%% file: main.tex
\documentclass[10pt, journal]{IEEEtran}
\usepackage{textcomp}
\usepackage[colorlinks,linkcolor=red,anchorcolor=blue,citecolor=blue]{hyperref}
\usepackage{tikz}
\usepackage{amsthm}
\usepackage{amsmath}
\usepackage[linesnumbered,ruled,vlined]{algorithm2e}
\usepackage{graphicx}
\usepackage{url}
\usetikzlibrary{arrows, positioning, calc}
\tikzstyle{vertex}=[draw,fill=black!15,circle,minimum size=18pt,inner sep=0pt]
\usepackage{amsmath,amssymb}
\usepackage{multicol}
\usepackage{multirow}
\usepackage{pgfplots}
\usepackage[utf8]{inputenc}
\usepackage{authblk}
\usepackage[FIGBOTCAP]{subfigure}
\usetikzlibrary{shapes.geometric}
\usepackage{adjustbox}
\usepackage{bm}
\usepackage[utf8]{inputenc}
\usepackage{amsthm}
\usetikzlibrary{decorations.pathreplacing,calligraphy}
\usepackage[english]{babel}
\usepackage{bm}
 \usetikzlibrary{patterns}
 \usetikzlibrary{spy}
\newtheorem{theorem}{Finding}
\usetikzlibrary{positioning}
\usetikzlibrary{shapes.arrows}
\usepackage{amssymb}
\usepackage{pgfplotstable}
\usepackage{float}
\usepackage{flushend}

\usepackage{comment}

\begin{document}

\bstctlcite{IEEEexample:BSTcontrol}

\title{Improving the Energy Efficiency of \\ High Throughput Computing: \\A Measurement-Based Case Study}
\author{Damu Ding\href{https://orcid.org/0000-0001-9692-7756}{\protect\includegraphics[scale=0.1]{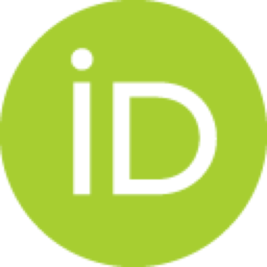}},  Xinpeng Hong\href{https://orcid.org/0000-0001-8525-6424}{\protect\includegraphics[scale=0.1]{orcid/icon.eps}}, Alastair Dewhurst\href{https://orcid.org/0000-0002-2062-8052}{\protect\includegraphics[scale=0.1]{orcid/icon.eps}},  James Walder, Daniel Schien\href{https://orcid.org/0000-0002-5290-7443}{\protect\includegraphics[scale=0.1]{orcid/icon.eps}}, David Greenwood\href{https://orcid.org/0000-0001-8632-6033}{\protect\includegraphics[scale=0.1]{orcid/icon.eps}}, Noa Zilberman\href{https://orcid.org/0000-0002-3655-2873}{\protect\includegraphics[scale=0.1]{orcid/icon.eps}}
    
\thanks{Affiliations as of the time of writing:
}
\thanks{Damu Ding, Xinpeng Hong, and Noa Zilberman were with University of Oxford, UK.
}
\thanks{Alastair Dewhurst and James Walder were with Science and Technology Facilities Council, Harwell Campus, UK.
}
\thanks{Daniel Schien was with University of Bristol, UK.
}
\thanks{David Greenwood was with University of Newcastle, UK.
}
}

\maketitle

\begin{abstract}
The significant energy consumed by data centers has become a concern both for costs and associated carbon emissions. 
In particular, the energy efficiency of servers is a key consideration for data center operators, and understanding servers' power consumption under different operating conditions is an important aspect of it. 
In this paper we present a measurement-based case study of high-throughput computing. We analyze power usage information of an operational data center, combined with focused measurements of power reduction techniques for a representative high-throughput workload.  
The study points out the obstacles encountered by data center operators in their efforts to minimize energy consumption and carbon emissions, and discusses the impact of server configuration adjustments on the energy consumption of processing jobs.  
We offer actionable recommendations for decreasing the energy usage of servers, while considering both performance and carbon emissions.
\end{abstract}

\section{Introduction}
The electricity consumption of data centers has been increasing over time with the growing demand for computing power, storage capacity, and data processing capabilities. According to the International Energy Agency (IEA), data centers’ global electricity consumption was 460TWh as of 2022, and this number is expected to continuously increase~\cite{iea2024}.
The increased energy usage leads to rising financial burdens and a negative impact on the environment.  
In recent years, numerous methods have been proposed to enhance data centers' energy efficiency, such as consolidating multiple servers into a single physical unit using virtualization\cite{varasteh2015server}, adjusting server power usage by workload \cite{lo2014towards}, improving cooling efficiency \cite{capozzoli2015cooling}, and utilizing energy-efficient hardware like Solid State Drives (SSD)\cite{shuja2014survey}.

Rather than upgrading data center infrastructure, which is associated with additional embodied carbon, a simpler approach to saving energy is improving computing-related energy efficiency. 
  In this work, we focus on \emph{High Throughput Computing} (HTC), which aims to maximize job throughput without prioritizing task-level latency~\cite{tsaregorodtsev2004dirac}. This is not like \emph{High Performance Computing} (HPC), which seeks to minimize time-to-solution (i.e., the average time to finish jobs) by distributing tasks across multiple workers at the cost of parallelization.
Both kinds of computing lead to high servers utilization under heavy workloads, but given the independent nature of jobs, HTC is operating under different constraints, such as the number of jobs to be finished within a given period of time. 

To balance the requirements of performance and energy efficiency, it is necessary to investigate the trade-off between power consumption and job execution performance, and to understand how they vary with different configurations. Furthermore, this needs to be assessed on accumulated energy-to-performance scale, rather than average power, to account for hidden overhead costs.  This measurement study is designed for this purpose, covering a crucial aspect of the broader effort to make data centers more sustainable and cost-efficient.

Although measurement-based analyses of data center server energy consumption have been conducted in the past, there has been a gap in recent studies, with the most recent analyses dating back to 2023~\cite{aroca2015measurement, arslan2023green, ciko2023going}. This means that there is a scarcity of empirical studies examining energy trends in current servers. As both workloads and servers have come a long way since, it is important to get a fresh look on the subject, and to provide the community with up to date data and new insights. Understanding the handling of configuration parameters and their impact on various variables may result in substantial cost and energy savings.

In this paper, we present a case study of improving the energy efficiency of high throughput computing, based on power measurements from a UK Science and Technology Facilities Council (STFC) data center, and focused power optimization measurements in a local cluster. 
As access to renewable energy varies between data center operators, our study provides an insight into some of the challenges faced by data center operators when trying to reduce the operational carbon emissions of their facilities. It further discusses how tuning server configurations affects the energy consumption of processing jobs, ranging from the number of active CPU cores and the number of jobs to 
performance boosting technologies. 
Finally, we provide practical recommendations for energy reduction of servers, with a combined view of both performance and energy efficiency.


In summary, this paper makes the following contributions:
\begin{itemize}
    \item We analyze a dataset of racks power consumption within a real data center and discuss power consumption trends. 
    \item We propose a simple but effective model to monitor the power and energy consumption of data center on service level, based on our power analysis.
    \item We measure the power, energy consumption, and job execution time in two kinds of data center servers and analyze their sensitivity to different configurations and operating scenarios.
    \item Based on our measurement results, we provide recommendations to balance the energy consumption and performance in data center servers for green computing.
\end{itemize}

The rest of the paper is organized as follows. Section \ref{sec:background} introduces background information and discusses related work. In Section \ref{sec:motivation}, we present a study of power consumption in a real data center.  Section \ref{sec:method} describes our energy consumption modelling. In Section \ref{sec:results}, we present the experimental setup of our testbed and report measurement results. Section \ref{sec:rec} provides recommendations for reducing data center energy consumption based on our measurement results.
 Finally, Section \ref{sec:conclusion} concludes the paper.

\section{Background and related work} \label{sec:background}
\subsection{Background} \label{sec:related}

This subsection provides the background of the concepts discussed in this paper.

\noindent\textbf{Definition of power and energy}
Power consumption $P$ is an instantaneous value that indicates the energy per unit time, and its unit is Watt.
Energy $E$ is the cumulative power consumption over a given period  of $T$ seconds. Thus, energy consumption can be formulated as:
\begin{equation}\label{eq:1}
    E = \bar{P} \cdot T
\end{equation}
where $\bar{P}$ is the average power consumption during period $T$. Note that the unit of power ($P$) is Watt (W), and 
the unit of energy ($E$) can be kWh (kilo Watt hour) or Joule (J). Since $1$ kWh is equal to $3.6 \times 10^6$ Joules, they are interchangeable. If not otherwise specified, kWh is the energy unit in the remainder of this paper.


\noindent\textbf{Energy consumption and carbon emissions}
Cloud data centers contribute significantly to environmental impact, particularly due to their substantial energy consumption~\cite{iea2024}.
Carbon intensity is a measure of the ratio between energy consumption and carbon emissions: when the carbon intensity is high, energy consumption results in higher $CO_2$ emissions. It is a property of the mix of energy sources that are being used. 
Data centers aim to reduce their carbon footprint by utilizing renewable energy sources (e.g., wind, hydro, or solar power), adopting low-carbon procurement standards, implementing measurements to prevent deforestation, and other sustainable practices. 
While some data centers (especially hyperscalers~\cite{iea}) can prioritize renewable energy procurement, others rely on the energy mix available from regional or national grids, depending on local regulations and infrastructure. Despite ongoing efforts, the average carbon emission of modern data center is still very high: according to the IEA~\cite{iea, malmodin2024ict}, data centers generated 330 Megatons of $CO_2$ equivalent in 2020, constituting 0.6\% of total global greenhouse gas (GHG) emissions and 0.9\% of energy-related GHG emissions. 

\noindent\textbf{Power Usage Effectiveness} Power Usage Effectiveness (\emph{PUE}) \cite{dayarathna2015data} is a metric to quantify the ratio between overall data center energy and IT equipment energy. Additional energy consumption beyond IT equipment energy is attributed to overheads such as cooling, and an energy-efficient data center should have a PUE close to 1. However, PUE does not consider the utilization of computational resources~\cite{brady2013case}.

\noindent\textbf{High Throughput Computing versus High Performance Computing}
High Performance Computing (HPC) and High Throughput Computing (HTC) are two computing paradigms that address different types of workloads with distinct characteristics. While both aim to achieve high performance, their metrics for performance are different. 
HPC aims to minimize time-to-solution for complex computational tasks (e.g., floating point operations (FLoPS)), while HTC focuses on maximizing the number of jobs completed within a given time frame~\cite{tsaregorodtsev2004dirac}. This dictates different types of optimizations and resource allocation schemes.
Large Hadron Collider beauty experiment (LHCb) is a typical scientific computing example of HTC~\cite{krawczyk2022ethernet}: it is a particle physics detector experiment that needs to maintain high real-time experimental data throughput in each server, and the server needs to send important filtered physical information to a collector within the servers' cluster for further processing.

\noindent\textbf{Cores, Threads and Jobs}
Many chip multiprocessors (CMPs) today have multiple processing cores, often referred to simply as \emph{cores}. Each core is an independent processing unit, with its own dedicated hardware resources, and running a program. All CPU cores are integrated into a single chip, typically residing on the same silicon die. Each core typically has one or two levels of dedicated cache, while higher-level caches and external memory are generally shared among multiple cores. Data center servers have tens to hundreds of cores, meaning that they can execute many programs in parallel. In addition, each core typically handles one hardware \emph{thread} at a time, although Intel and AMD CPUs allow two threads to be 'pinned' to a core. Moreover, a program may also support multiple threads through software level parallelism. In this paper, we refer only to the hardware-level threads. In addition, a thread is sometimes referred to as a \emph{logical core} or a virtual CPU (\emph{vCPU}). 

We consider each isolated running application as a \emph{job}. The application can be executed either directly on the physical server or in a virtualized environment (e.g., a virtual machine or a docker container). Using virtualization is the preferred option in most data centers, as it provides isolation between users, enables easy resource management and migration.

\noindent\textbf{DVFS and performance boosting}
Dynamic Voltage and Frequency Scaling (\emph{DVFS}) is a technique that can dynamically adjust the voltage alongside the frequency of a CPU. 
 In general, power grows with the square of voltage and linearly with frequency:
\begin{equation}
    P = C \cdot V^2 \cdot f
\end{equation}
where $C$ is the capacitance, $V$ is the supply voltage, and $f$ is the operating frequency.

DVFS monitors the activity of the CPU, and based on the measurements decreases voltage and frequency when the device is idle, and increases them when the utilization is high to maximize performance. This leads in turn to power savings in time of low activity. 

Performance-boosting technologies, such as Intel's Turbo Boost and AMD's Precision Boost, are forms of DVFS designed to temporarily increase CPU frequency to optimize job execution performance. These technologies boost the frequency for short bursts, typically during high-demand moments, but cannot sustain the increased frequency over long workloads. As such, they are focused on enhancing performance rather than saving energy.


\subsection{Related work} \label{sec:related}
This subsection presents recent related work on sustainable computing in data centers.

\noindent\textbf{Power and energy consumption models and measurements} Modelling the power consumption of data centers servers was the focus of many works~\cite{song2013unified, perumal2014power}. Most works modelled the power consumption as an additive model~\cite{jin2020review}, measuring the sum of the power consumption all of server components. 
However, these works rarely considered the combined impact of other parameters (e.g., the number of CPU cores, the number of jobs, DVFS) on power consumption. In this work, a non-linear model built on top of Baseline-active model~\cite{jin2020review} is used to investigate the effect of multiple parameters on the energy consumption of highly-utilized data center servers. 
Jordi et al.~\cite{aroca2015measurement} conducted measurement-based experiments on energy consumption of data center servers. 
They only considered the impact of frequency, number of cores and I/O. This work, beyond exploring energy consumption in newer generations of servers, explores more parameters and provides further sensitivity analysis.

\noindent\textbf{Server virtualization} 
Server virtualization (e.g., virtual machines and containers), which enables running multiple isolated applications on a single server, can help reduce energy usage if suitable virtualization is deployed. Jin et al. \cite{jin2012energy} empirically investigated the energy efficiency using server virtualization. It has been shown that an optimum number of virtual machines can maximize power savings. In \cite{zhang2020estimating}, Zhang et al. proposed a method to estimate the power consumption of containers and virtual machines for better resource scheduling and green computing. This work builds upon previous research to explore the benefits of server virtualization and containerization, aiming to identify the optimal strategy for minimizing the energy consumption of containers running on a single server. By doing so, it seeks to reduce the overall energy consumption of the data center.


\noindent\textbf{Power consumption adjustment based on workload} 
Power consumption can be reduced by adjusting the voltage and clock frequency through DVFS or by reducing the number of active cores in a server~\cite{lo2014towards}. Usually, DVFS is managed by keeping track of the CPU utilization. The frequency is increased when the utilization is high and becomes low when idle.
However, Google~\cite{kaffes2020leveraging} indicated that when the machines are heavily utilized, DVFS has limited effectiveness in terms of energy savings. 
Other similar solutions, like reducing the number of active cores, rely on slowing down the workload to reduce real-time power consumption, but this may significantly affect the user experience and result in even longer execution time. As we show in this paper, reducing the number of active cores has negative impacts on energy saving in data center servers.

\noindent\textbf{Additional approaches} 
There are many additional approaches to increase energy efficiency, such as improving cooling efficiency~\cite{capozzoli2015cooling} and using energy-efficient hardware~\cite{shuja2014survey}. 
For example, it has been shown that replacing air-cooled system with liquid-cooled system can help reduce the PUE from 1.48 to 1.14 \cite{capozzoli2015cooling}. 
It is also possible to save energy by using more energy-efficient components (e.g., Solid State Drive (SSD)) in data center servers. Hard Disk Drives (HDD) consume 85\% of overall energy when the server is idle \cite{narayanan2009migrating, sprecher2014recycling}, whereas in SSD the energy consumption is proportional to I/O operations per second. This means that upgrading storage devices to SSD can reduce energy usage~\cite{shuja2014survey}.
Unlike aforementioned approaches, the goal of this work is not to minimize the PUE or upgrade IT infrastructure, but to decrease the energy consumption of existing IT equipment, especially when highly utilized.

\section{A Study of Power Consumption at STFC RAL} \label{sec:motivation}

\begin{figure}[t]
    \centering
    \includegraphics[width=\linewidth]{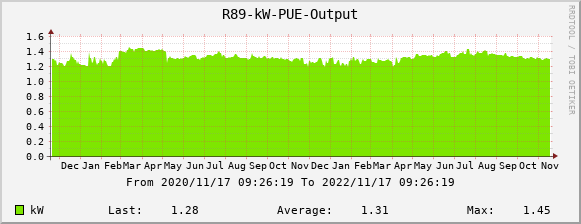}
    \vspace{-2em}
    \caption{PUE in STFC over the last 2 years}
    \label{fig:pue}
\end{figure}

\begin{figure}[t]
    \centering
    \includegraphics[width=\linewidth]{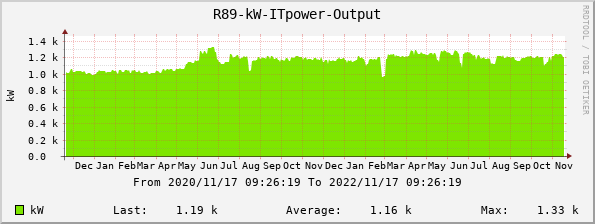}
        \vspace{-2em}
    \caption{IT power consumption in STFC over the last 2 years}
        \vspace{-1em}
    \label{fig:stfc}
\end{figure}

\begin{table*}[t]
\caption{Monitored Racks's Specifications and Median and $75^{th}$ Percentile Power Measurements. Server and core power are normalized.}
\vspace{-1em}
\label{table:rack}
\resizebox{\linewidth}{!}{
\begin{tabular}{ |c |c|c |c |c|c|c|c|c|c|c|c|c|}
\hline
\multirow{2}{*}{Rack ID} &  \multirow{2}{*}{Type}&  \multirow{2}{*}{\#Servers} &  \multirow{2}{*}{CPU} &  \multirow{2}{*}{\begin{tabular}{c}Logical cores \\per server\end{tabular}} & \multirow{2}{*}{\begin{tabular}{c}Memory \\ (GB)\end{tabular} } &\multirow{2}{*}{\begin{tabular}{c}Disk \\ (GB) \end{tabular}} & \multicolumn{2}{c|}{Power per rack (W)}&\multicolumn{2}{c|}{Power per server (W)}&\multicolumn{2}{c|}{Power per core (W)}\\
 \cline{8-13}
&&&&&&& Median& 75\% & Median & 75\%& Median & 75\%\\
 \hline
  113 & Dell16 & 48 & \begin{tabular}{c} Intel Xeon E5-2630 V4 2.30GHz\end{tabular} &40 & 160 &2400 & 3673 & 9747 & 76.5 & 203.1 & 1.91 & 5.08\\ 
 \hline
   124 & Dell17 & 48 & \begin{tabular}{c}Intel Xeon Gold 6130 2.10GHz \end{tabular}&64 & 192&1863& 18049 & 18369 & 376 & 382.7 & 5.88 & 5.98\\ 
 \hline
    \multirow{2}{*}{125} & Dell17  & 24 & \begin{tabular}{c} Intel Xeon Gold 6130 2.10GHz  \end{tabular}&64 & 192&1863& \multirow{2}{*}{17092} & \multirow{2}{*}{17980} & \multirow{2}{*}{356.1} & \multirow{2}{*}{374.59} & \multirow{2}{*}{5.93}& \multirow{2}{*}{6.24} \\ 
\cline{2-7}
   ~ &  XMA17 & 24 & \begin{tabular}{c} Intel Xeon Gold 5120 2.20GHz \end{tabular}&56 & 192&1863& &  &  & &  & \\ 
 \hline
169 & XMA20& 20 & \begin{tabular}{c}AMD EPYC 7452 2.35GHz \end{tabular}&128 & 512& 3576& 10578 &10713 & 528.9 & 535.7 & 4.13 & 4.19\\ 
 \hline
170 & XMA20 & 24 & \begin{tabular}{c}AMD EPYC 7452 2.35GHz \end{tabular}&128 & 512& 3576& 12578& 12712 & 524.1 & 529.7 & 4.09 & 4.14\\ 
 \hline
 177 &XMA21& 12 & \begin{tabular}{c}AMD EPYC 7763 2.45Ghz \end{tabular}&256& 1024&7680& 12394 & 12541 & 1032.9 & 1045.1 & 4.03 & 4.08\\ 
 \hline
 179 & XMA20 & 20 & \begin{tabular}{c}AMD EPYC 7452 2.35GHz\end{tabular}&128 & 512& 3576& 10669& 10753 & 533.5 & 537.6 & 4.17 & 4.2\\ 
 \hline
 180 &XMA20& 24 & \begin{tabular}{c}AMD EPYC 7452 2.35GHz\end{tabular}&128 & 512& 3576& 12804& 12902 & 533.5 & 537.6 & 4.17 & 4.2\\ 
 \hline
 224 &XMA21& 12 & \begin{tabular}{c}AMD EPYC 7763 2.45Ghz\end{tabular}&256& 1024&7680& 11390 & 11700 & 949.2 & 975 & 3.71 & 3.81\\ 
 \hline
  406 &Dell19& 48 & \begin{tabular}{c}AMD EPYC 7452 2.35GHz\end{tabular}&128 & 512&3576& 30295 & 30606 & 631.16 & 637.65 & 4.93 & 4.98\\ 
 \hline
  407 &Dell19& 48 & \begin{tabular}{c} AMD EPYC 7452 2.35GHz\end{tabular}&128 & 512&3576& 25037 & 25444 & 521.6 & 530.1 & 4.08 & 4.14\\ 
 \hline
\end{tabular}
}
\end{table*}

We consider R89, a data center operated by STFC~\cite{STFC} at the Rutherford Appleton Labs (RALs) as a real world use case for analysis purposes. The goal is to minimize the energy consumption of the data center. 
As a reference, from 2019 to 2020, their utilities generated 33,885 tonnes of $CO_2$~\cite{stfcreport}. 
Hence, diminishing the electricity consumption of servers in data centers is still valid, and can facilitate a reduction in  carbon emissions. 

The CERN particle collider is the largest high energy physics experimental site in the world, and STFC RAL provides Tier-1 computing infrastructure for CERN experiments, being responsible for processing around 10\% of CERN data. The majority of compute is required for Monte Carlo analysis, reconstruction and filtering. The type of compute is High Throughput Computing (HTC), meaning that many jobs by different uses are executed simultaneously within the data centers, with multiple users sharing the same server. This indicates that the data center servers have high utilization but low dependence between servers, even within the same rack. 

It is critical to study the energy usage of HTC because most scientific computing, like particle collider for nuclear research, is running HTC intensive workloads.
Almost 100\% of STFC's computing workloads are HTC, because Monte Carlo simulations require very large computational resources over a long period of time. Minimizing STFC's energy consumption can effectively reduce the carbon emissions.

STFC is continuously monitoring its PUE and power usage over time.  
As shown in Figure~\ref{fig:pue}, from 2020 to 2022, the PUE of STFC ranges from 1.28 to 1.45, and the small variations are caused by the seasons.
Figure~\ref{fig:stfc} shows the IT power usage of the STFC data center (without cooling and other overheads) over the last two years. It shows that the data center consumes about 1.2MW, and this power consumption does not vary significantly over time, except when new equipment is introduced. This power consumption is also coupled with high utilization of the computing resources, of over 92\%. 
Changes in the data center's overall energy consumption were mostly due to PUE changes, i.e., changes in cooling efficiency over the seasons, rather than changes in IT power consumption. 

The insight from these results is that for this data center there is little opportunity to reduce non-renewable energy carbon emissions or energy costs by shifting workloads over time, as the data center is continuously highly utilized, throughout the year and during all hours of the day.

\begin{figure*}[t]
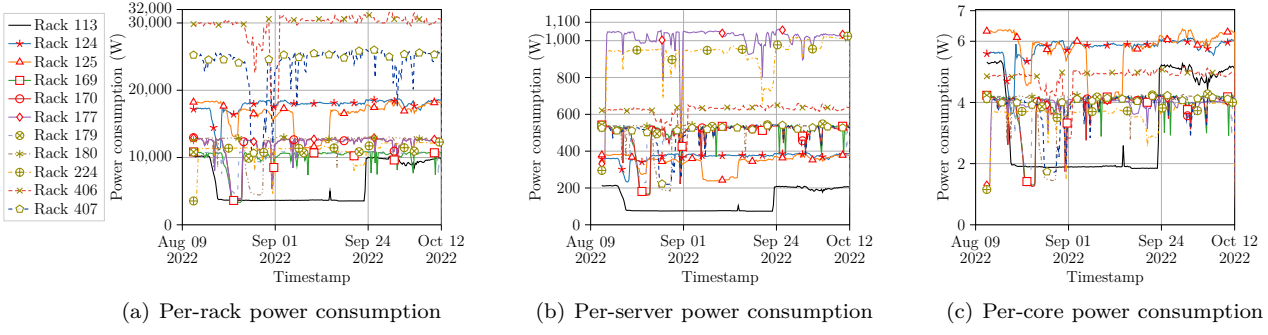

\centering
\subfigure[Per-rack power consumption ]{
\hspace{-3.5em}
{\scalebox{0.5}{
 \input{STFC/power-per-rack}
    \label{fig:stfc:rack}
    }
    }
    }
\subfigure[Per-server power consumption ]{
{\scalebox{0.5}{
 \input{STFC/power-per-server}
    \label{fig:stfc:server}
    }
    }
    }
\subfigure[Per-core power consumption]{
{\scalebox{0.5}{
 \input{STFC/power-per-core}
    \label{fig:stfc:core}
    }
    }
    }
\caption{Rack-level, server-level (normalized), and core-level (normalized) power consumption over time based on PDU measurements}
    \vspace{-1em}
\label{fig:stfc:power}
\end{figure*}

Servers in the data center are organized within racks, where each server is connected to two Power Distribution Units (\emph{PDU}), for redundancy purposes. Each rack typically hosts three PDUs. We collect a dataset of power measurement reports by the PDUs over two months. The measurement resolution is 6 hours, and the start and stop time of all PDUs are identical. Data is collected from 335 PDUs, but only 315 contain non-NULL values and are used for the analysis. The power consumed by PDUs varies significantly, from 2.2W to 6634W, but per-PDU the variance over time is small, with the average standard deviation being 9.1\%.

While the initial goal of our analysis was to observe variability of power consumption at server level and identify trends in power consumption, this could not be achieved as available measurements include only the total power consumption per PDU, rather than per socket, and as servers within a rack have their pair of power connections spread between three or four PDUs in a mixed manner. 
For this reason, we consider rack-level power consumption, as an aggregation of the power consumed by all PDUs within a given rack, and therefore all the computing equipment within that rack. 
 Once racks with partial PDU information (e.g., measurements for one PDU are missing) are omitted, we are left with 11 racks. The specifications of the racks are provided in Table~\ref{table:rack}. As can be seen, the dataset contains a mix of server specifications from different vendors, CPU generation, number of cores, and with a different number of servers per rack.


Figure.\ref{fig:stfc:power} shows the power consumption on rack level, for each of the 11 racks, based on the PDU measurements. We also capture the median and $75^{th}$ percentile in Table~\ref{table:rack}, with more information available in the accompanying repository \cite{cquandri}. 
As the figure shows, the power consumption of racks varies significantly between configurations: from 10kW to 30kW. Some equipment within the racks experiences failures from time to time, as demonstrated by significant power drops of certain racks. In particular, rack 113 had servers taken out of production between the middle of August 2022 and the end of September 2022.  


The power consumption of a rack is not representative of its computing power, as the number of servers and cores varies between racks. We define $server\ power=\frac{rack\ power}{\# servers}$ and $core\ power=\frac{rack\ power}{\# servers\times\#cores\ per\ server}$. Note that we refer to the number of \emph{logical} cores per server. 
The results are shown in Figure~\ref{fig:stfc:server}, Figure~\ref{fig:stfc:core} and Table~\ref{table:rack}.
As can be expected, servers with more cores consume more power. 


A surprising result is that the difference in power per logical core is small when considering different generations of AMD processors and different number of cores. For example, based on their thermal design power (TDP), AMD EPYC 7763 is supposed to be 10\% more power efficient per core than AMD EPYC 7452, while running at 4.2\% higher frequency.
However, our results show that racks with these two generations of servers can have differences in median and $75^{th}$ percentile as small as $1.5\%$, e.g. racks 170 and 177. This is as the results take into account the entire power consumed by the rack, including a server's power supply, memories, local storage and other overheads.

While it is commonly assumed that moving to newer, denser generation of CPUs improves the power efficiency and should therefore be preferred, our insight is that it may be better to delay the decommission of systems, as the overall power saving per core will be small. This is reinforced when considering embodied carbon emissions of servers.

\section{Modeling Energy Consumption} \label{sec:method}

Based on the previously discussed results, reducing the carbon emissions of the studied data center is not possible by a temporal workload shift. This leads us to explore carbon reduction through the reduction of energy consumption of workloads. 
Unlike a goal of purely reducing the power consumption of a server, our goal is to minimize the overall energy consumption of the data center over time, while optimizing for performance. As discussed next, these two goals are not the same. 
 To this end, this section outlines the modeling of energy usage by data center servers.

\subsection{Single server energy consumption model}
Baseline-Active (BA) model~\cite{jin2020review} is the most popular model of servers' power consumption. The baseline power $P_{baseline}$ is the power consumption of an idle server, which includes the power usages of CPU, memory, disk, and any other device, like a Network Interface Card (NIC). The baseline power usually does not vary significantly over time. The active power $P_{active}$ indicates the additional power consumption when running one or several jobs on the server. $P_{active}$ contains elements such as the power consumed by  memory and disk usage, network I/O as well as CPU utilization.  Therefore, the overall power consumption $P$ can be expressed as: 
\begin{equation}
    P = P_{baseline} + P_{active}
\end{equation}
In this paper, $P_{active}$ is extended to a non-linear model that considers two more parameters,  the number of utilized logical cores $N_{cores}$ (i.e., the concurrency level of the workload as described in \cite{shi2013cpt}) and the number of running jobs $N_{jobs}$ on the server, the power consumption can be denoted by:
\begin{equation}
    P = P_{baseline} + P_{unit}\cdot N_{cores} \cdot N_{jobs}
\end{equation}
where $P_{unit}$ is the average power consumption (over the sampling period) of a single logical core executing one thread of a job, where the complete job is running over $N_{cores}$.
Simply minimizing $P_{unit}$ (e.g., reducing CPU frequency) will reduce $P$, but may also increase execution time, leading to higher energy consumption. Therefore, a different method should be used to minimize energy consumption.

Being $\Bar{P}_{baseline}$ the average idle power consumption during period $T$, by Equation \ref{eq:1}, the total energy consumption is:
\begin{align}
    E &= \Bar{P}_{baseline}\cdot T  + \Bar{P}_{unit}\cdot N_{cores} \cdot N_{jobs} \cdot T \\
    & = E_{baseline}  + E_{unit}\cdot N_{cores} \cdot N_{jobs}
\end{align}
Where $E_{baseline}=\Bar{P}_{baseline}\cdot T$ and  $E_{unit}=\Bar{P}_{unit}\cdot T $. To minimize energy consumption both $E_{baseline}$ and $E_{unit}$ should be minimized. However, this is not sufficient, as $T$ depends on the number of cores used ($N_{cores}$) and potentially also the number of jobs ($N_{jobs}$).

The above equation can be expressed as:
\begin{equation}
    E = E_{unit}^{*}\cdot N_{cores} \cdot N_{jobs}
\end{equation}
where $E_{unit}^{*}$ is the energy consumption per core per job including the baseline. This model is limited, as it assumes that just one type of workload is running. If different types of workloads (denoted by $K_{workloads}$) are running, the model would be:

\begin{equation}
    E^{\dagger} = \sum_{i=1}^{K_{workloads}}{E_{unit_i}^{*}\cdot N_{cores_i} \cdot N_{jobs_i}}
\end{equation}


The model above provides a level of abstraction above CPU power states, though their effect is captured in Section~\ref{sec:exp}.

\subsection{Data center energy consumption modeling}
In STFC data center, depending on rack and server dimensions, each rack contains 12-48 servers. A top-of-rack switch is deployed on the rack, so that the servers can connect to the data center network. The PDUs are also connected, allowing operators to remote monitor and control the power and energy consumption of servers. 

If only a single type of a job runs on all servers in a data center, and all servers are the same, the total IT energy consumption of data centers can be denoted by:
\begin{equation}
    E_{tot} = E \cdot \sum_{i=1}^{N_{rack}}{N_{server}^{i}}
\end{equation}
where $E$ represents the total energy consumption of a single server, $N_{server}^{i}$ is the number of servers in rack $i$, and $N_{rack}$ is the number of racks in data center.

However, the reality is that the data center runs multiple types of jobs in parallel, and different types of servers have different energy consumption. Therefore, a more realistic model would be:
\begin{equation}
    E_{tot} =  \sum_{i=1}^{N_{rack}}{\sum_{j=1}^{N_{server}}{E^{\dagger}_{ij}}}
\end{equation}
Considering that the average PUE of STFC RAL from 2020 to 2022 is 1.31 (see Figure~\ref{fig:pue}), the overall energy consumption can be estimated as:
\begin{equation}
    E_{overall} =  1.31E_{tot}
\end{equation}
that is, the energy consumption of non-IT equipment (e.g., cooling) is around 31\% of that of IT equipment.

It is crucial to understand that the goal of this paper is to reduce energy consumption of IT equipment (i.e., $E_{tot}$) with an optimized scheduled job execution strategy, meaning minimizing $E^{\dagger}$. Therefore, reducing the energy consumption of non-IT equipment is beyond the scope of our work. In the following sections, we limit the discussion to a single workload, with the objective function of minimizing $E$, and the discussion is extended to the general case in Section~\ref{sec:rec}. 

\section{Empirical results } \label{sec:results}
In this section, we will present our experimental results in terms of power consumption, total energy consumption, and job execution time in two different data center servers. In the following experiments, the comparison between the Intel and AMD servers should not be interpreted as a direct like-for-like comparison, since the setup of each server is quite different and their workload for the same job is not equal. Instead, it is useful to consider trends and the power/energy consumption per job when excluding the baseline. 

\subsection{Experimental setup} \label{sec:exp}
\input{figure/testbed.tex}

To better understand energy-to-performance trade-offs, a focused set of measurements is run on a small research and development cluster. This subsection describes the setup of our measurements. Figure \ref{fig:arch} provides an overview of our physical testbed, focusing on the evaluated servers.

\subsubsection{Physical setup}
Two servers are used for the evaluation, one (referred to as \emph{Intel server} for the sake of conciseness) is DELL PowerEdge R740 using Intel Xeon Gold 6234 CPU, running at 3.30GHz, and the other one (called \emph{AMD server}) is ASUS  ESC4000A-E10 server equipped with a AMD EPYC 7302P 16-Core Processor. As shown in Table \ref{table:servers}, both servers have a total of 16 cores and 32 threads (logical cores); the Intel server is dual sockets, while the AMD server has one CPU socket. 
Both servers are quipped with DDR4 memory, and use SSD NVMe disk during the experiments. 
Both servers are equipped with a dual-port NVIDIA Mellanox ConnectX 5~\emph{NIC} (Network Interface Card). 
In addition, each server is equipped with HDD (idle), and one accelerator card: Alveo U280 in the Intel Server, and NVIDIA BlueField-2 in the AMD servers. Both accelerator cards are idle, and represent the increasingly common case of acceleration presence. They are further discussed in Section~\ref{sec:accelerators}.
 Each server is connected to two smart PDUs (Eaton ePDU EMAH06), that collect real-time power and energy consumption. 
By default, DVFS and performance boosting technologies are enabled. 
Note that the PDUs have IEC Class 1 (±1\%) billing grade accuracy. During our experiments, no other jobs are running in servers, and so no external interference. The read power is always stable.

\begin{table}
\caption{Servers Information}
\vspace{-1em}
\label{table:servers}
\resizebox{\linewidth}{!}{
\begin{tabular}{ |l |c |c |}
\hline
 Name & Intel & AMD \\ 
 \hline
 Type & DELL PowerEdge R470 &  ASUS  ESC4000A-E10\\
 \hline
  Motherboard & 0WXD1Y A01 &  KRPG-U8 60SB08F0-SB0A04 \\
 \hline
 CPU & Intel Xeon Gold 6234  &  AMD EPYC 7302P\\ 
 \hline
 CPU Frequency & Base 3.3GHz, Max 4.0GHz & Base 3.0GHz, Max 3.3GHz \\
 \hline
CPU Frequency Range & 1.2GHz-4GHz & 1.5GHz-3.3GHz  \\
 \hline
  Num. of Sockets & 2 & 1  \\
 \hline
  Cores per socket & 8 & 16  \\
 \hline
   Threads per core & 2 & 2  \\
 \hline
  Threads per server & 32 & 32  \\
 \hline
   Memory & 384GB DDR4-2933MT/s & 256GB DDR4-3200MT/s   \\
 \hline
  SSD capacity & 1.6TB  & 2TB  \\
 \hline
 NIC & \begin{tabular}{c} NVIDIA ConnectX-5\\ MCX516A-CCAT\end{tabular}  &  \begin{tabular}{c} NVIDIA ConnectX-5\\ MCX516A-CDAT\end{tabular} \\
 \hline
 Accelerator & AMD Alveo U280 FPGA & NVIDIA BlueField-2 DPU\\
 \hline
\end{tabular}
}
\end{table}

\subsubsection{Workload}
ATLAS~\cite{rimoldi2004simulation} is an experiment of CERN particle collider. 
 Workloads related to ATLAS consume 37.94\% of the Tier-1 batch farm, with simulations being about half of that. 
ATLAS simulations generate intensive computing loads with low concurrency, making it is a good  use case for analyzing the energy efficiency of servers running HTC. More details are in~\cite{cquandri}.

ATLAS simulations are running on servers within Docker~\cite{docker} containers. Each Docker container running anATLAS simulation is considered a \emph{job}. 
CVMFS~\cite{aguado2008cvmfs}, the CERN VM File System is running on the server too. 
The number of logical cores used by a workload can be configured. In our experiments, each physical core has two threads (logical cores) running, i.e., if the number of logical cores is set to 32, it means that there are 16 cores running 32 threads. CPU utilization refers to per-core utilization. For example, when 8 cores are configured, the CPU utilization during an ATLAS simulation can reach 800\%.

\subsubsection{Power and energy measurements}

The experimental setup uses \emph{SNMP} (Simple Network Management Protocol) to read the power consumption from the two PDUs connected to the server. Given power read $P_1$ from PDU 1  $P_2$ from PDU 2, the sum of two read values, $P_1 + P_2$, is the overall power consumption $P$ of the server. Similarly, we read the cumulative energy consumption $E$ from the start of a job. $E_{1}$ and $E_{2}$ are used to denote the energy consumption of PDU 1 and PDU 2, correspondingly. Therefore, the overall energy consumption of a job $E_{tot}$ is computed as the total energy consumption at the end $ E_{1}^{end} + E_{2}^{end} $  minus that at the beginning $ E_{1}^{start} + E_{2}^{start} $. The PDUs sample at 50Hz frequency, and we collect all values every 5 seconds. The monitoring and data collection is done from a different server, so there is no additional overhead on the monitored servers. Throughout this paper, the power and energy consumption of an idle server is denoted as \emph{baseline}.

\subsection{Baseline Power}\label{sec:accelerators}
We begin by measuring the power and energy consumption of an idle server. In addition, we measure the power consumption contribution of each accelerator card and NICs. While we measure the idle power of a NIC, the power consumed by a NIC changes when a job is running, as data is fetched and sent. All cards used have a connected 100G Direct Attach Cable (DAC). 
Figure~\ref{fig:intel:components} shows the idle power and cumulative energy of the Intel server over an hour, and the effect of the cards. The baseline power of the server is 251W with both cards, and 235W without the FPGA card. The NIC consumes 15W, and is a mandatory component. 
As shown in Figure~\ref{fig:amd:components}, the baseline power of the AMD server is 204W, including 58W contributed by the DPU and 38W by the NIC. 
 The higher NIC contribution was checked multiple times, and is in part due to the server consuming more power when the NIC is present and in part due to higher transceivers draw. Overall, it is observed that the baseline energy consumption of the Intel server is 0.25kWh, and the baseline of the AMD server is 0.2kWh. 

 To evaluate the maximum power consumption of each server, the Linux \emph{stress} tool is used. The Intel server consumes up to 460W, while the maximum measured power consumption of the AMD server is 310W. This means that an idle Intel server consumes 54\% of the maximum server power, and the idle AMD server consumes 66\% of its maximum power. These idle-to-maximum power ratios match similar power measurements in STFC RAL for production servers of the same generation. 
 
Note that GPUs were not part of STFC data center servers. For  information on energy consumption of GPUs, the readers can refer to the report of our UKRI Net-Zero partners~\cite{gpuconsumption}. 

\begin{theorem}
The energy consumption of idle servers is considerable, and can be more than a half of a fully utilized server. 
\end{theorem}
\begin{figure}[t]
\centering
\subfigure[Power consumption]{
{\scalebox{0.43}{
 \input{results/power_intel_components.tex}
    \label{fig:power:intel:components}
    }
    }
    }
\hspace{-1.5em}
\subfigure[Energy consumption]{
{\scalebox{0.43}{
 \input{results/energy_intel_components.tex}
    \label{fig:energy:intel:components}
    }
    }
    }
\caption{Intel Server Baseline Measurements}
    \vspace{-1em}
\label{fig:intel:components}
\end{figure}
\begin{figure}[t]
\centering
\subfigure[Power consumption]{
{\scalebox{0.43}{
 \input{results/power_amd_components.tex}
    \label{fig:power:amd:components}
    }
    }
    }
\hspace{-1.5em}
\subfigure[Energy consumption]{
{\scalebox{0.43}{
 \input{results/energy_amd_components.tex}
    \label{fig:energy:amd:components}
    }
    }
    }
\caption{AMD Server Baseline Measurements}
\vspace{-1em}
\label{fig:amd:components}
\end{figure}

\subsection{Impact of DVFS}\label{sec:dvfs}

This section explores the effect of DVFS on energy consumption. In all the experiments, we run a single job using its default configuration of 8 logical cores. 

We used \emph{cpupower}~\cite{cpupower} to force the CPU frequency of the Intel server to two different values, the highest 4GHz and the lowest 1.2GHz, and compared it to running with DVFS enabled. The results, shown in Figure~\ref{fig:power:intel:dvfs} demonstrate that using DVFS and setting to peak frequency have almost the same power profile. This is because the CPU utilization in our experiments is always high and the CPU frequency is always at peak when running jobs. When the frequency is low (i.e., 1.2GHz), even though the server runs at full utilization, the power consumption decreases to 100W, and the execution time triples. The energy consumption shown in Figure~\ref{fig:energy:intel:dvfs} tells us that the lowest CPU frequency is much less energy efficient than DVFS: DVFS only needs 0.25kWh to finish the job, while 1.2GHz requires more than 0.5kWh to execute the same job. Note that in this and following energy consumption figures we use filled markers (e.g., {\color{red}{$\blacksquare$}}) to indicate job completion time.

A similar experiment is executed on the AMD server, adjusting the CPU frequency to the minimum 1.5 GHz and the peak 3GHz. Figure~\ref{fig:amd:dvfs} shows a comparison between DVFS and fixed CPU frequencies. The power consumption in Figure~\ref{fig:power:amd:dvfs} shows similar trends to the Intel server, that is, DVFS and CPU frequency of 3GHz have a similar performance, and 1.5GHz only consumes 10W more than the baseline.  Figure~\ref{fig:energy:amd:dvfs} reveals running at a high frequency consumes about half the energy of running the same job at lower frequency. 

\begin{theorem}
    Running at low frequency reduces the momentary power consumption, but running at high frequency reduces the overall amounts of energy required to complete a job.
\end{theorem}
\begin{figure}[t]
\centering
\subfigure[Power consumption]{
{\scalebox{0.43}{
 \input{results/power_intel_dvfs.tex}
    \label{fig:power:intel:dvfs}
    }
    }
    }
\hspace{-1.5em}
\subfigure[Energy consumption]{
{\scalebox{0.43}{
 \input{results/energy_intel_dvfs.tex}
    \label{fig:energy:intel:dvfs}
    }
    }
    }
\caption{Intel server: The effect of CPU Frequency (1 job with 8 cores)}
\vspace{-1em}
\label{fig:intel:dvfs}
\end{figure}

\begin{figure}[t]
\centering
\subfigure[Power consumption]{
{\scalebox{0.43}{
 \input{results/power_amd_dvfs.tex}
    \label{fig:power:amd:dvfs}
    }
    }
    }
\hspace{-1.5em}
\subfigure[Energy consumption]{
{\scalebox{0.43}{
 \input{results/energy_amd_dvfs.tex}
    \label{fig:energy:amd:dvfs}
    }
    }
    }
\caption{AMD server: The effect of CPU Frequency (1 job with 8 cores)}
\vspace{-1em}
\label{fig:amd:dvfs}
\end{figure}



Figure~\ref{fig:intel:boost} shows a comparison of power and energy consumption when Turbo Boost is enabled or disabled in the Intel server.
In Figure~\ref{fig:power:intel:boost}, shows that disabling Turbo Boost saves around 40W of power consumption, but it takes around 500 seconds longer to execute the job. When observing the energy consumption in Figure~\ref{fig:energy:intel:boost}, it is evident that enabling boosting is more energy efficient, as for the same job execution, the total energy consumption with boosting enabled ({\color{red}{$\blacksquare$}}) is lower than that without boosting ({\color{red}{$\blacktriangle$}}).

Figure~\ref{fig:amd:boost} shows the equivalent experiment using the AMD server, when enabling or disabling Precision Boost.
Figure~\ref{fig:power:amd:boost} indicates that boosting has the same impact on AMD server as on Intel's, that is, the power consumption with boosting disabled is lower than that with boosting enabled, but the execution time is longer. However, in AMD server, less than 15W are saved when boosting is disabled.
This makes the differences in total energy consumption small: as shown in Figure \ref{fig:energy:amd:boost}, the curves of boosting enabled and boosting disabled almost overlap, but still with boosting enabled the server finishes the job earlier and saves energy.

\begin{theorem}
Enabling boosting can reduce job execution time with some additional power overhead, leading to lower total energy consumption.
\end{theorem}

Note that the above observation holds mainly due to the significant contribution of baseline power.

\begin{figure}[t]
\centering
\subfigure[Power consumption]{
{\scalebox{0.43}{
 \input{results/power_intel_boost.tex}
    \label{fig:power:intel:boost}
    }
    }
    }
\hspace{-2em}
\subfigure[Energy consumption]{
{\scalebox{0.43}{
 \input{results/energy_intel_boost.tex}
    \label{fig:energy:intel:boost}
    }
    }
    }
\caption{Intel server: Impact of boosting (1 job with 8 cores)}
\vspace{-1em}
\label{fig:intel:boost}
\end{figure}

\begin{figure}[t]
\centering
\subfigure[Power consumption]{
{\scalebox{0.43}{
 \input{results/power_amd_boost.tex}
    \label{fig:power:amd:boost}
    }
    }
    }
    \hspace{-2em}
\subfigure[Energy consumption]{
{\scalebox{0.43}{
 \input{results/energy_amd_boost.tex}
    \label{fig:energy:amd:boost}
    }
    }
    }
\caption{AMD server: Impact of boosting (1 job with 8 cores)}
\vspace{-1em}
\label{fig:amd:boost}
\end{figure}

\subsection{Impact of cores number and jobs number - Power}
\begin{figure}[t]
\centering
\subfigure[Intel]{
{\scalebox{0.43}{
 \input{results/power_intel.tex}
    \label{fig:power:intel}
    }
    }
    }
    \hspace{-2em}
\subfigure[AMD]{
{\scalebox{0.43}{
 \input{results/power_amd.tex}
    \label{fig:power:amd}
    }
    }
    }
\caption{Sensitivity analysis: Num. of cores (1 job) to power consumption }
\vspace{-1em}
\label{fig:power}
\end{figure}

\begin{figure}[t]
\centering
\subfigure[Intel]{
{\scalebox{0.43}{
 \input{results/power_intel_jobs.tex}
    \label{fig:power:intel:job}
    }
    }
    }
    \hspace{-2em}
\subfigure[AMD]{
{\scalebox{0.43}{
 \input{results/power_amd_jobs.tex}
    \label{fig:power:amd:job}
    }
    }
    }
\caption{Sensitivity analysis: Num. of jobs (8 cores) to power consumption }
\vspace{-1em}
\label{fig:power:job}
\end{figure}

In the following experiments, we explore how the number of jobs and the number of logical cores affect energy efficiency. 
Figure \ref{fig:power} reports the sensitivity of power consumption to the number of cores. When there are more cores assigned to a job, the execution time becomes lower, but the power consumption during execution is higher. As shown in Figure~\ref{fig:power:intel}, the baseline of Intel server is around 250W. 4 cores execution requires additional 40W, while 8 cores execution needs 90W more than baseline. However, when the number of cores is 16, the power consumption reaches 460W, the maximum. Increasing the number of cores does not significantly increase the power consumption during execution, but the execution time is lower. Figure~\ref{fig:power:amd} shows the power consumption of AMD server: similar to Intel server, the power consumption of AMD server increases as the number of cores increases but requires shorter execution time. 
When we increase the number of threads from 16 to 32, unlike Intel server, the power consumption continues increasing.

Figure \ref{fig:power:job} shows a comparison of power consumption when changing the number of jobs, while keeping the number of cores constant (8). From Figure \ref{fig:power:intel:job} it can be seen that in Intel server, the power consumption of 2 jobs ($\sim$200W more than baseline) is double of that of 1 job ($\sim$100W), for almost same execution time. On the other hand, when running 4 jobs, the maximum power consumption does not increase relative to 2 jobs, but job execution is longer. This is due to resources bottleneck, as they are fully utilized. Figure \ref{fig:power:amd:job} confirms that an increased number of jobs has the same effect in AMD server:  the 1 job simulation requires additional 50W over the baseline, and 2 job simulations require twice of that. It is worth noticing that each job on the AMD server consumes only half of the energy consumption on Intel server. Once the power consumption of 4 jobs execution reaches around 310W, the AMD server is at peak utilization and requires more time for completing all 4 jobs.


\subsection{Impact of cores number and jobs number - Total energy}
\begin{figure}[t]
\centering
\subfigure[Intel]{
{\scalebox{0.43}{
 \input{results/energy_intel.tex}
    \label{fig:energy:intel}
    }
    }
    }
    \hspace{-2em}
\subfigure[AMD]{
{\scalebox{0.43}{
 \input{results/energy_amd.tex}
    \label{fig:energy:amd}
    }
    }
    }
\caption{Sensitivity analysis: Num. of cores (1 job) to energy consumption }
\vspace{-1em}
\label{fig:energy}
\end{figure}

\begin{figure}[t]
\centering
\subfigure[Intel]{
{\scalebox{0.43}{
 \input{results/energy_intel_jobs.tex}
    \label{fig:energy:intel:job}
    }
    }
    }
\hspace{-2em}
\subfigure[AMD]{
{\scalebox{0.43}{
 \input{results/energy_amd_jobs.tex}
    \label{fig:energy:amd:job}
    }
    }
    }
\caption{Sensitivity analysis: Num. of jobs (8 cores) to energy consumption }
\vspace{-1em}
\label{fig:energy:job:2}
\end{figure}

Figure \ref{fig:energy} shows a sensitivity analysis of energy consumption to the number of cores. In Intel server (see Figure \ref{fig:energy:intel}), during job execution, the slope of the graph increase with the number of cores (up to 16 physical cores). However, the overall energy consumption of jobs executed on a higher number of cores is lower than that of jobs using less cores. 
 Figure~\ref{fig:energy:amd} reveals that the number of cores has the same effect on energy consumption in AMD server. When more cores are used, less energy is consumed by the same job. Interestingly, using 16 cores (32 threads) less than 0.1kWh is required to complete a job.

Figure \ref{fig:energy:job:2} illustrates energy consumption sensitivity to the number of jobs. As shown in Figure \ref{fig:energy:intel:job}, on Intel server, the energy consumption of 2 jobs is initially the same as 4 jobs (until the 2 jobs are complete), and 1 job consumes about half the energy of 4 jobs. However, that means that running 4 jobs is twice as efficient as running a single job, in terms of energy per job. 
 Figure \ref{fig:energy:amd:job} shows that similarly on AMD server, the total energy consumption increases as the number of jobs increases. 
These results are further analyzed for per-job energy efficiency in Section \ref{subsec:stas}.

\begin{theorem}
As long as CPU resources are not fully utilized, it is more energy efficient to increase the number
of jobs running on a CPU.
\end{theorem}

\subsection{Impact of CPU sockets}
\begin{figure}[t]
\centering
\subfigure[1 job]{
{\scalebox{0.42}{
 \input{results/power_intel_sockets.tex}
    \label{fig:power:intel:socket}
    }
    }
    }
        \hspace{-1em}
\subfigure[2 jobs]{
{\scalebox{0.42}{
 \input{results/power_intel_sockets_2jobs.tex}
    \label{fig:power:intel:socket:2}
    }
    }
    }
\caption{Sensitivity analysis: Use of sockets to power consumption}
\vspace{-1em}
\label{fig:power:job:2}
\end{figure}

\begin{figure}[t]
\centering
\subfigure[1 job]{
{\scalebox{0.42}{
 \input{results/energy_intel_sockets.tex}
    \label{fig:energy:intel:socket}
    }
    }
    }
    \hspace{-1em}
\subfigure[2 jobs]{
{\scalebox{0.42}{
 \input{results/energy_intel_sockets_2jobs.tex}
    \label{fig:energy:intel:socket:2}
    }
    }
    }
\caption{Sensitivity analysis: Use of sockets to energy consumption}
\vspace{-1em}
\label{fig:energy:job}
\end{figure}

So far, running the simulation on Intel server with the default 8 cores per job, the cores were always distributed between two CPU sockets.
In order to further understand the impact of CPU sockets usage on power and energy consumption, we used \emph{taskset}, 
a Linux shell command, to force all cores to run on a specific socket. 
In this experiment, we only consider 1 job and 2 jobs, as each job uses 8 logical cores, and only 16 are available on a single socket.

The difference in power consumption between using 1 or 2 sockets on Intel server is reported in Figure~\ref{fig:power:job:2}. For one job, Figure~\ref{fig:power:intel:socket} shows that the curves of both cases almost overlap, meaning that the distribution between sockets does not affect the power consumption considerably, but the job execution time is slightly shorter on a single socket. However, as shown in Figure \ref{fig:power:intel:socket:2}, when running two jobs and all 16 logical cores are forced on one socket, the power consumption is the same as that of one job, but the execution time is nearly as long. This also matches the power consumption results for 2 jobs and 4 jobs previously shown in Figure \ref{fig:power:intel:job}. 
 Note that initially the power consumption is high for 2 jobs on a single socket as child processes can be pinned to cores only after the job is fully running and children processes are forked.

The total energy consumption is analyzed in Figure \ref{fig:energy:job}. 
Figure~\ref{fig:energy:intel:socket} shows that for a single job the energy consumption of same socket is slightly lower than that of different sockets. Figure \ref{fig:energy:intel:socket:2} shows however that for 2 jobs, the execution on different sockets is more energy efficient. This is because forcing 16 logical cores on the same socket lowers the performance and increases execution time, leading to overall higher execution time.
\begin{theorem}
    If the total number of logical cores required by a set of jobs is less than the number of physical cores on a socket, it is preferred to use a single socket. 
\end{theorem}

\subsection{Impact of job scheduling}
\begin{figure}[t]
\centering
\subfigure[Power consumption]{
{\scalebox{0.43}{
 \input{results/power_intel_seq.tex}
    \label{fig:power:seq}
    }
    }
    }
    \hspace{-2em}
\subfigure[Total energy consumption]{
{\scalebox{0.43}{
 \input{results/energy_intel_seq.tex}
    \label{fig:energy:seq:sub}
    }
    }
    }
\caption{Staggered jobs start on Intel server ($T_{int} = 450s$)}
\vspace{-1em}
\label{fig:energy:seq}
\end{figure}
\begin{figure}[t]
\centering
\subfigure[Power consumption]{
{\scalebox{0.43}{
 \input{results/power_amd_seq.tex}
    \label{fig:power:seq:amd}
    }
    }
    }
\subfigure[Total energy consumption]{
{\scalebox{0.43}{
 \input{results/energy_amd_seq.tex}
    \label{fig:energy:seq:amd}
    }
    }
    }
\caption{Staggered jobs start on AMD server ($T_{int} = 450s$)}
\vspace{-1em}
\label{fig:amd:energy:seq}
\end{figure}

\begin{table*}[t]
 \caption{Summary of experiments}
\label{table3}
 \centering
 \vspace{-1em}
 \Large
\begin{adjustbox}{width=\textwidth}
\begin{tabular}{|c|c|c|c|c|c|c|c|c|c|c|c|c|c||c|c|}
\hline
\multirow{3}{*}{\textbf{Server}} &\multirow{3}{*}{\textbf{ \#Jobs}} & \multirow{3}{*}{\textbf{\#Cores}} & \multicolumn{2}{c|}{\begin{tabular}[x]{@{}c@{}} \textbf{Avg. power} \\ \textbf{consumption}\end{tabular}}   & \multicolumn{2}{c|}{\begin{tabular}[x]{@{}c@{}} \textbf{Total energy} \\ \textbf{consumption}\end{tabular}}   & \multicolumn{7}{c}{\textbf{Per job per core}}& \multicolumn{2}{|c|}{\textbf{Per job}}\\
\cline{4-16}
&&&  \multirow{2}{*}{\begin{tabular}[x]{@{}c@{}}  \textbf{per hour }\\ \textbf{(W)} \end{tabular}}  & \multirow{2}{*}{\begin{tabular}[x]{@{}c@{}} \textbf{during} \\ \textbf{execution time} \\ \textbf{(W)} \end{tabular}} 
&\multirow{2}{*}{\begin{tabular}[x]{@{}c@{}} \textbf{ per hour }\\ \textbf{(kWh)} \end{tabular}}  & \multirow{2}{*}{\begin{tabular}[x]{@{}c@{}}  \textbf{during} \\ \textbf{execution} \\ \textbf{time (kWh) }\end{tabular}} 
&  \multicolumn{4}{c|}{\textbf{Avg. power consumption}} & \multicolumn{5}{c|}{\textbf{Total energy consumption}} \\
\cline{8-16}
&&&&&&&\begin{tabular}[x]{@{}c@{}} \textbf{per hour }\\ \textbf{(W)} \textcolor{red}{$\ast$} \end{tabular} & \begin{tabular}[x]{@{}c@{}}  \textbf{during} \\ \textbf{execution} \\ \textbf{time} $\bar{P}_{unit}^{*}$ \\\textbf{(W) }\textcolor{red}{$\ast$}  \end{tabular} &  \begin{tabular}[x]{@{}c@{}}   \textbf{per hour} \\ \textbf{(W)} \textcolor{red}{$\dagger$} \end{tabular} & \begin{tabular}[x]{@{}c@{}}  \textbf{during} \\ \textbf{execution} \\ \textbf{time} $\bar{P}_{unit}$\\ \textbf{(W) }\textcolor{red}{$\dagger$}  \end{tabular}  &  \begin{tabular}[x]{@{}c@{}}  \textbf{per hour} \\\ \textbf{(}\bm{$\times 10^{-3}$} \\ \textbf{kWh)} \textcolor{red}{$\ast$} \end{tabular}  & \begin{tabular}[x]{@{}c@{}}  \textbf{during} \\ \textbf{execution} \\ \textbf{time} $E_{unit}$\\\ \textbf{(}\bm{$\times 10^{-3}$} \\ \textbf{kWh)} \textcolor{red}{$\dagger$}  \end{tabular} &  \begin{tabular}[x]{@{}c@{}}  \textbf{during} \\ \textbf{execution} \\ \textbf{time} $E_{unit}^{*}$\\\ \textbf{(}\bm{$\times 10^{-3}$} \\ \textbf{kWh)} \textcolor{red}{$\ast$}  \end{tabular}  & \begin{tabular}[x]{@{}c@{}}  \textbf{during} \\ \textbf{execution} \\ \textbf{time} $E_{unit}$\\\ \textbf{(}\bm{$\times 10^{-3}$} \\ \textbf{kWh)} \textcolor{red}{$\dagger$}  \end{tabular} &  \begin{tabular}[x]{@{}c@{}}  \textbf{during} \\ \textbf{execution} \\ \textbf{time} $E_{unit}^{*}$\\\ \textbf{(}\bm{$\times 10^{-3}$} \\ \textbf{kWh)} \textcolor{red}{$\ast$}  \end{tabular} \\
\hline
\begin{tabular}[x]{@{}c@{}}  Intel \\(Baseline)  \end{tabular}   & / & /  &252.7 & / &0.251 &/ & / & / & / & / & / & / & / & / & /\\
\hline
Intel  & 1 & 1  &267.58 & 268.49  & 0.265& 1.351 & 13.79 & 13.79 & 267.58 & 268.49 & 14.00 & 1351.00 & 80.22 & 1351 & 80.22\\
\hline
Intel  & 1 & 4  &302.55 & 303.88 & 0.304& 0.392 & 12.2 & 12.2 & 75.38 & 75.38 & 13.25 & 98.00 & 16.62 & 392.00 & 66.50\\
\hline
Intel  & 1 & 8  &326.90 & 350.22 & 0.324& 0.251 & 8.99  & 11.25& 40.57 & 42.84 & 9.13 & 31.38 & 9.45& 251.04 & 75.63\\
\hline
Intel  & 1 & 16  & 328.00 & 397.38 & 0.323& 0.202 & 4.58 & 8.88 & 20.37 & 24.67 &  3.63 & 12.63& 4.63& 202.08 & 74.03\\
\hline
Intel  & 1 & \begin{tabular}[x]{@{}c@{}}  16 \\ (32 \\ threads)   \end{tabular} &314.83 & 431.32 & 0.313& 0.147 & 3.88 & 11.16& 19.68 &26.96 & 3.00 & 9.19& 3.81 &147.04 & 61.01\\
\hline
Intel  & 2 & 8  & 394.77 & 456.66 & 0.394& 0.315 & 8.87 & 12.74& 21.74 & 28.54  & 8.06 & 19.69& 8.98&157.52 & 71.81\\
\hline
Intel  & 4 & 8  &462.21 & 461.35 & 0.462& 0.523 & 6.54 & 6.52& 12.88 &14.42 & 6.59 & 16.34& 7.44& 130.72 & 59.50\\
\hline
\begin{tabular}[x]{@{}c@{}}  AMD \\(Baseline)  \end{tabular} & / & /  &205.06 & / & 0.205& / & /&/ &/& / & / & / & / &/ & /\\
\hline
AMD  & 1 & 1 &211.39 & 211.06 & 0.210& 1.241 & 6.66 & 6.39& 211.39 & 211.06 & 4.00 & 1212.95 & 34.48&1212.95 & 34.48\\
\hline
AMD  & 1 & 4  &223.93 & 223.33 & 0.225& 0.354 & 4.21 & 4.21& 55.48  & 55.48 & 5.00 & 88.50 & 7.18 &351.24 & 28.73\\
\hline
AMD  & 1 & 8  &234.63 & 242.95 & 0.234& 0.189 & 3.70 & 4.74& 29.33 & 30.37 & 3.63 & 23.63& 3.67 &189.04 & 29.39\\
\hline
AMD  & 1 & 16 &235.33 & 274.87 & 0.235& 0.118 & 1.89 & 4.36& 14.71 & 17.18 & 1.875 & 7.38& 1.87 &118.08 & 29.86\\
\hline
AMD  & 1 &  \begin{tabular}[x]{@{}c@{}}  16 \\ (32 \\ threads)   \end{tabular}  &236.34 & 290.48 & 0.236& 0.104 & 2.00 & 5.34&14.77 & 18.16 & 1.94 & 6.56 & 1.92& 104.90 & 30.08\\
\hline
AMD  & 2 & 8 &264.49 & 280.37 & 0.266& 0.222 & 3.71 & 2.35& 16.53 &17.52 & 3.81 & 13.88 & 3.72 &111.04 & 29.75\\
\hline
AMD  & 4 & 8  &310.76 & 311.02 & 0.297& 0.310 & 2.86 & 2.86& 9.27 & 9.27 & 2.88& 9.69 & 3.56 &77.52 & 28.50\\
\hline
\multicolumn{11}{l}{\textcolor{red}{$\ast$} Baseline excluded \textcolor{red}{$\dagger$} Baseline included} \\
\end{tabular}
\end{adjustbox}
\vspace{-1em}
\label{table:summary}
\end{table*}

When starting a new job, there is a short period of initialization time, before children processes are forked. During this time, the power consumption of a job is low. In this subsection, we check the possibility of improving energy efficiency by taking advantage of this low-power period. 
Figure~\ref{fig:energy:seq} shows the case of a staggered deployment of jobs on Intel server. At the beginning of the experiment, the first job is started. Afterwards, every 450 seconds we launch a new job until the number of jobs reaches 4. 450 seconds is chosen as it is the approximate initialization time for the default 8-cores job configuration. Figure \ref{fig:power:seq}, shows the difference between jobs starting at the same time or staggered: when 4 jobs are started in parallel, the power consumption quickly reaches 460W, and the jobs finish after 4100s, which is faster than staggered jobs (as can be expected due to the delayed starts). The power consumption of staggered deployment reaches 460W when at least 2 jobs running.
Figure~\ref{fig:energy:seq:sub} provides a view of the overall energy consumption. It shows that for jobs triggered in parallel, the overall energy consume is somewhat lower, even if we consider CPU idle time at the end of a run.

Figure \ref{fig:amd:energy:seq} performs the same comparison on the AMD server. As shown in Figure \ref{fig:power:seq:amd}, given the same time interval of 450 seconds, the power consumption increases as the number of jobs increases, unlike on the first server where peak power was reaches with 2 jobs. 
4 jobs triggered in parallel can finish execution 1350 seconds earlier than jobs deployed in a staggered manner. In terms of total energy consumption, the curves in  Figure \ref{fig:energy:seq:amd} reveal that the consumed energy of both scenarios is quite close, and the energy consumption of parallel start is slightly lower at the end of the experiment (including idle time).

\begin{theorem}
   A parallel triggered start of jobs is slightly more energy efficient than a staggered start.
\end{theorem}

\subsection{Execution time comparison}

Figure \ref{fig:time} presents a comparison of job execution time between the two server. The execution time is measured with a terminal application called \emph{prmon} binding with the process ID of jobs. As shown in Figure~\ref{fig:time:cores}, as the number of logical cores (threads) assigned for a single job increases, the execution time decreases. When varying the number of jobs, as shown in Figure~\ref{fig:time:jobs}, the execution time of 1 job and 2 jobs quite similar, while 4 jobs takes more time. However, one considers the per-job execution time, 4 jobs running in parallel require less time than the other cases.


\begin{theorem}
Given sufficient CPU resources, running multiple jobs in parallel can reduce per-job execution time.
\end{theorem}

\begin{figure}[t]
\centering
\subfigure[Varying number of cores (1 job)]{
{\scalebox{0.44}{
 \input{results/time_cores.tex}
    \label{fig:time:cores}
    }
    }
    }
\subfigure[Varying number of jobs (8 cores)]{
{\scalebox{0.44}{
 \input{results/time_jobs.tex}
    \label{fig:time:jobs}
    }
    }
    }
\caption{A comparison of job execution time}
\vspace{-1em}
\label{fig:time}
\end{figure}

\subsection{Collected statistics} \label{subsec:stas}

A partial statistics collection of this work's experiments is provided in Table \ref{table:summary}. The table uses two computed metrics, average power consumption per hour, and average power consumption during execution time. Average power consumption per hour indicates the power efficiency including idle time at the end of a job, while average power consumption during execution time only considers the power consumption when the jobs are still running. The total energy consumption per hour and the total energy consumption during execution time, refers to the same two measurement scenarios applied to energy consumption. 

As the table shows, increasing the number of cores assigned to a job improves the energy efficiency.  Moreover, enabling more threads on the same cores (e.g., from 16 threads to 32 threads) can efficiently reduce the energy consumption. Another observation is that for both, given the same number of active cores in an experiment, 2 jobs with 8 cores each consume less energy per job than 1 job with 16 cores.

The next step is to subtract the baseline power from the results, and analyze per-core, per-job power consumption of each experiment. These two additional metrics are referred to as average power consumption per-job per-core per-hour, and average power consumption per-job per-core during execution time. The results indicate that when idle time exists, running a single job with more cores is preferred. Otherwise, running multiple jobs with the same overall number of cores is more power efficient. Still, the effect of baseline power cannot be neglected. With it, given the same number of cores, running multiple jobs in parallel is the best solution. 
 
Table~\ref{table:summary} also presents the total energy consumption per-job, per-core per-hour and during execution time. Note that the baseline of energy consumption during execution time is the energy consumed until the end of each job's execution, so the baseline is not the same for different cases. The result is noticeable: given the same total number of cores, running a single job is more energy efficient than running multiple jobs.

However, per-job per-core energy consumption is not enough to evaluate a job's energy efficiency. This is because this metric only indicates the energy consumption of each core, but without considering the impact of job execution time and per-job energy efficiency. Therefore, the table also presents the energy consumption per job, and this metric is evaluated together with per-job execution time in the next subsection.

\begin{theorem}
Regardless if servers baseline power is included ($\dagger$) or excluded ($\ast$), running a single job with more logical cores is better in terms of per-job per-core energy efficiency than multiple jobs using the same total number of logical cores for the ATLAS simulation workload.
\end{theorem}

\subsection{Energy-execution time trade-off}

In this subsection, we propose a key metric to evaluate energy efficiency for servers, that is, the energy-execution time trade-off. Intuitively, the best energy-execution time trade-off should present the lowest per-job execution time and total energy consumption. 
To explore the trade-off, we correlate per-job execution time and per-job total energy consumption as a linear regression model. The results of 1 job and 1 core are omitted from the figures of this subsection due to their extremely high execution time and energy consumption.

Figure~\ref{fig:time_energy_intel} shows the energy-execution time trade-off on Intel server. As shown in Figure~\ref{fig:time_eng:intel}, when considering the idle power consumption, running 4 jobs in parallel with 8 cores has the best performance: it takes only around 1000 seconds and 130Wh to execute one job. Running 1 job with 16 cores (32 threads) has comparable performance as 2 jobs with 8 cores, but 1 job with 16 cores (32 threads) has a slightly lower energy consumption. Since all points are close to the linear regression line, it is also obvious that higher execution time leads to worse energy-time performance. On the other hand, when the baseline is excluded (see Figure \ref{fig:time_eng_ex:intel}), the energy consumption of all job executions is mostly about 70Wh, with 10\% relative error to the linear regression line. 4 jobs with 8 cores still outperform other options because of its low job execution time. Moreover, 1 job with 16 physical cores (32 threads) can save approximately 10Wh of energy with respect to that of 2 jobs with 8 cores. 

 Figure~\ref{fig:time_energy_amd} illustrates the energy-execution time trade-off comparison for AMD server. 
    Figure~\ref{fig:time_eng:amd} presents energy-execution time trade-off when the baseline energy consumption is included. As can be expected, the energy consumption linearly increases with job execution time. 4-jobs execution with 8 cores each outperforms the other 5 scenarios both in terms of execution time per job and total energy consumption per job. One observation is that the performance of (i.) 1 job with 16 cores (32 threads), (ii.) 1 job with 16 cores, and (iii.) 2 jobs with 8 cores is very similar: they have close energy consumption as well as execution time, but 1 job execution with 16 cores (32 threads) is slightly better than the other two scenarios. If the baseline is excluded, the results shown in Figure~\ref{fig:time_eng_ex:amd} provide an important observation, as the energy consumption of all scenarios is quite flat and close to 30Wh, and the difference is only the execution time per job. This is an indicator of a good resource usage by the AMD CPU, meaning that the energy consumed by similar processing jobs sufficiently resourced is the same.

\begin{theorem}
A linear regression model can accurately approximate the energy consumption of a job. 
\end{theorem}

\begin{figure}[t]
\centering
\subfigure[Baseline included]{
{\scalebox{0.44}{
 \input{results/time_energy_intel.tex}
    \label{fig:time_eng:intel}
    }
    }
    }
\hspace{-2.5em}
\subfigure[Baseline excluded]{
{\scalebox{0.44}{
\input{results/time_energy_intel_ex_baseline.tex}
    \label{fig:time_eng_ex:intel}
    }
    }
    }
    \vspace{-1em}
\caption{Intel server: Energy-execution time trade-off comparison}
\label{fig:time_energy_intel}
\vspace{-1em}
\end{figure}
\begin{figure}[t]
\centering
\subfigure[Baseline included]{
{\scalebox{0.44}{
 \input{results/time_energy_amd.tex}
    \label{fig:time_eng:amd}
    }
    }
    }
\hspace{-2.5em}
\subfigure[Baseline excluded]{
{\scalebox{0.44}{
\input{results/time_energy_amd_ex_baseline.tex}
    \label{fig:time_eng_ex:amd}
    }
    }
    }
    \vspace{-1em}
\caption{AMD server: Energy-execution time trade-off comparison}
\label{fig:time_energy_amd}
\vspace{-1em}
\end{figure}

\section{Lessons Learned and Recommendations} \label{sec:rec}
Based on the experimental results, this section presents a series of lessons learned and recommendations, which suggest how jobs can be executed to improve energy efficiency for high-throughput workloads on data center servers.

IT power is often referred to as \emph{critical power}~\cite{barroso2018datacenter}, and while it excludes power required for cooling and other overheads, it includes the power overheads of computing itself. One of the interesting lessons of this work, evident both from the study of power consumption at STFC RAL (Section~\ref{sec:motivation}) and the focused measurements (Section~\ref{sec:exp}), is the significant effect of computing overheads. These take different forms, and the baseline power of a server only partially captures the effect.

One clear recommendation is to focus efforts on minimizing the baseline power of servers. Servers from different vendors with similar CPU, memory and storage specifications may have different baseline power, and data center operator should be attentive to that. Note that over-provisioning a server may increase its energy consumption over time.

A related consideration, based on the analysis in Section~\ref{sec:motivation}, is that upgrading to newer generations of CPUs may not (significantly) increase energy efficiency, as the normalized per-core power consumption (i.e., rack power divided by the number of cores in the rack) is not remarkably different. Given that significant carbon-emissions are part of manufacturing new servers (embodied carbon emissions) and decommissioning servers, data center operators may opt to delay replacement of servers. Note that this recommendation applies where CPUs have a mostly similar per-core computing performance, as faster processing time does reduce energy consumption per job. Still, this depends on the ratio of baseline server power and maximum server power.


Our measurement results show that lower execution time improves the energy efficiency of jobs, due to the baseline power. Combined with DVFS-related measurements, we reinforce the value of using DVFS and boosting even in highly utilized servers, for the studied high-throughput workload. 

It may be assumed that an intuitive way to reduce the energy consumption of a data center is to shut down idle service. There are two limitations to this approach. First, frequently turning off servers reduces their reliability~\cite{guenter2011managing}. Second, data centers such as STFC RAL are highly utilized and may not have idle operational machines. One potential trajectory may be to introduce finer grain electronic power control mechanisms that allow to turn off an unused core or an unused socket, without affecting reliability. These may also be applied to other parts of the server, e.g., shutting down unused accelerator cards.  

A way for data center operators to reduce their energy consumption is to throttle the servers by reducing CPU speed. While this reduces the amount of energy taken from the grid, this approach has a negative effect on the energy consumption per job, meaning that over time the energy cost of running a fixed set of experiments will increase. This approach can still be used on short time scales, e.g., if there is a limit to the power that can be drawn from the grid, and the PUE is higher than usual due to hot weather.

Another recommended method for data center operators to reduce energy consumption is by reducing utilization through policies and motivating users. For example, introducing \emph{spot jobs} that run only when utilization is below a certain level or when spare resources are available (e.g., to fully populate CPU cores). Users can also be incentivized to run their jobs only at times that energy is cheap or carbon intensity is low, e.g., providing a designated queue or a class for jobs that are environmentally friendly.




\section{Conclusion and future work}\label{sec:conclusion}
This paper presented an empirical study on the energy efficiency of data center servers running high-throughput computing workloads. The data analysis of power measurements of a real data center, showed the similarity and differences between different server families. 
 To further improve the energy efficiency of servers, we conduct intensive measurements, and obtain detailed statistics for power and energy usage of current servers. The findings reveal fundamental energy-performance trade-off in servers: minimizing both per-job energy consumption and  per-job execution time can assist in choosing the best energy saving strategy. Last, this work discussed lessons learned and recommendations to help data center operators, administrators, and engineers to better design and execute jobs in an energy-efficient way for sustainable computing.
Future work will explore effective mechanisms to reduce the idle power of servers. Moreover, we plan to explore power efficiency of other data center equipment.


\section*{Acknowledgement}
This work was funded by the UKRI Net-Zero Digital Research Infrastructure Scoping Project (NE/W007134/1).

%

\bibliographystyle{IEEEtran}
\bibliography{ref}

\end{document}

%% file: figure/testbed.tex
\begin{figure}[t]
\centering
\begin{adjustbox}{width=0.8\linewidth}
\begin{tikzpicture}[]

\node(intel) at (2.5, 2)[rectangle, draw, inner sep=1, line width=1mm,  minimum width = 4.5cm, minimum height=3cm, align=center]{};

\node(docker1) at (2.5, 3)[]{\includegraphics[scale = 0.5]{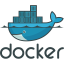}};

\node(d1) at (1.3, 2)[rectangle, fill=red]{Container 1};
\node(d2) at (1.3, 1.5)[rectangle, fill=red]{Container 2};
\node(d3) at (3.7, 2)[rectangle, fill=red]{Container 3};
\node(d3) at (3.7, 1.5)[rectangle, fill=red]{Container 4};
\node(p1) at (1.6, 3.7)[rectangle, draw, inner sep=1, line width=1mm,  minimum width = 0.5cm, minimum height=0.2cm,fill=black]{};
\node(p2) at (3.4, 3.7)[rectangle, draw, inner sep=1, line width=1mm,  minimum width = 0.5cm, minimum height=0.2cm,fill=black]{};

\node(server1) at (2.5, 1)[]{\textbf{Intel Server}};

\node(amd) at (8.5, 2)[rectangle, draw, inner sep=1, line width=1mm,  minimum width = 4.5cm, minimum height=3cm,align=center]{};
\node(dd1) at (7.3, 2)[rectangle, fill=red]{Container 1};
\node(dd2) at (7.3, 1.5)[rectangle, fill=red]{Container 2};
\node(dd3) at (9.7, 2)[rectangle, fill=red]{Container 3};
\node(dd3) at (9.7, 1.5)[rectangle, fill=red]{Container 4};

\node(docker2) at (8.5, 3)[]{\includegraphics[scale = 0.5]{figure/docker-icon.png}};
\node(server1) at (8.5, 1)[]{\textbf{AMD Server}};

\node(p3) at (7.6, 3.7)[rectangle, draw, inner sep=1, line width=1mm,  minimum width = 0.5cm, minimum height=0.2cm,fill=black]{};
\node(p4) at (9.4, 3.7)[rectangle, draw, inner sep=1, line width=1mm,  minimum width = 0.5cm, minimum height=0.2cm,fill=black]{};

\node(pdu1) at (2.5, 6)[]{\includegraphics[scale = 0.25]{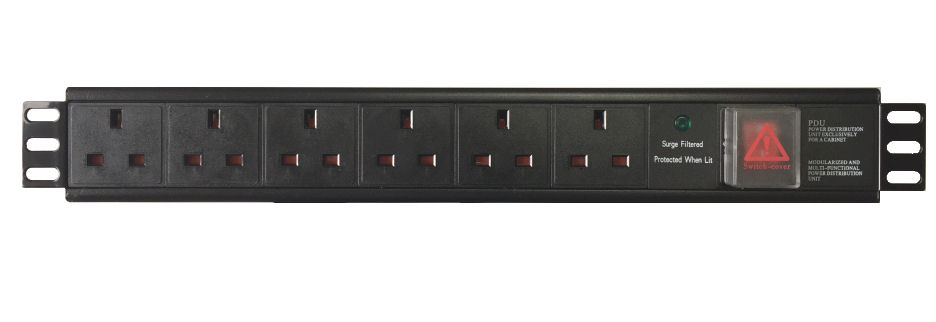} };
\node(pdu1x) at (2.5,6.5)[]{\textbf{PDU 1}};

\node(p1x) at (3.4, 5.7)[rectangle, draw, inner sep=1, line width=1mm,  minimum width = 0.5cm, minimum height=0.2cm,fill=black]{};
\node(p1xx) at (1.6, 5.7)[rectangle, draw, inner sep=1, line width=1mm,  minimum width = 0.5cm, minimum height=0.2cm,fill=black]{};

\node(pdu2) at (8.5, 6)[]{\includegraphics[scale = 0.25]{figure/PDU.jpeg} };
\node(pdu2x) at (8.5,6.5)[]{\textbf{PDU 2}};

\node(p2x) at (9.4, 5.7)[rectangle, draw, inner sep=1, line width=1mm,  minimum width = 0.5cm, minimum height=0.2cm,fill=black]{};
\node(p2xx) at (7.6, 5.7)[rectangle, draw, inner sep=1, line width=1mm,  minimum width = 0.5cm, minimum height=0.2cm,fill=black]{};

\draw[thick,-, dashed](p1) edge[bend left=40, draw=olive] node [left] {}  (p1xx);
\draw[thick,-,dashed](p2) edge[bend right=30, draw=olive] node [left] {}  (p2xx);

\draw[thick,-](p3) edge[bend left=30, draw=blue] node [left] {}  (p1x);
\draw[thick,-](p4) edge[bend right=40, draw=blue] node [left] {}  (p2x);

\node(power) at (5.5, 9)[]{\includegraphics[scale = 0.125]{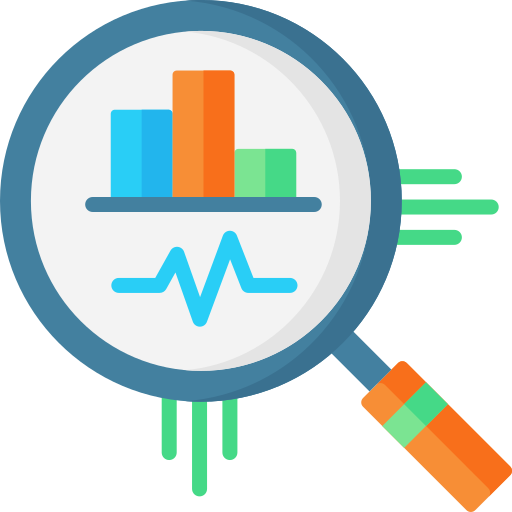}};

\node(powerx) at (8, 8.75)[inner sep=1, line width=1mm,  minimum width = 4cm, minimum height=1.2cm, align =center]{ \textcolor{orange}{\textbf{Energy consumption}} \\ \bm{$E_{tot} = E_{1}^{end} + E_{2}^{end} $ }\\ \bm{$- (E_{1}^{start} + E_{2}^{start})$}};

\node(powerx2) at (3.1, 8.75)[inner sep=1, line width=1mm,  minimum width = 4cm, minimum height=1.2cm, align =center]{\textcolor{orange}{\textbf{Power consumption}} \\ \bm{$P = P_{1}+P_{2}$}};

\draw [-, thick](pdu1x) |- (5.5,7.25) ;
\draw [-, thick](pdu2x) |- (5.5,7.25) ;
\draw [->, thick](5.5,7.25) -- (power);
\node(power) at (5.5,7)[]{\textbf{SNMP}};

\end{tikzpicture}
\end{adjustbox}
\caption{The physical testbed for power and energy measurements}
    \vspace{-1em}
\label{fig:arch}
\end{figure}

%% file: results/power_intel_components.tex
\begin{tikzpicture}[font=\Large]

\definecolor{color0}{rgb}{0.12156862745098,0.466666666666667,0.705882352941177}
\definecolor{color1}{rgb}{1,0.498039215686275,0.0549019607843137}
\definecolor{color2}{rgb}{0.172549019607843,0.627450980392157,0.172549019607843}
\definecolor{color3}{rgb}{0.83921568627451,0.152941176470588,0.156862745098039}
\definecolor{color4}{rgb}{0.580392156862745,0.403921568627451,0.741176470588235}
\definecolor{color5}{rgb}{0,0,0}

\begin{axis}[
legend cell align={left},
legend columns=1,
legend style={fill opacity=0.8, draw opacity=1, text opacity=1, at={(1.02,1.2)}, anchor=east, draw=white!80.0!black},
tick align=outside,
tick pos=left,
x grid style={white!69.01960784313725!black},
xlabel={Time (SEC)},
xmin=0, xmax=3600,
xtick style={color=black},
xtick={0,900,1800,2700,3600},
y grid style={white!69.01960784313725!black},
ylabel={Power consumption (W)},
ymin=100, ymax=300,
ytick={100,150,200,250, 300},
xmajorgrids,
ymajorgrids,
ytick style={color=black}
]
\addplot [thick, color5]
table {%
0 254
5 253
10 254
15 253
20 253
25 253
30 253
35 253
40 253
45 253
50 253
55 254
60 253
65 254
70 253
75 252
80 254
85 254
90 254
95 254
100 254
105 253
110 253
115 253
120 253
125 253
130 254
135 253
140 254
145 252
150 254
155 254
160 253
165 254
170 253
175 252
180 253
185 253
190 253
195 253
200 253
205 253
210 263
215 251
220 253
225 252
230 252
235 253
240 252
245 253
250 253
255 253
260 253
265 252
270 252
275 254
280 252
285 252
290 252
295 253
300 254
305 252
310 253
315 253
320 253
325 252
330 252
335 253
340 252
345 254
350 253
355 253
360 253
365 253
370 253
375 252
380 252
385 252
390 253
395 252
400 252
405 254
410 252
415 253
420 252
425 254
430 254
435 253
440 254
445 252
450 252
455 252
460 253
465 251
470 250
475 253
480 253
485 253
490 254
495 253
500 253
505 252
510 252
515 254
520 252
525 253
530 254
535 253
540 253
545 253
550 253
555 254
560 254
565 252
570 253
575 253
580 253
585 254
590 252
595 253
600 252
605 252
610 252
615 253
620 253
625 252
630 253
635 252
640 252
645 252
650 253
655 254
660 253
665 254
670 253
675 252
680 254
685 252
690 252
695 252
700 253
705 254
710 253
715 254
720 254
725 254
730 252
735 253
740 253
745 253
750 253
755 254
760 252
765 250
770 253
775 253
780 253
785 253
790 253
795 253
800 254
805 253
810 253
815 252
820 253
825 253
830 252
835 254
840 253
845 253
850 252
855 252
860 253
865 252
870 253
875 253
880 253
885 254
890 252
895 253
900 253
905 250
910 253
915 252
920 253
925 253
930 253
935 254
940 254
945 253
950 254
955 253
960 254
965 254
970 253
975 253
980 254
985 253
990 253
995 252
1000 253
1005 254
1010 253
1015 253
1020 253
1025 252
1030 254
1035 253
1040 253
1045 252
1050 254
1055 253
1060 252
1065 253
1070 252
1075 254
1080 253
1085 253
1090 252
1095 253
1100 252
1105 252
1110 253
1115 253
1120 254
1125 252
1130 254
1135 252
1140 253
1145 253
1150 252
1155 253
1160 254
1165 253
1170 251
1175 253
1180 252
1185 253
1190 253
1195 253
1200 253
1205 252
1210 252
1215 253
1220 252
1225 253
1230 254
1235 254
1240 253
1245 253
1250 253
1255 253
1260 252
1265 253
1270 254
1275 254
1280 254
1285 252
1290 254
1295 252
1300 252
1305 254
1310 252
1315 254
1320 254
1325 254
1330 253
1335 254
1340 252
1345 252
1350 253
1355 254
1360 254
1365 253
1370 253
1375 254
1380 263
1385 251
1390 254
1395 254
1400 253
1405 253
1410 254
1415 251
1420 251
1425 252
1430 251
1435 251
1440 251
1445 251
1450 251
1455 252
1460 251
1465 251
1470 251
1475 251
1480 252
1485 251
1490 252
1495 251
1500 251
1505 252
1510 251
1515 251
1520 251
1525 251
1530 251
1535 251
1540 251
1545 251
1550 251
1555 251
1560 251
1565 249
1570 251
1575 251
1580 251
1585 251
1590 252
1595 251
1600 252
1605 252
1610 251
1615 252
1620 251
1625 251
1630 251
1635 251
1640 252
1645 251
1650 252
1655 252
1660 254
1665 252
1670 252
1675 253
1680 253
1685 254
1690 252
1695 253
1700 252
1705 252
1710 252
1715 253
1720 253
1725 253
1730 252
1735 252
1740 252
1745 253
1750 252
1755 253
1760 254
1765 254
1770 253
1775 253
1780 254
1785 254
1790 253
1795 253
1800 252
1805 254
1810 252
1815 253
1820 253
1825 254
1830 252
1835 252
1840 253
1845 250
1850 252
1855 253
1860 252
1865 253
1870 252
1875 253
1880 253
1885 252
1890 253
1895 253
1900 253
1905 252
1910 252
1915 252
1920 252
1925 253
1930 252
1935 253
1940 254
1945 253
1950 253
1955 253
1960 253
1965 253
1970 253
1975 253
1980 252
1985 253
1990 250
1995 253
2000 253
2005 254
2010 253
2015 253
2020 253
2025 252
2030 252
2035 253
2040 253
2045 252
2050 253
2055 252
2060 253
2065 254
2070 253
2075 253
2080 252
2085 252
2090 253
2095 253
2100 252
2105 253
2110 253
2115 253
2120 252
2125 252
2130 253
2135 254
2140 253
2145 252
2150 252
2155 252
2160 254
2165 253
2170 252
2175 253
2180 253
2185 252
2190 253
2195 254
2200 254
2205 253
2210 254
2215 253
2220 254
2225 253
2230 254
2235 253
2240 253
2245 252
2250 252
2255 254
2260 253
2265 253
2270 254
2275 253
2280 253
2285 253
2290 254
2295 253
2300 253
2305 253
2310 253
2315 254
2320 254
2325 252
2330 254
2335 252
2340 253
2345 252
2350 253
2355 253
2360 253
2365 254
2370 254
2375 252
2380 252
2385 253
2390 263
2395 254
2400 254
2405 253
2410 253
2415 254
2420 252
2425 251
2430 253
2435 254
2440 254
2445 252
2450 254
2455 254
2460 253
2465 253
2470 253
2475 253
2480 254
2485 254
2490 253
2495 253
2500 253
2505 253
2510 254
2515 253
2520 254
2525 253
2530 254
2535 254
2540 253
2545 254
2550 254
2555 254
2560 254
2565 253
2570 253
2575 252
2580 254
2585 254
2590 253
2595 254
2600 252
2605 254
2610 253
2615 253
2620 254
2625 253
2630 253
2635 253
2640 253
2645 252
2650 253
2655 253
2660 253
2665 253
2670 252
2675 253
2680 252
2685 253
2690 253
2695 252
2700 253
2705 253
2710 253
2715 253
2720 252
2725 254
2730 252
2735 253
2740 253
2745 254
2750 253
2755 253
2760 253
2765 254
2770 254
2775 252
2780 253
2785 253
2790 253
2795 253
2800 257
2805 251
2810 254
2815 253
2820 253
2825 252
2830 253
2835 252
2840 253
2845 253
2850 253
2855 253
2860 264
2865 251
2870 253
2875 254
2880 253
2885 253
2890 253
2895 253
2900 254
2905 254
2910 252
2915 254
2920 254
2925 253
2930 253
2935 253
2940 253
2945 253
2950 254
2955 253
2960 252
2965 254
2970 252
2975 254
2980 254
2985 254
2990 253
2995 253
3000 253
3005 253
3010 254
3015 253
3020 253
3025 253
3030 253
3035 253
3040 253
3045 253
3050 253
3055 254
3060 253
3065 253
3070 253
3075 253
3080 252
3085 251
3090 252
3095 252
3100 252
3105 251
3110 251
3115 252
3120 252
3125 251
3130 253
3135 252
3140 252
3145 252
3150 252
3155 252
3160 251
3165 251
3170 252
3175 252
3180 251
3185 252
3190 252
3195 252
3200 251
3205 251
3210 252
3215 251
3220 252
3225 252
3230 252
3235 252
3240 251
3245 252
3250 252
3255 252
3260 251
3265 252
3270 252
3275 252
3280 252
3285 252
3290 251
3295 252
3300 251
3305 252
3310 252
3315 251
3320 251
3325 251
3330 252
3335 252
3340 252
3345 252
3350 251
3355 251
3360 252
3365 252
3370 252
3375 252
3380 252
3385 251
3390 251
3395 252
3400 251
3405 251
3410 252
3415 251
3420 252
3425 252
3430 252
3435 252
3440 252
3445 251
3450 252
3455 252
3460 252
3465 251
3470 252
3475 252
3480 251
3485 252
3490 252
3495 251
3500 252
3505 252
3510 252
3515 252
3520 251
3525 252
3530 252
3535 252
3540 252
3545 251
3550 251
3555 252
3560 251
3565 251
3570 251
3575 251
3580 251
3585 251
3590 251
3595 252
3600 252
};
\addlegendentry{Baseline}
\addplot [thick, color0, mark=*, mark size=3, mark options={solid,fill=white,draw=red},  mark repeat={180}]
table{
0 236
5 236
10 235
15 236
20 235
25 236
30 236
35 236
40 236
45 236
50 236
55 235
60 236
65 236
70 235
75 236
80 235
85 236
90 236
95 236
100 236
105 235
110 235
115 236
120 236
125 236
130 235
135 236
140 236
145 235
150 235
155 235
160 235
165 236
170 236
175 236
180 236
185 235
190 236
195 235
200 235
205 235
210 235
215 236
220 236
225 236
230 236
235 235
240 233
245 236
250 236
255 236
260 236
265 236
270 235
275 235
280 236
285 236
290 235
295 236
300 236
305 235
310 235
315 236
320 235
325 236
330 235
335 235
340 235
345 236
350 236
355 236
360 236
365 235
370 235
375 235
380 236
385 236
390 236
395 235
400 236
405 235
410 236
415 235
420 234
425 235
430 235
435 235
440 236
445 236
450 235
455 236
460 235
465 235
470 236
475 235
480 236
485 235
490 235
495 235
500 235
505 236
510 236
515 235
520 235
525 235
530 236
535 236
540 236
545 235
550 234
555 236
560 236
565 236
570 236
575 236
580 236
585 235
590 235
595 235
600 235
605 235
610 234
615 235
620 235
625 236
630 236
635 234
640 236
645 236
650 235
655 236
660 235
665 234
670 235
675 236
680 236
685 235
690 235
695 235
700 235
705 235
710 235
715 235
720 235
725 236
730 236
735 236
740 235
745 236
750 235
755 236
760 235
765 236
770 235
775 235
780 235
785 236
790 236
795 235
800 235
805 236
810 236
815 236
820 236
825 236
830 236
835 235
840 235
845 236
850 235
855 236
860 236
865 235
870 235
875 236
880 236
885 236
890 236
895 235
900 236
905 235
910 236
915 235
920 235
925 235
930 236
935 235
940 235
945 235
950 236
955 235
960 235
965 235
970 235
975 236
980 235
985 236
990 235
995 235
1000 236
1005 236
1010 235
1015 236
1020 235
1025 235
1030 236
1035 236
1040 235
1045 236
1050 235
1055 236
1060 236
1065 235
1070 235
1075 236
1080 235
1085 236
1090 235
1095 235
1100 235
1105 235
1110 235
1115 235
1120 235
1125 234
1130 236
1135 236
1140 234
1145 236
1150 236
1155 234
1160 236
1165 235
1170 236
1175 236
1180 235
1185 236
1190 235
1195 235
1200 236
1205 236
1210 236
1215 236
1220 236
1225 236
1230 234
1235 235
1240 236
1245 235
1250 236
1255 235
1260 236
1265 235
1270 236
1275 236
1280 235
1285 236
1290 236
1295 236
1300 236
1305 235
1310 236
1315 235
1320 235
1325 236
1330 236
1335 235
1340 235
1345 236
1350 235
1355 236
1360 236
1365 236
1370 236
1375 236
1380 235
1385 235
1390 235
1395 236
1400 236
1405 236
1410 236
1415 235
1420 236
1425 236
1430 236
1435 236
1440 235
1445 236
1450 236
1455 235
1460 236
1465 236
1470 236
1475 236
1480 236
1485 235
1490 235
1495 236
1500 236
1505 235
1510 236
1515 236
1520 235
1525 235
1530 235
1535 236
1540 236
1545 236
1550 236
1555 236
1560 235
1565 236
1570 235
1575 236
1580 235
1585 236
1590 235
1595 236
1600 236
1605 235
1610 235
1615 236
1620 236
1625 235
1630 233
1635 235
1640 235
1645 236
1650 236
1655 235
1660 236
1665 235
1670 235
1675 236
1680 235
1685 236
1690 235
1695 235
1700 236
1705 236
1710 236
1715 235
1720 236
1725 236
1730 235
1735 236
1740 236
1745 236
1750 235
1755 236
1760 236
1765 236
1770 235
1775 235
1780 235
1785 236
1790 235
1795 236
1800 236
1805 236
1810 236
1815 236
1820 233
1825 236
1830 236
1835 235
1840 235
1845 236
1850 235
1855 235
1860 236
1865 236
1870 236
1875 235
1880 236
1885 236
1890 236
1895 236
1900 236
1905 236
1910 236
1915 236
1920 236
1925 236
1930 235
1935 236
1940 236
1945 236
1950 235
1955 235
1960 235
1965 236
1970 236
1975 236
1980 236
1985 235
1990 236
1995 235
2000 235
2005 235
2010 235
2015 235
2020 236
2025 235
2030 235
2035 236
2040 236
2045 236
2050 235
2055 236
2060 234
2065 236
2070 236
2075 236
2080 236
2085 236
2090 234
2095 235
2100 235
2105 235
2110 236
2115 236
2120 236
2125 236
2130 233
2135 235
2140 235
2145 235
2150 235
2155 236
2160 236
2165 235
2170 236
2175 236
2180 236
2185 235
2190 235
2195 236
2200 235
2205 236
2210 236
2215 235
2220 235
2225 236
2230 235
2235 236
2240 236
2245 236
2250 236
2255 236
2260 236
2265 235
2270 235
2275 236
2280 236
2285 236
2290 236
2295 235
2300 236
2305 235
2310 236
2315 236
2320 235
2325 235
2330 235
2335 235
2340 236
2345 234
2350 235
2355 236
2360 236
2365 234
2370 236
2375 235
2380 235
2385 235
2390 235
2395 235
2400 236
2405 236
2410 235
2415 235
2420 235
2425 236
2430 236
2435 235
2440 236
2445 235
2450 235
2455 235
2460 236
2465 235
2470 236
2475 235
2480 235
2485 235
2490 235
2495 235
2500 235
2505 235
2510 235
2515 234
2520 236
2525 236
2530 236
2535 235
2540 236
2545 236
2550 236
2555 236
2560 235
2565 236
2570 236
2575 235
2580 236
2585 235
2590 236
2595 236
2600 235
2605 236
2610 236
2615 236
2620 235
2625 236
2630 236
2635 235
2640 235
2645 236
2650 235
2655 236
2660 236
2665 236
2670 235
2675 235
2680 236
2685 236
2690 236
2695 236
2700 236
2705 236
2710 236
2715 235
2720 236
2725 235
2730 236
2735 236
2740 236
2745 236
2750 236
2755 236
2760 236
2765 235
2770 235
2775 235
2780 235
2785 236
2790 236
2795 235
2800 235
2805 236
2810 235
2815 235
2820 237
2825 234
2830 237
2835 236
2840 236
2845 236
2850 236
2855 236
2860 236
2865 235
2870 235
2875 236
2880 235
2885 235
2890 236
2895 235
2900 236
2905 236
2910 235
2915 235
2920 236
2925 236
2930 236
2935 236
2940 237
2945 236
2950 236
2955 236
2960 235
2965 233
2970 235
2975 235
2980 235
2985 235
2990 235
2995 236
3000 236
3005 236
3010 236
3015 236
3020 235
3025 235
3030 236
3035 235
3040 235
3045 235
3050 235
3055 233
3060 235
3065 236
3070 235
3075 236
3080 236
3085 235
3090 235
3095 236
3100 235
3105 236
3110 236
3115 236
3120 235
3125 235
3130 235
3135 236
3140 236
3145 236
3150 236
3155 236
3160 236
3165 236
3170 236
3175 235
3180 236
3185 235
3190 235
3195 236
3200 236
3205 236
3210 236
3215 235
3220 235
3225 236
3230 236
3235 235
3240 236
3245 236
3250 236
3255 236
3260 236
3265 236
3270 236
3275 236
3280 235
3285 235
3290 235
3295 236
3300 236
3305 236
3310 236
3315 235
3320 235
3325 235
3330 235
3335 235
3340 236
3345 236
3350 235
3355 236
3360 236
3365 236
3370 236
3375 235
3380 236
3385 236
3390 236
3395 235
3400 235
3405 235
3410 236
3415 236
3420 236
3425 235
3430 236
3435 235
3440 236
3445 236
3450 235
3455 236
3460 236
3465 236
3470 235
3475 236
3480 236
3485 236
3490 235
3495 235
3500 236
3505 236
3510 236
3515 235
3520 235
3525 236
3530 236
3535 236
3540 236
3545 236
3550 236
3555 236
3560 235
3565 236
3570 236
3575 236
3580 236
3585 236
3590 236
3595 236
3600 235
};
\addlegendentry{Baseline without FPGA}
\addplot [thick, color1, mark=triangle*, mark size=3, mark options={solid,fill=white,draw=red},  mark repeat={180}]
table{
0 219
5 220
10 220
15 220
20 220
25 219
30 220
35 219
40 219
45 219
50 220
55 220
60 219
65 220
70 220
75 219
80 219
85 219
90 220
95 219
100 219
105 220
110 221
115 219
120 220
125 219
130 220
135 219
140 220
145 221
150 219
155 220
160 220
165 220
170 219
175 220
180 220
185 219
190 219
195 220
200 220
205 220
210 220
215 220
220 219
225 220
230 220
235 219
240 219
245 220
250 220
255 220
260 219
265 219
270 220
275 219
280 219
285 220
290 219
295 220
300 219
305 219
310 220
315 219
320 220
325 219
330 219
335 219
340 220
345 220
350 219
355 220
360 219
365 219
370 219
375 220
380 220
385 219
390 220
395 220
400 220
405 220
410 220
415 220
420 220
425 219
430 220
435 220
440 220
445 220
450 220
455 220
460 219
465 219
470 220
475 219
480 219
485 219
490 219
495 219
500 220
505 219
510 220
515 219
520 219
525 220
530 219
535 220
540 219
545 220
550 220
555 219
560 219
565 219
570 221
575 219
580 219
585 219
590 219
595 220
600 219
605 219
610 220
615 219
620 219
625 219
630 220
635 219
640 219
645 220
650 219
655 219
660 220
665 219
670 219
675 219
680 219
685 219
690 219
695 220
700 219
705 220
710 220
715 219
720 229
725 217
730 219
735 220
740 219
745 220
750 220
755 220
760 220
765 220
770 219
775 219
780 220
785 220
790 220
795 220
800 219
805 219
810 219
815 220
820 219
825 219
830 219
835 219
840 219
845 219
850 219
855 220
860 219
865 220
870 219
875 220
880 220
885 219
890 220
895 220
900 219
905 219
910 220
915 220
920 219
925 220
930 219
935 220
940 220
945 219
950 220
955 219
960 220
965 219
970 218
975 219
980 220
985 220
990 219
995 220
1000 219
1005 219
1010 220
1015 219
1020 219
1025 219
1030 220
1035 220
1040 220
1045 219
1050 220
1055 219
1060 220
1065 220
1070 219
1075 220
1080 220
1085 220
1090 219
1095 219
1100 219
1105 220
1110 220
1115 219
1120 220
1125 219
1130 220
1135 219
1140 220
1145 219
1150 219
1155 220
1160 219
1165 220
1170 219
1175 219
1180 219
1185 219
1190 220
1195 220
1200 220
1205 219
1210 219
1215 219
1220 220
1225 220
1230 219
1235 220
1240 219
1245 219
1250 219
1255 220
1260 220
1265 219
1270 220
1275 219
1280 220
1285 219
1290 219
1295 219
1300 220
1305 219
1310 219
1315 220
1320 219
1325 220
1330 219
1335 219
1340 220
1345 219
1350 220
1355 219
1360 219
1365 220
1370 220
1375 220
1380 219
1385 220
1390 218
1395 220
1400 220
1405 219
1410 220
1415 219
1420 220
1425 219
1430 220
1435 220
1440 220
1445 219
1450 220
1455 219
1460 220
1465 219
1470 220
1475 219
1480 220
1485 219
1490 220
1495 219
1500 220
1505 220
1510 220
1515 219
1520 219
1525 219
1530 220
1535 219
1540 220
1545 219
1550 219
1555 220
1560 219
1565 220
1570 218
1575 220
1580 220
1585 220
1590 219
1595 219
1600 220
1605 219
1610 219
1615 220
1620 220
1625 219
1630 219
1635 220
1640 219
1645 220
1650 219
1655 219
1660 219
1665 220
1670 219
1675 217
1680 219
1685 220
1690 219
1695 219
1700 220
1705 220
1710 220
1715 219
1720 219
1725 219
1730 219
1735 219
1740 219
1745 219
1750 219
1755 219
1760 220
1765 219
1770 219
1775 219
1780 219
1785 219
1790 219
1795 220
1800 219
1805 220
1810 219
1815 219
1820 219
1825 220
1830 219
1835 220
1840 219
1845 219
1850 219
1855 219
1860 219
1865 219
1870 219
1875 219
1880 220
1885 220
1890 219
1895 220
1900 220
1905 219
1910 220
1915 219
1920 220
1925 219
1930 219
1935 220
1940 219
1945 219
1950 220
1955 219
1960 219
1965 219
1970 220
1975 220
1980 219
1985 219
1990 219
1995 219
2000 220
2005 219
2010 220
2015 220
2020 220
2025 219
2030 220
2035 220
2040 220
2045 219
2050 219
2055 220
2060 218
2065 219
2070 219
2075 220
2080 219
2085 219
2090 218
2095 220
2100 219
2105 219
2110 220
2115 220
2120 219
2125 220
2130 220
2135 220
2140 220
2145 219
2150 219
2155 219
2160 220
2165 219
2170 219
2175 219
2180 219
2185 219
2190 220
2195 220
2200 219
2205 219
2210 220
2215 219
2220 220
2225 219
2230 220
2235 219
2240 219
2245 219
2250 219
2255 219
2260 219
2265 219
2270 219
2275 218
2280 220
2285 219
2290 217
2295 219
2300 219
2305 219
2310 219
2315 220
2320 219
2325 219
2330 220
2335 219
2340 219
2345 219
2350 220
2355 220
2360 219
2365 219
2370 220
2375 219
2380 219
2385 220
2390 220
2395 219
2400 219
2405 220
2410 220
2415 219
2420 219
2425 219
2430 220
2435 219
2440 219
2445 220
2450 219
2455 219
2460 219
2465 220
2470 219
2475 220
2480 220
2485 220
2490 219
2495 219
2500 219
2505 219
2510 220
2515 219
2520 220
2525 219
2530 219
2535 219
2540 219
2545 220
2550 219
2555 220
2560 219
2565 219
2570 220
2575 220
2580 219
2585 219
2590 219
2595 219
2600 219
2605 217
2610 219
2615 219
2620 219
2625 219
2630 220
2635 219
2640 219
2645 219
2650 219
2655 219
2660 219
2665 219
2670 219
2675 219
2680 220
2685 219
2690 219
2695 219
2700 219
2705 219
2710 219
2715 217
2720 219
2725 219
2730 219
2735 220
2740 219
2745 219
2750 220
2755 219
2760 219
2765 220
2770 220
2775 220
2780 220
2785 220
2790 220
2795 218
2800 219
2805 220
2810 219
2815 219
2820 219
2825 219
2830 219
2835 220
2840 219
2845 219
2850 220
2855 219
2860 219
2865 220
2870 220
2875 219
2880 219
2885 219
2890 219
2895 219
2900 220
2905 220
2910 219
2915 219
2920 220
2925 220
2930 219
2935 219
2940 219
2945 219
2950 219
2955 219
2960 219
2965 219
2970 219
2975 219
2980 220
2985 219
2990 219
2995 218
3000 219
3005 219
3010 219
3015 220
3020 217
3025 219
3030 220
3035 219
3040 219
3045 219
3050 219
3055 219
3060 220
3065 219
3070 219
3075 219
3080 220
3085 220
3090 219
3095 219
3100 220
3105 220
3110 219
3115 219
3120 219
3125 219
3130 220
3135 220
3140 220
3145 219
3150 220
3155 219
3160 220
3165 220
3170 220
3175 220
3180 219
3185 220
3190 220
3195 219
3200 219
3205 218
3210 220
3215 219
3220 219
3225 218
3230 219
3235 220
3240 217
3245 216
3250 219
3255 218
3260 219
3265 219
3270 219
3275 218
3280 220
3285 220
3290 219
3295 219
3300 219
3305 219
3310 219
3315 219
3320 220
3325 220
3330 220
3335 219
3340 219
3345 220
3350 220
3355 220
3360 219
3365 219
3370 220
3375 218
3380 220
3385 219
3390 219
3395 219
3400 220
3405 219
3410 220
3415 219
3420 220
3425 220
3430 219
3435 219
3440 218
3445 220
3450 218
3455 220
3460 219
3465 220
3470 219
3475 219
3480 219
3485 219
3490 220
3495 220
3500 219
3505 219
3510 219
3515 219
3520 219
3525 219
3530 219
3535 230
3540 217
3545 219
3550 219
3555 219
3560 219
3565 219
3570 219
3575 219
3580 219
3585 219
3590 219
3595 219
3600 219
};
\addlegendentry{Baseline without FPGA and NIC}

\end{axis}

\end{tikzpicture}

%% file: results/energy_intel_components.tex
\begin{tikzpicture}[font=\Large]

\definecolor{color0}{rgb}{0.12156862745098,0.466666666666667,0.705882352941177}
\definecolor{color1}{rgb}{1,0.498039215686275,0.0549019607843137}
\definecolor{color2}{rgb}{0.172549019607843,0.627450980392157,0.172549019607843}
\definecolor{color3}{rgb}{0.83921568627451,0.152941176470588,0.156862745098039}
\definecolor{color4}{rgb}{0.580392156862745,0.403921568627451,0.741176470588235}
\definecolor{color5}{rgb}{0,0,0}

\begin{axis}[
legend cell align={left},
legend columns=1,
legend style={fill opacity=0.8, draw opacity=1, text opacity=1, at={(1.05,1.2)}, anchor=east, draw=white!80.0!black},
tick align=outside,
tick pos=left,
x grid style={white!69.01960784313725!black},
xlabel={Time (SEC)},
xmin=0, xmax=3600,
xtick={0,900,1800,2700,3600},
xtick style={color=black},
y grid style={white!69.01960784313725!black},
ylabel={Total energy consumption (kWh)},
ymin=0, ymax=0.3,
ytick style={color=black},
xmajorgrids,
ymajorgrids,
ytick={0.05,0.1,0.15,0.2,0.25,0.3},
yticklabel style={
        /pgf/number format/fixed,
        /pgf/number format/precision=2
},
scaled y ticks=false,
 y label style={at={(axis description cs:-0.04,.5)}}
]
\addplot [thick, color5]
table [y expr=(\thisrowno{1}-678611)/1000] {%
0 678611
5 678611
10 678611
15 678611
20 678612
25 678612
30 678613
35 678613
40 678613
45 678614
50 678614
55 678614
60 678615
65 678615
70 678615
75 678616
80 678616
85 678616
90 678617
95 678617
100 678617
105 678618
110 678618
115 678619
120 678619
125 678619
130 678620
135 678620
140 678620
145 678621
150 678621
155 678621
160 678622
165 678622
170 678622
175 678623
180 678623
185 678623
190 678624
195 678624
200 678625
205 678625
210 678625
215 678626
220 678626
225 678626
230 678627
235 678627
240 678627
245 678628
250 678628
255 678628
260 678629
265 678629
270 678629
275 678630
280 678630
285 678630
290 678631
295 678631
300 678632
305 678632
310 678632
315 678633
320 678633
325 678633
330 678634
335 678634
340 678634
345 678635
350 678635
355 678635
360 678636
365 678636
370 678636
375 678637
380 678637
385 678637
390 678638
395 678638
400 678639
405 678639
410 678639
415 678640
420 678640
425 678640
430 678641
435 678641
440 678641
445 678642
450 678642
455 678642
460 678643
465 678643
470 678643
475 678644
480 678644
485 678644
490 678645
495 678645
500 678645
505 678646
510 678646
515 678647
520 678647
525 678647
530 678648
535 678648
540 678648
545 678649
550 678649
555 678649
560 678650
565 678650
570 678650
575 678651
580 678651
585 678651
590 678652
595 678652
600 678653
605 678653
610 678653
615 678654
620 678654
625 678654
630 678655
635 678655
640 678655
645 678656
650 678656
655 678656
660 678657
665 678657
670 678657
675 678658
680 678658
685 678659
690 678659
695 678659
700 678660
705 678660
710 678660
715 678661
720 678661
725 678661
730 678662
735 678662
740 678662
745 678663
750 678663
755 678663
760 678664
765 678664
770 678664
775 678665
780 678665
785 678666
790 678666
795 678666
800 678667
805 678667
810 678667
815 678668
820 678668
825 678668
830 678669
835 678669
840 678669
845 678670
850 678670
855 678670
860 678671
865 678671
870 678671
875 678672
880 678672
885 678672
890 678673
895 678673
900 678674
905 678674
910 678674
915 678675
920 678675
925 678675
930 678676
935 678676
940 678676
945 678677
950 678677
955 678677
960 678678
965 678678
970 678678
975 678679
980 678679
985 678679
990 678680
995 678680
1000 678680
1005 678681
1010 678681
1015 678682
1020 678682
1025 678682
1030 678683
1035 678683
1040 678683
1045 678684
1050 678684
1055 678684
1060 678685
1065 678685
1070 678685
1075 678686
1080 678686
1085 678686
1090 678687
1095 678687
1100 678688
1105 678688
1110 678688
1115 678689
1120 678689
1125 678689
1130 678690
1135 678690
1140 678690
1145 678691
1150 678691
1155 678691
1160 678692
1165 678692
1170 678692
1175 678693
1180 678693
1185 678694
1190 678694
1195 678694
1200 678695
1205 678695
1210 678695
1215 678696
1220 678696
1225 678696
1230 678697
1235 678697
1240 678697
1245 678698
1250 678698
1255 678698
1260 678699
1265 678699
1270 678699
1275 678700
1280 678700
1285 678701
1290 678701
1295 678701
1300 678702
1305 678702
1310 678702
1315 678703
1320 678703
1325 678703
1330 678704
1335 678704
1340 678704
1345 678705
1350 678705
1355 678705
1360 678706
1365 678706
1370 678706
1375 678707
1380 678707
1385 678707
1390 678708
1395 678708
1400 678709
1405 678709
1410 678709
1415 678710
1420 678710
1425 678710
1430 678711
1435 678711
1440 678711
1445 678712
1450 678712
1455 678712
1460 678713
1465 678713
1470 678713
1475 678714
1480 678714
1485 678714
1490 678715
1495 678715
1500 678716
1505 678716
1510 678716
1515 678717
1520 678717
1525 678717
1530 678718
1535 678718
1540 678718
1545 678719
1550 678719
1555 678719
1560 678720
1565 678720
1570 678720
1575 678721
1580 678721
1585 678721
1590 678722
1595 678722
1600 678723
1605 678723
1610 678723
1615 678724
1620 678724
1625 678724
1630 678725
1635 678725
1640 678725
1645 678726
1650 678726
1655 678726
1660 678727
1665 678727
1670 678727
1675 678728
1680 678728
1685 678729
1690 678729
1695 678729
1700 678730
1705 678730
1710 678730
1715 678731
1720 678731
1725 678731
1730 678732
1735 678732
1740 678732
1745 678733
1750 678733
1755 678733
1760 678734
1765 678734
1770 678734
1775 678735
1780 678735
1785 678736
1790 678736
1795 678736
1800 678737
1805 678737
1810 678737
1815 678738
1820 678738
1825 678738
1830 678739
1835 678739
1840 678739
1845 678740
1850 678740
1855 678740
1860 678741
1865 678741
1870 678742
1875 678742
1880 678742
1885 678743
1890 678743
1895 678743
1900 678744
1905 678744
1910 678744
1915 678745
1920 678745
1925 678745
1930 678746
1935 678746
1940 678746
1945 678747
1950 678747
1955 678747
1960 678748
1965 678748
1970 678748
1975 678749
1980 678749
1985 678750
1990 678750
1995 678750
2000 678751
2005 678751
2010 678751
2015 678752
2020 678752
2025 678752
2030 678753
2035 678753
2040 678753
2045 678754
2050 678754
2055 678754
2060 678755
2065 678755
2070 678755
2075 678756
2080 678756
2085 678757
2090 678757
2095 678757
2100 678758
2105 678758
2110 678758
2115 678759
2120 678759
2125 678759
2130 678760
2135 678760
2140 678760
2145 678761
2150 678761
2155 678761
2160 678762
2165 678762
2170 678763
2175 678763
2180 678763
2185 678764
2190 678764
2195 678764
2200 678765
2205 678765
2210 678765
2215 678766
2220 678766
2225 678766
2230 678767
2235 678767
2240 678767
2245 678768
2250 678768
2255 678768
2260 678769
2265 678769
2270 678769
2275 678770
2280 678770
2285 678771
2290 678771
2295 678771
2300 678772
2305 678772
2310 678772
2315 678773
2320 678773
2325 678773
2330 678774
2335 678774
2340 678774
2345 678775
2350 678775
2355 678776
2360 678776
2365 678776
2370 678777
2375 678777
2380 678777
2385 678778
2390 678778
2395 678778
2400 678779
2405 678779
2410 678779
2415 678780
2420 678780
2425 678780
2430 678781
2435 678781
2440 678782
2445 678782
2450 678782
2455 678783
2460 678783
2465 678783
2470 678784
2475 678784
2480 678784
2485 678785
2490 678785
2495 678785
2500 678786
2505 678786
2510 678786
2515 678787
2520 678787
2525 678788
2530 678788
2535 678788
2540 678789
2545 678789
2550 678789
2555 678790
2560 678790
2565 678790
2570 678791
2575 678791
2580 678791
2585 678792
2590 678792
2595 678792
2600 678793
2605 678793
2610 678793
2615 678794
2620 678794
2625 678795
2630 678795
2635 678795
2640 678796
2645 678796
2650 678796
2655 678797
2660 678797
2665 678797
2670 678798
2675 678798
2680 678798
2685 678799
2690 678799
2695 678799
2700 678800
2705 678800
2710 678800
2715 678801
2720 678801
2725 678802
2730 678802
2735 678802
2740 678803
2745 678803
2750 678803
2755 678804
2760 678804
2765 678804
2770 678805
2775 678805
2780 678805
2785 678806
2790 678806
2795 678807
2800 678807
2805 678807
2810 678808
2815 678808
2820 678808
2825 678809
2830 678809
2835 678809
2840 678810
2845 678810
2850 678810
2855 678811
2860 678811
2865 678811
2870 678812
2875 678812
2880 678812
2885 678813
2890 678813
2895 678813
2900 678814
2905 678814
2910 678815
2915 678815
2920 678815
2925 678816
2930 678816
2935 678816
2940 678817
2945 678817
2950 678817
2955 678818
2960 678818
2965 678818
2970 678819
2975 678819
2980 678819
2985 678820
2990 678820
2995 678821
3000 678821
3005 678821
3010 678822
3015 678822
3020 678822
3025 678823
3030 678823
3035 678823
3040 678824
3045 678824
3050 678824
3055 678825
3060 678825
3065 678825
3070 678826
3075 678826
3080 678827
3085 678827
3090 678827
3095 678828
3100 678828
3105 678828
3110 678829
3115 678829
3120 678829
3125 678830
3130 678830
3135 678830
3140 678831
3145 678831
3150 678831
3155 678832
3160 678832
3165 678832
3170 678833
3175 678833
3180 678833
3185 678834
3190 678834
3195 678835
3200 678835
3205 678835
3210 678836
3215 678836
3220 678836
3225 678837
3230 678837
3235 678837
3240 678838
3245 678838
3250 678838
3255 678839
3260 678839
3265 678839
3270 678840
3275 678840
3280 678841
3285 678841
3290 678841
3295 678842
3300 678842
3305 678842
3310 678843
3315 678843
3320 678843
3325 678844
3330 678844
3335 678844
3340 678845
3345 678845
3350 678845
3355 678846
3360 678846
3365 678846
3370 678847
3375 678847
3380 678847
3385 678848
3390 678848
3395 678849
3400 678849
3405 678849
3410 678850
3415 678850
3420 678850
3425 678851
3430 678851
3435 678851
3440 678852
3445 678852
3450 678852
3455 678853
3460 678853
3465 678853
3470 678854
3475 678854
3480 678855
3485 678855
3490 678855
3495 678856
3500 678856
3505 678856
3510 678857
3515 678857
3520 678857
3525 678858
3530 678858
3535 678858
3540 678859
3545 678859
3550 678859
3555 678860
3560 678860
3565 678861
3570 678861
3575 678861
3580 678862
3585 678862
3590 678862
3595 678863
3600 678863
};
\addlegendentry{Baseline}
\addplot [thick, color0, mark=*, mark size=3, mark options={solid,fill=white,draw=red},  mark repeat={180}]
table [y expr=(\thisrowno{1}-729766)/1000] {%
0 729766
5 729766
10 729767
15 729767
20 729767
25 729768
30 729768
35 729768
40 729768
45 729769
50 729769
55 729769
60 729770
65 729770
70 729770
75 729771
80 729771
85 729771
90 729772
95 729772
100 729772
105 729773
110 729773
115 729773
120 729774
125 729774
130 729774
135 729775
140 729775
145 729775
150 729776
155 729776
160 729776
165 729777
170 729777
175 729777
180 729778
185 729778
190 729778
195 729779
200 729779
205 729779
210 729779
215 729780
220 729780
225 729780
230 729781
235 729781
240 729781
245 729782
250 729782
255 729782
260 729783
265 729783
270 729783
275 729784
280 729784
285 729784
290 729785
295 729785
300 729785
305 729786
310 729786
315 729786
320 729787
325 729787
330 729787
335 729788
340 729788
345 729788
350 729789
355 729789
360 729789
365 729790
370 729790
375 729790
380 729791
385 729791
390 729791
395 729792
400 729792
405 729792
410 729793
415 729793
420 729793
425 729794
430 729794
435 729794
440 729795
445 729795
450 729795
455 729796
460 729796
465 729796
470 729796
475 729797
480 729797
485 729797
490 729798
495 729798
500 729798
505 729799
510 729799
515 729799
520 729800
525 729800
530 729800
535 729801
540 729801
545 729801
550 729802
555 729802
560 729802
565 729803
570 729803
575 729803
580 729804
585 729804
590 729804
595 729805
600 729805
605 729805
610 729806
615 729806
620 729806
625 729807
630 729807
635 729807
640 729808
645 729808
650 729808
655 729809
660 729809
665 729809
670 729810
675 729810
680 729810
685 729810
690 729811
695 729811
700 729811
705 729812
710 729812
715 729812
720 729813
725 729813
730 729813
735 729814
740 729814
745 729814
750 729815
755 729815
760 729815
765 729816
770 729816
775 729816
780 729817
785 729817
790 729817
795 729818
800 729818
805 729818
810 729819
815 729819
820 729819
825 729820
830 729820
835 729820
840 729821
845 729821
850 729821
855 729822
860 729822
865 729822
870 729823
875 729823
880 729823
885 729823
890 729824
895 729824
900 729824
905 729825
910 729825
915 729825
920 729826
925 729826
930 729826
935 729827
940 729827
945 729828
950 729828
955 729828
960 729829
965 729829
970 729829
975 729829
980 729830
985 729830
990 729830
995 729831
1000 729831
1005 729831
1010 729832
1015 729832
1020 729832
1025 729833
1030 729833
1035 729833
1040 729834
1045 729834
1050 729834
1055 729835
1060 729835
1065 729835
1070 729836
1075 729836
1080 729836
1085 729837
1090 729837
1095 729837
1100 729838
1105 729838
1110 729838
1115 729839
1120 729839
1125 729839
1130 729840
1135 729840
1140 729840
1145 729840
1150 729841
1155 729841
1160 729841
1165 729842
1170 729842
1175 729842
1180 729843
1185 729843
1190 729843
1195 729844
1200 729844
1205 729844
1210 729845
1215 729845
1220 729845
1225 729846
1230 729846
1235 729846
1240 729847
1245 729847
1250 729847
1255 729848
1260 729848
1265 729848
1270 729849
1275 729849
1280 729849
1285 729850
1290 729850
1295 729850
1300 729851
1305 729851
1310 729851
1315 729852
1320 729852
1325 729852
1330 729853
1335 729853
1340 729853
1345 729853
1350 729854
1355 729854
1360 729854
1365 729855
1370 729855
1375 729855
1380 729856
1385 729856
1390 729856
1395 729857
1400 729857
1405 729857
1410 729858
1415 729858
1420 729858
1425 729859
1430 729859
1435 729859
1440 729860
1445 729860
1450 729860
1455 729861
1460 729861
1465 729861
1470 729862
1475 729862
1480 729862
1485 729863
1490 729863
1495 729863
1500 729864
1505 729864
1510 729864
1515 729865
1520 729865
1525 729865
1530 729865
1535 729866
1540 729866
1545 729867
1550 729867
1555 729867
1560 729868
1565 729868
1570 729868
1575 729869
1580 729869
1585 729869
1590 729870
1595 729870
1600 729870
1605 729871
1610 729871
1615 729871
1620 729871
1625 729872
1630 729872
1635 729872
1640 729873
1645 729873
1650 729873
1655 729874
1660 729874
1665 729874
1670 729875
1675 729875
1680 729875
1685 729876
1690 729876
1695 729876
1700 729877
1705 729877
1710 729877
1715 729878
1720 729878
1725 729878
1730 729879
1735 729879
1740 729879
1745 729880
1750 729880
1755 729880
1760 729881
1765 729881
1770 729881
1775 729882
1780 729882
1785 729882
1790 729883
1795 729883
1800 729883
1805 729884
1810 729884
1815 729884
1820 729884
1825 729885
1830 729885
1835 729885
1840 729886
1845 729886
1850 729886
1855 729887
1860 729887
1865 729887
1870 729888
1875 729888
1880 729888
1885 729889
1890 729889
1895 729889
1900 729890
1905 729890
1910 729890
1915 729891
1920 729891
1925 729891
1930 729892
1935 729892
1940 729892
1945 729893
1950 729893
1955 729893
1960 729894
1965 729894
1970 729894
1975 729895
1980 729895
1985 729895
1990 729896
1995 729896
2000 729896
2005 729896
2010 729897
2015 729897
2020 729897
2025 729898
2030 729898
2035 729898
2040 729899
2045 729899
2050 729899
2055 729900
2060 729900
2065 729900
2070 729901
2075 729901
2080 729901
2085 729902
2090 729902
2095 729902
2100 729903
2105 729903
2110 729903
2115 729904
2120 729904
2125 729904
2130 729905
2135 729905
2140 729905
2145 729906
2150 729906
2155 729906
2160 729907
2165 729907
2170 729907
2175 729908
2180 729908
2185 729908
2190 729909
2195 729909
2200 729909
2205 729910
2210 729910
2215 729910
2220 729911
2225 729911
2230 729911
2235 729912
2240 729912
2245 729912
2250 729913
2255 729913
2260 729913
2265 729913
2270 729914
2275 729914
2280 729914
2285 729915
2290 729915
2295 729915
2300 729916
2305 729916
2310 729916
2315 729917
2320 729917
2325 729917
2330 729918
2335 729918
2340 729918
2345 729919
2350 729919
2355 729919
2360 729920
2365 729920
2370 729920
2375 729921
2380 729921
2385 729921
2390 729922
2395 729922
2400 729922
2405 729923
2410 729923
2415 729923
2420 729924
2425 729924
2430 729924
2435 729925
2440 729925
2445 729925
2450 729925
2455 729926
2460 729926
2465 729926
2470 729927
2475 729927
2480 729927
2485 729928
2490 729928
2495 729928
2500 729929
2505 729929
2510 729929
2515 729930
2520 729930
2525 729930
2530 729931
2535 729931
2540 729931
2545 729932
2550 729932
2555 729932
2560 729933
2565 729933
2570 729933
2575 729934
2580 729934
2585 729934
2590 729935
2595 729935
2600 729935
2605 729936
2610 729936
2615 729936
2620 729937
2625 729937
2630 729937
2635 729937
2640 729938
2645 729938
2650 729938
2655 729939
2660 729939
2665 729939
2670 729940
2675 729940
2680 729940
2685 729941
2690 729941
2695 729941
2700 729942
2705 729942
2710 729942
2715 729943
2720 729943
2725 729943
2730 729944
2735 729944
2740 729944
2745 729945
2750 729945
2755 729945
2760 729946
2765 729946
2770 729946
2775 729947
2780 729947
2785 729947
2790 729948
2795 729948
2800 729948
2805 729949
2810 729949
2815 729949
2820 729950
2825 729950
2830 729950
2835 729951
2840 729951
2845 729951
2850 729952
2855 729952
2860 729952
2865 729953
2870 729953
2875 729953
2880 729954
2885 729954
2890 729954
2895 729955
2900 729955
2905 729955
2910 729955
2915 729956
2920 729956
2925 729956
2930 729957
2935 729957
2940 729957
2945 729958
2950 729958
2955 729958
2960 729959
2965 729959
2970 729959
2975 729960
2980 729960
2985 729960
2990 729961
2995 729961
3000 729961
3005 729962
3010 729962
3015 729962
3020 729963
3025 729963
3030 729963
3035 729964
3040 729964
3045 729964
3050 729965
3055 729965
3060 729965
3065 729965
3070 729966
3075 729966
3080 729966
3085 729967
3090 729967
3095 729967
3100 729968
3105 729968
3110 729968
3115 729969
3120 729969
3125 729969
3130 729970
3135 729970
3140 729970
3145 729971
3150 729971
3155 729971
3160 729972
3165 729972
3170 729972
3175 729973
3180 729973
3185 729973
3190 729974
3195 729974
3200 729974
3205 729975
3210 729975
3215 729975
3220 729976
3225 729976
3230 729976
3235 729977
3240 729977
3245 729977
3250 729978
3255 729978
3260 729978
3265 729979
3270 729979
3275 729979
3280 729980
3285 729980
3290 729980
3295 729981
3300 729981
3305 729981
3310 729981
3315 729982
3320 729982
3325 729982
3330 729983
3335 729983
3340 729984
3345 729984
3350 729984
3355 729985
3360 729985
3365 729985
3370 729985
3375 729986
3380 729986
3385 729986
3390 729987
3395 729987
3400 729987
3405 729988
3410 729988
3415 729988
3420 729989
3425 729989
3430 729989
3435 729990
3440 729990
3445 729990
3450 729991
3455 729991
3460 729991
3465 729992
3470 729992
3475 729992
3480 729993
3485 729993
3490 729993
3495 729994
3500 729994
3505 729994
3510 729995
3515 729995
3520 729995
3525 729996
3530 729996
3535 729996
3540 729997
3545 729997
3550 729997
3555 729998
3560 729998
3565 729998
3570 729999
3575 729999
3580 729999
3585 729999
3590 730000
3595 730000
3600 730000
};
\addlegendentry{Baseline without FPGA}
\addplot [thick, color1, mark=triangle*, mark size=3, mark options={solid,fill=white,draw=red},  mark repeat={180}]
table[y expr=(\thisrowno{1}-730333)/1000]{
0 730333
5 730333
10 730334
15 730334
20 730334
25 730335
30 730335
35 730335
40 730336
45 730336
50 730336
55 730336
60 730337
65 730337
70 730337
75 730338
80 730338
85 730338
90 730339
95 730339
100 730339
105 730339
110 730340
115 730340
120 730340
125 730341
130 730341
135 730341
140 730342
145 730342
150 730342
155 730343
160 730343
165 730343
170 730343
175 730344
180 730344
185 730344
190 730345
195 730345
200 730345
205 730346
210 730346
215 730346
220 730346
225 730347
230 730347
235 730347
240 730348
245 730348
250 730348
255 730349
260 730349
265 730349
270 730350
275 730350
280 730350
285 730350
290 730351
295 730351
300 730351
305 730352
310 730352
315 730352
320 730352
325 730353
330 730353
335 730353
340 730354
345 730354
350 730354
355 730355
360 730355
365 730355
370 730356
375 730356
380 730356
385 730356
390 730357
395 730357
400 730357
405 730358
410 730358
415 730358
420 730359
425 730359
430 730359
435 730359
440 730360
445 730360
450 730360
455 730360
460 730361
465 730361
470 730361
475 730362
480 730362
485 730362
490 730363
495 730363
500 730363
505 730363
510 730364
515 730364
520 730364
525 730365
530 730365
535 730365
540 730366
545 730366
550 730366
555 730366
560 730367
565 730367
570 730367
575 730368
580 730368
585 730368
590 730369
595 730369
600 730369
605 730369
610 730370
615 730370
620 730370
625 730371
630 730371
635 730371
640 730372
645 730372
650 730372
655 730373
660 730373
665 730373
670 730373
675 730374
680 730374
685 730374
690 730375
695 730375
700 730375
705 730376
710 730376
715 730376
720 730376
725 730377
730 730377
735 730377
740 730378
745 730378
750 730378
755 730379
760 730379
765 730379
770 730380
775 730380
780 730380
785 730380
790 730381
795 730381
800 730381
805 730382
810 730382
815 730382
820 730383
825 730383
830 730383
835 730383
840 730384
845 730384
850 730384
855 730385
860 730385
865 730385
870 730386
875 730386
880 730386
885 730386
890 730387
895 730387
900 730387
905 730388
910 730388
915 730388
920 730389
925 730389
930 730389
935 730390
940 730390
945 730390
950 730390
955 730391
960 730391
965 730391
970 730392
975 730392
980 730392
985 730393
990 730393
995 730393
1000 730393
1005 730394
1010 730394
1015 730394
1020 730395
1025 730395
1030 730395
1035 730396
1040 730396
1045 730396
1050 730396
1055 730396
1060 730397
1065 730397
1070 730397
1075 730398
1080 730398
1085 730398
1090 730399
1095 730399
1100 730399
1105 730400
1110 730400
1115 730400
1120 730400
1125 730401
1130 730401
1135 730401
1140 730402
1145 730402
1150 730402
1155 730403
1160 730403
1165 730403
1170 730403
1175 730404
1180 730404
1185 730404
1190 730405
1195 730405
1200 730405
1205 730406
1210 730406
1215 730406
1220 730407
1225 730407
1230 730407
1235 730407
1240 730408
1245 730408
1250 730408
1255 730409
1260 730409
1265 730409
1270 730410
1275 730410
1280 730410
1285 730410
1290 730411
1295 730411
1300 730411
1305 730412
1310 730412
1315 730412
1320 730413
1325 730413
1330 730413
1335 730413
1340 730414
1345 730414
1350 730414
1355 730415
1360 730415
1365 730415
1370 730416
1375 730416
1380 730416
1385 730417
1390 730417
1395 730417
1400 730417
1405 730418
1410 730418
1415 730418
1420 730419
1425 730419
1430 730419
1435 730419
1440 730420
1445 730420
1450 730420
1455 730421
1460 730421
1465 730421
1470 730422
1475 730422
1480 730422
1485 730423
1490 730423
1495 730423
1500 730424
1505 730424
1510 730424
1515 730424
1520 730425
1525 730425
1530 730425
1535 730426
1540 730426
1545 730426
1550 730426
1555 730427
1560 730427
1565 730427
1570 730428
1575 730428
1580 730428
1585 730429
1590 730429
1595 730429
1600 730430
1605 730430
1610 730430
1615 730430
1620 730431
1625 730431
1630 730431
1635 730432
1640 730432
1645 730432
1650 730432
1655 730433
1660 730433
1665 730433
1670 730433
1675 730434
1680 730434
1685 730434
1690 730435
1695 730435
1700 730435
1705 730436
1710 730436
1715 730436
1720 730437
1725 730437
1730 730437
1735 730437
1740 730438
1745 730438
1750 730438
1755 730439
1760 730439
1765 730439
1770 730439
1775 730440
1780 730440
1785 730440
1790 730441
1795 730441
1800 730441
1805 730442
1810 730442
1815 730442
1820 730443
1825 730443
1830 730443
1835 730443
1840 730444
1845 730444
1850 730444
1855 730445
1860 730445
1865 730445
1870 730446
1875 730446
1880 730446
1885 730446
1890 730447
1895 730447
1900 730447
1905 730448
1910 730448
1915 730448
1920 730449
1925 730449
1930 730449
1935 730450
1940 730450
1945 730450
1950 730450
1955 730451
1960 730451
1965 730451
1970 730452
1975 730452
1980 730452
1985 730453
1990 730453
1995 730453
2000 730453
2005 730454
2010 730454
2015 730454
2020 730455
2025 730455
2030 730455
2035 730456
2040 730456
2045 730456
2050 730456
2055 730457
2060 730457
2065 730457
2070 730458
2075 730458
2080 730458
2085 730459
2090 730459
2095 730459
2100 730459
2105 730460
2110 730460
2115 730460
2120 730461
2125 730461
2130 730461
2135 730462
2140 730462
2145 730462
2150 730462
2155 730463
2160 730463
2165 730463
2170 730464
2175 730464
2180 730464
2185 730465
2190 730465
2195 730465
2200 730466
2205 730466
2210 730466
2215 730466
2220 730467
2225 730467
2230 730467
2235 730468
2240 730468
2245 730468
2250 730468
2255 730469
2260 730469
2265 730469
2270 730470
2275 730470
2280 730470
2285 730470
2290 730471
2295 730471
2300 730471
2305 730472
2310 730472
2315 730472
2320 730473
2325 730473
2330 730473
2335 730473
2340 730474
2345 730474
2350 730474
2355 730475
2360 730475
2365 730475
2370 730476
2375 730476
2380 730476
2385 730476
2390 730477
2395 730477
2400 730477
2405 730478
2410 730478
2415 730478
2420 730479
2425 730479
2430 730479
2435 730480
2440 730480
2445 730480
2450 730480
2455 730481
2460 730481
2465 730481
2470 730482
2475 730482
2480 730482
2485 730483
2490 730483
2495 730483
2500 730483
2505 730484
2510 730484
2515 730484
2520 730485
2525 730485
2530 730485
2535 730486
2540 730486
2545 730486
2550 730486
2555 730487
2560 730487
2565 730487
2570 730488
2575 730488
2580 730488
2585 730489
2590 730489
2595 730489
2600 730490
2605 730490
2610 730490
2615 730490
2620 730491
2625 730491
2630 730491
2635 730492
2640 730492
2645 730492
2650 730493
2655 730493
2660 730493
2665 730493
2670 730494
2675 730494
2680 730494
2685 730495
2690 730495
2695 730495
2700 730496
2705 730496
2710 730496
2715 730496
2720 730497
2725 730497
2730 730497
2735 730498
2740 730498
2745 730498
2750 730499
2755 730499
2760 730499
2765 730500
2770 730500
2775 730500
2780 730500
2785 730501
2790 730501
2795 730501
2800 730502
2805 730502
2810 730502
2815 730503
2820 730503
2825 730503
2830 730503
2835 730504
2840 730504
2845 730504
2850 730504
2855 730505
2860 730505
2865 730505
2870 730506
2875 730506
2880 730506
2885 730507
2890 730507
2895 730507
2900 730507
2905 730508
2910 730508
2915 730508
2920 730509
2925 730509
2930 730509
2935 730510
2940 730510
2945 730510
2950 730510
2955 730511
2960 730511
2965 730511
2970 730512
2975 730512
2980 730512
2985 730513
2990 730513
2995 730513
3000 730513
3005 730514
3010 730514
3015 730514
3020 730515
3025 730515
3030 730515
3035 730516
3040 730516
3045 730516
3050 730517
3055 730517
3060 730517
3065 730517
3070 730518
3075 730518
3080 730518
3085 730519
3090 730519
3095 730519
3100 730519
3105 730520
3110 730520
3115 730520
3120 730521
3125 730521
3130 730521
3135 730522
3140 730522
3145 730522
3150 730523
3155 730523
3160 730523
3165 730524
3170 730524
3175 730524
3180 730524
3185 730525
3190 730525
3195 730525
3200 730526
3205 730526
3210 730526
3215 730526
3220 730527
3225 730527
3230 730527
3235 730528
3240 730528
3245 730528
3250 730529
3255 730529
3260 730529
3265 730530
3270 730530
3275 730530
3280 730530
3285 730531
3290 730531
3295 730531
3300 730532
3305 730532
3310 730532
3315 730533
3320 730533
3325 730533
3330 730533
3335 730534
3340 730534
3345 730534
3350 730535
3355 730535
3360 730535
3365 730536
3370 730536
3375 730536
3380 730536
3385 730537
3390 730537
3395 730537
3400 730538
3405 730538
3410 730538
3415 730539
3420 730539
3425 730539
3430 730539
3435 730540
3440 730540
3445 730540
3450 730541
3455 730541
3460 730541
3465 730541
3470 730542
3475 730542
3480 730542
3485 730543
3490 730543
3495 730543
3500 730544
3505 730544
3510 730544
3515 730544
3520 730545
3525 730545
3530 730545
3535 730546
3540 730546
3545 730546
3550 730547
3555 730547
3560 730547
3565 730548
3570 730548
3575 730548
3580 730548
3585 730549
3590 730549
3595 730549
3600 730550
};
\addlegendentry{Baseline without FPGA and NIC}

\end{axis}

\end{tikzpicture}

%% file: results/power_amd_components.tex
\begin{tikzpicture}[font=\Large]

\definecolor{color0}{rgb}{0.12156862745098,0.466666666666667,0.705882352941177}
\definecolor{color1}{rgb}{1,0.498039215686275,0.0549019607843137}
\definecolor{color2}{rgb}{0.172549019607843,0.627450980392157,0.172549019607843}
\definecolor{color3}{rgb}{0.83921568627451,0.152941176470588,0.156862745098039}
\definecolor{color4}{rgb}{0.580392156862745,0.403921568627451,0.741176470588235}
\definecolor{color5}{rgb}{0,0,0}

\begin{axis}[
legend cell align={left},
legend columns=1,
legend style={fill opacity=0.8, draw opacity=1, text opacity=1, at={(1.02,1.2)}, anchor=east, draw=white!80.0!black},
tick align=outside,
tick pos=left,
x grid style={white!69.01960784313725!black},
xlabel={Time (SEC)},
xmin=0, xmax=3600,
xtick style={color=black},
xtick={0,900,1800,2700,3600},
y grid style={white!69.01960784313725!black},
ylabel={Power consumption (W)},
ymin=100, ymax=300,
xmajorgrids,
ymajorgrids,
ytick={100, 150,200,250, 300},
ytick style={color=black}
]
\addplot [thick, color5]
table {%
0 205
5 205
10 205
15 205
20 205
25 205
30 205
35 205
40 206
45 206
50 205
55 205
60 205
65 206
70 205
75 205
80 205
85 205
90 205
95 206
100 205
105 205
110 205
115 205
120 205
125 205
130 205
135 205
140 206
145 205
150 205
155 205
160 205
165 205
170 205
175 206
180 205
185 206
190 205
195 206
200 205
205 205
210 206
215 205
220 206
225 206
230 205
235 205
240 205
245 205
250 205
255 206
260 206
265 205
270 205
275 205
280 205
285 206
290 205
295 204
300 206
305 205
310 205
315 205
320 206
325 205
330 205
335 206
340 205
345 205
350 206
355 206
360 206
365 205
370 205
375 205
380 206
385 205
390 205
395 205
400 205
405 205
410 205
415 205
420 205
425 206
430 205
435 205
440 205
445 205
450 205
455 205
460 205
465 205
470 205
475 205
480 205
485 206
490 205
495 205
500 205
505 205
510 205
515 205
520 206
525 206
530 205
535 205
540 205
545 205
550 205
555 205
560 206
565 206
570 206
575 205
580 206
585 206
590 205
595 205
600 205
605 206
610 205
615 206
620 206
625 205
630 205
635 206
640 206
645 206
650 205
655 206
660 205
665 205
670 205
675 206
680 205
685 205
690 206
695 205
700 206
705 205
710 205
715 205
720 206
725 205
730 205
735 204
740 206
745 206
750 205
755 205
760 206
765 206
770 206
775 205
780 205
785 206
790 205
795 205
800 205
805 206
810 205
815 206
820 205
825 205
830 205
835 205
840 206
845 205
850 206
855 206
860 206
865 206
870 206
875 205
880 206
885 205
890 205
895 206
900 205
905 206
910 206
915 204
920 205
925 205
930 205
935 206
940 205
945 205
950 205
955 206
960 206
965 205
970 206
975 206
980 205
985 205
990 205
995 205
1000 205
1005 205
1010 205
1015 206
1020 206
1025 206
1030 206
1035 205
1040 205
1045 205
1050 205
1055 205
1060 205
1065 206
1070 205
1075 205
1080 205
1085 206
1090 206
1095 206
1100 205
1105 205
1110 206
1115 205
1120 205
1125 205
1130 205
1135 205
1140 205
1145 205
1150 205
1155 205
1160 205
1165 206
1170 205
1175 206
1180 206
1185 205
1190 205
1195 205
1200 206
1205 205
1210 206
1215 205
1220 205
1225 206
1230 205
1235 206
1240 206
1245 205
1250 206
1255 206
1260 206
1265 205
1270 206
1275 206
1280 205
1285 204
1290 206
1295 204
1300 206
1305 206
1310 206
1315 206
1320 205
1325 205
1330 205
1335 205
1340 205
1345 205
1350 206
1355 205
1360 205
1365 205
1370 205
1375 206
1380 206
1385 205
1390 206
1395 206
1400 205
1405 206
1410 205
1415 205
1420 206
1425 205
1430 205
1435 205
1440 205
1445 206
1450 206
1455 206
1460 206
1465 205
1470 206
1475 206
1480 205
1485 205
1490 205
1495 205
1500 205
1505 205
1510 205
1515 205
1520 205
1525 206
1530 205
1535 205
1540 205
1545 205
1550 205
1555 204
1560 204
1565 204
1570 205
1575 205
1580 205
1585 205
1590 205
1595 205
1600 205
1605 205
1610 205
1615 205
1620 205
1625 205
1630 205
1635 205
1640 205
1645 205
1650 206
1655 205
1660 205
1665 205
1670 205
1675 205
1680 204
1685 205
1690 206
1695 205
1700 206
1705 205
1710 205
1715 206
1720 206
1725 206
1730 205
1735 206
1740 205
1745 206
1750 205
1755 205
1760 205
1765 205
1770 205
1775 205
1780 206
1785 205
1790 205
1795 205
1800 205
1805 205
1810 206
1815 205
1820 205
1825 206
1830 205
1835 205
1840 205
1845 205
1850 205
1855 206
1860 206
1865 206
1870 205
1875 205
1880 205
1885 205
1890 206
1895 205
1900 206
1905 206
1910 206
1915 205
1920 206
1925 206
1930 205
1935 205
1940 205
1945 206
1950 206
1955 205
1960 205
1965 205
1970 205
1975 205
1980 205
1985 205
1990 205
1995 205
2000 205
2005 205
2010 205
2015 205
2020 205
2025 205
2030 205
2035 206
2040 206
2045 205
2050 205
2055 205
2060 206
2065 205
2070 205
2075 205
2080 205
2085 205
2090 206
2095 205
2100 205
2105 205
2110 205
2115 205
2120 205
2125 205
2130 205
2135 205
2140 205
2145 206
2150 205
2155 205
2160 206
2165 206
2170 205
2175 206
2180 205
2185 206
2190 206
2195 206
2200 206
2205 205
2210 206
2215 205
2220 206
2225 206
2230 206
2235 205
2240 206
2245 206
2250 205
2255 205
2260 205
2265 205
2270 205
2275 205
2280 206
2285 206
2290 206
2295 205
2300 206
2305 205
2310 205
2315 205
2320 206
2325 206
2330 206
2335 205
2340 205
2345 205
2350 206
2355 205
2360 205
2365 205
2370 206
2375 206
2380 206
2385 205
2390 205
2395 206
2400 206
2405 205
2410 205
2415 205
2420 205
2425 205
2430 206
2435 205
2440 206
2445 206
2450 206
2455 206
2460 206
2465 205
2470 205
2475 205
2480 206
2485 206
2490 205
2495 206
2500 205
2505 206
2510 205
2515 206
2520 206
2525 206
2530 206
2535 206
2540 205
2545 206
2550 206
2555 206
2560 205
2565 206
2570 205
2575 206
2580 205
2585 205
2590 206
2595 205
2600 205
2605 206
2610 205
2615 205
2620 206
2625 205
2630 205
2635 204
2640 205
2645 205
2650 206
2655 205
2660 205
2665 206
2670 205
2675 205
2680 205
2685 205
2690 206
2695 205
2700 206
2705 205
2710 205
2715 206
2720 205
2725 206
2730 206
2735 206
2740 205
2745 207
2750 204
2755 204
2760 204
2765 205
2770 204
2775 204
2780 204
2785 203
2790 205
2795 204
2800 204
2805 204
2810 204
2815 204
2820 204
2825 204
2830 204
2835 204
2840 204
2845 204
2850 204
2855 204
2860 204
2865 204
2870 204
2875 204
2880 204
2885 204
2890 204
2895 204
2900 204
2905 204
2910 204
2915 204
2920 204
2925 204
2930 204
2935 205
2940 204
2945 203
2950 204
2955 205
2960 203
2965 204
2970 204
2975 205
2980 204
2985 205
2990 205
2995 205
3000 205
3005 205
3010 204
3015 204
3020 205
3025 205
3030 204
3035 204
3040 204
3045 204
3050 204
3055 204
3060 204
3065 204
3070 204
3075 204
3080 204
3085 205
3090 204
3095 204
3100 205
3105 205
3110 204
3115 204
3120 204
3125 204
3130 205
3135 205
3140 204
3145 205
3150 205
3155 204
3160 204
3165 204
3170 204
3175 204
3180 205
3185 205
3190 205
3195 204
3200 204
3205 204
3210 204
3215 204
3220 204
3225 204
3230 204
3235 204
3240 204
3245 204
3250 204
3255 204
3260 204
3265 205
3270 204
3275 204
3280 204
3285 204
3290 205
3295 204
3300 204
3305 205
3310 204
3315 204
3320 205
3325 204
3330 204
3335 204
3340 204
3345 204
3350 204
3355 204
3360 204
3365 204
3370 204
3375 204
3380 204
3385 204
3390 204
3395 204
3400 205
3405 203
3410 204
3415 204
3420 203
3425 205
3430 204
3435 204
3440 204
3445 204
3450 204
3455 204
3460 204
3465 204
3470 204
3475 204
3480 204
3485 205
3490 204
3495 204
3500 204
3505 204
3510 205
3515 205
3520 205
3525 204
3530 204
3535 204
3540 205
3545 205
3550 204
3555 205
3560 205
3565 205
3570 204
3575 204
3580 204
3585 204
3590 204
3595 204
3600 204
};

\addlegendentry{Baseline}
\addplot [thick, color0, mark=*, mark size=3, mark options={solid,fill=white,draw=red},  mark repeat={180}]
table{0 165
5 166
10 165
15 166
20 166
25 166
30 166
35 166
40 166
45 165
50 166
55 166
60 166
65 166
70 166
75 166
80 165
85 166
90 166
95 166
100 166
105 165
110 166
115 165
120 165
125 166
130 165
135 166
140 165
145 166
150 165
155 166
160 165
165 165
170 166
175 166
180 166
185 165
190 166
195 166
200 165
205 166
210 166
215 165
220 165
225 165
230 166
235 166
240 166
245 166
250 166
255 165
260 166
265 166
270 166
275 166
280 166
285 166
290 166
295 166
300 166
305 166
310 166
315 165
320 166
325 166
330 166
335 166
340 165
345 166
350 166
355 165
360 166
365 166
370 166
375 166
380 165
385 166
390 166
395 165
400 166
405 166
410 166
415 166
420 166
425 166
430 166
435 166
440 166
445 166
450 166
455 165
460 166
465 166
470 165
475 166
480 165
485 165
490 165
495 166
500 166
505 166
510 165
515 166
520 166
525 165
530 166
535 166
540 166
545 165
550 166
555 166
560 166
565 166
570 166
575 166
580 165
585 166
590 166
595 165
600 166
605 166
610 166
615 165
620 166
625 167
630 169
635 165
640 166
645 165
650 166
655 165
660 166
665 166
670 166
675 165
680 166
685 166
690 165
695 166
700 165
705 166
710 166
715 165
720 165
725 166
730 166
735 166
740 166
745 166
750 165
755 165
760 166
765 165
770 166
775 165
780 166
785 165
790 166
795 166
800 166
805 166
810 165
815 166
820 166
825 166
830 165
835 166
840 165
845 166
850 166
855 166
860 166
865 166
870 166
875 166
880 166
885 166
890 165
895 166
900 166
905 166
910 166
915 166
920 166
925 166
930 166
935 166
940 165
945 166
950 166
955 165
960 166
965 166
970 166
975 166
980 166
985 166
990 166
995 166
1000 166
1005 166
1010 166
1015 165
1020 166
1025 166
1030 166
1035 166
1040 167
1045 166
1050 166
1055 166
1060 166
1065 166
1070 166
1075 166
1080 166
1085 166
1090 166
1095 165
1100 166
1105 166
1110 165
1115 165
1120 166
1125 166
1130 166
1135 166
1140 166
1145 165
1150 166
1155 165
1160 166
1165 166
1170 166
1175 166
1180 166
1185 165
1190 165
1195 166
1200 165
1205 165
1210 166
1215 166
1220 166
1225 166
1230 165
1235 166
1240 166
1245 166
1250 166
1255 166
1260 166
1265 166
1270 166
1275 166
1280 166
1285 166
1290 166
1295 166
1300 166
1305 165
1310 166
1315 166
1320 165
1325 166
1330 166
1335 165
1340 166
1345 166
1350 166
1355 166
1360 165
1365 166
1370 165
1375 165
1380 165
1385 166
1390 165
1395 165
1400 166
1405 166
1410 165
1415 165
1420 165
1425 166
1430 166
1435 166
1440 166
1445 166
1450 166
1455 166
1460 166
1465 166
1470 166
1475 165
1480 166
1485 165
1490 165
1495 166
1500 166
1505 166
1510 166
1515 166
1520 166
1525 165
1530 166
1535 166
1540 166
1545 166
1550 165
1555 166
1560 166
1565 166
1570 166
1575 165
1580 166
1585 165
1590 165
1595 165
1600 166
1605 165
1610 166
1615 166
1620 166
1625 166
1630 165
1635 165
1640 166
1645 166
1650 166
1655 166
1660 166
1665 166
1670 166
1675 166
1680 166
1685 166
1690 166
1695 166
1700 166
1705 167
1710 166
1715 166
1720 166
1725 166
1730 166
1735 166
1740 166
1745 165
1750 166
1755 166
1760 166
1765 166
1770 166
1775 166
1780 166
1785 166
1790 165
1795 166
1800 166
1805 166
1810 166
1815 165
1820 166
1825 165
1830 165
1835 165
1840 166
1845 165
1850 166
1855 165
1860 166
1865 166
1870 166
1875 166
1880 166
1885 165
1890 166
1895 166
1900 166
1905 166
1910 166
1915 166
1920 165
1925 165
1930 166
1935 166
1940 166
1945 166
1950 166
1955 166
1960 165
1965 165
1970 165
1975 166
1980 166
1985 165
1990 165
1995 166
2000 166
2005 166
2010 166
2015 166
2020 166
2025 166
2030 166
2035 166
2040 165
2045 165
2050 166
2055 166
2060 166
2065 166
2070 166
2075 166
2080 165
2085 166
2090 165
2095 166
2100 166
2105 166
2110 166
2115 166
2120 166
2125 166
2130 166
2135 165
2140 166
2145 166
2150 166
2155 166
2160 166
2165 165
2170 166
2175 166
2180 166
2185 166
2190 166
2195 166
2200 166
2205 166
2210 166
2215 165
2220 165
2225 166
2230 166
2235 166
2240 166
2245 166
2250 166
2255 166
2260 166
2265 166
2270 165
2275 166
2280 166
2285 166
2290 166
2295 166
2300 166
2305 166
2310 166
2315 165
2320 166
2325 166
2330 166
2335 166
2340 166
2345 166
2350 166
2355 166
2360 166
2365 166
2370 166
2375 165
2380 166
2385 166
2390 166
2395 166
2400 166
2405 166
2410 166
2415 166
2420 166
2425 166
2430 166
2435 166
2440 166
2445 166
2450 165
2455 166
2460 166
2465 166
2470 166
2475 166
2480 166
2485 166
2490 166
2495 166
2500 166
2505 166
2510 166
2515 166
2520 166
2525 166
2530 166
2535 166
2540 166
2545 166
2550 166
2555 166
2560 166
2565 166
2570 166
2575 166
2580 166
2585 166
2590 166
2595 166
2600 166
2605 166
2610 166
2615 166
2620 166
2625 166
2630 166
2635 166
2640 166
2645 166
2650 166
2655 166
2660 166
2665 166
2670 166
2675 166
2680 166
2685 166
2690 166
2695 166
2700 166
2705 166
2710 166
2715 166
2720 167
2725 166
2730 166
2735 166
2740 166
2745 166
2750 166
2755 166
2760 166
2765 166
2770 166
2775 166
2780 166
2785 166
2790 166
2795 166
2800 165
2805 166
2810 166
2815 166
2820 166
2825 166
2830 166
2835 166
2840 167
2845 166
2850 166
2855 166
2860 165
2865 166
2870 166
2875 166
2880 166
2885 166
2890 166
2895 166
2900 167
2905 166
2910 166
2915 166
2920 166
2925 166
2930 166
2935 166
2940 166
2945 166
2950 166
2955 166
2960 166
2965 166
2970 166
2975 166
2980 166
2985 166
2990 166
2995 166
3000 166
3005 166
3010 166
3015 167
3020 166
3025 166
3030 166
3035 166
3040 166
3045 166
3050 166
3055 166
3060 166
3065 166
3070 166
3075 166
3080 166
3085 166
3090 166
3095 166
3100 166
3105 166
3110 166
3115 166
3120 166
3125 166
3130 166
3135 166
3140 166
3145 166
3150 166
3155 166
3160 166
3165 166
3170 166
3175 166
3180 166
3185 166
3190 166
3195 167
3200 167
3205 166
3210 166
3215 166
3220 166
3225 166
3230 166
3235 166
3240 166
3245 166
3250 166
3255 167
3260 166
3265 166
3270 166
3275 166
3280 166
3285 166
3290 166
3295 166
3300 165
3305 166
3310 166
3315 166
3320 166
3325 166
3330 166
3335 166
3340 166
3345 166
3350 166
3355 166
3360 166
3365 166
3370 165
3375 166
3380 166
3385 166
3390 166
3395 166
3400 166
3405 165
3410 166
3415 166
3420 166
3425 166
3430 166
3435 166
3440 166
3445 166
3450 166
3455 166
3460 166
3465 166
3470 166
3475 166
3480 166
3485 166
3490 166
3495 166
3500 166
3505 166
3510 165
3515 166
3520 165
3525 166
3530 166
3535 166
3540 166
3545 166
3550 166
3555 166
3560 167
3565 166
3570 166
3575 166
3580 166
3585 166
3590 166
3595 166
3600 166
};
\addlegendentry{Baseline without NIC}
\addplot [thick, color1, mark=triangle*, mark size=3, mark options={solid,fill=white,draw=red},  mark repeat={180}]
table{
0 108
5 108
10 109
15 108
20 108
25 108
30 108
35 108
40 108
45 108
50 108
55 108
60 108
65 107
70 108
75 108
80 107
85 108
90 108
95 108
100 109
105 108
110 108
115 108
120 108
125 108
130 107
135 108
140 108
145 107
150 108
155 108
160 108
165 107
170 108
175 108
180 108
185 107
190 108
195 108
200 108
205 108
210 108
215 108
220 108
225 108
230 109
235 108
240 108
245 108
250 108
255 108
260 109
265 108
270 108
275 108
280 108
285 108
290 109
295 108
300 108
305 108
310 108
315 108
320 108
325 108
330 108
335 107
340 106
345 108
350 108
355 108
360 108
365 108
370 108
375 107
380 107
385 108
390 108
395 107
400 108
405 108
410 108
415 108
420 110
425 108
430 108
435 108
440 108
445 108
450 108
455 108
460 108
465 108
470 108
475 108
480 108
485 108
490 108
495 108
500 108
505 108
510 108
515 107
520 108
525 108
530 108
535 108
540 108
545 108
550 108
555 107
560 108
565 107
570 108
575 108
580 107
585 107
590 108
595 108
600 108
605 108
610 108
615 106
620 108
625 108
630 107
635 108
640 108
645 108
650 109
655 108
660 110
665 108
670 107
675 108
680 108
685 108
690 108
695 108
700 108
705 108
710 108
715 108
720 110
725 108
730 108
735 108
740 108
745 108
750 108
755 108
760 108
765 108
770 108
775 108
780 108
785 108
790 108
795 108
800 108
805 108
810 108
815 108
820 108
825 108
830 108
835 108
840 108
845 109
850 108
855 108
860 108
865 108
870 108
875 108
880 108
885 108
890 108
895 108
900 108
905 108
910 108
915 108
920 108
925 108
930 108
935 108
940 108
945 108
950 108
955 108
960 108
965 108
970 108
975 108
980 108
985 108
990 108
995 108
1000 108
1005 108
1010 108
1015 108
1020 108
1025 108
1030 108
1035 108
1040 108
1045 108
1050 108
1055 107
1060 108
1065 108
1070 108
1075 108
1080 108
1085 108
1090 108
1095 108
1100 108
1105 108
1110 108
1115 108
1120 108
1125 108
1130 108
1135 108
1140 108
1145 108
1150 109
1155 108
1160 108
1165 108
1170 108
1175 108
1180 108
1185 108
1190 108
1195 108
1200 108
1205 108
1210 108
1215 108
1220 108
1225 108
1230 108
1235 111
1240 108
1245 108
1250 108
1255 108
1260 107
1265 108
1270 108
1275 107
1280 108
1285 107
1290 108
1295 108
1300 108
1305 108
1310 108
1315 108
1320 108
1325 108
1330 108
1335 107
1340 108
1345 108
1350 108
1355 112
1360 108
1365 108
1370 108
1375 108
1380 108
1385 108
1390 108
1395 108
1400 108
1405 108
1410 108
1415 108
1420 109
1425 108
1430 108
1435 108
1440 108
1445 108
1450 108
1455 108
1460 108
1465 108
1470 108
1475 123
1480 108
1485 108
1490 108
1495 108
1500 108
1505 108
1510 108
1515 107
1520 107
1525 108
1530 108
1535 108
1540 108
1545 108
1550 108
1555 108
1560 108
1565 108
1570 108
1575 108
1580 108
1585 107
1590 108
1595 108
1600 108
1605 108
1610 108
1615 108
1620 108
1625 108
1630 108
1635 108
1640 108
1645 108
1650 108
1655 108
1660 108
1665 108
1670 108
1675 108
1680 108
1685 108
1690 108
1695 108
1700 107
1705 108
1710 108
1715 108
1720 108
1725 108
1730 108
1735 108
1740 108
1745 108
1750 108
1755 108
1760 106
1765 108
1770 108
1775 108
1780 108
1785 108
1790 109
1795 108
1800 107
1805 107
1810 108
1815 108
1820 108
1825 108
1830 108
1835 108
1840 108
1845 108
1850 108
1855 108
1860 108
1865 108
1870 108
1875 108
1880 108
1885 109
1890 108
1895 108
1900 108
1905 108
1910 108
1915 108
1920 108
1925 110
1930 108
1935 108
1940 108
1945 108
1950 108
1955 109
1960 108
1965 110
1970 108
1975 108
1980 108
1985 107
1990 106
1995 107
2000 108
2005 108
2010 107
2015 107
2020 108
2025 108
2030 108
2035 108
2040 108
2045 107
2050 108
2055 108
2060 108
2065 107
2070 108
2075 108
2080 107
2085 108
2090 108
2095 108
2100 108
2105 108
2110 108
2115 108
2120 108
2125 108
2130 108
2135 108
2140 108
2145 108
2150 108
2155 108
2160 108
2165 108
2170 108
2175 108
2180 108
2185 108
2190 108
2195 108
2200 108
2205 108
2210 108
2215 108
2220 108
2225 108
2230 108
2235 108
2240 108
2245 108
2250 108
2255 107
2260 108
2265 108
2270 108
2275 108
2280 107
2285 108
2290 108
2295 108
2300 107
2305 108
2310 108
2315 108
2320 108
2325 108
2330 108
2335 108
2340 108
2345 108
2350 108
2355 108
2360 108
2365 108
2370 108
2375 108
2380 108
2385 108
2390 108
2395 108
2400 108
2405 108
2410 108
2415 108
2420 107
2425 108
2430 108
2435 108
2440 108
2445 108
2450 108
2455 108
2460 107
2465 108
2470 107
2475 108
2480 108
2485 107
2490 107
2495 107
2500 109
2505 110
2510 108
2515 108
2520 108
2525 108
2530 108
2535 108
2540 108
2545 108
2550 108
2555 108
2560 108
2565 108
2570 108
2575 108
2580 108
2585 108
2590 108
2595 109
2600 108
2605 108
2610 108
2615 108
2620 108
2625 108
2630 108
2635 108
2640 108
2645 108
2650 108
2655 108
2660 108
2665 108
2670 108
2675 108
2680 108
2685 108
2690 108
2695 108
2700 108
2705 108
2710 108
2715 108
2720 108
2725 108
2730 108
2735 108
2740 108
2745 108
2750 109
2755 108
2760 108
2765 108
2770 109
2775 108
2780 108
2785 108
2790 108
2795 108
2800 108
2805 108
2810 108
2815 109
2820 108
2825 108
2830 108
2835 109
2840 108
2845 108
2850 108
2855 110
2860 108
2865 108
2870 108
2875 108
2880 108
2885 108
2890 108
2895 108
2900 108
2905 108
2910 108
2915 108
2920 108
2925 108
2930 108
2935 108
2940 108
2945 108
2950 109
2955 108
2960 108
2965 108
2970 108
2975 108
2980 108
2985 108
2990 109
2995 108
3000 108
3005 108
3010 109
3015 108
3020 108
3025 108
3030 108
3035 108
3040 109
3045 108
3050 110
3055 108
3060 108
3065 108
3070 108
3075 108
3080 108
3085 108
3090 108
3095 108
3100 108
3105 108
3110 108
3115 108
3120 108
3125 108
3130 108
3135 108
3140 108
3145 108
3150 114
3155 108
3160 108
3165 108
3170 108
3175 108
3180 108
3185 108
3190 108
3195 108
3200 108
3205 108
3210 108
3215 109
3220 108
3225 108
3230 108
3235 108
3240 108
3245 108
3250 108
3255 108
3260 108
3265 108
3270 108
3275 108
3280 108
3285 109
3290 108
3295 108
3300 108
3305 108
3310 108
3315 108
3320 109
3325 108
3330 108
3335 108
3340 109
3345 108
3350 108
3355 108
3360 108
3365 108
3370 108
3375 108
3380 108
3385 108
3390 108
3395 108
3400 108
3405 108
3410 108
3415 108
3420 108
3425 108
3430 108
3435 107
3440 107
3445 109
3450 108
3455 107
3460 108
3465 108
3470 108
3475 108
3480 108
3485 109
3490 108
3495 108
3500 109
3505 108
3510 109
3515 108
3520 108
3525 108
3530 108
3535 108
3540 108
3545 108
3550 108
3555 108
3560 108
3565 108
3570 108
3575 108
3580 108
3585 108
3590 108
3595 108
3600 108

};
\addlegendentry{Baseline without DPU and NIC}

\end{axis}

\end{tikzpicture}

%% file: results/energy_amd_components.tex
\begin{tikzpicture}[font=\Large]

\definecolor{color0}{rgb}{0.12156862745098,0.466666666666667,0.705882352941177}
\definecolor{color1}{rgb}{1,0.498039215686275,0.0549019607843137}
\definecolor{color2}{rgb}{0.172549019607843,0.627450980392157,0.172549019607843}
\definecolor{color3}{rgb}{0.83921568627451,0.152941176470588,0.156862745098039}
\definecolor{color4}{rgb}{0.580392156862745,0.403921568627451,0.741176470588235}
\definecolor{color5}{rgb}{0,0,0}

\begin{axis}[
legend cell align={left},
legend columns=1,
legend style={fill opacity=0.8, draw opacity=1, text opacity=1, at={(1.05,1.2)}, anchor=east, draw=white!80.0!black},
tick align=outside,
tick pos=left,
x grid style={white!69.01960784313725!black},
xmajorgrids,
ymajorgrids,
xlabel={Time (SEC)},
xmin=0, xmax=3600,
xtick={0,900,1800,2700,3600},
xtick style={color=black},
y grid style={white!69.01960784313725!black},
ylabel={Total energy consumption (kWh)},
ymin=0, ymax=0.3,
ytick style={color=black},
ytick={0.05,0.1,0.15,0.2,0.25,0.3},
yticklabel style={
        /pgf/number format/fixed,
        /pgf/number format/precision=2
},
scaled y ticks=false,
 y label style={at={(axis description cs:-0.04,.5)}}
]
\addplot [thick, color5]
table [y expr=(\thisrowno{1}-263550)/1000] {%
0 263550
5 263551
10 263551
15 263551
20 263551
25 263551
30 263552
35 263552
40 263552
45 263553
50 263553
55 263553
60 263553
65 263554
70 263554
75 263554
80 263554
85 263555
90 263555
95 263556
100 263556
105 263556
110 263556
115 263556
120 263557
125 263558
130 263558
135 263558
140 263558
145 263558
150 263558
155 263559
160 263560
165 263560
170 263560
175 263560
180 263560
185 263560
190 263561
195 263561
200 263562
205 263562
210 263562
215 263562
220 263563
225 263563
230 263563
235 263563
240 263564
245 263564
250 263565
255 263565
260 263565
265 263565
270 263565
275 263565
280 263567
285 263567
290 263567
295 263567
300 263567
305 263567
310 263567
315 263568
320 263569
325 263569
330 263569
335 263569
340 263570
345 263570
350 263571
355 263571
360 263571
365 263571
370 263571
375 263571
380 263573
385 263573
390 263573
395 263573
400 263573
405 263573
410 263574
415 263575
420 263575
425 263575
430 263575
435 263575
440 263575
445 263576
450 263576
455 263577
460 263577
465 263577
470 263577
475 263578
480 263578
485 263578
490 263578
495 263579
500 263579
505 263580
510 263580
515 263580
520 263580
525 263580
530 263580
535 263582
540 263582
545 263582
550 263582
555 263582
560 263582
565 263583
570 263583
575 263584
580 263584
585 263584
590 263584
595 263584
600 263585
605 263585
610 263585
615 263586
620 263586
625 263586
630 263587
635 263587
640 263587
645 263587
650 263587
655 263588
660 263589
665 263589
670 263589
675 263589
680 263589
685 263589
690 263590
695 263591
700 263591
705 263591
710 263591
715 263591
720 263591
725 263592
730 263593
735 263593
740 263593
745 263593
750 263593
755 263594
760 263594
765 263594
770 263595
775 263595
780 263595
785 263596
790 263596
795 263596
800 263596
805 263596
810 263597
815 263598
820 263598
825 263598
830 263598
835 263598
840 263598
845 263598
850 263600
855 263600
860 263600
865 263600
870 263600
875 263600
880 263601
885 263601
890 263602
895 263602
900 263602
905 263602
910 263603
915 263603
920 263603
925 263603
930 263603
935 263603
940 263604
945 263604
950 263605
955 263605
960 263605
965 263605
970 263605
975 263606
980 263607
985 263607
990 263607
995 263607
1000 263607
1005 263607
1010 263608
1015 263609
1020 263609
1025 263609
1030 263609
1035 263609
1040 263610
1045 263610
1050 263610
1055 263611
1060 263611
1065 263611
1070 263611
1075 263612
1080 263612
1085 263612
1090 263612
1095 263613
1100 263613
1105 263614
1110 263614
1115 263614
1120 263614
1125 263614
1130 263614
1135 263616
1140 263616
1145 263616
1150 263616
1155 263616
1160 263616
1165 263617
1170 263617
1175 263618
1180 263618
1185 263618
1190 263618
1195 263618
1200 263619
1205 263619
1210 263620
1215 263620
1220 263620
1225 263620
1230 263621
1235 263621
1240 263621
1245 263621
1250 263622
1255 263622
1260 263623
1265 263623
1270 263623
1275 263623
1280 263623
1285 263623
1290 263625
1295 263625
1300 263625
1305 263625
1310 263625
1315 263625
1320 263626
1325 263626
1330 263627
1335 263627
1340 263627
1345 263627
1350 263627
1355 263628
1360 263628
1365 263628
1370 263629
1375 263629
1380 263629
1385 263630
1390 263630
1395 263630
1400 263630
1405 263630
1410 263631
1415 263632
1420 263632
1425 263632
1430 263632
1435 263632
1440 263632
1445 263633
1450 263634
1455 263634
1460 263634
1465 263634
1470 263634
1475 263634
1480 263635
1485 263635
1490 263636
1495 263636
1500 263636
1505 263636
1510 263637
1515 263637
1520 263637
1525 263637
1530 263637
1535 263638
1540 263638
1545 263639
1550 263639
1555 263639
1560 263639
1565 263639
1570 263639
1575 263641
1580 263641
1585 263641
1590 263641
1595 263641
1600 263641
1605 263641
1610 263642
1615 263643
1620 263643
1625 263643
1630 263643
1635 263643
1640 263644
1645 263644
1650 263644
1655 263645
1660 263645
1665 263645
1670 263646
1675 263646
1680 263646
1685 263646
1690 263646
1695 263647
1700 263648
1705 263648
1710 263648
1715 263648
1720 263648
1725 263648
1730 263649
1735 263650
1740 263650
1745 263650
1750 263650
1755 263650
1760 263650
1765 263651
1770 263652
1775 263652
1780 263652
1785 263652
1790 263652
1795 263653
1800 263653
1805 263653
1810 263654
1815 263654
1820 263654
1825 263655
1830 263655
1835 263655
1840 263655
1845 263655
1850 263656
1855 263657
1860 263657
1865 263657
1870 263657
1875 263657
1880 263657
1885 263657
1890 263659
1895 263659
1900 263659
1905 263659
1910 263659
1915 263659
1920 263660
1925 263660
1930 263661
1935 263661
1940 263661
1945 263661
1950 263662
1955 263662
1960 263662
1965 263662
1970 263663
1975 263663
1980 263664
1985 263664
1990 263664
1995 263664
2000 263664
2005 263664
2010 263666
2015 263666
2020 263666
2025 263666
2030 263666
2035 263666
2040 263666
2045 263667
2050 263668
2055 263668
2060 263668
2065 263668
2070 263668
2075 263669
2080 263669
2085 263669
2090 263670
2095 263670
2100 263670
2105 263671
2110 263671
2115 263671
2120 263671
2125 263671
2130 263671
2135 263672
2140 263672
2145 263673
2150 263673
2155 263673
2160 263673
2165 263673
2170 263673
2175 263675
2180 263675
2185 263675
2190 263675
2195 263675
2200 263675
2205 263676
2210 263676
2215 263677
2220 263677
2225 263677
2230 263677
2235 263678
2240 263678
2245 263678
2250 263679
2255 263679
2260 263679
2265 263679
2270 263680
2275 263680
2280 263680
2285 263680
2290 263681
2295 263681
2300 263682
2305 263682
2310 263682
2315 263682
2320 263682
2325 263682
2330 263684
2335 263684
2340 263684
2345 263684
2350 263684
2355 263684
2360 263685
2365 263685
2370 263686
2375 263686
2380 263686
2385 263686
2390 263686
2395 263687
2400 263687
2405 263687
2410 263688
2415 263688
2420 263688
2425 263689
2430 263689
2435 263689
2440 263689
2445 263689
2450 263690
2455 263691
2460 263691
2465 263691
2470 263691
2475 263691
2480 263691
2485 263692
2490 263693
2495 263693
2500 263693
2505 263693
2510 263693
2515 263693
2520 263694
2525 263695
2530 263695
2535 263695
2540 263695
2545 263695
2550 263696
2555 263696
2560 263696
2565 263697
2570 263697
2575 263697
2580 263698
2585 263698
2590 263698
2595 263698
2600 263698
2605 263699
2610 263700
2615 263700
2620 263700
2625 263700
2630 263700
2635 263700
2640 263700
2645 263702
2650 263702
2655 263702
2660 263702
2665 263702
2670 263702
2675 263703
2680 263703
2685 263704
2690 263704
2695 263704
2700 263704
2705 263705
2710 263705
2715 263705
2720 263705
2725 263705
2730 263706
2735 263706
2740 263707
2745 263707
2750 263707
2755 263707
2760 263707
2765 263707
2770 263708
2775 263709
2780 263709
2785 263709
2790 263709
2795 263709
2800 263709
2805 263710
2810 263711
2815 263711
2820 263711
2825 263711
2830 263711
2835 263712
2840 263712
2845 263712
2850 263713
2855 263713
2860 263713
2865 263714
2870 263714
2875 263714
2880 263714
2885 263714
2890 263715
2895 263715
2900 263716
2905 263716
2910 263716
2915 263716
2920 263716
2925 263716
2930 263718
2935 263718
2940 263718
2945 263718
2950 263718
2955 263718
2960 263719
2965 263719
2970 263720
2975 263720
2980 263720
2985 263720
2990 263721
2995 263721
3000 263721
3005 263721
3010 263722
3015 263722
3020 263722
3025 263723
3030 263723
3035 263723
3040 263723
3045 263723
3050 263724
3055 263725
3060 263725
3065 263725
3070 263725
3075 263725
3080 263725
3085 263726
3090 263727
3095 263727
3100 263727
3105 263727
3110 263727
3115 263728
3120 263728
3125 263728
3130 263729
3135 263729
3140 263729
3145 263729
3150 263730
3155 263730
3160 263730
3165 263730
3170 263731
3175 263731
3180 263732
3185 263732
3190 263732
3195 263732
3200 263732
3205 263732
3210 263734
3215 263734
3220 263734
3225 263734
3230 263734
3235 263734
3240 263735
3245 263736
3250 263736
3255 263736
3260 263736
3265 263736
3270 263736
3275 263737
3280 263737
3285 263738
3290 263738
3295 263738
3300 263738
3305 263739
3310 263739
3315 263739
3320 263739
3325 263739
3330 263740
3335 263740
3340 263741
3345 263741
3350 263741
3355 263741
3360 263741
3365 263741
3370 263743
3375 263743
3380 263743
3385 263743
3390 263743
3395 263743
3400 263743
3405 263744
3410 263745
3415 263745
3420 263745
3425 263745
3430 263745
3435 263746
3440 263746
3445 263746
3450 263747
3455 263747
3460 263747
3465 263748
3470 263748
3475 263748
3480 263748
3485 263748
3490 263749
3495 263750
3500 263750
3505 263750
3510 263750
3515 263750
3520 263750
3525 263750
3530 263752
3535 263752
3540 263752
3545 263752
3550 263752
3555 263752
3560 263753
3565 263753
3570 263754
3575 263754
3580 263754
3585 263754
3590 263755
3595 263755
3600 263755
};
\addlegendentry{Baseline}
\addplot [thick, color0, mark=*, mark size=3, mark options={solid,fill=white,draw=red},  mark repeat={180}]
table [y expr=(\thisrowno{1}-497737)/1000] {%
0 497737
5 497737
10 497738
15 497738
20 497738
25 497738
30 497738
35 497739
40 497739
45 497740
50 497740
55 497740
60 497740
65 497740
70 497740
75 497740
80 497741
85 497742
90 497742
95 497742
100 497742
105 497742
110 497742
115 497743
120 497743
125 497743
130 497743
135 497743
140 497744
145 497744
150 497745
155 497745
160 497745
165 497745
170 497745
175 497745
180 497745
185 497746
190 497746
195 497747
200 497747
205 497747
210 497747
215 497747
220 497747
225 497748
230 497748
235 497748
240 497748
245 497749
250 497749
255 497749
260 497750
265 497750
270 497750
275 497750
280 497750
285 497750
290 497750
295 497751
300 497751
305 497752
310 497752
315 497752
320 497753
325 497753
330 497753
335 497753
340 497753
345 497753
350 497753
355 497754
360 497755
365 497755
370 497755
375 497755
380 497755
385 497755
390 497756
395 497756
400 497756
405 497756
410 497756
415 497757
420 497757
425 497758
430 497758
435 497758
440 497758
445 497758
450 497758
455 497758
460 497759
465 497760
470 497760
475 497760
480 497760
485 497760
490 497760
495 497760
500 497761
505 497761
510 497761
515 497761
520 497762
525 497762
530 497762
535 497763
540 497763
545 497763
550 497763
555 497763
560 497763
565 497763
570 497764
575 497765
580 497765
585 497765
590 497765
595 497765
600 497765
605 497766
610 497766
615 497766
620 497766
625 497766
630 497767
635 497767
640 497768
645 497768
650 497768
655 497768
660 497768
665 497768
670 497768
675 497768
680 497769
685 497770
690 497770
695 497770
700 497770
705 497770
710 497770
715 497771
720 497771
725 497771
730 497771
735 497772
740 497772
745 497772
750 497773
755 497773
760 497773
765 497773
770 497773
775 497773
780 497773
785 497774
790 497775
795 497775
800 497775
805 497775
810 497775
815 497775
820 497776
825 497776
830 497776
835 497776
840 497776
845 497777
850 497777
855 497778
860 497778
865 497778
870 497778
875 497778
880 497778
885 497778
890 497778
895 497779
900 497779
905 497780
910 497780
915 497781
920 497781
925 497781
930 497781
935 497781
940 497781
945 497781
950 497781
955 497782
960 497783
965 497783
970 497783
975 497783
980 497783
985 497783
990 497784
995 497784
1000 497784
1005 497784
1010 497784
1015 497785
1020 497785
1025 497786
1030 497786
1035 497786
1040 497786
1045 497786
1050 497786
1055 497786
1060 497787
1065 497788
1070 497788
1075 497788
1080 497788
1085 497788
1090 497788
1095 497789
1100 497789
1105 497789
1110 497789
1115 497789
1120 497790
1125 497790
1130 497790
1135 497791
1140 497791
1145 497791
1150 497791
1155 497791
1160 497791
1165 497791
1170 497792
1175 497793
1180 497793
1185 497793
1190 497793
1195 497793
1200 497793
1205 497794
1210 497794
1215 497794
1220 497794
1225 497794
1230 497795
1235 497795
1240 497796
1245 497796
1250 497796
1255 497796
1260 497796
1265 497796
1270 497796
1275 497797
1280 497798
1285 497798
1290 497798
1295 497798
1300 497798
1305 497798
1310 497799
1315 497799
1320 497799
1325 497799
1330 497799
1335 497800
1340 497800
1345 497800
1350 497801
1355 497801
1360 497801
1365 497801
1370 497801
1375 497801
1380 497801
1385 497802
1390 497803
1395 497803
1400 497803
1405 497803
1410 497803
1415 497803
1420 497804
1425 497804
1430 497804
1435 497804
1440 497804
1445 497805
1450 497805
1455 497806
1460 497806
1465 497806
1470 497806
1475 497806
1480 497806
1485 497806
1490 497807
1495 497807
1500 497807
1505 497808
1510 497808
1515 497809
1520 497809
1525 497809
1530 497809
1535 497809
1540 497809
1545 497809
1550 497810
1555 497810
1560 497811
1565 497811
1570 497811
1575 497811
1580 497811
1585 497811
1590 497812
1595 497812
1600 497812
1605 497812
1610 497812
1615 497813
1620 497813
1625 497814
1630 497814
1635 497814
1640 497814
1645 497814
1650 497814
1655 497814
1660 497815
1665 497816
1670 497816
1675 497816
1680 497816
1685 497816
1690 497816
1695 497817
1700 497817
1705 497817
1710 497817
1715 497817
1720 497818
1725 497818
1730 497819
1735 497819
1740 497819
1745 497819
1750 497819
1755 497819
1760 497819
1765 497820
1770 497820
1775 497821
1780 497821
1785 497821
1790 497821
1795 497821
1800 497821
1805 497822
1810 497822
1815 497822
1820 497822
1825 497822
1830 497823
1835 497823
1840 497824
1845 497824
1850 497824
1855 497824
1860 497824
1865 497824
1870 497824
1875 497825
1880 497826
1885 497826
1890 497826
1895 497826
1900 497826
1905 497826
1910 497827
1915 497827
1920 497827
1925 497827
1930 497827
1935 497828
1940 497828
1945 497829
1950 497829
1955 497829
1960 497829
1965 497829
1970 497829
1975 497829
1980 497830
1985 497830
1990 497831
1995 497831
2000 497831
2005 497831
2010 497831
2015 497831
2020 497832
2025 497832
2030 497832
2035 497832
2040 497832
2045 497833
2050 497833
2055 497834
2060 497834
2065 497834
2070 497834
2075 497834
2080 497834
2085 497834
2090 497835
2095 497835
2100 497836
2105 497836
2110 497836
2115 497837
2120 497837
2125 497837
2130 497837
2135 497837
2140 497837
2145 497837
2150 497838
2155 497839
2160 497839
2165 497839
2170 497839
2175 497839
2180 497839
2185 497840
2190 497840
2195 497840
2200 497840
2205 497840
2210 497841
2215 497841
2220 497842
2225 497842
2230 497842
2235 497842
2240 497842
2245 497842
2250 497842
2255 497842
2260 497843
2265 497844
2270 497844
2275 497844
2280 497844
2285 497844
2290 497844
2295 497845
2300 497845
2305 497845
2310 497845
2315 497845
2320 497846
2325 497846
2330 497847
2335 497847
2340 497847
2345 497847
2350 497847
2355 497847
2360 497847
2365 497848
2370 497849
2375 497849
2380 497849
2385 497849
2390 497849
2395 497849
2400 497850
2405 497850
2410 497850
2415 497850
2420 497850
2425 497851
2430 497851
2435 497852
2440 497852
2445 497852
2450 497852
2455 497852
2460 497852
2465 497852
2470 497853
2475 497853
2480 497854
2485 497854
2490 497854
2495 497854
2500 497854
2505 497854
2510 497855
2515 497855
2520 497855
2525 497855
2530 497855
2535 497856
2540 497856
2545 497857
2550 497857
2555 497857
2560 497857
2565 497857
2570 497857
2575 497857
2580 497858
2585 497858
2590 497859
2595 497859
2600 497859
2605 497859
2610 497859
2615 497860
2620 497860
2625 497860
2630 497860
2635 497860
2640 497861
2645 497861
2650 497862
2655 497862
2660 497862
2665 497862
2670 497862
2675 497862
2680 497862
2685 497862
2690 497863
2695 497863
2700 497864
2705 497864
2710 497864
2715 497865
2720 497865
2725 497865
2730 497865
2735 497865
2740 497865
2745 497865
2750 497866
2755 497867
2760 497867
2765 497867
2770 497867
2775 497867
2780 497867
2785 497868
2790 497868
2795 497868
2800 497868
2805 497868
2810 497869
2815 497869
2820 497870
2825 497870
2830 497870
2835 497870
2840 497870
2845 497870
2850 497870
2855 497871
2860 497871
2865 497872
2870 497872
2875 497872
2880 497872
2885 497872
2890 497872
2895 497873
2900 497873
2905 497873
2910 497873
2915 497873
2920 497874
2925 497874
2930 497875
2935 497875
2940 497875
2945 497875
2950 497875
2955 497875
2960 497875
2965 497876
2970 497877
2975 497877
2980 497877
2985 497877
2990 497877
2995 497877
3000 497878
3005 497878
3010 497878
3015 497878
3020 497878
3025 497879
3030 497879
3035 497880
3040 497880
3045 497880
3050 497880
3055 497880
3060 497880
3065 497880
3070 497881
3075 497881
3080 497882
3085 497882
3090 497882
3095 497882
3100 497882
3105 497882
3110 497883
3115 497883
3120 497883
3125 497883
3130 497883
3135 497884
3140 497884
3145 497885
3150 497885
3155 497885
3160 497885
3165 497885
3170 497885
3175 497885
3180 497886
3185 497887
3190 497887
3195 497887
3200 497887
3205 497887
3210 497887
3215 497888
3220 497888
3225 497888
3230 497888
3235 497888
3240 497889
3245 497889
3250 497890
3255 497890
3260 497890
3265 497890
3270 497890
3275 497890
3280 497890
3285 497891
3290 497891
3295 497891
3300 497892
3305 497892
3310 497893
3315 497893
3320 497893
3325 497893
3330 497893
3335 497893
3340 497893
3345 497894
3350 497894
3355 497895
3360 497895
3365 497895
3370 497895
3375 497895
3380 497895
3385 497896
3390 497896
3395 497896
3400 497896
3405 497897
3410 497897
3415 497897
3420 497898
3425 497898
3430 497898
3435 497898
3440 497898
3445 497898
3450 497898
3455 497899
3460 497900
3465 497900
3470 497900
3475 497900
3480 497900
3485 497900
3490 497901
3495 497901
3500 497901
3505 497901
3510 497901
3515 497902
3520 497902
3525 497903
3530 497903
3535 497903
3540 497903
3545 497903
3550 497903
3555 497903
3560 497904
3565 497904
3570 497905
3575 497905
3580 497905
3585 497905
3590 497905
3595 497905
3600 497906
};
\addlegendentry{Baseline without NIC}
\addplot [thick, color1, mark=triangle*, mark size=3, mark options={solid,fill=white,draw=red},  mark repeat={180}]
table [y expr=(\thisrowno{1}-498123)/1000] {
0 498123
5 498123
10 498123
15 498123
20 498123
25 498123
30 498123
35 498124
40 498124
45 498124
50 498124
55 498124
60 498124
65 498124
70 498125
75 498125
80 498125
85 498126
90 498126
95 498126
100 498126
105 498126
110 498126
115 498126
120 498126
125 498126
130 498126
135 498127
140 498127
145 498127
150 498127
155 498127
160 498128
165 498128
170 498128
175 498128
180 498128
185 498129
190 498129
195 498129
200 498129
205 498129
210 498129
215 498129
220 498129
225 498129
230 498129
235 498129
240 498130
245 498130
250 498130
255 498131
260 498131
265 498131
270 498131
275 498131
280 498131
285 498131
290 498132
295 498132
300 498132
305 498132
310 498132
315 498132
320 498132
325 498132
330 498132
335 498132
340 498133
345 498133
350 498134
355 498134
360 498134
365 498134
370 498134
375 498134
380 498134
385 498134
390 498135
395 498135
400 498135
405 498135
410 498135
415 498135
420 498135
425 498135
430 498135
435 498135
440 498135
445 498136
450 498136
455 498136
460 498136
465 498136
470 498136
475 498137
480 498138
485 498138
490 498138
495 498138
500 498138
505 498138
510 498138
515 498138
520 498138
525 498138
530 498139
535 498139
540 498139
545 498139
550 498139
555 498139
560 498139
565 498139
570 498140
575 498140
580 498141
585 498141
590 498141
595 498141
600 498141
605 498141
610 498141
615 498141
620 498141
625 498141
630 498141
635 498142
640 498142
645 498142
650 498142
655 498142
660 498142
665 498143
670 498143
675 498143
680 498143
685 498144
690 498144
695 498144
700 498144
705 498144
710 498144
715 498144
720 498144
725 498144
730 498144
735 498145
740 498145
745 498145
750 498145
755 498145
760 498146
765 498146
770 498146
775 498146
780 498146
785 498147
790 498147
795 498147
800 498147
805 498147
810 498147
815 498147
820 498147
825 498147
830 498147
835 498147
840 498148
845 498148
850 498148
855 498149
860 498149
865 498149
870 498149
875 498149
880 498149
885 498149
890 498150
895 498150
900 498150
905 498150
910 498150
915 498150
920 498150
925 498150
930 498150
935 498150
940 498151
945 498152
950 498152
955 498152
960 498152
965 498152
970 498152
975 498152
980 498152
985 498152
990 498153
995 498153
1000 498153
1005 498153
1010 498153
1015 498153
1020 498153
1025 498153
1030 498153
1035 498153
1040 498154
1045 498154
1050 498154
1055 498154
1060 498154
1065 498154
1070 498154
1075 498155
1080 498156
1085 498156
1090 498156
1095 498156
1100 498156
1105 498156
1110 498156
1115 498156
1120 498156
1125 498156
1130 498157
1135 498157
1140 498157
1145 498157
1150 498157
1155 498157
1160 498157
1165 498157
1170 498158
1175 498158
1180 498159
1185 498159
1190 498159
1195 498159
1200 498159
1205 498159
1210 498159
1215 498159
1220 498159
1225 498159
1230 498160
1235 498160
1240 498160
1245 498160
1250 498160
1255 498160
1260 498160
1265 498161
1270 498161
1275 498161
1280 498162
1285 498162
1290 498162
1295 498162
1300 498162
1305 498162
1310 498162
1315 498162
1320 498162
1325 498162
1330 498162
1335 498163
1340 498163
1345 498163
1350 498163
1355 498163
1360 498164
1365 498164
1370 498164
1375 498164
1380 498164
1385 498165
1390 498165
1395 498165
1400 498165
1405 498165
1410 498165
1415 498165
1420 498165
1425 498165
1430 498165
1435 498166
1440 498166
1445 498166
1450 498167
1455 498167
1460 498167
1465 498167
1470 498167
1475 498167
1480 498167
1485 498168
1490 498168
1495 498168
1500 498168
1505 498168
1510 498168
1515 498168
1520 498168
1525 498168
1530 498168
1535 498169
1540 498169
1545 498170
1550 498170
1555 498170
1560 498170
1565 498170
1570 498170
1575 498170
1580 498170
1585 498170
1590 498171
1595 498171
1600 498171
1605 498171
1610 498171
1615 498171
1620 498171
1625 498171
1630 498171
1635 498171
1640 498172
1645 498172
1650 498172
1655 498172
1660 498172
1665 498172
1670 498172
1675 498174
1680 498174
1685 498174
1690 498174
1695 498174
1700 498174
1705 498174
1710 498174
1715 498174
1720 498174
1725 498174
1730 498175
1735 498175
1740 498175
1745 498175
1750 498175
1755 498175
1760 498175
1765 498175
1770 498176
1775 498176
1780 498177
1785 498177
1790 498177
1795 498177
1800 498177
1805 498177
1810 498177
1815 498177
1820 498177
1825 498177
1830 498178
1835 498178
1840 498178
1845 498178
1850 498178
1855 498178
1860 498179
1865 498179
1870 498179
1875 498179
1880 498180
1885 498180
1890 498180
1895 498180
1900 498180
1905 498180
1910 498180
1915 498180
1920 498180
1925 498180
1930 498180
1935 498181
1940 498181
1945 498181
1950 498181
1955 498182
1960 498182
1965 498182
1970 498182
1975 498182
1980 498182
1985 498183
1990 498183
1995 498183
2000 498183
2005 498183
2010 498183
2015 498183
2020 498183
2025 498183
2030 498183
2035 498184
2040 498184
2045 498184
2050 498185
2055 498185
2060 498185
2065 498185
2070 498185
2075 498185
2080 498185
2085 498186
2090 498186
2095 498186
2100 498186
2105 498186
2110 498186
2115 498186
2120 498186
2125 498186
2130 498186
2135 498187
2140 498187
2145 498188
2150 498188
2155 498188
2160 498188
2165 498188
2170 498188
2175 498188
2180 498188
2185 498188
2190 498189
2195 498189
2200 498189
2205 498189
2210 498189
2215 498189
2220 498189
2225 498189
2230 498189
2235 498189
2240 498190
2245 498190
2250 498190
2255 498190
2260 498190
2265 498190
2270 498191
2275 498192
2280 498192
2285 498192
2290 498192
2295 498192
2300 498192
2305 498192
2310 498192
2315 498192
2320 498192
2325 498193
2330 498193
2335 498193
2340 498193
2345 498193
2350 498193
2355 498193
2360 498193
2365 498194
2370 498194
2375 498194
2380 498195
2385 498195
2390 498195
2395 498195
2400 498195
2405 498195
2410 498195
2415 498195
2420 498195
2425 498195
2430 498196
2435 498196
2440 498196
2445 498196
2450 498196
2455 498196
2460 498197
2465 498197
2470 498197
2475 498197
2480 498198
2485 498198
2490 498198
2495 498198
2500 498198
2505 498198
2510 498198
2515 498198
2520 498198
2525 498198
2530 498199
2535 498199
2540 498199
2545 498199
2550 498199
2555 498200
2560 498200
2565 498200
2570 498200
2575 498200
2580 498200
2585 498201
2590 498201
2595 498201
2600 498201
2605 498201
2610 498201
2615 498201
2620 498201
2625 498201
2630 498201
2635 498202
2640 498202
2645 498203
2650 498203
2655 498203
2660 498203
2665 498203
2670 498203
2675 498203
2680 498203
2685 498204
2690 498204
2695 498204
2700 498204
2705 498204
2710 498204
2715 498204
2720 498204
2725 498204
2730 498204
2735 498205
2740 498206
2745 498206
2750 498206
2755 498206
2760 498206
2765 498206
2770 498206
2775 498206
2780 498206
2785 498207
2790 498207
2795 498207
2800 498207
2805 498207
2810 498207
2815 498207
2820 498207
2825 498207
2830 498207
2835 498208
2840 498208
2845 498208
2850 498208
2855 498208
2860 498208
2865 498208
2870 498209
2875 498210
2880 498210
2885 498210
2890 498210
2895 498210
2900 498210
2905 498210
2910 498210
2915 498210
2920 498210
2925 498211
2930 498211
2935 498211
2940 498211
2945 498211
2950 498211
2955 498211
2960 498211
2965 498212
2970 498212
2975 498213
2980 498213
2985 498213
2990 498213
2995 498213
3000 498213
3005 498213
3010 498213
3015 498213
3020 498213
3025 498214
3030 498214
3035 498214
3040 498214
3045 498214
3050 498214
3055 498214
3060 498215
3065 498215
3070 498215
3075 498215
3080 498216
3085 498216
3090 498216
3095 498216
3100 498216
3105 498216
3110 498216
3115 498216
3120 498216
3125 498216
3130 498217
3135 498217
3140 498217
3145 498217
3150 498218
3155 498218
3160 498218
3165 498218
3170 498218
3175 498218
3180 498219
3185 498219
3190 498219
3195 498219
3200 498219
3205 498219
3210 498219
3215 498219
3220 498219
3225 498219
3230 498220
3235 498220
3240 498220
3245 498221
3250 498221
3255 498221
3260 498221
3265 498221
3270 498221
3275 498221
3280 498222
3285 498222
3290 498222
3295 498222
3300 498222
3305 498222
3310 498222
3315 498222
3320 498222
3325 498222
3330 498222
3335 498223
3340 498224
3345 498224
3350 498224
3355 498224
3360 498224
3365 498224
3370 498224
3375 498224
3380 498224
3385 498225
3390 498225
3395 498225
3400 498225
3405 498225
3410 498225
3415 498225
3420 498225
3425 498225
3430 498225
3435 498226
3440 498226
3445 498226
3450 498226
3455 498226
3460 498226
3465 498226
3470 498228
3475 498228
3480 498228
3485 498228
3490 498228
3495 498228
3500 498228
3505 498228
3510 498228
3515 498228
3520 498228
3525 498229
3530 498229
3535 498229
3540 498229
3545 498229
3550 498229
3555 498229
3560 498229
3565 498230
3570 498230
3575 498231
3580 498231
3585 498231
3590 498231
3595 498231
3600 498231
};
\addlegendentry{Baseline without DPU and NIC}

\end{axis}

\end{tikzpicture}

%% file: results/power_amd_dvfs.tex
\begin{tikzpicture}[font=\Large]

\definecolor{color0}{rgb}{0.12156862745098,0.466666666666667,0.705882352941177}
\definecolor{color1}{rgb}{1,0.498039215686275,0.0549019607843137}
\definecolor{color2}{rgb}{0.172549019607843,0.627450980392157,0.172549019607843}
\definecolor{color3}{rgb}{0.83921568627451,0.152941176470588,0.156862745098039}
\definecolor{color4}{rgb}{0.580392156862745,0.403921568627451,0.741176470588235}
\definecolor{color5}{rgb}{0,0,0}

\begin{axis}[
legend cell align={left},
legend columns=2,
legend style={fill opacity=0.8, draw opacity=1, text opacity=1, at={(1.02,1.18)}, anchor=east, draw=white!80.0!black},
tick align=outside,
tick pos=left,
x grid style={white!69.01960784313725!black},
xlabel={Time (SEC)},
xmin=0, xmax=6000,
xtick style={color=black},
xtick={0,1200,2400,3600,4800,6000},
y grid style={white!69.01960784313725!black},
ylabel={Power consumption (W)},
ymin=180, ymax=260,
ytick={180,200,220, 250},
xmajorgrids,
ymajorgrids,
ytick style={color=black}
]
\addplot [thick, color5, mark=.]
table{
0 205
5 205
10 204
15 205
20 205
25 204
30 205
35 204
40 204
45 204
50 205
55 205
60 204
65 205
70 204
75 205
80 204
85 205
90 205
95 205
100 205
105 204
110 204
115 205
120 205
125 204
130 205
135 205
140 204
145 204
150 205
155 205
160 205
165 204
170 205
175 204
180 204
185 205
190 205
195 204
200 204
205 206
210 206
215 206
220 206
225 206
230 206
235 206
240 205
245 206
250 205
255 206
260 205
265 206
270 205
275 206
280 206
285 205
290 205
295 205
300 205
305 205
310 205
315 206
320 205
325 205
330 205
335 205
340 206
345 205
350 205
355 205
360 205
365 205
370 206
375 206
380 205
385 205
390 205
395 206
400 205
405 206
410 206
415 206
420 206
425 206
430 206
435 206
440 206
445 205
450 208
455 205
460 205
465 206
470 206
475 206
480 205
485 205
490 206
495 205
500 205
505 205
510 205
515 205
520 205
525 206
530 205
535 206
540 205
545 206
550 205
555 205
560 205
565 205
570 205
575 206
580 206
585 206
590 205
595 206
600 205
605 206
610 206
615 206
620 205
625 206
630 206
635 206
640 205
645 206
650 206
655 206
660 205
665 206
670 206
675 205
680 206
685 206
690 205
695 205
700 205
705 205
710 205
715 205
720 205
725 206
730 205
735 206
740 207
745 207
750 206
755 205
760 205
765 205
770 206
775 206
780 205
785 206
790 206
795 206
800 205
805 206
810 205
815 205
820 206
825 205
830 206
835 205
840 205
845 206
850 206
855 206
860 205
865 206
870 205
875 206
880 205
885 206
890 205
895 206
900 206
905 206
910 205
915 206
920 206
925 205
930 205
935 205
940 206
945 206
950 205
955 206
960 205
965 206
970 206
975 206
980 205
985 206
990 206
995 206
1000 206
1005 205
1010 206
1015 205
1020 205
1025 206
1030 206
1035 205
1040 206
1045 206
1050 205
1055 205
1060 205
1065 205
1070 206
1075 205
1080 206
1085 206
1090 205
1095 205
1100 205
1105 205
1110 205
1115 205
1120 206
1125 205
1130 205
1135 206
1140 205
1145 206
1150 205
1155 205
1160 206
1165 206
1170 206
1175 205
1180 206
1185 205
1190 205
1195 205
1200 205
1205 205
1210 205
1215 205
1220 206
1225 206
1230 206
1235 205
1240 206
1245 206
1250 206
1255 205
1260 205
1265 206
1270 205
1275 205
1280 205
1285 205
1290 205
1295 205
1300 205
1305 205
1310 205
1315 205
1320 206
1325 206
1330 206
1335 206
1340 206
1345 205
1350 206
1355 206
1360 206
1365 205
1370 206
1375 205
1380 206
1385 206
1390 205
1395 205
1400 206
1405 206
1410 206
1415 206
1420 206
1425 205
1430 205
1435 206
1440 205
1445 206
1450 206
1455 205
1460 206
1465 206
1470 206
1475 205
1480 205
1485 206
1490 206
1495 206
1500 205
1505 206
1510 206
1515 205
1520 205
1525 205
1530 206
1535 206
1540 206
1545 206
1550 205
1555 206
1560 206
1565 206
1570 206
1575 206
1580 205
1585 206
1590 206
1595 205
1600 206
1605 206
1610 206
1615 205
1620 206
1625 206
1630 205
1635 206
1640 206
1645 207
1650 206
1655 206
1660 206
1665 205
1670 205
1675 205
1680 205
1685 206
1690 206
1695 206
1700 206
1705 205
1710 205
1715 206
1720 206
1725 206
1730 206
1735 205
1740 205
1745 206
1750 206
1755 206
1760 206
1765 205
1770 205
1775 205
1780 205
1785 205
1790 205
1795 205
1800 206
1805 205
1810 206
1815 205
1820 205
1825 206
1830 206
1835 205
1840 206
1845 206
1850 206
1855 206
1860 205
1865 205
1870 206
1875 205
1880 205
1885 205
1890 205
1895 206
1900 206
1905 205
1910 206
1915 205
1920 206
1925 205
1930 205
1935 206
1940 205
1945 205
1950 205
1955 204
1960 206
1965 206
1970 205
1975 205
1980 205
1985 205
1990 205
1995 205
2000 206
2005 205
2010 205
2015 205
2020 205
2025 205
2030 205
2035 205
2040 205
2045 205
2050 205
2055 205
2060 205
2065 206
2070 205
2075 206
2080 206
2085 205
2090 205
2095 205
2100 206
2105 205
2110 205
2115 206
2120 205
2125 205
2130 205
2135 205
2140 205
2145 206
2150 206
2155 205
2160 205
2165 205
2170 205
2175 206
2180 205
2185 206
2190 205
2195 205
2200 205
2205 205
2210 205
2215 205
2220 206
2225 205
2230 205
2235 206
2240 206
2245 206
2250 205
2255 205
2260 205
2265 205
2270 205
2275 205
2280 205
2285 205
2290 205
2295 205
2300 206
2305 206
2310 206
2315 205
2320 205
2325 205
2330 205
2335 205
2340 205
2345 205
2350 206
2355 206
2360 205
2365 205
2370 205
2375 204
2380 205
2385 206
2390 205
2395 205
2400 205
2405 205
2410 204
2415 205
2420 205
2425 205
2430 206
2435 206
2440 205
2445 206
2450 206
2455 205
2460 205
2465 205
2470 206
2475 205
2480 206
2485 205
2490 206
2495 205
2500 205
2505 206
2510 205
2515 205
2520 205
2525 206
2530 206
2535 205
2540 206
2545 206
2550 205
2555 206
2560 206
2565 205
2570 206
2575 205
2580 206
2585 205
2590 206
2595 205
2600 205
2605 206
2610 206
2615 206
2620 205
2625 206
2630 206
2635 205
2640 205
2645 205
2650 206
2655 205
2660 206
2665 206
2670 206
2675 205
2680 206
2685 206
2690 206
2695 206
2700 205
2705 206
2710 206
2715 206
2720 205
2725 206
2730 205
2735 205
2740 206
2745 205
2750 205
2755 205
2760 205
2765 206
2770 207
2775 205
2780 206
2785 205
2790 205
2795 206
2800 205
2805 200
2810 205
2815 205
2820 205
2825 205
2830 205
2835 205
2840 205
2845 206
2850 205
2855 206
2860 205
2865 205
2870 205
2875 204
2880 204
2885 204
2890 205
2895 204
2900 205
2905 205
2910 206
2915 205
2920 205
2925 205
2930 206
2935 205
2940 204
2945 205
2950 205
2955 206
2960 205
2965 205
2970 205
2975 205
2980 206
2985 205
2990 204
2995 206
3000 205
3005 205
3010 205
3015 204
3020 205
3025 205
3030 204
3035 204
3040 205
3045 205
3050 205
3055 205
3060 205
3065 205
3070 204
3075 205
3080 205
3085 205
3090 205
3095 205
3100 205
3105 205
3110 205
3115 205
3120 205
3125 205
3130 205
3135 205
3140 205
3145 207
3150 205
3155 205
3160 205
3165 205
3170 205
3175 205
3180 205
3185 204
3190 205
3195 205
3200 205
3205 205
3210 204
3215 205
3220 205
3225 206
3230 205
3235 205
3240 205
3245 205
3250 205
3255 205
3260 205
3265 205
3270 205
3275 204
3280 205
3285 205
3290 204
3295 204
3300 205
3305 205
3310 205
3315 205
3320 205
3325 206
3330 205
3335 205
3340 205
3345 204
3350 205
3355 205
3360 205
3365 205
3370 205
3375 205
3380 204
3385 205
3390 204
3395 205
3400 205
3405 205
3410 205
3415 204
3420 205
3425 205
3430 205
3435 205
3440 205
3445 208
3450 205
3455 205
3460 204
3465 205
3470 205
3475 205
3480 206
3485 205
3490 205
3495 205
3500 205
3505 205
3510 205
3515 204
3520 204
3525 205
3530 205
3535 205
3540 205
3545 204
3550 205
3555 205
3560 205
3565 206
3570 204
3575 205
3580 205
3585 205
3590 205
3595 205
3600 205
3605 205
3610 205
3615 205
3620 205
3625 205
3630 205
3635 204
3640 204
3645 205
3650 205
3655 204
3660 204
3665 205
3670 204
3675 205
3680 205
3685 205
3690 205
3695 205
3700 205
3705 205
3710 205
3715 205
3720 206
3725 204
3730 205
3735 204
3740 204
3745 206
3750 205
3755 205
3760 205
3765 205
3770 205
3775 205
3780 205
3785 205
3790 205
3795 204
3800 204
3805 205
3810 206
3815 205
3820 204
3825 205
3830 204
3835 205
3840 205
3845 204
3850 205
3855 204
3860 205
3865 205
3870 205
3875 205
3880 205
3885 205
3890 205
3895 205
3900 205
3905 204
3910 205
3915 205
3920 205
3925 205
3930 204
3935 205
3940 205
3945 204
3950 205
3955 205
3960 205
3965 205
3970 205
3975 205
3980 205
3985 204
3990 204
3995 205
4000 205
4005 204
4010 204
4015 205
4020 205
4025 205
4030 205
4035 204
4040 205
4045 205
4050 205
4055 205
4060 204
4065 205
4070 204
4075 205
4080 205
4085 205
4090 205
4095 205
4100 205
4105 205
4110 205
4115 205
4120 205
4125 205
4130 205
4135 205
4140 204
4145 205
4150 205
4155 205
4160 205
4165 205
4170 204
4175 205
4180 205
4185 205
4190 205
4195 205
4200 204
4205 205
4210 204
4215 205
4220 205
4225 205
4230 205
4235 205
4240 205
4245 204
4250 204
4255 205
4260 205
4265 205
4270 205
4275 205
4280 204
4285 204
4290 205
4295 205
4300 204
4305 206
4310 205
4315 205
4320 205
4325 205
4330 205
4335 205
4340 205
4345 205
4350 204
4355 205
4360 205
4365 205
4370 205
4375 205
4380 205
4385 205
4390 205
4395 205
4400 205
4405 205
4410 205
4415 205
4420 205
4425 204
4430 205
4435 206
4440 205
4445 205
4450 204
4455 205
4460 205
4465 205
4470 205
4475 205
4480 204
4485 205
4490 206
4495 204
4500 205
4505 205
4510 204
4515 204
4520 204
4525 205
4530 204
4535 205
4540 205
4545 205
4550 205
4555 205
4560 205
4565 205
4570 205
4575 206
4580 205
4585 205
4590 205
4595 205
4600 205
4605 205
4610 205
4615 205
4620 205
4625 205
4630 204
4635 205
4640 205
4645 205
4650 205
4655 205
4660 204
4665 204
4670 205
4675 205
4680 205
4685 205
4690 205
4695 205
4700 205
4705 205
4710 205
4715 205
4720 205
4725 205
4730 204
4735 204
4740 205
4745 205
4750 205
4755 205
4760 204
4765 205
4770 205
4775 205
4780 204
4785 205
4790 204
4795 205
4800 205
4805 205
4810 204
4815 204
4820 205
4825 205
4830 205
4835 204
4840 205
4845 204
4850 205
4855 205
4860 204
4865 206
4870 205
4875 205
4880 204
4885 204
4890 204
4895 205
4900 205
4905 205
4910 205
4915 205
4920 205
4925 205
4930 205
4935 205
4940 204
4945 205
4950 205
4955 205
4960 205
4965 205
4970 205
4975 205
4980 205
4985 205
4990 206
4995 205
5000 205
5005 205
5010 205
5015 205
5020 204
5025 205
5030 205
5035 205
5040 205
5045 205
5050 204
5055 204
5060 205
5065 204
5070 205
5075 205
5080 206
5085 205
5090 205
5095 205
5100 205
5105 205
5110 205
5115 205
5120 205
5125 206
5130 205
5135 205
5140 205
5145 205
5150 205
5155 205
5160 205
5165 204
5170 205
5175 204
5180 204
5185 205
5190 204
5195 204
5200 205
5205 205
5210 204
5215 205
5220 205
5225 205
5230 205
5235 205
5240 205
5245 204
5250 205
5255 204
5260 205
5265 205
5270 205
5275 205
5280 204
5285 205
5290 205
5295 205
5300 205
5305 205
5310 204
5315 205
5320 204
5325 204
5330 205
5335 205
5340 205
5345 205
5350 204
5355 205
5360 205
5365 205
5370 205
5375 204
5380 205
5385 205
5390 205
5395 205
5400 204
5405 205
5410 205
5415 205
5420 204
5425 205
5430 204
5435 205
5440 205
5445 205
5450 205
5455 205
5460 205
5465 204
5470 205
5475 204
5480 205
5485 205
5490 205
5495 204
5500 204
5505 205
5510 205
5515 204
5520 204
5525 205
5530 205
5535 206
5540 205
5545 204
5550 204
5555 206
5560 205
5565 204
5570 205
5575 204
5580 205
5585 205
5590 205
5595 205
5600 206
5605 205
5610 205
5615 205
5620 205
5625 205
5630 205
5635 205
5640 205
5645 205
5650 204
5655 205
5660 205
5665 205
5670 205
5675 204
5680 204
5685 205
5690 205
5695 205
5700 206
5705 205
5710 205
5715 205
5720 205
5725 205
5730 204
5735 205
5740 205
5745 205
5750 205
5755 205
5760 204
5765 205
5770 205
5775 205
5780 205
5785 205
5790 204
5795 205
5800 205
5805 206
5810 205
5815 205
5820 205
5825 205
5830 204
5835 206
5840 206
5845 205
5850 205
5855 205
5860 204
5865 205
5870 206
5875 205
5880 205
5885 205
5890 205
5895 205
5900 205
5905 205
5910 204
5915 205
5920 206
5925 205
5930 205
5935 205
5940 205
5945 206
5950 205
5955 205
5960 205
5965 205
5970 205
5975 204
5980 205
5985 205
5990 204
5995 205
6000 205
6005 204
6010 205
6015 205
6020 205
6025 205
6030 205
6035 205
6040 205
6045 205
6050 205
6055 205
6060 205
6065 205
6070 204
6075 206
6080 206
6085 204
6090 205
6095 204
6100 204
6105 205
6110 205
6115 205
6120 205
6125 206
6130 205
6135 205
6140 206
6145 205
6150 206
6155 206
6160 204
6165 205
6170 205
6175 205
6180 205
6185 205
6190 206
6195 205
6200 204
6205 205
6210 205
6215 205
6220 205
6225 205
6230 205
6235 205
6240 205
6245 205
6250 205
6255 204
6260 204
6265 204
6270 203
6275 204
6280 204
6285 204
6290 204
6295 204
6300 204
6305 203
6310 203
6315 204
6320 204
6325 204
6330 204
6335 204
6340 203
6345 204
6350 204
6355 204
6360 203
6365 205
6370 203
6375 204
6380 204
6385 204
6390 204
6395 204
6400 204
6405 204
6410 204
6415 203
6420 205
6425 205
6430 204
6435 204
6440 205
6445 204
6450 204
6455 204
6460 204
6465 204
6470 204
6475 203
6480 204
6485 204
6490 204
6495 203
6500 204
6505 204
6510 204
6515 204
6520 204
6525 204
6530 204
6535 203
6540 204
6545 204
6550 204
6555 204
6560 204
6565 204
6570 203
6575 204
6580 204
6585 205
6590 203
6595 203
6600 204
6605 204
6610 203
6615 204
6620 205
6625 203
6630 204
6635 204
6640 204
6645 204
6650 204
6655 204
6660 203
6665 204
6670 203
6675 203
6680 203
6685 204
6690 204
6695 204
6700 204
6705 204
6710 204
6715 204
6720 204
6725 204
6730 204
6735 204
6740 206
6745 204
6750 204
6755 204
6760 204
6765 203
6770 204
6775 204
6780 204
6785 203
6790 203
6795 203
6800 204
6805 203
6810 204
6815 204
6820 204
6825 204
6830 203
6835 204
6840 204
6845 204
6850 204
6855 203
6860 204
6865 204
6870 204
6875 204
6880 203
6885 204
6890 204
6895 204
6900 203
6905 204
6910 204
6915 204
6920 204
6925 204
6930 204
6935 204
6940 203
6945 203
6950 204
6955 203
6960 205
6965 203
6970 204
6975 204
6980 204
6985 204
6990 204
6995 204
7000 204
7005 204
7010 204
7015 204
7020 204
7025 203
7030 204
7035 203
7040 207
7045 203
7050 203
7055 204
7060 204
7065 204
7070 204
7075 203
7080 203
7085 204
7090 203
7095 204
7100 204
7105 204
7110 203
7115 204
7120 204
7125 204
7130 204
7135 203
7140 204
7145 204
7150 205
7155 204
7160 204
7165 204
7170 204
7175 204
7180 204
7185 204
7190 203
7195 203
7200 204
};
\addlegendentry{Baseline}
\addplot [thick, color3, mark=o, mark size=3, mark options={solid,fill=white,draw=red},  mark repeat={160}]
table{
0 210
5 211
10 211
15 211
20 211
25 206
30 210
35 204
40 205
45 205
50 206
55 205
60 205
65 204
70 205
75 204
80 207
85 205
90 205
95 205
100 205
105 208
110 210
115 211
120 211
125 211
130 210
135 208
140 209
145 208
150 208
155 210
160 211
165 210
170 211
175 211
180 210
185 223
190 241
195 241
200 241
205 241
210 242
215 242
220 242
225 242
230 242
235 241
240 241
245 242
250 241
255 242
260 241
265 242
270 241
275 241
280 241
285 242
290 242
295 242
300 242
305 242
310 242
315 242
320 243
325 243
330 242
335 243
340 242
345 242
350 243
355 243
360 243
365 243
370 242
375 242
380 247
385 243
390 242
395 243
400 243
405 243
410 242
415 242
420 243
425 243
430 244
435 243
440 243
445 243
450 244
455 243
460 243
465 243
470 243
475 244
480 244
485 243
490 243
495 244
500 243
505 244
510 243
515 244
520 243
525 243
530 244
535 244
540 243
545 244
550 243
555 243
560 243
565 244
570 244
575 244
580 243
585 244
590 244
595 245
600 246
605 246
610 244
615 246
620 245
625 245
630 245
635 245
640 245
645 245
650 246
655 244
660 245
665 246
670 245
675 245
680 247
685 245
690 245
695 244
700 245
705 245
710 246
715 245
720 244
725 245
730 246
735 245
740 246
745 245
750 245
755 245
760 245
765 245
770 245
775 244
780 246
785 245
790 246
795 245
800 246
805 245
810 246
815 245
820 245
825 245
830 245
835 246
840 245
845 240
850 244
855 244
860 244
865 244
870 245
875 244
880 245
885 244
890 244
895 245
900 245
905 245
910 244
915 245
920 245
925 244
930 244
935 244
940 245
945 244
950 244
955 245
960 245
965 244
970 244
975 244
980 244
985 245
990 245
995 244
1000 244
1005 244
1010 245
1015 245
1020 244
1025 245
1030 244
1035 245
1040 244
1045 245
1050 244
1055 244
1060 244
1065 244
1070 245
1075 245
1080 245
1085 245
1090 245
1095 244
1100 245
1105 244
1110 245
1115 245
1120 245
1125 245
1130 244
1135 244
1140 245
1145 245
1150 245
1155 245
1160 244
1165 245
1170 245
1175 245
1180 245
1185 245
1190 245
1195 244
1200 245
1205 244
1210 243
1215 245
1220 245
1225 245
1230 245
1235 245
1240 245
1245 245
1250 243
1255 243
1260 242
1265 243
1270 242
1275 242
1280 244
1285 243
1290 242
1295 242
1300 244
1305 246
1310 244
1315 244
1320 245
1325 244
1330 245
1335 245
1340 245
1345 244
1350 244
1355 244
1360 245
1365 244
1370 244
1375 245
1380 245
1385 245
1390 245
1395 245
1400 245
1405 244
1410 244
1415 245
1420 245
1425 245
1430 246
1435 244
1440 245
1445 246
1450 245
1455 245
1460 245
1465 245
1470 245
1475 244
1480 245
1485 244
1490 244
1495 245
1500 245
1505 245
1510 245
1515 244
1520 245
1525 245
1530 245
1535 245
1540 245
1545 245
1550 245
1555 244
1560 244
1565 245
1570 244
1575 244
1580 244
1585 244
1590 244
1595 245
1600 245
1605 244
1610 245
1615 244
1620 245
1625 245
1630 245
1635 244
1640 242
1645 242
1650 242
1655 242
1660 243
1665 242
1670 243
1675 242
1680 243
1685 243
1690 243
1695 244
1700 244
1705 243
1710 243
1715 242
1720 242
1725 244
1730 243
1735 245
1740 245
1745 244
1750 245
1755 244
1760 244
1765 245
1770 244
1775 244
1780 245
1785 244
1790 244
1795 245
1800 243
1805 244
1810 244
1815 244
1820 245
1825 245
1830 245
1835 245
1840 244
1845 244
1850 244
1855 245
1860 245
1865 244
1870 245
1875 246
1880 245
1885 244
1890 245
1895 245
1900 245
1905 244
1910 245
1915 244
1920 245
1925 244
1930 244
1935 244
1940 245
1945 244
1950 245
1955 245
1960 245
1965 244
1970 245
1975 245
1980 244
1985 244
1990 245
1995 244
2000 244
2005 245
2010 245
2015 246
2020 246
2025 244
2030 244
2035 244
2040 245
2045 245
2050 245
2055 245
2060 245
2065 245
2070 245
2075 246
2080 244
2085 244
2090 244
2095 244
2100 245
2105 245
2110 245
2115 245
2120 245
2125 245
2130 245
2135 245
2140 245
2145 245
2150 245
2155 245
2160 244
2165 245
2170 245
2175 246
2180 245
2185 245
2190 245
2195 245
2200 244
2205 246
2210 245
2215 244
2220 245
2225 245
2230 245
2235 244
2240 244
2245 245
2250 246
2255 245
2260 245
2265 245
2270 244
2275 245
2280 246
2285 245
2290 245
2295 245
2300 245
2305 244
2310 246
2315 245
2320 245
2325 244
2330 245
2335 245
2340 245
2345 245
2350 245
2355 245
2360 245
2365 244
2370 245
2375 245
2380 246
2385 245
2390 245
2395 244
2400 245
2405 245
2410 245
2415 245
2420 245
2425 245
2430 245
2435 246
2440 243
2445 242
2450 243
2455 243
2460 243
2465 244
2470 243
2475 243
2480 242
2485 243
2490 243
2495 243
2500 243
2505 243
2510 243
2515 242
2520 242
2525 242
2530 242
2535 243
2540 245
2545 244
2550 244
2555 245
2560 244
2565 245
2570 244
2575 244
2580 245
2585 245
2590 244
2595 244
2600 244
2605 244
2610 245
2615 245
2620 244
2625 244
2630 244
2635 245
2640 245
2645 244
2650 245
2655 244
2660 244
2665 245
2670 245
2675 244
2680 244
2685 244
2690 245
2695 244
2700 244
2705 245
2710 245
2715 245
2720 245
2725 245
2730 245
2735 244
2740 227
2745 227
2750 217
2755 212
2760 213
2765 212
2770 212
2775 212
2780 211
2785 206
2790 205
2795 204
2800 206
2805 205
2810 205
2815 205
2820 205
2825 205
2830 205
2835 205
2840 206
2845 205
2850 205
2855 204
2860 205
2865 205
2870 204
2875 204
2880 205
2885 205
2890 204
2895 204
2900 205
2905 204
2910 204
2915 204
2920 204
2925 204
2930 205
2935 205
2940 205
2945 204
2950 204
2955 204
2960 205
2965 205
2970 204
2975 204
2980 205
2985 204
2990 205
2995 204
3000 204
3005 205
3010 205
3015 204
3020 205
3025 205
3030 205
3035 205
3040 205
3045 205
3050 204
3055 205
3060 204
3065 203
3070 204
3075 204
3080 204
3085 204
3090 204
3095 204
3100 205
3105 204
3110 204
3115 203
3120 204
3125 205
3130 204
3135 204
3140 204
3145 204
3150 204
3155 204
3160 204
3165 204
3170 204
3175 205
3180 204
3185 204
3190 204
3195 203
3200 205
3205 205
3210 204
3215 203
3220 204
3225 204
3230 204
3235 204
3240 204
3245 204
3250 204
3255 203
3260 204
3265 204
3270 203
3275 204
3280 204
3285 204
3290 204
3295 204
3300 204
3305 204
3310 203
3315 204
3320 204
3325 204
3330 204
3335 204
3340 204
3345 204
3350 203
3355 204
3360 204
3365 204
3370 204
3375 207
3380 204
3385 204
3390 204
3395 204
3400 204
3405 204
3410 205
3415 204
3420 204
3425 204
3430 204
3435 204
3440 203
3445 204
3450 204
3455 204
3460 203
3465 203
3470 204
3475 204
3480 204
3485 204
3490 204
3495 204
3500 204
3505 205
3510 205
3515 204
3520 203
3525 204
3530 204
3535 203
3540 204
3545 204
3550 204
3555 204
3560 204
3565 204
3570 204
3575 204
3580 204
3585 203
3590 203
3595 204
3600 204
3605 204
3610 204
3615 204
3620 205
3625 204
3630 204
3635 204
3640 204
3645 204
3650 204
3655 205
3660 204
3665 204
3670 204
3675 204
3680 204
3685 204
3690 204
3695 204
3700 205
3705 204
3710 205
3715 204
3720 204
3725 204
3730 203
3735 205
3740 204
3745 204
3750 203
3755 204
3760 205
3765 204
3770 204
3775 204
3780 204
3785 204
3790 204
3795 204
3800 204
3805 204
3810 204
3815 204
3820 204
3825 204
3830 204
3835 204
3840 204
3845 204
3850 204
3855 204
3860 204
3865 203
3870 204
3875 204
3880 205
3885 204
3890 205
3895 203
3900 204
3905 204
3910 204
3915 204
3920 205
3925 204
3930 205
3935 205
3940 205
3945 204
3950 204
3955 204
3960 204
3965 204
3970 204
3975 205
3980 205
3985 205
3990 205
3995 204
4000 204
4005 205
4010 204
4015 204
4020 204
4025 204
4030 205
4035 205
4040 204
4045 205
4050 204
4055 205
4060 205
4065 205
4070 205
4075 205
4080 205
4085 205
4090 205
4095 204
4100 204
4105 204
4110 204
4115 205
4120 204
4125 204
4130 205
4135 204
4140 204
4145 204
4150 205
4155 204
4160 204
4165 204
4170 204
4175 204
4180 204
4185 205
4190 204
4195 204
4200 205
4205 204
4210 204
4215 203
4220 204
4225 204
4230 204
4235 204
4240 205
4245 204
4250 204
4255 204
4260 205
4265 205
4270 204
4275 207
4280 204
4285 204
4290 205
4295 204
4300 204
4305 204
4310 205
4315 204
4320 205
4325 205
4330 204
4335 205
4340 204
4345 204
4350 204
4355 205
4360 204
4365 204
4370 204
4375 205
4380 205
4385 205
4390 205
4395 204
4400 205
4405 204
4410 205
4415 205
4420 205
4425 204
4430 204
4435 205
4440 204
4445 205
4450 204
4455 204
4460 204
4465 205
4470 204
4475 204
4480 204
4485 205
4490 204
4495 205
4500 205
4505 204
4510 204
4515 204
4520 204
4525 205
4530 205
4535 204
4540 204
4545 204
4550 205
4555 204
4560 205
4565 204
4570 205
4575 207
4580 204
4585 205
4590 204
4595 205
4600 205
4605 204
4610 204
4615 205
4620 204
4625 204
4630 205
4635 204
4640 205
4645 205
4650 204
4655 205
4660 205
4665 204
4670 203
4675 204
4680 204
4685 205
4690 204
4695 204
4700 205
4705 204
4710 205
4715 204
4720 204
4725 205
4730 204
4735 205
4740 204
4745 204
4750 205
4755 205
4760 205
4765 204
4770 204
4775 204
4780 204
4785 205
4790 204
4795 205
4800 204
4805 205
4810 204
4815 204
4820 204
4825 204
4830 204
4835 204
4840 204
4845 204
4850 205
4855 205
4860 204
4865 205
4870 204
4875 210
4880 204
4885 204
4890 204
4895 204
4900 204
4905 204
4910 204
4915 205
4920 204
4925 205
4930 205
4935 205
4940 204
4945 205
4950 204
4955 204
4960 205
4965 204
4970 205
4975 204
4980 205
4985 205
4990 204
4995 204
5000 204
5005 204
5010 204
5015 204
5020 205
5025 205
5030 204
5035 204
5040 204
5045 205
5050 204
5055 204
5060 205
5065 205
5070 204
5075 204
5080 205
5085 204
5090 204
5095 204
5100 205
5105 204
5110 204
5115 204
5120 205
5125 204
5130 205
5135 204
5140 204
5145 205
5150 205
5155 205
5160 205
5165 205
5170 205
5175 205
5180 204
5185 204
5190 205
5195 205
5200 204
5205 204
5210 204
5215 204
5220 204
5225 205
5230 204
5235 205
5240 204
5245 205
5250 205
5255 204
5260 204
5265 205
5270 205
5275 205
5280 205
5285 204
5290 204
5295 204
5300 204
5305 204
5310 204
5315 205
5320 205
5325 204
5330 205
5335 205
5340 204
5345 204
5350 204
5355 204
5360 204
5365 204
5370 204
5375 204
5380 204
5385 204
5390 204
5395 205
5400 205
5405 205
5410 204
5415 205
5420 204
5425 204
5430 204
5435 205
5440 204
5445 204
5450 205
5455 205
5460 205
5465 205
5470 205
5475 206
5480 204
5485 204
5490 205
5495 204
5500 204
5505 205
5510 204
5515 204
5520 204
5525 204
5530 204
5535 204
5540 205
5545 204
5550 204
5555 204
5560 205
5565 204
5570 205
5575 205
5580 204
5585 204
5590 204
5595 204
5600 204
5605 205
5610 204
5615 205
5620 205
5625 204
5630 204
5635 204
5640 204
5645 205
5650 205
5655 204
5660 205
5665 205
5670 204
5675 204
5680 205
5685 204
5690 204
5695 204
5700 205
5705 204
5710 205
5715 204
5720 204
5725 205
5730 204
5735 204
5740 205
5745 205
5750 204
5755 205
5760 204
5765 204
5770 204
5775 205
5780 205
5785 204
5790 204
5795 204
5800 205
5805 204
5810 204
5815 204
5820 205
5825 205
5830 204
5835 205
5840 205
5845 205
5850 204
5855 205
5860 205
5865 204
5870 204
5875 205
5880 205
5885 205
5890 205
5895 204
5900 204
5905 204
5910 204
5915 205
5920 205
5925 204
5930 204
5935 205
5940 204
5945 205
5950 205
5955 204
5960 204
5965 205
5970 204
5975 205
5980 205
5985 205
5990 204
5995 204
6000 204
};
\addlegendentry{DVFS}
\addplot [thick, color1, mark=triangle*, mark size=3, mark options={solid,fill=white,draw=red},  mark repeat={180}]
table{0 205
5 209
10 210
15 208
20 209
25 208
30 209
35 208
40 207
45 208
50 207
55 208
60 209
65 208
70 206
75 206
80 206
85 206
90 206
95 206
100 206
105 206
110 206
115 206
120 211
125 207
130 206
135 206
140 206
145 206
150 206
155 206
160 208
165 206
170 206
175 206
180 207
185 206
190 207
195 207
200 206
205 206
210 207
215 207
220 206
225 206
230 207
235 209
240 208
245 210
250 209
255 209
260 208
265 208
270 209
275 208
280 209
285 209
290 207
295 208
300 208
305 208
310 208
315 208
320 208
325 208
330 208
335 208
340 209
345 208
350 209
355 208
360 208
365 208
370 208
375 208
380 207
385 207
390 208
395 208
400 208
405 215
410 214
415 214
420 214
425 214
430 214
435 214
440 214
445 214
450 215
455 214
460 215
465 215
470 215
475 214
480 214
485 214
490 214
495 215
500 214
505 214
510 215
515 214
520 215
525 215
530 215
535 214
540 215
545 214
550 214
555 214
560 214
565 214
570 214
575 214
580 214
585 215
590 214
595 215
600 215
605 214
610 214
615 215
620 214
625 215
630 215
635 214
640 215
645 215
650 215
655 214
660 214
665 214
670 214
675 214
680 215
685 215
690 214
695 214
700 215
705 214
710 214
715 214
720 214
725 215
730 215
735 215
740 214
745 213
750 214
755 215
760 215
765 215
770 215
775 215
780 215
785 215
790 214
795 214
800 214
805 215
810 214
815 214
820 215
825 214
830 214
835 214
840 215
845 214
850 215
855 214
860 214
865 215
870 214
875 214
880 214
885 215
890 214
895 214
900 215
905 215
910 215
915 214
920 215
925 214
930 214
935 215
940 214
945 214
950 214
955 214
960 214
965 215
970 214
975 214
980 214
985 214
990 215
995 215
1000 215
1005 215
1010 214
1015 214
1020 214
1025 214
1030 214
1035 215
1040 215
1045 214
1050 215
1055 214
1060 214
1065 215
1070 214
1075 214
1080 215
1085 214
1090 214
1095 214
1100 214
1105 214
1110 215
1115 214
1120 215
1125 214
1130 215
1135 214
1140 215
1145 215
1150 215
1155 215
1160 214
1165 214
1170 214
1175 215
1180 214
1185 214
1190 214
1195 214
1200 215
1205 214
1210 214
1215 215
1220 214
1225 214
1230 215
1235 214
1240 214
1245 215
1250 215
1255 214
1260 214
1265 214
1270 214
1275 214
1280 214
1285 215
1290 215
1295 214
1300 214
1305 216
1310 214
1315 214
1320 215
1325 214
1330 215
1335 215
1340 215
1345 214
1350 214
1355 215
1360 214
1365 215
1370 215
1375 214
1380 214
1385 214
1390 215
1395 215
1400 214
1405 214
1410 214
1415 214
1420 216
1425 214
1430 215
1435 215
1440 215
1445 214
1450 215
1455 215
1460 214
1465 215
1470 215
1475 215
1480 215
1485 214
1490 215
1495 215
1500 215
1505 215
1510 214
1515 214
1520 215
1525 215
1530 215
1535 215
1540 215
1545 214
1550 215
1555 215
1560 215
1565 215
1570 215
1575 215
1580 214
1585 215
1590 215
1595 214
1600 215
1605 214
1610 215
1615 214
1620 215
1625 215
1630 215
1635 215
1640 214
1645 215
1650 214
1655 215
1660 218
1665 214
1670 215
1675 215
1680 215
1685 216
1690 214
1695 215
1700 215
1705 215
1710 214
1715 215
1720 215
1725 215
1730 215
1735 215
1740 215
1745 215
1750 215
1755 215
1760 214
1765 215
1770 214
1775 214
1780 214
1785 214
1790 215
1795 215
1800 214
1805 215
1810 215
1815 215
1820 215
1825 215
1830 216
1835 215
1840 215
1845 215
1850 214
1855 215
1860 215
1865 215
1870 214
1875 214
1880 215
1885 214
1890 215
1895 215
1900 214
1905 214
1910 214
1915 215
1920 215
1925 215
1930 214
1935 214
1940 214
1945 215
1950 215
1955 214
1960 215
1965 214
1970 215
1975 214
1980 214
1985 215
1990 214
1995 214
2000 215
2005 214
2010 214
2015 215
2020 215
2025 214
2030 214
2035 215
2040 215
2045 214
2050 214
2055 215
2060 215
2065 214
2070 214
2075 214
2080 215
2085 214
2090 214
2095 215
2100 215
2105 215
2110 215
2115 214
2120 215
2125 214
2130 215
2135 215
2140 214
2145 215
2150 214
2155 215
2160 215
2165 215
2170 214
2175 215
2180 214
2185 214
2190 215
2195 214
2200 215
2205 214
2210 214
2215 214
2220 215
2225 215
2230 214
2235 215
2240 214
2245 215
2250 215
2255 214
2260 214
2265 215
2270 215
2275 214
2280 214
2285 215
2290 215
2295 214
2300 215
2305 215
2310 215
2315 215
2320 214
2325 215
2330 214
2335 214
2340 214
2345 215
2350 216
2355 214
2360 214
2365 215
2370 215
2375 215
2380 215
2385 214
2390 215
2395 214
2400 214
2405 214
2410 216
2415 214
2420 214
2425 214
2430 215
2435 215
2440 214
2445 214
2450 215
2455 215
2460 215
2465 215
2470 216
2475 214
2480 215
2485 215
2490 214
2495 215
2500 215
2505 214
2510 214
2515 214
2520 215
2525 214
2530 215
2535 216
2540 215
2545 215
2550 214
2555 215
2560 215
2565 214
2570 215
2575 214
2580 215
2585 216
2590 214
2595 215
2600 215
2605 215
2610 214
2615 214
2620 215
2625 215
2630 215
2635 215
2640 215
2645 215
2650 214
2655 215
2660 215
2665 214
2670 215
2675 215
2680 214
2685 214
2690 215
2695 215
2700 214
2705 214
2710 215
2715 214
2720 214
2725 215
2730 214
2735 215
2740 215
2745 215
2750 214
2755 213
2760 214
2765 214
2770 215
2775 214
2780 214
2785 214
2790 215
2795 214
2800 215
2805 215
2810 214
2815 214
2820 215
2825 214
2830 215
2835 214
2840 214
2845 215
2850 214
2855 214
2860 215
2865 214
2870 215
2875 215
2880 215
2885 214
2890 215
2895 214
2900 215
2905 214
2910 215
2915 214
2920 214
2925 214
2930 214
2935 215
2940 214
2945 215
2950 215
2955 214
2960 214
2965 214
2970 214
2975 214
2980 214
2985 215
2990 215
2995 215
3000 214
3005 215
3010 214
3015 215
3020 214
3025 214
3030 215
3035 215
3040 215
3045 214
3050 214
3055 214
3060 214
3065 215
3070 215
3075 215
3080 214
3085 215
3090 215
3095 214
3100 214
3105 215
3110 214
3115 215
3120 215
3125 215
3130 214
3135 215
3140 215
3145 214
3150 214
3155 215
3160 214
3165 215
3170 215
3175 215
3180 215
3185 214
3190 215
3195 215
3200 214
3205 215
3210 214
3215 215
3220 215
3225 214
3230 214
3235 215
3240 214
3245 215
3250 214
3255 215
3260 215
3265 214
3270 214
3275 215
3280 214
3285 215
3290 214
3295 214
3300 215
3305 214
3310 214
3315 215
3320 215
3325 215
3330 214
3335 215
3340 214
3345 214
3350 214
3355 215
3360 214
3365 215
3370 214
3375 214
3380 215
3385 215
3390 214
3395 215
3400 214
3405 214
3410 215
3415 215
3420 215
3425 215
3430 214
3435 215
3440 215
3445 215
3450 215
3455 215
3460 214
3465 215
3470 215
3475 215
3480 214
3485 215
3490 215
3495 215
3500 215
3505 215
3510 215
3515 214
3520 215
3525 214
3530 214
3535 215
3540 215
3545 214
3550 215
3555 213
3560 214
3565 215
3570 214
3575 214
3580 215
3585 215
3590 215
3595 214
3600 215
3605 215
3610 215
3615 214
3620 215
3625 214
3630 215
3635 215
3640 215
3645 215
3650 215
3655 215
3660 215
3665 214
3670 214
3675 214
3680 214
3685 214
3690 214
3695 214
3700 215
3705 214
3710 215
3715 214
3720 214
3725 214
3730 215
3735 215
3740 215
3745 215
3750 214
3755 214
3760 214
3765 215
3770 215
3775 215
3780 215
3785 214
3790 215
3795 215
3800 215
3805 215
3810 215
3815 215
3820 215
3825 214
3830 214
3835 214
3840 215
3845 215
3850 215
3855 215
3860 214
3865 215
3870 214
3875 215
3880 214
3885 215
3890 215
3895 215
3900 214
3905 215
3910 215
3915 215
3920 214
3925 214
3930 215
3935 214
3940 214
3945 214
3950 214
3955 214
3960 215
3965 215
3970 215
3975 214
3980 215
3985 214
3990 214
3995 215
4000 215
4005 214
4010 215
4015 215
4020 215
4025 215
4030 215
4035 215
4040 215
4045 215
4050 215
4055 214
4060 214
4065 215
4070 214
4075 215
4080 214
4085 215
4090 215
4095 215
4100 215
4105 214
4110 215
4115 215
4120 215
4125 214
4130 214
4135 214
4140 214
4145 215
4150 214
4155 214
4160 215
4165 215
4170 215
4175 214
4180 215
4185 215
4190 214
4195 215
4200 215
4205 214
4210 215
4215 215
4220 214
4225 215
4230 215
4235 214
4240 214
4245 214
4250 215
4255 214
4260 214
4265 215
4270 214
4275 214
4280 215
4285 215
4290 214
4295 215
4300 215
4305 214
4310 215
4315 214
4320 215
4325 215
4330 215
4335 215
4340 215
4345 215
4350 215
4355 215
4360 215
4365 215
4370 215
4375 215
4380 214
4385 215
4390 215
4395 215
4400 215
4405 215
4410 214
4415 214
4420 215
4425 214
4430 214
4435 215
4440 214
4445 214
4450 214
4455 215
4460 215
4465 215
4470 215
4475 215
4480 215
4485 214
4490 214
4495 214
4500 215
4505 214
4510 215
4515 215
4520 214
4525 214
4530 215
4535 214
4540 214
4545 215
4550 214
4555 214
4560 214
4565 214
4570 215
4575 214
4580 215
4585 215
4590 214
4595 214
4600 214
4605 215
4610 214
4615 215
4620 215
4625 215
4630 214
4635 215
4640 215
4645 215
4650 215
4655 216
4660 214
4665 215
4670 215
4675 215
4680 215
4685 214
4690 215
4695 214
4700 215
4705 215
4710 214
4715 214
4720 214
4725 215
4730 214
4735 215
4740 215
4745 214
4750 214
4755 214
4760 215
4765 215
4770 214
4775 214
4780 215
4785 214
4790 214
4795 215
4800 215
4805 213
4810 214
4815 215
4820 215
4825 214
4830 214
4835 214
4840 215
4845 215
4850 214
4855 214
4860 214
4865 215
4870 215
4875 215
4880 215
4885 214
4890 214
4895 215
4900 215
4905 214
4910 214
4915 214
4920 214
4925 215
4930 215
4935 215
4940 215
4945 215
4950 215
4955 214
4960 215
4965 215
4970 215
4975 214
4980 214
4985 214
4990 214
4995 214
5000 214
5005 214
5010 215
5015 214
5020 214
5025 214
5030 215
5035 215
5040 214
5045 215
5050 214
5055 215
5060 215
5065 215
5070 214
5075 215
5080 215
5085 215
5090 215
5095 215
5100 215
5105 215
5110 215
5115 214
5120 214
5125 214
5130 215
5135 214
5140 214
5145 215
5150 214
5155 215
5160 215
5165 215
5170 215
5175 215
5180 215
5185 214
5190 215
5195 214
5200 215
5205 214
5210 214
5215 215
5220 214
5225 215
5230 215
5235 215
5240 215
5245 214
5250 215
5255 215
5260 215
5265 215
5270 215
5275 215
5280 214
5285 215
5290 215
5295 215
5300 215
5305 214
5310 215
5315 215
5320 215
5325 215
5330 215
5335 215
5340 214
5345 215
5350 215
5355 215
5360 214
5365 214
5370 215
5375 214
5380 215
5385 215
5390 215
5395 215
5400 215
5405 215
5410 215
5415 215
5420 214
5425 214
5430 215
5435 214
5440 214
5445 215
5450 215
5455 214
5460 214
5465 214
5470 215
5475 215
5480 215
5485 215
5490 215
5495 215
5500 215
5505 215
5510 215
5515 215
5520 214
5525 215
5530 215
5535 214
5540 215
5545 215
5550 214
5555 215
5560 215
5565 215
5570 215
5575 215
5580 215
5585 215
5590 215
5595 215
5600 215
5605 214
5610 214
5615 214
5620 215
5625 215
5630 215
5635 214
5640 215
5645 215
5650 215
5655 214
5660 214
5665 214
5670 214
5675 214
5680 215
5685 215
5690 215
5695 214
5700 215
5705 215
5710 214
5715 215
5720 214
5725 214
5730 214
5735 214
5740 214
5745 215
5750 214
5755 215
5760 214
5765 214
5770 214
5775 214
5780 215
5785 215
5790 215
5795 215
5800 214
5805 215
5810 215
5815 215
5820 214
5825 215
5830 214
5835 215
5840 214
5845 215
5850 214
5855 215
5860 215
5865 215
5870 214
5875 215
5880 214
5885 215
5890 215
5895 215
5900 215
5905 215
5910 212
5915 212
5920 211
5925 211
5930 211
5935 210
5940 210
5945 209
5950 209
5955 209
5960 209
5965 209
5970 209
5975 209
5980 209
5985 208
5990 209
5995 209
6000 208
6005 208
6010 209
6015 206
6020 206
6025 206
};
\addlegendentry{1.5GHz}
\addplot [thick, color2, mark=square*, mark size=3, mark options={solid,fill=white,draw=red},  mark repeat={180}]
table{
0 208
5 216
10 214
15 214
20 214
25 214
30 215
35 208
40 208
45 208
50 208
55 208
60 208
65 209
70 209
75 208
80 208
85 208
90 208
95 208
100 209
105 211
110 212
115 212
120 214
125 213
130 214
135 212
140 211
145 211
150 211
155 211
160 214
165 212
170 213
175 213
180 216
185 212
190 242
195 242
200 242
205 243
210 242
215 242
220 242
225 243
230 243
235 243
240 243
245 243
250 243
255 243
260 243
265 244
270 243
275 244
280 243
285 243
290 244
295 244
300 243
305 244
310 243
315 244
320 243
325 244
330 244
335 244
340 243
345 244
350 244
355 244
360 244
365 244
370 244
375 244
380 243
385 243
390 244
395 244
400 244
405 244
410 244
415 244
420 244
425 244
430 244
435 244
440 244
445 244
450 244
455 245
460 244
465 245
470 244
475 245
480 244
485 245
490 244
495 245
500 246
505 245
510 246
515 244
520 245
525 245
530 245
535 245
540 244
545 245
550 244
555 245
560 245
565 248
570 244
575 244
580 245
585 244
590 246
595 244
600 245
605 244
610 244
615 245
620 245
625 245
630 244
635 245
640 245
645 245
650 244
655 244
660 245
665 244
670 245
675 245
680 245
685 245
690 244
695 245
700 245
705 245
710 244
715 245
720 245
725 245
730 245
735 245
740 245
745 246
750 244
755 245
760 244
765 244
770 245
775 245
780 244
785 245
790 245
795 244
800 245
805 245
810 245
815 244
820 245
825 245
830 244
835 245
840 245
845 245
850 245
855 244
860 245
865 244
870 245
875 244
880 245
885 245
890 245
895 245
900 245
905 245
910 245
915 245
920 245
925 246
930 245
935 245
940 246
945 244
950 245
955 245
960 244
965 244
970 244
975 245
980 245
985 245
990 245
995 245
1000 245
1005 245
1010 245
1015 245
1020 246
1025 245
1030 245
1035 244
1040 244
1045 245
1050 245
1055 245
1060 245
1065 245
1070 245
1075 245
1080 244
1085 245
1090 245
1095 244
1100 245
1105 245
1110 245
1115 244
1120 245
1125 245
1130 246
1135 245
1140 244
1145 245
1150 245
1155 245
1160 244
1165 245
1170 245
1175 245
1180 244
1185 245
1190 246
1195 244
1200 245
1205 245
1210 246
1215 245
1220 244
1225 246
1230 245
1235 244
1240 245
1245 245
1250 245
1255 245
1260 244
1265 244
1270 245
1275 244
1280 245
1285 244
1290 244
1295 245
1300 245
1305 245
1310 245
1315 245
1320 245
1325 245
1330 244
1335 244
1340 245
1345 245
1350 245
1355 245
1360 245
1365 244
1370 245
1375 244
1380 244
1385 245
1390 245
1395 244
1400 244
1405 244
1410 244
1415 245
1420 245
1425 245
1430 245
1435 244
1440 245
1445 246
1450 244
1455 245
1460 244
1465 245
1470 245
1475 245
1480 245
1485 245
1490 245
1495 244
1500 245
1505 245
1510 244
1515 245
1520 245
1525 245
1530 245
1535 244
1540 245
1545 244
1550 245
1555 244
1560 245
1565 244
1570 245
1575 245
1580 244
1585 244
1590 245
1595 245
1600 244
1605 245
1610 246
1615 245
1620 245
1625 245
1630 245
1635 245
1640 244
1645 245
1650 245
1655 245
1660 244
1665 244
1670 244
1675 245
1680 244
1685 244
1690 245
1695 244
1700 245
1705 244
1710 244
1715 244
1720 244
1725 244
1730 244
1735 245
1740 245
1745 245
1750 244
1755 245
1760 245
1765 245
1770 245
1775 245
1780 245
1785 244
1790 244
1795 244
1800 245
1805 243
1810 244
1815 245
1820 244
1825 245
1830 245
1835 245
1840 244
1845 245
1850 245
1855 245
1860 245
1865 245
1870 245
1875 245
1880 244
1885 244
1890 244
1895 244
1900 244
1905 244
1910 245
1915 244
1920 245
1925 245
1930 244
1935 244
1940 244
1945 245
1950 244
1955 245
1960 244
1965 245
1970 244
1975 244
1980 245
1985 245
1990 244
1995 244
2000 245
2005 245
2010 244
2015 245
2020 244
2025 245
2030 245
2035 245
2040 244
2045 245
2050 244
2055 244
2060 245
2065 246
2070 245
2075 245
2080 245
2085 245
2090 245
2095 245
2100 244
2105 245
2110 245
2115 244
2120 245
2125 245
2130 244
2135 244
2140 245
2145 245
2150 245
2155 245
2160 244
2165 245
2170 246
2175 245
2180 245
2185 245
2190 244
2195 245
2200 245
2205 245
2210 245
2215 245
2220 245
2225 246
2230 244
2235 244
2240 244
2245 245
2250 244
2255 245
2260 244
2265 244
2270 245
2275 245
2280 247
2285 246
2290 245
2295 245
2300 244
2305 245
2310 245
2315 245
2320 244
2325 246
2330 245
2335 244
2340 245
2345 245
2350 244
2355 245
2360 245
2365 245
2370 245
2375 245
2380 245
2385 245
2390 244
2395 244
2400 244
2405 244
2410 245
2415 245
2420 245
2425 244
2430 245
2435 244
2440 244
2445 244
2450 244
2455 244
2460 244
2465 245
2470 245
2475 244
2480 246
2485 244
2490 245
2495 245
2500 245
2505 244
2510 244
2515 245
2520 245
2525 245
2530 245
2535 245
2540 244
2545 245
2550 245
2555 244
2560 245
2565 244
2570 244
2575 244
2580 245
2585 245
2590 245
2595 245
2600 244
2605 244
2610 245
2615 245
2620 244
2625 245
2630 245
2635 245
2640 244
2645 245
2650 244
2655 244
2660 244
2665 245
2670 244
2675 244
2680 245
2685 245
2690 244
2695 244
2700 245
2705 244
2710 244
2715 244
2720 246
2725 246
2730 245
2735 244
2740 244
2745 246
2750 244
2755 231
2760 226
2765 224
2770 219
2775 214
2780 214
2785 215
2790 214
2795 214
2800 214
2805 211
2810 207
2815 208
2820 208
2825 208
2830 208
2835 208
2840 208
2845 207
2850 207
2855 207
2860 207
2865 208
2870 208
2875 206
2880 207
2885 208
2890 207
2895 207
2900 207
2905 207
2910 207
2915 207
2920 207
2925 207
2930 207
2935 207
2940 207
2945 207
2950 207
2955 207
2960 207
2965 207
2970 207
2975 207
2980 208
2985 207
2990 207
2995 207
3000 207
3005 207
3010 207
3015 206
3020 207
3025 207
3030 207
3035 207
3040 207
3045 206
3050 208
3055 206
3060 207
3065 207
3070 208
3075 207
3080 207
3085 207
3090 206
3095 206
3100 207
3105 207
3110 207
3115 206
3120 206
3125 207
3130 207
3135 207
3140 207
3145 207
3150 207
3155 207
3160 207
3165 207
3170 207
3175 207
3180 206
3185 207
3190 206
3195 206
3200 207
3205 207
3210 207
3215 207
3220 207
3225 207
3230 206
3235 207
3240 207
3245 206
3250 206
3255 206
3260 207
3265 206
3270 207
3275 206
3280 206
3285 206
3290 206
3295 206
3300 206
3305 207
3310 207
3315 206
3320 206
3325 206
3330 207
3335 207
3340 208
3345 206
3350 208
3355 207
3360 207
3365 208
3370 207
3375 207
3380 207
3385 206
3390 206
3395 207
3400 207
3405 206
3410 206
3415 206
3420 207
3425 206
3430 206
3435 207
3440 207
3445 206
3450 207
3455 207
3460 207
3465 206
3470 207
3475 207
3480 207
3485 208
3490 206
3495 206
3500 207
3505 206
3510 208
3515 206
3520 208
3525 207
3530 207
3535 207
3540 206
3545 207
3550 207
3555 207
3560 206
3565 207
3570 206
3575 207
3580 207
3585 207
3590 206
3595 206
3600 207
3605 207
3610 207
3615 206
3620 207
3625 207
3630 207
3635 206
3640 207
3645 206
3650 206
3655 207
3660 206
3665 206
3670 207
3675 207
3680 207
3685 207
3690 207
3695 207
3700 207
3705 207
3710 206
3715 207
3720 207
3725 207
3730 207
3735 207
3740 206
3745 207
3750 207
3755 207
3760 207
3765 207
3770 207
3775 207
3780 207
3785 207
3790 207
3795 207
3800 207
3805 207
3810 207
3815 207
3820 207
3825 206
3830 206
3835 206
3840 207
3845 206
3850 207
3855 207
3860 207
3865 206
3870 207
3875 207
3880 207
3885 206
3890 206
3895 207
3900 206
3905 207
3910 206
3915 208
3920 206
3925 207
3930 206
3935 207
3940 207
3945 207
3950 207
3955 206
3960 207
3965 207
3970 207
3975 207
3980 207
3985 207
3990 207
3995 207
4000 207
4005 207
4010 207
4015 207
4020 207
4025 207
4030 207
4035 207
4040 207
4045 207
4050 207
4055 207
4060 207
4065 207
4070 206
4075 207
4080 207
4085 206
4090 207
4095 207
4100 207
4105 206
4110 206
4115 207
4120 207
4125 206
4130 208
4135 206
4140 207
4145 207
4150 206
4155 207
4160 207
4165 208
4170 207
4175 207
4180 207
4185 207
4190 207
4195 206
4200 207
4205 207
4210 207
4215 206
4220 207
4225 207
4230 207
4235 207
4240 207
4245 207
4250 207
4255 207
4260 207
4265 207
4270 208
4275 206
4280 208
4285 207
4290 207
4295 207
4300 207
4305 207
4310 207
4315 206
4320 206
4325 207
4330 207
4335 207
4340 207
4345 207
4350 206
4355 206
4360 207
4365 207
4370 207
4375 207
4380 207
4385 207
4390 207
4395 206
4400 206
4405 207
4410 207
4415 207
4420 207
4425 207
4430 207
4435 206
4440 207
4445 207
4450 208
4455 207
4460 207
4465 207
4470 207
4475 208
4480 207
4485 207
4490 207
4495 208
4500 206
4505 208
4510 206
4515 207
4520 206
4525 207
4530 206
4535 206
4540 207
4545 206
4550 206
4555 207
4560 206
4565 207
4570 206
4575 206
4580 207
4585 207
4590 207
4595 206
4600 207
4605 207
4610 206
4615 207
4620 206
4625 206
4630 207
4635 207
4640 206
4645 207
4650 207
4655 207
4660 207
4665 207
4670 207
4675 206
4680 207
4685 206
4690 207
4695 206
4700 206
4705 207
4710 206
4715 207
4720 207
4725 207
4730 207
4735 207
4740 207
4745 207
4750 207
4755 207
4760 207
4765 207
4770 207
4775 207
4780 206
4785 207
4790 207
4795 207
4800 207
4805 207
4810 207
4815 207
4820 207
4825 207
4830 207
4835 207
4840 207
4845 206
4850 206
4855 206
4860 207
4865 207
4870 206
4875 207
4880 206
4885 206
4890 206
4895 207
4900 207
4905 207
4910 207
4915 207
4920 207
4925 207
4930 207
4935 207
4940 206
4945 207
4950 207
4955 207
4960 207
4965 206
4970 207
4975 207
4980 207
4985 207
4990 207
4995 206
5000 207
5005 207
5010 207
5015 207
5020 207
5025 207
5030 207
5035 206
5040 207
5045 206
5050 206
5055 207
5060 206
5065 212
5070 206
5075 207
5080 206
5085 207
5090 206
5095 206
5100 207
5105 207
5110 207
5115 206
5120 207
5125 207
5130 206
5135 207
5140 207
5145 207
5150 206
5155 206
5160 207
5165 207
5170 207
5175 206
5180 206
5185 208
5190 207
5195 207
5200 206
5205 207
5210 206
5215 207
5220 207
5225 207
5230 207
5235 206
5240 207
5245 207
5250 206
5255 206
5260 207
5265 207
5270 207
5275 206
5280 207
5285 206
5290 206
5295 207
5300 206
5305 207
5310 207
5315 206
5320 206
5325 207
5330 206
5335 206
5340 207
5345 207
5350 206
5355 207
5360 206
5365 206
5370 206
5375 206
5380 207
5385 207
5390 207
5395 207
5400 207
5405 206
5410 207
5415 207
5420 206
5425 206
5430 206
5435 207
5440 207
5445 206
5450 207
5455 206
5460 206
5465 207
5470 207
5475 206
5480 206
5485 206
5490 207
5495 207
5500 206
5505 207
5510 207
5515 206
5520 207
5525 207
5530 206
5535 206
5540 207
5545 207
5550 207
5555 206
5560 206
5565 206
5570 207
5575 207
5580 206
5585 207
5590 207
5595 206
5600 206
5605 206
5610 206
5615 207
5620 207
5625 207
5630 206
5635 206
5640 206
5645 207
5650 206
5655 206
5660 206
5665 208
5670 207
5675 206
5680 207
5685 207
5690 206
5695 207
5700 207
5705 206
5710 206
5715 207
5720 206
5725 207
5730 207
5735 207
5740 207
5745 206
5750 206
5755 207
5760 206
5765 207
5770 206
5775 206
5780 207
5785 206
5790 207
5795 206
5800 206
5805 206
5810 206
5815 207
5820 206
5825 206
5830 207
5835 207
5840 207
5845 206
5850 206
5855 206
5860 206
5865 207
5870 206
5875 206
5880 207
5885 207
5890 207
5895 206
5900 206
5905 207
5910 207
5915 207
5920 206
5925 207
5930 206
5935 206
5940 207
5945 207
5950 206
5955 206
5960 207
5965 207
5970 207
5975 207
5980 207
5985 207
5990 207
5995 206
};
\addlegendentry{3GHz}

\end{axis}

\end{tikzpicture}

%% file: results/power_intel_boost.tex
\begin{tikzpicture}[font=\Large]

\definecolor{color0}{rgb}{0.12156862745098,0.466666666666667,0.705882352941177}
\definecolor{color1}{rgb}{1,0.498039215686275,0.0549019607843137}
\definecolor{color2}{rgb}{0.172549019607843,0.627450980392157,0.172549019607843}
\definecolor{color3}{rgb}{0.83921568627451,0.152941176470588,0.156862745098039}
\definecolor{color4}{rgb}{0.580392156862745,0.403921568627451,0.741176470588235}
\definecolor{color5}{rgb}{0,0,0}

\begin{axis}[
legend cell align={left},
legend columns=2,
legend style={fill opacity=0.8, draw opacity=1, text opacity=1, at={(1.02,1.18)}, anchor=east, draw=white!80.0!black},
tick align=outside,
tick pos=left,
x grid style={white!69.01960784313725!black},
xlabel={Time (SEC)},
xmin=0, xmax=3600,
xtick style={color=black},
xtick={0,900,1800,2700,3600},
y grid style={white!69.01960784313725!black},
ylabel={Power consumption (W)},
ymin=150, ymax=380,
ytick={150,200,250,300,350},
xmajorgrids,
ymajorgrids,
ytick style={color=black}
]
\addplot [thick, color5, mark=.]
table {%
0 254
5 262
10 252
15 253
20 253
25 253
30 253
35 254
40 253
45 253
50 254
55 254
60 253
65 254
70 254
75 254
80 254
85 254
90 254
95 254
100 254
105 254
110 253
115 254
120 254
125 254
130 254
135 253
140 253
145 253
150 253
155 254
160 254
165 253
170 254
175 254
180 253
185 254
190 254
195 254
200 253
205 254
210 254
215 254
220 253
225 254
230 253
235 253
240 253
245 252
250 253
255 253
260 254
265 254
270 254
275 253
280 254
285 253
290 253
295 253
300 254
305 253
310 254
315 253
320 253
325 253
330 254
335 254
340 253
345 253
350 253
355 254
360 253
365 253
370 254
375 254
380 254
385 254
390 254
395 254
400 253
405 253
410 253
415 254
420 253
425 254
430 253
435 253
440 254
445 254
450 254
455 254
460 253
465 254
470 253
475 253
480 252
485 254
490 254
495 253
500 253
505 254
510 254
515 254
520 253
525 253
530 254
535 254
540 253
545 254
550 253
555 253
560 253
565 253
570 254
575 253
580 254
585 253
590 254
595 254
600 253
605 254
610 253
615 254
620 254
625 253
630 254
635 253
640 253
645 254
650 254
655 253
660 254
665 253
670 253
675 254
680 253
685 254
690 254
695 254
700 254
705 253
710 254
715 254
720 253
725 254
730 253
735 253
740 253
745 254
750 254
755 254
760 254
765 254
770 254
775 253
780 254
785 254
790 254
795 253
800 254
805 253
810 253
815 253
820 254
825 254
830 253
835 253
840 253
845 253
850 253
855 253
860 254
865 253
870 253
875 252
880 254
885 254
890 253
895 253
900 254
905 253
910 254
915 254
920 253
925 254
930 254
935 254
940 253
945 254
950 253
955 254
960 254
965 253
970 254
975 254
980 254
985 252
990 253
995 253
1000 253
1005 253
1010 253
1015 253
1020 252
1025 254
1030 253
1035 254
1040 252
1045 254
1050 254
1055 254
1060 253
1065 254
1070 254
1075 254
1080 253
1085 253
1090 253
1095 254
1100 254
1105 253
1110 253
1115 254
1120 253
1125 254
1130 254
1135 253
1140 253
1145 253
1150 253
1155 254
1160 254
1165 254
1170 254
1175 254
1180 253
1185 253
1190 254
1195 253
1200 253
1205 254
1210 254
1215 254
1220 254
1225 254
1230 251
1235 253
1240 254
1245 253
1250 254
1255 253
1260 254
1265 254
1270 254
1275 254
1280 254
1285 253
1290 253
1295 253
1300 253
1305 254
1310 254
1315 254
1320 254
1325 254
1330 254
1335 253
1340 255
1345 254
1350 254
1355 253
1360 254
1365 254
1370 254
1375 253
1380 254
1385 254
1390 253
1395 254
1400 255
1405 253
1410 254
1415 254
1420 254
1425 254
1430 253
1435 254
1440 254
1445 253
1450 253
1455 253
1460 253
1465 253
1470 253
1475 253
1480 254
1485 254
1490 254
1495 253
1500 254
1505 253
1510 252
1515 254
1520 254
1525 254
1530 253
1535 254
1540 253
1545 254
1550 253
1555 254
1560 252
1565 254
1570 254
1575 254
1580 254
1585 254
1590 253
1595 254
1600 251
1605 254
1610 253
1615 254
1620 253
1625 254
1630 253
1635 254
1640 254
1645 253
1650 253
1655 254
1660 254
1665 253
1670 253
1675 253
1680 253
1685 254
1690 254
1695 254
1700 254
1705 253
1710 253
1715 254
1720 254
1725 252
1730 252
1735 253
1740 254
1745 253
1750 254
1755 253
1760 253
1765 253
1770 254
1775 252
1780 253
1785 253
1790 252
1795 253
1800 253
1805 253
1810 253
1815 253
1820 254
1825 252
1830 254
1835 253
1840 253
1845 253
1850 252
1855 252
1860 253
1865 253
1870 253
1875 254
1880 253
1885 252
1890 253
1895 253
1900 253
1905 252
1910 254
1915 252
1920 254
1925 253
1930 254
1935 253
1940 253
1945 254
1950 253
1955 254
1960 253
1965 253
1970 254
1975 254
1980 252
1985 253
1990 254
1995 253
2000 253
2005 253
2010 254
2015 253
2020 253
2025 253
2030 253
2035 253
2040 254
2045 254
2050 253
2055 254
2060 253
2065 253
2070 254
2075 253
2080 253
2085 254
2090 253
2095 254
2100 254
2105 253
2110 254
2115 254
2120 252
2125 254
2130 254
2135 253
2140 253
2145 253
2150 254
2155 254
2160 254
2165 253
2170 251
2175 253
2180 253
2185 254
2190 254
2195 253
2200 253
2205 254
2210 253
2215 254
2220 254
2225 254
2230 253
2235 254
2240 253
2245 254
2250 253
2255 254
2260 254
2265 253
2270 254
2275 254
2280 253
2285 253
2290 254
2295 254
2300 253
2305 253
2310 254
2315 253
2320 253
2325 253
2330 253
2335 254
2340 254
2345 254
2350 254
2355 253
2360 253
2365 253
2370 253
2375 254
2380 254
2385 253
2390 253
2395 254
2400 253
2405 253
2410 253
2415 253
2420 254
2425 253
2430 253
2435 253
2440 253
2445 254
2450 254
2455 254
2460 254
2465 251
2470 254
2475 253
2480 254
2485 253
2490 253
2495 254
2500 254
2505 253
2510 254
2515 254
2520 254
2525 253
2530 253
2535 251
2540 252
2545 253
2550 254
2555 253
2560 253
2565 254
2570 253
2575 253
2580 253
2585 253
2590 253
2595 253
2600 253
2605 253
2610 253
2615 253
2620 253
2625 261
2630 252
2635 254
2640 253
2645 254
2650 253
2655 254
2660 253
2665 253
2670 254
2675 254
2680 253
2685 253
2690 253
2695 253
2700 253
2705 254
2710 254
2715 254
2720 254
2725 253
2730 253
2735 253
2740 253
2745 254
2750 254
2755 253
2760 254
2765 252
2770 254
2775 254
2780 254
2785 254
2790 254
2795 253
2800 254
2805 253
2810 253
2815 253
2820 253
2825 253
2830 254
2835 254
2840 254
2845 254
2850 254
2855 254
2860 253
2865 254
2870 254
2875 253
2880 254
2885 251
2890 253
2895 254
2900 253
2905 253
2910 254
2915 253
2920 249
2925 252
2930 253
2935 252
2940 251
2945 252
2950 252
2955 253
2960 252
2965 252
2970 252
2975 252
2980 252
2985 251
2990 252
2995 252
3000 251
3005 252
3010 252
3015 252
3020 251
3025 251
3030 252
3035 251
3040 252
3045 252
3050 251
3055 252
3060 252
3065 252
3070 252
3075 252
3080 252
3085 252
3090 253
3095 251
3100 252
3105 252
3110 251
3115 252
3120 252
3125 251
3130 252
3135 253
3140 252
3145 252
3150 251
3155 251
3160 252
3165 252
3170 252
3175 251
3180 252
3185 251
3190 251
3195 253
3200 251
3205 251
3210 251
3215 252
3220 251
3225 252
3230 252
3235 252
3240 252
3245 251
3250 252
3255 251
3260 251
3265 252
3270 252
3275 251
3280 252
3285 251
3290 251
3295 251
3300 251
3305 252
3310 251
3315 253
3320 251
3325 251
3330 251
3335 252
3340 251
3345 251
3350 252
3355 252
3360 252
3365 252
3370 252
3375 252
3380 252
3385 253
3390 251
3395 253
3400 251
3405 252
3410 252
3415 252
3420 252
3425 251
3430 251
3435 253
3440 252
3445 251
3450 252
3455 251
3460 251
3465 251
3470 253
3475 253
3480 254
3485 253
3490 254
3495 254
3500 253
3505 254
3510 253
3515 254
3520 254
3525 254
3530 254
3535 253
3540 254
3545 253
3550 254
3555 254
3560 254
3565 253
3570 253
3575 254
3580 253
3585 254
3590 254
3595 253
3600 252
};
\addlegendentry{Baseline}
\addplot [thick, color2, mark=square*, mark size=3, mark options={solid,fill=white,draw=red},  mark repeat={180}]
table {%
0 254
5 263
10 266
15 269
20 270
25 261
30 266
35 254
40 253
45 253
50 255
55 254
60 253
65 254
70 254
75 253
80 253
85 253
90 255
95 254
100 256
105 256
110 267
115 268
120 268
125 266
130 264
135 262
140 263
145 263
150 255
155 264
160 255
165 255
170 254
175 255
180 254
185 256
190 255
195 255
200 268
205 266
210 266
215 258
220 268
225 260
230 353
235 355
240 355
245 356
250 355
255 356
260 357
265 357
270 357
275 357
280 357
285 356
290 358
295 359
300 359
305 358
310 361
315 359
320 362
325 362
330 362
335 360
340 362
345 361
350 362
355 361
360 362
365 362
370 362
375 362
380 363
385 363
390 363
395 362
400 362
405 363
410 364
415 364
420 363
425 365
430 364
435 365
440 364
445 365
450 365
455 366
460 366
465 366
470 365
475 365
480 365
485 366
490 366
495 366
500 366
505 361
510 365
515 363
520 365
525 363
530 364
535 362
540 364
545 364
550 363
555 365
560 363
565 365
570 364
575 365
580 364
585 363
590 364
595 365
600 363
605 365
610 362
615 376
620 366
625 367
630 365
635 367
640 366
645 366
650 368
655 366
660 366
665 365
670 367
675 367
680 366
685 367
690 366
695 365
700 368
705 368
710 368
715 366
720 368
725 366
730 367
735 366
740 367
745 367
750 367
755 367
760 367
765 368
770 365
775 369
780 368
785 367
790 368
795 367
800 367
805 368
810 367
815 368
820 367
825 368
830 366
835 367
840 368
845 368
850 367
855 367
860 366
865 368
870 367
875 366
880 368
885 367
890 367
895 367
900 367
905 367
910 366
915 367
920 366
925 368
930 367
935 367
940 367
945 368
950 366
955 367
960 366
965 366
970 366
975 364
980 367
985 367
990 366
995 368
1000 366
1005 368
1010 368
1015 367
1020 367
1025 368
1030 368
1035 366
1040 367
1045 367
1050 369
1055 367
1060 368
1065 368
1070 368
1075 367
1080 368
1085 367
1090 368
1095 367
1100 367
1105 367
1110 369
1115 367
1120 367
1125 367
1130 368
1135 367
1140 367
1145 377
1150 365
1155 368
1160 367
1165 367
1170 369
1175 368
1180 369
1185 368
1190 368
1195 369
1200 368
1205 368
1210 369
1215 367
1220 368
1225 369
1230 368
1235 368
1240 369
1245 368
1250 368
1255 368
1260 369
1265 368
1270 368
1275 367
1280 368
1285 367
1290 368
1295 368
1300 369
1305 367
1310 369
1315 368
1320 368
1325 367
1330 367
1335 368
1340 366
1345 368
1350 368
1355 368
1360 367
1365 368
1370 366
1375 368
1380 367
1385 366
1390 365
1395 366
1400 366
1405 367
1410 366
1415 367
1420 366
1425 367
1430 366
1435 367
1440 365
1445 366
1450 366
1455 367
1460 366
1465 367
1470 366
1475 366
1480 367
1485 366
1490 368
1495 367
1500 368
1505 367
1510 366
1515 366
1520 368
1525 365
1530 367
1535 367
1540 367
1545 365
1550 367
1555 366
1560 364
1565 366
1570 366
1575 367
1580 366
1585 368
1590 368
1595 367
1600 367
1605 367
1610 366
1615 367
1620 367
1625 366
1630 367
1635 367
1640 367
1645 366
1650 366
1655 367
1660 367
1665 367
1670 368
1675 368
1680 366
1685 367
1690 367
1695 366
1700 368
1705 367
1710 366
1715 366
1720 368
1725 367
1730 368
1735 367
1740 368
1745 368
1750 367
1755 367
1760 367
1765 368
1770 368
1775 367
1780 367
1785 368
1790 367
1795 367
1800 367
1805 368
1810 366
1815 368
1820 368
1825 368
1830 367
1835 368
1840 367
1845 368
1850 367
1855 369
1860 368
1865 366
1870 367
1875 367
1880 368
1885 368
1890 368
1895 368
1900 367
1905 368
1910 367
1915 367
1920 367
1925 366
1930 366
1935 368
1940 367
1945 368
1950 367
1955 368
1960 367
1965 367
1970 368
1975 370
1980 367
1985 367
1990 367
1995 367
2000 368
2005 369
2010 368
2015 367
2020 368
2025 367
2030 367
2035 367
2040 367
2045 368
2050 367
2055 367
2060 368
2065 366
2070 366
2075 366
2080 367
2085 368
2090 367
2095 366
2100 368
2105 366
2110 367
2115 366
2120 368
2125 367
2130 366
2135 366
2140 367
2145 366
2150 368
2155 367
2160 366
2165 367
2170 367
2175 367
2180 367
2185 367
2190 366
2195 366
2200 366
2205 366
2210 367
2215 368
2220 366
2225 368
2230 368
2235 367
2240 368
2245 368
2250 367
2255 367
2260 368
2265 368
2270 368
2275 368
2280 368
2285 367
2290 368
2295 367
2300 368
2305 368
2310 369
2315 367
2320 368
2325 367
2330 369
2335 368
2340 369
2345 367
2350 368
2355 368
2360 368
2365 368
2370 367
2375 367
2380 367
2385 368
2390 367
2395 368
2400 369
2405 367
2410 368
2415 368
2420 369
2425 368
2430 368
2435 369
2440 367
2445 368
2450 368
2455 368
2460 368
2465 369
2470 369
2475 367
2480 370
2485 345
2490 314
2495 313
2500 290
2505 273
2510 274
2515 272
2520 273
2525 267
2530 270
2535 259
2540 259
2545 258
2550 259
2555 256
2560 256
2565 256
2570 257
2575 260
2580 256
2585 257
2590 256
2595 256
2600 255
2605 255
2610 256
2615 256
2620 255
2625 255
2630 256
2635 256
2640 255
2645 257
2650 255
2655 255
2660 255
2665 255
2670 255
2675 255
2680 255
2685 255
2690 254
2695 256
2700 256
2705 254
2710 255
2715 254
2720 255
2725 255
2730 255
2735 254
2740 256
2745 254
2750 255
2755 255
2760 256
2765 252
2770 254
2775 255
2780 254
2785 254
2790 254
2795 256
2800 254
2805 254
2810 255
2815 255
2820 254
2825 255
2830 253
2835 255
2840 253
2845 255
2850 254
2855 254
2860 253
2865 254
2870 253
2875 257
2880 254
2885 254
2890 254
2895 254
2900 255
2905 253
2910 254
2915 254
2920 254
2925 253
2930 253
2935 253
2940 255
2945 254
2950 255
2955 253
2960 255
2965 254
2970 254
2975 254
2980 254
2985 253
2990 254
2995 254
3000 254
3005 255
3010 253
3015 254
3020 253
3025 253
3030 253
3035 253
3040 253
3045 251
3050 253
3055 255
3060 253
3065 254
3070 253
3075 255
3080 254
3085 254
3090 254
3095 254
3100 254
3105 253
3110 255
3115 253
3120 253
3125 253
3130 253
3135 253
3140 254
3145 253
3150 254
3155 253
3160 255
3165 254
3170 254
3175 253
3180 253
3185 254
3190 254
3195 254
3200 253
3205 253
3210 253
3215 253
3220 254
3225 253
3230 254
3235 254
3240 253
3245 253
3250 254
3255 255
3260 253
3265 254
3270 254
3275 254
3280 254
3285 254
3290 254
3295 253
3300 253
3305 253
3310 255
3315 254
3320 254
3325 253
3330 253
3335 254
3340 253
3345 253
3350 253
3355 254
3360 255
3365 253
3370 254
3375 255
3380 254
3385 254
3390 254
3395 253
3400 254
3405 254
3410 254
3415 253
3420 253
3425 254
3430 253
3435 255
3440 253
3445 254
3450 253
3455 253
3460 254
3465 253
3470 253
3475 253
3480 253
3485 254
3490 253
3495 254
3500 253
3505 253
3510 253
3515 254
3520 254
3525 253
3530 254
3535 253
3540 254
3545 253
3550 253
3555 254
3560 254
3565 253
3570 254
3575 253
3580 253
3585 253
3590 253
3595 253
};

\addlegendentry{Boosting enabled}
\addplot [thick, color1, mark=triangle*, mark size=3, mark options={solid,fill=white,draw=red},  mark repeat={180}]
table{0 257
5 264
10 266
15 265
20 265
25 265
30 265
35 265
40 257
45 256
50 267
55 255
60 256
65 256
70 257
75 256
80 257
85 257
90 257
95 256
100 256
105 256
110 256
115 256
120 256
125 255
130 255
135 256
140 256
145 256
150 255
155 257
160 256
165 255
170 256
175 255
180 256
185 256
190 258
195 264
200 263
205 264
210 264
215 264
220 263
225 262
230 261
235 261
240 260
245 261
250 262
255 263
260 263
265 264
270 260
275 264
280 308
285 317
290 318
295 318
300 318
305 317
310 317
315 318
320 318
325 319
330 318
335 319
340 319
345 318
350 318
355 320
360 320
365 319
370 319
375 320
380 319
385 319
390 320
395 319
400 320
405 320
410 319
415 319
420 319
425 319
430 319
435 320
440 321
445 319
450 320
455 319
460 319
465 320
470 320
475 320
480 320
485 319
490 320
495 320
500 321
505 319
510 321
515 321
520 319
525 320
530 321
535 319
540 321
545 321
550 321
555 320
560 320
565 321
570 320
575 320
580 320
585 319
590 321
595 321
600 320
605 320
610 321
615 321
620 320
625 321
630 321
635 320
640 321
645 320
650 321
655 321
660 321
665 321
670 319
675 320
680 319
685 320
690 319
695 321
700 320
705 319
710 320
715 320
720 319
725 323
730 319
735 321
740 319
745 320
750 320
755 320
760 320
765 321
770 320
775 320
780 320
785 319
790 321
795 320
800 321
805 320
810 321
815 320
820 320
825 320
830 319
835 321
840 321
845 320
850 321
855 320
860 319
865 320
870 320
875 319
880 320
885 320
890 320
895 319
900 320
905 320
910 320
915 320
920 320
925 320
930 331
935 317
940 321
945 319
950 320
955 320
960 320
965 320
970 320
975 320
980 321
985 320
990 321
995 319
1000 320
1005 320
1010 320
1015 320
1020 319
1025 321
1030 320
1035 319
1040 319
1045 320
1050 320
1055 320
1060 320
1065 320
1070 320
1075 320
1080 319
1085 321
1090 320
1095 320
1100 319
1105 320
1110 321
1115 320
1120 321
1125 320
1130 320
1135 321
1140 319
1145 321
1150 320
1155 319
1160 319
1165 320
1170 320
1175 321
1180 320
1185 319
1190 321
1195 321
1200 319
1205 320
1210 320
1215 318
1220 320
1225 319
1230 321
1235 319
1240 320
1245 322
1250 320
1255 321
1260 320
1265 321
1270 321
1275 320
1280 320
1285 321
1290 320
1295 320
1300 320
1305 321
1310 322
1315 320
1320 320
1325 323
1330 320
1335 321
1340 321
1345 321
1350 320
1355 320
1360 320
1365 320
1370 321
1375 320
1380 321
1385 320
1390 321
1395 320
1400 321
1405 321
1410 321
1415 321
1420 321
1425 322
1430 322
1435 330
1440 318
1445 320
1450 321
1455 321
1460 321
1465 321
1470 321
1475 321
1480 322
1485 322
1490 322
1495 322
1500 321
1505 321
1510 321
1515 321
1520 321
1525 322
1530 320
1535 320
1540 322
1545 321
1550 322
1555 321
1560 320
1565 322
1570 322
1575 321
1580 321
1585 321
1590 321
1595 320
1600 320
1605 322
1610 320
1615 321
1620 321
1625 322
1630 321
1635 321
1640 321
1645 319
1650 320
1655 321
1660 321
1665 321
1670 322
1675 321
1680 320
1685 319
1690 320
1695 322
1700 321
1705 321
1710 322
1715 321
1720 320
1725 322
1730 321
1735 321
1740 322
1745 321
1750 321
1755 321
1760 320
1765 321
1770 321
1775 320
1780 321
1785 319
1790 321
1795 321
1800 321
1805 320
1810 321
1815 321
1820 321
1825 319
1830 321
1835 320
1840 321
1845 321
1850 321
1855 321
1860 320
1865 320
1870 320
1875 321
1880 322
1885 322
1890 322
1895 320
1900 322
1905 320
1910 320
1915 321
1920 321
1925 321
1930 321
1935 320
1940 320
1945 320
1950 321
1955 320
1960 320
1965 321
1970 320
1975 320
1980 321
1985 320
1990 321
1995 321
2000 320
2005 321
2010 321
2015 319
2020 320
2025 320
2030 319
2035 321
2040 320
2045 320
2050 321
2055 321
2060 321
2065 320
2070 320
2075 320
2080 321
2085 320
2090 321
2095 320
2100 320
2105 321
2110 321
2115 319
2120 319
2125 319
2130 320
2135 320
2140 320
2145 320
2150 321
2155 320
2160 321
2165 320
2170 320
2175 321
2180 320
2185 321
2190 321
2195 319
2200 321
2205 321
2210 321
2215 321
2220 320
2225 321
2230 321
2235 321
2240 320
2245 320
2250 320
2255 321
2260 321
2265 320
2270 321
2275 321
2280 320
2285 321
2290 320
2295 321
2300 321
2305 321
2310 321
2315 320
2320 321
2325 322
2330 320
2335 320
2340 321
2345 321
2350 321
2355 321
2360 321
2365 319
2370 321
2375 320
2380 321
2385 321
2390 320
2395 320
2400 322
2405 321
2410 320
2415 321
2420 321
2425 321
2430 321
2435 321
2440 321
2445 321
2450 321
2455 321
2460 320
2465 320
2470 321
2475 321
2480 320
2485 321
2490 320
2495 322
2500 320
2505 322
2510 322
2515 322
2520 321
2525 321
2530 321
2535 321
2540 321
2545 320
2550 321
2555 321
2560 320
2565 320
2570 321
2575 318
2580 321
2585 321
2590 321
2595 322
2600 321
2605 320
2610 320
2615 320
2620 320
2625 320
2630 320
2635 320
2640 319
2645 320
2650 319
2655 320
2660 322
2665 321
2670 320
2675 321
2680 321
2685 320
2690 320
2695 320
2700 319
2705 319
2710 320
2715 320
2720 321
2725 321
2730 319
2735 319
2740 331
2745 318
2750 321
2755 321
2760 321
2765 320
2770 320
2775 320
2780 321
2785 320
2790 320
2795 320
2800 320
2805 320
2810 320
2815 320
2820 321
2825 320
2830 320
2835 320
2840 320
2845 321
2850 320
2855 320
2860 321
2865 320
2870 321
2875 320
2880 320
2885 321
2890 321
2895 320
2900 319
2905 321
2910 321
2915 321
2920 320
2925 320
2930 320
2935 321
2940 321
2945 320
2950 319
2955 320
2960 317
2965 320
2970 320
2975 319
2980 321
2985 321
2990 297
2995 290
3000 288
3005 280
3010 266
3015 265
3020 265
3025 264
3030 265
3035 265
3040 264
3045 257
3050 257
3055 256
3060 257
3065 257
3070 256
3075 257
3080 256
3085 257
3090 255
3095 255
3100 256
3105 255
3110 256
3115 255
3120 255
3125 256
3130 255
3135 255
3140 256
3145 255
3150 256
3155 255
3160 255
3165 256
3170 256
3175 256
3180 254
3185 256
3190 255
3195 256
3200 254
3205 255
3210 256
3215 254
3220 256
3225 256
3230 255
3235 255
3240 254
3245 254
3250 254
3255 254
3260 255
3265 255
3270 255
3275 255
3280 254
3285 255
3290 254
3295 255
3300 254
3305 255
3310 255
3315 254
3320 255
3325 254
3330 254
3335 255
3340 254
3345 255
3350 254
3355 255
3360 254
3365 255
3370 255
3375 255
3380 255
3385 255
3390 254
3395 255
3400 254
3405 255
3410 254
3415 255
3420 255
3425 254
3430 254
3435 254
3440 255
3445 255
3450 255
3455 254
3460 255
3465 254
3470 255
3475 255
3480 255
3485 256
3490 255
3495 256
3500 256
3505 256
3510 254
3515 254
3520 255
3525 255
3530 256
3535 255
3540 256
3545 255
3550 256
3555 255
3560 255
3565 255
3570 255
3575 256
3580 256
3585 255
3590 254
3595 254
3600 253
};
\addlegendentry{Boosting disabled}

\end{axis}

\end{tikzpicture}

%% file: results/energy_intel_boost.tex
\begin{tikzpicture}[font=\Large,spy using outlines= {circle, magnification=2, connect spies}]

\definecolor{color0}{rgb}{0.12156862745098,0.466666666666667,0.705882352941177}
\definecolor{color1}{rgb}{1,0.498039215686275,0.0549019607843137}
\definecolor{color2}{rgb}{0.172549019607843,0.627450980392157,0.172549019607843}
\definecolor{color3}{rgb}{0.83921568627451,0.152941176470588,0.156862745098039}
\definecolor{color4}{rgb}{0.580392156862745,0.403921568627451,0.741176470588235}
\definecolor{color5}{rgb}{0,0,0}

\begin{axis}[
legend cell align={left},
legend columns=2,
legend style={fill opacity=0.8, draw opacity=1, text opacity=1, at={(1.05,1.15)}, anchor=east, draw=white!80.0!black},
tick align=outside,
tick pos=left,
x grid style={white!69.01960784313725!black},
xlabel={Time (SEC)},
xmin=0, xmax=3600,
xtick={0,900,1800,2700,3600},
xtick style={color=black},
y grid style={white!69.01960784313725!black},
xmajorgrids,
ymajorgrids,
ylabel={Total energy consumption (kWh)},
ymin=0, ymax=0.35,
ytick style={color=black},
ytick={0.05,0.1,0.15,0.2,0.25,0.3,0.35},
yticklabel style={
        /pgf/number format/fixed,
        /pgf/number format/precision=2
},
 y label style={at={(axis description cs:-0.04,.5)}}
]
\addplot [semithick, color5]
table [y expr=(\thisrowno{1}-678611)/1000] {%
0 678611
5 678611
10 678611
15 678611
20 678612
25 678612
30 678613
35 678613
40 678613
45 678614
50 678614
55 678614
60 678615
65 678615
70 678615
75 678616
80 678616
85 678616
90 678617
95 678617
100 678617
105 678618
110 678618
115 678619
120 678619
125 678619
130 678620
135 678620
140 678620
145 678621
150 678621
155 678621
160 678622
165 678622
170 678622
175 678623
180 678623
185 678623
190 678624
195 678624
200 678625
205 678625
210 678625
215 678626
220 678626
225 678626
230 678627
235 678627
240 678627
245 678628
250 678628
255 678628
260 678629
265 678629
270 678629
275 678630
280 678630
285 678630
290 678631
295 678631
300 678632
305 678632
310 678632
315 678633
320 678633
325 678633
330 678634
335 678634
340 678634
345 678635
350 678635
355 678635
360 678636
365 678636
370 678636
375 678637
380 678637
385 678637
390 678638
395 678638
400 678639
405 678639
410 678639
415 678640
420 678640
425 678640
430 678641
435 678641
440 678641
445 678642
450 678642
455 678642
460 678643
465 678643
470 678643
475 678644
480 678644
485 678644
490 678645
495 678645
500 678645
505 678646
510 678646
515 678647
520 678647
525 678647
530 678648
535 678648
540 678648
545 678649
550 678649
555 678649
560 678650
565 678650
570 678650
575 678651
580 678651
585 678651
590 678652
595 678652
600 678653
605 678653
610 678653
615 678654
620 678654
625 678654
630 678655
635 678655
640 678655
645 678656
650 678656
655 678656
660 678657
665 678657
670 678657
675 678658
680 678658
685 678659
690 678659
695 678659
700 678660
705 678660
710 678660
715 678661
720 678661
725 678661
730 678662
735 678662
740 678662
745 678663
750 678663
755 678663
760 678664
765 678664
770 678664
775 678665
780 678665
785 678666
790 678666
795 678666
800 678667
805 678667
810 678667
815 678668
820 678668
825 678668
830 678669
835 678669
840 678669
845 678670
850 678670
855 678670
860 678671
865 678671
870 678671
875 678672
880 678672
885 678672
890 678673
895 678673
900 678674
905 678674
910 678674
915 678675
920 678675
925 678675
930 678676
935 678676
940 678676
945 678677
950 678677
955 678677
960 678678
965 678678
970 678678
975 678679
980 678679
985 678679
990 678680
995 678680
1000 678680
1005 678681
1010 678681
1015 678682
1020 678682
1025 678682
1030 678683
1035 678683
1040 678683
1045 678684
1050 678684
1055 678684
1060 678685
1065 678685
1070 678685
1075 678686
1080 678686
1085 678686
1090 678687
1095 678687
1100 678688
1105 678688
1110 678688
1115 678689
1120 678689
1125 678689
1130 678690
1135 678690
1140 678690
1145 678691
1150 678691
1155 678691
1160 678692
1165 678692
1170 678692
1175 678693
1180 678693
1185 678694
1190 678694
1195 678694
1200 678695
1205 678695
1210 678695
1215 678696
1220 678696
1225 678696
1230 678697
1235 678697
1240 678697
1245 678698
1250 678698
1255 678698
1260 678699
1265 678699
1270 678699
1275 678700
1280 678700
1285 678701
1290 678701
1295 678701
1300 678702
1305 678702
1310 678702
1315 678703
1320 678703
1325 678703
1330 678704
1335 678704
1340 678704
1345 678705
1350 678705
1355 678705
1360 678706
1365 678706
1370 678706
1375 678707
1380 678707
1385 678707
1390 678708
1395 678708
1400 678709
1405 678709
1410 678709
1415 678710
1420 678710
1425 678710
1430 678711
1435 678711
1440 678711
1445 678712
1450 678712
1455 678712
1460 678713
1465 678713
1470 678713
1475 678714
1480 678714
1485 678714
1490 678715
1495 678715
1500 678716
1505 678716
1510 678716
1515 678717
1520 678717
1525 678717
1530 678718
1535 678718
1540 678718
1545 678719
1550 678719
1555 678719
1560 678720
1565 678720
1570 678720
1575 678721
1580 678721
1585 678721
1590 678722
1595 678722
1600 678723
1605 678723
1610 678723
1615 678724
1620 678724
1625 678724
1630 678725
1635 678725
1640 678725
1645 678726
1650 678726
1655 678726
1660 678727
1665 678727
1670 678727
1675 678728
1680 678728
1685 678729
1690 678729
1695 678729
1700 678730
1705 678730
1710 678730
1715 678731
1720 678731
1725 678731
1730 678732
1735 678732
1740 678732
1745 678733
1750 678733
1755 678733
1760 678734
1765 678734
1770 678734
1775 678735
1780 678735
1785 678736
1790 678736
1795 678736
1800 678737
1805 678737
1810 678737
1815 678738
1820 678738
1825 678738
1830 678739
1835 678739
1840 678739
1845 678740
1850 678740
1855 678740
1860 678741
1865 678741
1870 678742
1875 678742
1880 678742
1885 678743
1890 678743
1895 678743
1900 678744
1905 678744
1910 678744
1915 678745
1920 678745
1925 678745
1930 678746
1935 678746
1940 678746
1945 678747
1950 678747
1955 678747
1960 678748
1965 678748
1970 678748
1975 678749
1980 678749
1985 678750
1990 678750
1995 678750
2000 678751
2005 678751
2010 678751
2015 678752
2020 678752
2025 678752
2030 678753
2035 678753
2040 678753
2045 678754
2050 678754
2055 678754
2060 678755
2065 678755
2070 678755
2075 678756
2080 678756
2085 678757
2090 678757
2095 678757
2100 678758
2105 678758
2110 678758
2115 678759
2120 678759
2125 678759
2130 678760
2135 678760
2140 678760
2145 678761
2150 678761
2155 678761
2160 678762
2165 678762
2170 678763
2175 678763
2180 678763
2185 678764
2190 678764
2195 678764
2200 678765
2205 678765
2210 678765
2215 678766
2220 678766
2225 678766
2230 678767
2235 678767
2240 678767
2245 678768
2250 678768
2255 678768
2260 678769
2265 678769
2270 678769
2275 678770
2280 678770
2285 678771
2290 678771
2295 678771
2300 678772
2305 678772
2310 678772
2315 678773
2320 678773
2325 678773
2330 678774
2335 678774
2340 678774
2345 678775
2350 678775
2355 678776
2360 678776
2365 678776
2370 678777
2375 678777
2380 678777
2385 678778
2390 678778
2395 678778
2400 678779
2405 678779
2410 678779
2415 678780
2420 678780
2425 678780
2430 678781
2435 678781
2440 678782
2445 678782
2450 678782
2455 678783
2460 678783
2465 678783
2470 678784
2475 678784
2480 678784
2485 678785
2490 678785
2495 678785
2500 678786
2505 678786
2510 678786
2515 678787
2520 678787
2525 678788
2530 678788
2535 678788
2540 678789
2545 678789
2550 678789
2555 678790
2560 678790
2565 678790
2570 678791
2575 678791
2580 678791
2585 678792
2590 678792
2595 678792
2600 678793
2605 678793
2610 678793
2615 678794
2620 678794
2625 678795
2630 678795
2635 678795
2640 678796
2645 678796
2650 678796
2655 678797
2660 678797
2665 678797
2670 678798
2675 678798
2680 678798
2685 678799
2690 678799
2695 678799
2700 678800
2705 678800
2710 678800
2715 678801
2720 678801
2725 678802
2730 678802
2735 678802
2740 678803
2745 678803
2750 678803
2755 678804
2760 678804
2765 678804
2770 678805
2775 678805
2780 678805
2785 678806
2790 678806
2795 678807
2800 678807
2805 678807
2810 678808
2815 678808
2820 678808
2825 678809
2830 678809
2835 678809
2840 678810
2845 678810
2850 678810
2855 678811
2860 678811
2865 678811
2870 678812
2875 678812
2880 678812
2885 678813
2890 678813
2895 678813
2900 678814
2905 678814
2910 678815
2915 678815
2920 678815
2925 678816
2930 678816
2935 678816
2940 678817
2945 678817
2950 678817
2955 678818
2960 678818
2965 678818
2970 678819
2975 678819
2980 678819
2985 678820
2990 678820
2995 678821
3000 678821
3005 678821
3010 678822
3015 678822
3020 678822
3025 678823
3030 678823
3035 678823
3040 678824
3045 678824
3050 678824
3055 678825
3060 678825
3065 678825
3070 678826
3075 678826
3080 678827
3085 678827
3090 678827
3095 678828
3100 678828
3105 678828
3110 678829
3115 678829
3120 678829
3125 678830
3130 678830
3135 678830
3140 678831
3145 678831
3150 678831
3155 678832
3160 678832
3165 678832
3170 678833
3175 678833
3180 678833
3185 678834
3190 678834
3195 678835
3200 678835
3205 678835
3210 678836
3215 678836
3220 678836
3225 678837
3230 678837
3235 678837
3240 678838
3245 678838
3250 678838
3255 678839
3260 678839
3265 678839
3270 678840
3275 678840
3280 678841
3285 678841
3290 678841
3295 678842
3300 678842
3305 678842
3310 678843
3315 678843
3320 678843
3325 678844
3330 678844
3335 678844
3340 678845
3345 678845
3350 678845
3355 678846
3360 678846
3365 678846
3370 678847
3375 678847
3380 678847
3385 678848
3390 678848
3395 678849
3400 678849
3405 678849
3410 678850
3415 678850
3420 678850
3425 678851
3430 678851
3435 678851
3440 678852
3445 678852
3450 678852
3455 678853
3460 678853
3465 678853
3470 678854
3475 678854
3480 678855
3485 678855
3490 678855
3495 678856
3500 678856
3505 678856
3510 678857
3515 678857
3520 678857
3525 678858
3530 678858
3535 678858
3540 678859
3545 678859
3550 678859
3555 678860
3560 678860
3565 678861
3570 678861
3575 678861
3580 678862
3585 678862
3590 678862
3595 678863
3600 678863
};
\addlegendentry{Baseline}
\addplot [semithick, color2, mark=square*, mark size=3, mark options={solid,fill=white,draw=red},  mark repeat={160}]
table[y expr=(\thisrowno{1}-368832)/1000] {%
0 368832
5 368832
10 368833
15 368833
20 368833
25 368834
30 368834
35 368834
40 368835
45 368835
50 368835
55 368836
60 368836
65 368836
70 368837
75 368837
80 368838
85 368838
90 368838
95 368839
100 368839
105 368839
110 368840
115 368840
120 368840
125 368841
130 368841
135 368842
140 368842
145 368842
150 368843
155 368843
160 368843
165 368844
170 368844
175 368845
180 368845
185 368846
190 368846
195 368846
200 368847
205 368847
210 368848
215 368849
220 368849
225 368849
230 368850
235 368850
240 368851
245 368852
250 368852
255 368852
260 368853
265 368853
270 368854
275 368854
280 368855
285 368855
290 368856
295 368856
300 368857
305 368857
310 368858
315 368859
320 368859
325 368860
330 368860
335 368861
340 368861
345 368862
350 368862
355 368863
360 368863
365 368864
370 368864
375 368865
380 368865
385 368866
390 368866
395 368867
400 368867
405 368868
410 368868
415 368869
420 368869
425 368870
430 368870
435 368871
440 368871
445 368872
450 368872
455 368873
460 368873
465 368874
470 368874
475 368875
480 368875
485 368876
490 368876
495 368877
500 368877
505 368878
510 368878
515 368879
520 368879
525 368880
530 368880
535 368880
540 368881
545 368881
550 368882
555 368882
560 368883
565 368883
570 368884
575 368885
580 368885
585 368886
590 368886
595 368887
600 368887
605 368888
610 368888
615 368889
620 368889
625 368890
630 368890
635 368891
640 368891
645 368892
650 368892
655 368893
660 368893
665 368894
670 368894
675 368895
680 368895
685 368896
690 368896
695 368897
700 368897
705 368898
710 368898
715 368899
720 368899
725 368900
730 368900
735 368901
740 368901
745 368902
750 368902
755 368903
760 368903
765 368904
770 368904
775 368905
780 368905
785 368906
790 368906
795 368907
800 368907
805 368908
810 368908
815 368909
820 368909
825 368910
830 368911
835 368911
840 368912
845 368912
850 368913
855 368913
860 368914
865 368914
870 368915
875 368915
880 368916
885 368916
890 368917
895 368917
900 368918
905 368918
910 368919
915 368919
920 368920
925 368920
930 368921
935 368921
940 368922
945 368922
950 368923
955 368923
960 368924
965 368924
970 368925
975 368925
980 368926
985 368926
990 368927
995 368927
1000 368928
1005 368928
1010 368929
1015 368929
1020 368930
1025 368930
1030 368931
1035 368931
1040 368932
1045 368932
1050 368933
1055 368934
1060 368934
1065 368935
1070 368935
1075 368936
1080 368936
1085 368937
1090 368937
1095 368938
1100 368938
1105 368939
1110 368939
1115 368940
1120 368940
1125 368941
1130 368941
1135 368942
1140 368942
1145 368943
1150 368943
1155 368944
1160 368944
1165 368945
1170 368945
1175 368946
1180 368946
1185 368947
1190 368947
1195 368948
1200 368948
1205 368949
1210 368949
1215 368950
1220 368950
1225 368951
1230 368951
1235 368952
1240 368952
1245 368953
1250 368953
1255 368954
1260 368954
1265 368955
1270 368955
1275 368956
1280 368956
1285 368957
1290 368958
1295 368958
1300 368959
1305 368959
1310 368960
1315 368960
1320 368961
1325 368961
1330 368962
1335 368962
1340 368963
1345 368963
1350 368964
1355 368964
1360 368965
1365 368965
1370 368966
1375 368966
1380 368967
1385 368967
1390 368968
1395 368968
1400 368969
1405 368969
1410 368970
1415 368970
1420 368971
1425 368971
1430 368972
1435 368972
1440 368973
1445 368973
1450 368974
1455 368974
1460 368975
1465 368975
1470 368976
1475 368976
1480 368977
1485 368977
1490 368978
1495 368978
1500 368979
1505 368979
1510 368980
1515 368980
1520 368981
1525 368982
1530 368982
1535 368983
1540 368983
1545 368984
1550 368984
1555 368985
1560 368985
1565 368986
1570 368986
1575 368987
1580 368987
1585 368988
1590 368988
1595 368989
1600 368989
1605 368990
1610 368990
1615 368991
1620 368991
1625 368992
1630 368992
1635 368993
1640 368993
1645 368994
1650 368994
1655 368995
1660 368995
1665 368996
1670 368996
1675 368997
1680 368997
1685 368998
1690 368998
1695 368999
1700 368999
1705 369000
1710 369000
1715 369001
1720 369001
1725 369002
1730 369002
1735 369003
1740 369003
1745 369004
1750 369004
1755 369005
1760 369005
1765 369006
1770 369006
1775 369007
1780 369007
1785 369008
1790 369008
1795 369009
1800 369009
1805 369010
1810 369010
1815 369011
1820 369011
1825 369012
1830 369012
1835 369013
1840 369014
1845 369014
1850 369014
1855 369015
1860 369016
1865 369016
1870 369017
1875 369017
1880 369018
1885 369018
1890 369019
1895 369019
1900 369020
1905 369020
1910 369021
1915 369021
1920 369022
1925 369022
1930 369023
1935 369023
1940 369024
1945 369024
1950 369025
1955 369025
1960 369026
1965 369026
1970 369027
1975 369027
1980 369028
1985 369028
1990 369029
1995 369029
2000 369030
2005 369030
2010 369031
2015 369031
2020 369032
2025 369032
2030 369033
2035 369033
2040 369034
2045 369034
2050 369035
2055 369035
2060 369036
2065 369036
2070 369037
2075 369037
2080 369038
2085 369038
2090 369039
2095 369040
2100 369040
2105 369041
2110 369041
2115 369041
2120 369042
2125 369043
2130 369043
2135 369044
2140 369044
2145 369045
2150 369045
2155 369046
2160 369046
2165 369047
2170 369047
2175 369048
2180 369048
2185 369049
2190 369049
2195 369050
2200 369050
2205 369051
2210 369051
2215 369052
2220 369052
2225 369053
2230 369053
2235 369054
2240 369054
2245 369055
2250 369055
2255 369056
2260 369056
2265 369057
2270 369057
2275 369058
2280 369058
2285 369059
2290 369059
2295 369060
2300 369060
2305 369061
2310 369061
2315 369062
2320 369062
2325 369063
2330 369063
2335 369064
2340 369064
2345 369065
2350 369065
2355 369066
2360 369066
2365 369067
2370 369067
2375 369068
2380 369068
2385 369069
2390 369070
2395 369070
2400 369070
2405 369071
2410 369071
2415 369072
2420 369073
2425 369073
2430 369074
2435 369074
2440 369075
2445 369075
2450 369075
2455 369076
2460 369076
2465 369077
2470 369077
2475 369077
2480 369078
2485 369078
2490 369079
2495 369079
2500 369079
2505 369080
2510 369080
2515 369080
2520 369081
2525 369081
2530 369081
2535 369082
2540 369082
2545 369082
2550 369083
2555 369083
2560 369083
2565 369084
2570 369084
2575 369085
2580 369085
2585 369085
2590 369086
2595 369086
2600 369086
2605 369087
2610 369087
2615 369087
2620 369088
2625 369088
2630 369088
2635 369089
2640 369089
2645 369089
2650 369090
2655 369090
2660 369091
2665 369091
2670 369091
2675 369092
2680 369092
2685 369092
2690 369093
2695 369093
2700 369093
2705 369094
2710 369094
2715 369094
2720 369095
2725 369095
2730 369095
2735 369096
2740 369096
2745 369097
2750 369097
2755 369097
2760 369098
2765 369098
2770 369098
2775 369099
2780 369099
2785 369099
2790 369100
2795 369100
2800 369100
2805 369101
2810 369101
2815 369101
2820 369102
2825 369102
2830 369102
2835 369103
2840 369103
2845 369104
2850 369104
2855 369104
2860 369105
2865 369105
2870 369105
2875 369106
2880 369106
2885 369106
2890 369107
2895 369107
2900 369107
2905 369108
2910 369108
2915 369108
2920 369109
2925 369109
2930 369109
2935 369110
2940 369110
2945 369110
2950 369111
2955 369111
2960 369111
2965 369112
2970 369112
2975 369113
2980 369113
2985 369113
2990 369114
2995 369114
3000 369114
3005 369115
3010 369115
3015 369115
3020 369116
3025 369116
3030 369116
3035 369117
3040 369117
3045 369117
3050 369118
3055 369118
3060 369118
3065 369119
3070 369119
3075 369119
3080 369120
3085 369120
3090 369121
3095 369121
3100 369121
3105 369122
3110 369122
3115 369122
3120 369123
3125 369123
3130 369123
3135 369124
3140 369124
3145 369124
3150 369125
3155 369125
3160 369125
3165 369126
3170 369126
3175 369126
3180 369127
3185 369127
3190 369128
3195 369128
3200 369128
3205 369129
3210 369129
3215 369129
3220 369130
3225 369130
3230 369130
3235 369131
3240 369131
3245 369131
3250 369132
3255 369132
3260 369132
3265 369133
3270 369133
3275 369134
3280 369134
3285 369134
3290 369135
3295 369135
3300 369135
3305 369136
3310 369136
3315 369136
3320 369137
3325 369137
3330 369137
3335 369138
3340 369138
3345 369138
3350 369139
3355 369139
3360 369139
3365 369140
3370 369140
3375 369141
3380 369141
3385 369141
3390 369142
3395 369142
3400 369142
3405 369143
3410 369143
3415 369143
3420 369144
3425 369144
3430 369144
3435 369145
3440 369145
3445 369145
3450 369146
3455 369146
3460 369146
3465 369147
3470 369147
3475 369148
3480 369148
3485 369148
3490 369149
3495 369149
3500 369149
3505 369150
3510 369150
3515 369150
3520 369151
3525 369151
3530 369151
3535 369152
3540 369152
3545 369153
3550 369153
3555 369153
3560 369154
3565 369154
3570 369154
3575 369155
3580 369155
3585 369155
3590 369156
3595 369156
3600 369156
};
\addlegendentry{Boosting enabled}
\addplot [semithick, color1, mark=triangle*, mark size=3, mark options={solid,fill=white,draw=red},  mark repeat={180}]
table[y expr=(\thisrowno{1}-666249)/1000] {
0 666249
5 666250
10 666250
15 666251
20 666251
25 666251
30 666252
35 666252
40 666252
45 666253
50 666253
55 666253
60 666254
65 666254
70 666254
75 666255
80 666255
85 666256
90 666256
95 666256
100 666257
105 666257
110 666258
115 666258
120 666258
125 666259
130 666259
135 666259
140 666260
145 666260
150 666260
155 666261
160 666261
165 666261
170 666262
175 666262
180 666262
185 666263
190 666263
195 666264
200 666264
205 666264
210 666265
215 666265
220 666265
225 666266
230 666266
235 666266
240 666267
245 666267
250 666268
255 666268
260 666268
265 666269
270 666269
275 666269
280 666270
285 666270
290 666271
295 666271
300 666271
305 666272
310 666272
315 666273
320 666273
325 666274
330 666274
335 666275
340 666275
345 666275
350 666276
355 666276
360 666277
365 666277
370 666278
375 666278
380 666278
385 666279
390 666279
395 666280
400 666280
405 666281
410 666281
415 666282
420 666282
425 666282
430 666283
435 666283
440 666284
445 666284
450 666285
455 666285
460 666286
465 666286
470 666286
475 666287
480 666287
485 666288
490 666288
495 666289
500 666289
505 666290
510 666290
515 666290
520 666291
525 666291
530 666292
535 666292
540 666293
545 666293
550 666293
555 666294
560 666294
565 666295
570 666295
575 666296
580 666296
585 666297
590 666297
595 666297
600 666298
605 666298
610 666299
615 666299
620 666300
625 666300
630 666301
635 666301
640 666301
645 666302
650 666302
655 666303
660 666303
665 666304
670 666304
675 666304
680 666305
685 666305
690 666306
695 666306
700 666306
705 666307
710 666307
715 666308
720 666308
725 666309
730 666309
735 666310
740 666310
745 666310
750 666311
755 666311
760 666312
765 666312
770 666313
775 666313
780 666314
785 666314
790 666314
795 666315
800 666315
805 666316
810 666316
815 666317
820 666317
825 666318
830 666318
835 666318
840 666319
845 666319
850 666320
855 666320
860 666321
865 666321
870 666322
875 666322
880 666322
885 666323
890 666323
895 666324
900 666324
905 666325
910 666325
915 666326
920 666326
925 666326
930 666327
935 666327
940 666328
945 666328
950 666329
955 666329
960 666330
965 666330
970 666330
975 666331
980 666331
985 666332
990 666332
995 666333
1000 666333
1005 666334
1010 666334
1015 666334
1020 666335
1025 666335
1030 666336
1035 666336
1040 666337
1045 666337
1050 666337
1055 666338
1060 666338
1065 666339
1070 666339
1075 666340
1080 666340
1085 666341
1090 666341
1095 666341
1100 666342
1105 666342
1110 666343
1115 666343
1120 666344
1125 666344
1130 666345
1135 666345
1140 666345
1145 666346
1150 666346
1155 666347
1160 666347
1165 666348
1170 666348
1175 666349
1180 666349
1185 666349
1190 666350
1195 666350
1200 666351
1205 666351
1210 666352
1215 666352
1220 666352
1225 666353
1230 666353
1235 666354
1240 666354
1245 666355
1250 666355
1255 666356
1260 666356
1265 666357
1270 666357
1275 666357
1280 666358
1285 666358
1290 666359
1295 666359
1300 666360
1305 666360
1310 666360
1315 666361
1320 666361
1325 666362
1330 666362
1335 666363
1340 666363
1345 666364
1350 666364
1355 666364
1360 666365
1365 666365
1370 666366
1375 666366
1380 666367
1385 666367
1390 666368
1395 666368
1400 666369
1405 666369
1410 666369
1415 666370
1420 666370
1425 666371
1430 666371
1435 666372
1440 666372
1445 666372
1450 666373
1455 666373
1460 666374
1465 666374
1470 666375
1475 666375
1480 666376
1485 666376
1490 666377
1495 666377
1500 666377
1505 666378
1510 666378
1515 666379
1520 666379
1525 666380
1530 666380
1535 666380
1540 666381
1545 666381
1550 666382
1555 666382
1560 666383
1565 666383
1570 666384
1575 666384
1580 666384
1585 666385
1590 666385
1595 666386
1600 666386
1605 666387
1610 666387
1615 666388
1620 666388
1625 666388
1630 666389
1635 666389
1640 666390
1645 666390
1650 666391
1655 666391
1660 666392
1665 666392
1670 666392
1675 666393
1680 666393
1685 666394
1690 666394
1695 666395
1700 666395
1705 666396
1710 666396
1715 666396
1720 666397
1725 666397
1730 666398
1735 666398
1740 666399
1745 666399
1750 666400
1755 666400
1760 666401
1765 666401
1770 666401
1775 666402
1780 666402
1785 666403
1790 666403
1795 666404
1800 666404
1805 666404
1810 666405
1815 666405
1820 666406
1825 666406
1830 666407
1835 666407
1840 666408
1845 666408
1850 666408
1855 666409
1860 666409
1865 666410
1870 666410
1875 666411
1880 666411
1885 666412
1890 666412
1895 666412
1900 666413
1905 666413
1910 666414
1915 666414
1920 666415
1925 666415
1930 666415
1935 666416
1940 666416
1945 666417
1950 666417
1955 666418
1960 666418
1965 666419
1970 666419
1975 666419
1980 666420
1985 666420
1990 666421
1995 666421
2000 666422
2005 666422
2010 666423
2015 666423
2020 666423
2025 666424
2030 666424
2035 666425
2040 666425
2045 666426
2050 666426
2055 666427
2060 666427
2065 666427
2070 666428
2075 666428
2080 666429
2085 666429
2090 666430
2095 666430
2100 666430
2105 666431
2110 666431
2115 666432
2120 666432
2125 666433
2130 666433
2135 666434
2140 666434
2145 666435
2150 666435
2155 666435
2160 666436
2165 666436
2170 666437
2175 666437
2180 666438
2185 666438
2190 666438
2195 666439
2200 666439
2205 666440
2210 666440
2215 666441
2220 666441
2225 666442
2230 666442
2235 666443
2240 666443
2245 666443
2250 666444
2255 666444
2260 666445
2265 666445
2270 666446
2275 666446
2280 666447
2285 666447
2290 666447
2295 666448
2300 666448
2305 666449
2310 666449
2315 666450
2320 666450
2325 666451
2330 666451
2335 666451
2340 666452
2345 666452
2350 666453
2355 666453
2360 666454
2365 666454
2370 666455
2375 666455
2380 666455
2385 666456
2390 666456
2395 666457
2400 666457
2405 666458
2410 666458
2415 666458
2420 666459
2425 666459
2430 666460
2435 666460
2440 666461
2445 666461
2450 666462
2455 666462
2460 666462
2465 666463
2470 666463
2475 666464
2480 666464
2485 666465
2490 666465
2495 666465
2500 666466
2505 666466
2510 666467
2515 666467
2520 666468
2525 666468
2530 666469
2535 666469
2540 666469
2545 666470
2550 666470
2555 666471
2560 666471
2565 666472
2570 666472
2575 666473
2580 666473
2585 666473
2590 666474
2595 666474
2600 666475
2605 666475
2610 666476
2615 666476
2620 666477
2625 666477
2630 666477
2635 666478
2640 666478
2645 666479
2650 666479
2655 666480
2660 666480
2665 666481
2670 666481
2675 666481
2680 666482
2685 666482
2690 666483
2695 666483
2700 666484
2705 666484
2710 666484
2715 666485
2720 666485
2725 666486
2730 666486
2735 666487
2740 666487
2745 666488
2750 666488
2755 666488
2760 666489
2765 666489
2770 666490
2775 666490
2780 666491
2785 666491
2790 666492
2795 666492
2800 666492
2805 666493
2810 666493
2815 666494
2820 666494
2825 666495
2830 666495
2835 666496
2840 666496
2845 666497
2850 666497
2855 666497
2860 666498
2865 666498
2870 666499
2875 666499
2880 666500
2885 666500
2890 666500
2895 666501
2900 666501
2905 666502
2910 666502
2915 666503
2920 666503
2925 666504
2930 666504
2935 666504
2940 666505
2945 666505
2950 666506
2955 666506
2960 666507
2965 666507
2970 666508
2975 666508
2980 666508
2985 666509
2990 666509
2995 666510
3000 666510
3005 666510
3010 666511
3015 666511
3020 666512
3025 666512
3030 666512
3035 666513
3040 666513
3045 666513
3050 666514
3055 666514
3060 666515
3065 666515
3070 666515
3075 666516
3080 666516
3085 666516
3090 666517
3095 666517
3100 666517
3105 666518
3110 666518
3115 666518
3120 666519
3125 666519
3130 666520
3135 666520
3140 666520
3145 666521
3150 666521
3155 666521
3160 666522
3165 666522
3170 666522
3175 666523
3180 666523
3185 666523
3190 666524
3195 666524
3200 666524
3205 666525
3210 666525
3215 666525
3220 666526
3225 666526
3230 666527
3235 666527
3240 666527
3245 666528
3250 666528
3255 666528
3260 666529
3265 666529
3270 666529
3275 666530
3280 666530
3285 666530
3290 666531
3295 666531
3300 666531
3305 666532
3310 666532
3315 666532
3320 666533
3325 666533
3330 666534
3335 666534
3340 666534
3345 666535
3350 666535
3355 666535
3360 666536
3365 666536
3370 666536
3375 666537
3380 666537
3385 666538
3390 666538
3395 666538
3400 666539
3405 666539
3410 666539
3415 666540
3420 666540
3425 666540
3430 666541
3435 666541
3440 666541
3445 666542
3450 666542
3455 666542
3460 666543
3465 666543
3470 666543
3475 666544
3480 666544
3485 666545
3490 666545
3495 666545
3500 666546
3505 666546
3510 666546
3515 666547
3520 666547
3525 666547
3530 666548
3535 666548
3540 666548
3545 666549
3550 666549
3555 666549
3560 666550
3565 666550
3570 666550
3575 666551
3580 666551
3585 666552
3590 666552
3595 666552
3600 666553
};
\addlegendentry{Boosting disabled}

\addplot [semithick, color2,mark=square*, mark size=3, mark options={solid,fill=red,draw=red}]
table[y expr=(\thisrowno{1}-368832)/1000]{
2530 369081
};
\addplot [semithick, color1, mark=triangle*, mark size=3, mark options={solid,fill=red,draw=red}]
table[y expr=(\thisrowno{1}-666249)/1000]{
3040 666513
};
\end{axis}
\node[text width=8cm] at (3.5,-1.5) {Filled markers indicate the end of jobs};

\end{tikzpicture}

%% file: results/power_amd_boost.tex
\begin{tikzpicture}[font=\Large]

\definecolor{color0}{rgb}{0.12156862745098,0.466666666666667,0.705882352941177}
\definecolor{color1}{rgb}{1,0.498039215686275,0.0549019607843137}
\definecolor{color2}{rgb}{0.172549019607843,0.627450980392157,0.172549019607843}
\definecolor{color3}{rgb}{0.83921568627451,0.152941176470588,0.156862745098039}
\definecolor{color4}{rgb}{0.580392156862745,0.403921568627451,0.741176470588235}
\definecolor{color5}{rgb}{0,0,0}

\begin{axis}[
legend cell align={left},
legend columns=2,
legend style={fill opacity=0.8, draw opacity=1, text opacity=1, at={(1.02,1.18)}, anchor=east, draw=white!80.0!black},
tick align=outside,
tick pos=left,
x grid style={white!69.01960784313725!black},
xlabel={Time (SEC)},
xmin=0, xmax=3600,
xtick style={color=black},
xtick={0,900,1800,2700,3600},
y grid style={white!69.01960784313725!black},
ylabel={Power consumption (W)},
ymin=150, ymax=260,
xmajorgrids,
ymajorgrids,
ytick={150,200,230,250},
ytick style={color=black}
]
\addplot [thick, color5, mark=.]
table {%
0 205
5 205
10 205
15 205
20 205
25 205
30 205
35 205
40 206
45 206
50 205
55 205
60 205
65 206
70 205
75 205
80 205
85 205
90 205
95 206
100 205
105 205
110 205
115 205
120 205
125 205
130 205
135 205
140 206
145 205
150 205
155 205
160 205
165 205
170 205
175 206
180 205
185 206
190 205
195 206
200 205
205 205
210 206
215 205
220 206
225 206
230 205
235 205
240 205
245 205
250 205
255 206
260 206
265 205
270 205
275 205
280 205
285 206
290 205
295 204
300 206
305 205
310 205
315 205
320 206
325 205
330 205
335 206
340 205
345 205
350 206
355 206
360 206
365 205
370 205
375 205
380 206
385 205
390 205
395 205
400 205
405 205
410 205
415 205
420 205
425 206
430 205
435 205
440 205
445 205
450 205
455 205
460 205
465 205
470 205
475 205
480 205
485 206
490 205
495 205
500 205
505 205
510 205
515 205
520 206
525 206
530 205
535 205
540 205
545 205
550 205
555 205
560 206
565 206
570 206
575 205
580 206
585 206
590 205
595 205
600 205
605 206
610 205
615 206
620 206
625 205
630 205
635 206
640 206
645 206
650 205
655 206
660 205
665 205
670 205
675 206
680 205
685 205
690 206
695 205
700 206
705 205
710 205
715 205
720 206
725 205
730 205
735 204
740 206
745 206
750 205
755 205
760 206
765 206
770 206
775 205
780 205
785 206
790 205
795 205
800 205
805 206
810 205
815 206
820 205
825 205
830 205
835 205
840 206
845 205
850 206
855 206
860 206
865 206
870 206
875 205
880 206
885 205
890 205
895 206
900 205
905 206
910 206
915 204
920 205
925 205
930 205
935 206
940 205
945 205
950 205
955 206
960 206
965 205
970 206
975 206
980 205
985 205
990 205
995 205
1000 205
1005 205
1010 205
1015 206
1020 206
1025 206
1030 206
1035 205
1040 205
1045 205
1050 205
1055 205
1060 205
1065 206
1070 205
1075 205
1080 205
1085 206
1090 206
1095 206
1100 205
1105 205
1110 206
1115 205
1120 205
1125 205
1130 205
1135 205
1140 205
1145 205
1150 205
1155 205
1160 205
1165 206
1170 205
1175 206
1180 206
1185 205
1190 205
1195 205
1200 206
1205 205
1210 206
1215 205
1220 205
1225 206
1230 205
1235 206
1240 206
1245 205
1250 206
1255 206
1260 206
1265 205
1270 206
1275 206
1280 205
1285 204
1290 206
1295 204
1300 206
1305 206
1310 206
1315 206
1320 205
1325 205
1330 205
1335 205
1340 205
1345 205
1350 206
1355 205
1360 205
1365 205
1370 205
1375 206
1380 206
1385 205
1390 206
1395 206
1400 205
1405 206
1410 205
1415 205
1420 206
1425 205
1430 205
1435 205
1440 205
1445 206
1450 206
1455 206
1460 206
1465 205
1470 206
1475 206
1480 205
1485 205
1490 205
1495 205
1500 205
1505 205
1510 205
1515 205
1520 205
1525 206
1530 205
1535 205
1540 205
1545 205
1550 205
1555 204
1560 204
1565 204
1570 205
1575 205
1580 205
1585 205
1590 205
1595 205
1600 205
1605 205
1610 205
1615 205
1620 205
1625 205
1630 205
1635 205
1640 205
1645 205
1650 206
1655 205
1660 205
1665 205
1670 205
1675 205
1680 204
1685 205
1690 206
1695 205
1700 206
1705 205
1710 205
1715 206
1720 206
1725 206
1730 205
1735 206
1740 205
1745 206
1750 205
1755 205
1760 205
1765 205
1770 205
1775 205
1780 206
1785 205
1790 205
1795 205
1800 205
1805 205
1810 206
1815 205
1820 205
1825 206
1830 205
1835 205
1840 205
1845 205
1850 205
1855 206
1860 206
1865 206
1870 205
1875 205
1880 205
1885 205
1890 206
1895 205
1900 206
1905 206
1910 206
1915 205
1920 206
1925 206
1930 205
1935 205
1940 205
1945 206
1950 206
1955 205
1960 205
1965 205
1970 205
1975 205
1980 205
1985 205
1990 205
1995 205
2000 205
2005 205
2010 205
2015 205
2020 205
2025 205
2030 205
2035 206
2040 206
2045 205
2050 205
2055 205
2060 206
2065 205
2070 205
2075 205
2080 205
2085 205
2090 206
2095 205
2100 205
2105 205
2110 205
2115 205
2120 205
2125 205
2130 205
2135 205
2140 205
2145 206
2150 205
2155 205
2160 206
2165 206
2170 205
2175 206
2180 205
2185 206
2190 206
2195 206
2200 206
2205 205
2210 206
2215 205
2220 206
2225 206
2230 206
2235 205
2240 206
2245 206
2250 205
2255 205
2260 205
2265 205
2270 205
2275 205
2280 206
2285 206
2290 206
2295 205
2300 206
2305 205
2310 205
2315 205
2320 206
2325 206
2330 206
2335 205
2340 205
2345 205
2350 206
2355 205
2360 205
2365 205
2370 206
2375 206
2380 206
2385 205
2390 205
2395 206
2400 206
2405 205
2410 205
2415 205
2420 205
2425 205
2430 206
2435 205
2440 206
2445 206
2450 206
2455 206
2460 206
2465 205
2470 205
2475 205
2480 206
2485 206
2490 205
2495 206
2500 205
2505 206
2510 205
2515 206
2520 206
2525 206
2530 206
2535 206
2540 205
2545 206
2550 206
2555 206
2560 205
2565 206
2570 205
2575 206
2580 205
2585 205
2590 206
2595 205
2600 205
2605 206
2610 205
2615 205
2620 206
2625 205
2630 205
2635 204
2640 205
2645 205
2650 206
2655 205
2660 205
2665 206
2670 205
2675 205
2680 205
2685 205
2690 206
2695 205
2700 206
2705 205
2710 205
2715 206
2720 205
2725 206
2730 206
2735 206
2740 205
2745 207
2750 204
2755 204
2760 204
2765 205
2770 204
2775 204
2780 204
2785 203
2790 205
2795 204
2800 204
2805 204
2810 204
2815 204
2820 204
2825 204
2830 204
2835 204
2840 204
2845 204
2850 204
2855 204
2860 204
2865 204
2870 204
2875 204
2880 204
2885 204
2890 204
2895 204
2900 204
2905 204
2910 204
2915 204
2920 204
2925 204
2930 204
2935 205
2940 204
2945 203
2950 204
2955 205
2960 203
2965 204
2970 204
2975 205
2980 204
2985 205
2990 205
2995 205
3000 205
3005 205
3010 204
3015 204
3020 205
3025 205
3030 204
3035 204
3040 204
3045 204
3050 204
3055 204
3060 204
3065 204
3070 204
3075 204
3080 204
3085 205
3090 204
3095 204
3100 205
3105 205
3110 204
3115 204
3120 204
3125 204
3130 205
3135 205
3140 204
3145 205
3150 205
3155 204
3160 204
3165 204
3170 204
3175 204
3180 205
3185 205
3190 205
3195 204
3200 204
3205 204
3210 204
3215 204
3220 204
3225 204
3230 204
3235 204
3240 204
3245 204
3250 204
3255 204
3260 204
3265 205
3270 204
3275 204
3280 204
3285 204
3290 205
3295 204
3300 204
3305 205
3310 204
3315 204
3320 205
3325 204
3330 204
3335 204
3340 204
3345 204
3350 204
3355 204
3360 204
3365 204
3370 204
3375 204
3380 204
3385 204
3390 204
3395 204
3400 205
3405 203
3410 204
3415 204
3420 203
3425 205
3430 204
3435 204
3440 204
3445 204
3450 204
3455 204
3460 204
3465 204
3470 204
3475 204
3480 204
3485 205
3490 204
3495 204
3500 204
3505 204
3510 205
3515 205
3520 205
3525 204
3530 204
3535 204
3540 205
3545 205
3550 204
3555 205
3560 205
3565 205
3570 204
3575 204
3580 204
3585 204
3590 204
3595 204
3600 204
};
\addlegendentry{Baseline}
\addplot [thick, color2, mark=square*, mark size=3, mark options={solid,fill=white,draw=red},  mark repeat={180}]
table{
0 205
5 208
10 212
15 212
20 212
25 213
30 212
35 213
40 207
45 206
50 207
55 206
60 206
65 206
70 206
75 206
80 206
85 206
90 206
95 206
100 206
105 206
110 207
115 210
120 213
125 213
130 213
135 213
140 213
145 212
150 211
155 210
160 212
165 211
170 212
175 212
180 212
185 214
190 212
195 213
200 243
205 242
210 242
215 242
220 242
225 243
230 245
235 243
240 243
245 244
250 243
255 243
260 243
265 243
270 243
275 244
280 243
285 243
290 243
295 243
300 243
305 243
310 244
315 244
320 244
325 244
330 244
335 244
340 244
345 242
350 244
355 244
360 244
365 244
370 243
375 243
380 245
385 246
390 246
395 245
400 245
405 246
410 246
415 246
420 245
425 246
430 246
435 246
440 247
445 246
450 246
455 246
460 246
465 246
470 246
475 246
480 246
485 246
490 246
495 246
500 246
505 246
510 246
515 246
520 246
525 246
530 247
535 246
540 246
545 246
550 246
555 246
560 246
565 246
570 246
575 246
580 247
585 246
590 246
595 246
600 246
605 246
610 246
615 247
620 247
625 246
630 247
635 245
640 246
645 247
650 246
655 247
660 246
665 246
670 247
675 247
680 246
685 246
690 246
695 246
700 246
705 246
710 246
715 247
720 246
725 246
730 247
735 247
740 246
745 246
750 246
755 247
760 247
765 246
770 246
775 247
780 246
785 246
790 246
795 246
800 247
805 246
810 246
815 247
820 246
825 246
830 250
835 247
840 246
845 246
850 247
855 246
860 246
865 246
870 246
875 247
880 246
885 246
890 246
895 246
900 246
905 247
910 246
915 247
920 246
925 247
930 246
935 246
940 246
945 247
950 246
955 247
960 247
965 246
970 247
975 246
980 246
985 247
990 247
995 247
1000 246
1005 247
1010 248
1015 245
1020 244
1025 244
1030 244
1035 245
1040 245
1045 245
1050 245
1055 245
1060 245
1065 245
1070 246
1075 245
1080 244
1085 244
1090 245
1095 244
1100 245
1105 244
1110 245
1115 245
1120 244
1125 245
1130 245
1135 245
1140 245
1145 246
1150 245
1155 245
1160 245
1165 245
1170 245
1175 245
1180 244
1185 245
1190 245
1195 245
1200 244
1205 244
1210 244
1215 245
1220 245
1225 245
1230 244
1235 245
1235 245
1240 245
1245 245
1250 245
1255 245
1260 245
1265 245
1270 244
1275 245
1280 245
1285 245
1290 245
1295 245
1300 244
1305 245
1310 244
1315 245
1320 245
1325 245
1330 245
1335 245
1340 244
1345 245
1350 245
1355 244
1360 244
1365 245
1370 245
1375 245
1380 245
1385 244
1390 244
1395 244
1400 245
1405 245
1410 244
1415 244
1420 244
1425 245
1430 245
1435 246
1440 246
1445 247
1450 246
1455 247
1460 247
1465 246
1470 247
1475 247
1480 246
1485 246
1490 246
1495 246
1500 247
1505 247
1510 247
1515 246
1520 247
1525 247
1530 246
1535 246
1540 247
1545 246
1550 247
1555 246
1560 246
1565 246
1570 246
1575 246
1580 246
1585 246
1590 246
1595 246
1600 246
1605 247
1610 246
1615 246
1620 246
1625 247
1630 246
1635 247
1640 247
1645 246
1650 246
1655 247
1660 247
1665 246
1670 247
1675 246
1680 246
1685 247
1690 247
1695 246
1700 246
1705 246
1710 246
1715 246
1720 246
1725 247
1730 246
1735 246
1740 247
1745 246
1750 247
1755 246
1760 247
1765 246
1770 246
1775 246
1780 246
1785 245
1790 244
1795 245
1800 245
1805 245
1810 246
1815 244
1820 245
1825 246
1830 246
1835 246
1840 246
1845 246
1850 246
1855 246
1860 246
1865 247
1870 247
1875 247
1880 247
1885 247
1890 247
1895 247
1900 247
1905 247
1910 248
1915 247
1920 247
1925 247
1930 247
1935 248
1940 247
1945 247
1950 247
1955 247
1960 246
1965 245
1970 245
1975 245
1980 245
1985 245
1990 245
1995 245
2000 246
2005 246
2010 245
2015 246
2020 245
2025 244
2030 245
2035 245
2040 244
2045 245
2050 244
2055 245
2060 244
2065 245
2070 247
2075 246
2080 246
2085 246
2090 247
2095 247
2100 247
2105 247
2110 247
2115 248
2120 247
2125 247
2130 247
2135 248
2140 248
2145 247
2150 247
2155 247
2160 248
2165 248
2170 247
2175 248
2180 247
2185 247
2190 247
2195 247
2200 247
2205 247
2210 247
2215 246
2220 248
2225 248
2230 248
2235 247
2240 247
2245 247
2250 247
2255 247
2260 247
2265 248
2270 246
2275 247
2280 247
2285 247
2290 248
2295 247
2300 247
2305 247
2310 247
2315 247
2320 247
2325 249
2330 247
2335 247
2340 247
2345 247
2350 248
2355 247
2360 248
2365 247
2370 245
2375 245
2380 245
2385 246
2390 245
2395 245
2400 245
2405 245
2410 245
2415 245
2420 246
2425 245
2430 245
2435 245
2440 245
2445 245
2450 246
2455 245
2460 245
2465 245
2470 246
2475 246
2480 245
2485 247
2490 248
2495 248
2500 247
2505 247
2510 248
2515 248
2520 247
2525 248
2530 246
2535 248
2540 248
2545 248
2550 247
2555 247
2560 247
2565 248
2570 247
2575 247
2580 248
2585 247
2590 247
2595 247
2600 247
2605 247
2610 247
2615 247
2620 247
2625 248
2630 247
2635 248
2640 247
2645 247
2650 248
2655 247
2660 247
2665 247
2670 247
2675 247
2680 247
2685 247
2690 247
2695 248
2700 247
2705 248
2710 248
2715 246
2720 247
2725 247
2730 247
2735 247
2740 247
2745 247
2750 231
2755 229
2760 223
2765 215
2770 214
2775 213
2780 213
2785 214
2790 215
2795 213
2800 209
2805 207
2810 207
2815 207
2820 207
2825 207
2830 207
2835 207
2840 207
2845 207
2850 206
2855 206
2860 206
2865 206
2870 207
2875 207
2880 207
2885 206
2890 206
2895 207
2900 206
2905 207
2910 206
2915 206
2920 206
2925 206
2930 206
2935 207
2940 206
2945 207
2950 206
2955 206
2960 206
2965 206
2970 207
2975 206
2980 206
2985 206
2990 206
2995 206
3000 206
3005 206
3010 206
3015 206
3020 206
3025 205
3030 206
3035 206
3040 206
3045 206
3050 206
3055 206
3060 206
3065 205
3070 206
3075 206
3080 206
3085 206
3090 206
3095 206
3100 206
3105 206
3110 205
3115 206
3120 207
3125 205
3130 206
3135 206
3140 206
3145 206
3150 206
3155 206
3160 206
3165 205
3170 207
3175 205
3180 206
3185 206
3190 206
3195 206
3200 206
3205 206
3210 206
3215 206
3220 205
3225 206
3230 206
3235 206
3240 205
3245 206
3250 205
3255 205
3260 206
3265 206
3270 206
3275 206
3280 205
3285 205
3290 206
3295 205
3300 206
3305 206
3310 206
3315 206
3320 206
3325 206
3330 206
3335 206
3340 206
3345 205
3350 206
3355 206
3360 206
3365 205
3370 206
3375 205
3380 206
3385 206
3390 206
3395 206
3400 206
3405 206
3410 206
3415 206
3420 205
3425 205
3430 205
3435 207
3440 205
3445 205
3450 206
3455 206
3460 206
3465 206
3470 206
3475 206
3480 206
3485 205
3490 206
3495 206
3500 205
3505 205
3510 206
3515 206
3520 206
3525 206
3530 206
3535 206
3540 205
3545 206
3550 206
3555 206
3560 206
3565 206
3570 206
3575 205
3580 206
3585 206
3590 206
3595 207
3600 206
3605 206
3610 206
3615 206
3620 206
3625 206
3630 206
3635 206
3640 205
3645 206
3650 206
3655 206
3660 206
3665 205
3670 206
3675 206
3680 205
3685 206
3690 206
3695 205
3700 206
3705 206
3710 205
3715 206
3720 206
3725 206
3730 206
3735 206
3740 206
3745 206
3750 206
3755 206
3760 205
3765 206
3770 207
3775 206
3780 206
3785 205
3790 205
3795 206
3800 205
3805 206
3810 206
3815 206
3820 206
3825 209
3830 205
3835 207
3840 206
3845 206
3850 206
3855 205
3860 206
3865 206
3870 206
3875 206
3880 206
3885 205
3890 206
3895 205
3900 205
3905 206
3910 206
3915 206
3920 206
3925 205
3930 206
3935 206
3940 206
3945 206
3950 205
3955 206
3960 205
3965 206
3970 205
3975 206
3980 205
3985 205
3990 205
3995 206
4000 206
4005 206
4010 206
4015 206
4020 206
4025 206
4030 206
4035 206
4040 206
4045 206
4050 205
4055 205
4060 206
4065 205
4070 206
4075 206
4080 205
4085 206
4090 206
4095 206
4100 205
4105 206
4110 206
4115 205
4120 205
4125 206
4130 206
4135 205
4140 206
4145 205
4150 205
4155 205
4160 205
4165 205
4170 206
4175 206
4180 206
4185 206
4190 205
4195 205
4200 206
4205 206
4210 206
4215 206
4220 206
4225 206
4230 206
4235 206
4240 206
4245 205
4250 206
4255 205
4260 205
4265 206
4270 206
4275 206
4280 206
4285 206
4290 205
4295 206
4300 205
4305 206
4310 205
4315 206
4320 206
4325 206
4330 206
4335 205
4340 206
4345 206
4350 206
4355 206
4360 206
4365 205
4370 206
4375 206
4380 206
4385 206
4390 205
4395 206
4400 205
4405 205
4410 206
4415 205
4420 206
4425 207
4430 206
4435 206
4440 206
4445 206
4450 206
4455 206
4460 206
4465 206
4470 206
4475 206
4480 206
4485 206
4490 206
4495 206
4500 206
4505 206
4510 206
4515 206
4520 206
4525 205
4530 206
4535 206
4540 206
4545 206
4550 206
4555 206
4560 206
4565 206
4570 206
4575 206
4580 205
4585 206
4590 206
4595 206
4600 205
4605 205
4610 206
4615 206
4620 206
4625 206
4630 206
4635 206
4640 206
4645 206
4650 206
4655 206
4660 206
4665 206
4670 205
4675 206
4680 206
4685 206
4690 206
4695 206
4700 206
4705 206
4710 206
4715 206
4720 206
4725 208
4730 206
4735 206
4740 206
4745 205
4750 206
4755 206
4760 206
4765 207
4770 207
4775 205
4780 206
4785 206
4790 206
4795 206
4800 206
4805 206
4810 206
4815 206
4820 206
4825 206
4830 206
4835 206
4840 206
4845 206
4850 206
4855 206
4860 206
4865 206
4870 206
4875 206
4880 205
4885 207
4890 206
4895 206
4900 206
4905 205
4910 206
4915 206
4920 206
4925 206
4930 206
4935 205
4940 206
4945 206
4950 206
4955 206
4960 206
4965 206
4970 206
4975 206
4980 206
4985 206
4990 206
4995 206
5000 205
5005 206
5010 206
5015 206
5020 205
5025 208
5030 205
5035 206
5040 205
5045 206
5050 206
5055 206
5060 206
5065 206
5070 206
5075 206
5080 207
5085 206
5090 206
5095 206
5100 206
5105 206
5110 206
5115 206
5120 206
5125 206
5130 206
5135 206
5140 206
5145 206
5150 206
5155 205
5160 205
5165 206
5170 206
5175 206
5180 206
5185 206
5190 206
5195 205
5200 206
5205 206
5210 206
5215 206
5220 206
5225 205
5230 206
5235 206
5240 206
5245 206
5250 206
5255 206
5260 206
5265 206
5270 206
5275 207
5280 206
5285 206
5290 206
5295 206
5300 206
5305 206
5310 206
5315 206
5320 206
5325 207
5330 206
5335 206
5340 206
5345 205
5350 206
5355 206
5360 207
5365 205
5370 207
5375 206
5380 206
5385 206
5390 206
5395 205
5400 206
5405 206
5410 206
5415 206
5420 206
5425 206
5430 207
5435 206
5440 206
5445 205
5450 206
5455 206
5460 206
5465 206
5470 207
5475 206
5480 206
5485 205
5490 206
5495 206
5500 206
5505 205
5510 206
5515 206
5520 206
5525 205
5530 206
5535 205
5540 207
5545 205
5550 206
5555 206
5560 205
5565 206
5570 206
5575 205
5580 206
5585 206
5590 206
5595 205
5600 206
5605 206
5610 206
5615 206
5620 206
5625 205
5630 206
5635 206
5640 206
5645 206
5650 206
5655 206
5660 206
5665 205
5670 206
5675 206
5680 206
5685 206
5690 206
5695 206
5700 206
5705 206
5710 206
5715 206
5720 205
5725 206
5730 207
5735 206
5740 206
5745 205
5750 205
5755 206
5760 207
5765 206
5770 206
5775 205
5780 205
5785 206
5790 206
5795 206
5800 206
5805 205
5810 205
5815 206
5820 206
5825 206
5830 206
5835 205
5840 206
5845 206
5850 206
5855 205
5860 206
5865 206
5870 206
5875 205
5880 206
5885 206
5890 205
5895 205
5900 206
5905 206
5910 205
5915 206
5920 205
5925 206
5930 206
5935 206
5940 205
5945 206
5950 205
5955 206
5960 205
5965 205
5970 206
5975 207
5980 207
5985 206
5990 205
5995 206
};

\addlegendentry{Boosting enabled}
\addplot [thick, color1, mark=triangle*, mark size=3, mark options={solid,fill=white,draw=red},  mark repeat={180}]
table{
0 206
5 210
10 210
15 210
20 210
25 211
30 211
35 212
40 207
45 207
50 206
55 206
60 207
65 207
70 206
75 207
80 207
85 206
90 206
95 206
100 207
105 207
110 206
115 211
120 211
125 211
130 211
135 212
140 211
145 210
150 210
155 209
160 209
165 210
170 211
175 213
180 210
185 210
190 206
195 214
200 211
205 232
210 233
215 233
220 233
225 233
230 233
235 234
240 234
245 232
250 233
255 234
260 233
265 232
270 233
275 233
280 233
285 233
290 233
295 233
300 233
305 234
310 233
315 232
320 233
325 233
330 233
335 234
340 234
345 234
350 234
355 233
360 234
365 233
370 234
375 234
380 234
385 234
390 234
395 234
400 233
405 234
410 234
415 233
420 233
425 234
430 234
435 234
440 233
445 233
450 233
455 234
460 235
465 233
470 233
475 234
480 234
485 234
490 234
495 234
500 234
505 234
510 234
515 234
520 234
525 234
530 234
535 234
540 233
545 234
550 233
555 233
560 233
565 232
570 232
575 233
580 232
585 232
590 232
595 234
600 233
605 232
610 232
615 232
620 233
625 233
630 233
635 232
640 232
645 233
650 233
655 233
660 233
665 233
670 233
675 233
680 235
685 234
690 234
695 233
700 234
705 234
710 234
715 234
720 233
725 235
730 234
735 234
740 233
745 232
750 234
755 234
760 234
765 234
770 234
775 234
780 234
785 234
790 234
795 234
800 233
805 234
810 234
815 234
820 234
825 234
830 234
835 234
840 234
845 233
850 233
855 234
860 234
865 233
870 234
875 234
880 234
885 234
890 234
895 234
900 234
905 234
910 234
915 234
920 234
925 234
930 234
935 234
940 234
945 233
950 233
955 234
960 234
965 234
970 234
975 233
980 235
985 233
990 234
995 234
1000 234
1005 233
1010 234
1015 234
1020 234
1025 234
1030 235
1035 234
1040 234
1045 234
1050 234
1055 234
1060 233
1065 234
1070 234
1075 234
1080 234
1085 234
1090 234
1095 234
1100 234
1105 234
1110 234
1115 234
1120 234
1125 234
1130 234
1135 234
1140 233
1145 238
1150 234
1155 234
1160 234
1165 234
1170 234
1175 234
1180 234
1185 233
1190 233
1195 234
1200 233
1205 234
1210 234
1215 234
1220 234
1225 235
1230 234
1235 234
1240 234
1245 234
1250 234
1255 234
1260 234
1265 234
1270 233
1275 234
1280 234
1285 234
1290 234
1295 233
1300 234
1305 234
1310 234
1315 233
1320 233
1325 233
1330 232
1335 233
1340 232
1345 233
1350 232
1355 232
1360 232
1365 233
1370 234
1375 234
1380 234
1385 233
1390 234
1395 233
1400 234
1405 234
1410 234
1415 233
1420 234
1425 234
1430 234
1435 234
1440 234
1445 234
1450 234
1455 234
1460 234
1465 235
1470 234
1475 234
1480 233
1485 234
1490 234
1495 234
1500 233
1505 233
1510 233
1515 232
1520 233
1525 232
1530 232
1535 232
1540 232
1545 233
1550 232
1555 233
1560 234
1565 234
1570 234
1575 233
1580 234
1585 234
1590 234
1595 234
1600 234
1605 234
1610 234
1615 234
1620 233
1625 234
1630 234
1635 234
1640 233
1645 234
1650 234
1655 234
1660 234
1665 234
1670 234
1675 233
1680 233
1685 233
1690 234
1695 234
1700 234
1705 233
1710 234
1715 233
1720 234
1725 234
1730 234
1735 234
1740 234
1745 235
1750 234
1755 233
1760 234
1765 234
1770 234
1775 233
1780 233
1785 234
1790 234
1795 234
1800 234
1805 233
1810 234
1815 234
1820 233
1825 234
1830 234
1835 234
1840 233
1845 233
1850 234
1855 233
1860 234
1865 234
1870 235
1875 233
1880 234
1885 233
1890 234
1895 234
1900 235
1905 234
1910 235
1915 234
1920 233
1925 234
1930 234
1935 233
1940 235
1945 234
1950 234
1955 234
1960 234
1965 234
1970 233
1975 234
1980 234
1985 234
1990 233
1995 234
2000 234
2005 234
2010 234
2015 233
2020 234
2025 234
2030 234
2035 234
2040 234
2045 235
2050 233
2055 234
2060 234
2065 234
2070 234
2075 234
2080 234
2085 234
2090 234
2095 234
2100 234
2105 234
2110 234
2115 234
2120 234
2125 234
2130 235
2135 233
2140 233
2145 232
2150 232
2155 233
2160 233
2165 232
2170 232
2175 233
2180 232
2185 233
2190 233
2195 233
2200 233
2205 233
2210 233
2215 232
2220 233
2225 234
2230 234
2235 233
2240 233
2245 233
2250 234
2255 235
2260 234
2265 234
2270 234
2275 234
2280 234
2285 234
2290 235
2295 234
2300 234
2305 234
2310 234
2315 234
2320 234
2325 234
2330 234
2335 234
2340 233
2345 235
2350 234
2355 234
2360 234
2365 234
2370 233
2375 234
2380 234
2385 234
2390 235
2395 234
2400 235
2405 234
2410 234
2415 234
2420 234
2425 234
2430 234
2435 233
2440 234
2445 235
2450 234
2455 234
2460 234
2465 235
2470 234
2475 234
2480 234
2485 234
2490 234
2495 234
2500 234
2505 234
2510 234
2515 234
2520 234
2525 234
2530 234
2535 234
2540 235
2545 235
2550 234
2555 234
2560 235
2565 234
2570 234
2575 234
2580 234
2585 234
2590 234
2595 234
2600 234
2605 234
2610 234
2615 234
2620 234
2625 234
2630 235
2635 234
2640 234
2645 236
2650 235
2655 234
2660 234
2665 235
2670 234
2675 233
2680 234
2685 235
2690 234
2695 234
2700 234
2705 234
2710 234
2715 234
2720 234
2725 234
2730 234
2735 235
2740 235
2745 234
2750 234
2755 234
2760 234
2765 234
2770 234
2775 234
2780 234
2785 234
2790 234
2795 235
2800 234
2805 234
2810 233
2815 234
2820 234
2825 235
2830 234
2835 234
2840 235
2845 235
2850 234
2855 235
2860 235
2865 234
2870 234
2875 234
2880 235
2885 234
2890 234
2895 234
2900 234
2905 234
2910 233
2915 234
2920 234
2925 234
2930 234
2935 234
2940 233
2945 234
2950 233
2955 233
2960 233
2965 233
2970 235
2975 234
2980 234
2985 234
2990 234
2995 234
3000 234
3005 234
3010 233
3015 234
3020 234
3025 233
3030 233
3035 232
3040 233
3045 220
3050 217
3055 216
3060 214
3065 211
3070 212
3075 212
3080 212
3085 211
3090 211
3095 212
3100 206
3105 207
3110 206
3115 206
3120 206
3125 207
3130 207
3135 206
3140 206
3145 206
3150 206
3155 206
3160 207
3165 207
3170 206
3175 206
3180 207
3185 206
3190 206
3195 206
3200 207
3205 206
3210 206
3215 206
3220 206
3225 206
3230 207
3235 206
3240 206
3245 207
3250 206
3255 206
3260 206
3265 206
3270 206
3275 206
3280 206
3285 206
3290 206
3295 206
3300 206
3305 206
3310 206
3315 206
3320 206
3325 206
3330 206
3335 206
3340 206
3345 206
3350 206
3355 206
3360 206
3365 206
3370 206
3375 206
3380 206
3385 205
3390 206
3395 206
3400 205
3405 206
3410 205
3415 206
3420 206
3425 206
3430 207
3435 205
3440 206
3445 206
3450 206
3455 206
3460 207
3465 206
3470 206
3475 206
3480 206
3485 206
3490 206
3495 206
3500 206
3505 206
3510 206
3515 206
3520 206
3525 206
3530 206
3535 206
3540 206
3545 206
3550 206
3555 206
3560 206
3565 205
3570 206
3575 206
3580 206
3585 206
3590 206
3595 206
3600 206
3605 205
3610 206
3615 206
3620 206
3625 205
3630 205
3635 206
3640 207
3645 206
3650 205
3655 205
3660 206
3665 206
3670 206
3675 206
3680 206
3685 206
3690 206
3695 206
3700 206
3705 205
3710 207
3715 206
3720 206
3725 205
3730 206
3735 206
3740 205
3745 206
3750 206
3755 206
3760 206
3765 205
3770 206
3775 206
3780 207
3785 206
3790 206
3795 206
3800 206
3805 206
3810 206
3815 205
3820 206
3825 206
3830 206
3835 206
3840 206
3845 206
3850 206
3855 205
3860 206
3865 205
3870 206
3875 206
3880 206
3885 206
3890 205
3895 206
3900 206
3905 206
3910 206
3915 206
3920 206
3925 206
3930 205
3935 206
3940 206
3945 206
3950 205
3955 206
3960 206
3965 206
3970 206
3975 206
3980 206
3985 205
3990 206
3995 206
4000 206
4005 206
4010 206
4015 206
4020 205
4025 206
4030 205
4035 206
4040 206
4045 206
4050 206
4055 206
4060 206
4065 206
4070 206
4075 205
4080 205
4085 205
4090 206
4095 206
4100 206
4105 206
4110 205
4115 206
4120 206
4125 206
4130 206
4135 206
4140 206
4145 206
4150 206
4155 206
4160 206
4165 206
4170 205
4175 206
4180 206
4185 207
4190 206
4195 206
4200 206
4205 205
4210 206
4215 205
4220 205
4225 206
4230 205
4235 205
4240 206
4245 206
4250 206
4255 206
4260 205
4265 205
4270 205
4275 205
4280 205
4285 206
4290 206
4295 206
4300 205
4305 206
4310 205
4315 206
4320 206
4325 205
4330 205
4335 206
4340 206
4345 205
4350 206
4355 206
4360 205
4365 205
4370 206
4375 205
4380 206
4385 205
4390 206
4395 205
4400 206
4405 205
4410 206
4415 206
4420 206
4425 206
4430 206
4435 206
4440 205
4445 206
4450 207
4455 205
4460 206
4465 206
4470 205
4475 205
4480 206
4485 206
4490 206
4495 206
4500 205
4505 206
4510 205
4515 205
4520 205
4525 205
4530 205
4535 205
4540 205
4545 206
4550 205
4555 206
4560 205
4565 206
4570 206
4575 205
4580 206
4585 206
4590 206
4595 206
4600 206
4605 205
4610 206
4615 206
4620 205
4625 206
4630 206
4635 205
4640 206
4645 205
4650 205
4655 205
4660 206
4665 206
4670 205
4675 206
4680 206
4685 206
4690 205
4695 205
4700 205
4705 205
4710 206
4715 205
4720 206
4725 206
4730 205
4735 205
4740 205
4745 205
4750 205
4755 205
4760 206
4765 205
4770 205
4775 206
4780 205
4785 205
4790 206
4795 205
4800 206
4805 205
4810 206
4815 205
4820 205
4825 205
4830 206
4835 206
4840 205
4845 206
4850 206
4855 205
4860 205
4865 206
4870 206
4875 206
4880 205
4885 205
4890 206
4895 205
4900 206
4905 205
4910 205
4915 205
4920 206
4925 206
4930 205
4935 206
4940 206
4945 205
4950 205
4955 205
4960 207
4965 205
4970 206
4975 205
4980 206
4985 205
4990 206
4995 206
5000 206
5005 206
5010 206
5015 205
5020 206
5025 206
5030 206
5035 206
5040 206
5045 206
5050 205
5055 206
5060 206
5065 205
5070 207
5075 206
5080 206
5085 206
5090 206
5095 206
5100 206
5105 205
5110 205
5115 205
5120 205
5125 205
5130 206
5135 205
5140 205
5145 205
5150 205
5155 206
5160 205
5165 205
5170 205
5175 206
5180 206
5185 205
5190 205
5195 206
5200 206
5205 205
5210 205
5215 205
5220 205
5225 206
5230 205
5235 205
5240 205
5245 205
5250 206
5255 206
5260 205
5265 205
5270 205
5275 205
5280 206
5285 205
5290 205
5295 206
5300 206
5305 206
5310 205
5315 206
5320 205
5325 206
5330 206
5335 206
5340 206
5345 205
5350 205
5355 205
5360 206
5365 205
5370 206
5375 207
5380 205
5385 206
5390 205
5395 205
5400 205
5405 206
5410 206
5415 206
5420 205
5425 205
5430 205
5435 205
5440 206
5445 205
5450 205
5455 206
5460 205
5465 205
5470 206
5475 205
5480 205
5485 206
5490 206
5495 205
5500 206
5505 206
5510 206
5515 206
5520 205
5525 206
5530 205
5535 205
5540 205
5545 206
5550 206
5555 206
5560 206
5565 205
5570 206
5575 206
5580 206
5585 206
5590 206
5595 206
5600 206
5605 206
5610 206
5615 206
5620 206
5625 205
5630 205
5635 206
5640 210
5645 205
5650 205
5655 206
5660 205
5665 205
5670 206
5675 206
5680 206
5685 205
5690 206
5695 205
5700 206
5705 205
5710 206
5715 206
5720 205
5725 206
5730 206
5735 206
5740 205
5745 206
5750 206
5755 205
5760 205
5765 206
5770 206
5775 205
5780 206
5785 206
5790 206
5795 206
5800 205
5805 206
5810 205
5815 206
5820 205
5825 205
5830 206
5835 205
5840 206
5845 206
5850 206
5855 206
5860 205
5865 206
5870 205
5875 205
5880 206
5885 205
5890 206
5895 206
5900 206
5905 206
5910 206
5915 205
5920 206
5925 205
5930 206
5935 205
5940 205
5945 206
5950 205
5955 206
5960 206
5965 205
5970 206
5975 205
5980 206
5985 205
5990 205
5995 205

};
\addlegendentry{Boosting disabled}

\end{axis}

\end{tikzpicture}

%% file: results/energy_amd_boost.tex
\begin{tikzpicture}[font=\Large,spy using outlines= {circle, magnification=2, connect spies}]

\definecolor{color0}{rgb}{0.12156862745098,0.466666666666667,0.705882352941177}
\definecolor{color1}{rgb}{1,0.498039215686275,0.0549019607843137}
\definecolor{color2}{rgb}{0.172549019607843,0.627450980392157,0.172549019607843}
\definecolor{color3}{rgb}{0.83921568627451,0.152941176470588,0.156862745098039}
\definecolor{color4}{rgb}{0.580392156862745,0.403921568627451,0.741176470588235}
\definecolor{color5}{rgb}{0,0,0}

\begin{axis}[
legend cell align={left},
legend columns=2,
legend style={fill opacity=0.8, draw opacity=1, text opacity=1, at={(1.05,1.15)}, anchor=east, draw=white!80.0!black},
tick align=outside,
tick pos=left,
x grid style={white!69.01960784313725!black},
xmajorgrids,
y grid style={white!69.01960784313725!black},
ymajorgrids,
xlabel={Time (SEC)},
xmin=0, xmax=3600,
xtick={0,900,1800,2700,3600},
xtick style={color=black},
y grid style={white!69.01960784313725!black},
ylabel={Total energy consumption (kWh)},
ymin=0, ymax=0.25,
ytick style={color=black},
ytick={0.1,0.15,0.2,0.25},
 y label style={at={(axis description cs:-0.04,.5)}}
]
\addplot [semithick, color5]
table[y expr=(\thisrowno{1}-350880)/1000] {%
0 350880
5 350880
10 350880
15 350880
20 350880
25 350880
30 350881
35 350882
40 350882
45 350882
50 350882
55 350882
60 350883
65 350883
70 350883
75 350884
80 350884
85 350884
90 350884
95 350885
100 350885
105 350885
110 350885
115 350886
120 350886
125 350887
130 350887
135 350887
140 350887
145 350887
150 350888
155 350889
160 350889
165 350889
170 350889
175 350889
180 350889
185 350889
190 350891
195 350891
200 350891
205 350891
210 350891
215 350891
220 350892
225 350892
230 350893
235 350893
240 350893
245 350893
250 350894
255 350894
260 350894
265 350894
270 350895
275 350895
280 350895
285 350896
290 350896
295 350896
300 350896
305 350896
310 350897
315 350898
320 350898
325 350898
330 350899
335 350899
340 350899
345 350900
350 350900
355 350900
360 350900
365 350900
370 350901
375 350901
380 350902
385 350902
390 350902
395 350902
400 350902
405 350902
410 350904
415 350904
420 350904
425 350904
430 350904
435 350904
440 350905
445 350905
450 350906
455 350906
460 350906
465 350906
470 350907
475 350907
480 350907
485 350907
490 350908
495 350908
500 350908
505 350909
510 350909
515 350909
520 350909
525 350910
530 350910
535 350911
540 350911
545 350911
550 350911
555 350911
560 350911
565 350913
570 350913
575 350913
580 350913
585 350913
590 350913
595 350913
600 350914
605 350915
610 350915
615 350915
620 350915
625 350915
630 350916
635 350916
640 350916
645 350917
650 350917
655 350917
660 350918
665 350918
670 350918
675 350918
680 350918
685 350919
690 350920
695 350920
700 350920
705 350920
710 350920
715 350920
720 350920
725 350922
730 350922
735 350922
740 350922
745 350922
750 350922
755 350923
760 350923
765 350924
770 350924
775 350924
780 350924
785 350925
790 350925
795 350925
800 350925
805 350926
810 350926
815 350927
820 350927
825 350927
830 350927
835 350927
840 350928
845 350928
850 350929
855 350929
860 350929
865 350929
870 350929
875 350929
880 350931
885 350931
890 350931
895 350931
900 350931
905 350931
910 350932
915 350932
920 350932
925 350932
930 350933
935 350933
940 350933
945 350934
950 350934
955 350934
960 350934
965 350934
970 350935
975 350936
980 350936
985 350936
990 350936
995 350936
1000 350936
1005 350936
1010 350938
1015 350938
1020 350938
1025 350938
1030 350938
1035 350938
1040 350939
1045 350940
1050 350940
1055 350940
1060 350940
1065 350940
1070 350941
1075 350941
1080 350941
1085 350942
1090 350942
1095 350942
1100 350943
1105 350943
1110 350943
1115 350943
1120 350943
1125 350944
1130 350944
1135 350945
1140 350945
1145 350945
1150 350945
1155 350945
1160 350945
1165 350947
1170 350947
1175 350947
1180 350947
1185 350947
1190 350947
1195 350948
1200 350948
1205 350949
1210 350949
1215 350949
1220 350949
1225 350949
1230 350950
1235 350950
1240 350950
1245 350951
1250 350951
1255 350951
1260 350952
1265 350952
1270 350952
1275 350952
1280 350952
1285 350953
1290 350954
1295 350954
1300 350954
1305 350954
1310 350954
1315 350954
1320 350955
1325 350956
1330 350956
1335 350956
1340 350956
1345 350956
1350 350956
1355 350957
1360 350958
1365 350958
1370 350958
1375 350958
1380 350958
1385 350959
1390 350959
1395 350959
1400 350960
1405 350960
1410 350960
1415 350961
1420 350961
1425 350961
1430 350961
1435 350961
1440 350962
1445 350963
1450 350963
1455 350963
1460 350963
1465 350963
1470 350963
1475 350963
1480 350965
1485 350965
1490 350965
1495 350965
1500 350965
1505 350965
1510 350966
1515 350966
1520 350966
1525 350967
1530 350967
1535 350967
1540 350968
1545 350968
1550 350968
1555 350968
1560 350968
1565 350969
1570 350969
1575 350970
1580 350970
1585 350970
1590 350970
1595 350970
1600 350970
1605 350972
1610 350972
1615 350972
1620 350972
1625 350972
1630 350972
1635 350972
1640 350973
1645 350974
1650 350974
1655 350974
1660 350974
1665 350974
1670 350975
1675 350975
1680 350975
1685 350976
1690 350976
1695 350976
1700 350977
1705 350977
1710 350977
1715 350977
1720 350977
1725 350978
1730 350979
1735 350979
1740 350979
1745 350979
1750 350979
1755 350979
1760 350979
1765 350981
1770 350981
1775 350981
1780 350981
1785 350981
1790 350981
1795 350982
1800 350982
1805 350983
1810 350983
1815 350983
1820 350983
1825 350984
1830 350984
1835 350984
1840 350985
1845 350985
1850 350985
1855 350986
1860 350986
1865 350986
1870 350986
1875 350986
1880 350987
1885 350987
1890 350988
1895 350988
1900 350988
1905 350988
1910 350988
1915 350988
1920 350990
1925 350990
1930 350990
1935 350990
1940 350990
1945 350990
1950 350991
1955 350991
1960 350992
1965 350992
1970 350992
1975 350992
1980 350992
1985 350993
1990 350993
1995 350993
2000 350994
2005 350994
2010 350994
2015 350995
2020 350995
2025 350995
2030 350995
2035 350995
2040 350996
2045 350997
2050 350997
2055 350997
2060 350997
2065 350997
2070 350997
2075 350997
2080 350999
2085 350999
2090 350999
2095 350999
2100 350999
2105 350999
2110 351000
2115 351000
2120 351000
2125 351001
2130 351001
2135 351001
2140 351002
2145 351002
2150 351002
2155 351002
2160 351002
2165 351003
2170 351004
2175 351004
2180 351004
2185 351004
2190 351004
2195 351004
2200 351004
2205 351006
2210 351006
2215 351006
2220 351006
2225 351006
2230 351006
2235 351007
2240 351007
2245 351008
2250 351008
2255 351008
2260 351008
2265 351009
2270 351009
2275 351009
2280 351009
2285 351010
2290 351010
2295 351010
2300 351011
2305 351011
2310 351011
2315 351011
2320 351012
2325 351012
2330 351013
2335 351013
2340 351013
2345 351013
2350 351013
2355 351013
2360 351015
2365 351015
2370 351015
2375 351015
2380 351015
2385 351015
2390 351016
2395 351016
2400 351017
2405 351017
2410 351017
2415 351017
2420 351017
2425 351018
2430 351018
2435 351018
2440 351019
2445 351019
2450 351019
2455 351020
2460 351020
2465 351020
2470 351020
2475 351020
2480 351021
2485 351022
2490 351022
2495 351022
2500 351022
2505 351022
2510 351022
2515 351022
2520 351024
2525 351024
2530 351024
2535 351024
2540 351024
2545 351024
2550 351025
2555 351026
2560 351026
2565 351026
2570 351026
2575 351026
2580 351027
2585 351027
2590 351027
2595 351027
2600 351028
2605 351028
2610 351028
2615 351029
2620 351029
2625 351029
2630 351029
2635 351030
2640 351030
2645 351031
2650 351031
2655 351031
2660 351031
2665 351031
2670 351031
2675 351033
2680 351033
2685 351033
2690 351033
2695 351033
2700 351033
2705 351034
2710 351034
2715 351034
2720 351034
2725 351035
2730 351035
2735 351035
2740 351036
2745 351036
2750 351036
2755 351036
2760 351036
2765 351037
2770 351038
2775 351038
2780 351038
2785 351038
2790 351038
2795 351038
2800 351038
2805 351040
2810 351040
2815 351040
2820 351040
2825 351040
2830 351040
2835 351041
2840 351042
2845 351042
2850 351042
2855 351042
2860 351042
2865 351043
2870 351043
2875 351043
2880 351044
2885 351044
2890 351044
2895 351045
2900 351045
2905 351045
2910 351045
2915 351045
2920 351046
2925 351046
2930 351047
2935 351047
2940 351047
2945 351047
2950 351047
2955 351047
2960 351049
2965 351049
2970 351049
2975 351049
2980 351049
2985 351049
2990 351050
2995 351050
3000 351051
3005 351051
3010 351051
3015 351051
3020 351052
3025 351052
3030 351052
3035 351052
3040 351053
3045 351053
3050 351053
3055 351054
3060 351054
3065 351054
3070 351054
3075 351054
3080 351055
3085 351056
3090 351056
3095 351056
3100 351056
3105 351056
3110 351056
3115 351058
3120 351058
3125 351058
3130 351058
3135 351058
3140 351058
3145 351058
3150 351059
3155 351060
3160 351060
3165 351060
3170 351060
3175 351060
3180 351061
3185 351061
3190 351061
3195 351062
3200 351062
3205 351062
3210 351063
3215 351063
3220 351063
3225 351063
3230 351063
3235 351064
3240 351064
3245 351065
3250 351065
3255 351065
3260 351065
3265 351065
3270 351065
3275 351067
3280 351067
3285 351067
3290 351067
3295 351067
3300 351067
3305 351068
3310 351068
3315 351068
3320 351068
3325 351069
3330 351069
3335 351070
3340 351070
3345 351070
3350 351070
3355 351070
3360 351070
3365 351071
3370 351072
3375 351072
3380 351072
3385 351072
3390 351072
3395 351072
3400 351074
3405 351074
3410 351074
3415 351074
3420 351074
3425 351074
3430 351075
3435 351075
3440 351076
3445 351076
3450 351076
3455 351076
3460 351076
3465 351077
3470 351077
3475 351077
3480 351078
3485 351078
3490 351078
3495 351079
3500 351079
3505 351079
3510 351079
3515 351079
3520 351080
3525 351081
3530 351081
3535 351081
3540 351081
3545 351081
3550 351081
3555 351082
3560 351083
3565 351083
3570 351083
3575 351083
3580 351083
3585 351083
3590 351084
3595 351085
3600 351085
3605 351085
3610 351085
3615 351085
3620 351086
3625 351086
3630 351086
3635 351087
3640 351087
3645 351087
3650 351088
3655 351088
3660 351088
3665 351088
3670 351088
3675 351089
3680 351090
3685 351090
3690 351090
3695 351090
3700 351090
3705 351090
3710 351090
3715 351092
3720 351092
3725 351092
3730 351092
3735 351092
3740 351092
3745 351093
3750 351094
3755 351094
3760 351094
3765 351094
3770 351094
3775 351095
3780 351095
3785 351095
3790 351096
3795 351096
3800 351096
3805 351096
3810 351097
3815 351097
3820 351097
3825 351097
3830 351098
3835 351098
3840 351099
3845 351099
3850 351099
3855 351099
3860 351099
3865 351099
3870 351101
3875 351101
3880 351101
3885 351101
3890 351101
3895 351101
3900 351102
3905 351102
3910 351102
3915 351102
3920 351103
3925 351103
3930 351103
3935 351104
3940 351104
3945 351104
3950 351104
3955 351104
3960 351105
3965 351106
3970 351106
3975 351106
3980 351106
3985 351106
3990 351106
3995 351106
4000 351108
4005 351108
4010 351108
4015 351108
4020 351108
4025 351108
4030 351109
4035 351109
4040 351110
4045 351110
4050 351110
4055 351110
4060 351111
4065 351111
4070 351111
4075 351112
4080 351112
4085 351112
4090 351112
4095 351113
4100 351113
4105 351113
4110 351113
4115 351114
4120 351114
4125 351115
4130 351115
4135 351115
4140 351115
4145 351115
4150 351115
4155 351117
4160 351117
4165 351117
4170 351117
4175 351117
4180 351117
4185 351118
4190 351118
4195 351119
4200 351119
4205 351119
4210 351119
4215 351119
4220 351120
4225 351120
4230 351121
4235 351121
4240 351121
4245 351121
4250 351122
4255 351122
4260 351122
4265 351122
4270 351123
4275 351123
4280 351124
4285 351124
4290 351124
4295 351124
4300 351124
4305 351124
4310 351125
4315 351126
4320 351126
4325 351126
4330 351126
4335 351126
4340 351126
4345 351127
4350 351128
4355 351128
4360 351128
4365 351128
4370 351128
4375 351129
4380 351129
4385 351130
4390 351130
4395 351130
4400 351130
4405 351130
4410 351131
4415 351131
4420 351131
4425 351132
4430 351132
4435 351132
4440 351133
4445 351133
4450 351133
4455 351133
4460 351133
4465 351134
4470 351134
4475 351135
4480 351135
4485 351135
4490 351135
4495 351135
4500 351135
4505 351136
4510 351136
4515 351136
4520 351137
4525 351137
4530 351137
4535 351137
4540 351138
4545 351138
4550 351138
4555 351138
4560 351139
4565 351140
4570 351140
4575 351140
4580 351140
4585 351140
4590 351140
4595 351142
4600 351142
4605 351142
4610 351142
4615 351142
4620 351142
4625 351142
4630 351143
4635 351144
4640 351144
4645 351144
4650 351144
4655 351144
4660 351145
4665 351145
4670 351145
4675 351146
4680 351146
4685 351146
4690 351146
4695 351147
4700 351147
4705 351147
4710 351147
4715 351148
4720 351148
4725 351149
4730 351149
4735 351149
4740 351149
4745 351149
4750 351150
4755 351150
4760 351151
4765 351151
4770 351151
4775 351151
4780 351151
4785 351151
4790 351153
4795 351153
4800 351153
4805 351153
4810 351153
4815 351153
4820 351154
4825 351154
4830 351155
4835 351155
4840 351155
4845 351155
4850 351155
4855 351156
4860 351156
4865 351157
4870 351157
4875 351157
4880 351157
4885 351158
4890 351158
4895 351158
4900 351158
4905 351159
4910 351159
4915 351159
4920 351160
4925 351160
4930 351160
4935 351160
4940 351161
4945 351161
4950 351162
4955 351162
4960 351162
4965 351162
4970 351162
4975 351162
4980 351164
4985 351164
4990 351164
4995 351164
5000 351164
5005 351164
5010 351164
5015 351165
5020 351166
5025 351166
5030 351166
5035 351166
5040 351166
5045 351167
5050 351167
5055 351167
5060 351168
5065 351168
5070 351168
5075 351168
5080 351169
5085 351169
5090 351169
5095 351170
5100 351170
5105 351170
5110 351171
5115 351171
5120 351172
5125 351172
5130 351172
5135 351172
5140 351172
5145 351173
5150 351173
5155 351174
5160 351174
5165 351175
5170 351175
5175 351175
5180 351175
5185 351175
5190 351175
5195 351177
5200 351177
5205 351177
5210 351177
5215 351177
5220 351177
5225 351177
5230 351178
5235 351179
5240 351179
5245 351179
5250 351179
5255 351179
5260 351180
5265 351180
5270 351181
5275 351181
5280 351181
5285 351181
5290 351181
5295 351182
5300 351182
5305 351182
5310 351183
5315 351183
5320 351183
5325 351184
5330 351184
5335 351184
};
\addlegendentry{Baseline}
\addplot [semithick, color2, mark=square*, mark size=3, mark options={solid,fill=white,draw=red},  mark repeat={160}]
table[y expr=(\thisrowno{1}-272285)/1000] {%
0 272285
5 272285
10 272285
15 272286
20 272286
25 272287
30 272287
35 272287
40 272287
45 272287
50 272288
55 272289
60 272289
65 272289
70 272289
75 272289
80 272289
85 272290
90 272291
95 272291
100 272291
105 272291
110 272291
115 272292
120 272292
125 272293
130 272293
135 272293
140 272293
145 272293
150 272294
155 272294
160 272294
165 272295
170 272295
175 272295
180 272296
185 272296
190 272296
195 272296
200 272297
205 272297
210 272298
215 272298
220 272298
225 272298
230 272299
235 272300
240 272300
245 272300
250 272300
255 272301
260 272301
265 272301
270 272301
275 272302
280 272302
285 272303
290 272303
295 272303
300 272304
305 272304
310 272304
315 272305
320 272305
325 272305
330 272306
335 272306
340 272306
345 272306
350 272307
355 272308
360 272308
365 272308
370 272308
375 272308
380 272309
385 272310
390 272310
395 272310
400 272310
405 272310
410 272312
415 272312
420 272312
425 272312
430 272312
435 272312
440 272314
445 272314
450 272314
455 272314
460 272314
465 272315
470 272316
475 272316
480 272316
485 272316
490 272316
495 272317
500 272317
505 272318
510 272318
515 272318
520 272319
525 272319
530 272319
535 272320
540 272320
545 272320
550 272321
555 272321
560 272321
565 272322
570 272322
575 272323
580 272323
585 272323
590 272323
595 272324
600 272324
605 272325
610 272325
615 272325
620 272325
625 272325
630 272327
635 272327
640 272327
645 272327
650 272327
655 272327
660 272329
665 272329
670 272329
675 272329
680 272329
685 272330
690 272331
695 272331
700 272331
705 272331
710 272331
715 272332
720 272333
725 272333
730 272333
735 272333
740 272334
745 272334
750 272335
755 272335
760 272335
765 272335
770 272336
775 272336
780 272336
785 272337
790 272337
795 272338
800 272338
805 272338
810 272338
815 272339
820 272339
825 272340
830 272340
835 272340
840 272340
845 272340
850 272341
855 272342
860 272342
865 272342
870 272343
875 272343
880 272343
885 272344
890 272344
895 272344
900 272345
905 272345
910 272345
915 272346
920 272346
925 272347
930 272347
935 272347
940 272347
945 272348
950 272348
955 272349
960 272349
965 272349
970 272349
975 272350
980 272351
985 272351
990 272351
995 272351
1000 272351
1005 272351
1010 272353
1015 272353
1020 272353
1025 272353
1030 272353
1035 272354
1040 272355
1045 272355
1050 272355
1055 272355
1060 272355
1065 272356
1070 272357
1075 272357
1080 272357
1085 272357
1090 272358
1095 272358
1100 272359
1105 272359
1110 272359
1115 272360
1120 272360
1125 272360
1130 272360
1135 272361
1140 272361
1145 272362
1150 272362
1155 272362
1160 272362
1165 272363
1170 272364
1175 272364
1180 272364
1185 272364
1190 272364
1195 272365
1200 272366
1205 272366
1210 272366
1215 272366
1220 272366
1225 272368
1230 272368
1235 272368
1240 272368
1245 272368
1250 272368
1255 272370
1260 272370
1265 272370
1270 272370
1275 272370
1280 272371
1285 272371
1290 272372
1295 272372
1300 272372
1305 272373
1310 272373
1315 272373
1320 272374
1325 272374
1330 272374
1335 272375
1340 272375
1345 272375
1350 272376
1355 272376
1360 272377
1365 272377
1370 272377
1375 272377
1380 272378
1385 272378
1390 272379
1395 272379
1400 272379
1405 272379
1410 272379
1415 272381
1420 272381
1425 272381
1430 272381
1435 272381
1440 272381
1445 272382
1450 272383
1455 272383
1460 272383
1465 272383
1470 272384
1475 272384
1480 272385
1485 272385
1490 272385
1495 272386
1500 272386
1505 272386
1510 272386
1515 272387
1520 272387
1525 272388
1530 272388
1535 272388
1540 272388
1545 272389
1550 272390
1555 272390
1560 272390
1565 272390
1570 272390
1575 272392
1580 272392
1585 272392
1590 272392
1595 272392
1600 272392
1605 272394
1610 272394
1615 272394
1620 272394
1625 272394
1630 272395
1635 272395
1640 272396
1645 272396
1650 272396
1655 272396
1660 272397
1665 272397
1670 272398
1675 272398
1680 272398
1685 272399
1690 272399
1695 272399
1700 272400
1705 272400
1710 272400
1715 272401
1720 272401
1725 272401
1730 272402
1735 272402
1740 272403
1745 272403
1750 272403
1755 272403
1760 272404
1765 272404
1770 272405
1775 272405
1780 272405
1785 272405
1790 272405
1795 272407
1800 272407
1805 272407
1810 272407
1815 272407
1820 272407
1825 272409
1830 272409
1835 272409
1840 272409
1845 272409
1850 272410
1855 272411
1860 272411
1865 272411
1870 272411
1875 272411
1880 272412
1885 272413
1890 272413
1895 272413
1900 272413
1905 272414
1910 272414
1915 272414
1920 272415
1925 272415
1930 272415
1935 272416
1940 272416
1945 272416
1950 272417
1955 272417
1960 272418
1965 272418
1970 272418
1975 272418
1980 272419
1985 272420
1990 272420
1995 272420
2000 272420
2005 272420
2010 272421
2015 272422
2020 272422
2025 272422
2030 272422
2035 272422
2040 272423
2045 272423
2050 272424
2055 272424
2060 272424
2065 272424
2070 272425
2075 272425
2080 272426
2085 272426
2090 272426
2095 272427
2100 272427
2105 272427
2110 272428
2115 272428
2120 272429
2125 272429
2130 272429
2135 272429
2140 272430
2145 272430
2150 272431
2155 272431
2160 272431
2165 272431
2170 272431
2175 272433
2180 272433
2185 272433
2190 272433
2195 272433
2200 272433
2205 272435
2210 272435
2215 272435
2220 272435
2225 272435
2230 272436
2235 272437
2240 272437
2245 272437
2250 272437
2255 272438
2260 272438
2265 272439
2270 272439
2275 272439
2280 272439
2285 272440
2290 272440
2295 272441
2300 272441
2305 272441
2310 272442
2315 272442
2320 272442
2325 272442
2330 272443
2335 272443
2340 272444
2345 272444
2350 272444
2355 272444
2360 272445
2365 272446
2370 272446
2375 272446
2380 272446
2385 272446
2390 272447
2395 272448
2400 272448
2405 272448
2410 272448
2415 272448
2420 272450
2425 272450
2430 272450
2435 272450
2440 272450
2445 272451
2450 272451
2455 272452
2460 272452
2465 272452
2470 272452
2475 272453
2480 272453
2485 272454
2490 272454
2495 272454
2500 272455
2505 272455
2510 272455
2515 272456
2520 272456
2525 272456
2530 272457
2535 272457
2540 272457
2545 272458
2550 272458
2555 272459
2560 272459
2565 272459
2570 272459
2575 272460
2580 272460
2585 272461
2590 272461
2595 272461
2600 272461
2605 272461
2610 272463
2615 272463
2620 272463
2625 272463
2630 272463
2635 272464
2640 272464
2645 272464
2650 272465
2655 272465
2660 272465
2665 272466
2670 272466
2675 272466
2680 272467
2685 272467
2690 272467
2695 272468
2700 272468
2705 272468
2710 272469
2715 272469
2720 272470
2725 272470
2730 272470
2735 272470
2740 272471
2745 272471
2750 272472
2755 272472
2760 272472
2765 272472
2770 272472
2775 272473
2780 272474
2785 272474
2790 272474
2795 272474
2800 272474
2805 272474
2810 272475
2815 272476
2820 272476
2825 272476
2830 272476
2835 272476
2840 272477
2845 272477
2850 272478
2855 272478
2860 272478
2865 272478
2870 272478
2875 272479
2880 272479
2885 272479
2890 272480
2895 272480
2900 272480
2905 272481
2910 272481
2915 272481
2920 272481
2925 272481
2930 272482
2935 272482
2940 272483
2945 272483
2950 272483
2955 272483
2960 272483
2965 272484
2970 272485
2975 272485
2980 272485
2985 272485
2990 272485
2995 272485
3000 272486
3005 272487
3010 272487
3015 272487
3020 272487
3025 272487
3030 272488
3035 272488
3040 272488
3045 272489
3050 272489
3055 272489
3060 272489
3065 272490
3070 272490
3075 272490
3080 272491
3085 272491
3090 272491
3095 272492
3100 272492
3105 272492
3110 272492
3115 272492
3120 272493
3125 272494
3130 272494
3135 272494
3140 272494
3145 272494
3150 272494
3155 272494
3160 272496
3165 272496
3170 272496
3175 272496
3180 272496
3185 272496
3190 272497
3195 272497
3200 272498
3205 272498
3210 272498
3215 272498
3220 272499
3225 272499
3230 272499
3235 272499
3240 272499
3245 272499
3250 272500
3255 272500
3260 272500
3265 272501
3270 272501
3275 272501
3280 272501
3285 272502
3290 272502
3295 272502
3300 272503
3305 272503
3310 272503
3315 272503
3320 272503
3325 272504
3330 272505
3335 272505
3340 272505
3345 272505
3350 272505
3355 272505
3360 272506
3365 272507
3370 272507
3375 272507
3380 272507
3385 272507
3390 272507
3395 272508
3400 272508
3405 272509
3410 272509
3415 272509
3420 272509
3425 272510
3430 272510
3435 272510
3440 272511
3445 272511
3450 272511
3455 272512
3460 272512
3465 272512
3470 272512
3475 272512
3480 272513
3485 272514
3490 272514
3495 272514
3500 272514
3505 272514
3510 272514
3515 272514
3520 272516
3525 272516
3530 272516
3535 272516
3540 272516
3545 272516
3550 272517
3555 272517
3560 272518
3565 272518
3570 272518
3575 272518
3580 272519
3585 272519
3590 272519
3595 272519
3600 272520
};
\addlegendentry{Boosting enabled}
\addplot [semithick, color1, mark=triangle*, mark size=3, mark options={solid,fill=white,draw=red},  mark repeat={180}]
table[y expr=(\thisrowno{1}-449640)/1000] {%
0 449640
5 449640
10 449640
15 449640
20 449641
25 449642
30 449642
35 449642
40 449642
45 449642
50 449642
55 449643
60 449644
65 449644
70 449644
75 449644
80 449644
85 449645
90 449645
95 449645
100 449646
105 449646
110 449646
115 449647
120 449647
125 449647
130 449647
135 449647
140 449648
145 449649
150 449649
155 449649
160 449649
165 449649
170 449649
175 449651
180 449651
185 449651
190 449651
195 449651
200 449651
205 449652
210 449653
215 449653
220 449653
225 449653
230 449653
235 449654
240 449655
245 449655
250 449655
255 449655
260 449657
265 449657
270 449657
275 449658
280 449658
285 449658
290 449658
295 449659
300 449659
305 449660
310 449660
315 449660
320 449660
325 449661
330 449662
335 449662
340 449662
345 449662
350 449662
355 449663
360 449664
365 449664
370 449664
375 449664
380 449664
385 449664
390 449666
395 449666
400 449666
405 449666
410 449666
415 449667
420 449668
425 449668
430 449668
435 449668
440 449669
445 449669
450 449670
455 449670
460 449670
465 449670
470 449671
475 449671
480 449672
485 449672
490 449672
495 449673
500 449673
505 449673
510 449673
515 449674
520 449674
525 449675
530 449675
535 449675
540 449675
545 449676
550 449677
555 449677
560 449677
565 449677
570 449677
575 449678
580 449679
585 449679
590 449679
595 449679
600 449679
605 449681
610 449681
615 449681
620 449681
625 449681
630 449682
635 449682
640 449683
645 449683
650 449683
655 449683
660 449684
665 449684
670 449685
675 449685
680 449685
685 449686
690 449686
695 449686
700 449687
705 449687
710 449687
715 449688
720 449688
725 449688
730 449689
735 449689
740 449690
745 449690
750 449690
755 449690
760 449690
765 449691
770 449692
775 449692
780 449692
785 449692
790 449692
795 449694
800 449694
805 449694
810 449694
815 449694
820 449695
825 449696
830 449696
835 449696
840 449696
845 449696
850 449697
855 449697
860 449698
865 449698
870 449698
875 449699
880 449699
885 449699
890 449700
895 449700
900 449700
905 449701
910 449701
915 449701
920 449701
925 449702
930 449703
935 449703
940 449703
945 449703
950 449703
955 449704
960 449705
965 449705
970 449705
975 449705
980 449705
985 449707
990 449707
995 449707
1000 449707
1005 449707
1010 449708
1015 449708
1020 449709
1025 449709
1030 449709
1035 449709
1040 449710
1045 449710
1050 449711
1055 449711
1060 449711
1065 449712
1070 449712
1075 449712
1080 449713
1085 449713
1090 449713
1095 449714
1100 449714
1105 449714
1110 449715
1115 449715
1120 449716
1125 449716
1130 449716
1135 449716
1140 449716
1145 449717
1150 449718
1155 449718
1160 449718
1165 449718
1170 449718
1175 449720
1180 449720
1185 449720
1190 449720
1195 449720
1200 449720
1205 449722
1210 449722
1215 449722
1220 449722
1225 449722
1230 449723
1235 449724
1240 449724
1245 449724
1250 449724
1255 449725
1260 449725
1265 449726
1270 449726
1275 449726
1280 449726
1285 449727
1290 449727
1295 449727
1300 449728
1305 449728
1310 449728
1315 449729
1320 449729
1325 449729
1330 449730
1335 449730
1340 449731
1345 449731
1350 449731
1355 449731
1360 449732
1365 449732
1370 449733
1375 449733
1380 449733
1385 449733
1390 449733
1395 449735
1400 449735
1405 449735
1410 449735
1415 449735
1420 449736
1425 449737
1430 449737
1435 449737
1440 449737
1445 449737
1450 449738
1455 449738
1460 449739
1465 449739
1470 449739
1475 449740
1480 449740
1485 449740
1490 449741
1495 449741
1500 449742
1505 449742
1510 449742
1515 449742
1520 449743
1525 449743
1530 449744
1535 449744
1540 449744
1545 449744
1550 449744
1555 449746
1560 449746
1565 449746
1570 449746
1575 449746
1580 449746
1585 449748
1590 449748
1595 449748
1600 449748
1605 449748
1610 449749
1615 449750
1620 449750
1625 449750
1630 449750
1635 449751
1640 449751
1645 449751
1650 449752
1655 449752
1660 449752
1665 449753
1670 449753
1675 449753
1680 449754
1685 449754
1690 449755
1695 449755
1700 449755
1705 449755
1710 449756
1715 449756
1720 449757
1725 449757
1730 449757
1735 449757
1740 449758
1745 449759
1750 449759
1755 449759
1760 449759
1765 449759
1770 449759
1775 449761
1780 449761
1785 449761
1790 449761
1795 449761
1800 449762
1805 449763
1810 449763
1815 449763
1820 449763
1825 449764
1830 449764
1835 449765
1840 449765
1845 449765
1850 449765
1855 449766
1860 449766
1865 449767
1870 449767
1875 449767
1880 449768
1885 449768
1890 449768
1895 449768
1900 449769
1905 449769
1910 449770
1915 449770
1920 449770
1925 449770
1930 449771
1935 449772
1940 449772
1945 449772
1950 449772
1955 449772
1960 449774
1965 449774
1970 449774
1975 449774
1980 449774
1985 449774
1990 449775
1995 449776
2000 449776
2005 449776
2010 449776
2015 449777
2020 449777
2025 449778
2030 449778
2035 449778
2040 449778
2045 449779
2050 449779
2055 449779
2060 449780
2065 449780
2070 449780
2075 449781
2080 449781
2085 449781
2090 449782
2095 449782
2100 449783
2105 449783
2110 449783
2115 449783
2120 449784
2125 449785
2130 449785
2135 449785
2140 449785
2145 449785
2150 449786
2155 449787
2160 449787
2165 449787
2170 449787
2175 449787
2180 449788
2185 449789
2190 449789
2195 449789
2200 449789
2205 449789
2210 449790
2215 449791
2220 449791
2225 449791
2230 449791
2235 449792
2240 449792
2245 449793
2250 449793
2255 449793
2260 449794
2265 449794
2270 449794
2275 449795
2280 449795
2285 449795
2290 449796
2295 449796
2300 449796
2305 449796
2310 449797
2315 449798
2320 449798
2325 449798
2330 449798
2335 449798
2340 449799
2345 449800
2350 449800
2355 449800
2360 449800
2365 449800
2370 449802
2375 449802
2380 449802
2385 449802
2390 449802
2395 449802
2400 449804
2405 449804
2410 449804
2415 449804
2420 449804
2425 449805
2430 449806
2435 449806
2440 449806
2445 449806
2450 449807
2455 449807
2460 449807
2465 449808
2470 449808
2475 449808
2480 449809
2485 449809
2490 449809
2495 449810
2500 449810
2505 449811
2510 449811
2515 449811
2520 449811
2525 449812
2530 449812
2535 449813
2540 449813
2545 449813
2550 449813
2555 449813
2560 449815
2565 449815
2570 449815
2575 449815
2580 449815
2585 449816
2590 449817
2595 449817
2600 449817
2605 449817
2610 449817
2615 449818
2620 449819
2625 449819
2630 449819
2635 449819
2640 449820
2645 449820
2650 449820
2655 449821
2660 449821
2665 449821
2670 449822
2675 449822
2680 449822
2685 449822
2690 449823
2695 449824
2700 449824
2705 449824
2710 449824
2715 449824
2720 449825
2725 449826
2730 449826
2735 449826
2740 449826
2745 449826
2750 449828
2755 449828
2760 449828
2765 449828
2770 449828
2775 449828
2780 449829
2785 449830
2790 449830
2795 449830
2800 449830
2805 449830
2810 449831
2815 449831
2820 449831
2825 449832
2830 449832
2835 449832
2840 449832
2845 449833
2850 449833
2855 449833
2860 449833
2865 449834
2870 449834
2875 449835
2880 449835
2885 449835
2890 449835
2895 449835
2900 449836
2905 449837
2910 449837
2915 449837
2920 449837
2925 449837
2930 449837
2935 449838
2940 449839
2945 449839
2950 449839
2955 449839
2960 449839
2965 449839
2970 449840
2975 449840
2980 449841
2985 449841
2990 449841
2995 449841
3000 449842
3005 449842
3010 449842
3015 449842
3020 449843
3025 449843
3030 449844
3035 449844
3040 449844
3045 449844
3050 449844
3055 449844
3060 449846
3065 449846
3070 449846
3075 449846
3080 449846
3085 449846
3090 449847
3095 449848
3100 449848
3105 449848
3110 449848
3115 449848
3120 449848
3125 449849
3130 449849
3135 449850
3140 449850
3145 449850
3150 449850
3155 449851
3160 449851
3165 449851
3170 449851
3175 449852
3180 449852
3185 449853
3190 449853
3195 449853
3200 449853
3205 449853
3210 449853
3215 449855
3220 449855
3225 449855
3230 449855
3235 449855
3240 449855
3245 449855
3250 449856
3255 449856
3260 449857
3265 449857
3270 449858
3275 449858
3280 449858
3285 449858
3290 449858
3295 449858
3300 449860
3305 449860
3310 449860
3315 449860
3320 449860
3325 449860
3330 449860
3335 449861
3340 449862
3345 449862
3350 449862
3355 449862
3360 449862
3365 449863
3370 449863
3375 449864
3380 449864
3385 449864
3390 449864
3395 449865
3400 449865
3405 449865
3410 449865
3415 449866
3420 449866
3425 449867
3430 449867
3435 449867
3440 449867
3445 449867
3450 449867
3455 449868
3460 449869
3465 449869
3470 449869
3475 449869
3480 449869
3485 449869
3490 449870
3495 449871
3500 449871
3505 449871
3510 449871
3515 449871
3520 449872
3525 449872
3530 449872
3535 449873
3540 449873
3545 449873
3550 449874
3555 449874
3560 449874
3565 449874
3570 449874
3575 449875
3580 449876
3585 449876
3590 449876
3595 449876
3600 449876
3605 449876
3610 449877
3615 449878
3620 449878
3625 449878
3630 449878
3635 449878
3640 449878
3645 449879
3650 449880
3655 449880
3660 449880
3665 449880
3670 449880
3675 449881
3680 449881
3685 449881
3690 449882
3695 449882
3700 449882
3705 449883
3710 449883
3715 449883
3720 449883
3725 449883
3730 449884
3735 449884
3740 449885
3745 449885
3750 449885
3755 449885
3760 449885
3765 449885
3770 449887
3775 449887
3780 449887
3785 449887
3790 449887
3795 449887
3800 449888
3805 449888
3810 449889
3815 449889
3820 449889
3825 449889
3830 449890
3835 449890
3840 449890
3845 449890
3850 449890
3855 449890
3860 449891
3865 449891
3870 449892
3875 449892
3880 449892
3885 449892
3890 449892
3895 449893
3900 449894
3905 449894
3910 449894
3915 449894
3920 449894
3925 449894
3930 449894
3935 449896
3940 449896
3945 449896
3950 449896
3955 449896
3960 449896
3965 449897
3970 449897
3975 449898
3980 449898
3985 449898
3990 449898
3995 449899
4000 449899
4005 449899
4010 449899
4015 449900
4020 449900
4025 449901
4030 449901
4035 449901
4040 449901
4045 449901
4050 449901
4055 449902
4060 449903
4065 449903
4070 449903
4075 449903
4080 449903
4085 449903
4090 449905
4095 449905
4100 449905
4105 449905
4110 449905
4115 449905
4120 449906
4125 449906
4130 449906
4135 449907
4140 449907
4145 449907
4150 449908
4155 449908
4160 449908
4165 449908
4170 449909
4175 449909
4180 449909
4185 449910
4190 449910
4195 449910
4200 449910
4205 449910
4210 449911
4215 449912
4220 449912
4225 449912
4230 449912
4235 449912
4240 449912
4245 449913
4250 449914
4255 449914
4260 449914
4265 449914
4270 449914
4275 449915
4280 449915
4285 449915
4290 449916
4295 449916
4300 449916
4305 449916
4310 449917
4315 449917
4320 449917
4325 449917
4330 449918
4335 449918
4340 449919
4345 449919
4350 449919
4355 449919
4360 449919
4365 449919
4370 449921
4375 449921
4380 449921
4385 449921
4390 449921
4395 449921
4400 449922
4405 449922
4410 449923
4415 449923
4420 449923
4425 449923
4430 449924
4435 449924
4440 449924
4445 449924
4450 449924
4455 449925
4460 449925
4465 449925
4470 449926
4475 449926
4480 449926
4485 449926
4490 449926
4495 449927
4500 449928
4505 449928
4510 449928
4515 449928
4520 449928
4525 449928
4530 449929
4535 449930
4540 449930
4545 449930
4550 449930
4555 449930
4560 449931
4565 449931
4570 449931
4575 449932
4580 449932
4585 449932
4590 449932
4595 449933
4600 449933
4605 449933
4610 449933
4615 449934
4620 449934
4625 449935
4630 449935
4635 449935
4640 449935
4645 449935
4650 449935
4655 449937
4660 449937
4665 449937
4670 449937
4675 449937
4680 449937
4685 449938
4690 449939
4695 449939
4700 449939
4705 449939
4710 449939
4715 449939
4720 449940
4725 449940
4730 449941
4735 449941
4740 449941
4745 449941
4750 449942
4755 449942
4760 449942
4765 449942
4770 449943
4775 449943
4780 449944
4785 449944
4790 449944
4795 449944
4800 449944
4805 449944
4810 449946
4815 449946
4820 449946
4825 449946
4830 449946
4835 449946
4840 449946
4845 449947
4850 449948
4855 449948
4860 449948
4865 449948
4870 449948
4875 449949
4880 449949
4885 449949
4890 449950
4895 449950
4900 449950
4905 449951
4910 449951
4915 449951
4920 449951
4925 449952
4930 449952
4935 449953
4940 449953
4945 449953
4950 449953
4955 449953
4960 449953
4965 449954
4970 449955
4975 449955
4980 449955
4985 449955
4990 449955
4995 449955
5000 449956
5005 449957
5010 449957
5015 449957
5020 449957
5025 449957
5030 449958
5035 449958
5040 449958
5045 449958
5050 449958
5055 449959
5060 449959
5065 449960
5070 449960
5075 449960
5080 449960
5085 449960
5090 449960
5095 449961
5100 449962
5105 449962
5110 449962
5115 449962
5120 449962
5125 449962
5130 449964
5135 449964
5140 449964
5145 449964
5150 449964
5155 449964
5160 449965
5165 449965
5170 449966
5175 449966
5180 449966
5185 449966
5190 449967
5195 449967
5200 449967
5205 449967
5210 449968
5215 449968
5220 449969
5225 449969
5230 449969
5235 449969
5240 449969
5245 449969
5250 449970
5255 449971
5260 449971
5265 449971
5270 449971
5275 449971
5280 449971
5285 449972
5290 449973
5295 449973
5300 449973
5305 449973
5310 449973
5315 449974
5320 449974
5325 449974
5330 449975
5335 449975
5340 449975
5345 449976
5350 449976
5355 449976
5360 449976
5365 449977
5370 449977
5375 449977
5380 449978
5385 449978
5390 449978
5395 449978
5400 449978
5405 449979
5410 449980
5415 449980
5420 449980
5425 449980
5430 449980
5435 449980
5440 449981
5445 449982
5450 449982
5455 449982
5460 449982
5465 449982
5470 449983
5475 449983
5480 449983
5485 449984
5490 449984
5495 449984
5500 449984
5505 449985
5510 449985
5515 449985
5520 449985
5525 449986
5530 449986
5535 449987
5540 449987
5545 449987
5550 449987
5555 449987
5560 449987
5565 449989
5570 449989
5575 449989
5580 449989
5585 449989
5590 449989
5595 449990
5600 449991
5605 449991
5610 449991
5615 449991
5620 449991
5625 449991
5630 449992
5635 449992
5640 449992
5645 449992
5650 449993
5655 449993
5660 449993
5665 449994
5670 449994
5675 449994
5680 449994
5685 449994
5690 449995
5695 449995
5700 449996
5705 449996
5710 449996
5715 449996
5720 449996
5725 449996
5730 449998
5735 449998
5740 449998
5745 449998
5750 449998
5755 449998
5760 449999
5765 449999
5770 450000
5775 450000
5780 450000
5785 450000
5790 450001
5795 450001
5800 450001
5805 450001
5810 450002
5815 450002
5820 450003
5825 450003
5830 450003
5835 450003
5840 450003
5845 450003
5850 450004
5855 450005
5860 450005
5865 450005
5870 450005
5875 450005
5880 450005
5885 450006
5890 450007
5895 450007
5900 450007
5905 450007
5910 450007
5915 450008
5920 450008
5925 450009
5930 450009
5935 450009
5940 450009
5945 450009
5950 450010
5955 450010
5960 450010
5965 450011
5970 450011
5975 450011
5980 450012
5985 450012
5990 450012
5995 450012
6000 450012
};
\addlegendentry{Boosting disabled}

\addplot [semithick, color2,mark=square*, mark size=3, mark options={solid,fill=red,draw=red}]
table[y expr=(\thisrowno{1}-272285)/1000]{
2765 272472
};
\addplot [semithick, color1, mark=triangle*, mark size=3, mark options={solid,fill=red,draw=red}]
table[y expr=(\thisrowno{1}-449640)/1000]{
3095 449848
};
\end{axis}
\node[text width=8cm] at (3.5,-1.5) {Filled markers indicate the end of jobs};

\end{tikzpicture}

%% file: results/power_intel.tex
\begin{tikzpicture}[font=\Large]

\definecolor{color0}{rgb}{0.12156862745098,0.466666666666667,0.705882352941177}
\definecolor{color1}{rgb}{1,0.498039215686275,0.0549019607843137}
\definecolor{color2}{rgb}{0.172549019607843,0.627450980392157,0.172549019607843}
\definecolor{color3}{rgb}{0.83921568627451,0.152941176470588,0.156862745098039}
\definecolor{color4}{rgb}{0.580392156862745,0.403921568627451,0.741176470588235}
\definecolor{color5}{rgb}{0,0,0}

\begin{axis}[
legend cell align={left},
legend columns=2,
legend style={fill opacity=0.8, draw opacity=1, text opacity=1, at={(1.02,1.15)}, anchor=east, draw=white!80.0!black},
tick align=outside,
tick pos=left,
x grid style={white!69.01960784313725!black},
xlabel={Time (SEC)},
xmin=0, xmax=3600,
xtick style={color=black},
xtick={0,900,1800,2700,3600},
y grid style={white!69.01960784313725!black},
ylabel={Power consumption (W)},
ymin=150, ymax=500,
ytick={150,200,250, 300,350,400,450},
xmajorgrids,
ymajorgrids,
ytick style={color=black}
]
\addplot [thick, color5]
table {%
0 254
5 253
10 254
15 253
20 253
25 253
30 253
35 253
40 253
45 253
50 253
55 254
60 253
65 254
70 253
75 252
80 254
85 254
90 254
95 254
100 254
105 253
110 253
115 253
120 253
125 253
130 254
135 253
140 254
145 252
150 254
155 254
160 253
165 254
170 253
175 252
180 253
185 253
190 253
195 253
200 253
205 253
210 263
215 251
220 253
225 252
230 252
235 253
240 252
245 253
250 253
255 253
260 253
265 252
270 252
275 254
280 252
285 252
290 252
295 253
300 254
305 252
310 253
315 253
320 253
325 252
330 252
335 253
340 252
345 254
350 253
355 253
360 253
365 253
370 253
375 252
380 252
385 252
390 253
395 252
400 252
405 254
410 252
415 253
420 252
425 254
430 254
435 253
440 254
445 252
450 252
455 252
460 253
465 251
470 250
475 253
480 253
485 253
490 254
495 253
500 253
505 252
510 252
515 254
520 252
525 253
530 254
535 253
540 253
545 253
550 253
555 254
560 254
565 252
570 253
575 253
580 253
585 254
590 252
595 253
600 252
605 252
610 252
615 253
620 253
625 252
630 253
635 252
640 252
645 252
650 253
655 254
660 253
665 254
670 253
675 252
680 254
685 252
690 252
695 252
700 253
705 254
710 253
715 254
720 254
725 254
730 252
735 253
740 253
745 253
750 253
755 254
760 252
765 250
770 253
775 253
780 253
785 253
790 253
795 253
800 254
805 253
810 253
815 252
820 253
825 253
830 252
835 254
840 253
845 253
850 252
855 252
860 253
865 252
870 253
875 253
880 253
885 254
890 252
895 253
900 253
905 250
910 253
915 252
920 253
925 253
930 253
935 254
940 254
945 253
950 254
955 253
960 254
965 254
970 253
975 253
980 254
985 253
990 253
995 252
1000 253
1005 254
1010 253
1015 253
1020 253
1025 252
1030 254
1035 253
1040 253
1045 252
1050 254
1055 253
1060 252
1065 253
1070 252
1075 254
1080 253
1085 253
1090 252
1095 253
1100 252
1105 252
1110 253
1115 253
1120 254
1125 252
1130 254
1135 252
1140 253
1145 253
1150 252
1155 253
1160 254
1165 253
1170 251
1175 253
1180 252
1185 253
1190 253
1195 253
1200 253
1205 252
1210 252
1215 253
1220 252
1225 253
1230 254
1235 254
1240 253
1245 253
1250 253
1255 253
1260 252
1265 253
1270 254
1275 254
1280 254
1285 252
1290 254
1295 252
1300 252
1305 254
1310 252
1315 254
1320 254
1325 254
1330 253
1335 254
1340 252
1345 252
1350 253
1355 254
1360 254
1365 253
1370 253
1375 254
1380 263
1385 251
1390 254
1395 254
1400 253
1405 253
1410 254
1415 251
1420 251
1425 252
1430 251
1435 251
1440 251
1445 251
1450 251
1455 252
1460 251
1465 251
1470 251
1475 251
1480 252
1485 251
1490 252
1495 251
1500 251
1505 252
1510 251
1515 251
1520 251
1525 251
1530 251
1535 251
1540 251
1545 251
1550 251
1555 251
1560 251
1565 249
1570 251
1575 251
1580 251
1585 251
1590 252
1595 251
1600 252
1605 252
1610 251
1615 252
1620 251
1625 251
1630 251
1635 251
1640 252
1645 251
1650 252
1655 252
1660 254
1665 252
1670 252
1675 253
1680 253
1685 254
1690 252
1695 253
1700 252
1705 252
1710 252
1715 253
1720 253
1725 253
1730 252
1735 252
1740 252
1745 253
1750 252
1755 253
1760 254
1765 254
1770 253
1775 253
1780 254
1785 254
1790 253
1795 253
1800 252
1805 254
1810 252
1815 253
1820 253
1825 254
1830 252
1835 252
1840 253
1845 250
1850 252
1855 253
1860 252
1865 253
1870 252
1875 253
1880 253
1885 252
1890 253
1895 253
1900 253
1905 252
1910 252
1915 252
1920 252
1925 253
1930 252
1935 253
1940 254
1945 253
1950 253
1955 253
1960 253
1965 253
1970 253
1975 253
1980 252
1985 253
1990 250
1995 253
2000 253
2005 254
2010 253
2015 253
2020 253
2025 252
2030 252
2035 253
2040 253
2045 252
2050 253
2055 252
2060 253
2065 254
2070 253
2075 253
2080 252
2085 252
2090 253
2095 253
2100 252
2105 253
2110 253
2115 253
2120 252
2125 252
2130 253
2135 254
2140 253
2145 252
2150 252
2155 252
2160 254
2165 253
2170 252
2175 253
2180 253
2185 252
2190 253
2195 254
2200 254
2205 253
2210 254
2215 253
2220 254
2225 253
2230 254
2235 253
2240 253
2245 252
2250 252
2255 254
2260 253
2265 253
2270 254
2275 253
2280 253
2285 253
2290 254
2295 253
2300 253
2305 253
2310 253
2315 254
2320 254
2325 252
2330 254
2335 252
2340 253
2345 252
2350 253
2355 253
2360 253
2365 254
2370 254
2375 252
2380 252
2385 253
2390 263
2395 254
2400 254
2405 253
2410 253
2415 254
2420 252
2425 251
2430 253
2435 254
2440 254
2445 252
2450 254
2455 254
2460 253
2465 253
2470 253
2475 253
2480 254
2485 254
2490 253
2495 253
2500 253
2505 253
2510 254
2515 253
2520 254
2525 253
2530 254
2535 254
2540 253
2545 254
2550 254
2555 254
2560 254
2565 253
2570 253
2575 252
2580 254
2585 254
2590 253
2595 254
2600 252
2605 254
2610 253
2615 253
2620 254
2625 253
2630 253
2635 253
2640 253
2645 252
2650 253
2655 253
2660 253
2665 253
2670 252
2675 253
2680 252
2685 253
2690 253
2695 252
2700 253
2705 253
2710 253
2715 253
2720 252
2725 254
2730 252
2735 253
2740 253
2745 254
2750 253
2755 253
2760 253
2765 254
2770 254
2775 252
2780 253
2785 253
2790 253
2795 253
2800 257
2805 251
2810 254
2815 253
2820 253
2825 252
2830 253
2835 252
2840 253
2845 253
2850 253
2855 253
2860 264
2865 251
2870 253
2875 254
2880 253
2885 253
2890 253
2895 253
2900 254
2905 254
2910 252
2915 254
2920 254
2925 253
2930 253
2935 253
2940 253
2945 253
2950 254
2955 253
2960 252
2965 254
2970 252
2975 254
2980 254
2985 254
2990 253
2995 253
3000 253
3005 253
3010 254
3015 253
3020 253
3025 253
3030 253
3035 253
3040 253
3045 253
3050 253
3055 254
3060 253
3065 253
3070 253
3075 253
3080 252
3085 251
3090 252
3095 252
3100 252
3105 251
3110 251
3115 252
3120 252
3125 251
3130 253
3135 252
3140 252
3145 252
3150 252
3155 252
3160 251
3165 251
3170 252
3175 252
3180 251
3185 252
3190 252
3195 252
3200 251
3205 251
3210 252
3215 251
3220 252
3225 252
3230 252
3235 252
3240 251
3245 252
3250 252
3255 252
3260 251
3265 252
3270 252
3275 252
3280 252
3285 252
3290 251
3295 252
3300 251
3305 252
3310 252
3315 251
3320 251
3325 251
3330 252
3335 252
3340 252
3345 252
3350 251
3355 251
3360 252
3365 252
3370 252
3375 252
3380 252
3385 251
3390 251
3395 252
3400 251
3405 251
3410 252
3415 251
3420 252
3425 252
3430 252
3435 252
3440 252
3445 251
3450 252
3455 252
3460 252
3465 251
3470 252
3475 252
3480 251
3485 252
3490 252
3495 251
3500 252
3505 252
3510 252
3515 252
3520 251
3525 252
3530 252
3535 252
3540 252
3545 251
3550 251
3555 252
3560 251
3565 251
3570 251
3575 251
3580 251
3585 251
3590 251
3595 252
3600 252
};
\addlegendentry{Baseline}
\addplot [thick, color0, mark=star, mark size=3, mark options={solid,fill=white,draw=red},  mark repeat={180}]
table {%
0 254
5 263
10 263
15 265
20 268
25 269
30 265
35 257
40 255
45 253
50 254
55 253
60 253
65 253
70 254
75 254
80 253
85 254
90 254
95 253
100 254
105 253
110 256
115 267
120 267
125 267
130 268
135 262
140 263
145 262
150 263
155 262
160 266
165 267
170 266
175 259
180 267
185 266
190 266
195 266
200 265
205 266
210 265
215 267
220 267
225 266
230 265
235 265
240 265
245 265
250 266
255 266
260 267
265 267
270 266
275 265
280 267
285 266
290 266
295 267
300 266
305 266
310 268
315 267
320 267
325 266
330 266
335 267
340 266
345 267
350 266
355 267
360 267
365 266
370 266
375 266
380 266
385 266
390 267
395 266
400 266
405 267
410 267
415 266
420 266
425 266
430 268
435 267
440 267
445 266
450 267
455 267
460 266
465 266
470 267
475 267
480 267
485 266
490 266
495 266
500 265
505 265
510 265
515 266
520 265
525 266
530 266
535 266
540 267
545 267
550 266
555 267
560 266
565 266
570 267
575 266
580 266
585 267
590 266
595 266
600 267
605 266
610 268
615 266
620 267
625 267
630 267
635 266
640 267
645 266
650 267
655 267
660 266
665 266
670 267
675 266
680 266
685 266
690 267
695 267
700 266
705 267
710 266
715 266
720 267
725 266
730 267
735 267
740 266
745 267
750 266
755 266
760 267
765 267
770 266
775 266
780 266
785 267
790 265
795 266
800 266
805 267
810 266
815 266
820 266
825 267
830 267
835 267
840 267
845 266
850 267
855 267
860 267
865 267
870 268
875 267
880 267
885 267
890 266
895 266
900 267
905 267
910 266
915 267
920 266
925 266
930 266
935 267
940 266
945 266
950 267
955 267
960 266
965 267
970 266
975 267
980 266
985 266
990 266
995 267
1000 266
1005 266
1010 266
1015 267
1020 267
1025 267
1030 267
1035 267
1040 266
1045 268
1050 266
1055 266
1060 267
1065 266
1070 266
1075 267
1080 266
1085 267
1090 267
1095 266
1100 267
1105 266
1110 268
1115 267
1120 267
1125 267
1130 267
1135 267
1140 268
1145 269
1150 268
1155 267
1160 266
1165 266
1170 267
1175 267
1180 266
1185 267
1190 266
1195 267
1200 267
1205 267
1210 266
1215 266
1220 266
1225 268
1230 266
1235 267
1240 267
1245 267
1250 266
1255 267
1260 266
1265 266
1270 267
1275 267
1280 267
1285 267
1290 268
1295 267
1300 266
1305 266
1310 268
1315 267
1320 266
1325 267
1330 266
1335 266
1340 268
1345 266
1350 267
1355 266
1360 267
1365 266
1370 267
1375 267
1380 266
1385 266
1390 267
1395 266
1400 267
1405 267
1410 268
1415 267
1420 267
1425 266
1430 266
1435 267
1440 267
1445 266
1450 267
1455 266
1460 266
1465 268
1470 267
1475 267
1480 267
1485 267
1490 267
1495 268
1500 266
1505 266
1510 267
1515 267
1520 267
1525 266
1530 268
1535 266
1540 266
1545 267
1550 268
1555 267
1560 267
1565 267
1570 269
1575 267
1580 267
1585 266
1590 268
1595 266
1600 268
1605 267
1610 266
1615 267
1620 266
1625 267
1630 267
1635 267
1640 267
1645 267
1650 266
1655 266
1660 268
1665 267
1670 266
1675 267
1680 266
1685 266
1690 267
1695 266
1700 267
1705 266
1710 266
1715 267
1720 267
1725 267
1730 266
1735 266
1740 267
1745 271
1750 266
1755 266
1760 268
1765 267
1770 268
1775 267
1780 267
1785 266
1790 267
1795 267
1800 266
1805 267
1810 266
1815 266
1820 266
1825 266
1830 267
1835 267
1840 267
1845 266
1850 266
1855 266
1860 266
1865 267
1870 266
1875 266
1880 267
1885 266
1890 266
1895 267
1900 266
1905 267
1910 267
1915 266
1920 266
1925 267
1930 267
1935 267
1940 267
1945 267
1950 267
1955 266
1960 267
1965 267
1970 267
1975 267
1980 266
1985 267
1990 266
1995 266
2000 266
2005 267
2010 266
2015 267
2020 267
2025 267
2030 267
2035 266
2040 267
2045 268
2050 267
2055 266
2060 266
2065 267
2070 268
2075 266
2080 267
2085 266
2090 266
2095 267
2100 267
2105 267
2110 266
2115 266
2120 267
2125 267
2130 266
2135 267
2140 267
2145 267
2150 266
2155 267
2160 267
2165 267
2170 268
2175 267
2180 266
2185 266
2190 268
2195 266
2200 266
2205 266
2210 267
2215 266
2220 267
2225 267
2230 266
2235 266
2240 267
2245 267
2250 267
2255 268
2260 267
2265 266
2270 267
2275 268
2280 266
2285 266
2290 267
2295 267
2300 267
2305 266
2310 266
2315 266
2320 267
2325 267
2330 267
2335 267
2340 266
2345 272
2350 266
2355 266
2360 267
2365 267
2370 267
2375 266
2380 267
2385 267
2390 268
2395 267
2400 268
2405 267
2410 266
2415 267
2420 267
2425 266
2430 268
2435 267
2440 267
2445 267
2450 266
2455 267
2460 267
2465 266
2470 266
2475 267
2480 266
2485 267
2490 267
2495 268
2500 266
2505 266
2510 267
2515 267
2520 268
2525 268
2530 266
2535 267
2540 268
2545 266
2550 268
2555 266
2560 267
2565 268
2570 267
2575 267
2580 267
2585 266
2590 266
2595 268
2600 268
2605 267
2610 268
2615 268
2620 267
2625 267
2630 267
2635 266
2640 267
2645 269
2650 267
2655 267
2660 267
2665 267
2670 267
2675 267
2680 267
2685 268
2690 266
2695 268
2700 266
2705 267
2710 267
2715 268
2720 266
2725 266
2730 266
2735 268
2740 268
2745 267
2750 267
2755 267
2760 266
2765 268
2770 267
2775 266
2780 268
2785 268
2790 267
2795 267
2800 268
2805 266
2810 266
2815 268
2820 267
2825 268
2830 267
2835 266
2840 267
2845 268
2850 267
2855 268
2860 266
2865 268
2870 266
2875 267
2880 266
2885 267
2890 267
2895 268
2900 267
2905 266
2910 267
2915 266
2920 266
2925 267
2930 266
2935 266
2940 268
2945 268
2950 267
2955 266
2960 267
2965 267
2970 267
2975 267
2980 268
2985 266
2990 268
2995 266
3000 267
3005 267
3010 267
3015 267
3020 267
3025 267
3030 266
3035 266
3040 266
3045 268
3050 267
3055 267
3060 267
3065 266
3070 266
3075 266
3080 267
3085 266
3090 267
3095 267
3100 267
3105 267
3110 266
3115 266
3120 267
3125 266
3130 266
3135 267
3140 266
3145 266
3150 268
3155 267
3160 266
3165 267
3170 267
3175 267
3180 267
3185 266
3190 268
3195 267
3200 267
3205 267
3210 267
3215 267
3220 267
3225 267
3230 267
3235 266
3240 267
3245 268
3250 268
3255 267
3260 266
3265 267
3270 266
3275 266
3280 267
3285 268
3290 266
3295 267
3300 266
3305 268
3310 267
3315 267
3320 266
3325 268
3330 268
3335 266
3340 267
3345 268
3350 268
3355 268
3360 267
3365 267
3370 266
3375 268
3380 267
3385 267
3390 266
3395 268
3400 268
3405 267
3410 267
3415 267
3420 268
3425 267
3430 267
3435 268
3440 267
3445 268
3450 266
3455 266
3460 267
3465 266
3470 268
3475 268
3480 276
3485 277
3490 277
3495 276
3500 278
3505 277
3510 277
3515 271
3520 268
3525 266
3530 267
3535 267
3540 268
3545 267
3550 268
3555 266
3560 268
3565 267
3570 268
3575 268
3580 266
3585 267
3590 267
3595 267
3600 267
};
\addlegendentry{1 core}
\addplot [thick, color1, mark=triangle*, mark size=3, mark options={solid,fill=white,draw=red},  mark repeat={180}]
table {%
0 255
5 262
10 255
15 255
20 255
25 255
30 256
35 254
40 255
45 256
50 256
55 267
60 271
65 270
70 255
75 256
80 254
85 256
90 255
95 255
100 257
105 256
110 254
115 256
120 255
125 255
130 257
135 254
140 254
145 254
150 254
155 254
160 255
165 255
170 254
175 254
180 254
185 257
190 255
195 255
200 254
205 256
210 255
215 254
220 256
225 267
230 266
235 266
240 267
245 267
250 255
255 255
260 259
265 264
270 263
275 261
280 255
285 255
290 254
295 263
300 256
305 254
310 255
315 254
320 254
325 254
330 256
335 254
340 254
345 255
350 255
355 254
360 256
365 255
370 255
375 259
380 267
385 268
390 262
395 271
400 267
405 256
410 305
415 306
420 307
425 306
430 304
435 305
440 305
445 304
450 305
455 306
460 305
465 305
470 305
475 306
480 307
485 305
490 306
495 307
500 305
505 306
510 307
515 306
520 307
525 308
530 308
535 307
540 306
545 308
550 307
555 306
560 306
565 308
570 308
575 307
580 306
585 307
590 308
595 308
600 307
605 307
610 309
615 307
620 308
625 308
630 307
635 308
640 308
645 307
650 308
655 307
660 308
665 307
670 307
675 308
680 308
685 309
690 308
695 311
700 307
705 307
710 308
715 309
720 308
725 308
730 309
735 307
740 310
745 309
750 308
755 308
760 308
765 309
770 309
775 308
780 308
785 309
790 308
795 309
800 308
805 309
810 308
815 309
820 310
825 307
830 310
835 310
840 308
845 309
850 308
855 307
860 309
865 308
870 308
875 308
880 308
885 309
890 309
895 307
900 307
905 308
910 308
915 307
920 309
925 308
930 307
935 307
940 308
945 308
950 309
955 308
960 307
965 308
970 308
975 309
980 308
985 309
990 309
995 314
1000 308
1005 307
1010 308
1015 308
1020 308
1025 307
1030 309
1035 307
1040 307
1045 309
1050 307
1055 308
1060 309
1065 308
1070 308
1075 308
1080 309
1085 308
1090 309
1095 307
1100 308
1105 308
1110 308
1115 307
1120 308
1125 307
1130 308
1135 307
1140 308
1145 308
1150 308
1155 307
1160 307
1165 309
1170 308
1175 307
1180 308
1185 307
1190 307
1195 309
1200 308
1205 307
1210 309
1215 307
1220 308
1225 308
1230 308
1235 317
1240 319
1245 317
1250 319
1255 319
1260 318
1265 315
1270 309
1275 308
1280 308
1285 309
1290 308
1295 313
1300 310
1305 309
1310 302
1315 307
1320 307
1325 307
1330 307
1335 308
1340 307
1345 307
1350 306
1355 307
1360 308
1365 308
1370 307
1375 307
1380 308
1385 307
1390 307
1395 307
1400 308
1405 308
1410 308
1415 306
1420 308
1425 306
1430 306
1435 308
1440 307
1445 307
1450 308
1455 307
1460 307
1465 308
1470 306
1475 306
1480 306
1485 306
1490 307
1495 306
1500 306
1505 307
1510 308
1515 307
1520 307
1525 305
1530 306
1535 306
1540 307
1545 306
1550 305
1555 307
1560 306
1565 305
1570 306
1575 306
1580 307
1585 307
1590 306
1595 310
1600 306
1605 306
1610 306
1615 306
1620 305
1625 307
1630 307
1635 306
1640 306
1645 306
1650 305
1655 306
1660 306
1665 306
1670 306
1675 307
1680 306
1685 307
1690 306
1695 306
1700 306
1705 305
1710 307
1715 307
1720 306
1725 304
1730 305
1735 306
1740 306
1745 306
1750 307
1755 304
1760 306
1765 305
1770 307
1775 305
1780 305
1785 306
1790 305
1795 306
1800 306
1805 308
1810 305
1815 305
1820 305
1825 307
1830 307
1835 307
1840 306
1845 305
1850 305
1855 305
1860 307
1865 306
1870 305
1875 306
1880 305
1885 307
1890 305
1895 305
1900 306
1905 307
1910 307
1915 306
1920 305
1925 306
1930 306
1935 307
1940 305
1945 305
1950 305
1955 306
1960 306
1965 305
1970 306
1975 306
1980 306
1985 307
1990 307
1995 307
2000 306
2005 305
2010 306
2015 306
2020 306
2025 307
2030 308
2035 307
2040 307
2045 308
2050 306
2055 308
2060 307
2065 307
2070 307
2075 307
2080 306
2085 306
2090 308
2095 307
2100 306
2105 308
2110 306
2115 306
2120 307
2125 306
2130 307
2135 308
2140 307
2145 306
2150 306
2155 306
2160 306
2165 307
2170 307
2175 307
2180 307
2185 307
2190 306
2195 309
2200 308
2205 307
2210 307
2215 307
2220 306
2225 307
2230 306
2235 307
2240 307
2245 306
2250 306
2255 308
2260 308
2265 307
2270 306
2275 307
2280 308
2285 307
2290 307
2295 308
2300 307
2305 307
2310 307
2315 307
2320 307
2325 307
2330 307
2335 305
2340 307
2345 306
2350 306
2355 307
2360 306
2365 305
2370 307
2375 306
2380 307
2385 306
2390 307
2395 306
2400 306
2405 306
2410 307
2415 307
2420 306
2425 306
2430 307
2435 305
2440 306
2445 308
2450 307
2455 308
2460 307
2465 308
2470 307
2475 308
2480 308
2485 306
2490 307
2495 306
2500 307
2505 307
2510 307
2515 306
2520 307
2525 308
2530 308
2535 308
2540 306
2545 307
2550 307
2555 306
2560 308
2565 307
2570 306
2575 308
2580 307
2585 307
2590 308
2595 307
2600 306
2605 307
2610 307
2615 308
2620 307
2625 307
2630 307
2635 306
2640 308
2645 307
2650 308
2655 307
2660 307
2665 307
2670 308
2675 306
2680 306
2685 307
2690 309
2695 308
2700 308
2705 308
2710 307
2715 307
2720 308
2725 308
2730 308
2735 306
2740 306
2745 307
2750 306
2755 307
2760 307
2765 308
2770 307
2775 307
2780 307
2785 306
2790 307
2795 307
2800 308
2805 306
2810 306
2815 308
2820 308
2825 307
2830 308
2835 308
2840 307
2845 307
2850 307
2855 307
2860 308
2865 307
2870 308
2875 307
2880 307
2885 308
2890 307
2895 306
2900 308
2905 308
2910 307
2915 308
2920 307
2925 307
2930 308
2935 307
2940 306
2945 308
2950 308
2955 307
2960 308
2965 308
2970 307
2975 307
2980 307
2985 308
2990 307
2995 307
3000 308
3005 307
3010 308
3015 308
3020 308
3025 308
3030 307
3035 307
3040 308
3045 307
3050 308
3055 308
3060 309
3065 308
3070 307
3075 308
3080 307
3085 308
3090 307
3095 307
3100 309
3105 308
3110 308
3115 307
3120 307
3125 307
3130 307
3135 307
3140 308
3145 308
3150 307
3155 307
3160 307
3165 307
3170 308
3175 307
3180 308
3185 308
3190 307
3195 307
3200 308
3205 308
3210 308
3215 308
3220 308
3225 308
3230 308
3235 306
3240 308
3245 306
3250 308
3255 308
3260 308
3265 308
3270 309
3275 307
3280 308
3285 308
3290 308
3295 308
3300 307
3305 308
3310 308
3315 308
3320 307
3325 308
3330 307
3335 307
3340 306
3345 307
3350 307
3355 306
3360 307
3365 307
3370 306
3375 305
3380 306
3385 307
3390 306
3395 306
3400 306
3405 305
3410 306
3415 307
3420 306
3425 306
3430 307
3435 306
3440 305
3445 305
3450 306
3455 305
3460 306
3465 307
3470 306
3475 307
3480 306
3485 307
3490 307
3495 306
3500 306
3505 305
3510 305
3515 306
3520 305
3525 307
3530 305
3535 307
3540 306
3545 305
3550 306
3555 307
3560 305
3565 305
3570 306
3575 305
3580 306
3585 307
3590 306
3595 307
3600 307
};
\addlegendentry{4 cores}
\addplot [thick, color2, mark=square*, mark size=3, mark options={solid,fill=white,draw=red},  mark repeat={180}]
table {%
0 253
5 264
10 265
15 272
20 272
0 264
5 263
10 266
0 254
5 263
10 266
15 269
20 270
25 261
30 266
35 254
40 253
45 253
50 255
55 254
60 253
65 254
70 254
75 253
80 253
85 253
90 255
95 254
100 256
105 256
110 267
115 268
120 268
125 266
130 264
135 262
140 263
145 263
150 255
155 264
160 255
165 255
170 254
175 255
180 254
185 256
190 255
195 255
200 268
205 266
210 266
215 258
220 268
225 260
230 353
235 355
240 355
245 356
250 355
255 356
260 357
265 357
270 357
275 357
280 357
285 356
290 358
295 359
300 359
305 358
310 361
315 359
320 362
325 362
330 362
335 360
340 362
345 361
350 362
355 361
360 362
365 362
370 362
375 362
380 363
385 363
390 363
395 362
400 362
405 363
410 364
415 364
420 363
425 365
430 364
435 365
440 364
445 365
450 365
455 366
460 366
465 366
470 365
475 365
480 365
485 366
490 366
495 366
500 366
505 361
510 365
515 363
520 365
525 363
530 364
535 362
540 364
545 364
550 363
555 365
560 363
565 365
570 364
575 365
580 364
585 363
590 364
595 365
600 363
605 365
610 362
615 376
620 366
625 367
630 365
635 367
640 366
645 366
650 368
655 366
660 366
665 365
670 367
675 367
680 366
685 367
690 366
695 365
700 368
705 368
710 368
715 366
720 368
725 366
730 367
735 366
740 367
745 367
750 367
755 367
760 367
765 368
770 365
775 369
780 368
785 367
790 368
795 367
800 367
805 368
810 367
815 368
820 367
825 368
830 366
835 367
840 368
845 368
850 367
855 367
860 366
865 368
870 367
875 366
880 368
885 367
890 367
895 367
900 367
905 367
910 366
915 367
920 366
925 368
930 367
935 367
940 367
945 368
950 366
955 367
960 366
965 366
970 366
975 364
980 367
985 367
990 366
995 368
1000 366
1005 368
1010 368
1015 367
1020 367
1025 368
1030 368
1035 366
1040 367
1045 367
1050 369
1055 367
1060 368
1065 368
1070 368
1075 367
1080 368
1085 367
1090 368
1095 367
1100 367
1105 367
1110 369
1115 367
1120 367
1125 367
1130 368
1135 367
1140 367
1145 377
1150 365
1155 368
1160 367
1165 367
1170 369
1175 368
1180 369
1185 368
1190 368
1195 369
1200 368
1205 368
1210 369
1215 367
1220 368
1225 369
1230 368
1235 368
1240 369
1245 368
1250 368
1255 368
1260 369
1265 368
1270 368
1275 367
1280 368
1285 367
1290 368
1295 368
1300 369
1305 367
1310 369
1315 368
1320 368
1325 367
1330 367
1335 368
1340 366
1345 368
1350 368
1355 368
1360 367
1365 368
1370 366
1375 368
1380 367
1385 366
1390 365
1395 366
1400 366
1405 367
1410 366
1415 367
1420 366
1425 367
1430 366
1435 367
1440 365
1445 366
1450 366
1455 367
1460 366
1465 367
1470 366
1475 366
1480 367
1485 366
1490 368
1495 367
1500 368
1505 367
1510 366
1515 366
1520 368
1525 365
1530 367
1535 367
1540 367
1545 365
1550 367
1555 366
1560 364
1565 366
1570 366
1575 367
1580 366
1585 368
1590 368
1595 367
1600 367
1605 367
1610 366
1615 367
1620 367
1625 366
1630 367
1635 367
1640 367
1645 366
1650 366
1655 367
1660 367
1665 367
1670 368
1675 368
1680 366
1685 367
1690 367
1695 366
1700 368
1705 367
1710 366
1715 366
1720 368
1725 367
1730 368
1735 367
1740 368
1745 368
1750 367
1755 367
1760 367
1765 368
1770 368
1775 367
1780 367
1785 368
1790 367
1795 367
1800 367
1805 368
1810 366
1815 368
1820 368
1825 368
1830 367
1835 368
1840 367
1845 368
1850 367
1855 369
1860 368
1865 366
1870 367
1875 367
1880 368
1885 368
1890 368
1895 368
1900 367
1905 368
1910 367
1915 367
1920 367
1925 366
1930 366
1935 368
1940 367
1945 368
1950 367
1955 368
1960 367
1965 367
1970 368
1975 370
1980 367
1985 367
1990 367
1995 367
2000 368
2005 369
2010 368
2015 367
2020 368
2025 367
2030 367
2035 367
2040 367
2045 368
2050 367
2055 367
2060 368
2065 366
2070 366
2075 366
2080 367
2085 368
2090 367
2095 366
2100 368
2105 366
2110 367
2115 366
2120 368
2125 367
2130 366
2135 366
2140 367
2145 366
2150 368
2155 367
2160 366
2165 367
2170 367
2175 367
2180 367
2185 367
2190 366
2195 366
2200 366
2205 366
2210 367
2215 368
2220 366
2225 368
2230 368
2235 367
2240 368
2245 368
2250 367
2255 367
2260 368
2265 368
2270 368
2275 368
2280 368
2285 367
2290 368
2295 367
2300 368
2305 368
2310 369
2315 367
2320 368
2325 367
2330 369
2335 368
2340 369
2345 367
2350 368
2355 368
2360 368
2365 368
2370 367
2375 367
2380 367
2385 368
2390 367
2395 368
2400 369
2405 367
2410 368
2415 368
2420 369
2425 368
2430 368
2435 369
2440 367
2445 368
2450 368
2455 368
2460 368
2465 369
2470 369
2475 367
2480 370
2485 345
2490 314
2495 313
2500 290
2505 273
2510 274
2515 272
2520 273
2525 267
2530 270
2535 259
2540 259
2545 258
2550 259
2555 256
2560 256
2565 256
2570 257
2575 260
2580 256
2585 257
2590 256
2595 256
2600 255
2605 255
2610 256
2615 256
2620 255
2625 255
2630 256
2635 256
2640 255
2645 257
2650 255
2655 255
2660 255
2665 255
2670 255
2675 255
2680 255
2685 255
2690 254
2695 256
2700 256
2705 254
2710 255
2715 254
2720 255
2725 255
2730 255
2735 254
2740 256
2745 254
2750 255
2755 255
2760 256
2765 252
2770 254
2775 255
2780 254
2785 254
2790 254
2795 256
2800 254
2805 254
2810 255
2815 255
2820 254
2825 255
2830 253
2835 255
2840 253
2845 255
2850 254
2855 254
2860 253
2865 254
2870 253
2875 257
2880 254
2885 254
2890 254
2895 254
2900 255
2905 253
2910 254
2915 254
2920 254
2925 253
2930 253
2935 253
2940 255
2945 254
2950 255
2955 253
2960 255
2965 254
2970 254
2975 254
2980 254
2985 253
2990 254
2995 254
3000 254
3005 255
3010 253
3015 254
3020 253
3025 253
3030 253
3035 253
3040 253
3045 251
3050 253
3055 255
3060 253
3065 254
3070 253
3075 255
3080 254
3085 254
3090 254
3095 254
3100 254
3105 253
3110 255
3115 253
3120 253
3125 253
3130 253
3135 253
3140 254
3145 253
3150 254
3155 253
3160 255
3165 254
3170 254
3175 253
3180 253
3185 254
3190 254
3195 254
3200 253
3205 253
3210 253
3215 253
3220 254
3225 253
3230 254
3235 254
3240 253
3245 253
3250 254
3255 255
3260 253
3265 254
3270 254
3275 254
3280 254
3285 254
3290 254
3295 253
3300 253
3305 253
3310 255
3315 254
3320 254
3325 253
3330 253
3335 254
3340 253
3345 253
3350 253
3355 254
3360 255
3365 253
3370 254
3375 255
3380 254
3385 254
3390 254
3395 253
3400 254
3405 254
3410 254
3415 253
3420 253
3425 254
3430 253
3435 255
3440 253
3445 254
3450 253
3455 253
3460 254
3465 253
3470 253
3475 253
3480 253
3485 254
3490 253
3495 254
3500 253
3505 253
3510 253
3515 254
3520 254
3525 253
3530 254
3535 253
3540 254
3545 253
3550 253
3555 254
3560 254
3565 253
3570 254
3575 253
3580 253
3585 253
3590 253
3595 253
};
\addlegendentry{8 cores}
\addplot [thick, color3, mark=o, mark size=3, mark options={solid,fill=white,draw=red},  mark repeat={180}]
table {%
0 256
5 256
10 266
15 269
20 273
25 275
30 260
35 269
40 257
45 256
50 255
55 255
60 256
65 255
70 257
75 256
80 256
85 255
90 256
95 256
100 256
105 256
110 264
115 270
120 270
125 269
130 270
135 264
140 265
145 264
150 267
155 267
160 270
165 271
170 269
175 270
180 322
185 404
190 465
195 468
200 470
205 471
210 473
215 475
220 474
225 475
230 474
235 474
240 476
245 476
250 475
255 476
260 476
265 475
270 474
275 472
280 473
285 472
290 472
295 470
300 470
305 472
310 471
315 471
320 469
325 472
330 473
335 474
340 473
345 473
350 474
355 474
360 474
365 473
370 474
375 474
380 473
385 474
390 474
395 474
400 474
405 475
410 474
415 474
420 473
425 473
430 475
435 475
440 474
445 475
450 475
455 472
460 475
465 474
470 475
475 475
480 475
485 474
490 474
495 475
500 476
505 474
510 476
515 475
520 474
525 476
530 475
535 476
540 475
545 476
550 475
555 476
560 476
565 476
570 476
575 476
580 476
585 475
590 475
595 476
600 477
605 475
610 477
615 476
620 475
625 476
630 477
635 477
640 476
645 478
650 476
655 477
660 476
665 477
670 477
675 477
680 477
685 477
690 477
695 477
700 477
705 478
710 478
715 477
720 478
725 474
730 477
735 478
740 477
745 478
750 478
755 477
760 478
765 478
770 477
775 478
780 478
785 477
790 478
795 477
800 478
805 477
810 477
815 478
820 478
825 477
830 477
835 477
840 477
845 476
850 476
855 476
860 477
865 477
870 477
875 476
880 476
885 476
890 475
895 476
900 476
905 477
910 477
915 477
920 477
925 476
930 476
935 476
940 475
945 475
950 475
955 475
960 475
965 475
970 475
975 476
980 476
985 475
990 476
995 475
1000 475
1005 475
1010 474
1015 474
1020 475
1025 474
1030 475
1035 474
1040 475
1045 474
1050 475
1055 474
1060 475
1065 475
1070 475
1075 474
1080 475
1085 475
1090 475
1095 474
1100 475
1105 474
1110 475
1115 475
1120 475
1125 475
1130 474
1135 474
1140 474
1145 474
1150 474
1155 474
1160 474
1165 474
1170 475
1175 474
1180 474
1185 475
1190 475
1195 475
1200 474
1205 475
1210 474
1215 474
1220 475
1225 476
1230 476
1235 475
1240 474
1245 476
1250 475
1255 476
1260 475
1265 476
1270 475
1275 477
1280 475
1285 475
1290 476
1295 476
1300 476
1305 477
1310 476
1315 477
1320 477
1325 476
1330 477
1335 476
1340 476
1345 474
1350 423
1355 364
1360 334
1365 330
1370 315
1375 314
1380 300
1385 299
1390 282
1395 260
1400 272
1405 260
1410 259
1415 259
1420 258
1425 258
1430 258
1435 258
1440 259
1445 258
1450 257
1455 258
1460 259
1465 258
1470 257
1475 257
1480 257
1485 257
1490 258
1495 258
1500 258
1505 257
1510 257
1515 257
1520 257
1525 257
1530 256
1535 257
1540 256
1545 256
1550 256
1555 267
1560 254
1565 257
1570 257
1575 257
1580 257
1585 257
1590 257
1595 257
1600 257
1605 256
1610 256
1615 257
1620 257
1625 257
1630 256
1635 255
1640 256
1645 256
1650 256
1655 257
1660 257
1665 256
1670 255
1675 256
1680 256
1685 255
1690 255
1695 256
1700 255
1705 255
1710 255
1715 256
1720 255
1725 256
1730 256
1735 256
1740 255
1745 256
1750 256
1755 256
1760 255
1765 256
1770 255
1775 256
1780 255
1785 255
1790 255
1795 256
1800 256
1805 256
1810 255
1815 255
1820 256
1825 256
1830 256
1835 255
1840 255
1845 255
1850 256
1855 255
1860 256
1865 255
1870 255
1875 256
1880 255
1885 255
1890 256
1895 255
1900 255
1905 256
1910 253
1915 253
1920 255
1925 255
1930 256
1935 255
1940 256
1945 256
1950 256
1955 256
1960 256
1965 256
1970 255
1975 256
1980 256
1985 256
1990 255
1995 256
2000 255
2005 256
2010 256
2015 256
2020 256
2025 256
2030 256
2035 256
2040 256
2045 256
2050 255
2055 255
2060 256
2065 256
2070 255
2075 255
2080 256
2085 255
2090 256
2095 256
2100 256
2105 255
2110 255
2115 255
2120 256
2125 255
2130 256
2135 255
2140 256
2145 255
2150 256
2155 255
2160 256
2165 256
2170 255
2175 256
2180 256
2185 256
2190 256
2195 255
2200 256
2205 256
2210 255
2215 256
2220 255
2225 256
2230 256
2235 256
2240 255
2245 256
2250 256
2255 255
2260 255
2265 255
2270 256
2275 255
2280 256
2285 255
2290 255
2295 255
2300 255
2305 256
2310 256
2315 256
2320 255
2325 255
2330 256
2335 255
2340 256
2345 255
2350 256
2355 256
2360 255
2365 255
2370 255
2375 255
2380 255
2385 255
2390 255
2395 255
2400 256
2405 256
2410 255
2415 256
2420 255
2425 255
2430 255
2435 255
2440 255
2445 255
2450 255
2455 255
2460 255
2465 255
2470 255
2475 255
2480 255
2485 255
2490 255
2495 256
2500 255
2505 255
2510 255
2515 255
2520 253
2525 255
2530 256
2535 256
2540 256
2545 255
2550 255
2555 256
2560 256
2565 256
2570 255
2575 256
2580 256
2585 255
2590 255
2595 254
2600 256
2605 256
2610 255
2615 255
2620 255
2625 255
2630 256
2635 256
2640 255
2645 255
2650 255
2655 256
2660 256
2665 255
2670 255
2675 256
2680 255
2685 255
2690 256
2695 256
2700 254
2705 255
2710 255
2715 255
2720 255
2725 255
2730 256
2735 256
2740 255
2745 255
2750 255
2755 256
2760 256
2765 256
2770 255
2775 256
2780 256
2785 255
2790 255
2795 255
2800 256
2805 254
2810 256
2815 255
2820 256
2825 255
2830 256
2835 256
2840 256
2845 255
2850 256
2855 256
2860 255
2865 255
2870 255
2875 256
2880 256
2885 255
2890 256
2895 255
2900 256
2905 255
2910 255
2915 255
2920 255
2925 256
2930 255
2935 256
2940 256
2945 255
2950 256
2955 255
2960 256
2965 256
2970 256
2975 256
2980 256
2985 256
2990 256
2995 256
3000 256
3005 255
3010 256
3015 255
3020 256
3025 256
3030 255
3035 256
3040 256
3045 255
3050 256
3055 255
3060 255
3065 255
3070 256
3075 255
3080 256
3085 256
3090 256
3095 256
3100 256
3105 255
3110 256
3115 256
3120 255
3125 256
3130 256
3135 256
3140 255
3145 256
3150 255
3155 256
3160 255
3165 255
3170 256
3175 255
3180 255
3185 256
3190 256
3195 256
3200 256
3205 255
3210 256
3215 255
3220 256
3225 255
3230 255
3235 255
3240 256
3245 255
3250 256
3255 255
3260 256
3265 256
3270 256
3275 256
3280 256
3285 256
3290 255
3295 256
3300 256
3305 256
3310 256
3315 256
3320 256
3325 256
3330 256
3335 256
3340 256
3345 255
3350 255
3355 255
3360 255
3365 255
3370 256
3375 255
3380 255
3385 255
3390 256
3395 256
3400 255
3405 255
3410 255
3415 255
3420 255
3425 253
3430 255
3435 255
3440 256
3445 256
3450 256
3455 255
3460 256
3465 255
3470 256
3475 255
3480 255
3485 255
3490 255
3495 255
3500 255
3505 255
3510 255
3515 255
3520 255
3525 256
3530 255
3535 255
3540 256
3545 257
3550 256
3555 258
3560 256
3565 255
3570 256
3575 256
3580 255
3585 256
3590 255
3595 256
3600 256
3605 256
3610 255
3615 256
3620 256
3625 255
3630 256
3635 256
3640 255
3645 255
3650 255
3655 256
3660 256
3665 256
3670 255
3675 255
3680 256
3685 256
3690 255
3695 256
3700 255
3705 256
3710 255
3715 256
3720 256
3725 256
3730 256
3735 255
3740 255
3745 254
3750 254
3755 256
3760 256
3765 256
3770 256
3775 256
3780 255
3785 256
3790 255
3795 255
3800 255
3805 256
3810 256
3815 256
3820 256
3825 256
3830 256
3835 256
3840 256
3845 256
3850 256
3855 260
3860 255
3865 255
3870 256
3875 256
3880 256
3885 256
3890 255
3895 255
3900 256
3905 255
3910 255
3915 256
3920 256
3925 256
3930 255
3935 255
3940 255
3945 256
3950 256
3955 256
3960 256
3965 255
3970 256
3975 255
3980 255
3985 255
3990 256
3995 256
4000 255
4005 256
4010 256
4015 256
4020 255
4025 256
4030 255
4035 256
4040 256
4045 256
4050 255
4055 256
4060 256
4065 256
4070 256
4075 256
4080 255
4085 255
4090 256
4095 256
4100 255
4105 255
4110 256
4115 256
4120 256
4125 256
4130 256
4135 256
4140 256
4145 256
4150 256
4155 256
4160 256
4165 255
4170 256
4175 256
4180 256
4185 255
4190 256
4195 255
4200 256
4205 254
4210 256
4215 256
4220 256
4225 255
4230 256
4235 255
4240 255
4245 255
4250 256
4255 256
4260 256
4265 256
4270 256
4275 255
4280 256
4285 256
4290 256
4295 256
4300 256
4305 256
4310 255
4315 256
4320 255
4325 256
4330 256
4335 255
4340 255
4345 256
4350 255
4355 255
4360 253
4365 254
4370 256
4375 255
4380 255
4385 256
4390 256
4395 255
4400 256
4405 256
4410 254
4415 255
4420 256
4425 255
4430 255
4435 256
4440 256
4445 256
4450 256
4455 256
4460 256
4465 256
4470 255
4475 256
4480 255
4485 255
4490 256
4495 256
4500 256
4505 256
4510 255
4515 256
4520 255
4525 256
4530 256
4535 256
4540 256
4545 255
4550 255
4555 255
4560 256
4565 256
4570 256
4575 256
4580 256
4585 255
4590 255
4595 255
4600 256
4605 255
4610 256
4615 256
4620 256
4625 256
4630 256
4635 255
4640 256
4645 255
4650 255
4655 256
4660 256
4665 255
4670 256
4675 255
4680 256
4685 256
4690 256
4695 256
4700 255
4705 254
4710 255
4715 255
4720 255
4725 255
4730 255
4735 256
4740 255
4745 256
4750 255
4755 254
4760 254
4765 256
4770 255
4775 255
4780 255
4785 255
4790 256
4795 254
4800 256
4805 256
4810 255
4815 256
4820 255
4825 255
4830 256
4835 256
4840 256
4845 256
4850 255
4855 255
4860 254
4865 256
4870 255
4875 254
4880 256
4885 254
4890 255
4895 256
4900 255
4905 255
4910 255
4915 256
4920 256
4925 254
4930 255
4935 255
4940 256
4945 256
4950 255
4955 256
4960 256
4965 256
4970 254
4975 256
4980 255
4985 256
4990 256
4995 255
5000 256
5005 256
5010 256
5015 255
5020 255
5025 254
5030 255
5035 256
5040 255
5045 256
5050 255
5055 255
5060 256
5065 256
5070 255
5075 255
5080 256
5085 256
5090 256
5095 256
5100 256
5105 255
5110 256
5115 256
5120 256
5125 255
5130 255
5135 255
5140 256
5145 256
5150 256
5155 256
5160 256
5165 254
5170 255
5175 256
5180 256
5185 256
5190 255
5195 256
5200 253
5205 255
5210 256
5215 255
5220 255
5225 256
5230 256
5235 255
5240 256
5245 256
5250 255
5255 255
5260 256
5265 256
5270 256
5275 256
5280 255
5285 256
5290 256
5295 256
5300 256
5305 256
5310 256
5315 256
5320 256
5325 255
5330 256
5335 256
5340 256
5345 255
5350 256
5355 256
5360 256
5365 255
5370 256
5375 256
5380 255
5385 255
5390 256
5395 255

};
\addlegendentry{16 cores}
\addplot [thick, color4, mark=diamond*, mark size=3, mark options={solid,fill=white,draw=red},  mark repeat={160}]
table {%
5 254
10 265
15 267
20 270
25 271
30 267
35 267
40 255
45 254
50 255
55 254
60 255
65 253
70 255
75 254
80 254
85 255
90 255
95 254
100 254
105 255
110 258
115 267
120 268
125 268
130 266
135 264
140 264
145 262
150 263
155 265
160 266
165 268
170 266
175 266
180 267
185 275
190 455
195 464
200 466
205 466
210 466
215 466
220 467
225 468
230 467
235 469
240 469
245 469
250 470
255 472
260 472
265 472
270 472
275 472
280 474
285 474
290 475
295 473
300 472
305 473
310 473
315 472
320 474
325 474
330 473
335 472
340 472
345 474
350 473
355 474
360 473
365 473
370 474
375 474
380 483
385 470
390 472
395 473
400 471
405 473
410 473
415 473
420 472
425 472
430 473
435 472
440 472
445 473
450 472
455 473
460 472
465 472
470 473
475 471
480 473
485 473
490 472
495 473
500 472
505 472
510 473
515 473
520 473
525 473
530 473
535 473
540 473
545 473
550 474
555 472
560 473
565 472
570 472
575 474
580 473
585 474
590 474
595 474
600 473
605 474
610 471
615 474
620 475
625 474
630 474
635 474
640 475
645 474
650 473
655 472
660 472
665 473
670 474
675 472
680 472
685 472
690 473
695 472
700 472
705 473
710 471
715 471
720 472
725 471
730 471
735 472
740 472
745 471
750 472
755 473
760 473
765 473
770 472
775 473
780 472
785 472
790 473
795 472
800 472
805 473
810 472
815 473
820 473
825 474
830 473
835 474
840 472
845 473
850 473
855 474
860 474
865 474
870 474
875 474
880 474
885 472
890 473
895 474
900 475
905 474
910 474
915 475
920 475
925 475
930 475
935 475
940 474
945 474
950 474
955 475
960 473
965 474
970 472
975 475
980 474
985 474
990 474
995 473
1000 474
1005 473
1010 475
1015 474
1020 475
1025 474
1030 475
1035 475
1040 474
1045 475
1050 474
1055 474
1060 474
1065 474
1070 476
1075 475
1080 475
1085 474
1090 476
1095 474
1100 475
1105 474
1110 476
1115 474
1120 476
1125 475
1130 474
1135 474
1140 475
1145 474
1150 475
1155 476
1160 474
1165 461
1170 393
1175 376
1180 337
1185 315
1190 312
1195 309
1200 286
1205 284
1210 272
1215 265
1220 271
1225 258
1230 257
1235 256
1240 255
1245 257
1250 257
1255 256
1260 257
1265 256
1270 256
1275 255
1280 257
1285 256
1290 256
1295 256
1300 257
1305 256
1310 256
1315 255
1320 256
1325 255
1330 256
1335 254
1340 256
1345 255
1350 256
1355 256
1360 255
1365 256
1370 256
1375 256
1380 255
1385 255
1390 254
1395 255
1400 255
1405 255
1410 255
1415 254
1420 254
1425 255
1430 254
1435 254
1440 255
1445 255
1450 256
1455 254
1460 255
1465 253
1470 255
1475 255
1480 255
1485 253
1490 255
1495 263
1500 252
1505 253
1510 253
1515 253
1520 255
1525 254
1530 254
1535 253
1540 255
1545 254
1550 255
1555 254
1560 254
1565 254
1570 255
1575 254
1580 254
1585 253
1590 253
1595 255
1600 254
1605 255
1610 254
1615 253
1620 254
1625 255
1630 253
1635 253
1640 254
1645 255
1650 254
1655 254
1660 254
1665 253
1670 254
1675 253
1680 253
1685 254
1690 254
1695 253
1700 253
1705 254
1710 255
1715 254
1720 254
1725 254
1730 254
1735 254
1740 254
1745 254
1750 254
1755 254
1760 255
1765 254
1770 253
1775 254
1780 254
1785 253
1790 254
1795 254
1800 254
1805 255
1810 254
1815 253
1820 254
1825 253
1830 254
1835 254
1840 253
1845 253
1850 253
1855 255
1860 253
1865 255
1870 254
1875 254
1880 254
1885 254
1890 254
1895 254
1900 254
1905 254
1910 254
1915 254
1920 254
1925 253
1930 253
1935 253
1940 254
1945 254
1950 254
1955 254
1960 254
1965 254
1970 254
1975 254
1980 254
1985 254
1990 254
1995 254
2000 254
2005 255
2010 254
2015 253
2020 254
2025 253
2030 254
2035 253
2040 253
2045 254
2050 253
2055 253
2060 254
2065 253
2070 254
2075 254
2080 254
2085 255
2090 254
2095 254
2100 254
2105 254
2110 253
2115 253
2120 254
2125 255
2130 253
2135 254
2140 255
2145 254
2150 254
2155 253
2160 254
2165 253
2170 253
2175 254
2180 255
2185 254
2190 254
2195 254
2200 254
2205 254
2210 253
2215 253
2220 254
2225 253
2230 254
2235 254
2240 253
2245 254
2250 254
2255 253
2260 254
2265 254
2270 253
2275 254
2280 254
2285 253
2290 254
2295 255
2300 253
2305 254
2310 254
2315 255
2320 255
2325 255
2330 254
2335 254
2340 251
2345 254
2350 253
2355 254
2360 254
2365 254
2370 254
2375 253
2380 254
2385 254
2390 255
2395 254
2400 254
2405 253
2410 254
2415 253
2420 254
2425 254
2430 254
2435 253
2440 255
2445 254
2450 254
2455 254
2460 254
2465 253
2470 254
2475 253
2480 253
2485 255
2490 254
2495 255
2500 254
2505 253
2510 254
2515 253
2520 254
2525 255
2530 254
2535 254
2540 254
2545 253
2550 255
2555 254
2560 253
2565 253
2570 253
2575 254
2580 253
2585 254
2590 253
2595 254
2600 254
2605 253
2610 254
2615 253
2620 254
2625 255
2630 254
2635 254
2640 255
2645 254
2650 253
2655 253
2660 254
2665 254
2670 251
2675 254
2680 253
2685 253
2690 254
2695 253
2700 253
2705 253
2710 253
2715 254
2720 253
2725 254
2730 254
2735 254
2740 254
2745 254
2750 253
2755 254
2760 254
2765 255
2770 253
2775 254
2780 253
2785 253
2790 253
2795 253
2800 253
2805 253
2810 254
2815 254
2820 254
2825 253
2830 258
2835 254
2840 254
2845 254
2850 254
2855 255
2860 254
2865 252
2870 253
2875 254
2880 253
2885 254
2890 253
2895 253
2900 254
2905 254
2910 253
2915 254
2920 254
2925 253
2930 254
2935 255
2940 253
2945 254
2950 254
2955 254
2960 254
2965 254
2970 253
2975 254
2980 254
2985 254
2990 254
2995 253
3000 254
3005 254
3010 254
3015 254
3020 253
3025 253
3030 253
3035 253
3040 254
3045 254
3050 254
3055 253
3060 254
3065 254
3070 253
3075 253
3080 254
3085 253
3090 255
3095 254
3100 254
3105 253
3110 253
3115 254
3120 254
3125 254
3130 258
3135 253
3140 253
3145 254
3150 254
3155 254
3160 254
3165 254
3170 253
3175 253
3180 254
3185 255
3190 253
3195 254
3200 253
3205 253
3210 252
3215 254
3220 254
3225 254
3230 254
3235 254
3240 263
3245 251
3250 254
3255 253
3260 253
3265 253
3270 253
3275 253
3280 254
3285 254
3290 254
3295 253
3300 255
3305 253
3310 254
3315 254
3320 253
3325 254
3330 254
3335 253
3340 251
3345 254
3350 253
3355 255
3360 254
3365 254
3370 254
3375 253
3380 253
3385 254
3390 254
3395 254
3400 254
3405 255
3410 253
3415 255
3420 254
3425 254
3430 254
3435 254
3440 254
3445 254
3450 254
3455 254
3460 254
3465 254
3470 253
3475 254
3480 254
3485 253
3490 254
3495 253
3500 253
3505 254
3510 255
3515 254
3520 255
3525 253
3530 254
3535 254
3540 254
3545 253
3550 254
3555 254
3560 254
3565 253
3570 254
3575 254
3580 264
3585 251
3590 254
3595 254
3600 254
};
\addlegendentry{16 cores (32 threads)}
\end{axis}

\end{tikzpicture}

%% file: results/power_amd.tex
\begin{tikzpicture}[font=\Large]

\definecolor{color0}{rgb}{0.12156862745098,0.466666666666667,0.705882352941177}
\definecolor{color1}{rgb}{1,0.498039215686275,0.0549019607843137}
\definecolor{color2}{rgb}{0.172549019607843,0.627450980392157,0.172549019607843}
\definecolor{color3}{rgb}{0.83921568627451,0.152941176470588,0.156862745098039}
\definecolor{color4}{rgb}{0.580392156862745,0.403921568627451,0.741176470588235}
\definecolor{color5}{rgb}{0,0,0}

\begin{axis}[
legend cell align={left},
legend columns=2,
legend style={fill opacity=0.8, draw opacity=1, text opacity=1, at={(1.02,1.18)}, anchor=east, draw=white!80.0!black},
tick align=outside,
tick pos=left,
x grid style={white!69.01960784313725!black},
xlabel={Time (SEC)},
xmin=0, xmax=3600,
xtick style={color=black},
xtick={0,900,1800,2700,3600},
y grid style={white!69.01960784313725!black},
ylabel={Power consumption (W)},
ymin=150, ymax=350,
ytick={150,200,250,300,350},
xmajorgrids,
ymajorgrids,
ytick style={color=black}
]
\addplot [thick, color5, mark=.]
table {%
0 205
5 205
10 205
15 205
20 205
25 205
30 205
35 205
40 206
45 206
50 205
55 205
60 205
65 206
70 205
75 205
80 205
85 205
90 205
95 206
100 205
105 205
110 205
115 205
120 205
125 205
130 205
135 205
140 206
145 205
150 205
155 205
160 205
165 205
170 205
175 206
180 205
185 206
190 205
195 206
200 205
205 205
210 206
215 205
220 206
225 206
230 205
235 205
240 205
245 205
250 205
255 206
260 206
265 205
270 205
275 205
280 205
285 206
290 205
295 204
300 206
305 205
310 205
315 205
320 206
325 205
330 205
335 206
340 205
345 205
350 206
355 206
360 206
365 205
370 205
375 205
380 206
385 205
390 205
395 205
400 205
405 205
410 205
415 205
420 205
425 206
430 205
435 205
440 205
445 205
450 205
455 205
460 205
465 205
470 205
475 205
480 205
485 206
490 205
495 205
500 205
505 205
510 205
515 205
520 206
525 206
530 205
535 205
540 205
545 205
550 205
555 205
560 206
565 206
570 206
575 205
580 206
585 206
590 205
595 205
600 205
605 206
610 205
615 206
620 206
625 205
630 205
635 206
640 206
645 206
650 205
655 206
660 205
665 205
670 205
675 206
680 205
685 205
690 206
695 205
700 206
705 205
710 205
715 205
720 206
725 205
730 205
735 204
740 206
745 206
750 205
755 205
760 206
765 206
770 206
775 205
780 205
785 206
790 205
795 205
800 205
805 206
810 205
815 206
820 205
825 205
830 205
835 205
840 206
845 205
850 206
855 206
860 206
865 206
870 206
875 205
880 206
885 205
890 205
895 206
900 205
905 206
910 206
915 204
920 205
925 205
930 205
935 206
940 205
945 205
950 205
955 206
960 206
965 205
970 206
975 206
980 205
985 205
990 205
995 205
1000 205
1005 205
1010 205
1015 206
1020 206
1025 206
1030 206
1035 205
1040 205
1045 205
1050 205
1055 205
1060 205
1065 206
1070 205
1075 205
1080 205
1085 206
1090 206
1095 206
1100 205
1105 205
1110 206
1115 205
1120 205
1125 205
1130 205
1135 205
1140 205
1145 205
1150 205
1155 205
1160 205
1165 206
1170 205
1175 206
1180 206
1185 205
1190 205
1195 205
1200 206
1205 205
1210 206
1215 205
1220 205
1225 206
1230 205
1235 206
1240 206
1245 205
1250 206
1255 206
1260 206
1265 205
1270 206
1275 206
1280 205
1285 204
1290 206
1295 204
1300 206
1305 206
1310 206
1315 206
1320 205
1325 205
1330 205
1335 205
1340 205
1345 205
1350 206
1355 205
1360 205
1365 205
1370 205
1375 206
1380 206
1385 205
1390 206
1395 206
1400 205
1405 206
1410 205
1415 205
1420 206
1425 205
1430 205
1435 205
1440 205
1445 206
1450 206
1455 206
1460 206
1465 205
1470 206
1475 206
1480 205
1485 205
1490 205
1495 205
1500 205
1505 205
1510 205
1515 205
1520 205
1525 206
1530 205
1535 205
1540 205
1545 205
1550 205
1555 204
1560 204
1565 204
1570 205
1575 205
1580 205
1585 205
1590 205
1595 205
1600 205
1605 205
1610 205
1615 205
1620 205
1625 205
1630 205
1635 205
1640 205
1645 205
1650 206
1655 205
1660 205
1665 205
1670 205
1675 205
1680 204
1685 205
1690 206
1695 205
1700 206
1705 205
1710 205
1715 206
1720 206
1725 206
1730 205
1735 206
1740 205
1745 206
1750 205
1755 205
1760 205
1765 205
1770 205
1775 205
1780 206
1785 205
1790 205
1795 205
1800 205
1805 205
1810 206
1815 205
1820 205
1825 206
1830 205
1835 205
1840 205
1845 205
1850 205
1855 206
1860 206
1865 206
1870 205
1875 205
1880 205
1885 205
1890 206
1895 205
1900 206
1905 206
1910 206
1915 205
1920 206
1925 206
1930 205
1935 205
1940 205
1945 206
1950 206
1955 205
1960 205
1965 205
1970 205
1975 205
1980 205
1985 205
1990 205
1995 205
2000 205
2005 205
2010 205
2015 205
2020 205
2025 205
2030 205
2035 206
2040 206
2045 205
2050 205
2055 205
2060 206
2065 205
2070 205
2075 205
2080 205
2085 205
2090 206
2095 205
2100 205
2105 205
2110 205
2115 205
2120 205
2125 205
2130 205
2135 205
2140 205
2145 206
2150 205
2155 205
2160 206
2165 206
2170 205
2175 206
2180 205
2185 206
2190 206
2195 206
2200 206
2205 205
2210 206
2215 205
2220 206
2225 206
2230 206
2235 205
2240 206
2245 206
2250 205
2255 205
2260 205
2265 205
2270 205
2275 205
2280 206
2285 206
2290 206
2295 205
2300 206
2305 205
2310 205
2315 205
2320 206
2325 206
2330 206
2335 205
2340 205
2345 205
2350 206
2355 205
2360 205
2365 205
2370 206
2375 206
2380 206
2385 205
2390 205
2395 206
2400 206
2405 205
2410 205
2415 205
2420 205
2425 205
2430 206
2435 205
2440 206
2445 206
2450 206
2455 206
2460 206
2465 205
2470 205
2475 205
2480 206
2485 206
2490 205
2495 206
2500 205
2505 206
2510 205
2515 206
2520 206
2525 206
2530 206
2535 206
2540 205
2545 206
2550 206
2555 206
2560 205
2565 206
2570 205
2575 206
2580 205
2585 205
2590 206
2595 205
2600 205
2605 206
2610 205
2615 205
2620 206
2625 205
2630 205
2635 204
2640 205
2645 205
2650 206
2655 205
2660 205
2665 206
2670 205
2675 205
2680 205
2685 205
2690 206
2695 205
2700 206
2705 205
2710 205
2715 206
2720 205
2725 206
2730 206
2735 206
2740 205
2745 207
2750 204
2755 204
2760 204
2765 205
2770 204
2775 204
2780 204
2785 203
2790 205
2795 204
2800 204
2805 204
2810 204
2815 204
2820 204
2825 204
2830 204
2835 204
2840 204
2845 204
2850 204
2855 204
2860 204
2865 204
2870 204
2875 204
2880 204
2885 204
2890 204
2895 204
2900 204
2905 204
2910 204
2915 204
2920 204
2925 204
2930 204
2935 205
2940 204
2945 203
2950 204
2955 205
2960 203
2965 204
2970 204
2975 205
2980 204
2985 205
2990 205
2995 205
3000 205
3005 205
3010 204
3015 204
3020 205
3025 205
3030 204
3035 204
3040 204
3045 204
3050 204
3055 204
3060 204
3065 204
3070 204
3075 204
3080 204
3085 205
3090 204
3095 204
3100 205
3105 205
3110 204
3115 204
3120 204
3125 204
3130 205
3135 205
3140 204
3145 205
3150 205
3155 204
3160 204
3165 204
3170 204
3175 204
3180 205
3185 205
3190 205
3195 204
3200 204
3205 204
3210 204
3215 204
3220 204
3225 204
3230 204
3235 204
3240 204
3245 204
3250 204
3255 204
3260 204
3265 205
3270 204
3275 204
3280 204
3285 204
3290 205
3295 204
3300 204
3305 205
3310 204
3315 204
3320 205
3325 204
3330 204
3335 204
3340 204
3345 204
3350 204
3355 204
3360 204
3365 204
3370 204
3375 204
3380 204
3385 204
3390 204
3395 204
3400 205
3405 203
3410 204
3415 204
3420 203
3425 205
3430 204
3435 204
3440 204
3445 204
3450 204
3455 204
3460 204
3465 204
3470 204
3475 204
3480 204
3485 205
3490 204
3495 204
3500 204
3505 204
3510 205
3515 205
3520 205
3525 204
3530 204
3535 204
3540 205
3545 205
3550 204
3555 205
3560 205
3565 205
3570 204
3575 204
3580 204
3585 204
3590 204
3595 204
3600 204
};
\addlegendentry{Baseline}
\addplot [thick, color0, mark=star, mark size=3, mark options={solid,fill=white,draw=red},  mark repeat={180}]
table {%
0 211
5 212
10 212
15 212
20 208
25 211
30 205
35 206
40 206
45 208
50 205
55 205
60 206
65 205
70 206
75 206
80 207
85 206
90 206
95 206
100 206
105 206
110 206
115 207
120 206
125 205
130 206
135 207
140 206
145 206
150 208
155 206
160 206
165 207
170 212
175 211
180 211
185 213
190 212
195 210
200 211
205 210
210 210
215 211
220 210
225 211
230 211
235 212
240 206
245 211
250 211
255 212
260 212
265 211
270 212
275 212
280 211
285 212
290 211
295 212
300 212
305 213
310 212
315 213
320 211
325 211
330 212
335 211
340 212
345 212
350 213
355 213
360 212
365 213
370 212
375 212
380 212
385 212
390 212
395 212
400 212
405 211
410 212
415 212
420 212
425 212
430 212
435 211
440 211
445 212
450 211
455 211
460 211
465 212
470 212
475 212
480 212
485 212
490 212
495 211
500 211
505 211
510 211
515 212
520 212
525 211
530 211
535 211
540 212
545 211
550 211
555 212
560 211
565 211
570 212
575 211
580 211
585 212
590 212
595 211
600 211
605 211
610 212
615 211
620 211
625 212
630 212
635 212
640 211
645 211
650 211
655 211
660 211
665 211
670 212
675 212
680 212
685 212
690 212
695 212
700 212
705 212
710 211
715 212
720 212
725 212
730 211
735 211
740 212
745 212
750 211
755 212
760 212
765 211
770 211
775 211
780 212
785 211
790 210
795 212
800 211
805 211
810 211
815 212
820 213
825 212
830 213
835 212
840 212
845 212
850 212
855 213
860 211
865 212
870 211
875 212
880 211
885 212
890 212
895 211
900 212
905 212
910 212
915 211
920 211
925 212
930 213
935 212
940 212
945 211
950 212
955 212
960 213
965 212
970 211
975 213
980 211
985 212
990 212
995 212
1000 212
1005 212
1010 212
1015 212
1020 213
1025 212
1030 212
1035 212
1040 211
1045 212
1050 212
1055 213
1060 212
1065 210
1070 211
1075 211
1080 211
1085 212
1090 211
1095 211
1100 211
1105 211
1110 211
1115 212
1120 211
1125 212
1130 212
1135 212
1140 211
1145 211
1150 212
1155 211
1160 211
1165 211
1170 212
1175 212
1180 211
1185 212
1190 211
1195 212
1200 212
1205 212
1210 211
1215 211
1220 212
1225 212
1230 212
1235 213
1240 212
1245 212
1250 213
1255 212
1260 211
1265 212
1270 211
1275 211
1280 211
1285 211
1290 212
1295 211
1300 211
1305 212
1310 211
1315 211
1320 211
1325 211
1330 212
1335 211
1340 212
1345 211
1350 212
1355 213
1360 213
1365 212
1370 211
1375 212
1380 211
1385 212
1390 211
1395 212
1400 212
1405 212
1410 211
1415 212
1420 211
1425 211
1430 212
1435 212
1440 212
1445 212
1450 212
1455 212
1460 211
1465 212
1470 211
1475 211
1480 213
1485 212
1490 212
1495 212
1500 212
1505 212
1510 212
1515 212
1520 212
1525 212
1530 212
1535 211
1540 211
1545 212
1550 212
1555 212
1560 212
1565 212
1570 212
1575 212
1580 212
1585 212
1590 211
1595 212
1600 212
1605 212
1610 212
1615 211
1620 211
1625 212
1630 212
1635 212
1640 212
1645 212
1650 212
1655 211
1660 212
1665 211
1670 211
1675 212
1680 212
1685 212
1690 212
1695 212
1700 211
1705 212
1710 212
1715 211
1720 212
1725 212
1730 212
1735 212
1740 213
1745 212
1750 212
1755 212
1760 212
1765 211
1770 212
1775 211
1780 212
1785 212
1790 211
1795 211
1800 212
1805 211
1810 211
1815 211
1820 211
1825 211
1830 212
1835 211
1840 212
1845 212
1850 212
1855 212
1860 212
1865 211
1870 212
1875 213
1880 212
1885 213
1890 213
1895 212
1900 212
1905 212
1910 212
1915 212
1920 211
1925 212
1930 212
1935 212
1940 212
1945 212
1950 212
1955 212
1960 212
1965 212
1970 211
1975 212
1980 212
1985 211
1990 211
1995 211
2000 211
2005 211
2010 211
2015 212
2020 212
2025 212
2030 211
2035 211
2040 211
2045 212
2050 212
2055 212
2060 213
2065 212
2070 212
2075 212
2080 212
2085 213
2090 213
2095 212
2100 212
2105 212
2110 212
2115 213
2120 213
2125 212
2130 211
2135 212
2140 211
2145 211
2150 211
2155 212
2160 212
2165 212
2170 212
2175 213
2180 212
2185 212
2190 211
2195 210
2200 212
2205 211
2210 212
2215 212
2220 211
2225 211
2230 211
2235 211
2240 212
2245 211
2250 211
2255 212
2260 212
2265 212
2270 212
2275 212
2280 211
2285 212
2290 212
2295 212
2300 211
2305 212
2310 212
2315 211
2320 211
2325 212
2330 212
2335 213
2340 212
2345 213
2350 212
2355 212
2360 211
2365 212
2370 211
2375 212
2380 213
2385 211
2390 211
2395 211
2400 212
2405 211
2410 212
2415 212
2420 212
2425 212
2430 212
2435 211
2440 212
2445 212
2450 212
2455 211
2460 212
2465 212
2470 212
2475 212
2480 212
2485 212
2490 212
2495 212
2500 212
2505 212
2510 212
2515 213
2520 213
2525 211
2530 212
2535 211
2540 211
2545 211
2550 211
2555 211
2560 211
2565 211
2570 211
2575 212
2580 212
2585 211
2590 212
2595 211
2600 211
2605 211
2610 212
2615 211
2620 211
2625 211
2630 212
2635 212
2640 212
2645 211
2650 212
2655 211
2660 212
2665 212
2670 212
2675 212
2680 211
2685 212
2690 212
2695 211
2700 212
2705 211
2710 211
2715 212
2720 212
2725 212
2730 211
2735 212
2740 212
2745 211
2750 211
2755 211
2760 212
2765 211
2770 211
2775 212
2780 212
2785 211
2790 212
2795 212
2800 212
2805 212
2810 211
2815 211
2820 212
2825 212
2830 212
2835 212
2840 211
2845 211
2850 212
2855 211
2860 212
2865 212
2870 211
2875 212
2880 213
2885 212
2890 212
2895 212
2900 212
2905 211
2910 213
2915 212
2920 212
2925 211
2930 211
2935 212
2940 212
2945 210
2950 212
2955 212
2960 212
2965 211
2970 212
2975 212
2980 211
2985 212
2990 211
2995 212
3000 211
3005 211
3010 211
3015 212
3020 212
3025 212
3030 213
3035 212
3040 212
3045 211
3050 211
3055 212
3060 212
3065 212
3070 211
3075 212
3080 213
3085 213
3090 212
3095 211
3100 212
3105 212
3110 212
3115 212
3120 213
3125 213
3130 212
3135 211
3140 212
3145 212
3150 212
3155 212
3160 212
3165 212
3170 211
3175 212
3180 212
3185 212
3190 211
3195 211
3200 212
3205 212
3210 213
3215 212
3220 211
3225 211
3230 212
3235 212
3240 211
3245 212
3250 211
3255 212
3260 211
3265 212
3270 211
3275 212
3280 211
3285 211
3290 213
3295 213
3300 212
3305 212
3310 212
3315 211
3320 212
3325 210
3330 211
3335 212
3340 212
3345 211
3350 212
3355 211
3360 211
3365 212
3370 210
3375 212
3380 211
3385 212
3390 212
3395 213
3400 211
3405 211
3410 211
3415 211
3420 211
3425 212
3430 212
3435 211
3440 211
3445 213
3450 211
3455 212
3460 212
3465 212
3470 211
3475 212
3480 212
3485 211
3490 212
3495 212
3500 211
3505 212
3510 212
3515 212
3520 212
3525 212
3530 212
3535 211
3540 212
3545 211
3550 212
3555 212
3560 211
3565 212
3570 212
3575 212
3580 212
3585 211
3590 211
3595 211
3600 212
};
\addlegendentry{1 core}
\addplot [thick, color1, mark=triangle*, mark size=3, mark options={solid,fill=white,draw=red},  mark repeat={180}]
table {%
0 210
5 210
10 211
15 211
20 211
25 205
30 209
35 210
40 205
45 206
50 205
55 205
60 205
65 205
70 205
75 205
80 205
85 206
90 205
95 205
100 205
105 205
110 205
115 206
120 205
125 205
130 205
135 206
140 205
145 206
150 205
155 205
160 205
165 205
170 205
175 204
180 205
185 205
190 205
195 205
200 206
205 205
210 206
215 205
220 206
225 205
230 204
235 204
240 205
245 205
250 205
255 206
260 205
265 205
270 205
275 205
280 205
285 205
290 205
295 206
300 207
305 207
310 206
315 205
320 205
325 206
330 205
335 204
340 206
345 206
350 205
355 205
360 205
365 206
370 205
375 205
380 206
385 204
390 206
395 205
400 206
405 206
410 205
415 206
420 205
425 204
430 208
435 211
440 211
445 212
450 211
455 210
460 209
465 209
470 210
475 209
480 209
485 209
490 210
495 211
500 210
505 211
510 211
515 224
520 223
525 225
530 225
535 224
540 224
545 224
550 225
555 224
560 225
565 225
570 224
575 224
580 224
585 224
590 225
595 225
600 225
605 226
610 224
615 224
620 226
625 225
630 225
635 226
640 225
645 225
650 225
655 225
660 225
665 224
670 224
675 224
680 224
685 224
690 224
695 224
700 224
705 225
710 224
715 225
720 225
725 225
730 224
735 226
740 224
745 225
750 226
755 225
760 225
765 225
770 225
775 226
780 226
785 226
790 225
795 226
800 226
805 226
810 227
815 226
820 226
825 225
830 227
835 227
840 226
845 226
850 227
855 226
860 226
865 227
870 226
875 227
880 226
885 226
890 226
895 226
900 224
905 226
910 224
915 223
920 224
925 224
930 224
935 224
940 224
945 224
950 224
955 225
960 225
965 224
970 224
975 225
980 224
985 224
990 224
995 224
1000 224
1005 224
1010 225
1015 224
1020 226
1025 225
1030 225
1035 224
1040 226
1045 227
1050 226
1055 226
1060 226
1065 226
1070 226
1075 227
1080 226
1085 226
1090 225
1095 226
1100 227
1105 226
1110 226
1115 226
1120 226
1125 226
1130 226
1135 226
1140 226
1145 227
1150 227
1155 227
1160 224
1165 224
1170 225
1175 225
1180 224
1185 224
1190 224
1195 224
1200 225
1205 226
1210 224
1215 224
1220 224
1225 224
1230 223
1235 224
1240 224
1245 224
1250 224
1255 224
1260 224
1265 224
1270 224
1275 224
1280 224
1285 225
1290 223
1295 224
1300 224
1305 224
1310 224
1315 224
1320 224
1325 224
1330 224
1335 223
1340 223
1345 224
1350 224
1355 224
1360 224
1365 223
1370 223
1375 223
1380 224
1385 224
1390 224
1395 224
1400 224
1405 224
1410 223
1415 224
1420 224
1425 224
1430 224
1435 224
1440 225
1445 224
1450 224
1455 225
1460 223
1465 224
1470 224
1475 223
1480 223
1485 224
1490 224
1495 224
1500 223
1505 225
1510 223
1515 224
1520 223
1525 224
1530 224
1535 224
1540 224
1545 223
1550 224
1555 224
1560 224
1565 225
1570 224
1575 224
1580 224
1585 224
1590 223
1595 225
1600 225
1605 224
1610 224
1615 224
1620 224
1625 225
1630 224
1635 225
1640 224
1645 224
1650 224
1655 224
1660 224
1665 225
1670 225
1675 225
1680 224
1685 225
1690 224
1695 224
1700 224
1705 224
1710 224
1715 224
1720 224
1725 224
1730 224
1735 223
1740 224
1745 224
1750 224
1755 224
1760 224
1765 224
1770 224
1775 224
1780 224
1785 224
1790 224
1795 224
1800 224
1805 226
1810 225
1815 225
1820 225
1825 226
1830 225
1835 225
1840 226
1845 225
1850 225
1855 226
1860 225
1865 225
1870 224
1875 226
1880 226
1885 226
1890 225
1895 225
1900 226
1905 226
1910 226
1915 226
1920 226
1925 226
1930 225
1935 226
1940 226
1945 226
1950 226
1955 226
1960 226
1965 225
1970 226
1975 225
1980 226
1985 226
1990 225
1995 225
2000 226
2005 226
2010 226
2015 226
2020 226
2025 226
2030 226
2035 226
2040 224
2045 225
2050 224
2055 225
2060 225
2065 224
2070 224
2075 224
2080 226
2085 226
2090 225
2095 225
2100 225
2105 226
2110 224
2115 225
2120 224
2125 224
2130 225
2135 224
2140 225
2145 225
2150 226
2155 224
2160 225
2165 225
2170 225
2175 225
2180 225
2185 225
2190 225
2195 224
2200 226
2205 224
2210 226
2215 225
2220 225
2225 224
2230 225
2235 224
2240 225
2245 226
2250 225
2255 225
2260 225
2265 225
2270 225
2275 225
2280 225
2285 224
2290 219
2295 225
2300 224
2305 225
2310 224
2315 224
2320 225
2325 224
2330 223
2335 224
2340 224
2345 224
2350 224
2355 224
2360 224
2365 224
2370 224
2375 224
2380 224
2385 224
2390 225
2395 225
2400 224
2405 223
2410 224
2415 224
2420 225
2425 224
2430 224
2435 224
2440 224
2445 224
2450 224
2455 225
2460 224
2465 224
2470 225
2475 225
2480 224
2485 225
2490 224
2495 224
2500 224
2505 224
2510 224
2515 224
2520 223
2525 224
2530 224
2535 225
2540 225
2545 225
2550 225
2555 224
2560 224
2565 224
2570 224
2575 224
2580 224
2585 224
2590 224
2595 224
2600 224
2605 224
2610 225
2615 224
2620 224
2625 224
2630 224
2635 224
2640 224
2645 224
2650 225
2655 225
2660 224
2665 225
2670 224
2675 225
2680 224
2685 224
2690 225
2695 224
2700 223
2705 224
2710 224
2715 224
2720 224
2725 224
2730 224
2735 224
2740 223
2745 224
2750 224
2755 224
2760 224
2765 224
2770 224
2775 223
2780 224
2785 223
2790 223
2795 224
2800 225
2805 224
2810 224
2815 224
2820 224
2825 224
2830 224
2835 224
2840 224
2845 225
2850 224
2855 224
2860 224
2865 224
2870 224
2875 224
2880 224
2885 224
2890 223
2895 225
2900 223
2905 224
2910 224
2915 224
2920 225
2925 224
2930 225
2935 225
2940 225
2945 226
2950 226
2955 225
2960 226
2965 225
2970 225
2975 226
2980 225
2985 225
2990 225
2995 225
3000 225
3005 225
3010 225
3015 225
3020 225
3025 224
3030 224
3035 224
3040 224
3045 224
3050 224
3055 224
3060 224
3065 224
3070 224
3075 223
3080 224
3085 226
3090 225
3095 224
3100 223
3105 224
3110 222
3115 223
3120 223
3125 225
3130 225
3135 226
3140 225
3145 225
3150 224
3155 225
3160 225
3165 226
3170 225
3175 224
3180 224
3185 224
3190 224
3195 224
3200 224
3205 224
3210 224
3215 225
3220 224
3225 224
3230 224
3235 224
3240 224
3245 225
3250 224
3255 224
3260 224
3265 225
3270 224
3275 224
3280 224
3285 224
3290 225
3295 224
3300 224
3305 224
3310 224
3315 224
3320 224
3325 224
3330 224
3335 224
3340 224
3345 224
3350 224
3355 223
3360 223
3365 224
3370 224
3375 224
3380 224
3385 223
3390 224
3395 224
3400 223
3405 224
3410 224
3415 224
3420 224
3425 224
3430 224
3435 223
3440 223
3445 224
3450 224
3455 223
3460 224
3465 224
3470 223
3475 223
3480 224
3485 224
3490 224
3495 225
3500 224
3505 224
3510 224
3515 224
3520 224
3525 224
3530 224
3535 224
3540 223
3545 224
3550 225
3555 224
3560 224
3565 224
3570 224
3575 224
3580 224
3585 225
3590 224
3595 224
3600 223
};
\addlegendentry{4 cores}
\addplot [thick, color2, mark=square*, mark size=3, mark options={solid,fill=white,draw=red},  mark repeat={180}]
table {%
5 210
10 211
15 212
20 212
25 211
30 212
35 207
40 206
45 206
50 206
55 205
60 205
65 206
70 206
75 206
80 206
85 209
90 206
95 207
100 205
105 207
110 212
115 213
120 213
125 213
130 213
135 210
140 211
145 211
150 211
155 211
160 211
165 211
170 211
175 211
180 214
185 211
190 245
195 242
200 242
205 242
210 242
215 243
220 242
225 243
230 243
235 242
240 244
245 244
250 245
255 245
260 243
265 242
270 243
275 243
280 243
285 243
290 243
295 243
300 243
305 243
310 242
315 244
320 244
325 244
330 244
335 244
340 244
345 245
350 244
355 244
360 245
365 243
370 244
375 244
380 244
385 244
390 243
395 244
400 244
405 244
410 244
415 245
420 246
425 246
430 245
435 246
440 246
445 246
450 245
455 244
460 244
465 245
470 244
475 244
480 243
485 244
490 244
495 244
500 244
505 244
510 244
515 244
520 245
525 246
530 246
535 246
540 246
545 245
550 245
555 246
560 247
565 246
570 245
575 245
580 245
585 246
590 246
595 246
600 246
605 246
610 247
615 246
620 246
625 246
630 246
635 246
640 246
645 245
650 245
655 244
660 244
665 245
670 244
675 244
680 245
685 245
690 245
695 245
700 244
705 245
710 245
715 245
720 245
725 245
730 245
735 245
740 245
745 244
750 245
755 244
760 245
765 244
770 245
775 245
780 245
785 245
790 245
795 245
800 245
805 245
810 245
815 245
820 245
825 244
830 244
835 245
840 245
845 245
850 245
855 245
860 245
865 244
870 245
875 245
880 244
885 244
890 245
895 245
900 244
905 245
910 244
915 244
920 245
925 245
930 244
935 245
940 246
945 246
950 246
955 246
960 246
965 246
970 247
975 246
980 246
985 246
990 247
995 246
1000 246
1005 247
1010 246
1015 247
1020 246
1025 246
1030 246
1035 246
1040 246
1045 246
1050 247
1055 246
1060 246
1065 246
1070 246
1075 247
1080 247
1085 247
1090 247
1095 246
1100 246
1105 246
1110 246
1115 246
1120 246
1125 246
1130 246
1135 245
1140 246
1145 245
1150 247
1155 247
1160 246
1165 247
1170 247
1175 246
1180 247
1185 246
1190 246
1195 246
1200 245
1205 244
1210 244
1215 247
1220 247
1225 246
1230 247
1235 246
1240 247
1245 247
1250 247
1255 248
1260 247
1265 247
1270 247
1275 248
1280 247
1285 246
1290 247
1295 247
1300 247
1305 247
1310 247
1315 247
1320 247
1325 247
1330 247
1335 245
1340 246
1345 247
1350 247
1355 246
1360 247
1365 246
1370 247
1375 246
1380 247
1385 248
1390 246
1395 246
1400 247
1405 247
1410 247
1415 248
1420 247
1425 247
1430 247
1435 246
1440 245
1445 245
1450 245
1455 245
1460 246
1465 247
1470 248
1475 248
1480 247
1485 247
1490 245
1495 245
1500 245
1505 248
1510 247
1515 247
1520 247
1525 247
1530 248
1535 247
1540 247
1545 246
1550 247
1555 247
1560 247
1565 248
1570 248
1575 246
1580 248
1585 247
1590 247
1595 248
1600 247
1605 247
1610 248
1615 248
1620 247
1625 247
1630 247
1635 245
1640 246
1645 245
1650 245
1655 246
1660 245
1665 245
1670 245
1675 245
1680 244
1685 246
1690 245
1695 245
1700 245
1705 246
1710 245
1715 245
1720 245
1725 245
1730 244
1735 245
1740 245
1745 245
1750 245
1755 245
1760 246
1765 244
1770 245
1775 245
1780 245
1785 245
1790 244
1795 245
1800 245
1805 245
1810 246
1815 245
1820 244
1825 245
1830 245
1835 245
1840 246
1845 246
1850 244
1855 245
1860 246
1865 245
1870 245
1875 246
1880 246
1885 246
1890 245
1895 245
1900 248
1905 247
1910 247
1915 247
1920 247
1925 247
1930 247
1935 247
1940 247
1945 247
1950 247
1955 247
1960 248
1965 248
1970 248
1975 246
1980 247
1985 247
1990 247
1995 247
2000 247
2005 247
2010 247
2015 248
2020 248
2025 247
2030 247
2035 248
2040 247
2045 247
2050 247
2055 248
2060 246
2065 247
2070 246
2075 247
2080 246
2085 248
2090 247
2095 247
2100 247
2105 247
2110 248
2115 247
2120 248
2125 247
2130 248
2135 247
2140 246
2145 249
2150 248
2155 247
2160 247
2165 247
2170 248
2175 247
2180 248
2185 248
2190 247
2195 246
2200 248
2205 248
2210 248
2215 247
2220 247
2225 249
2230 248
2235 248
2240 247
2245 247
2250 246
2255 247
2260 247
2265 248
2270 247
2275 247
2280 248
2285 248
2290 248
2295 247
2300 248
2305 247
2310 247
2315 247
2320 247
2325 248
2330 247
2335 248
2340 248
2345 247
2350 247
2355 247
2360 246
2365 247
2370 247
2375 247
2380 247
2385 248
2390 247
2395 246
2400 246
2405 247
2410 247
2415 247
2420 247
2425 247
2430 246
2435 247
2440 248
2445 248
2450 247
2455 247
2460 247
2465 247
2470 247
2475 247
2480 248
2485 247
2490 247
2495 247
2500 248
2505 245
2510 245
2515 245
2520 246
2525 245
2530 245
2535 246
2540 245
2545 246
2550 245
2555 245
2560 245
2565 245
2570 246
2575 246
2580 245
2585 244
2590 246
2595 245
2600 245
2605 245
2610 245
2615 245
2620 247
2625 247
2630 248
2635 247
2640 247
2645 247
2650 247
2655 247
2660 247
2665 248
2670 247
2675 247
2680 247
2685 246
2690 248
2695 247
2700 248
2705 246
2710 247
2715 247
2720 247
2725 247
2730 247
2735 246
2740 237
2745 226
2750 227
2755 218
2760 213
2765 214
2770 214
2775 214
2780 214
2785 210
2790 213
2795 207
2800 207
2805 208
2810 206
2815 207
2820 207
2825 206
2830 206
2835 206
2840 206
2845 206
2850 206
2855 207
2860 206
2865 206
2870 206
2875 207
2880 206
2885 206
2890 206
2895 207
2900 206
2905 206
2910 206
2915 207
2920 206
2925 206
2930 207
2935 206
2940 208
2945 206
2950 207
2955 206
2960 207
2965 206
2970 206
2975 206
2980 206
2985 206
2990 207
2995 206
3000 206
3005 206
3010 206
3015 206
3020 205
3025 206
3030 206
3035 206
3040 206
3045 206
3050 206
3055 206
3060 206
3065 207
3070 206
3075 206
3080 209
3085 205
3090 205
3095 206
3100 206
3105 206
3110 207
3115 206
3120 207
3125 207
3130 206
3135 206
3140 205
3145 206
3150 206
3155 205
3160 207
3165 206
3170 206
3175 206
3180 206
3185 206
3190 206
3195 206
3200 205
3205 205
3210 205
3215 206
3220 206
3225 206
3230 206
3235 205
3240 206
3245 205
3250 205
3255 205
3260 206
3265 205
3270 205
3275 205
3280 206
3285 206
3290 205
3295 206
3300 206
3305 205
3310 206
3315 206
3320 205
3325 206
3330 206
3335 205
3340 205
3345 206
3350 206
3355 206
3360 206
3365 205
3370 206
3375 206
3380 208
3385 206
3390 206
3395 206
3400 206
3405 206
3410 206
3415 206
3420 205
3425 206
3430 206
3435 206
3440 206
3445 206
3450 206
3455 206
3460 205
3465 206
3470 206
3475 206
3480 205
3485 206
3490 206
3495 205
3500 205
3505 205
3510 205
3515 206
3520 206
3525 206
3530 206
3535 206
3540 206
3545 206
3550 206
3555 205
3560 207
3565 205
3570 206
3575 205
3580 205
3585 205
3590 204
3595 206
3600 206
};
\addlegendentry{8 cores}
\addplot [thick, color3, mark=o, mark size=3, mark options={solid,fill=white,draw=red},  mark repeat={180}]
table {%
0 207
5 206
10 206
15 212
20 211
25 213
30 213
35 214
40 210
45 213
50 208
55 207
60 206
65 208
70 207
75 207
80 207
85 207
90 206
95 207
100 207
105 207
110 207
115 206
120 208
125 212
130 212
135 212
140 213
145 213
150 211
155 210
160 210
165 210
170 210
175 210
180 212
185 211
190 211
195 216
200 213
205 243
210 277
215 278
220 279
225 279
230 276
235 281
240 280
245 280
250 280
255 281
260 281
265 281
270 282
275 282
280 282
285 282
290 283
295 283
300 283
305 283
310 284
315 285
320 286
325 286
330 286
335 286
340 286
345 287
350 286
355 287
360 287
365 286
370 287
375 287
380 286
385 287
390 287
395 287
400 287
405 287
410 287
415 286
420 287
425 287
430 288
435 287
440 287
445 287
450 287
455 287
460 287
465 287
470 287
475 288
480 288
485 286
490 287
495 288
500 287
505 287
510 288
515 288
520 286
525 287
530 287
535 288
540 287
545 287
550 287
555 288
560 290
565 286
570 287
575 287
580 287
585 287
590 287
595 287
600 287
605 287
610 287
615 287
620 289
625 288
630 287
635 287
640 287
645 288
650 287
655 287
660 288
665 288
670 287
675 289
680 287
685 287
690 287
695 288
700 288
705 287
710 288
715 287
720 289
725 288
730 288
735 288
740 288
745 287
750 287
755 287
760 287
765 287
770 289
775 287
780 287
785 287
790 288
795 287
800 287
805 287
810 287
815 287
820 287
825 287
830 287
835 288
840 287
845 287
850 287
855 287
860 288
865 288
870 287
875 287
880 287
885 287
890 287
895 287
900 287
905 287
910 287
915 288
920 287
925 287
930 287
935 287
940 287
945 288
950 286
955 287
960 287
965 287
970 287
975 287
980 287
985 287
990 287
995 287
1000 287
1005 287
1010 287
1015 287
1020 284
1025 288
1030 287
1035 287
1040 287
1045 287
1050 286
1055 287
1060 287
1065 287
1070 287
1075 287
1080 287
1085 287
1090 287
1095 287
1100 287
1105 287
1110 287
1115 287
1120 287
1125 287
1130 287
1135 287
1140 287
1145 287
1150 286
1155 286
1160 286
1165 286
1170 287
1175 286
1180 286
1185 286
1190 286
1195 286
1200 287
1205 287
1210 286
1215 286
1220 286
1225 286
1230 286
1235 286
1240 287
1245 287
1250 287
1255 287
1260 286
1265 287
1270 286
1275 287
1280 286
1285 287
1290 287
1295 287
1300 287
1305 287
1310 286
1315 286
1320 287
1325 288
1330 287
1335 287
1340 286
1345 287
1350 287
1355 288
1360 287
1365 287
1370 287
1375 287
1380 286
1385 287
1390 286
1395 287
1400 288
1405 287
1410 287
1415 287
1420 286
1425 287
1430 286
1435 286
1440 287
1445 285
1450 287
1455 286
1460 286
1465 287
1470 287
1475 288
1480 287
1485 287
1490 287
1495 286
1500 273
1505 265
1510 236
1515 235
1520 231
1525 227
1530 226
1535 221
1540 219
1545 218
1550 216
1555 214
1560 212
1565 207
1570 208
1575 206
1580 207
1585 206
1590 208
1595 207
1600 208
1605 207
1610 207
1615 207
1620 207
1625 207
1630 206
1635 207
1640 206
1645 206
1650 206
1655 206
1660 206
1665 206
1670 207
1675 206
1680 206
1685 206
1690 207
1695 206
1700 207
1705 206
1710 206
1715 206
1720 207
1725 205
1730 207
1735 206
1740 207
1745 205
1750 206
1755 206
1760 206
1765 206
1770 206
1775 206
1780 206
1785 206
1790 205
1795 206
1800 205
1805 206
1810 207
1815 207
1820 206
1825 207
1830 206
1835 206
1840 206
1845 206
1850 205
1855 206
1860 205
1865 206
1870 206
1875 206
1880 205
1885 206
1890 206
1895 206
1900 206
1905 205
1910 206
1915 205
1920 206
1925 207
1930 206
1935 206
1940 207
1945 206
1950 206
1955 207
1960 205
1965 205
1970 206
1975 206
1980 205
1985 206
1990 206
1995 206
2000 207
2005 206
2010 206
2015 206
2020 205
2025 207
2030 206
2035 206
2040 205
2045 205
2050 206
2055 206
2060 205
2065 206
2070 205
2075 206
2080 206
2085 206
2090 205
2095 206
2100 206
2105 206
2110 207
2115 206
2120 205
2125 205
2130 206
2135 206
2140 206
2145 205
2150 206
2155 205
2160 206
2165 205
2170 206
2175 205
2180 206
2185 205
2190 206
2195 205
2200 206
2205 206
2210 205
2215 206
2220 205
2225 206
2230 205
2235 206
2240 206
2245 206
2250 206
2255 206
2260 206
2265 206
2270 205
2275 207
2280 206
2285 207
2290 205
2295 205
2300 205
2305 205
2310 205
2315 206
2320 207
2325 205
2330 206
2335 206
2340 205
2345 205
2350 206
2355 205
2360 206
2365 206
2370 206
2375 206
2380 206
2385 205
2390 205
2395 206
2400 205
2405 206
2410 206
2415 207
2420 206
2425 206
2430 205
2435 206
2440 206
2445 206
2450 205
2455 205
2460 205
2465 205
2470 205
2475 207
2480 206
2485 206
2490 206
2495 206
2500 205
2505 206
2510 205
2515 206
2520 206
2525 206
2530 206
2535 206
2540 206
2545 205
2550 205
2555 205
2560 205
2565 206
2570 205
2575 206
2580 206
2585 205
2590 206
2595 206
2600 205
2605 206
2610 206
2615 206
2620 206
2625 205
2630 206
2635 206
2640 205
2645 205
2650 205
2655 206
2660 205
2665 207
2670 206
2675 205
2680 206
2685 206
2690 206
2695 206
2700 206
2705 206
2710 206
2715 205
2720 205
2725 205
2730 205
2735 206
2740 206
2745 205
2750 207
2755 206
2760 206
2765 205
2770 205
2775 206
2780 206
2785 205
2790 205
2795 206
2800 206
2805 205
2810 206
2815 205
2820 204
2825 208
2830 206
2835 206
2840 205
2845 206
2850 206
2855 205
2860 205
2865 206
2870 205
2875 206
2880 206
2885 205
2890 205
2895 205
2900 205
2905 205
2910 206
2915 206
2920 206
2925 205
2930 205
2935 205
2940 205
2945 205
2950 206
2955 206
2960 206
2965 205
2970 206
2975 205
2980 205
2985 205
2990 206
2995 204
3000 206
3005 206
3010 205
3015 205
3020 206
3025 206
3030 205
3035 206
3040 206
3045 206
3050 205
3055 205
3060 205
3065 205
3070 205
3075 207
3080 206
3085 205
3090 206
3095 206
3100 205
3105 205
3110 206
3115 205
3120 207
3125 206
3130 205
3135 205
3140 205
3145 205
3150 205
3155 204
3160 206
3165 205
3170 206
3175 205
3180 207
3185 206
3190 206
3195 205
3200 206
3205 206
3210 205
3215 205
3220 206
3225 206
3230 205
3235 206
3240 205
3245 206
3250 205
3255 205
3260 205
3265 206
3270 205
3275 206
3280 206
3285 206
3290 205
3295 206
3300 205
3305 205
3310 206
3315 204
3320 206
3325 206
3330 206
3335 206
3340 205
3345 206
3350 205
3355 205
3360 206
3365 205
3370 206
3375 205
3380 206
3385 206
3390 206
3395 205
3400 206
3405 206
3410 206
3415 205
3420 205
3425 208
3430 205
3435 206
3440 205
3445 205
3450 205
3455 205
3460 205
3465 206
3470 205
3475 205
3480 206
3485 206
3490 205
3495 205
3500 205
3505 206
3510 206
3515 205
3520 207
3525 205
3530 206
3535 205
3540 205
3545 205
3550 205
3555 205
3560 206
3565 205
3570 206
3575 205
3580 206
3585 205
3590 206
3595 205
3600 206
};
\addlegendentry{16 cores}
\addplot [thick, color4, mark=diamond*, mark size=3, mark options={solid,fill=white,draw=red},  mark repeat={160}]
table {%
0 210
5 210
10 211
15 211
20 212
25 205
30 211
35 207
40 206
45 205
50 206
55 205
60 206
65 205
70 206
75 207
80 205
85 205
90 206
95 207
100 205
105 207
110 205
115 206
120 206
125 207
130 205
135 206
140 206
145 206
150 211
155 211
160 212
165 211
170 212
175 210
180 210
185 210
190 210
195 210
200 209
205 211
210 212
215 211
220 206
225 213
230 242
235 260
240 284
245 294
250 301
255 303
260 301
265 303
270 303
275 303
280 303
285 304
290 305
295 308
300 308
305 309
310 310
315 311
320 312
325 311
330 312
335 312
340 312
345 312
350 312
355 312
360 312
365 315
370 316
375 315
380 315
385 316
390 313
395 314
400 315
405 315
410 315
415 315
420 315
425 315
430 314
435 313
440 315
445 315
450 315
455 315
460 315
465 313
470 314
475 315
480 315
485 315
490 313
495 315
500 315
505 316
510 315
515 315
520 316
525 315
530 315
535 315
540 315
545 315
550 314
555 315
560 315
565 315
570 315
575 315
580 315
585 316
590 317
595 316
600 315
605 315
610 315
615 315
620 314
625 313
630 315
635 315
640 314
645 317
650 315
655 315
660 315
665 315
670 315
675 315
680 314
685 315
690 315
695 312
700 315
705 315
710 312
715 313
720 313
725 315
730 315
735 315
740 315
745 315
750 312
755 315
760 315
765 314
770 313
775 315
780 315
785 315
790 315
795 315
800 312
805 315
810 315
815 315
820 315
825 316
830 315
835 313
840 315
845 315
850 315
855 316
860 315
865 315
870 315
875 315
880 315
885 315
890 315
895 315
900 315
905 315
910 315
915 313
920 315
925 315
930 315
935 315
940 315
945 316
950 315
955 315
960 315
965 313
970 313
975 315
980 316
985 315
990 315
995 315
1000 315
1005 316
1010 315
1015 315
1020 315
1025 316
1030 315
1035 317
1040 313
1045 315
1050 315
1055 317
1060 315
1065 316
1070 315
1075 316
1080 315
1085 315
1090 316
1095 316
1100 315
1105 315
1110 316
1115 315
1120 315
1125 315
1130 315
1135 315
1140 315
1145 313
1150 315
1155 315
1160 315
1165 315
1170 315
1175 315
1180 314
1185 314
1190 315
1195 316
1200 315
1205 313
1210 316
1215 315
1220 315
1225 303
1230 295
1235 290
1240 267
1245 252
1250 245
1255 227
1260 226
1265 226
1270 222
1275 223
1280 219
1285 214
1290 214
1295 213
1300 213
1305 209
1310 207
1315 207
1320 207
1325 208
1330 207
1335 207
1340 206
1345 207
1350 206
1355 207
1360 206
1365 206
1370 206
1375 206
1380 207
1385 207
1390 206
1395 206
1400 207
1405 206
1410 207
1415 207
1420 206
1425 206
1430 206
1435 206
1440 206
1445 207
1450 207
1455 205
1460 207
1465 206
1470 206
1475 207
1480 205
1485 207
1490 206
1495 207
1500 206
1505 206
1510 206
1515 206
1520 206
1525 206
1530 206
1535 206
1540 206
1545 206
1550 206
1555 206
1560 206
1565 206
1570 206
1575 206
1580 206
1585 205
1590 205
1595 206
1600 207
1605 206
1610 206
1615 205
1620 206
1625 206
1630 207
1635 205
1640 206
1645 206
1650 206
1655 206
1660 206
1665 206
1670 206
1675 206
1680 206
1685 207
1690 205
1695 206
1700 206
1705 206
1710 205
1715 205
1720 205
1725 206
1730 206
1735 206
1740 206
1745 206
1750 205
1755 206
1760 206
1765 206
1770 206
1775 206
1780 206
1785 206
1790 206
1795 206
1800 206
1805 206
1810 206
1815 205
1820 206
1825 205
1830 206
1835 206
1840 207
1845 205
1850 206
1855 205
1860 206
1865 205
1870 205
1875 206
1880 206
1885 206
1890 205
1895 206
1900 206
1905 206
1910 206
1915 206
1920 205
1925 205
1930 205
1935 206
1940 205
1945 206
1950 205
1955 206
1960 206
1965 206
1970 205
1975 206
1980 205
1985 206
1990 206
1995 205
2000 206
2005 205
2010 206
2015 207
2020 206
2025 205
2030 206
2035 206
2040 206
2045 205
2050 205
2055 206
2060 206
2065 206
2070 206
2075 206
2080 206
2085 207
2090 205
2095 205
2100 206
2105 206
2110 205
2115 206
2120 206
2125 205
2130 205
2135 206
2140 205
2145 206
2150 206
2155 206
2160 206
2165 205
2170 206
2175 205
2180 206
2185 205
2190 206
2195 205
2200 206
2205 205
2210 206
2215 206
2220 205
2225 206
2230 205
2235 205
2240 206
2245 205
2250 206
2255 206
2260 206
2265 206
2270 205
2275 207
2280 205
2285 206
2290 205
2295 205
2300 205
2305 206
2310 205
2315 206
2320 206
2325 206
2330 206
2335 206
2340 205
2345 206
2350 205
2355 206
2360 206
2365 206
2370 206
2375 206
2380 206
2385 205
2390 205
2395 206
2400 205
2405 205
2410 206
2415 205
2420 206
2425 206
2430 205
2435 206
2440 206
2445 206
2450 206
2455 206
2460 206
2465 205
2470 206
2475 206
2480 205
2485 206
2490 205
2495 205
2500 205
2505 206
2510 206
2515 206
2520 205
2525 206
2530 206
2535 205
2540 205
2545 206
2550 206
2555 206
2560 207
2565 206
2570 206
2575 205
2580 206
2585 206
2590 206
2595 206
2600 205
2605 206
2610 206
2615 205
2620 206
2625 206
2630 206
2635 205
2640 206
2645 206
2650 206
2655 205
2660 207
2665 206
2670 205
2675 207
2680 206
2685 206
2690 206
2695 205
2700 205
2705 206
2710 206
2715 206
2720 206
2725 206
2730 205
2735 206
2740 205
2745 205
2750 206
2755 205
2760 206
2765 205
2770 206
2775 206
2780 206
2785 205
2790 205
2795 205
2800 205
2805 206
2810 205
2815 205
2820 206
2825 206
2830 206
2835 205
2840 206
2845 205
2850 205
2855 206
2860 205
2865 206
2870 206
2875 206
2880 206
2885 206
2890 205
2895 207
2900 205
2905 206
2910 207
2915 206
2920 206
2925 206
2930 206
2935 206
2940 206
2945 206
2950 206
2955 207
2960 206
2965 206
2970 205
2975 206
2980 205
2985 206
2990 206
2995 205
3000 205
3005 205
3010 206
3015 206
3020 205
3025 205
3030 206
3035 205
3040 205
3045 205
3050 205
3055 206
3060 205
3065 206
3070 207
3075 206
3080 206
3085 205
3090 205
3095 206
3100 205
3105 205
3110 205
3115 205
3120 206
3125 206
3130 205
3135 206
3140 206
3145 205
3150 205
3155 206
3160 206
3165 206
3170 206
3175 206
3180 206
3185 206
3190 206
3195 205
3200 206
3205 205
3210 206
3215 205
3220 206
3225 206
3230 206
3235 206
3240 207
3245 205
3250 205
3255 205
3260 206
3265 205
3270 205
3275 205
3280 205
3285 207
3290 206
3295 206
3300 205
3305 206
3310 206
3315 205
3320 206
3325 206
3330 206
3335 207
3340 205
3345 205
3350 206
3355 205
3360 206
3365 206
3370 206
3375 206
3380 206
3385 206
3390 205
3395 205
3400 206
3405 206
3410 206
3415 205
3420 207
3425 206
3430 206
3435 205
3440 206
3445 205
3450 205
3455 205
3460 206
3465 206
3470 206
3475 206
3480 206
3485 205
3490 205
3495 206
3500 206
3505 207
3510 206
3515 206
3520 206
3525 206
3530 206
3535 206
3540 206
3545 207
3550 205
3555 206
3560 206
3565 206
3570 206
3575 205
3580 206
3585 206
3590 206
3595 206
3600 206
};
\addlegendentry{16 cores (32 threads)}
\end{axis}

\end{tikzpicture}

%% file: results/power_intel_jobs.tex
\begin{tikzpicture}[font=\Large]

\definecolor{color0}{rgb}{0.12156862745098,0.466666666666667,0.705882352941177}
\definecolor{color1}{rgb}{1,0.498039215686275,0.0549019607843137}
\definecolor{color2}{rgb}{0.172549019607843,0.627450980392157,0.172549019607843}
\definecolor{color3}{rgb}{0.83921568627451,0.152941176470588,0.156862745098039}
\definecolor{color4}{rgb}{0.580392156862745,0.403921568627451,0.741176470588235}
\definecolor{color5}{rgb}{0,0,0}

\begin{axis}[
legend cell align={left},
legend columns=4,
legend style={fill opacity=0.8, draw opacity=1, text opacity=1, at={(1.05,1.12)}, anchor=east, draw=white!80.0!black},
tick align=outside,
tick pos=left,
x grid style={white!69.01960784313725!black},
xlabel={Time (SEC)},
xmin=0, xmax=5400,
xtick style={color=black},
xtick={0,1000,2000,3000,4000,5000},
y grid style={white!69.01960784313725!black},
ylabel={Power consumption (W)},
ymin=150, ymax=500,
ytick={150,200,250,300,350,400,450,500},
xmajorgrids,
ymajorgrids,
ytick style={color=black}
]
\addplot [thick, color5]
table {%
0 254
5 262
10 252
15 253
20 253
25 253
30 253
35 254
40 253
45 253
50 254
55 254
60 253
65 254
70 254
75 254
80 254
85 254
90 254
95 254
100 254
105 254
110 253
115 254
120 254
125 254
130 254
135 253
140 253
145 253
150 253
155 254
160 254
165 253
170 254
175 254
180 253
185 254
190 254
195 254
200 253
205 254
210 254
215 254
220 253
225 254
230 253
235 253
240 253
245 252
250 253
255 253
260 254
265 254
270 254
275 253
280 254
285 253
290 253
295 253
300 254
305 253
310 254
315 253
320 253
325 253
330 254
335 254
340 253
345 253
350 253
355 254
360 253
365 253
370 254
375 254
380 254
385 254
390 254
395 254
400 253
405 253
410 253
415 254
420 253
425 254
430 253
435 253
440 254
445 254
450 254
455 254
460 253
465 254
470 253
475 253
480 252
485 254
490 254
495 253
500 253
505 254
510 254
515 254
520 253
525 253
530 254
535 254
540 253
545 254
550 253
555 253
560 253
565 253
570 254
575 253
580 254
585 253
590 254
595 254
600 253
605 254
610 253
615 254
620 254
625 253
630 254
635 253
640 253
645 254
650 254
655 253
660 254
665 253
670 253
675 254
680 253
685 254
690 254
695 254
700 254
705 253
710 254
715 254
720 253
725 254
730 253
735 253
740 253
745 254
750 254
755 254
760 254
765 254
770 254
775 253
780 254
785 254
790 254
795 253
800 254
805 253
810 253
815 253
820 254
825 254
830 253
835 253
840 253
845 253
850 253
855 253
860 254
865 253
870 253
875 252
880 254
885 254
890 253
895 253
900 254
905 253
910 254
915 254
920 253
925 254
930 254
935 254
940 253
945 254
950 253
955 254
960 254
965 253
970 254
975 254
980 254
985 252
990 253
995 253
1000 253
1005 253
1010 253
1015 253
1020 252
1025 254
1030 253
1035 254
1040 252
1045 254
1050 254
1055 254
1060 253
1065 254
1070 254
1075 254
1080 253
1085 253
1090 253
1095 254
1100 254
1105 253
1110 253
1115 254
1120 253
1125 254
1130 254
1135 253
1140 253
1145 253
1150 253
1155 254
1160 254
1165 254
1170 254
1175 254
1180 253
1185 253
1190 254
1195 253
1200 253
1205 254
1210 254
1215 254
1220 254
1225 254
1230 251
1235 253
1240 254
1245 253
1250 254
1255 253
1260 254
1265 254
1270 254
1275 254
1280 254
1285 253
1290 253
1295 253
1300 253
1305 254
1310 254
1315 254
1320 254
1325 254
1330 254
1335 253
1340 255
1345 254
1350 254
1355 253
1360 254
1365 254
1370 254
1375 253
1380 254
1385 254
1390 253
1395 254
1400 255
1405 253
1410 254
1415 254
1420 254
1425 254
1430 253
1435 254
1440 254
1445 253
1450 253
1455 253
1460 253
1465 253
1470 253
1475 253
1480 254
1485 254
1490 254
1495 253
1500 254
1505 253
1510 252
1515 254
1520 254
1525 254
1530 253
1535 254
1540 253
1545 254
1550 253
1555 254
1560 252
1565 254
1570 254
1575 254
1580 254
1585 254
1590 253
1595 254
1600 251
1605 254
1610 253
1615 254
1620 253
1625 254
1630 253
1635 254
1640 254
1645 253
1650 253
1655 254
1660 254
1665 253
1670 253
1675 253
1680 253
1685 254
1690 254
1695 254
1700 254
1705 253
1710 253
1715 254
1720 254
1725 252
1730 252
1735 253
1740 254
1745 253
1750 254
1755 253
1760 253
1765 253
1770 254
1775 252
1780 253
1785 253
1790 252
1795 253
1800 253
1805 253
1810 253
1815 253
1820 254
1825 252
1830 254
1835 253
1840 253
1845 253
1850 252
1855 252
1860 253
1865 253
1870 253
1875 254
1880 253
1885 252
1890 253
1895 253
1900 253
1905 252
1910 254
1915 252
1920 254
1925 253
1930 254
1935 253
1940 253
1945 254
1950 253
1955 254
1960 253
1965 253
1970 254
1975 254
1980 252
1985 253
1990 254
1995 253
2000 253
2005 253
2010 254
2015 253
2020 253
2025 253
2030 253
2035 253
2040 254
2045 254
2050 253
2055 254
2060 253
2065 253
2070 254
2075 253
2080 253
2085 254
2090 253
2095 254
2100 254
2105 253
2110 254
2115 254
2120 252
2125 254
2130 254
2135 253
2140 253
2145 253
2150 254
2155 254
2160 254
2165 253
2170 251
2175 253
2180 253
2185 254
2190 254
2195 253
2200 253
2205 254
2210 253
2215 254
2220 254
2225 254
2230 253
2235 254
2240 253
2245 254
2250 253
2255 254
2260 254
2265 253
2270 254
2275 254
2280 253
2285 253
2290 254
2295 254
2300 253
2305 253
2310 254
2315 253
2320 253
2325 253
2330 253
2335 254
2340 254
2345 254
2350 254
2355 253
2360 253
2365 253
2370 253
2375 254
2380 254
2385 253
2390 253
2395 254
2400 253
2405 253
2410 253
2415 253
2420 254
2425 253
2430 253
2435 253
2440 253
2445 254
2450 254
2455 254
2460 254
2465 251
2470 254
2475 253
2480 254
2485 253
2490 253
2495 254
2500 254
2505 253
2510 254
2515 254
2520 254
2525 253
2530 253
2535 251
2540 252
2545 253
2550 254
2555 253
2560 253
2565 254
2570 253
2575 253
2580 253
2585 253
2590 253
2595 253
2600 253
2605 253
2610 253
2615 253
2620 253
2625 261
2630 252
2635 254
2640 253
2645 254
2650 253
2655 254
2660 253
2665 253
2670 254
2675 254
2680 253
2685 253
2690 253
2695 253
2700 253
2705 254
2710 254
2715 254
2720 254
2725 253
2730 253
2735 253
2740 253
2745 254
2750 254
2755 253
2760 254
2765 252
2770 254
2775 254
2780 254
2785 254
2790 254
2795 253
2800 254
2805 253
2810 253
2815 253
2820 253
2825 253
2830 254
2835 254
2840 254
2845 254
2850 254
2855 254
2860 253
2865 254
2870 254
2875 253
2880 254
2885 251
2890 253
2895 254
2900 253
2905 253
2910 254
2915 253
2920 249
2925 252
2930 253
2935 252
2940 251
2945 252
2950 252
2955 253
2960 252
2965 252
2970 252
2975 252
2980 252
2985 251
2990 252
2995 252
3000 251
3005 252
3010 252
3015 252
3020 251
3025 251
3030 252
3035 251
3040 252
3045 252
3050 251
3055 252
3060 252
3065 252
3070 252
3075 252
3080 252
3085 252
3090 253
3095 251
3100 252
3105 252
3110 251
3115 252
3120 252
3125 251
3130 252
3135 253
3140 252
3145 252
3150 251
3155 251
3160 252
3165 252
3170 252
3175 251
3180 252
3185 251
3190 251
3195 253
3200 251
3205 251
3210 251
3215 252
3220 251
3225 252
3230 252
3235 252
3240 252
3245 251
3250 252
3255 251
3260 251
3265 252
3270 252
3275 251
3280 252
3285 251
3290 251
3295 251
3300 251
3305 252
3310 251
3315 253
3320 251
3325 251
3330 251
3335 252
3340 251
3345 251
3350 252
3355 252
3360 252
3365 252
3370 252
3375 252
3380 252
3385 253
3390 251
3395 253
3400 251
3405 252
3410 252
3415 252
3420 252
3425 251
3430 251
3435 253
3440 252
3445 251
3450 252
3455 251
3460 251
3465 251
3470 253
3475 253
3480 254
3485 253
3490 254
3495 254
3500 253
3505 254
3510 253
3515 254
3520 254
3525 254
3530 254
3535 253
3540 254
3545 253
3550 254
3555 254
3560 254
3565 253
3570 253
3575 254
3580 253
3585 254
3590 254
3595 253
3600 252
3605 253
3610 253
3615 254
3620 253
3625 255
3630 254
3635 254
3640 253
3645 253
3650 254
3655 253
3660 254
3665 254
3670 253
3675 252
3680 252
3685 254
3690 253
3695 253
3700 254
3705 253
3710 254
3715 253
3720 253
3725 253
3730 253
3735 252
3740 252
3745 254
3750 253
3755 250
3760 253
3765 253
3770 253
3775 254
3780 253
3785 253
3790 253
3795 254
3800 252
3805 254
3810 254
3815 253
3820 253
3825 253
3830 253
3835 254
3840 253
3845 252
3850 254
3855 253
3860 253
3865 253
3870 253
3875 253
3880 253
3885 253
3890 253
3895 254
3900 252
3905 254
3910 253
3915 253
3920 253
3925 253
3930 254
3935 254
3940 253
3945 253
3950 252
3955 253
3960 253
3965 252
3970 253
3975 254
3980 253
3985 254
3990 252
3995 253
4000 254
4005 253
4010 254
4015 253
4020 253
4025 254
4030 253
4035 254
4040 254
4045 254
4050 253
4055 254
4060 253
4065 254
4070 254
4075 254
4080 254
4085 254
4090 253
4095 254
4100 253
4105 253
4110 254
4115 253
4120 253
4125 254
4130 253
4135 253
4140 254
4145 254
4150 254
4155 253
4160 253
4165 254
4170 253
4175 253
4180 253
4185 254
4190 254
4195 253
4200 253
4205 254
4210 253
4215 253
4220 254
4225 253
4230 253
4235 253
4240 253
4245 254
4250 253
4255 253
4260 254
4265 253
4270 254
4275 254
4280 253
4285 253
4290 253
4295 254
4300 254
4305 254
4310 254
4315 253
4320 254
4325 254
4330 254
4335 253
4340 263
4345 253
4350 254
4355 254
4360 254
4365 254
4370 254
4375 254
4380 254
4385 254
4390 253
4395 253
4400 254
4405 253
4410 254
4415 254
4420 253
4425 253
4430 254
4435 253
4440 253
4445 254
4450 254
4455 253
4460 254
4465 253
4470 254
4475 253
4480 254
4485 253
4490 254
4495 254
4500 253
4505 253
4510 254
4515 253
4520 254
4525 253
4530 253
4535 253
4540 254
4545 254
4550 253
4555 253
4560 253
4565 253
4570 253
4575 254
4580 254
4585 253
4590 254
4595 254
4600 254
4605 254
4610 253
4615 254
4620 253
4625 253
4630 253
4635 253
4640 254
4645 253
4650 254
4655 254
4660 254
4665 253
4670 254
4675 254
4680 254
4685 253
4690 253
4695 254
4700 254
4705 254
4710 254
4715 254
4720 254
4725 253
4730 254
4735 253
4740 254
4745 253
4750 253
4755 254
4760 254
4765 253
4770 254
4775 254
4780 253
4785 254
4790 253
4795 253
4800 254
4805 253
4810 254
4815 254
4820 254
4825 254
4830 254
4835 253
4840 254
4845 254
4850 254
4855 253
4860 254
4865 253
4870 254
4875 253
4880 254
4885 253
4890 254
4895 254
4900 254
4905 253
4910 253
4915 254
4920 253
4925 254
4930 254
4935 254
4940 253
4945 254
4950 253
4955 254
4960 254
4965 252
4970 252
4975 254
4980 254
4985 254
4990 254
4995 254
5000 253
5005 254
5010 253
5015 254
5020 253
5025 254
5030 254
5035 254
5040 254
5045 252
5050 255
5055 253
5060 253
5065 253
5070 254
5075 254
5080 253
5085 254
5090 251
5095 253
5100 254
5105 254
5110 254
5115 254
5120 253
5125 254
5130 254
5135 254
5140 254
5145 254
5150 254
5155 254
5160 254
5165 253
5170 254
5175 253
5180 254
5185 254
5190 253
5195 254
5200 253
5205 253
5210 253
5215 252
5220 253
5225 254
5230 254
5235 253
5240 253
5245 253
5250 253
5255 253
5260 253
5265 253
5270 253
5275 252
5280 254
5285 254
5290 252
5295 253
5300 253
5305 254
5310 253
5315 254
5320 254
5325 254
5330 253
5335 254
5340 254
5345 253
5350 253
5355 254
5360 254
5365 253
5370 253
5375 253
5380 254
5385 254
5390 253
5395 253
};
\addlegendentry{Baseline}
\addplot [thick, color0,, mark=triangle*, mark size=3, mark options={solid,fill=white,draw=red},  mark repeat={180}]
table{0 254
5 261
10 264
15 266
20 271
25 272
30 264
35 265
40 254
45 255
50 254
55 255
60 254
65 254
70 256
75 253
80 255
85 254
90 253
95 254
100 253
105 257
110 256
115 266
120 268
125 268
130 267
135 266
140 263
145 264
150 263
155 262
160 269
165 268
170 267
175 272
180 266
185 352
190 356
195 358
200 357
205 357
210 357
215 357
220 357
225 359
230 359
235 359
240 359
245 359
250 358
255 357
260 359
265 359
270 358
275 359
280 358
285 359
290 360
295 359
300 360
305 361
310 370
315 371
320 371
325 370
330 367
335 369
340 367
345 371
350 363
355 363
360 364
365 364
370 363
375 363
380 364
385 363
390 363
395 363
400 363
405 362
410 363
415 363
420 363
425 364
430 364
435 364
440 364
445 363
450 364
455 364
460 365
465 365
470 363
475 363
480 365
485 364
490 365
495 365
500 365
505 365
510 365
515 365
520 365
525 365
530 364
535 364
540 364
545 365
550 365
555 362
560 364
565 374
570 363
575 364
580 365
585 364
590 365
595 364
600 365
605 365
610 365
615 365
620 364
625 365
630 365
635 365
640 363
645 363
650 364
655 364
660 365
665 365
670 364
675 363
680 365
685 365
690 365
695 364
700 364
705 363
710 363
715 365
720 364
725 363
730 365
735 363
740 364
745 364
750 363
755 365
760 363
765 364
770 365
775 365
780 364
785 365
790 364
795 365
800 366
805 365
810 363
815 365
820 362
825 364
830 365
835 364
840 365
845 365
850 364
855 366
860 365
865 365
870 366
875 366
880 364
885 365
890 365
895 366
900 364
905 365
910 365
915 366
920 364
925 365
930 364
935 364
940 365
945 365
950 365
955 365
960 365
965 365
970 364
975 362
980 364
985 365
990 366
995 366
1000 366
1005 366
1010 366
1015 365
1020 366
1025 366
1030 367
1035 366
1040 366
1045 366
1050 366
1055 367
1060 366
1065 366
1070 366
1075 365
1080 365
1085 366
1090 367
1095 366
1100 366
1105 366
1110 367
1115 366
1120 366
1125 367
1130 366
1135 365
1140 366
1145 365
1150 367
1155 366
1160 366
1165 366
1170 366
1175 366
1180 377
1185 375
1190 367
1195 367
1200 368
1205 367
1210 366
1215 367
1220 365
1225 365
1230 366
1235 366
1240 366
1245 366
1250 365
1255 366
1260 366
1265 366
1270 366
1275 365
1280 366
1285 366
1290 366
1295 366
1300 366
1305 366
1310 366
1315 365
1320 366
1325 366
1330 366
1335 367
1340 367
1345 366
1350 365
1355 366
1360 365
1365 367
1370 366
1375 367
1380 367
1385 367
1390 366
1395 367
1400 367
1405 366
1410 363
1415 364
1420 366
1425 365
1430 367
1435 365
1440 367
1445 366
1450 366
1455 366
1460 367
1465 367
1470 367
1475 367
1480 366
1485 365
1490 367
1495 367
1500 367
1505 367
1510 367
1515 368
1520 368
1525 366
1530 366
1535 366
1540 368
1545 367
1550 367
1555 368
1560 376
1565 377
1570 377
1575 377
1580 377
1585 377
1590 376
1595 377
1600 368
1605 370
1610 368
1615 368
1620 367
1625 368
1630 367
1635 366
1640 367
1645 366
1650 365
1655 366
1660 366
1665 367
1670 366
1675 365
1680 367
1685 366
1690 366
1695 365
1700 365
1705 366
1710 366
1715 366
1720 366
1725 365
1730 367
1735 367
1740 365
1745 367
1750 366
1755 366
1760 366
1765 367
1770 367
1775 367
1780 366
1785 367
1790 366
1795 367
1800 367
1805 366
1810 367
1815 367
1820 366
1825 368
1830 367
1835 367
1840 367
1845 367
1850 366
1855 366
1860 365
1865 366
1870 366
1875 365
1880 366
1885 366
1890 367
1895 366
1900 367
1905 367
1910 367
1915 366
1920 366
1925 366
1930 366
1935 367
1940 365
1945 367
1950 367
1955 367
1960 366
1965 366
1970 366
1975 367
1980 367
1985 366
1990 366
1995 366
2000 366
2005 367
2010 367
2015 366
2020 365
2025 366
2030 366
2035 366
2040 367
2045 367
2050 368
2055 365
2060 366
2065 366
2070 366
2075 367
2080 368
2085 368
2090 367
2095 367
2100 367
2105 367
2110 365
2115 366
2120 367
2125 366
2130 366
2135 367
2140 367
2145 366
2150 367
2155 366
2160 366
2165 367
2170 366
2175 366
2180 366
2185 366
2190 367
2195 367
2200 366
2205 367
2210 366
2215 366
2220 366
2225 367
2230 367
2235 367
2240 366
2245 366
2250 366
2255 366
2260 358
2265 365
2270 367
2275 366
2280 364
2285 367
2290 365
2295 374
2300 366
2305 366
2310 367
2315 366
2320 366
2325 367
2330 366
2335 365
2340 365
2345 366
2350 366
2355 366
2360 367
2365 362
2370 365
2375 366
2380 365
2385 366
2390 368
2395 365
2400 365
2405 367
2410 365
2415 365
2420 366
2425 365
2430 366
2435 367
2440 313
2445 313
2450 299
2455 272
2460 272
2465 271
2470 270
2475 270
2480 268
2485 269
2490 260
2495 256
2500 256
2505 256
2510 256
2515 256
2520 256
2525 255
2530 256
2535 255
2540 257
2545 255
2550 255
2555 255
2560 255
2565 255
2570 255
2575 256
2580 254
2585 254
2590 254
2595 254
2600 254
2605 254
2610 254
2615 254
2620 254
2625 255
2630 255
2635 254
2640 254
2645 253
2650 256
2655 253
2660 254
2665 253
2670 253
2675 253
2680 251
2685 253
2690 254
2695 254
2700 255
2705 253
2710 254
2715 253
2720 254
2725 253
2730 254
2735 254
2740 253
2745 254
2750 253
2755 254
2760 253
2765 254
2770 254
2775 253
2780 253
2785 254
2790 263
2795 254
2800 254
2805 254
2810 253
2815 253
2820 254
2825 254
2830 253
2835 253
2840 253
2845 254
2850 253
2855 254
2860 254
2865 253
2870 253
2875 253
2880 253
2885 253
2890 253
2895 254
2900 253
2905 253
2910 253
2915 254
2920 253
2925 254
2930 253
2935 253
2940 253
2945 253
2950 257
2955 254
2960 253
2965 253
2970 253
2975 253
2980 253
2985 253
2990 251
2995 253
3000 254
3005 253
3010 254
3015 254
3020 254
3025 254
3030 253
3035 253
3040 254
3045 253
3050 253
3055 254
3060 253
3065 253
3070 254
3075 253
3080 254
3085 254
3090 253
3095 253
3100 254
3105 254
3110 254
3115 254
3120 254
3125 253
3130 254
3135 254
3140 254
3145 254
3150 253
3155 254
3160 254
3165 253
3170 254
3175 254
3180 253
3185 253
3190 254
3195 253
3200 253
3205 254
3210 253
3215 254
3220 254
3225 253
3230 254
3235 253
3240 254
3245 254
3250 255
3255 254
3260 253
3265 253
3270 251
3275 254
3280 254
3285 254
3290 254
3295 255
3300 254
3305 253
3310 253
3315 254
3320 253
3325 253
3330 253
3335 254
3340 254
3345 253
3350 254
3355 254
3360 253
3365 253
3370 253
3375 253
3380 254
3385 253
3390 253
3395 254
3400 253
3405 253
3410 262
3415 262
3420 262
3425 263
3430 263
3435 263
3440 263
3445 262
3450 253
3455 254
3460 254
3465 251
3470 254
3475 254
3480 254
3485 253
3490 254
3495 254
3500 253
3505 254
3510 253
3515 253
3520 254
3525 253
3530 254
3535 254
3540 254
3545 254
3550 258
3555 253
3560 254
3565 253
3570 253
3575 254
3580 253
3585 254
3590 253
3595 253
3600 254
3605 254
3610 254
3615 253
3620 253
3625 253
3630 253
3635 253
3640 254
3645 254
3650 252
3655 253
3660 254
3665 253
3670 253
3675 254
3680 254
3685 253
3690 253
3695 254
3700 254
3705 254
3710 254
3715 254
3720 253
3725 254
3730 254
3735 254
3740 254
3745 254
3750 253
3755 254
3760 253
3765 254
3770 254
3775 254
3780 253
3785 253
3790 254
3795 254
3800 254
3805 253
3810 254
3815 254
3820 253
3825 253
3830 253
3835 253
3840 254
3845 254
3850 254
3855 253
3860 253
3865 253
3870 253
3875 253
3880 254
3885 253
3890 254
3895 253
3900 253
3905 251
3910 253
3915 254
3920 253
3925 254
3930 254
3935 254
3940 254
3945 254
3950 253
3955 254
3960 254
3965 253
3970 254
3975 253
3980 254
3985 254
3990 254
3995 253
4000 253
4005 254
4010 254
4015 254
4020 253
4025 254
4030 253
4035 253
4040 254
4045 254
4050 253
4055 253
4060 253
4065 254
4070 253
4075 254
4080 254
4085 254
4090 253
4095 253
4100 254
4105 253
4110 254
4115 253
4120 254
4125 254
4130 254
4135 254
4140 254
4145 253
4150 255
4155 253
4160 253
4165 253
4170 254
4175 254
4180 254
4185 253
4190 253
4195 253
4200 254
4205 254
4210 253
4215 253
4220 254
4225 253
4230 254
4235 254
4240 254
4245 254
4250 254
4255 254
4260 253
4265 254
4270 254
4275 253
4280 253
4285 254
4290 254
4295 254
4300 254
4305 254
4310 253
4315 253
4320 253
4325 254
4330 254
4335 253
4340 254
4345 254
4350 254
4355 254
4360 254
4365 254
4370 254
4375 254
4380 254
4385 254
4390 254
4395 254
4400 253
4405 254
4410 254
4415 254
4420 254
4425 253
4430 254
4435 254
4440 254
4445 254
4450 254
4455 254
4460 253
4465 254
4470 253
4475 253
4480 253
4485 253
4490 253
4495 254
4500 254
4505 254
4510 254
4515 254
4520 254
4525 254
4530 253
4535 253
4540 254
4545 253
4550 254
4555 254
4560 254
4565 254
4570 254
4575 254
4580 253
4585 253
4590 253
4595 253
4600 254
4605 254
4610 253
4615 253
4620 254
4625 254
4630 253
4635 254
4640 253
4645 253
4650 254
4655 253
4660 253
4665 254
4670 255
4675 254
4680 254
4685 253
4690 253
4695 254
4700 253
4705 254
4710 254
4715 253
4720 254
4725 254
4730 254
4735 253
4740 254
4745 254
4750 253
4755 254
4760 254
4765 254
4770 253
4775 254
4780 254
4785 253
4790 254
4795 254
4800 253
4805 254
4810 254
4815 254
4820 254
4825 253
4830 254
4835 254
4840 253
4845 254
4850 254
4855 253
4860 253
4865 253
4870 254
4875 254
4880 254
4885 253
4890 254
4895 254
4900 254
4905 254
4910 254
4915 254
4920 264
4925 251
4930 254
4935 253
4940 254
4945 254
4950 253
4955 253
4960 253
4965 253
4970 253
4975 254
4980 253
4985 254
4990 254
4995 254
5000 253
5005 253
5010 254
5015 253
5020 254
5025 254
5030 254
5035 253
5040 254
5045 253
5050 256
5055 253
5060 254
5065 254
5070 254
5075 254
5080 253
5085 254
5090 254
5095 254
5100 254
5105 254
5110 253
5115 254
5120 253
5125 254
5130 254
5135 253
5140 254
5145 253
5150 253
5155 254
5160 254
5165 254
5170 253
5175 254
5180 253
5185 254
5190 253
5195 254
5200 253
5205 254
5210 254
5215 254
5220 254
5225 254
5230 254
5235 254
5240 251
5245 253
5250 254
5255 254
5260 253
5265 253
5270 253
5275 254
5280 254
5285 253
5290 254
5295 253
5300 253
5305 253
5310 254
5315 253
5320 253
5325 254
5330 253
5335 254
5340 254
5345 254
5350 254
5355 253
5360 254
5365 253
5370 254
5375 254
5380 254
5385 254
5390 254
5395 254
5400 254
};
\addlegendentry{1 job}
\addplot [thick, color1,, mark=o, mark size=3, mark options={solid,fill=white,draw=red},  mark repeat={180}]
table {%
0 255
5 256
10 277
15 279
20 282
25 289
30 268
35 278
40 257
45 255
50 256
55 256
60 256
65 255
70 256
75 256
80 256
85 256
90 256
95 256
100 257
105 253
110 256
115 256
120 270
125 281
130 282
135 283
140 279
145 274
150 274
155 274
160 275
165 277
170 282
175 284
180 279
185 288
190 285
195 463
200 463
205 464
210 468
215 467
220 468
225 470
230 470
235 470
240 470
245 472
250 473
255 472
260 473
265 471
270 471
275 471
280 471
285 471
290 472
295 472
300 471
305 471
310 471
315 471
320 471
325 471
330 471
335 471
340 471
345 472
350 472
355 471
360 472
365 472
370 471
375 472
380 471
385 472
390 472
395 472
400 472
405 472
410 471
415 473
420 473
425 473
430 473
435 472
440 472
445 474
450 473
455 472
460 474
465 471
470 474
475 475
480 474
485 474
490 474
495 474
500 474
505 474
510 475
515 475
520 475
525 474
530 475
535 474
540 474
545 474
550 474
555 475
560 475
565 474
570 475
575 474
580 475
585 474
590 475
595 475
600 474
605 475
610 475
615 475
620 475
625 475
630 474
635 475
640 475
645 475
650 475
655 474
660 475
665 475
670 474
675 474
680 473
685 474
690 474
695 474
700 474
705 475
710 474
715 475
720 474
725 474
730 474
735 474
740 474
745 475
750 474
755 474
760 475
765 475
770 474
775 474
780 475
785 474
790 475
795 473
800 474
805 474
810 475
815 475
820 474
825 474
830 474
835 475
840 474
845 475
850 475
855 474
860 475
865 474
870 472
875 474
880 474
885 475
890 474
895 475
900 474
905 475
910 474
915 475
920 476
925 475
930 476
935 475
940 476
945 474
950 476
955 475
960 475
965 475
970 475
975 476
980 475
985 475
990 476
995 476
1000 476
1005 475
1010 476
1015 476
1020 476
1025 476
1030 475
1035 475
1040 476
1045 477
1050 477
1055 476
1060 476
1065 476
1070 476
1075 477
1080 477
1085 477
1090 475
1095 476
1100 476
1105 476
1110 477
1115 476
1120 477
1125 477
1130 477
1135 477
1140 477
1145 477
1150 477
1155 476
1160 477
1165 477
1170 477
1175 477
1180 478
1185 478
1190 478
1195 477
1200 477
1205 477
1210 478
1215 477
1220 478
1225 477
1230 478
1235 477
1240 478
1245 477
1250 477
1255 477
1260 478
1265 477
1270 477
1275 478
1280 477
1285 477
1290 478
1295 477
1300 478
1305 476
1310 478
1315 478
1320 478
1325 477
1330 477
1335 477
1340 477
1345 477
1350 478
1355 477
1360 477
1365 478
1370 478
1375 477
1380 478
1385 478
1390 477
1395 477
1400 478
1405 478
1410 478
1415 477
1420 478
1425 477
1430 479
1435 478
1440 478
1445 478
1450 477
1455 477
1460 478
1465 478
1470 479
1475 478
1480 478
1485 479
1490 478
1495 478
1500 477
1505 477
1510 477
1515 477
1520 477
1525 477
1530 477
1535 477
1540 477
1545 478
1550 478
1555 477
1560 478
1565 478
1570 478
1575 478
1580 479
1585 478
1590 478
1595 478
1600 478
1605 477
1610 478
1615 478
1620 477
1625 478
1630 478
1635 477
1640 478
1645 477
1650 477
1655 478
1660 478
1665 477
1670 478
1675 477
1680 478
1685 478
1690 477
1695 477
1700 477
1705 477
1710 477
1715 477
1720 476
1725 476
1730 476
1735 476
1740 476
1745 476
1750 476
1755 475
1760 476
1765 476
1770 476
1775 475
1780 475
1785 476
1790 475
1795 475
1800 476
1805 475
1810 476
1815 476
1820 476
1825 476
1830 476
1835 475
1840 475
1845 475
1850 475
1855 476
1860 475
1865 474
1870 474
1875 474
1880 476
1885 475
1890 473
1895 472
1900 474
1905 475
1910 475
1915 474
1920 474
1925 475
1930 475
1935 474
1940 475
1945 474
1950 474
1955 474
1960 473
1965 473
1970 474
1975 473
1980 474
1985 473
1990 473
1995 473
2000 474
2005 473
2010 472
2015 473
2020 473
2025 473
2030 474
2035 473
2040 473
2045 472
2050 474
2055 474
2060 473
2065 472
2070 474
2075 473
2080 473
2085 472
2090 473
2095 474
2100 474
2105 474
2110 473
2115 473
2120 473
2125 472
2130 474
2135 473
2140 473
2145 473
2150 473
2155 472
2160 474
2165 473
2170 473
2175 474
2180 474
2185 475
2190 474
2195 474
2200 474
2205 474
2210 474
2215 475
2220 474
2225 474
2230 475
2235 474
2240 474
2245 474
2250 474
2255 474
2260 476
2265 475
2270 474
2275 475
2280 475
2285 475
2290 475
2295 475
2300 476
2305 476
2310 475
2315 474
2320 475
2325 474
2330 475
2335 474
2340 476
2345 475
2350 476
2355 475
2360 476
2365 475
2370 475
2375 475
2380 475
2385 476
2390 475
2395 476
2400 475
2405 476
2410 476
2415 475
2420 474
2425 476
2430 472
2435 434
2440 429
2445 400
2450 384
2455 382
2460 380
2465 380
2470 374
2475 372
2480 335
2485 315
2490 316
2495 273
2500 272
2505 272
2510 271
2515 271
2520 270
2525 258
2530 258
2535 257
2540 258
2545 257
2550 258
2555 258
2560 258
2565 258
2570 258
2575 258
2580 257
2585 257
2590 257
2595 256
2600 257
2605 257
2610 256
2615 257
2620 256
2625 257
2630 256
2635 256
2640 257
2645 256
2650 256
2655 257
2660 256
2665 256
2670 257
2675 257
2680 256
2685 256
2690 256
2695 256
2700 255
2705 257
2710 255
2715 257
2720 255
2725 255
2730 256
2735 255
2740 256
2745 256
2750 255
2755 255
2760 255
2765 256
2770 256
2775 255
2780 255
2785 255
2790 256
2795 256
2800 255
2805 255
2810 255
2815 256
2820 255
2825 256
2830 256
2835 255
2840 256
2845 256
2850 255
2855 255
2860 256
2865 256
2870 256
2875 256
2880 256
2885 255
2890 255
2895 256
2900 259
2905 253
2910 254
2915 255
2920 255
2925 255
2930 255
2935 255
2940 256
2945 256
2950 255
2955 256
2960 256
2965 255
2970 256
2975 256
2980 256
2985 256
2990 255
2995 256
3000 255
3005 255
3010 256
3015 256
3020 255
3025 256
3030 256
3035 255
3040 256
3045 256
3050 255
3055 255
3060 254
3065 254
3070 256
3075 255
3080 254
3085 254
3090 255
3095 255
3100 254
3105 254
3110 255
3115 254
3120 254
3125 254
3130 255
3135 255
3140 254
3145 255
3150 255
3155 255
3160 255
3165 255
3170 255
3175 254
3180 254
3185 255
3190 255
3195 255
3200 255
3205 255
3210 255
3215 255
3220 254
3225 256
3230 255
3235 255
3240 254
3245 254
3250 254
3255 255
3260 255
3265 252
3270 254
3275 255
3280 255
3285 254
3290 254
3295 255
3300 255
3305 254
3310 255
3315 256
3320 254
3325 254
3330 255
3335 255
3340 254
3345 255
3350 255
3355 254
3360 254
3365 254
3370 255
3375 254
3380 254
3385 255
3390 254
3395 254
3400 254
3405 255
3410 254
3415 255
3420 254
3425 256
3430 255
3435 255
3440 255
3445 255
3450 255
3455 254
3460 255
3465 256
3470 255
3475 254
3480 254
3485 255
3490 255
3495 254
3500 260
3505 255
3510 255
3515 254
3520 256
3525 254
3530 255
3535 254
3540 255
3545 256
3550 254
3555 247
3560 253
3565 253
3570 253
3575 253
3580 253
3585 253
3590 253
3595 253
3600 253
3605 253
3610 253
3615 253
3620 253
3625 253
3630 253
3635 253
3640 253
3645 253
3650 253
3655 254
3660 253
3665 253
3670 253
3675 253
3680 263
3685 251
3690 255
3695 254
3700 254
3705 254
3710 254
3715 254
3720 254
3725 254
3730 254
3735 255
3740 254
3745 254
3750 255
3755 254
3760 254
3765 254
3770 254
3775 254
3780 254
3785 254
3790 254
3795 255
3800 256
3805 254
3810 254
3815 254
3820 255
3825 254
3830 255
3835 254
3840 255
3845 254
3850 254
3855 254
3860 255
3865 254
3870 254
3875 254
3880 254
3885 254
3890 254
3895 254
3900 254
3905 254
3910 254
3915 254
3920 254
3925 254
3930 254
3935 254
3940 255
3945 252
3950 254
3955 254
3960 255
3965 254
3970 255
3975 255
3980 255
3985 256
3990 255
3995 255
4000 255
4005 254
4010 254
4015 254
4020 255
4025 255
4030 255
4035 255
4040 255
4045 254
4050 254
4055 254
4060 254
4065 255
4070 254
4075 254
4080 254
4085 255
4090 254
4095 255
4100 255
4105 254
4110 255
4115 254
4120 254
4125 256
4130 255
4135 255
4140 255
4145 255
4150 256
4155 254
4160 254
4165 254
4170 255
4175 254
4180 254
4185 255
4190 255
4195 254
4200 255
4205 254
4210 255
4215 255
4220 255
4225 254
4230 255
4235 254
4240 254
4245 256
4250 254
4255 254
4260 255
4265 255
4270 255
4275 254
4280 255
4285 254
4290 254
4295 253
4300 253
4305 253
4310 254
4315 254
4320 254
4325 253
4330 254
4335 254
4340 254
4345 253
4350 254
4355 254
4360 253
4365 253
4370 254
4375 253
4380 253
4385 254
4390 253
4395 253
4400 253
4405 253
4410 251
4415 253
4420 254
4425 253
4430 253
4435 253
4440 254
4445 253
4450 254
4455 254
4460 253
4465 253
4470 253
4475 254
4480 254
4485 254
4490 254
4495 254
4500 254
4505 253
4510 253
4515 253
4520 253
4525 254
4530 253
4535 253
4540 253
4545 253
4550 253
4555 254
4560 253
4565 254
4570 253
4575 253
4580 253
4585 254
4590 253
4595 253
4600 254
4605 253
4610 253
4615 253
4620 254
4625 254
4630 253
4635 253
4640 254
4645 253
4650 252
4655 252
4660 253
4665 253
4670 254
4675 254
4680 253
4685 254
4690 254
4695 253
4700 253
4705 254
4710 254
4715 253
4720 253
4725 253
4730 254
4735 253
4740 252
4745 254
4750 253
4755 253
4760 252
4765 252
4770 254
4775 254
4780 254
4785 254
4790 254
4795 254
4800 253
4805 253
4810 254
4815 252
4820 253
4825 254
4830 253
4835 254
4840 254
4845 254
4850 253
4855 254
4860 254
4865 254
4870 254
4875 253
4880 253
4885 252
4890 254
4895 254
4900 254
4905 254
4910 252
4915 253
4920 254
4925 253
4930 254
4935 254
4940 253
4945 253
4950 253
4955 253
4960 253
4965 252
4970 254
4975 253
4980 254
4985 254
4990 254
4995 253
5000 254
5005 254
5010 253
5015 254
5020 254
5025 253
5030 253
5035 253
5040 253
5045 253
5050 254
5055 254
5060 254
5065 254
5070 254
5075 254
5080 253
5085 253
5090 253
5095 253
5100 253
5105 254
5110 254
5115 254
5120 253
5125 253
5130 253
5135 254
5140 253
5145 253
5150 254
5155 253
5160 254
5165 253
5170 254
5175 254
5180 253
5185 254
5190 253
5195 254
5200 254
5205 253
5210 254
5215 254
5220 254
5225 253
5230 253
5235 253
5240 254
5245 253
5250 254
5255 252
5260 253
5265 254
5270 254
5275 254
5280 250
5285 254
5290 253
5295 263
5300 251
5305 253
5310 253
5315 252
5320 253
5325 251
5330 254
5335 253
5340 254
5345 253
5350 252
5355 253
5360 253
5365 254
5370 254
5375 253
5380 253
5385 253
5390 254
5395 254

};
\addlegendentry{2 jobs}
\addplot [thick, color2, mark=square*, mark size=3, mark options={solid,fill=white,draw=red},  mark repeat={180}]
table {%
0 253
5 262
10 287
15 302
20 311
25 323
30 311
35 301
40 260
45 254
50 254
55 253
60 255
65 254
70 254
75 253
80 253
85 255
90 253
95 253
100 255
105 254
110 265
115 254
120 305
125 306
130 305
135 306
140 305
145 292
150 294
155 289
160 289
165 312
170 307
175 307
180 309
185 305
190 469
195 462
200 463
205 465
210 465
215 466
220 466
225 467
230 468
235 466
240 468
245 467
250 468
255 469
260 470
265 471
270 470
275 470
280 469
285 469
290 469
295 469
300 469
305 469
310 469
315 469
320 469
325 469
330 469
335 469
340 469
345 469
350 469
355 469
360 469
365 470
370 468
375 471
380 470
385 472
390 470
395 470
400 471
405 470
410 471
415 470
420 471
425 472
430 471
435 470
440 471
445 470
450 471
455 471
460 472
465 472
470 471
475 471
480 472
485 472
490 472
495 471
500 472
505 472
510 471
515 472
520 472
525 471
530 472
535 472
540 472
545 472
550 472
555 472
560 472
565 472
570 472
575 473
580 472
585 473
590 473
595 473
600 473
605 473
610 473
615 471
620 473
625 473
630 474
635 473
640 474
645 474
650 475
655 474
660 475
665 475
670 473
675 475
680 475
685 474
690 474
695 474
700 475
705 475
710 474
715 475
720 474
725 474
730 475
735 473
740 473
745 473
750 473
755 473
760 473
765 473
770 474
775 473
780 474
785 473
790 473
795 473
800 474
805 473
810 474
815 473
820 474
825 473
830 474
835 473
840 473
845 473
850 472
855 473
860 472
865 473
870 472
875 472
880 472
885 472
890 472
895 473
900 472
905 471
910 472
915 471
920 471
925 471
930 471
935 471
940 471
945 472
950 471
955 471
960 471
965 469
970 471
975 471
980 471
985 471
990 471
995 471
1000 472
1005 472
1010 471
1015 470
1020 470
1025 471
1030 471
1035 470
1040 472
1045 470
1050 471
1055 470
1060 470
1065 472
1070 468
1075 471
1080 471
1085 470
1090 472
1095 470
1100 471
1105 472
1110 471
1115 472
1120 472
1125 471
1130 471
1135 472
1140 473
1145 472
1150 472
1155 472
1160 472
1165 472
1170 472
1175 472
1180 472
1185 472
1190 472
1195 472
1200 473
1205 473
1210 470
1215 469
1220 469
1225 469
1230 469
1235 469
1240 469
1245 469
1250 469
1255 469
1260 469
1265 470
1270 470
1275 469
1280 469
1285 469
1290 469
1295 468
1300 470
1305 469
1310 469
1315 468
1320 468
1325 472
1330 473
1335 472
1340 472
1345 472
1350 471
1355 471
1360 471
1365 472
1370 472
1375 471
1380 471
1385 472
1390 471
1395 471
1400 471
1405 471
1410 472
1415 471
1420 471
1425 472
1430 471
1435 471
1440 471
1445 473
1450 471
1455 471
1460 472
1465 471
1470 471
1475 471
1480 471
1485 471
1490 471
1495 471
1500 471
1505 471
1510 471
1515 471
1520 472
1525 473
1530 473
1535 473
1540 472
1545 473
1550 472
1555 473
1560 473
1565 473
1570 473
1575 473
1580 473
1585 474
1590 473
1595 473
1600 473
1605 474
1610 472
1615 472
1620 470
1625 470
1630 470
1635 469
1640 469
1645 471
1650 470
1655 473
1660 474
1665 474
1670 473
1675 474
1680 474
1685 473
1690 473
1695 473
1700 473
1705 473
1710 474
1715 473
1720 473
1725 474
1730 474
1735 473
1740 473
1745 473
1750 473
1755 474
1760 473
1765 473
1770 473
1775 474
1780 474
1785 473
1790 473
1795 473
1800 474
1805 473
1810 473
1815 473
1820 474
1825 473
1830 473
1835 474
1840 473
1845 473
1850 473
1855 473
1860 473
1865 474
1870 473
1875 473
1880 473
1885 473
1890 474
1895 473
1900 473
1905 473
1910 474
1915 471
1920 474
1925 474
1930 473
1935 474
1940 474
1945 474
1950 474
1955 475
1960 475
1965 475
1970 475
1975 474
1980 475
1985 475
1990 475
1995 474
2000 474
2005 474
2010 475
2015 475
2020 474
2025 474
2030 474
2035 475
2040 475
2045 475
2050 474
2055 475
2060 474
2065 474
2070 474
2075 474
2080 474
2085 474
2090 474
2095 475
2100 474
2105 475
2110 474
2115 475
2120 474
2125 474
2130 475
2135 474
2140 476
2145 475
2150 474
2155 474
2160 475
2165 476
2170 474
2175 474
2180 474
2185 474
2190 474
2195 474
2200 474
2205 474
2210 474
2215 475
2220 473
2225 474
2230 473
2235 473
2240 474
2245 473
2250 474
2255 472
2260 473
2265 472
2270 472
2275 473
2280 473
2285 474
2290 473
2295 473
2300 473
2305 472
2310 472
2315 474
2320 472
2325 473
2330 473
2335 472
2340 472
2345 472
2350 472
2355 472
2360 472
2365 472
2370 471
2375 471
2380 472
2385 471
2390 472
2395 471
2400 471
2405 472
2410 471
2415 471
2420 471
2425 471
2430 472
2435 471
2440 471
2445 472
2450 472
2455 471
2460 471
2465 471
2470 471
2475 472
2480 472
2485 472
2490 472
2495 472
2500 472
2505 473
2510 473
2515 472
2520 472
2525 474
2530 472
2535 472
2540 472
2545 473
2550 474
2555 472
2560 473
2565 473
2570 473
2575 474
2580 473
2585 473
2590 473
2595 474
2600 473
2605 474
2610 473
2615 474
2620 474
2625 474
2630 474
2635 474
2640 474
2645 474
2650 474
2655 475
2660 474
2665 474
2670 474
2675 474
2680 475
2685 474
2690 474
2695 474
2700 474
2705 475
2710 475
2715 475
2720 475
2725 472
2730 474
2735 474
2740 476
2745 474
2750 474
2755 475
2760 474
2765 475
2770 475
2775 474
2780 474
2785 475
2790 474
2795 473
2800 473
2805 473
2810 474
2815 473
2820 474
2825 474
2830 472
2835 473
2840 473
2845 473
2850 473
2855 472
2860 473
2865 472
2870 473
2875 473
2880 473
2885 473
2890 472
2895 472
2900 473
2905 472
2910 472
2915 472
2920 472
2925 472
2930 472
2935 471
2940 471
2945 471
2950 471
2955 471
2960 472
2965 471
2970 472
2975 471
2980 472
2985 471
2990 472
2995 471
3000 472
3005 471
3010 471
3015 471
3020 471
3025 471
3030 471
3035 471
3040 472
3045 471
3050 472
3055 472
3060 471
3065 471
3070 472
3075 471
3080 471
3085 471
3090 472
3095 472
3100 472
3105 472
3110 473
3115 472
3120 472
3125 472
3130 473
3135 473
3140 472
3145 472
3150 473
3155 473
3160 473
3165 472
3170 472
3175 472
3180 473
3185 474
3190 473
3195 473
3200 472
3205 473
3210 473
3215 473
3220 473
3225 472
3230 473
3235 473
3240 473
3245 473
3250 474
3255 474
3260 474
3265 475
3270 474
3275 474
3280 474
3285 474
3290 474
3295 474
3300 474
3305 474
3310 474
3315 474
3320 474
3325 475
3330 474
3335 474
3340 475
3345 474
3350 475
3355 474
3360 474
3365 474
3370 475
3375 475
3380 475
3385 475
3390 474
3395 474
3400 469
3405 470
3410 471
3415 470
3420 471
3425 472
3430 470
3435 472
3440 470
3445 475
3450 474
3455 476
3460 474
3465 472
3470 475
3475 475
3480 475
3485 475
3490 473
3495 475
3500 473
3505 474
3510 473
3515 474
3520 473
3525 474
3530 474
3535 474
3540 473
3545 473
3550 474
3555 473
3560 473
3565 474
3570 473
3575 472
3580 472
3585 473
3590 474
3595 473
3600 473
3605 473
3610 472
3615 473
3620 472
3625 473
3630 473
3635 472
3640 474
3645 472
3650 472
3655 472
3660 472
3665 472
3670 472
3675 472
3680 473
3685 473
3690 472
3695 472
3700 472
3705 472
3710 472
3715 472
3720 472
3725 472
3730 472
3735 472
3740 472
3745 473
3750 472
3755 473
3760 473
3765 472
3770 470
3775 472
3780 472
3785 473
3790 474
3795 473
3800 474
3805 473
3810 474
3815 474
3820 473
3825 474
3830 473
3835 473
3840 473
3845 475
3850 475
3855 474
3860 474
3865 474
3870 474
3875 474
3880 475
3885 473
3890 474
3895 475
3900 474
3905 474
3910 475
3915 474
3920 474
3925 474
3930 473
3935 472
3940 475
3945 474
3950 474
3955 475
3960 474
3965 473
3970 474
3975 473
3980 474
3985 474
3990 473
3995 473
4000 473
4005 474
4010 472
4015 468
4020 430
4025 398
4030 398
4035 395
4040 421
4045 408
4050 385
4055 317
4060 288
4065 287
4070 287
4075 290
4080 288
4085 274
4090 267
4095 263
4100 260
4105 258
4110 261
4115 260
4120 260
4125 261
4130 260
4135 259
4140 259
4145 261
4150 261
4155 260
4160 263
4165 262
4170 259
4175 261
4180 261
4185 262
4190 261
4195 263
4200 262
4205 261
4210 260
4215 261
4220 261
4225 262
4230 259
4235 261
4240 262
4245 260
4250 260
4255 263
4260 262
4265 260
4270 259
4275 258
4280 257
4285 257
4290 254
4295 257
4300 257
4305 257
4310 260
4315 258
4320 257
4325 256
4330 259
4335 258
4340 258
4345 258
4350 259
4355 259
4360 259
4365 256
4370 259
4375 257
4380 259
4385 259
4390 259
4395 258
4400 258
4405 258
4410 258
4415 257
4420 257
4425 260
4430 259
4435 259
4440 259
4445 260
4450 259
4455 259
4460 258
4465 259
4470 257
4475 258
4480 258
4485 258
4490 259
4495 260
4500 258
4505 260
4510 258
4515 258
4520 257
4525 259
4530 259
4535 258
4540 257
4545 259
4550 258
4555 258
4560 259
4565 259
4570 259
4575 259
4580 259
4585 259
4590 258
4595 259
4600 259
4605 258
4610 259
4615 259
4620 258
4625 260
4630 259
4635 258
4640 259
4645 259
4650 260
4655 257
4660 259
4665 258
4670 258
4675 260
4680 258
4685 258
4690 259
4695 258
4700 259
4705 259
4710 259
4715 258
4720 258
4725 259
4730 259
4735 258
4740 258
4745 258
4750 259
4755 259
4760 258
4765 259
4770 259
4775 258
4780 258
4785 258
4790 259
4795 259
4800 259
4805 258
4810 260
4815 258
4820 258
4825 259
4830 259
4835 258
4840 258
4845 258
4850 258
4855 259
4860 259
4865 258
4870 258
4875 259
4880 256
4885 258
4890 258
4895 258
4900 258
4905 259
4910 258
4915 261
4920 260
4925 259
4930 258
4935 259
4940 258
4945 258
4950 258
4955 258
4960 259
4965 262
4970 259
4975 259
4980 259
4985 259
4990 259
4995 259
5000 258
5005 257
5010 259
5015 259
5020 259
5025 258
5030 259
5035 258
5040 259
5045 258
5050 258
5055 259
5060 259
5065 257
5070 258
5075 259
5080 259
5085 259
5090 259
5095 259
5100 260
5105 257
5110 258
5115 259
5120 259
5125 258
5130 258
5135 259
5140 258
5145 257
5150 257
5155 258
5160 258
5165 257
5170 258
5175 257
5180 259
5185 258
5190 258
5195 258
5200 259
5205 257
5210 259
5215 259
5220 258
5225 258
5230 258
5235 256
5240 258
5245 260
5250 258
5255 258
5260 258
5265 258
5270 258
5275 258
5280 258
5285 257
5290 258
5295 258
5300 257
5305 257
5310 257
5315 257
5320 258
5325 256
5330 257
5335 255
5340 257
5345 258
5350 258
5355 257
5360 257
5365 258
5370 257
5375 258
5380 258
5385 258
5390 258
5395 259
};
\addlegendentry{4 jobs}
\end{axis}

\end{tikzpicture}

%% file: results/power_amd_jobs.tex
\begin{tikzpicture}[font=\Large]

\definecolor{color0}{rgb}{0.12156862745098,0.466666666666667,0.705882352941177}
\definecolor{color1}{rgb}{1,0.498039215686275,0.0549019607843137}
\definecolor{color2}{rgb}{0.172549019607843,0.627450980392157,0.172549019607843}
\definecolor{color3}{rgb}{0.83921568627451,0.152941176470588,0.156862745098039}
\definecolor{color4}{rgb}{0.580392156862745,0.403921568627451,0.741176470588235}
\definecolor{color5}{rgb}{0,0,0}

\begin{axis}[
legend cell align={left},
legend columns=4,
legend style={fill opacity=0.8, draw opacity=1, text opacity=1, at={(1.05,1.12)}, anchor=east, draw=white!80.0!black},
tick align=outside,
tick pos=left,
x grid style={white!69.01960784313725!black},
xlabel={Time (SEC)},
xmin=0, xmax=5400,
xtick={0,1000,2000,3000,4000,5000},
xtick style={color=black},
y grid style={white!69.01960784313725!black},
ylabel={Power consumption (W)},
ymin=150, ymax=350,
ytick={150,200,250,300,350},
xmajorgrids,
ymajorgrids,
ytick style={color=black}
]
\addplot [thick, color5, mark=.]
table{
0 205
5 205
10 204
15 205
20 205
25 204
30 205
35 204
40 204
45 204
50 205
55 205
60 204
65 205
70 204
75 205
80 204
85 205
90 205
95 205
100 205
105 204
110 204
115 205
120 205
125 204
130 205
135 205
140 204
145 204
150 205
155 205
160 205
165 204
170 205
175 204
180 204
185 205
190 205
195 204
200 204
205 206
210 206
215 206
220 206
225 206
230 206
235 206
240 205
245 206
250 205
255 206
260 205
265 206
270 205
275 206
280 206
285 205
290 205
295 205
300 205
305 205
310 205
315 206
320 205
325 205
330 205
335 205
340 206
345 205
350 205
355 205
360 205
365 205
370 206
375 206
380 205
385 205
390 205
395 206
400 205
405 206
410 206
415 206
420 206
425 206
430 206
435 206
440 206
445 205
450 208
455 205
460 205
465 206
470 206
475 206
480 205
485 205
490 206
495 205
500 205
505 205
510 205
515 205
520 205
525 206
530 205
535 206
540 205
545 206
550 205
555 205
560 205
565 205
570 205
575 206
580 206
585 206
590 205
595 206
600 205
605 206
610 206
615 206
620 205
625 206
630 206
635 206
640 205
645 206
650 206
655 206
660 205
665 206
670 206
675 205
680 206
685 206
690 205
695 205
700 205
705 205
710 205
715 205
720 205
725 206
730 205
735 206
740 207
745 207
750 206
755 205
760 205
765 205
770 206
775 206
780 205
785 206
790 206
795 206
800 205
805 206
810 205
815 205
820 206
825 205
830 206
835 205
840 205
845 206
850 206
855 206
860 205
865 206
870 205
875 206
880 205
885 206
890 205
895 206
900 206
905 206
910 205
915 206
920 206
925 205
930 205
935 205
940 206
945 206
950 205
955 206
960 205
965 206
970 206
975 206
980 205
985 206
990 206
995 206
1000 206
1005 205
1010 206
1015 205
1020 205
1025 206
1030 206
1035 205
1040 206
1045 206
1050 205
1055 205
1060 205
1065 205
1070 206
1075 205
1080 206
1085 206
1090 205
1095 205
1100 205
1105 205
1110 205
1115 205
1120 206
1125 205
1130 205
1135 206
1140 205
1145 206
1150 205
1155 205
1160 206
1165 206
1170 206
1175 205
1180 206
1185 205
1190 205
1195 205
1200 205
1205 205
1210 205
1215 205
1220 206
1225 206
1230 206
1235 205
1240 206
1245 206
1250 206
1255 205
1260 205
1265 206
1270 205
1275 205
1280 205
1285 205
1290 205
1295 205
1300 205
1305 205
1310 205
1315 205
1320 206
1325 206
1330 206
1335 206
1340 206
1345 205
1350 206
1355 206
1360 206
1365 205
1370 206
1375 205
1380 206
1385 206
1390 205
1395 205
1400 206
1405 206
1410 206
1415 206
1420 206
1425 205
1430 205
1435 206
1440 205
1445 206
1450 206
1455 205
1460 206
1465 206
1470 206
1475 205
1480 205
1485 206
1490 206
1495 206
1500 205
1505 206
1510 206
1515 205
1520 205
1525 205
1530 206
1535 206
1540 206
1545 206
1550 205
1555 206
1560 206
1565 206
1570 206
1575 206
1580 205
1585 206
1590 206
1595 205
1600 206
1605 206
1610 206
1615 205
1620 206
1625 206
1630 205
1635 206
1640 206
1645 207
1650 206
1655 206
1660 206
1665 205
1670 205
1675 205
1680 205
1685 206
1690 206
1695 206
1700 206
1705 205
1710 205
1715 206
1720 206
1725 206
1730 206
1735 205
1740 205
1745 206
1750 206
1755 206
1760 206
1765 205
1770 205
1775 205
1780 205
1785 205
1790 205
1795 205
1800 206
1805 205
1810 206
1815 205
1820 205
1825 206
1830 206
1835 205
1840 206
1845 206
1850 206
1855 206
1860 205
1865 205
1870 206
1875 205
1880 205
1885 205
1890 205
1895 206
1900 206
1905 205
1910 206
1915 205
1920 206
1925 205
1930 205
1935 206
1940 205
1945 205
1950 205
1955 204
1960 206
1965 206
1970 205
1975 205
1980 205
1985 205
1990 205
1995 205
2000 206
2005 205
2010 205
2015 205
2020 205
2025 205
2030 205
2035 205
2040 205
2045 205
2050 205
2055 205
2060 205
2065 206
2070 205
2075 206
2080 206
2085 205
2090 205
2095 205
2100 206
2105 205
2110 205
2115 206
2120 205
2125 205
2130 205
2135 205
2140 205
2145 206
2150 206
2155 205
2160 205
2165 205
2170 205
2175 206
2180 205
2185 206
2190 205
2195 205
2200 205
2205 205
2210 205
2215 205
2220 206
2225 205
2230 205
2235 206
2240 206
2245 206
2250 205
2255 205
2260 205
2265 205
2270 205
2275 205
2280 205
2285 205
2290 205
2295 205
2300 206
2305 206
2310 206
2315 205
2320 205
2325 205
2330 205
2335 205
2340 205
2345 205
2350 206
2355 206
2360 205
2365 205
2370 205
2375 204
2380 205
2385 206
2390 205
2395 205
2400 205
2405 205
2410 204
2415 205
2420 205
2425 205
2430 206
2435 206
2440 205
2445 206
2450 206
2455 205
2460 205
2465 205
2470 206
2475 205
2480 206
2485 205
2490 206
2495 205
2500 205
2505 206
2510 205
2515 205
2520 205
2525 206
2530 206
2535 205
2540 206
2545 206
2550 205
2555 206
2560 206
2565 205
2570 206
2575 205
2580 206
2585 205
2590 206
2595 205
2600 205
2605 206
2610 206
2615 206
2620 205
2625 206
2630 206
2635 205
2640 205
2645 205
2650 206
2655 205
2660 206
2665 206
2670 206
2675 205
2680 206
2685 206
2690 206
2695 206
2700 205
2705 206
2710 206
2715 206
2720 205
2725 206
2730 205
2735 205
2740 206
2745 205
2750 205
2755 205
2760 205
2765 206
2770 207
2775 205
2780 206
2785 205
2790 205
2795 206
2800 205
2805 200
2810 205
2815 205
2820 205
2825 205
2830 205
2835 205
2840 205
2845 206
2850 205
2855 206
2860 205
2865 205
2870 205
2875 204
2880 204
2885 204
2890 205
2895 204
2900 205
2905 205
2910 206
2915 205
2920 205
2925 205
2930 206
2935 205
2940 204
2945 205
2950 205
2955 206
2960 205
2965 205
2970 205
2975 205
2980 206
2985 205
2990 204
2995 206
3000 205
3005 205
3010 205
3015 204
3020 205
3025 205
3030 204
3035 204
3040 205
3045 205
3050 205
3055 205
3060 205
3065 205
3070 204
3075 205
3080 205
3085 205
3090 205
3095 205
3100 205
3105 205
3110 205
3115 205
3120 205
3125 205
3130 205
3135 205
3140 205
3145 207
3150 205
3155 205
3160 205
3165 205
3170 205
3175 205
3180 205
3185 204
3190 205
3195 205
3200 205
3205 205
3210 204
3215 205
3220 205
3225 206
3230 205
3235 205
3240 205
3245 205
3250 205
3255 205
3260 205
3265 205
3270 205
3275 204
3280 205
3285 205
3290 204
3295 204
3300 205
3305 205
3310 205
3315 205
3320 205
3325 206
3330 205
3335 205
3340 205
3345 204
3350 205
3355 205
3360 205
3365 205
3370 205
3375 205
3380 204
3385 205
3390 204
3395 205
3400 205
3405 205
3410 205
3415 204
3420 205
3425 205
3430 205
3435 205
3440 205
3445 208
3450 205
3455 205
3460 204
3465 205
3470 205
3475 205
3480 206
3485 205
3490 205
3495 205
3500 205
3505 205
3510 205
3515 204
3520 204
3525 205
3530 205
3535 205
3540 205
3545 204
3550 205
3555 205
3560 205
3565 206
3570 204
3575 205
3580 205
3585 205
3590 205
3595 205
3600 205
3605 205
3610 205
3615 205
3620 205
3625 205
3630 205
3635 204
3640 204
3645 205
3650 205
3655 204
3660 204
3665 205
3670 204
3675 205
3680 205
3685 205
3690 205
3695 205
3700 205
3705 205
3710 205
3715 205
3720 206
3725 204
3730 205
3735 204
3740 204
3745 206
3750 205
3755 205
3760 205
3765 205
3770 205
3775 205
3780 205
3785 205
3790 205
3795 204
3800 204
3805 205
3810 206
3815 205
3820 204
3825 205
3830 204
3835 205
3840 205
3845 204
3850 205
3855 204
3860 205
3865 205
3870 205
3875 205
3880 205
3885 205
3890 205
3895 205
3900 205
3905 204
3910 205
3915 205
3920 205
3925 205
3930 204
3935 205
3940 205
3945 204
3950 205
3955 205
3960 205
3965 205
3970 205
3975 205
3980 205
3985 204
3990 204
3995 205
4000 205
4005 204
4010 204
4015 205
4020 205
4025 205
4030 205
4035 204
4040 205
4045 205
4050 205
4055 205
4060 204
4065 205
4070 204
4075 205
4080 205
4085 205
4090 205
4095 205
4100 205
4105 205
4110 205
4115 205
4120 205
4125 205
4130 205
4135 205
4140 204
4145 205
4150 205
4155 205
4160 205
4165 205
4170 204
4175 205
4180 205
4185 205
4190 205
4195 205
4200 204
4205 205
4210 204
4215 205
4220 205
4225 205
4230 205
4235 205
4240 205
4245 204
4250 204
4255 205
4260 205
4265 205
4270 205
4275 205
4280 204
4285 204
4290 205
4295 205
4300 204
4305 206
4310 205
4315 205
4320 205
4325 205
4330 205
4335 205
4340 205
4345 205
4350 204
4355 205
4360 205
4365 205
4370 205
4375 205
4380 205
4385 205
4390 205
4395 205
4400 205
4405 205
4410 205
4415 205
4420 205
4425 204
4430 205
4435 206
4440 205
4445 205
4450 204
4455 205
4460 205
4465 205
4470 205
4475 205
4480 204
4485 205
4490 206
4495 204
4500 205
4505 205
4510 204
4515 204
4520 204
4525 205
4530 204
4535 205
4540 205
4545 205
4550 205
4555 205
4560 205
4565 205
4570 205
4575 206
4580 205
4585 205
4590 205
4595 205
4600 205
4605 205
4610 205
4615 205
4620 205
4625 205
4630 204
4635 205
4640 205
4645 205
4650 205
4655 205
4660 204
4665 204
4670 205
4675 205
4680 205
4685 205
4690 205
4695 205
4700 205
4705 205
4710 205
4715 205
4720 205
4725 205
4730 204
4735 204
4740 205
4745 205
4750 205
4755 205
4760 204
4765 205
4770 205
4775 205
4780 204
4785 205
4790 204
4795 205
4800 205
4805 205
4810 204
4815 204
4820 205
4825 205
4830 205
4835 204
4840 205
4845 204
4850 205
4855 205
4860 204
4865 206
4870 205
4875 205
4880 204
4885 204
4890 204
4895 205
4900 205
4905 205
4910 205
4915 205
4920 205
4925 205
4930 205
4935 205
4940 204
4945 205
4950 205
4955 205
4960 205
4965 205
4970 205
4975 205
4980 205
4985 205
4990 206
4995 205
5000 205
5005 205
5010 205
5015 205
5020 204
5025 205
5030 205
5035 205
5040 205
5045 205
5050 204
5055 204
5060 205
5065 204
5070 205
5075 205
5080 206
5085 205
5090 205
5095 205
5100 205
5105 205
5110 205
5115 205
5120 205
5125 206
5130 205
5135 205
5140 205
5145 205
5150 205
5155 205
5160 205
5165 204
5170 205
5175 204
5180 204
5185 205
5190 204
5195 204
5200 205
5205 205
5210 204
5215 205
5220 205
5225 205
5230 205
5235 205
5240 205
5245 204
5250 205
5255 204
5260 205
5265 205
5270 205
5275 205
5280 204
5285 205
5290 205
5295 205
5300 205
5305 205
5310 204
5315 205
5320 204
5325 204
5330 205
5335 205
5340 205
5345 205
5350 204
5355 205
5360 205
5365 205
5370 205
5375 204
5380 205
5385 205
5390 205
5395 205
5400 204
};
\addlegendentry{Baseline}
\addplot [thick, color0, mark=triangle*, mark size=3, mark options={solid,fill=white,draw=red},  mark repeat={180}]
table{
0 210
5 211
10 211
15 211
20 211
25 206
30 210
35 204
40 205
45 205
50 206
55 205
60 205
65 204
70 205
75 204
80 207
85 205
90 205
95 205
100 205
105 208
110 210
115 211
120 211
125 211
130 210
135 208
140 209
145 208
150 208
155 210
160 211
165 210
170 211
175 211
180 210
185 223
190 241
195 241
200 241
205 241
210 242
215 242
220 242
225 242
230 242
235 241
240 241
245 242
250 241
255 242
260 241
265 242
270 241
275 241
280 241
285 242
290 242
295 242
300 242
305 242
310 242
315 242
320 243
325 243
330 242
335 243
340 242
345 242
350 243
355 243
360 243
365 243
370 242
375 242
380 247
385 243
390 242
395 243
400 243
405 243
410 242
415 242
420 243
425 243
430 244
435 243
440 243
445 243
450 244
455 243
460 243
465 243
470 243
475 244
480 244
485 243
490 243
495 244
500 243
505 244
510 243
515 244
520 243
525 243
530 244
535 244
540 243
545 244
550 243
555 243
560 243
565 244
570 244
575 244
580 243
585 244
590 244
595 245
600 246
605 246
610 244
615 246
620 245
625 245
630 245
635 245
640 245
645 245
650 246
655 244
660 245
665 246
670 245
675 245
680 247
685 245
690 245
695 244
700 245
705 245
710 246
715 245
720 244
725 245
730 246
735 245
740 246
745 245
750 245
755 245
760 245
765 245
770 245
775 244
780 246
785 245
790 246
795 245
800 246
805 245
810 246
815 245
820 245
825 245
830 245
835 246
840 245
845 240
850 244
855 244
860 244
865 244
870 245
875 244
880 245
885 244
890 244
895 245
900 245
905 245
910 244
915 245
920 245
925 244
930 244
935 244
940 245
945 244
950 244
955 245
960 245
965 244
970 244
975 244
980 244
985 245
990 245
995 244
1000 244
1005 244
1010 245
1015 245
1020 244
1025 245
1030 244
1035 245
1040 244
1045 245
1050 244
1055 244
1060 244
1065 244
1070 245
1075 245
1080 245
1085 245
1090 245
1095 244
1100 245
1105 244
1110 245
1115 245
1120 245
1125 245
1130 244
1135 244
1140 245
1145 245
1150 245
1155 245
1160 244
1165 245
1170 245
1175 245
1180 245
1185 245
1190 245
1195 244
1200 245
1205 244
1210 243
1215 245
1220 245
1225 245
1230 245
1235 245
1240 245
1245 245
1250 243
1255 243
1260 242
1265 243
1270 242
1275 242
1280 244
1285 243
1290 242
1295 242
1300 244
1305 246
1310 244
1315 244
1320 245
1325 244
1330 245
1335 245
1340 245
1345 244
1350 244
1355 244
1360 245
1365 244
1370 244
1375 245
1380 245
1385 245
1390 245
1395 245
1400 245
1405 244
1410 244
1415 245
1420 245
1425 245
1430 246
1435 244
1440 245
1445 246
1450 245
1455 245
1460 245
1465 245
1470 245
1475 244
1480 245
1485 244
1490 244
1495 245
1500 245
1505 245
1510 245
1515 244
1520 245
1525 245
1530 245
1535 245
1540 245
1545 245
1550 245
1555 244
1560 244
1565 245
1570 244
1575 244
1580 244
1585 244
1590 244
1595 245
1600 245
1605 244
1610 245
1615 244
1620 245
1625 245
1630 245
1635 244
1640 242
1645 242
1650 242
1655 242
1660 243
1665 242
1670 243
1675 242
1680 243
1685 243
1690 243
1695 244
1700 244
1705 243
1710 243
1715 242
1720 242
1725 244
1730 243
1735 245
1740 245
1745 244
1750 245
1755 244
1760 244
1765 245
1770 244
1775 244
1780 245
1785 244
1790 244
1795 245
1800 243
1805 244
1810 244
1815 244
1820 245
1825 245
1830 245
1835 245
1840 244
1845 244
1850 244
1855 245
1860 245
1865 244
1870 245
1875 246
1880 245
1885 244
1890 245
1895 245
1900 245
1905 244
1910 245
1915 244
1920 245
1925 244
1930 244
1935 244
1940 245
1945 244
1950 245
1955 245
1960 245
1965 244
1970 245
1975 245
1980 244
1985 244
1990 245
1995 244
2000 244
2005 245
2010 245
2015 246
2020 246
2025 244
2030 244
2035 244
2040 245
2045 245
2050 245
2055 245
2060 245
2065 245
2070 245
2075 246
2080 244
2085 244
2090 244
2095 244
2100 245
2105 245
2110 245
2115 245
2120 245
2125 245
2130 245
2135 245
2140 245
2145 245
2150 245
2155 245
2160 244
2165 245
2170 245
2175 246
2180 245
2185 245
2190 245
2195 245
2200 244
2205 246
2210 245
2215 244
2220 245
2225 245
2230 245
2235 244
2240 244
2245 245
2250 246
2255 245
2260 245
2265 245
2270 244
2275 245
2280 246
2285 245
2290 245
2295 245
2300 245
2305 244
2310 246
2315 245
2320 245
2325 244
2330 245
2335 245
2340 245
2345 245
2350 245
2355 245
2360 245
2365 244
2370 245
2375 245
2380 246
2385 245
2390 245
2395 244
2400 245
2405 245
2410 245
2415 245
2420 245
2425 245
2430 245
2435 246
2440 243
2445 242
2450 243
2455 243
2460 243
2465 244
2470 243
2475 243
2480 242
2485 243
2490 243
2495 243
2500 243
2505 243
2510 243
2515 242
2520 242
2525 242
2530 242
2535 243
2540 245
2545 244
2550 244
2555 245
2560 244
2565 245
2570 244
2575 244
2580 245
2585 245
2590 244
2595 244
2600 244
2605 244
2610 245
2615 245
2620 244
2625 244
2630 244
2635 245
2640 245
2645 244
2650 245
2655 244
2660 244
2665 245
2670 245
2675 244
2680 244
2685 244
2690 245
2695 244
2700 244
2705 245
2710 245
2715 245
2720 245
2725 245
2730 245
2735 244
2740 227
2745 227
2750 217
2755 212
2760 213
2765 212
2770 212
2775 212
2780 211
2785 206
2790 205
2795 204
2800 206
2805 205
2810 205
2815 205
2820 205
2825 205
2830 205
2835 205
2840 206
2845 205
2850 205
2855 204
2860 205
2865 205
2870 204
2875 204
2880 205
2885 205
2890 204
2895 204
2900 205
2905 204
2910 204
2915 204
2920 204
2925 204
2930 205
2935 205
2940 205
2945 204
2950 204
2955 204
2960 205
2965 205
2970 204
2975 204
2980 205
2985 204
2990 205
2995 204
3000 204
3005 205
3010 205
3015 204
3020 205
3025 205
3030 205
3035 205
3040 205
3045 205
3050 204
3055 205
3060 204
3065 203
3070 204
3075 204
3080 204
3085 204
3090 204
3095 204
3100 205
3105 204
3110 204
3115 203
3120 204
3125 205
3130 204
3135 204
3140 204
3145 204
3150 204
3155 204
3160 204
3165 204
3170 204
3175 205
3180 204
3185 204
3190 204
3195 203
3200 205
3205 205
3210 204
3215 203
3220 204
3225 204
3230 204
3235 204
3240 204
3245 204
3250 204
3255 203
3260 204
3265 204
3270 203
3275 204
3280 204
3285 204
3290 204
3295 204
3300 204
3305 204
3310 203
3315 204
3320 204
3325 204
3330 204
3335 204
3340 204
3345 204
3350 203
3355 204
3360 204
3365 204
3370 204
3375 207
3380 204
3385 204
3390 204
3395 204
3400 204
3405 204
3410 205
3415 204
3420 204
3425 204
3430 204
3435 204
3440 203
3445 204
3450 204
3455 204
3460 203
3465 203
3470 204
3475 204
3480 204
3485 204
3490 204
3495 204
3500 204
3505 205
3510 205
3515 204
3520 203
3525 204
3530 204
3535 203
3540 204
3545 204
3550 204
3555 204
3560 204
3565 204
3570 204
3575 204
3580 204
3585 203
3590 203
3595 204
3600 204
3605 204
3610 204
3615 204
3620 205
3625 204
3630 204
3635 204
3640 204
3645 204
3650 204
3655 205
3660 204
3665 204
3670 204
3675 204
3680 204
3685 204
3690 204
3695 204
3700 205
3705 204
3710 205
3715 204
3720 204
3725 204
3730 203
3735 205
3740 204
3745 204
3750 203
3755 204
3760 205
3765 204
3770 204
3775 204
3780 204
3785 204
3790 204
3795 204
3800 204
3805 204
3810 204
3815 204
3820 204
3825 204
3830 204
3835 204
3840 204
3845 204
3850 204
3855 204
3860 204
3865 203
3870 204
3875 204
3880 205
3885 204
3890 205
3895 203
3900 204
3905 204
3910 204
3915 204
3920 205
3925 204
3930 205
3935 205
3940 205
3945 204
3950 204
3955 204
3960 204
3965 204
3970 204
3975 205
3980 205
3985 205
3990 205
3995 204
4000 204
4005 205
4010 204
4015 204
4020 204
4025 204
4030 205
4035 205
4040 204
4045 205
4050 204
4055 205
4060 205
4065 205
4070 205
4075 205
4080 205
4085 205
4090 205
4095 204
4100 204
4105 204
4110 204
4115 205
4120 204
4125 204
4130 205
4135 204
4140 204
4145 204
4150 205
4155 204
4160 204
4165 204
4170 204
4175 204
4180 204
4185 205
4190 204
4195 204
4200 205
4205 204
4210 204
4215 203
4220 204
4225 204
4230 204
4235 204
4240 205
4245 204
4250 204
4255 204
4260 205
4265 205
4270 204
4275 207
4280 204
4285 204
4290 205
4295 204
4300 204
4305 204
4310 205
4315 204
4320 205
4325 205
4330 204
4335 205
4340 204
4345 204
4350 204
4355 205
4360 204
4365 204
4370 204
4375 205
4380 205
4385 205
4390 205
4395 204
4400 205
4405 204
4410 205
4415 205
4420 205
4425 204
4430 204
4435 205
4440 204
4445 205
4450 204
4455 204
4460 204
4465 205
4470 204
4475 204
4480 204
4485 205
4490 204
4495 205
4500 205
4505 204
4510 204
4515 204
4520 204
4525 205
4530 205
4535 204
4540 204
4545 204
4550 205
4555 204
4560 205
4565 204
4570 205
4575 207
4580 204
4585 205
4590 204
4595 205
4600 205
4605 204
4610 204
4615 205
4620 204
4625 204
4630 205
4635 204
4640 205
4645 205
4650 204
4655 205
4660 205
4665 204
4670 203
4675 204
4680 204
4685 205
4690 204
4695 204
4700 205
4705 204
4710 205
4715 204
4720 204
4725 205
4730 204
4735 205
4740 204
4745 204
4750 205
4755 205
4760 205
4765 204
4770 204
4775 204
4780 204
4785 205
4790 204
4795 205
4800 204
4805 205
4810 204
4815 204
4820 204
4825 204
4830 204
4835 204
4840 204
4845 204
4850 205
4855 205
4860 204
4865 205
4870 204
4875 210
4880 204
4885 204
4890 204
4895 204
4900 204
4905 204
4910 204
4915 205
4920 204
4925 205
4930 205
4935 205
4940 204
4945 205
4950 204
4955 204
4960 205
4965 204
4970 205
4975 204
4980 205
4985 205
4990 204
4995 204
5000 204
5005 204
5010 204
5015 204
5020 205
5025 205
5030 204
5035 204
5040 204
5045 205
5050 204
5055 204
5060 205
5065 205
5070 204
5075 204
5080 205
5085 204
5090 204
5095 204
5100 205
5105 204
5110 204
5115 204
5120 205
5125 204
5130 205
5135 204
5140 204
5145 205
5150 205
5155 205
5160 205
5165 205
5170 205
5175 205
5180 204
5185 204
5190 205
5195 205
5200 204
5205 204
5210 204
5215 204
5220 204
5225 205
5230 204
5235 205
5240 204
5245 205
5250 205
5255 204
5260 204
5265 205
5270 205
5275 205
5280 205
5285 204
5290 204
5295 204
5300 204
5305 204
5310 204
5315 205
5320 205
5325 204
5330 205
5335 205
5340 204
5345 204
5350 204
5355 204
5360 204
5365 204
5370 204
5375 204
5380 204
5385 204
5390 204
5395 205
5400 205
5405 205
5410 204
5415 205
5420 204
5425 204
5430 204
5435 205
5440 204
5445 204
5450 205
5455 205
5460 205
5465 205
5470 205
5475 206
5480 204
5485 204
5490 205
5495 204
5500 204
5505 205
5510 204
5515 204
5520 204
5525 204
5530 204
5535 204
5540 205
5545 204
5550 204
5555 204
5560 205
5565 204
5570 205
5575 205
5580 204
5585 204
5590 204
5595 204
5600 204
5605 205
5610 204
5615 205
5620 205
5625 204
5630 204
5635 204
5640 204
5645 205
5650 205
5655 204
5660 205
5665 205
5670 204
5675 204
5680 205
5685 204
5690 204
5695 204
5700 205
5705 204
5710 205
5715 204
5720 204
5725 205
5730 204
5735 204
5740 205
5745 205
5750 204
5755 205
5760 204
5765 204
5770 204
5775 205
5780 205
5785 204
5790 204
5795 204
5800 205
5805 204
5810 204
5815 204
5820 205
5825 205
5830 204
5835 205
5840 205
5845 205
5850 204
5855 205
5860 205
5865 204
5870 204
5875 205
5880 205
5885 205
5890 205
5895 204
5900 204
5905 204
5910 204
5915 205
5920 205
5925 204
5930 204
5935 205
5940 204
5945 205
5950 205
5955 204
5960 204
5965 205
5970 204
5975 205
5980 205
5985 205
5990 204
5995 204
6000 204
};
\addlegendentry{1 job}
\addplot [thick, color1, mark=o, mark size=3, mark options={solid,fill=white,draw=red},  mark repeat={180}]
table{
0 216
5 215
10 215
15 216
20 216
25 217
30 208
35 206
40 205
45 206
50 205
55 206
60 206
65 205
70 206
75 206
80 206
85 205
90 205
95 206
100 207
105 208
110 205
115 207
120 206
125 208
130 214
135 216
140 217
145 216
150 216
155 216
160 214
165 213
170 212
175 212
180 213
185 213
190 215
195 216
200 213
205 217
210 217
215 278
220 278
225 280
230 279
235 279
240 280
245 280
250 281
255 280
260 281
265 281
270 281
275 281
280 281
285 282
290 282
295 282
300 283
305 283
310 283
315 283
320 282
325 283
330 284
335 284
340 285
345 284
350 285
355 285
360 285
365 286
370 287
375 287
380 287
385 286
390 287
395 286
400 287
405 287
410 287
415 286
420 286
425 287
430 286
435 287
440 287
445 286
450 286
455 286
460 286
465 287
470 287
475 287
480 286
485 287
490 286
495 286
500 286
505 286
510 286
515 286
520 286
525 286
530 286
535 286
540 287
545 286
550 287
555 287
560 287
565 287
570 286
575 287
580 286
585 287
590 287
595 287
600 287
605 286
610 287
615 288
620 286
625 286
630 286
635 287
640 287
645 286
650 285
655 287
660 287
665 287
670 287
675 287
680 286
685 287
690 287
695 287
700 287
705 286
710 287
715 287
720 287
725 286
730 286
735 286
740 287
745 286
750 287
755 288
760 284
765 287
770 286
775 285
780 286
785 286
790 285
795 287
800 287
805 286
810 286
815 286
820 286
825 286
830 287
835 288
840 287
845 287
850 287
855 286
860 286
865 286
870 286
875 287
880 286
885 285
890 286
895 286
900 287
905 285
910 285
915 287
920 287
925 286
930 287
935 285
940 286
945 286
950 286
955 287
960 286
965 286
970 286
975 285
980 286
985 286
990 287
995 285
1000 286
1005 286
1010 286
1015 285
1020 285
1025 287
1030 286
1035 287
1040 286
1045 287
1050 286
1055 286
1060 285
1065 287
1070 287
1075 286
1080 287
1085 286
1090 287
1095 287
1100 286
1105 286
1110 286
1115 286
1120 286
1125 287
1130 286
1135 287
1140 286
1145 286
1150 287
1155 287
1160 287
1165 286
1170 286
1175 287
1180 287
1185 287
1190 285
1195 286
1200 286
1205 287
1210 286
1215 286
1220 287
1225 286
1230 285
1235 285
1240 286
1245 286
1250 287
1255 287
1260 286
1265 287
1270 286
1275 287
1280 287
1285 287
1290 287
1295 286
1300 287
1305 286
1310 286
1315 285
1320 286
1325 286
1330 288
1335 287
1340 286
1345 286
1350 287
1355 288
1360 287
1365 286
1370 285
1375 286
1380 285
1385 286
1390 287
1395 286
1400 287
1405 287
1410 287
1415 287
1420 286
1425 286
1430 286
1435 286
1440 287
1445 286
1450 287
1455 286
1460 285
1465 286
1470 286
1475 286
1480 285
1485 286
1490 286
1495 286
1500 286
1505 286
1510 287
1515 285
1520 286
1525 286
1530 286
1535 286
1540 287
1545 285
1550 285
1555 285
1560 285
1565 286
1570 285
1575 286
1580 285
1585 286
1590 285
1595 287
1600 286
1605 285
1610 286
1615 285
1620 286
1625 286
1630 286
1635 285
1640 286
1645 285
1650 285
1655 286
1660 286
1665 285
1670 285
1675 285
1680 286
1685 285
1690 286
1695 285
1700 286
1705 285
1710 285
1715 285
1720 285
1725 286
1730 287
1735 286
1740 287
1745 286
1750 286
1755 287
1760 286
1765 286
1770 286
1775 286
1780 286
1785 286
1790 287
1795 287
1800 286
1805 286
1810 286
1815 285
1820 286
1825 286
1830 286
1835 286
1840 287
1845 286
1850 286
1855 285
1860 285
1865 285
1870 285
1875 285
1880 285
1885 285
1890 286
1895 286
1900 286
1905 286
1910 286
1915 286
1920 286
1925 286
1930 286
1935 286
1940 286
1945 286
1950 285
1955 287
1960 287
1965 286
1970 285
1975 285
1980 286
1985 285
1990 285
1995 285
2000 285
2005 286
2010 286
2015 286
2020 286
2025 286
2030 285
2035 285
2040 285
2045 286
2050 287
2055 286
2060 285
2065 286
2070 286
2075 286
2080 285
2085 285
2090 285
2095 286
2100 286
2105 285
2110 286
2115 286
2120 287
2125 285
2130 285
2135 286
2140 286
2145 285
2150 286
2155 286
2160 285
2165 286
2170 286
2175 286
2180 286
2185 286
2190 285
2195 285
2200 285
2205 286
2210 286
2215 285
2220 286
2225 286
2230 285
2235 286
2240 286
2245 287
2250 285
2255 285
2260 286
2265 286
2270 286
2275 286
2280 286
2285 286
2290 285
2295 285
2300 285
2305 285
2310 285
2315 285
2320 286
2325 286
2330 286
2335 287
2340 287
2345 286
2350 285
2355 286
2360 286
2365 286
2370 285
2375 285
2380 286
2385 286
2390 287
2395 287
2400 286
2405 286
2410 286
2415 288
2420 286
2425 286
2430 286
2435 286
2440 286
2445 286
2450 285
2455 286
2460 287
2465 286
2470 285
2475 286
2480 286
2485 286
2490 287
2495 287
2500 287
2505 287
2510 286
2515 285
2520 287
2525 286
2530 285
2535 286
2540 286
2545 286
2550 286
2555 286
2560 286
2565 286
2570 285
2575 287
2580 286
2585 285
2590 285
2595 286
2600 286
2605 286
2610 286
2615 286
2620 285
2625 287
2630 285
2635 286
2640 285
2645 286
2650 286
2655 285
2660 286
2665 286
2670 286
2675 285
2680 286
2685 285
2690 285
2695 285
2700 286
2705 285
2710 285
2715 285
2720 286
2725 286
2730 285
2735 285
2740 286
2745 285
2750 287
2755 285
2760 285
2765 285
2770 286
2775 286
2780 285
2785 285
2790 281
2795 260
2800 242
2805 239
2810 228
2815 218
2820 219
2825 219
2830 218
2835 215
2840 214
2845 210
2850 206
2855 209
2860 206
2865 206
2870 207
2875 206
2880 206
2885 205
2890 206
2895 206
2900 206
2905 206
2910 206
2915 205
2920 205
2925 206
2930 206
2935 205
2940 205
2945 205
2950 206
2955 205
2960 206
2965 205
2970 205
2975 205
2980 205
2985 205
2990 205
2995 205
3000 205
3005 205
3010 205
3015 206
3020 204
3025 205
3030 205
3035 205
3040 205
3045 205
3050 205
3055 204
3060 205
3065 204
3070 205
3075 205
3080 205
3085 205
3090 205
3095 205
3100 205
3105 205
3110 204
3115 204
3120 205
3125 205
3130 204
3135 204
3140 205
3145 205
3150 204
3155 204
3160 204
3165 204
3170 205
3175 205
3180 204
3185 205
3190 205
3195 204
3200 205
3205 205
3210 204
3215 204
3220 205
3225 204
3230 205
3235 204
3240 205
3245 204
3250 204
3255 204
3260 204
3265 205
3270 204
3275 204
3280 204
3285 205
3290 204
3295 205
3300 204
3305 204
3310 204
3315 204
3320 205
3325 205
3330 204
3335 205
3340 204
3345 205
3350 204
3355 204
3360 205
3365 205
3370 205
3375 204
3380 204
3385 205
3390 205
3395 204
3400 204
3405 205
3410 205
3415 205
3420 204
3425 204
3430 205
3435 205
3440 205
3445 205
3450 204
3455 205
3460 204
3465 204
3470 204
3475 204
3480 204
3485 204
3490 204
3495 205
3500 205
3505 204
3510 204
3515 204
3520 204
3525 204
3530 205
3535 204
3540 205
3545 204
3550 205
3555 205
3560 204
3565 205
3570 203
3575 204
3580 205
3585 204
3590 204
3595 204
3600 204
3605 204
3610 205
3615 204
3620 205
3625 205
3630 203
3635 204
3640 204
3645 204
3650 204
3655 204
3660 204
3665 205
3670 204
3675 204
3680 204
3685 205
3690 204
3695 204
3700 204
3705 204
3710 204
3715 205
3720 204
3725 205
3730 204
3735 204
3740 205
3745 205
3750 204
3755 207
3760 204
3765 204
3770 204
3775 204
3780 204
3785 205
3790 204
3795 204
3800 205
3805 204
3810 205
3815 205
3820 204
3825 205
3830 204
3835 204
3840 204
3845 204
3850 204
3855 205
3860 204
3865 203
3870 205
3875 205
3880 204
3885 205
3890 205
3895 205
3900 204
3905 204
3910 205
3915 204
3920 204
3925 205
3930 204
3935 204
3940 204
3945 204
3950 204
3955 204
3960 204
3965 204
3970 204
3975 204
3980 204
3985 205
3990 204
3995 204
4000 204
4005 204
4010 204
4015 204
4020 204
4025 204
4030 204
4035 205
4040 205
4045 204
4050 205
4055 207
4060 204
4065 204
4070 204
4075 205
4080 205
4085 204
4090 205
4095 204
4100 204
4105 204
4110 205
4115 204
4120 206
4125 205
4130 205
4135 204
4140 204
4145 204
4150 204
4155 204
4160 205
4165 205
4170 204
4175 204
4180 204
4185 205
4190 204
4195 204
4200 205
4205 204
4210 204
4215 204
4220 204
4225 204
4230 204
4235 204
4240 204
4245 204
4250 204
4255 205
4260 205
4265 205
4270 204
4275 204
4280 204
4285 204
4290 204
4295 204
4300 204
4305 205
4310 204
4315 204
4320 205
4325 204
4330 204
4335 205
4340 205
4345 204
4350 204
4355 205
4360 205
4365 204
4370 205
4375 204
4380 204
4385 205
4390 204
4395 204
4400 205
4405 205
4410 205
4415 204
4420 205
4425 205
4430 205
4435 205
4440 204
4445 204
4450 204
4455 204
4460 204
4465 205
4470 205
4475 204
4480 205
4485 205
4490 205
4495 204
4500 205
4505 204
4510 204
4515 204
4520 205
4525 204
4530 207
4535 204
4540 204
4545 204
4550 204
4555 205
4560 204
4565 204
4570 205
4575 205
4580 205
4585 205
4590 204
4595 204
4600 204
4605 204
4610 205
4615 205
4620 205
4625 205
4630 204
4635 204
4640 204
4645 204
4650 204
4655 207
4660 205
4665 205
4670 204
4675 204
4680 204
4685 205
4690 205
4695 204
4700 205
4705 205
4710 204
4715 205
4720 204
4725 205
4730 205
4735 204
4740 205
4745 204
4750 204
4755 204
4760 204
4765 204
4770 204
4775 205
4780 205
4785 204
4790 205
4795 204
4800 204
4805 204
4810 204
4815 204
4820 204
4825 204
4830 204
4835 204
4840 204
4845 205
4850 205
4855 205
4860 205
4865 204
4870 205
4875 204
4880 204
4885 204
4890 204
4895 205
4900 205
4905 204
4910 205
4915 205
4920 205
4925 205
4930 204
4935 204
4940 204
4945 204
4950 204
4955 205
4960 204
4965 204
4970 204
4975 204
4980 204
4985 204
4990 204
4995 204
5000 204
5005 204
5010 204
5015 204
5020 204
5025 204
5030 204
5035 204
5040 204
5045 204
5050 204
5055 204
5060 205
5065 204
5070 204
5075 205
5080 204
5085 205
5090 204
5095 204
5100 204
5105 205
5110 205
5115 204
5120 205
5125 204
5130 204
5135 204
5140 205
5145 204
5150 204
5155 204
5160 204
5165 205
5170 204
5175 204
5180 205
5185 204
5190 204
5195 204
5200 205
5205 204
5210 204
5215 205
5220 204
5225 204
5230 204
5235 205
5240 204
5245 204
5250 204
5255 205
5260 205
5265 205
5270 204
5275 204
5280 205
5285 204
5290 205
5295 204
5300 204
5305 204
5310 205
5315 204
5320 204
5325 204
5330 204
5335 204
5340 206
5345 204
5350 204
5355 204
5360 204
5365 204
5370 204
5375 204
5380 205
5385 205
5390 204
5395 205
5400 205
5405 204
5410 204
5415 205
5420 204
5425 205
5430 204
5435 204
5440 204
5445 204
5450 205
5455 205
5460 205
5465 204
5470 204
5475 204
5480 204
5485 204
5490 204
5495 204
5500 204
5505 205
5510 205
5515 204
5520 204
5525 205
5530 204
5535 204
5540 204
5545 204
5550 204
5555 204
5560 204
5565 204
5570 205
5575 205
5580 204
5585 204
5590 204
5595 205
5600 204
5605 204
5610 205
5615 204
5620 205
5625 204
5630 205
5635 204
5640 204
5645 204
5650 204
5655 205
5660 204
5665 204
5670 204
5675 204
5680 204
5685 205
5690 204
5695 204
5700 204
5705 204
5710 204
5715 204
5720 204
5725 205
5730 205
5735 204
5740 204
5745 204
5750 204
5755 204
5760 204
5765 204
5770 205
5775 204
5780 204
5785 204
5790 204
5795 204
5800 204
5805 204
5810 204
5815 204
5820 204
5825 204
5830 204
5835 205
5840 204
5845 205
5850 205
5855 205
5860 204
5865 205
5870 205
5875 204
5880 205
5885 204
5890 205
5895 204
5900 204
5905 204
5910 205
5915 204
5920 204
5925 204
5930 204
5935 205
5940 204
5945 204
5950 205
5955 204
5960 204
5965 205
5970 204
5975 205
5980 205
5985 204
5990 204
5995 204
6000 204
};
\addlegendentry{2 jobs}
\addplot [thick, color2, mark=square*, mark size=3, mark options={solid,fill=white,draw=red},  mark repeat={180}]
table {%
0 206
5 214
10 223
15 225
20 227
25 227
30 223
35 224
40 209
45 206
50 206
55 206
60 207
65 206
70 207
75 205
80 206
85 206
90 206
95 207
100 208
105 208
110 208
115 223
120 224
125 224
130 225
135 224
140 219
145 217
150 216
155 216
160 215
165 216
170 222
175 222
180 218
185 227
190 224
195 298
200 303
205 303
210 303
215 303
220 304
225 305
230 306
235 308
240 309
245 310
250 312
255 311
260 312
265 312
270 313
275 313
280 313
285 313
290 318
295 316
300 316
305 316
310 316
315 316
320 316
325 313
330 315
335 316
340 317
345 316
350 316
355 315
360 315
365 314
370 313
375 313
380 314
385 315
390 314
395 315
400 315
405 314
410 314
415 314
420 315
425 315
430 315
435 315
440 315
445 315
450 315
455 314
460 314
465 315
470 312
475 315
480 315
485 314
490 315
495 315
500 314
505 312
510 315
515 315
520 315
525 315
530 316
535 315
540 315
545 314
550 315
555 315
560 315
565 315
570 315
575 315
580 315
585 314
590 315
595 316
600 315
605 316
610 315
615 315
620 313
625 316
630 316
635 316
640 314
645 315
650 315
655 315
660 315
665 315
670 315
675 315
680 314
685 315
690 316
695 315
700 315
705 316
710 315
715 315
720 316
725 315
730 315
735 314
740 315
745 315
750 315
755 315
760 316
765 315
770 316
775 314
780 315
785 314
790 315
795 315
800 315
805 315
810 315
815 315
820 314
825 315
830 315
835 315
840 314
845 315
850 315
855 315
860 315
865 315
870 315
875 316
880 315
885 314
890 316
895 316
900 316
905 316
910 316
915 316
920 316
925 316
930 316
935 316
940 316
945 316
950 316
955 316
960 315
965 315
970 316
975 316
980 316
985 316
990 316
995 315
1000 316
1005 316
1010 316
1015 315
1020 316
1025 315
1030 316
1035 316
1040 315
1045 316
1050 315
1055 316
1060 316
1065 316
1070 316
1075 316
1080 315
1085 315
1090 316
1095 316
1100 316
1105 316
1110 316
1115 316
1120 316
1125 316
1130 316
1135 316
1140 316
1145 316
1150 316
1155 316
1160 315
1165 316
1170 316
1175 315
1180 315
1185 316
1190 316
1195 316
1200 316
1205 316
1210 316
1215 316
1220 316
1225 316
1230 316
1235 316
1240 316
1245 316
1250 316
1255 317
1260 316
1265 316
1270 316
1275 315
1280 315
1285 316
1290 316
1295 316
1300 316
1305 315
1310 316
1315 315
1320 315
1325 316
1330 316
1335 316
1340 315
1345 315
1350 315
1355 315
1360 316
1365 314
1370 316
1375 315
1380 313
1385 315
1390 315
1395 315
1400 316
1405 315
1410 316
1415 316
1420 315
1425 316
1430 315
1435 316
1440 313
1445 313
1450 315
1455 316
1460 315
1465 316
1470 316
1475 315
1480 315
1485 314
1490 316
1495 320
1500 315
1505 316
1510 318
1515 316
1520 317
1525 317
1530 316
1535 317
1540 315
1545 315
1550 317
1555 317
1560 316
1565 317
1570 316
1575 317
1580 316
1585 317
1590 317
1595 317
1600 316
1605 316
1610 316
1615 317
1620 316
1625 317
1630 317
1635 316
1640 316
1645 318
1650 317
1655 316
1660 317
1665 318
1670 318
1675 316
1680 315
1685 317
1690 319
1695 316
1700 317
1705 318
1710 317
1715 318
1720 318
1725 317
1730 317
1735 317
1740 317
1745 318
1750 318
1755 318
1760 317
1765 316
1770 317
1775 317
1780 318
1785 317
1790 318
1795 318
1800 317
1805 318
1810 316
1815 318
1820 317
1825 318
1830 318
1835 317
1840 317
1845 318
1850 317
1855 318
1860 318
1865 318
1870 317
1875 318
1880 316
1885 317
1890 318
1895 318
1900 317
1905 317
1910 317
1915 316
1920 317
1925 317
1930 317
1935 317
1940 316
1945 318
1950 317
1955 317
1960 318
1965 316
1970 317
1975 316
1980 316
1985 317
1990 316
1995 318
2000 316
2005 318
2010 318
2015 317
2020 317
2025 316
2030 317
2035 316
2040 318
2045 317
2050 318
2055 317
2060 318
2065 317
2070 318
2075 318
2080 317
2085 318
2090 316
2095 316
2100 318
2105 317
2110 318
2115 317
2120 317
2125 317
2130 317
2135 317
2140 316
2145 316
2150 318
2155 317
2160 316
2165 318
2170 317
2175 317
2180 318
2185 318
2190 316
2195 317
2200 317
2205 317
2210 316
2215 316
2220 317
2225 316
2230 316
2235 316
2240 317
2245 316
2250 317
2255 316
2260 316
2265 316
2270 316
2275 317
2280 316
2285 317
2290 316
2295 316
2300 316
2305 316
2310 318
2315 317
2320 318
2325 318
2330 317
2335 317
2340 319
2345 316
2350 317
2355 317
2360 318
2365 318
2370 318
2375 317
2380 318
2385 319
2390 318
2395 318
2400 318
2405 318
2410 318
2415 318
2420 319
2425 318
2430 318
2435 318
2440 319
2445 318
2450 318
2455 319
2460 319
2465 318
2470 317
2475 319
2480 318
2485 318
2490 319
2495 319
2500 318
2505 319
2510 318
2515 318
2520 318
2525 319
2530 318
2535 319
2540 318
2545 319
2550 318
2555 319
2560 319
2565 318
2570 318
2575 318
2580 318
2585 319
2590 319
2595 318
2600 319
2605 318
2610 318
2615 319
2620 319
2625 319
2630 318
2635 319
2640 318
2645 319
2650 318
2655 318
2660 318
2665 318
2670 318
2675 318
2680 317
2685 318
2690 318
2695 318
2700 318
2705 318
2710 318
2715 318
2720 318
2725 318
2730 318
2735 318
2740 317
2745 318
2750 318
2755 318
2760 318
2765 319
2770 318
2775 317
2780 317
2785 318
2790 317
2795 317
2800 316
2805 318
2810 318
2815 317
2820 318
2825 318
2830 316
2835 316
2840 317
2845 317
2850 318
2855 317
2860 316
2865 316
2870 316
2875 317
2880 317
2885 316
2890 316
2895 317
2900 317
2905 316
2910 317
2915 316
2920 317
2925 317
2930 318
2935 317
2940 316
2945 317
2950 317
2955 318
2960 317
2965 317
2970 316
2975 316
2980 317
2985 317
2990 316
2995 316
3000 317
3005 316
3010 317
3015 317
3020 317
3025 318
3030 318
3035 317
3040 318
3045 318
3050 317
3055 317
3060 318
3065 318
3070 317
3075 318
3080 318
3085 317
3090 316
3095 317
3100 318
3105 318
3110 317
3115 318
3120 318
3125 317
3130 316
3135 317
3140 318
3145 318
3150 316
3155 318
3160 318
3165 317
3170 316
3175 318
3180 318
3185 319
3190 318
3195 318
3200 317
3205 318
3210 318
3215 319
3220 318
3225 317
3230 317
3235 317
3240 318
3245 317
3250 318
3255 317
3260 318
3265 318
3270 317
3275 317
3280 317
3285 318
3290 316
3295 318
3300 316
3305 317
3310 317
3315 317
3320 316
3325 317
3330 317
3335 316
3340 316
3345 317
3350 316
3355 317
3360 316
3365 317
3370 316
3375 316
3380 316
3385 315
3390 316
3395 316
3400 316
3405 316
3410 315
3415 316
3420 313
3425 316
3430 316
3435 316
3440 316
3445 316
3450 315
3455 316
3460 316
3465 317
3470 315
3475 317
3480 316
3485 315
3490 316
3495 316
3500 316
3505 316
3510 315
3515 316
3520 315
3525 315
3530 315
3535 316
3540 315
3545 315
3550 316
3555 316
3560 316
3565 316
3570 316
3575 316
3580 316
3585 316
3590 316
3595 316
3600 316
3605 316
3610 316
3615 316
3620 316
3625 315
3630 316
3635 316
3640 317
3645 317
3650 316
3655 316
3660 316
3665 316
3670 315
3675 316
3680 316
3685 316
3690 316
3695 316
3700 315
3705 317
3710 316
3715 316
3720 316
3725 316
3730 316
3735 316
3740 316
3745 316
3750 318
3755 317
3760 317
3765 317
3770 317
3775 317
3780 318
3785 317
3790 318
3795 317
3800 319
3805 318
3810 318
3815 318
3820 317
3825 317
3830 317
3835 317
3840 318
3845 318
3850 318
3855 318
3860 318
3865 318
3870 318
3875 317
3880 318
3885 318
3890 318
3895 318
3900 316
3905 318
3910 317
3915 318
3920 318
3925 318
3930 319
3935 319
3940 319
3945 318
3950 319
3955 318
3960 318
3965 318
3970 319
3975 318
3980 318
3985 319
3990 319
3995 319
4000 319
4005 320
4010 319
4015 319
4020 318
4025 320
4030 318
4035 319
4040 318
4045 318
4050 319
4055 319
4060 319
4065 318
4070 317
4075 318
4080 317
4085 318
4090 317
4095 318
4100 318
4105 317
4110 317
4115 318
4120 319
4125 318
4130 318
4135 318
4140 318
4145 318
4150 317
4155 314
4160 314
4165 312
4170 312
4175 307
4180 308
4185 304
4190 302
4195 300
4200 295
4205 293
4210 289
4215 282
4220 276
4225 275
4230 261
4235 239
4240 230
4245 226
4250 220
4255 217
4260 216
4265 217
4270 213
4275 212
4280 212
4285 207
4290 206
4295 205
4300 205
4305 206
4310 206
4315 206
4320 205
4325 206
4330 206
4335 206
4340 205
4345 206
4350 205
4355 205
4360 205
4365 204
4370 206
4375 205
4380 205
4385 205
4390 205
4395 205
4400 205
4405 205
4410 205
4415 205
4420 205
4425 204
4430 205
4435 205
4440 205
4445 204
4450 205
4455 205
4460 205
4465 205
4470 205
4475 206
4480 206
4485 205
4490 205
4495 205
4500 204
4505 205
4510 205
4515 205
4520 204
4525 204
4530 204
4535 204
4540 204
4545 204
4550 204
4555 204
4560 204
4565 203
4570 204
4575 204
4580 204
4585 204
4590 204
4595 204
4600 204
4605 204
4610 204
4615 204
4620 204
4625 203
4630 204
4635 204
4640 204
4645 205
4650 204
4655 204
4660 204
4665 204
4670 203
4675 204
4680 204
4685 204
4690 204
4695 204
4700 204
4705 204
4710 203
4715 204
4720 204
4725 203
4730 204
4735 204
4740 205
4745 204
4750 203
4755 204
4760 204
4765 203
4770 204
4775 203
4780 204
4785 204
4790 204
4795 204
4800 204
4805 203
4810 203
4815 204
4820 204
4825 204
4830 204
4835 203
4840 204
4845 203
4850 204
4855 203
4860 203
4865 203
4870 203
4875 203
4880 203
4885 204
4890 204
4895 204
4900 203
4905 204
4910 203
4915 204
4920 204
4925 204
4930 203
4935 203
4940 203
4945 204
4950 204
4955 204
4960 203
4965 204
4970 204
4975 203
4980 203
4985 204
4990 203
4995 203
5000 204
5005 204
5010 204
5015 203
5020 204
5025 204
5030 204
5035 204
5040 204
5045 204
5050 203
5055 204
5060 204
5065 203
5070 204
5075 204
5080 203
5085 204
5090 204
5095 204
5100 204
5105 204
5110 203
5115 204
5120 204
5125 203
5130 203
5135 204
5140 204
5145 204
5150 203
5155 203
5160 203
5165 204
5170 204
5175 204
5180 204
5185 204
5190 204
5195 204
5200 203
5205 203
5210 202
5215 204
5220 203
5225 204
5230 204
5235 204
5240 205
5245 203
5250 204
5255 204
5260 204
5265 204
5270 203
5275 204
5280 203
5285 204
5290 204
5295 204
5300 204
5305 203
5310 205
5315 204
5320 204
5325 204
5330 204
5335 203
5340 203
5345 203
5350 204
5355 204
5360 204
5365 204
5370 204
5375 203
5380 203
5385 205
5390 204
5395 203
5400 204
5405 204
5410 203
5415 204
5420 203
5425 203
5430 204
5435 204
5440 204
5445 203
5450 204
5455 204
5460 204
5465 204
5470 205
5475 204
5480 204
5485 204
5490 204
5495 204
5500 203
5505 204
5510 203
5515 203
5520 204
5525 205
5530 203
5535 204
5540 204
5545 204
5550 203
5555 203
5560 204
5565 204
5570 204
5575 204
5580 203
5585 204
5590 204
5595 204
5600 204
5605 204
5610 206
5615 203
5620 204
5625 204
5630 204
5635 203
5640 203
5645 203
5650 202
5655 204
5660 204
5665 205
5670 204
5675 204
5680 203
5685 204
5690 203
5695 203
5700 203
5705 204
5710 203
5715 203
5720 203
5725 203
5730 204
5735 203
5740 203
5745 203
5750 204
5755 204
5760 203
5765 203
5770 203
5775 203
5780 203
5785 204
5790 204
5795 204
5800 203
};
\addlegendentry{4 jobs}
\end{axis}

\end{tikzpicture}

%% file: results/power_intel_sockets.tex
\begin{tikzpicture}[font=\Large]

\definecolor{color0}{rgb}{0.12156862745098,0.466666666666667,0.705882352941177}
\definecolor{color1}{rgb}{1,0.498039215686275,0.0549019607843137}
\definecolor{color2}{rgb}{0.172549019607843,0.627450980392157,0.172549019607843}
\definecolor{color3}{rgb}{0.83921568627451,0.152941176470588,0.156862745098039}
\definecolor{color4}{rgb}{0.580392156862745,0.403921568627451,0.741176470588235}
\definecolor{color5}{rgb}{0,0,0}

\begin{axis}[
legend cell align={left},
legend columns=2,
legend style={fill opacity=0.8, draw opacity=1, text opacity=1, at={(1.05,1.15)}, anchor=east, draw=white!80.0!black},
tick align=outside,
tick pos=left,
x grid style={white!69.01960784313725!black},
xlabel={Time (SEC)},
xmin=0, xmax=3600,
xtick={0,900,1800,2700,3600},
xtick style={color=black},
y grid style={white!69.01960784313725!black},
ylabel={Power consumption (W)},
ymin=150, ymax=500,
ytick={150,200,250,300,350,400,450,500},
xmajorgrids,
ymajorgrids,
ytick style={color=black}
]
\addplot [thick, color5, mark=.]
table {%
0 254
5 253
10 254
15 253
20 253
25 253
30 253
35 253
40 253
45 253
50 253
55 254
60 253
65 254
70 253
75 252
80 254
85 254
90 254
95 254
100 254
105 253
110 253
115 253
120 253
125 253
130 254
135 253
140 254
145 252
150 254
155 254
160 253
165 254
170 253
175 252
180 253
185 253
190 253
195 253
200 253
205 253
210 263
215 251
220 253
225 252
230 252
235 253
240 252
245 253
250 253
255 253
260 253
265 252
270 252
275 254
280 252
285 252
290 252
295 253
300 254
305 252
310 253
315 253
320 253
325 252
330 252
335 253
340 252
345 254
350 253
355 253
360 253
365 253
370 253
375 252
380 252
385 252
390 253
395 252
400 252
405 254
410 252
415 253
420 252
425 254
430 254
435 253
440 254
445 252
450 252
455 252
460 253
465 251
470 250
475 253
480 253
485 253
490 254
495 253
500 253
505 252
510 252
515 254
520 252
525 253
530 254
535 253
540 253
545 253
550 253
555 254
560 254
565 252
570 253
575 253
580 253
585 254
590 252
595 253
600 252
605 252
610 252
615 253
620 253
625 252
630 253
635 252
640 252
645 252
650 253
655 254
660 253
665 254
670 253
675 252
680 254
685 252
690 252
695 252
700 253
705 254
710 253
715 254
720 254
725 254
730 252
735 253
740 253
745 253
750 253
755 254
760 252
765 250
770 253
775 253
780 253
785 253
790 253
795 253
800 254
805 253
810 253
815 252
820 253
825 253
830 252
835 254
840 253
845 253
850 252
855 252
860 253
865 252
870 253
875 253
880 253
885 254
890 252
895 253
900 253
905 250
910 253
915 252
920 253
925 253
930 253
935 254
940 254
945 253
950 254
955 253
960 254
965 254
970 253
975 253
980 254
985 253
990 253
995 252
1000 253
1005 254
1010 253
1015 253
1020 253
1025 252
1030 254
1035 253
1040 253
1045 252
1050 254
1055 253
1060 252
1065 253
1070 252
1075 254
1080 253
1085 253
1090 252
1095 253
1100 252
1105 252
1110 253
1115 253
1120 254
1125 252
1130 254
1135 252
1140 253
1145 253
1150 252
1155 253
1160 254
1165 253
1170 251
1175 253
1180 252
1185 253
1190 253
1195 253
1200 253
1205 252
1210 252
1215 253
1220 252
1225 253
1230 254
1235 254
1240 253
1245 253
1250 253
1255 253
1260 252
1265 253
1270 254
1275 254
1280 254
1285 252
1290 254
1295 252
1300 252
1305 254
1310 252
1315 254
1320 254
1325 254
1330 253
1335 254
1340 252
1345 252
1350 253
1355 254
1360 254
1365 253
1370 253
1375 254
1380 263
1385 251
1390 254
1395 254
1400 253
1405 253
1410 254
1415 251
1420 251
1425 252
1430 251
1435 251
1440 251
1445 251
1450 251
1455 252
1460 251
1465 251
1470 251
1475 251
1480 252
1485 251
1490 252
1495 251
1500 251
1505 252
1510 251
1515 251
1520 251
1525 251
1530 251
1535 251
1540 251
1545 251
1550 251
1555 251
1560 251
1565 249
1570 251
1575 251
1580 251
1585 251
1590 252
1595 251
1600 252
1605 252
1610 251
1615 252
1620 251
1625 251
1630 251
1635 251
1640 252
1645 251
1650 252
1655 252
1660 254
1665 252
1670 252
1675 253
1680 253
1685 254
1690 252
1695 253
1700 252
1705 252
1710 252
1715 253
1720 253
1725 253
1730 252
1735 252
1740 252
1745 253
1750 252
1755 253
1760 254
1765 254
1770 253
1775 253
1780 254
1785 254
1790 253
1795 253
1800 252
1805 254
1810 252
1815 253
1820 253
1825 254
1830 252
1835 252
1840 253
1845 250
1850 252
1855 253
1860 252
1865 253
1870 252
1875 253
1880 253
1885 252
1890 253
1895 253
1900 253
1905 252
1910 252
1915 252
1920 252
1925 253
1930 252
1935 253
1940 254
1945 253
1950 253
1955 253
1960 253
1965 253
1970 253
1975 253
1980 252
1985 253
1990 250
1995 253
2000 253
2005 254
2010 253
2015 253
2020 253
2025 252
2030 252
2035 253
2040 253
2045 252
2050 253
2055 252
2060 253
2065 254
2070 253
2075 253
2080 252
2085 252
2090 253
2095 253
2100 252
2105 253
2110 253
2115 253
2120 252
2125 252
2130 253
2135 254
2140 253
2145 252
2150 252
2155 252
2160 254
2165 253
2170 252
2175 253
2180 253
2185 252
2190 253
2195 254
2200 254
2205 253
2210 254
2215 253
2220 254
2225 253
2230 254
2235 253
2240 253
2245 252
2250 252
2255 254
2260 253
2265 253
2270 254
2275 253
2280 253
2285 253
2290 254
2295 253
2300 253
2305 253
2310 253
2315 254
2320 254
2325 252
2330 254
2335 252
2340 253
2345 252
2350 253
2355 253
2360 253
2365 254
2370 254
2375 252
2380 252
2385 253
2390 263
2395 254
2400 254
2405 253
2410 253
2415 254
2420 252
2425 251
2430 253
2435 254
2440 254
2445 252
2450 254
2455 254
2460 253
2465 253
2470 253
2475 253
2480 254
2485 254
2490 253
2495 253
2500 253
2505 253
2510 254
2515 253
2520 254
2525 253
2530 254
2535 254
2540 253
2545 254
2550 254
2555 254
2560 254
2565 253
2570 253
2575 252
2580 254
2585 254
2590 253
2595 254
2600 252
2605 254
2610 253
2615 253
2620 254
2625 253
2630 253
2635 253
2640 253
2645 252
2650 253
2655 253
2660 253
2665 253
2670 252
2675 253
2680 252
2685 253
2690 253
2695 252
2700 253
2705 253
2710 253
2715 253
2720 252
2725 254
2730 252
2735 253
2740 253
2745 254
2750 253
2755 253
2760 253
2765 254
2770 254
2775 252
2780 253
2785 253
2790 253
2795 253
2800 257
2805 251
2810 254
2815 253
2820 253
2825 252
2830 253
2835 252
2840 253
2845 253
2850 253
2855 253
2860 264
2865 251
2870 253
2875 254
2880 253
2885 253
2890 253
2895 253
2900 254
2905 254
2910 252
2915 254
2920 254
2925 253
2930 253
2935 253
2940 253
2945 253
2950 254
2955 253
2960 252
2965 254
2970 252
2975 254
2980 254
2985 254
2990 253
2995 253
3000 253
3005 253
3010 254
3015 253
3020 253
3025 253
3030 253
3035 253
3040 253
3045 253
3050 253
3055 254
3060 253
3065 253
3070 253
3075 253
3080 252
3085 251
3090 252
3095 252
3100 252
3105 251
3110 251
3115 252
3120 252
3125 251
3130 253
3135 252
3140 252
3145 252
3150 252
3155 252
3160 251
3165 251
3170 252
3175 252
3180 251
3185 252
3190 252
3195 252
3200 251
3205 251
3210 252
3215 251
3220 252
3225 252
3230 252
3235 252
3240 251
3245 252
3250 252
3255 252
3260 251
3265 252
3270 252
3275 252
3280 252
3285 252
3290 251
3295 252
3300 251
3305 252
3310 252
3315 251
3320 251
3325 251
3330 252
3335 252
3340 252
3345 252
3350 251
3355 251
3360 252
3365 252
3370 252
3375 252
3380 252
3385 251
3390 251
3395 252
3400 251
3405 251
3410 252
3415 251
3420 252
3425 252
3430 252
3435 252
3440 252
3445 251
3450 252
3455 252
3460 252
3465 251
3470 252
3475 252
3480 251
3485 252
3490 252
3495 251
3500 252
3505 252
3510 252
3515 252
3520 251
3525 252
3530 252
3535 252
3540 252
3545 251
3550 251
3555 252
3560 251
3565 251
3570 251
3575 251
3580 251
3585 251
3590 251
3595 252
3600 252
};
\addlegendentry{Baseline}
\addplot [thick, color0, mark=triangle*, mark size=3, mark options={solid,fill=white,draw=red},  mark repeat={180}]
table {%
0 254
5 263
10 266
15 269
20 270
25 261
30 266
35 254
40 253
45 253
50 255
55 254
60 253
65 254
70 254
75 253
80 253
85 253
90 255
95 254
100 256
105 256
110 267
115 268
120 268
125 266
130 264
135 262
140 263
145 263
150 255
155 264
160 255
165 255
170 254
175 255
180 254
185 256
190 255
195 255
200 268
205 266
210 266
215 258
220 268
225 260
230 353
235 355
240 355
245 356
250 355
255 356
260 357
265 357
270 357
275 357
280 357
285 356
290 358
295 359
300 359
305 358
310 361
315 359
320 362
325 362
330 362
335 360
340 362
345 361
350 362
355 361
360 362
365 362
370 362
375 362
380 363
385 363
390 363
395 362
400 362
405 363
410 364
415 364
420 363
425 365
430 364
435 365
440 364
445 365
450 365
455 366
460 366
465 366
470 365
475 365
480 365
485 366
490 366
495 366
500 366
505 361
510 365
515 363
520 365
525 363
530 364
535 362
540 364
545 364
550 363
555 365
560 363
565 365
570 364
575 365
580 364
585 363
590 364
595 365
600 363
605 365
610 362
615 376
620 366
625 367
630 365
635 367
640 366
645 366
650 368
655 366
660 366
665 365
670 367
675 367
680 366
685 367
690 366
695 365
700 368
705 368
710 368
715 366
720 368
725 366
730 367
735 366
740 367
745 367
750 367
755 367
760 367
765 368
770 365
775 369
780 368
785 367
790 368
795 367
800 367
805 368
810 367
815 368
820 367
825 368
830 366
835 367
840 368
845 368
850 367
855 367
860 366
865 368
870 367
875 366
880 368
885 367
890 367
895 367
900 367
905 367
910 366
915 367
920 366
925 368
930 367
935 367
940 367
945 368
950 366
955 367
960 366
965 366
970 366
975 364
980 367
985 367
990 366
995 368
1000 366
1005 368
1010 368
1015 367
1020 367
1025 368
1030 368
1035 366
1040 367
1045 367
1050 369
1055 367
1060 368
1065 368
1070 368
1075 367
1080 368
1085 367
1090 368
1095 367
1100 367
1105 367
1110 369
1115 367
1120 367
1125 367
1130 368
1135 367
1140 367
1145 377
1150 365
1155 368
1160 367
1165 367
1170 369
1175 368
1180 369
1185 368
1190 368
1195 369
1200 368
1205 368
1210 369
1215 367
1220 368
1225 369
1230 368
1235 368
1240 369
1245 368
1250 368
1255 368
1260 369
1265 368
1270 368
1275 367
1280 368
1285 367
1290 368
1295 368
1300 369
1305 367
1310 369
1315 368
1320 368
1325 367
1330 367
1335 368
1340 366
1345 368
1350 368
1355 368
1360 367
1365 368
1370 366
1375 368
1380 367
1385 366
1390 365
1395 366
1400 366
1405 367
1410 366
1415 367
1420 366
1425 367
1430 366
1435 367
1440 365
1445 366
1450 366
1455 367
1460 366
1465 367
1470 366
1475 366
1480 367
1485 366
1490 368
1495 367
1500 368
1505 367
1510 366
1515 366
1520 368
1525 365
1530 367
1535 367
1540 367
1545 365
1550 367
1555 366
1560 364
1565 366
1570 366
1575 367
1580 366
1585 368
1590 368
1595 367
1600 367
1605 367
1610 366
1615 367
1620 367
1625 366
1630 367
1635 367
1640 367
1645 366
1650 366
1655 367
1660 367
1665 367
1670 368
1675 368
1680 366
1685 367
1690 367
1695 366
1700 368
1705 367
1710 366
1715 366
1720 368
1725 367
1730 368
1735 367
1740 368
1745 368
1750 367
1755 367
1760 367
1765 368
1770 368
1775 367
1780 367
1785 368
1790 367
1795 367
1800 367
1805 368
1810 366
1815 368
1820 368
1825 368
1830 367
1835 368
1840 367
1845 368
1850 367
1855 369
1860 368
1865 366
1870 367
1875 367
1880 368
1885 368
1890 368
1895 368
1900 367
1905 368
1910 367
1915 367
1920 367
1925 366
1930 366
1935 368
1940 367
1945 368
1950 367
1955 368
1960 367
1965 367
1970 368
1975 370
1980 367
1985 367
1990 367
1995 367
2000 368
2005 369
2010 368
2015 367
2020 368
2025 367
2030 367
2035 367
2040 367
2045 368
2050 367
2055 367
2060 368
2065 366
2070 366
2075 366
2080 367
2085 368
2090 367
2095 366
2100 368
2105 366
2110 367
2115 366
2120 368
2125 367
2130 366
2135 366
2140 367
2145 366
2150 368
2155 367
2160 366
2165 367
2170 367
2175 367
2180 367
2185 367
2190 366
2195 366
2200 366
2205 366
2210 367
2215 368
2220 366
2225 368
2230 368
2235 367
2240 368
2245 368
2250 367
2255 367
2260 368
2265 368
2270 368
2275 368
2280 368
2285 367
2290 368
2295 367
2300 368
2305 368
2310 369
2315 367
2320 368
2325 367
2330 369
2335 368
2340 369
2345 367
2350 368
2355 368
2360 368
2365 368
2370 367
2375 367
2380 367
2385 368
2390 367
2395 368
2400 369
2405 367
2410 368
2415 368
2420 369
2425 368
2430 368
2435 369
2440 367
2445 368
2450 368
2455 368
2460 368
2465 369
2470 369
2475 367
2480 370
2485 345
2490 314
2495 313
2500 290
2505 273
2510 274
2515 272
2520 273
2525 267
2530 270
2535 259
2540 259
2545 258
2550 259
2555 256
2560 256
2565 256
2570 257
2575 260
2580 256
2585 257
2590 256
2595 256
2600 255
2605 255
2610 256
2615 256
2620 255
2625 255
2630 256
2635 256
2640 255
2645 257
2650 255
2655 255
2660 255
2665 255
2670 255
2675 255
2680 255
2685 255
2690 254
2695 256
2700 256
2705 254
2710 255
2715 254
2720 255
2725 255
2730 255
2735 254
2740 256
2745 254
2750 255
2755 255
2760 256
2765 252
2770 254
2775 255
2780 254
2785 254
2790 254
2795 256
2800 254
2805 254
2810 255
2815 255
2820 254
2825 255
2830 253
2835 255
2840 253
2845 255
2850 254
2855 254
2860 253
2865 254
2870 253
2875 257
2880 254
2885 254
2890 254
2895 254
2900 255
2905 253
2910 254
2915 254
2920 254
2925 253
2930 253
2935 253
2940 255
2945 254
2950 255
2955 253
2960 255
2965 254
2970 254
2975 254
2980 254
2985 253
2990 254
2995 254
3000 254
3005 255
3010 253
3015 254
3020 253
3025 253
3030 253
3035 253
3040 253
3045 251
3050 253
3055 255
3060 253
3065 254
3070 253
3075 255
3080 254
3085 254
3090 254
3095 254
3100 254
3105 253
3110 255
3115 253
3120 253
3125 253
3130 253
3135 253
3140 254
3145 253
3150 254
3155 253
3160 255
3165 254
3170 254
3175 253
3180 253
3185 254
3190 254
3195 254
3200 253
3205 253
3210 253
3215 253
3220 254
3225 253
3230 254
3235 254
3240 253
3245 253
3250 254
3255 255
3260 253
3265 254
3270 254
3275 254
3280 254
3285 254
3290 254
3295 253
3300 253
3305 253
3310 255
3315 254
3320 254
3325 253
3330 253
3335 254
3340 253
3345 253
3350 253
3355 254
3360 255
3365 253
3370 254
3375 255
3380 254
3385 254
3390 254
3395 253
3400 254
3405 254
3410 254
3415 253
3420 253
3425 254
3430 253
3435 255
3440 253
3445 254
3450 253
3455 253
3460 254
3465 253
3470 253
3475 253
3480 253
3485 254
3490 253
3495 254
3500 253
3505 253
3510 253
3515 254
3520 254
3525 253
3530 254
3535 253
3540 254
3545 253
3550 253
3555 254
3560 254
3565 253
3570 254
3575 253
3580 253
3585 253
3590 253
3595 253
};
\addlegendentry{Different sockets}
\addplot [thick, color1, mark=square*, mark size=3, mark options={solid,fill=white,draw=red},  mark repeat={180}]
table {%
0 275
5 264
10 267
15 257
20 255
25 255
30 255
35 256
40 255
45 256
50 255
55 255
60 254
65 255
70 255
75 255
80 256
85 260
90 268
95 268
100 269
105 267
110 264
115 263
120 263
125 263
130 266
135 267
140 268
145 268
150 274
155 266
160 357
165 358
170 359
175 359
180 360
185 360
190 361
195 361
200 361
205 358
210 355
215 361
220 361
225 361
230 361
235 362
240 359
245 360
250 362
255 361
260 360
265 361
270 361
275 362
280 361
285 361
290 360
295 361
300 361
305 361
310 361
315 360
320 360
325 360
330 359
335 361
340 361
345 361
350 359
355 360
360 360
365 361
370 362
375 360
380 360
385 357
390 361
395 360
400 361
405 360
410 360
415 360
420 360
425 361
430 360
435 361
440 361
445 361
450 361
455 360
460 361
465 360
470 361
475 361
480 361
485 361
490 361
495 361
500 361
505 361
510 361
515 360
520 361
525 362
530 362
535 360
540 360
545 361
550 362
555 362
560 361
565 362
570 361
575 361
580 361
585 362
590 360
595 361
600 362
605 361
610 362
615 362
620 362
625 367
630 371
635 369
640 370
645 369
650 370
655 370
660 370
665 367
670 362
675 362
680 362
685 362
690 361
695 362
700 361
705 362
710 361
715 362
720 362
725 362
730 362
735 361
740 363
745 362
750 361
755 361
760 362
765 361
770 361
775 361
780 361
785 361
790 362
795 363
800 361
805 362
810 362
815 361
820 362
825 363
830 363
835 361
840 361
845 361
850 362
855 361
860 362
865 362
870 362
875 362
880 362
885 363
890 363
895 362
900 361
905 362
910 361
915 362
920 362
925 362
930 361
935 362
940 363
945 362
950 363
955 362
960 362
965 362
970 363
975 363
980 364
985 362
990 363
995 363
1000 362
1005 362
1010 363
1015 363
1020 362
1025 363
1030 363
1035 363
1040 364
1045 362
1050 362
1055 361
1060 360
1065 362
1070 362
1075 363
1080 362
1085 363
1090 362
1095 362
1100 363
1105 365
1110 360
1115 363
1120 363
1125 363
1130 363
1135 363
1140 362
1145 362
1150 362
1155 363
1160 362
1165 363
1170 362
1175 362
1180 362
1185 363
1190 364
1195 364
1200 362
1205 362
1210 362
1215 364
1220 364
1225 362
1230 362
1235 362
1240 363
1245 362
1250 363
1255 364
1260 362
1265 363
1270 362
1275 363
1280 362
1285 362
1290 364
1295 362
1300 362
1305 361
1310 363
1315 362
1320 362
1325 361
1330 361
1335 361
1340 363
1345 360
1350 359
1355 361
1360 361
1365 361
1370 361
1375 362
1380 361
1385 362
1390 362
1395 362
1400 361
1405 361
1410 361
1415 361
1420 361
1425 361
1430 362
1435 362
1440 361
1445 361
1450 361
1455 361
1460 361
1465 361
1470 361
1475 361
1480 361
1485 361
1490 360
1495 361
1500 362
1505 361
1510 361
1515 360
1520 362
1525 363
1530 359
1535 362
1540 360
1545 361
1550 361
1555 362
1560 362
1565 362
1570 361
1575 362
1580 360
1585 360
1590 362
1595 362
1600 362
1605 361
1610 361
1615 361
1620 361
1625 361
1630 360
1635 361
1640 361
1645 361
1650 362
1655 362
1660 361
1665 361
1670 362
1675 362
1680 363
1685 362
1690 361
1695 362
1700 361
1705 361
1710 362
1715 363
1720 362
1725 362
1730 362
1735 362
1740 362
1745 363
1750 362
1755 357
1760 360
1765 361
1770 360
1775 360
1780 359
1785 360
1790 360
1795 361
1800 361
1805 361
1810 359
1815 360
1820 360
1825 361
1830 360
1835 361
1840 359
1845 360
1850 362
1855 360
1860 359
1865 362
1870 360
1875 361
1880 359
1885 360
1890 361
1895 360
1900 362
1905 360
1910 359
1915 361
1920 360
1925 360
1930 361
1935 360
1940 360
1945 361
1950 361
1955 360
1960 361
1965 360
1970 360
1975 360
1980 360
1985 361
1990 360
1995 360
2000 361
2005 360
2010 360
2015 361
2020 360
2025 360
2030 360
2035 360
2040 360
2045 362
2050 360
2055 361
2060 360
2065 367
2070 360
2075 362
2080 360
2085 361
2090 361
2095 361
2100 361
2105 361
2110 361
2115 361
2120 361
2125 361
2130 362
2135 361
2140 361
2145 361
2150 361
2155 360
2160 362
2165 362
2170 361
2175 361
2180 360
2185 361
2190 361
2195 362
2200 360
2205 361
2210 360
2215 361
2220 361
2225 362
2230 362
2235 360
2240 361
2245 360
2250 360
2255 361
2260 361
2265 361
2270 361
2275 361
2280 362
2285 361
2290 361
2295 361
2300 361
2305 360
2310 360
2315 360
2320 360
2325 361
2330 361
2335 361
2340 361
2345 361
2350 362
2355 361
2360 360
2365 366
2370 361
2375 361
2380 357
2385 326
2390 311
2395 297
2400 270
2405 268
2410 269
2415 267
2420 264
2425 267
2430 256
2435 254
2440 253
2445 254
2450 254
2455 253
2460 254
2465 254
2470 253
2475 254
2480 252
2485 254
2490 253
2495 254
2500 254
2505 253
2510 254
2515 253
2520 253
2525 252
2530 253
2535 252
2540 253
2545 253
2550 253
2555 253
2560 252
2565 252
2570 253
2575 252
2580 252
2585 253
2590 252
2595 253
2600 253
2605 253
2610 252
2615 252
2620 252
2625 253
2630 252
2635 255
2640 253
2645 253
2650 253
2655 263
2660 251
2665 258
2670 253
2675 253
2680 252
2685 253
2690 252
2695 252
2700 253
2705 253
2710 253
2715 253
2720 253
2725 253
2730 253
2735 253
2740 251
2745 252
2750 252
2755 252
2760 252
2765 253
2770 252
2775 252
2780 252
2785 252
2790 251
2795 251
2800 252
2805 252
2810 251
2815 252
2820 251
2825 252
2830 252
2835 252
2840 252
2845 251
2850 251
2855 251
2860 251
2865 252
2870 251
2875 251
2880 252
2885 251
2890 251
2895 252
2900 252
2905 251
2910 251
2915 251
2920 252
2925 252
2930 251
2935 252
2940 251
2945 252
2950 252
2955 252
2960 252
2965 255
2970 251
2975 251
2980 252
2985 252
2990 251
2995 252
3000 252
3005 252
3010 251
3015 253
3020 251
3025 252
3030 252
3035 251
3040 252
3045 251
3050 252
3055 251
3060 253
3065 254
3070 252
3075 253
3080 253
3085 254
3090 253
3095 253
3100 253
3105 254
3110 253
3115 253
3120 253
3125 253
3130 253
3135 254
3140 253
3145 252
3150 254
3155 253
3160 252
3165 254
3170 253
3175 254
3180 253
3185 254
3190 253
3195 254
3200 253
3205 252
3210 253
3215 253
3220 254
3225 253
3230 253
3235 253
3240 253
3245 254
3250 253
3255 254
3260 253
3265 253
3270 253
3275 254
3280 253
3285 254
3290 254
3295 254
3300 253
3305 253
3310 253
3315 254
3320 254
3325 254
3330 253
3335 254
3340 253
3345 252
3350 253
3355 253
3360 253
3365 253
3370 253
3375 253
3380 253
3385 253
3390 253
3395 253
3400 253
3405 253
3410 253
3415 253
3420 253
3425 253
3430 253
3435 253
3440 253
3445 253
3450 253
3455 254
3460 253
3465 253
3470 254
3475 253
3480 253
3485 253
3490 253
3495 253
3500 253
3505 253
3510 254
3515 253
3520 254
3525 251
3530 252
3535 253
3540 254
3545 253
3550 254
3555 253
3560 253
3565 253
3570 253
3575 253
3580 253
3585 253
3590 253
3595 253
3600 253
};
\addlegendentry{Same socket}
\end{axis}

\end{tikzpicture}

%% file: results/power_intel_sockets_2jobs.tex
\begin{tikzpicture}[font=\Large]

\definecolor{color0}{rgb}{0.12156862745098,0.466666666666667,0.705882352941177}
\definecolor{color1}{rgb}{1,0.498039215686275,0.0549019607843137}
\definecolor{color2}{rgb}{0.172549019607843,0.627450980392157,0.172549019607843}
\definecolor{color3}{rgb}{0.83921568627451,0.152941176470588,0.156862745098039}
\definecolor{color4}{rgb}{0.580392156862745,0.403921568627451,0.741176470588235}
\definecolor{color5}{rgb}{0,0,0}

\begin{axis}[
legend cell align={left},
legend columns=2,
legend style={fill opacity=0.8, draw opacity=1, text opacity=1, at={(1.05,1.15)}, anchor=east, draw=white!80.0!black},
tick align=outside,
tick pos=left,
x grid style={white!69.01960784313725!black},
xlabel={Time (SEC)},
xmin=0, xmax=5400,
xtick style={color=black},
y grid style={white!69.01960784313725!black},
ylabel={Power consumption (W)},
ymin=150, ymax=500,
ytick={150,200,250,300,350,400,450,500},
xmajorgrids,
ymajorgrids,
ytick style={color=black}
]
\addplot [thick, color5, mark=.]
table {%
0 254
5 262
10 252
15 253
20 253
25 253
30 253
35 254
40 253
45 253
50 254
55 254
60 253
65 254
70 254
75 254
80 254
85 254
90 254
95 254
100 254
105 254
110 253
115 254
120 254
125 254
130 254
135 253
140 253
145 253
150 253
155 254
160 254
165 253
170 254
175 254
180 253
185 254
190 254
195 254
200 253
205 254
210 254
215 254
220 253
225 254
230 253
235 253
240 253
245 252
250 253
255 253
260 254
265 254
270 254
275 253
280 254
285 253
290 253
295 253
300 254
305 253
310 254
315 253
320 253
325 253
330 254
335 254
340 253
345 253
350 253
355 254
360 253
365 253
370 254
375 254
380 254
385 254
390 254
395 254
400 253
405 253
410 253
415 254
420 253
425 254
430 253
435 253
440 254
445 254
450 254
455 254
460 253
465 254
470 253
475 253
480 252
485 254
490 254
495 253
500 253
505 254
510 254
515 254
520 253
525 253
530 254
535 254
540 253
545 254
550 253
555 253
560 253
565 253
570 254
575 253
580 254
585 253
590 254
595 254
600 253
605 254
610 253
615 254
620 254
625 253
630 254
635 253
640 253
645 254
650 254
655 253
660 254
665 253
670 253
675 254
680 253
685 254
690 254
695 254
700 254
705 253
710 254
715 254
720 253
725 254
730 253
735 253
740 253
745 254
750 254
755 254
760 254
765 254
770 254
775 253
780 254
785 254
790 254
795 253
800 254
805 253
810 253
815 253
820 254
825 254
830 253
835 253
840 253
845 253
850 253
855 253
860 254
865 253
870 253
875 252
880 254
885 254
890 253
895 253
900 254
905 253
910 254
915 254
920 253
925 254
930 254
935 254
940 253
945 254
950 253
955 254
960 254
965 253
970 254
975 254
980 254
985 252
990 253
995 253
1000 253
1005 253
1010 253
1015 253
1020 252
1025 254
1030 253
1035 254
1040 252
1045 254
1050 254
1055 254
1060 253
1065 254
1070 254
1075 254
1080 253
1085 253
1090 253
1095 254
1100 254
1105 253
1110 253
1115 254
1120 253
1125 254
1130 254
1135 253
1140 253
1145 253
1150 253
1155 254
1160 254
1165 254
1170 254
1175 254
1180 253
1185 253
1190 254
1195 253
1200 253
1205 254
1210 254
1215 254
1220 254
1225 254
1230 251
1235 253
1240 254
1245 253
1250 254
1255 253
1260 254
1265 254
1270 254
1275 254
1280 254
1285 253
1290 253
1295 253
1300 253
1305 254
1310 254
1315 254
1320 254
1325 254
1330 254
1335 253
1340 255
1345 254
1350 254
1355 253
1360 254
1365 254
1370 254
1375 253
1380 254
1385 254
1390 253
1395 254
1400 255
1405 253
1410 254
1415 254
1420 254
1425 254
1430 253
1435 254
1440 254
1445 253
1450 253
1455 253
1460 253
1465 253
1470 253
1475 253
1480 254
1485 254
1490 254
1495 253
1500 254
1505 253
1510 252
1515 254
1520 254
1525 254
1530 253
1535 254
1540 253
1545 254
1550 253
1555 254
1560 252
1565 254
1570 254
1575 254
1580 254
1585 254
1590 253
1595 254
1600 251
1605 254
1610 253
1615 254
1620 253
1625 254
1630 253
1635 254
1640 254
1645 253
1650 253
1655 254
1660 254
1665 253
1670 253
1675 253
1680 253
1685 254
1690 254
1695 254
1700 254
1705 253
1710 253
1715 254
1720 254
1725 252
1730 252
1735 253
1740 254
1745 253
1750 254
1755 253
1760 253
1765 253
1770 254
1775 252
1780 253
1785 253
1790 252
1795 253
1800 253
1805 253
1810 253
1815 253
1820 254
1825 252
1830 254
1835 253
1840 253
1845 253
1850 252
1855 252
1860 253
1865 253
1870 253
1875 254
1880 253
1885 252
1890 253
1895 253
1900 253
1905 252
1910 254
1915 252
1920 254
1925 253
1930 254
1935 253
1940 253
1945 254
1950 253
1955 254
1960 253
1965 253
1970 254
1975 254
1980 252
1985 253
1990 254
1995 253
2000 253
2005 253
2010 254
2015 253
2020 253
2025 253
2030 253
2035 253
2040 254
2045 254
2050 253
2055 254
2060 253
2065 253
2070 254
2075 253
2080 253
2085 254
2090 253
2095 254
2100 254
2105 253
2110 254
2115 254
2120 252
2125 254
2130 254
2135 253
2140 253
2145 253
2150 254
2155 254
2160 254
2165 253
2170 251
2175 253
2180 253
2185 254
2190 254
2195 253
2200 253
2205 254
2210 253
2215 254
2220 254
2225 254
2230 253
2235 254
2240 253
2245 254
2250 253
2255 254
2260 254
2265 253
2270 254
2275 254
2280 253
2285 253
2290 254
2295 254
2300 253
2305 253
2310 254
2315 253
2320 253
2325 253
2330 253
2335 254
2340 254
2345 254
2350 254
2355 253
2360 253
2365 253
2370 253
2375 254
2380 254
2385 253
2390 253
2395 254
2400 253
2405 253
2410 253
2415 253
2420 254
2425 253
2430 253
2435 253
2440 253
2445 254
2450 254
2455 254
2460 254
2465 251
2470 254
2475 253
2480 254
2485 253
2490 253
2495 254
2500 254
2505 253
2510 254
2515 254
2520 254
2525 253
2530 253
2535 251
2540 252
2545 253
2550 254
2555 253
2560 253
2565 254
2570 253
2575 253
2580 253
2585 253
2590 253
2595 253
2600 253
2605 253
2610 253
2615 253
2620 253
2625 261
2630 252
2635 254
2640 253
2645 254
2650 253
2655 254
2660 253
2665 253
2670 254
2675 254
2680 253
2685 253
2690 253
2695 253
2700 253
2705 254
2710 254
2715 254
2720 254
2725 253
2730 253
2735 253
2740 253
2745 254
2750 254
2755 253
2760 254
2765 252
2770 254
2775 254
2780 254
2785 254
2790 254
2795 253
2800 254
2805 253
2810 253
2815 253
2820 253
2825 253
2830 254
2835 254
2840 254
2845 254
2850 254
2855 254
2860 253
2865 254
2870 254
2875 253
2880 254
2885 251
2890 253
2895 254
2900 253
2905 253
2910 254
2915 253
2920 249
2925 252
2930 253
2935 252
2940 251
2945 252
2950 252
2955 253
2960 252
2965 252
2970 252
2975 252
2980 252
2985 251
2990 252
2995 252
3000 251
3005 252
3010 252
3015 252
3020 251
3025 251
3030 252
3035 251
3040 252
3045 252
3050 251
3055 252
3060 252
3065 252
3070 252
3075 252
3080 252
3085 252
3090 253
3095 251
3100 252
3105 252
3110 251
3115 252
3120 252
3125 251
3130 252
3135 253
3140 252
3145 252
3150 251
3155 251
3160 252
3165 252
3170 252
3175 251
3180 252
3185 251
3190 251
3195 253
3200 251
3205 251
3210 251
3215 252
3220 251
3225 252
3230 252
3235 252
3240 252
3245 251
3250 252
3255 251
3260 251
3265 252
3270 252
3275 251
3280 252
3285 251
3290 251
3295 251
3300 251
3305 252
3310 251
3315 253
3320 251
3325 251
3330 251
3335 252
3340 251
3345 251
3350 252
3355 252
3360 252
3365 252
3370 252
3375 252
3380 252
3385 253
3390 251
3395 253
3400 251
3405 252
3410 252
3415 252
3420 252
3425 251
3430 251
3435 253
3440 252
3445 251
3450 252
3455 251
3460 251
3465 251
3470 253
3475 253
3480 254
3485 253
3490 254
3495 254
3500 253
3505 254
3510 253
3515 254
3520 254
3525 254
3530 254
3535 253
3540 254
3545 253
3550 254
3555 254
3560 254
3565 253
3570 253
3575 254
3580 253
3585 254
3590 254
3595 253
3600 252
3605 253
3610 253
3615 254
3620 253
3625 255
3630 254
3635 254
3640 253
3645 253
3650 254
3655 253
3660 254
3665 254
3670 253
3675 252
3680 252
3685 254
3690 253
3695 253
3700 254
3705 253
3710 254
3715 253
3720 253
3725 253
3730 253
3735 252
3740 252
3745 254
3750 253
3755 250
3760 253
3765 253
3770 253
3775 254
3780 253
3785 253
3790 253
3795 254
3800 252
3805 254
3810 254
3815 253
3820 253
3825 253
3830 253
3835 254
3840 253
3845 252
3850 254
3855 253
3860 253
3865 253
3870 253
3875 253
3880 253
3885 253
3890 253
3895 254
3900 252
3905 254
3910 253
3915 253
3920 253
3925 253
3930 254
3935 254
3940 253
3945 253
3950 252
3955 253
3960 253
3965 252
3970 253
3975 254
3980 253
3985 254
3990 252
3995 253
4000 254
4005 253
4010 254
4015 253
4020 253
4025 254
4030 253
4035 254
4040 254
4045 254
4050 253
4055 254
4060 253
4065 254
4070 254
4075 254
4080 254
4085 254
4090 253
4095 254
4100 253
4105 253
4110 254
4115 253
4120 253
4125 254
4130 253
4135 253
4140 254
4145 254
4150 254
4155 253
4160 253
4165 254
4170 253
4175 253
4180 253
4185 254
4190 254
4195 253
4200 253
4205 254
4210 253
4215 253
4220 254
4225 253
4230 253
4235 253
4240 253
4245 254
4250 253
4255 253
4260 254
4265 253
4270 254
4275 254
4280 253
4285 253
4290 253
4295 254
4300 254
4305 254
4310 254
4315 253
4320 254
4325 254
4330 254
4335 253
4340 263
4345 253
4350 254
4355 254
4360 254
4365 254
4370 254
4375 254
4380 254
4385 254
4390 253
4395 253
4400 254
4405 253
4410 254
4415 254
4420 253
4425 253
4430 254
4435 253
4440 253
4445 254
4450 254
4455 253
4460 254
4465 253
4470 254
4475 253
4480 254
4485 253
4490 254
4495 254
4500 253
4505 253
4510 254
4515 253
4520 254
4525 253
4530 253
4535 253
4540 254
4545 254
4550 253
4555 253
4560 253
4565 253
4570 253
4575 254
4580 254
4585 253
4590 254
4595 254
4600 254
4605 254
4610 253
4615 254
4620 253
4625 253
4630 253
4635 253
4640 254
4645 253
4650 254
4655 254
4660 254
4665 253
4670 254
4675 254
4680 254
4685 253
4690 253
4695 254
4700 254
4705 254
4710 254
4715 254
4720 254
4725 253
4730 254
4735 253
4740 254
4745 253
4750 253
4755 254
4760 254
4765 253
4770 254
4775 254
4780 253
4785 254
4790 253
4795 253
4800 254
4805 253
4810 254
4815 254
4820 254
4825 254
4830 254
4835 253
4840 254
4845 254
4850 254
4855 253
4860 254
4865 253
4870 254
4875 253
4880 254
4885 253
4890 254
4895 254
4900 254
4905 253
4910 253
4915 254
4920 253
4925 254
4930 254
4935 254
4940 253
4945 254
4950 253
4955 254
4960 254
4965 252
4970 252
4975 254
4980 254
4985 254
4990 254
4995 254
5000 253
5005 254
5010 253
5015 254
5020 253
5025 254
5030 254
5035 254
5040 254
5045 252
5050 255
5055 253
5060 253
5065 253
5070 254
5075 254
5080 253
5085 254
5090 251
5095 253
5100 254
5105 254
5110 254
5115 254
5120 253
5125 254
5130 254
5135 254
5140 254
5145 254
5150 254
5155 254
5160 254
5165 253
5170 254
5175 253
5180 254
5185 254
5190 253
5195 254
5200 253
5205 253
5210 253
5215 252
5220 253
5225 254
5230 254
5235 253
5240 253
5245 253
5250 253
5255 253
5260 253
5265 253
5270 253
5275 252
5280 254
5285 254
5290 252
5295 253
5300 253
5305 254
5310 253
5315 254
5320 254
5325 254
5330 253
5335 254
5340 254
5345 253
5350 253
5355 254
5360 254
5365 253
5370 253
5375 253
5380 254
5385 254
5390 253
5395 253
};
\addlegendentry{Baseline}
\addplot [thick, color0, mark=triangle*, mark size=3, mark options={solid,fill=white,draw=red},  mark repeat={180}]
table {%
0 255
5 256
10 277
15 279
20 282
25 289
30 268
35 278
40 257
45 255
50 256
55 256
60 256
65 255
70 256
75 256
80 256
85 256
90 256
95 256
100 257
105 253
110 256
115 256
120 270
125 281
130 282
135 283
140 279
145 274
150 274
155 274
160 275
165 277
170 282
175 284
180 279
185 288
190 285
195 463
200 463
205 464
210 468
215 467
220 468
225 470
230 470
235 470
240 470
245 472
250 473
255 472
260 473
265 471
270 471
275 471
280 471
285 471
290 472
295 472
300 471
305 471
310 471
315 471
320 471
325 471
330 471
335 471
340 471
345 472
350 472
355 471
360 472
365 472
370 471
375 472
380 471
385 472
390 472
395 472
400 472
405 472
410 471
415 473
420 473
425 473
430 473
435 472
440 472
445 474
450 473
455 472
460 474
465 471
470 474
475 475
480 474
485 474
490 474
495 474
500 474
505 474
510 475
515 475
520 475
525 474
530 475
535 474
540 474
545 474
550 474
555 475
560 475
565 474
570 475
575 474
580 475
585 474
590 475
595 475
600 474
605 475
610 475
615 475
620 475
625 475
630 474
635 475
640 475
645 475
650 475
655 474
660 475
665 475
670 474
675 474
680 473
685 474
690 474
695 474
700 474
705 475
710 474
715 475
720 474
725 474
730 474
735 474
740 474
745 475
750 474
755 474
760 475
765 475
770 474
775 474
780 475
785 474
790 475
795 473
800 474
805 474
810 475
815 475
820 474
825 474
830 474
835 475
840 474
845 475
850 475
855 474
860 475
865 474
870 472
875 474
880 474
885 475
890 474
895 475
900 474
905 475
910 474
915 475
920 476
925 475
930 476
935 475
940 476
945 474
950 476
955 475
960 475
965 475
970 475
975 476
980 475
985 475
990 476
995 476
1000 476
1005 475
1010 476
1015 476
1020 476
1025 476
1030 475
1035 475
1040 476
1045 477
1050 477
1055 476
1060 476
1065 476
1070 476
1075 477
1080 477
1085 477
1090 475
1095 476
1100 476
1105 476
1110 477
1115 476
1120 477
1125 477
1130 477
1135 477
1140 477
1145 477
1150 477
1155 476
1160 477
1165 477
1170 477
1175 477
1180 478
1185 478
1190 478
1195 477
1200 477
1205 477
1210 478
1215 477
1220 478
1225 477
1230 478
1235 477
1240 478
1245 477
1250 477
1255 477
1260 478
1265 477
1270 477
1275 478
1280 477
1285 477
1290 478
1295 477
1300 478
1305 476
1310 478
1315 478
1320 478
1325 477
1330 477
1335 477
1340 477
1345 477
1350 478
1355 477
1360 477
1365 478
1370 478
1375 477
1380 478
1385 478
1390 477
1395 477
1400 478
1405 478
1410 478
1415 477
1420 478
1425 477
1430 479
1435 478
1440 478
1445 478
1450 477
1455 477
1460 478
1465 478
1470 479
1475 478
1480 478
1485 479
1490 478
1495 478
1500 477
1505 477
1510 477
1515 477
1520 477
1525 477
1530 477
1535 477
1540 477
1545 478
1550 478
1555 477
1560 478
1565 478
1570 478
1575 478
1580 479
1585 478
1590 478
1595 478
1600 478
1605 477
1610 478
1615 478
1620 477
1625 478
1630 478
1635 477
1640 478
1645 477
1650 477
1655 478
1660 478
1665 477
1670 478
1675 477
1680 478
1685 478
1690 477
1695 477
1700 477
1705 477
1710 477
1715 477
1720 476
1725 476
1730 476
1735 476
1740 476
1745 476
1750 476
1755 475
1760 476
1765 476
1770 476
1775 475
1780 475
1785 476
1790 475
1795 475
1800 476
1805 475
1810 476
1815 476
1820 476
1825 476
1830 476
1835 475
1840 475
1845 475
1850 475
1855 476
1860 475
1865 474
1870 474
1875 474
1880 476
1885 475
1890 473
1895 472
1900 474
1905 475
1910 475
1915 474
1920 474
1925 475
1930 475
1935 474
1940 475
1945 474
1950 474
1955 474
1960 473
1965 473
1970 474
1975 473
1980 474
1985 473
1990 473
1995 473
2000 474
2005 473
2010 472
2015 473
2020 473
2025 473
2030 474
2035 473
2040 473
2045 472
2050 474
2055 474
2060 473
2065 472
2070 474
2075 473
2080 473
2085 472
2090 473
2095 474
2100 474
2105 474
2110 473
2115 473
2120 473
2125 472
2130 474
2135 473
2140 473
2145 473
2150 473
2155 472
2160 474
2165 473
2170 473
2175 474
2180 474
2185 475
2190 474
2195 474
2200 474
2205 474
2210 474
2215 475
2220 474
2225 474
2230 475
2235 474
2240 474
2245 474
2250 474
2255 474
2260 476
2265 475
2270 474
2275 475
2280 475
2285 475
2290 475
2295 475
2300 476
2305 476
2310 475
2315 474
2320 475
2325 474
2330 475
2335 474
2340 476
2345 475
2350 476
2355 475
2360 476
2365 475
2370 475
2375 475
2380 475
2385 476
2390 475
2395 476
2400 475
2405 476
2410 476
2415 475
2420 474
2425 476
2430 472
2435 434
2440 429
2445 400
2450 384
2455 382
2460 380
2465 380
2470 374
2475 372
2480 335
2485 315
2490 316
2495 273
2500 272
2505 272
2510 271
2515 271
2520 270
2525 258
2530 258
2535 257
2540 258
2545 257
2550 258
2555 258
2560 258
2565 258
2570 258
2575 258
2580 257
2585 257
2590 257
2595 256
2600 257
2605 257
2610 256
2615 257
2620 256
2625 257
2630 256
2635 256
2640 257
2645 256
2650 256
2655 257
2660 256
2665 256
2670 257
2675 257
2680 256
2685 256
2690 256
2695 256
2700 255
2705 257
2710 255
2715 257
2720 255
2725 255
2730 256
2735 255
2740 256
2745 256
2750 255
2755 255
2760 255
2765 256
2770 256
2775 255
2780 255
2785 255
2790 256
2795 256
2800 255
2805 255
2810 255
2815 256
2820 255
2825 256
2830 256
2835 255
2840 256
2845 256
2850 255
2855 255
2860 256
2865 256
2870 256
2875 256
2880 256
2885 255
2890 255
2895 256
2900 259
2905 253
2910 254
2915 255
2920 255
2925 255
2930 255
2935 255
2940 256
2945 256
2950 255
2955 256
2960 256
2965 255
2970 256
2975 256
2980 256
2985 256
2990 255
2995 256
3000 255
3005 255
3010 256
3015 256
3020 255
3025 256
3030 256
3035 255
3040 256
3045 256
3050 255
3055 255
3060 254
3065 254
3070 256
3075 255
3080 254
3085 254
3090 255
3095 255
3100 254
3105 254
3110 255
3115 254
3120 254
3125 254
3130 255
3135 255
3140 254
3145 255
3150 255
3155 255
3160 255
3165 255
3170 255
3175 254
3180 254
3185 255
3190 255
3195 255
3200 255
3205 255
3210 255
3215 255
3220 254
3225 256
3230 255
3235 255
3240 254
3245 254
3250 254
3255 255
3260 255
3265 252
3270 254
3275 255
3280 255
3285 254
3290 254
3295 255
3300 255
3305 254
3310 255
3315 256
3320 254
3325 254
3330 255
3335 255
3340 254
3345 255
3350 255
3355 254
3360 254
3365 254
3370 255
3375 254
3380 254
3385 255
3390 254
3395 254
3400 254
3405 255
3410 254
3415 255
3420 254
3425 256
3430 255
3435 255
3440 255
3445 255
3450 255
3455 254
3460 255
3465 256
3470 255
3475 254
3480 254
3485 255
3490 255
3495 254
3500 260
3505 255
3510 255
3515 254
3520 256
3525 254
3530 255
3535 254
3540 255
3545 256
3550 254
3555 247
3560 253
3565 253
3570 253
3575 253
3580 253
3585 253
3590 253
3595 253
3600 253
3605 253
3610 253
3615 253
3620 253
3625 253
3630 253
3635 253
3640 253
3645 253
3650 253
3655 254
3660 253
3665 253
3670 253
3675 253
3680 263
3685 251
3690 255
3695 254
3700 254
3705 254
3710 254
3715 254
3720 254
3725 254
3730 254
3735 255
3740 254
3745 254
3750 255
3755 254
3760 254
3765 254
3770 254
3775 254
3780 254
3785 254
3790 254
3795 255
3800 256
3805 254
3810 254
3815 254
3820 255
3825 254
3830 255
3835 254
3840 255
3845 254
3850 254
3855 254
3860 255
3865 254
3870 254
3875 254
3880 254
3885 254
3890 254
3895 254
3900 254
3905 254
3910 254
3915 254
3920 254
3925 254
3930 254
3935 254
3940 255
3945 252
3950 254
3955 254
3960 255
3965 254
3970 255
3975 255
3980 255
3985 256
3990 255
3995 255
4000 255
4005 254
4010 254
4015 254
4020 255
4025 255
4030 255
4035 255
4040 255
4045 254
4050 254
4055 254
4060 254
4065 255
4070 254
4075 254
4080 254
4085 255
4090 254
4095 255
4100 255
4105 254
4110 255
4115 254
4120 254
4125 256
4130 255
4135 255
4140 255
4145 255
4150 256
4155 254
4160 254
4165 254
4170 255
4175 254
4180 254
4185 255
4190 255
4195 254
4200 255
4205 254
4210 255
4215 255
4220 255
4225 254
4230 255
4235 254
4240 254
4245 256
4250 254
4255 254
4260 255
4265 255
4270 255
4275 254
4280 255
4285 254
4290 254
4295 253
4300 253
4305 253
4310 254
4315 254
4320 254
4325 253
4330 254
4335 254
4340 254
4345 253
4350 254
4355 254
4360 253
4365 253
4370 254
4375 253
4380 253
4385 254
4390 253
4395 253
4400 253
4405 253
4410 251
4415 253
4420 254
4425 253
4430 253
4435 253
4440 254
4445 253
4450 254
4455 254
4460 253
4465 253
4470 253
4475 254
4480 254
4485 254
4490 254
4495 254
4500 254
4505 253
4510 253
4515 253
4520 253
4525 254
4530 253
4535 253
4540 253
4545 253
4550 253
4555 254
4560 253
4565 254
4570 253
4575 253
4580 253
4585 254
4590 253
4595 253
4600 254
4605 253
4610 253
4615 253
4620 254
4625 254
4630 253
4635 253
4640 254
4645 253
4650 252
4655 252
4660 253
4665 253
4670 254
4675 254
4680 253
4685 254
4690 254
4695 253
4700 253
4705 254
4710 254
4715 253
4720 253
4725 253
4730 254
4735 253
4740 252
4745 254
4750 253
4755 253
4760 252
4765 252
4770 254
4775 254
4780 254
4785 254
4790 254
4795 254
4800 253
4805 253
4810 254
4815 252
4820 253
4825 254
4830 253
4835 254
4840 254
4845 254
4850 253
4855 254
4860 254
4865 254
4870 254
4875 253
4880 253
4885 252
4890 254
4895 254
4900 254
4905 254
4910 252
4915 253
4920 254
4925 253
4930 254
4935 254
4940 253
4945 253
4950 253
4955 253
4960 253
4965 252
4970 254
4975 253
4980 254
4985 254
4990 254
4995 253
5000 254
5005 254
5010 253
5015 254
5020 254
5025 253
5030 253
5035 253
5040 253
5045 253
5050 254
5055 254
5060 254
5065 254
5070 254
5075 254
5080 253
5085 253
5090 253
5095 253
5100 253
5105 254
5110 254
5115 254
5120 253
5125 253
5130 253
5135 254
5140 253
5145 253
5150 254
5155 253
5160 254
5165 253
5170 254
5175 254
5180 253
5185 254
5190 253
5195 254
5200 254
5205 253
5210 254
5215 254
5220 254
5225 253
5230 253
5235 253
5240 254
5245 253
5250 254
5255 252
5260 253
5265 254
5270 254
5275 254
5280 250
5285 254
5290 253
5295 263
5300 251
5305 253
5310 253
5315 252
5320 253
5325 251
5330 254
5335 253
5340 254
5345 253
5350 252
5355 253
5360 253
5365 254
5370 254
5375 253
5380 253
5385 253
5390 254
5395 254

};
\addlegendentry{Different sockets}
\addplot [thick, color1, mark=square*, mark size=3, mark options={solid,fill=white,draw=red},  mark repeat={180}]
table {%
0 256
5 259
10 276
15 279
20 287
25 291
30 265
35 280
40 256
45 256
50 256
55 256
60 255
65 256
70 256
75 256
80 255
85 257
90 257
95 255
100 257
105 261
110 270
115 281
120 289
125 279
130 282
135 276
140 274
145 274
150 273
155 276
160 281
165 283
170 273
175 296
180 285
185 460
190 462
195 463
200 366
205 465
210 467
215 468
220 369
225 367
230 367
235 365
240 365
245 365
250 365
255 366
260 366
265 365
270 365
275 365
280 365
285 366
290 366
295 365
300 366
305 365
310 366
315 365
320 366
325 366
330 364
335 365
340 365
345 366
350 366
355 367
360 365
365 366
370 365
375 365
380 366
385 365
390 364
395 365
400 365
405 366
410 364
415 366
420 365
425 366
430 366
435 365
440 365
445 365
450 366
455 366
460 366
465 364
470 365
475 365
480 365
485 365
490 366
495 365
500 365
505 365
510 365
515 365
520 365
525 365
530 365
535 365
540 365
545 366
550 365
555 366
560 364
565 366
570 366
575 366
580 366
585 366
590 366
595 365
600 365
605 366
610 365
615 364
620 365
625 365
630 365
635 366
640 366
645 366
650 365
655 365
660 366
665 366
670 366
675 366
680 367
685 365
690 367
695 366
700 365
705 366
710 363
715 365
720 367
725 365
730 365
735 365
740 365
745 365
750 366
755 366
760 365
765 366
770 366
775 367
780 366
785 366
790 367
795 365
800 366
805 366
810 366
815 365
820 365
825 365
830 365
835 365
840 365
845 365
850 365
855 364
860 366
865 365
870 366
875 368
880 364
885 365
890 365
895 365
900 365
905 365
910 364
915 365
920 364
925 365
930 365
935 364
940 365
945 365
950 365
955 365
960 365
965 365
970 364
975 364
980 365
985 364
990 365
995 365
1000 365
1005 365
1010 365
1015 365
1020 365
1025 365
1030 365
1035 365
1040 364
1045 364
1050 365
1055 365
1060 365
1065 364
1070 364
1075 365
1080 365
1085 366
1090 364
1095 364
1100 364
1105 365
1110 364
1115 364
1120 365
1125 365
1130 365
1135 365
1140 364
1145 364
1150 364
1155 365
1160 364
1165 364
1170 365
1175 370
1180 365
1185 366
1190 366
1195 365
1200 366
1205 366
1210 365
1215 366
1220 366
1225 366
1230 365
1235 365
1240 366
1245 366
1250 365
1255 365
1260 366
1265 365
1270 365
1275 365
1280 366
1285 365
1290 366
1295 365
1300 366
1305 365
1310 365
1315 365
1320 366
1325 366
1330 366
1335 365
1340 365
1345 363
1350 365
1355 366
1360 366
1365 365
1370 365
1375 365
1380 366
1385 366
1390 366
1395 366
1400 365
1405 366
1410 366
1415 366
1420 366
1425 365
1430 365
1435 366
1440 366
1445 365
1450 365
1455 364
1460 366
1465 365
1470 365
1475 368
1480 364
1485 365
1490 364
1495 365
1500 364
1505 365
1510 364
1515 364
1520 366
1525 366
1530 364
1535 365
1540 364
1545 365
1550 365
1555 364
1560 365
1565 364
1570 365
1575 365
1580 365
1585 364
1590 365
1595 364
1600 364
1605 364
1610 364
1615 365
1620 365
1625 364
1630 365
1635 364
1640 366
1645 364
1650 364
1655 365
1660 365
1665 365
1670 365
1675 364
1680 365
1685 365
1690 365
1695 365
1700 365
1705 365
1710 365
1715 365
1720 365
1725 364
1730 364
1735 365
1740 366
1745 364
1750 366
1755 365
1760 366
1765 364
1770 365
1775 365
1780 365
1785 364
1790 365
1795 364
1800 366
1805 365
1810 366
1815 366
1820 365
1825 365
1830 365
1835 365
1840 366
1845 366
1850 365
1855 365
1860 366
1865 366
1870 366
1875 365
1880 366
1885 366
1890 365
1895 367
1900 365
1905 367
1910 366
1915 365
1920 365
1925 366
1930 366
1935 365
1940 365
1945 365
1950 366
1955 366
1960 365
1965 366
1970 366
1975 366
1980 365
1985 365
1990 366
1995 366
2000 365
2005 367
2010 367
2015 365
2020 365
2025 365
2030 366
2035 366
2040 366
2045 366
2050 366
2055 367
2060 365
2065 365
2070 366
2075 367
2080 367
2085 366
2090 367
2095 365
2100 367
2105 367
2110 367
2115 367
2120 366
2125 366
2130 365
2135 365
2140 367
2145 365
2150 365
2155 366
2160 366
2165 366
2170 366
2175 365
2180 366
2185 366
2190 366
2195 365
2200 365
2205 366
2210 366
2215 365
2220 366
2225 366
2230 365
2235 365
2240 366
2245 366
2250 365
2255 365
2260 364
2265 366
2270 364
2275 365
2280 365
2285 364
2290 365
2295 364
2300 364
2305 365
2310 365
2315 365
2320 366
2325 364
2330 365
2335 365
2340 364
2345 364
2350 365
2355 365
2360 365
2365 365
2370 365
2375 364
2380 364
2385 364
2390 365
2395 364
2400 365
2405 364
2410 366
2415 365
2420 366
2425 365
2430 366
2435 365
2440 366
2445 366
2450 366
2455 366
2460 366
2465 366
2470 365
2475 365
2480 366
2485 366
2490 365
2495 366
2500 366
2505 366
2510 366
2515 366
2520 365
2525 365
2530 365
2535 365
2540 366
2545 366
2550 366
2555 365
2560 366
2565 366
2570 365
2575 365
2580 366
2585 366
2590 365
2595 363
2600 366
2605 365
2610 366
2615 365
2620 366
2625 366
2630 365
2635 365
2640 366
2645 366
2650 364
2655 366
2660 366
2665 366
2670 366
2675 363
2680 364
2685 366
2690 365
2695 366
2700 366
2705 366
2710 364
2715 366
2720 366
2725 365
2730 365
2735 365
2740 364
2745 364
2750 364
2755 365
2760 366
2765 364
2770 364
2775 364
2780 365
2785 366
2790 364
2795 365
2800 364
2805 365
2810 365
2815 365
2820 365
2825 362
2830 365
2835 365
2840 365
2845 364
2850 364
2855 365
2860 364
2865 365
2870 365
2875 364
2880 365
2885 366
2890 365
2895 365
2900 365
2905 365
2910 366
2915 365
2920 365
2925 366
2930 365
2935 366
2940 362
2945 366
2950 365
2955 365
2960 364
2965 366
2970 366
2975 366
2980 365
2985 365
2990 366
2995 366
3000 366
3005 366
3010 365
3015 365
3020 366
3025 366
3030 365
3035 365
3040 366
3045 366
3050 365
3055 366
3060 366
3065 366
3070 376
3075 363
3080 365
3085 366
3090 365
3095 366
3100 366
3105 365
3110 367
3115 365
3120 366
3125 366
3130 366
3135 366
3140 366
3145 365
3150 365
3155 365
3160 366
3165 366
3170 366
3175 365
3180 364
3185 363
3190 366
3195 364
3200 364
3205 365
3210 365
3215 366
3220 365
3225 365
3230 364
3235 365
3240 364
3245 365
3250 365
3255 365
3260 365
3265 365
3270 365
3275 364
3280 365
3285 366
3290 364
3295 365
3300 364
3305 364
3310 364
3315 364
3320 364
3325 364
3330 365
3335 364
3340 365
3345 365
3350 365
3355 364
3360 365
3365 364
3370 364
3375 364
3380 365
3385 364
3390 364
3395 365
3400 365
3405 365
3410 364
3415 364
3420 365
3425 364
3430 364
3435 364
3440 365
3445 364
3450 365
3455 363
3460 363
3465 363
3470 365
3475 364
3480 364
3485 363
3490 364
3495 365
3500 363
3505 364
3510 364
3515 365
3520 365
3525 364
3530 364
3535 364
3540 364
3545 363
3550 364
3555 363
3560 365
3565 365
3570 364
3575 363
3580 363
3585 364
3590 364
3595 365
3600 365
3605 365
3610 364
3615 365
3620 364
3625 365
3630 365
3635 365
3640 364
3645 364
3650 364
3655 365
3660 364
3665 365
3670 364
3675 364
3680 364
3685 365
3690 365
3695 365
3700 364
3705 365
3710 365
3715 365
3720 364
3725 365
3730 364
3735 365
3740 365
3745 365
3750 364
3755 365
3760 365
3765 365
3770 365
3775 365
3780 364
3785 365
3790 365
3795 365
3800 364
3805 364
3810 364
3815 365
3820 365
3825 364
3830 364
3835 365
3840 365
3845 365
3850 366
3855 364
3860 365
3865 365
3870 365
3875 364
3880 365
3885 366
3890 364
3895 365
3900 366
3905 365
3910 365
3915 364
3920 365
3925 366
3930 366
3935 365
3940 365
3945 364
3950 364
3955 365
3960 365
3965 366
3970 365
3975 366
3980 364
3985 366
3990 351
3995 349
4000 323
4005 290
4010 288
4015 287
4020 285
4025 284
4030 271
4035 271
4040 259
4045 258
4050 258
4055 258
4060 257
4065 267
4070 256
4075 257
4080 257
4085 258
4090 257
4095 256
4100 257
4105 256
4110 257
4115 257
4120 256
4125 257
4130 256
4135 257
4140 257
4145 257
4150 256
4155 257
4160 257
4165 256
4170 257
4175 256
4180 255
4185 256
4190 256
4195 256
4200 255
4205 256
4210 255
4215 256
4220 256
4225 253
4230 256
4235 256
4240 256
4245 255
4250 256
4255 255
4260 256
4265 256
4270 256
4275 256
4280 256
4285 256
4290 256
4295 255
4300 255
4305 256
4310 256
4315 256
4320 255
4325 256
4330 255
4335 256
4340 255
4345 255
4350 256
4355 255
4360 256
4365 255
4370 255
4375 256
4380 255
4385 256
4390 256
4395 256
4400 256
4405 255
4410 256
4415 256
4420 254
4425 256
4430 256
4435 255
4440 255
4445 256
4450 256
4455 256
4460 256
4465 256
4470 255
4475 253
4480 254
4485 255
4490 256
4495 256
4500 255
4505 254
4510 256
4515 256
4520 256
4525 255
4530 255
4535 256
4540 256
4545 256
4550 256
4555 256
4560 256
4565 256
4570 254
4575 255
4580 256
4585 255
4590 255
4595 255
4600 255
4605 256
4610 256
4615 254
4620 256
4625 255
4630 255
4635 255
4640 255
4645 256
4650 255
4655 256
4660 256
4665 256
4670 256
4675 256
4680 256
4685 256
4690 255
4695 255
4700 256
4705 255
4710 255
4715 255
4720 255
4725 256
4730 254
4735 256
4740 256
4745 255
4750 255
4755 254
4760 256
4765 256
4770 255
4775 256
4780 256
4785 256
4790 255
4795 256
4800 256
4805 256
4810 256
4815 255
4820 255
4825 255
4830 256
4835 256
4840 255
4845 255
4850 256
4855 256
4860 256
4865 256
4870 255
4875 256
4880 255
4885 255
4890 256
4895 256
4900 256
4905 256
4910 256
4915 255
4920 255
4925 255
4930 256
4935 256
4940 255
4945 255
4950 255
4955 256
4960 255
4965 255
4970 255
4975 255
4980 255
4985 256
4990 254
4995 255
5000 255
5005 256
5010 256
5015 255
5020 255
5025 256
5030 255
5035 256
5040 256
5045 255
5050 256
5055 256
5060 255
5065 256
5070 258
5075 255
5080 255
5085 256
5090 256
5095 255
5100 256
5105 254
5110 255
5115 255
5120 255
5125 256
5130 254
5135 256
5140 256
5145 254
5150 256
5155 255
5160 256
5165 256
5170 256
5175 255
5180 256
5185 255
5190 256
5195 254
5200 255
5205 254
5210 254
5215 255
5220 255
5225 256
5230 254
5235 255
5240 254
5245 255
5250 256
5255 255
5260 256
5265 256
5270 255
5275 255
5280 256
5285 255
5290 255
5295 255
5300 255
5305 256
5310 256
5315 255
5320 256
5325 256
5330 255
5335 256
5340 255
5345 256
5350 256
5355 256
5360 256
5365 255
5370 256
5375 255
5380 255
5385 256
5390 255
5395 255

};
\addlegendentry{Same socket}
\end{axis}

\end{tikzpicture}

%% file: results/energy_intel_sockets.tex
\begin{tikzpicture}[font=\Large]

\definecolor{color0}{rgb}{0.12156862745098,0.466666666666667,0.705882352941177}
\definecolor{color1}{rgb}{1,0.498039215686275,0.0549019607843137}
\definecolor{color2}{rgb}{0.172549019607843,0.627450980392157,0.172549019607843}
\definecolor{color3}{rgb}{0.83921568627451,0.152941176470588,0.156862745098039}
\definecolor{color4}{rgb}{0.580392156862745,0.403921568627451,0.741176470588235}
\definecolor{color5}{rgb}{0,0,0}

\begin{axis}[
legend cell align={left},
legend columns=2,
legend style={fill opacity=0.8, draw opacity=1, text opacity=1, at={(1.05,1.15)}, anchor=east, draw=white!80.0!black},
tick align=outside,
tick pos=left,
x grid style={white!69.01960784313725!black},
xlabel={Time (SEC)},
xmin=0, xmax=3600,
xtick={0,900,1800,2700,3600},
xtick style={color=black},
y grid style={white!69.01960784313725!black},
ylabel={Total energy consumption (kWh)},
ymin=0, ymax=0.55,
xmajorgrids,
ymajorgrids,
ytick style={color=black},
ytick={0.05,0.1,0.15,0.2,0.25,0.3,0.35,0.4,0.45,0.5,0.55},
yticklabel style={
        /pgf/number format/fixed,
        /pgf/number format/precision=2
},
 y label style={at={(axis description cs:-0.05,.5)}}
]
\addplot [semithick, color5]
table [y expr=(\thisrowno{1}-678611)/1000] {%
0 678611
5 678611
10 678611
15 678611
20 678612
25 678612
30 678613
35 678613
40 678613
45 678614
50 678614
55 678614
60 678615
65 678615
70 678615
75 678616
80 678616
85 678616
90 678617
95 678617
100 678617
105 678618
110 678618
115 678619
120 678619
125 678619
130 678620
135 678620
140 678620
145 678621
150 678621
155 678621
160 678622
165 678622
170 678622
175 678623
180 678623
185 678623
190 678624
195 678624
200 678625
205 678625
210 678625
215 678626
220 678626
225 678626
230 678627
235 678627
240 678627
245 678628
250 678628
255 678628
260 678629
265 678629
270 678629
275 678630
280 678630
285 678630
290 678631
295 678631
300 678632
305 678632
310 678632
315 678633
320 678633
325 678633
330 678634
335 678634
340 678634
345 678635
350 678635
355 678635
360 678636
365 678636
370 678636
375 678637
380 678637
385 678637
390 678638
395 678638
400 678639
405 678639
410 678639
415 678640
420 678640
425 678640
430 678641
435 678641
440 678641
445 678642
450 678642
455 678642
460 678643
465 678643
470 678643
475 678644
480 678644
485 678644
490 678645
495 678645
500 678645
505 678646
510 678646
515 678647
520 678647
525 678647
530 678648
535 678648
540 678648
545 678649
550 678649
555 678649
560 678650
565 678650
570 678650
575 678651
580 678651
585 678651
590 678652
595 678652
600 678653
605 678653
610 678653
615 678654
620 678654
625 678654
630 678655
635 678655
640 678655
645 678656
650 678656
655 678656
660 678657
665 678657
670 678657
675 678658
680 678658
685 678659
690 678659
695 678659
700 678660
705 678660
710 678660
715 678661
720 678661
725 678661
730 678662
735 678662
740 678662
745 678663
750 678663
755 678663
760 678664
765 678664
770 678664
775 678665
780 678665
785 678666
790 678666
795 678666
800 678667
805 678667
810 678667
815 678668
820 678668
825 678668
830 678669
835 678669
840 678669
845 678670
850 678670
855 678670
860 678671
865 678671
870 678671
875 678672
880 678672
885 678672
890 678673
895 678673
900 678674
905 678674
910 678674
915 678675
920 678675
925 678675
930 678676
935 678676
940 678676
945 678677
950 678677
955 678677
960 678678
965 678678
970 678678
975 678679
980 678679
985 678679
990 678680
995 678680
1000 678680
1005 678681
1010 678681
1015 678682
1020 678682
1025 678682
1030 678683
1035 678683
1040 678683
1045 678684
1050 678684
1055 678684
1060 678685
1065 678685
1070 678685
1075 678686
1080 678686
1085 678686
1090 678687
1095 678687
1100 678688
1105 678688
1110 678688
1115 678689
1120 678689
1125 678689
1130 678690
1135 678690
1140 678690
1145 678691
1150 678691
1155 678691
1160 678692
1165 678692
1170 678692
1175 678693
1180 678693
1185 678694
1190 678694
1195 678694
1200 678695
1205 678695
1210 678695
1215 678696
1220 678696
1225 678696
1230 678697
1235 678697
1240 678697
1245 678698
1250 678698
1255 678698
1260 678699
1265 678699
1270 678699
1275 678700
1280 678700
1285 678701
1290 678701
1295 678701
1300 678702
1305 678702
1310 678702
1315 678703
1320 678703
1325 678703
1330 678704
1335 678704
1340 678704
1345 678705
1350 678705
1355 678705
1360 678706
1365 678706
1370 678706
1375 678707
1380 678707
1385 678707
1390 678708
1395 678708
1400 678709
1405 678709
1410 678709
1415 678710
1420 678710
1425 678710
1430 678711
1435 678711
1440 678711
1445 678712
1450 678712
1455 678712
1460 678713
1465 678713
1470 678713
1475 678714
1480 678714
1485 678714
1490 678715
1495 678715
1500 678716
1505 678716
1510 678716
1515 678717
1520 678717
1525 678717
1530 678718
1535 678718
1540 678718
1545 678719
1550 678719
1555 678719
1560 678720
1565 678720
1570 678720
1575 678721
1580 678721
1585 678721
1590 678722
1595 678722
1600 678723
1605 678723
1610 678723
1615 678724
1620 678724
1625 678724
1630 678725
1635 678725
1640 678725
1645 678726
1650 678726
1655 678726
1660 678727
1665 678727
1670 678727
1675 678728
1680 678728
1685 678729
1690 678729
1695 678729
1700 678730
1705 678730
1710 678730
1715 678731
1720 678731
1725 678731
1730 678732
1735 678732
1740 678732
1745 678733
1750 678733
1755 678733
1760 678734
1765 678734
1770 678734
1775 678735
1780 678735
1785 678736
1790 678736
1795 678736
1800 678737
1805 678737
1810 678737
1815 678738
1820 678738
1825 678738
1830 678739
1835 678739
1840 678739
1845 678740
1850 678740
1855 678740
1860 678741
1865 678741
1870 678742
1875 678742
1880 678742
1885 678743
1890 678743
1895 678743
1900 678744
1905 678744
1910 678744
1915 678745
1920 678745
1925 678745
1930 678746
1935 678746
1940 678746
1945 678747
1950 678747
1955 678747
1960 678748
1965 678748
1970 678748
1975 678749
1980 678749
1985 678750
1990 678750
1995 678750
2000 678751
2005 678751
2010 678751
2015 678752
2020 678752
2025 678752
2030 678753
2035 678753
2040 678753
2045 678754
2050 678754
2055 678754
2060 678755
2065 678755
2070 678755
2075 678756
2080 678756
2085 678757
2090 678757
2095 678757
2100 678758
2105 678758
2110 678758
2115 678759
2120 678759
2125 678759
2130 678760
2135 678760
2140 678760
2145 678761
2150 678761
2155 678761
2160 678762
2165 678762
2170 678763
2175 678763
2180 678763
2185 678764
2190 678764
2195 678764
2200 678765
2205 678765
2210 678765
2215 678766
2220 678766
2225 678766
2230 678767
2235 678767
2240 678767
2245 678768
2250 678768
2255 678768
2260 678769
2265 678769
2270 678769
2275 678770
2280 678770
2285 678771
2290 678771
2295 678771
2300 678772
2305 678772
2310 678772
2315 678773
2320 678773
2325 678773
2330 678774
2335 678774
2340 678774
2345 678775
2350 678775
2355 678776
2360 678776
2365 678776
2370 678777
2375 678777
2380 678777
2385 678778
2390 678778
2395 678778
2400 678779
2405 678779
2410 678779
2415 678780
2420 678780
2425 678780
2430 678781
2435 678781
2440 678782
2445 678782
2450 678782
2455 678783
2460 678783
2465 678783
2470 678784
2475 678784
2480 678784
2485 678785
2490 678785
2495 678785
2500 678786
2505 678786
2510 678786
2515 678787
2520 678787
2525 678788
2530 678788
2535 678788
2540 678789
2545 678789
2550 678789
2555 678790
2560 678790
2565 678790
2570 678791
2575 678791
2580 678791
2585 678792
2590 678792
2595 678792
2600 678793
2605 678793
2610 678793
2615 678794
2620 678794
2625 678795
2630 678795
2635 678795
2640 678796
2645 678796
2650 678796
2655 678797
2660 678797
2665 678797
2670 678798
2675 678798
2680 678798
2685 678799
2690 678799
2695 678799
2700 678800
2705 678800
2710 678800
2715 678801
2720 678801
2725 678802
2730 678802
2735 678802
2740 678803
2745 678803
2750 678803
2755 678804
2760 678804
2765 678804
2770 678805
2775 678805
2780 678805
2785 678806
2790 678806
2795 678807
2800 678807
2805 678807
2810 678808
2815 678808
2820 678808
2825 678809
2830 678809
2835 678809
2840 678810
2845 678810
2850 678810
2855 678811
2860 678811
2865 678811
2870 678812
2875 678812
2880 678812
2885 678813
2890 678813
2895 678813
2900 678814
2905 678814
2910 678815
2915 678815
2920 678815
2925 678816
2930 678816
2935 678816
2940 678817
2945 678817
2950 678817
2955 678818
2960 678818
2965 678818
2970 678819
2975 678819
2980 678819
2985 678820
2990 678820
2995 678821
3000 678821
3005 678821
3010 678822
3015 678822
3020 678822
3025 678823
3030 678823
3035 678823
3040 678824
3045 678824
3050 678824
3055 678825
3060 678825
3065 678825
3070 678826
3075 678826
3080 678827
3085 678827
3090 678827
3095 678828
3100 678828
3105 678828
3110 678829
3115 678829
3120 678829
3125 678830
3130 678830
3135 678830
3140 678831
3145 678831
3150 678831
3155 678832
3160 678832
3165 678832
3170 678833
3175 678833
3180 678833
3185 678834
3190 678834
3195 678835
3200 678835
3205 678835
3210 678836
3215 678836
3220 678836
3225 678837
3230 678837
3235 678837
3240 678838
3245 678838
3250 678838
3255 678839
3260 678839
3265 678839
3270 678840
3275 678840
3280 678841
3285 678841
3290 678841
3295 678842
3300 678842
3305 678842
3310 678843
3315 678843
3320 678843
3325 678844
3330 678844
3335 678844
3340 678845
3345 678845
3350 678845
3355 678846
3360 678846
3365 678846
3370 678847
3375 678847
3380 678847
3385 678848
3390 678848
3395 678849
3400 678849
3405 678849
3410 678850
3415 678850
3420 678850
3425 678851
3430 678851
3435 678851
3440 678852
3445 678852
3450 678852
3455 678853
3460 678853
3465 678853
3470 678854
3475 678854
3480 678855
3485 678855
3490 678855
3495 678856
3500 678856
3505 678856
3510 678857
3515 678857
3520 678857
3525 678858
3530 678858
3535 678858
3540 678859
3545 678859
3550 678859
3555 678860
3560 678860
3565 678861
3570 678861
3575 678861
3580 678862
3585 678862
3590 678862
3595 678863
3600 678863
};
\addlegendentry{Baseline}
\addplot [semithick, color2, mark=triangle*, mark size=3, mark options={solid,fill=white,draw=red},  mark repeat={180}]
table[y expr=(\thisrowno{1}-368832)/1000] {%
0 368832
5 368832
10 368833
15 368833
20 368833
25 368834
30 368834
35 368834
40 368835
45 368835
50 368835
55 368836
60 368836
65 368836
70 368837
75 368837
80 368838
85 368838
90 368838
95 368839
100 368839
105 368839
110 368840
115 368840
120 368840
125 368841
130 368841
135 368842
140 368842
145 368842
150 368843
155 368843
160 368843
165 368844
170 368844
175 368845
180 368845
185 368846
190 368846
195 368846
200 368847
205 368847
210 368848
215 368849
220 368849
225 368849
230 368850
235 368850
240 368851
245 368852
250 368852
255 368852
260 368853
265 368853
270 368854
275 368854
280 368855
285 368855
290 368856
295 368856
300 368857
305 368857
310 368858
315 368859
320 368859
325 368860
330 368860
335 368861
340 368861
345 368862
350 368862
355 368863
360 368863
365 368864
370 368864
375 368865
380 368865
385 368866
390 368866
395 368867
400 368867
405 368868
410 368868
415 368869
420 368869
425 368870
430 368870
435 368871
440 368871
445 368872
450 368872
455 368873
460 368873
465 368874
470 368874
475 368875
480 368875
485 368876
490 368876
495 368877
500 368877
505 368878
510 368878
515 368879
520 368879
525 368880
530 368880
535 368880
540 368881
545 368881
550 368882
555 368882
560 368883
565 368883
570 368884
575 368885
580 368885
585 368886
590 368886
595 368887
600 368887
605 368888
610 368888
615 368889
620 368889
625 368890
630 368890
635 368891
640 368891
645 368892
650 368892
655 368893
660 368893
665 368894
670 368894
675 368895
680 368895
685 368896
690 368896
695 368897
700 368897
705 368898
710 368898
715 368899
720 368899
725 368900
730 368900
735 368901
740 368901
745 368902
750 368902
755 368903
760 368903
765 368904
770 368904
775 368905
780 368905
785 368906
790 368906
795 368907
800 368907
805 368908
810 368908
815 368909
820 368909
825 368910
830 368911
835 368911
840 368912
845 368912
850 368913
855 368913
860 368914
865 368914
870 368915
875 368915
880 368916
885 368916
890 368917
895 368917
900 368918
905 368918
910 368919
915 368919
920 368920
925 368920
930 368921
935 368921
940 368922
945 368922
950 368923
955 368923
960 368924
965 368924
970 368925
975 368925
980 368926
985 368926
990 368927
995 368927
1000 368928
1005 368928
1010 368929
1015 368929
1020 368930
1025 368930
1030 368931
1035 368931
1040 368932
1045 368932
1050 368933
1055 368934
1060 368934
1065 368935
1070 368935
1075 368936
1080 368936
1085 368937
1090 368937
1095 368938
1100 368938
1105 368939
1110 368939
1115 368940
1120 368940
1125 368941
1130 368941
1135 368942
1140 368942
1145 368943
1150 368943
1155 368944
1160 368944
1165 368945
1170 368945
1175 368946
1180 368946
1185 368947
1190 368947
1195 368948
1200 368948
1205 368949
1210 368949
1215 368950
1220 368950
1225 368951
1230 368951
1235 368952
1240 368952
1245 368953
1250 368953
1255 368954
1260 368954
1265 368955
1270 368955
1275 368956
1280 368956
1285 368957
1290 368958
1295 368958
1300 368959
1305 368959
1310 368960
1315 368960
1320 368961
1325 368961
1330 368962
1335 368962
1340 368963
1345 368963
1350 368964
1355 368964
1360 368965
1365 368965
1370 368966
1375 368966
1380 368967
1385 368967
1390 368968
1395 368968
1400 368969
1405 368969
1410 368970
1415 368970
1420 368971
1425 368971
1430 368972
1435 368972
1440 368973
1445 368973
1450 368974
1455 368974
1460 368975
1465 368975
1470 368976
1475 368976
1480 368977
1485 368977
1490 368978
1495 368978
1500 368979
1505 368979
1510 368980
1515 368980
1520 368981
1525 368982
1530 368982
1535 368983
1540 368983
1545 368984
1550 368984
1555 368985
1560 368985
1565 368986
1570 368986
1575 368987
1580 368987
1585 368988
1590 368988
1595 368989
1600 368989
1605 368990
1610 368990
1615 368991
1620 368991
1625 368992
1630 368992
1635 368993
1640 368993
1645 368994
1650 368994
1655 368995
1660 368995
1665 368996
1670 368996
1675 368997
1680 368997
1685 368998
1690 368998
1695 368999
1700 368999
1705 369000
1710 369000
1715 369001
1720 369001
1725 369002
1730 369002
1735 369003
1740 369003
1745 369004
1750 369004
1755 369005
1760 369005
1765 369006
1770 369006
1775 369007
1780 369007
1785 369008
1790 369008
1795 369009
1800 369009
1805 369010
1810 369010
1815 369011
1820 369011
1825 369012
1830 369012
1835 369013
1840 369014
1845 369014
1850 369014
1855 369015
1860 369016
1865 369016
1870 369017
1875 369017
1880 369018
1885 369018
1890 369019
1895 369019
1900 369020
1905 369020
1910 369021
1915 369021
1920 369022
1925 369022
1930 369023
1935 369023
1940 369024
1945 369024
1950 369025
1955 369025
1960 369026
1965 369026
1970 369027
1975 369027
1980 369028
1985 369028
1990 369029
1995 369029
2000 369030
2005 369030
2010 369031
2015 369031
2020 369032
2025 369032
2030 369033
2035 369033
2040 369034
2045 369034
2050 369035
2055 369035
2060 369036
2065 369036
2070 369037
2075 369037
2080 369038
2085 369038
2090 369039
2095 369040
2100 369040
2105 369041
2110 369041
2115 369041
2120 369042
2125 369043
2130 369043
2135 369044
2140 369044
2145 369045
2150 369045
2155 369046
2160 369046
2165 369047
2170 369047
2175 369048
2180 369048
2185 369049
2190 369049
2195 369050
2200 369050
2205 369051
2210 369051
2215 369052
2220 369052
2225 369053
2230 369053
2235 369054
2240 369054
2245 369055
2250 369055
2255 369056
2260 369056
2265 369057
2270 369057
2275 369058
2280 369058
2285 369059
2290 369059
2295 369060
2300 369060
2305 369061
2310 369061
2315 369062
2320 369062
2325 369063
2330 369063
2335 369064
2340 369064
2345 369065
2350 369065
2355 369066
2360 369066
2365 369067
2370 369067
2375 369068
2380 369068
2385 369069
2390 369070
2395 369070
2400 369070
2405 369071
2410 369071
2415 369072
2420 369073
2425 369073
2430 369074
2435 369074
2440 369075
2445 369075
2450 369075
2455 369076
2460 369076
2465 369077
2470 369077
2475 369077
2480 369078
2485 369078
2490 369079
2495 369079
2500 369079
2505 369080
2510 369080
2515 369080
2520 369081
2525 369081
2530 369081
2535 369082
2540 369082
2545 369082
2550 369083
2555 369083
2560 369083
2565 369084
2570 369084
2575 369085
2580 369085
2585 369085
2590 369086
2595 369086
2600 369086
2605 369087
2610 369087
2615 369087
2620 369088
2625 369088
2630 369088
2635 369089
2640 369089
2645 369089
2650 369090
2655 369090
2660 369091
2665 369091
2670 369091
2675 369092
2680 369092
2685 369092
2690 369093
2695 369093
2700 369093
2705 369094
2710 369094
2715 369094
2720 369095
2725 369095
2730 369095
2735 369096
2740 369096
2745 369097
2750 369097
2755 369097
2760 369098
2765 369098
2770 369098
2775 369099
2780 369099
2785 369099
2790 369100
2795 369100
2800 369100
2805 369101
2810 369101
2815 369101
2820 369102
2825 369102
2830 369102
2835 369103
2840 369103
2845 369104
2850 369104
2855 369104
2860 369105
2865 369105
2870 369105
2875 369106
2880 369106
2885 369106
2890 369107
2895 369107
2900 369107
2905 369108
2910 369108
2915 369108
2920 369109
2925 369109
2930 369109
2935 369110
2940 369110
2945 369110
2950 369111
2955 369111
2960 369111
2965 369112
2970 369112
2975 369113
2980 369113
2985 369113
2990 369114
2995 369114
3000 369114
3005 369115
3010 369115
3015 369115
3020 369116
3025 369116
3030 369116
3035 369117
3040 369117
3045 369117
3050 369118
3055 369118
3060 369118
3065 369119
3070 369119
3075 369119
3080 369120
3085 369120
3090 369121
3095 369121
3100 369121
3105 369122
3110 369122
3115 369122
3120 369123
3125 369123
3130 369123
3135 369124
3140 369124
3145 369124
3150 369125
3155 369125
3160 369125
3165 369126
3170 369126
3175 369126
3180 369127
3185 369127
3190 369128
3195 369128
3200 369128
3205 369129
3210 369129
3215 369129
3220 369130
3225 369130
3230 369130
3235 369131
3240 369131
3245 369131
3250 369132
3255 369132
3260 369132
3265 369133
3270 369133
3275 369134
3280 369134
3285 369134
3290 369135
3295 369135
3300 369135
3305 369136
3310 369136
3315 369136
3320 369137
3325 369137
3330 369137
3335 369138
3340 369138
3345 369138
3350 369139
3355 369139
3360 369139
3365 369140
3370 369140
3375 369141
3380 369141
3385 369141
3390 369142
3395 369142
3400 369142
3405 369143
3410 369143
3415 369143
3420 369144
3425 369144
3430 369144
3435 369145
3440 369145
3445 369145
3450 369146
3455 369146
3460 369146
3465 369147
3470 369147
3475 369148
3480 369148
3485 369148
3490 369149
3495 369149
3500 369149
3505 369150
3510 369150
3515 369150
3520 369151
3525 369151
3530 369151
3535 369152
3540 369152
3545 369153
3550 369153
3555 369153
3560 369154
3565 369154
3570 369154
3575 369155
3580 369155
3585 369155
3590 369156
3595 369156
3600 369156
3605 369157
3610 369157
3615 369157
3620 369158
3625 369158
3630 369159
3635 369159
3640 369159
3645 369160
3650 369160
3655 369160
3660 369161
3665 369161
3670 369161
3675 369162
3680 369162
3685 369162
3690 369163
3695 369163
3700 369163
3705 369164
3710 369164
3715 369164
3720 369165
3725 369165
3730 369166
3735 369166
3740 369166
3745 369167
3750 369167
3755 369167
3760 369168
3765 369168
3770 369168
3775 369169
3780 369169
3785 369169
3790 369170
3795 369170
3800 369170
3805 369171
3810 369171
3815 369172
3820 369172
3825 369172
3830 369173
3835 369173
3840 369173
3845 369174
3850 369174
3855 369174
3860 369175
3865 369175
3870 369175
3875 369176
3880 369176
3885 369176
3890 369177
3895 369177
3900 369177
3905 369178
3910 369178
3915 369178
3920 369179
3925 369179
3930 369180
3935 369180
3940 369180
3945 369181
3950 369181
3955 369181
3960 369182
3965 369182
3970 369182
3975 369183
3980 369183
3985 369183
3990 369184
3995 369184
4000 369184
4005 369185
4010 369185
4015 369185
4020 369186
4025 369186
4030 369186
4035 369187
4040 369187
4045 369188
4050 369188
4055 369188
4060 369189
4065 369189
4070 369189
4075 369190
4080 369190
4085 369190
4090 369191
4095 369191
4100 369191
4105 369192
4110 369192
4115 369192
4120 369193
4125 369193
4130 369194
4135 369194
4140 369194
4145 369195
4150 369195
4155 369195
4160 369196
4165 369196
4170 369196
4175 369197
4180 369197
4185 369197
4190 369198
4195 369198
4200 369198
4205 369199
4210 369199
4215 369200
4220 369200
4225 369200
4230 369201
4235 369201
4240 369201
4245 369202
4250 369202
4255 369202
4260 369203
4265 369203
4270 369203
4275 369204
4280 369204
4285 369204
4290 369205
4295 369205
4300 369205
4305 369206
4310 369206
4315 369207
4320 369207
4325 369207
4330 369208
4335 369208
4340 369208
4345 369209
4350 369209
4355 369209
4360 369210
4365 369210
4370 369210
4375 369211
4380 369211
4385 369211
4390 369212
4395 369212
4400 369212
4405 369213
4410 369213
4415 369213
4420 369214
4425 369214
4430 369215
4435 369215
4440 369215
4445 369216
4450 369216
4455 369216
4460 369217
4465 369217
4470 369217
4475 369218
4480 369218
4485 369218
4490 369219
4495 369219
4500 369219
4505 369220
4510 369220
4515 369220
4520 369221
4525 369221
4530 369221
4535 369222
4540 369222
4545 369223
4550 369223
4555 369223
4560 369224
4565 369224
4570 369224
4575 369225
4580 369225
4585 369225
4590 369226
4595 369226
4600 369226
4605 369227
4610 369227
4615 369227
4620 369228
4625 369228
4630 369229
4635 369229
4640 369229
4645 369230
4650 369230
4655 369230
4660 369231
4665 369231
4670 369231
4675 369232
4680 369232
4685 369232
4690 369233
4695 369233
4700 369233
4705 369234
4710 369234
4715 369234
4720 369235
4725 369235
4730 369236
4735 369236
4740 369236
4745 369237
4750 369237
4755 369237
4760 369238
4765 369238
4770 369239
4775 369239
4780 369239
4785 369240
4790 369240
4795 369240
4800 369241
4805 369241
4810 369241
4815 369242
4820 369242
4825 369242
4830 369243
4835 369243
4840 369243
4845 369244
4850 369244
4855 369244
4860 369245
4865 369245
4870 369245
4875 369246
4880 369246
4885 369247
4890 369247
4895 369247
4900 369248
4905 369248
4910 369248
4915 369249
4920 369249
4925 369249
4930 369250
4935 369250
4940 369250
4945 369251
4950 369251
4955 369251
4960 369252
4965 369252
4970 369252
4975 369253
4980 369253
4985 369253
4990 369254
4995 369254
5000 369255
5005 369255
5010 369255
5015 369256
5020 369256
5025 369256
5030 369257
5035 369257
5040 369257
5045 369258
5050 369258
5055 369258
5060 369259
5065 369259
5070 369259
5075 369260
5080 369260
5085 369260
5090 369261
5095 369261
5100 369262
5105 369262
5110 369262
5115 369263
5120 369263
5125 369263
};
\addlegendentry{Different sockets}
\addplot[semithick, color3, mark=square*, mark size=3, mark options={solid,fill=white,draw=red,fill opacity=0.4},  mark repeat={180}]
table[y expr=(\thisrowno{1}-459130)/1000] {%
0 459130
5 459131
10 459131
15 459131
20 459132
25 459132
30 459132
35 459133
40 459133
45 459134
50 459134
55 459134
60 459135
65 459135
70 459135
75 459136
80 459136
85 459136
90 459137
95 459137
100 459138
105 459138
110 459139
115 459139
120 459139
125 459140
130 459140
135 459140
140 459141
145 459141
150 459142
155 459142
160 459142
165 459143
170 459143
175 459144
180 459144
185 459145
190 459145
195 459146
200 459146
205 459147
210 459147
215 459148
220 459148
225 459149
230 459149
235 459150
240 459150
245 459151
250 459151
255 459152
260 459152
265 459153
270 459153
275 459154
280 459154
285 459155
290 459155
295 459156
300 459156
305 459157
310 459157
315 459158
320 459158
325 459159
330 459159
335 459160
340 459160
345 459161
350 459161
355 459162
360 459162
365 459163
370 459163
375 459164
380 459164
385 459165
390 459165
395 459166
400 459166
405 459167
410 459167
415 459168
420 459168
425 459169
430 459169
435 459170
440 459170
445 459171
450 459171
455 459172
460 459172
465 459173
470 459173
475 459174
480 459174
485 459175
490 459175
495 459176
500 459176
505 459177
510 459177
515 459178
520 459178
525 459179
530 459179
535 459180
540 459180
545 459181
550 459181
555 459182
560 459182
565 459183
570 459183
575 459184
580 459184
585 459185
590 459185
595 459186
600 459186
605 459187
610 459187
615 459188
620 459188
625 459189
630 459189
635 459190
640 459190
645 459191
650 459191
655 459192
660 459192
665 459193
670 459193
675 459194
680 459194
685 459195
690 459195
695 459196
700 459196
705 459197
710 459197
715 459198
720 459198
725 459199
730 459199
735 459200
740 459200
745 459201
750 459201
755 459202
760 459202
765 459203
770 459203
775 459204
780 459204
785 459205
790 459205
795 459206
800 459206
805 459207
810 459207
815 459208
820 459208
825 459209
830 459209
835 459210
840 459210
845 459211
850 459211
855 459212
860 459212
865 459213
870 459213
875 459214
880 459214
885 459215
890 459215
895 459216
900 459216
905 459217
910 459217
915 459218
920 459218
925 459219
930 459219
935 459220
940 459220
945 459221
950 459221
955 459222
960 459222
965 459223
970 459223
975 459224
980 459224
985 459225
990 459225
995 459226
1000 459226
1005 459227
1010 459227
1015 459228
1020 459228
1025 459229
1030 459229
1035 459230
1040 459230
1045 459231
1050 459231
1055 459232
1060 459232
1065 459233
1070 459233
1075 459234
1080 459234
1085 459235
1090 459235
1095 459236
1100 459236
1105 459237
1110 459237
1115 459238
1120 459238
1125 459239
1130 459239
1135 459240
1140 459240
1145 459241
1150 459241
1155 459242
1160 459242
1165 459243
1170 459243
1175 459244
1180 459244
1185 459245
1190 459245
1195 459246
1200 459246
1205 459247
1210 459247
1215 459248
1220 459248
1225 459249
1230 459249
1235 459250
1240 459250
1245 459251
1250 459251
1255 459252
1260 459252
1265 459253
1270 459253
1275 459254
1280 459254
1285 459255
1290 459255
1295 459256
1300 459256
1305 459257
1310 459257
1315 459258
1320 459258
1325 459259
1330 459259
1335 459260
1340 459260
1345 459261
1350 459261
1355 459262
1360 459262
1365 459263
1370 459263
1375 459264
1380 459264
1385 459265
1390 459265
1395 459266
1400 459266
1405 459267
1410 459267
1415 459268
1420 459268
1425 459269
1430 459269
1435 459270
1440 459270
1445 459271
1450 459271
1455 459272
1460 459272
1465 459273
1470 459273
1475 459274
1480 459274
1485 459275
1490 459275
1495 459276
1500 459276
1505 459277
1510 459277
1515 459278
1520 459278
1525 459279
1530 459279
1535 459280
1540 459280
1545 459281
1550 459281
1555 459282
1560 459282
1565 459283
1570 459283
1575 459284
1580 459284
1585 459285
1590 459285
1595 459286
1600 459286
1605 459287
1610 459287
1615 459288
1620 459288
1625 459289
1630 459289
1635 459290
1640 459290
1645 459291
1650 459292
1655 459292
1660 459292
1665 459293
1670 459294
1675 459294
1680 459294
1685 459295
1690 459296
1695 459296
1700 459296
1705 459297
1710 459297
1715 459298
1720 459298
1725 459299
1730 459299
1735 459300
1740 459300
1745 459301
1750 459301
1755 459302
1760 459302
1765 459303
1770 459303
1775 459304
1780 459304
1785 459305
1790 459305
1795 459306
1800 459306
1805 459307
1810 459307
1815 459308
1820 459308
1825 459309
1830 459309
1835 459310
1840 459310
1845 459311
1850 459311
1855 459312
1860 459312
1865 459313
1870 459313
1875 459314
1880 459314
1885 459315
1890 459316
1895 459316
1900 459317
1905 459317
1910 459318
1915 459318
1920 459319
1925 459319
1930 459320
1935 459320
1940 459321
1945 459321
1950 459322
1955 459322
1960 459323
1965 459323
1970 459324
1975 459324
1980 459325
1985 459325
1990 459326
1995 459326
2000 459327
2005 459327
2010 459328
2015 459328
2020 459329
2025 459329
2030 459330
2035 459330
2040 459331
2045 459331
2050 459332
2055 459332
2060 459333
2065 459333
2070 459334
2075 459334
2080 459335
2085 459335
2090 459336
2095 459336
2100 459337
2105 459337
2110 459338
2115 459338
2120 459339
2125 459339
2130 459340
2135 459340
2140 459341
2145 459341
2150 459342
2155 459342
2160 459343
2165 459343
2170 459344
2175 459344
2180 459345
2185 459345
2190 459346
2195 459346
2200 459347
2205 459347
2210 459348
2215 459348
2220 459349
2225 459349
2230 459350
2235 459350
2240 459351
2245 459351
2250 459352
2255 459352
2260 459353
2265 459353
2270 459354
2275 459354
2280 459355
2285 459355
2290 459356
2295 459356
2300 459357
2305 459357
2310 459358
2315 459358
2320 459359
2325 459359
2330 459360
2335 459360
2340 459361
2345 459361
2350 459362
2355 459362
2360 459363
2365 459363
2370 459364
2375 459364
2380 459365
2385 459365
2390 459366
2395 459366
2400 459366
2405 459367
2410 459367
2415 459368
2420 459368
2425 459368
2430 459369
2435 459369
2440 459369
2445 459370
2450 459370
2455 459370
2460 459371
2465 459371
2470 459371
2475 459372
2480 459372
2485 459372
2490 459373
2495 459373
2500 459373
2505 459374
2510 459374
2515 459374
2520 459375
2525 459375
2530 459375
2535 459376
2540 459376
2545 459376
2550 459377
2555 459377
2560 459377
2565 459378
2570 459378
2575 459379
2580 459379
2585 459379
2590 459380
2595 459380
2600 459380
2605 459381
2610 459381
2615 459381
2620 459382
2625 459382
2630 459382
2635 459383
2640 459383
2645 459383
2650 459384
2655 459384
2660 459384
2665 459385
2670 459385
2675 459385
2680 459386
2685 459386
2690 459387
2695 459387
2700 459387
2705 459388
2710 459388
2715 459388
2720 459389
2725 459389
2730 459389
2735 459390
2740 459390
2745 459390
2750 459391
2755 459391
2760 459391
2765 459392
2770 459392
2775 459392
2780 459393
2785 459393
2790 459394
2795 459394
2800 459394
2805 459395
2810 459395
2815 459395
2820 459396
2825 459396
2830 459396
2835 459397
2840 459397
2845 459397
2850 459398
2855 459398
2860 459398
2865 459399
2870 459399
2875 459399
2880 459400
2885 459400
2890 459400
2895 459401
2900 459401
2905 459401
2910 459402
2915 459402
2920 459403
2925 459403
2930 459403
2935 459404
2940 459404
2945 459404
2950 459405
2955 459405
2960 459405
2965 459406
2970 459406
2975 459406
2980 459407
2985 459407
2990 459407
2995 459408
3000 459408
3005 459408
3010 459409
3015 459409
3020 459410
3025 459410
3030 459410
3035 459410
3040 459411
3045 459411
3050 459412
3055 459412
3060 459412
3065 459413
3070 459413
3075 459413
3080 459414
3085 459414
3090 459415
3095 459415
3100 459415
3105 459416
3110 459416
3115 459416
3120 459417
3125 459417
3130 459418
3135 459418
3140 459418
3145 459418
3150 459419
3155 459419
3160 459420
3165 459420
3170 459420
3175 459421
3180 459421
3185 459421
3190 459422
3195 459422
3200 459422
3205 459423
3210 459423
3215 459423
3220 459424
3225 459424
3230 459424
3235 459425
3240 459425
3245 459426
3250 459426
3255 459426
3260 459427
3265 459427
3270 459427
3275 459428
3280 459428
3285 459428
3290 459429
3295 459429
3300 459429
3305 459430
3310 459430
3315 459430
3320 459431
3325 459431
3330 459431
3335 459432
3340 459432
3345 459433
3350 459433
3355 459433
3360 459434
3365 459434
3370 459434
3375 459435
3380 459435
3385 459435
3390 459436
3395 459436
3400 459436
3405 459437
3410 459437
3415 459437
3420 459438
3425 459438
3430 459438
3435 459439
3440 459439
3445 459440
3450 459440
3455 459440
3460 459441
3465 459441
3470 459441
3475 459442
3480 459442
3485 459442
3490 459443
3495 459443
3500 459443
3505 459444
3510 459444
3515 459444
3520 459445
3525 459445
3530 459445
3535 459446
3540 459446
3545 459446
3550 459447
3555 459447
3560 459448
3565 459448
3570 459448
3575 459449
3580 459449
3585 459449
3590 459450
3595 459450
3600 459450
};
\addlegendentry{Same socket}
\addplot [semithick, color2,mark=triangle*, mark size=3, mark options={solid,fill=red,draw=red}]
table[y expr=(\thisrowno{1}-459130)/1000]{
2535 459376
};
\addplot [semithick, color1, mark=square*, mark size=3, mark options={solid,fill=red,draw=red}]
table[y expr=(\thisrowno{1}-368832)/1000]{
2435 369074
};
\end{axis}
\node[text width=8cm] at (3,-1.5) {Filled markers indicate the end of jobs};

\end{tikzpicture}

%% file: results/energy_intel_sockets_2jobs.tex
\begin{tikzpicture}[font=\Large]

\definecolor{color0}{rgb}{0.12156862745098,0.466666666666667,0.705882352941177}
\definecolor{color1}{rgb}{1,0.498039215686275,0.0549019607843137}
\definecolor{color2}{rgb}{0.172549019607843,0.627450980392157,0.172549019607843}
\definecolor{color3}{rgb}{0.83921568627451,0.152941176470588,0.156862745098039}
\definecolor{color4}{rgb}{0.580392156862745,0.403921568627451,0.741176470588235}
\definecolor{color5}{rgb}{0,0,0}

\begin{axis}[
legend cell align={left},
legend columns=2,
legend style={fill opacity=0.8, draw opacity=1, text opacity=1, at={(1.05,1.15)}, anchor=east, draw=white!80.0!black},
tick align=outside,
tick pos=left,
x grid style={white!69.01960784313725!black},
xlabel={Time (SEC)},
xmin=0, xmax=5400,
xtick style={color=black},
y grid style={white!69.01960784313725!black},
ylabel={Total energy consumption (kWh)},
ymin=0, ymax=0.55,
ytick style={color=black},
xmajorgrids,
ymajorgrids,
ytick={0.05,0.1,0.15,0.2,0.25,0.3,0.35,0.4,0.45,0.5, 0.55},
yticklabel style={
        /pgf/number format/fixed,
        /pgf/number format/precision=2
},
 y label style={at={(axis description cs:-0.05,.5)}}
]
\addplot [semithick, color5]
table [y expr=(\thisrowno{1}-521738)/1000] {%
0 521738
5 521738
10 521739
15 521739
20 521739
25 521740
30 521740
35 521740
40 521741
45 521741
50 521741
55 521742
60 521742
65 521743
70 521743
75 521743
80 521744
85 521744
90 521744
95 521745
100 521745
105 521745
110 521746
115 521746
120 521746
125 521747
130 521747
135 521747
140 521748
145 521748
150 521748
155 521749
160 521749
165 521749
170 521750
175 521750
180 521751
185 521751
190 521751
195 521752
200 521752
205 521752
210 521753
215 521753
220 521753
225 521754
230 521754
235 521754
240 521755
245 521755
250 521755
255 521756
260 521756
265 521757
270 521757
275 521757
280 521758
285 521758
290 521758
295 521759
300 521759
305 521759
310 521760
315 521760
320 521760
325 521761
330 521761
335 521761
340 521762
345 521762
350 521762
355 521763
360 521763
365 521763
370 521764
375 521764
380 521765
385 521765
390 521765
395 521766
400 521766
405 521766
410 521767
415 521767
420 521767
425 521768
430 521768
435 521768
440 521769
445 521769
450 521769
455 521770
460 521770
465 521771
470 521771
475 521771
480 521772
485 521772
490 521772
495 521773
500 521773
505 521773
510 521774
515 521774
520 521774
525 521775
530 521775
535 521775
540 521776
545 521776
550 521776
555 521777
560 521777
565 521778
570 521778
575 521778
580 521779
585 521779
590 521779
595 521780
600 521780
605 521780
610 521781
615 521781
620 521781
625 521782
630 521782
635 521782
640 521783
645 521783
650 521784
655 521784
660 521784
665 521785
670 521785
675 521785
680 521786
685 521786
690 521786
695 521787
700 521787
705 521787
710 521788
715 521788
720 521788
725 521789
730 521789
735 521790
740 521790
745 521790
750 521791
755 521791
760 521791
765 521792
770 521792
775 521792
780 521793
785 521793
790 521793
795 521794
800 521794
805 521794
810 521795
815 521795
820 521795
825 521796
830 521796
835 521796
840 521797
845 521797
850 521798
855 521798
860 521798
865 521799
870 521799
875 521799
880 521800
885 521800
890 521800
895 521801
900 521801
905 521801
910 521802
915 521802
920 521802
925 521803
930 521803
935 521804
940 521804
945 521804
950 521805
955 521805
960 521805
965 521806
970 521806
975 521806
980 521807
985 521807
990 521807
995 521808
1000 521808
1005 521808
1010 521809
1015 521809
1020 521809
1025 521810
1030 521810
1035 521810
1040 521811
1045 521811
1050 521812
1055 521812
1060 521812
1065 521813
1070 521813
1075 521813
1080 521814
1085 521814
1090 521814
1095 521815
1100 521815
1105 521815
1110 521816
1115 521816
1120 521816
1125 521817
1130 521817
1135 521817
1140 521818
1145 521818
1150 521819
1155 521819
1160 521819
1165 521820
1170 521820
1175 521820
1180 521821
1185 521821
1190 521821
1195 521822
1200 521822
1205 521823
1210 521823
1215 521823
1220 521824
1225 521824
1230 521824
1235 521825
1240 521825
1245 521825
1250 521826
1255 521826
1260 521826
1265 521827
1270 521827
1275 521827
1280 521828
1285 521828
1290 521828
1295 521829
1300 521829
1305 521830
1310 521830
1315 521830
1320 521831
1325 521831
1330 521831
1335 521832
1340 521832
1345 521832
1350 521833
1355 521833
1360 521833
1365 521834
1370 521834
1375 521834
1380 521835
1385 521835
1390 521835
1395 521836
1400 521836
1405 521837
1410 521837
1415 521837
1420 521838
1425 521838
1430 521838
1435 521839
1440 521839
1445 521839
1450 521840
1455 521840
1460 521840
1465 521841
1470 521841
1475 521841
1480 521842
1485 521842
1490 521842
1495 521843
1500 521843
1505 521843
1510 521844
1515 521844
1520 521845
1525 521845
1530 521845
1535 521846
1540 521846
1545 521846
1550 521847
1555 521847
1560 521847
1565 521848
1570 521848
1575 521848
1580 521849
1585 521849
1590 521849
1595 521850
1600 521850
1605 521851
1610 521851
1615 521851
1620 521852
1625 521852
1630 521852
1635 521853
1640 521853
1645 521853
1650 521854
1655 521854
1660 521854
1665 521855
1670 521855
1675 521855
1680 521856
1685 521856
1690 521856
1695 521857
1700 521857
1705 521858
1710 521858
1715 521858
1720 521859
1725 521859
1730 521859
1735 521860
1740 521860
1745 521860
1750 521861
1755 521861
1760 521861
1765 521862
1770 521862
1775 521863
1780 521863
1785 521863
1790 521864
1795 521864
1800 521864
1805 521865
1810 521865
1815 521865
1820 521866
1825 521866
1830 521866
1835 521867
1840 521867
1845 521867
1850 521868
1855 521868
1860 521868
1865 521869
1870 521869
1875 521869
1880 521870
1885 521870
1890 521871
1895 521871
1900 521871
1905 521872
1910 521872
1915 521872
1920 521873
1925 521873
1930 521873
1935 521874
1940 521874
1945 521874
1950 521875
1955 521875
1960 521875
1965 521876
1970 521876
1975 521876
1980 521877
1985 521877
1990 521878
1995 521878
2000 521878
2005 521879
2010 521879
2015 521879
2020 521880
2025 521880
2030 521880
2035 521881
2040 521881
2045 521881
2050 521882
2055 521882
2060 521882
2065 521883
2070 521883
2075 521883
2080 521884
2085 521884
2090 521885
2095 521885
2100 521885
2105 521886
2110 521886
2115 521886
2120 521887
2125 521887
2130 521887
2135 521888
2140 521888
2145 521888
2150 521889
2155 521889
2160 521889
2165 521890
2170 521890
2175 521890
2180 521891
2185 521891
2190 521892
2195 521892
2200 521892
2205 521893
2210 521893
2215 521893
2220 521894
2225 521894
2230 521894
2235 521895
2240 521895
2245 521895
2250 521896
2255 521896
2260 521896
2265 521897
2270 521897
2275 521898
2280 521898
2285 521898
2290 521899
2295 521899
2300 521899
2305 521900
2310 521900
2315 521900
2320 521901
2325 521901
2330 521901
2335 521902
2340 521902
2345 521902
2350 521903
2355 521903
2360 521904
2365 521904
2370 521904
2375 521905
2380 521905
2385 521905
2390 521906
2395 521906
2400 521906
2405 521907
2410 521907
2415 521907
2420 521908
2425 521908
2430 521908
2435 521909
2440 521909
2445 521909
2450 521910
2455 521910
2460 521911
2465 521911
2470 521911
2475 521912
2480 521912
2485 521912
2490 521913
2495 521913
2500 521913
2505 521914
2510 521914
2515 521914
2520 521915
2525 521915
2530 521915
2535 521916
2540 521916
2545 521916
2550 521917
2555 521917
2560 521918
2565 521918
2570 521918
2575 521919
2580 521919
2585 521919
2590 521920
2595 521920
2600 521920
2605 521921
2610 521921
2615 521921
2620 521922
2625 521922
2630 521922
2635 521923
2640 521923
2645 521924
2650 521924
2655 521924
2660 521925
2665 521925
2670 521925
2675 521926
2680 521926
2685 521926
2690 521927
2695 521927
2700 521927
2705 521928
2710 521928
2715 521928
2720 521929
2725 521929
2730 521929
2735 521930
2740 521930
2745 521931
2750 521931
2755 521931
2760 521932
2765 521932
2770 521932
2775 521933
2780 521933
2785 521933
2790 521934
2795 521934
2800 521934
2805 521935
2810 521935
2815 521935
2820 521936
2825 521936
2830 521936
2835 521937
2840 521937
2845 521938
2850 521938
2855 521938
2860 521939
2865 521939
2870 521939
2875 521940
2880 521940
2885 521940
2890 521941
2895 521941
2900 521941
2905 521942
2910 521942
2915 521942
2920 521943
2925 521943
2930 521943
2935 521944
2940 521944
2945 521944
2950 521945
2955 521945
2960 521946
2965 521946
2970 521946
2975 521947
2980 521947
2985 521947
2990 521948
2995 521948
3000 521948
3005 521949
3010 521949
3015 521950
3020 521950
3025 521950
3030 521951
3035 521951
3040 521951
3045 521952
3050 521952
3055 521952
3060 521953
3065 521953
3070 521953
3075 521954
3080 521954
3085 521954
3090 521955
3095 521955
3100 521955
3105 521956
3110 521956
3115 521957
3120 521957
3125 521957
3130 521958
3135 521958
3140 521958
3145 521959
3150 521959
3155 521959
3160 521960
3165 521960
3170 521960
3175 521961
3180 521961
3185 521961
3190 521962
3195 521962
3200 521962
3205 521963
3210 521963
3215 521964
3220 521964
3225 521964
3230 521965
3235 521965
3240 521965
3245 521966
3250 521966
3255 521966
3260 521967
3265 521967
3270 521967
3275 521968
3280 521968
3285 521968
3290 521969
3295 521969
3300 521969
3305 521970
3310 521970
3315 521970
3320 521971
3325 521971
3330 521971
3335 521972
3340 521972
3345 521973
3350 521973
3355 521973
3360 521974
3365 521974
3370 521974
3375 521975
3380 521975
3385 521975
3390 521976
3395 521976
3400 521976
3405 521977
3410 521977
3415 521977
3420 521978
3425 521978
3430 521978
3435 521979
3440 521979
3445 521979
3450 521980
3455 521980
3460 521980
3465 521981
3470 521981
3475 521982
3480 521982
3485 521982
3490 521983
3495 521983
3500 521983
3505 521984
3510 521984
3515 521984
3520 521985
3525 521985
3530 521985
3535 521986
3540 521986
3545 521986
3550 521987
3555 521987
3560 521988
3565 521988
3570 521989
3575 521989
3580 521989
3585 521990
3590 521990
3595 521990
3600 521991
3605 521991
3610 521991
3615 521992
3620 521992
3625 521992
3630 521993
3635 521993
3640 521993
3645 521994
3650 521994
3655 521995
3660 521995
3665 521995
3670 521996
3675 521996
3680 521996
3685 521997
3690 521997
3695 521997
3700 521998
3705 521998
3710 521998
3715 521999
3720 521999
3725 521999
3730 522000
3735 522000
3740 522000
3745 522001
3750 522001
3755 522001
3760 522002
3765 522002
3770 522003
3775 522003
3780 522003
3785 522004
3790 522004
3795 522004
3800 522005
3805 522005
3810 522005
3815 522006
3820 522006
3825 522006
3830 522007
3835 522007
3840 522007
3845 522008
3850 522008
3855 522008
3860 522009
3865 522009
3870 522010
3875 522010
3880 522010
3885 522011
3890 522011
3895 522011
3900 522012
3905 522012
3910 522012
3915 522013
3920 522013
3925 522013
3930 522014
3935 522014
3940 522014
3945 522015
3950 522015
3955 522015
3960 522016
3965 522016
3970 522016
3975 522017
3980 522017
3985 522018
3990 522018
3995 522018
4000 522019
4005 522019
4010 522019
4015 522020
4020 522020
4025 522020
4030 522021
4035 522021
4040 522021
4045 522022
4050 522022
4055 522022
4060 522023
4065 522023
4070 522024
4075 522024
4080 522024
4085 522025
4090 522025
4095 522025
4100 522026
4105 522026
4110 522026
4115 522027
4120 522027
4125 522027
4130 522028
4135 522028
4140 522028
4145 522029
4150 522029
4155 522030
4160 522030
4165 522030
4170 522031
4175 522031
4180 522031
4185 522032
4190 522032
4195 522032
4200 522033
4205 522033
4210 522033
4215 522034
4220 522034
4225 522035
4230 522035
4235 522035
4240 522036
4245 522036
4250 522036
4255 522037
4260 522037
4265 522037
4270 522038
4275 522038
4280 522038
4285 522039
4290 522039
4295 522039
4300 522040
4305 522040
4310 522040
4315 522041
4320 522041
4325 522042
4330 522042
4335 522042
4340 522043
4345 522043
4350 522043
4355 522044
4360 522044
4365 522044
4370 522045
4375 522045
4380 522045
4385 522046
4390 522046
4395 522046
4400 522047
4405 522047
4410 522047
4415 522048
4420 522048
4425 522048
4430 522049
4435 522049
4440 522050
4445 522050
4450 522050
4455 522051
4460 522051
4465 522051
4470 522052
4475 522052
4480 522052
4485 522053
4490 522053
4495 522053
4500 522054
4505 522054
4510 522054
4515 522055
4520 522055
4525 522056
4530 522056
4535 522056
4540 522057
4545 522057
4550 522057
4555 522058
4560 522058
4565 522058
4570 522059
4575 522059
4580 522059
4585 522060
4590 522060
4595 522060
4600 522061
4605 522061
4610 522061
4615 522062
4620 522062
4625 522063
4630 522063
4635 522063
4640 522064
4645 522064
4650 522064
4655 522065
4660 522065
4665 522065
4670 522066
4675 522066
4680 522066
4685 522067
4690 522067
4695 522067
4700 522068
4705 522068
4710 522068
4715 522069
4720 522069
4725 522069
4730 522070
4735 522070
4740 522071
4745 522071
4750 522071
4755 522072
4760 522072
4765 522072
4770 522073
4775 522073
4780 522073
4785 522074
4790 522074
4795 522074
4800 522075
4805 522075
4810 522075
4815 522076
4820 522076
4825 522077
4830 522077
4835 522077
4840 522078
4845 522078
4850 522078
4855 522079
4860 522079
4865 522079
4870 522080
4875 522080
4880 522080
4885 522081
4890 522081
4895 522081
4900 522082
4905 522082
4910 522083
4915 522083
4920 522083
4925 522084
4930 522084
4935 522084
4940 522085
4945 522085
4950 522085
4955 522086
4960 522086
4965 522086
4970 522087
4975 522087
4980 522087
4985 522088
4990 522088
4995 522089
5000 522089
5005 522089
5010 522090
5015 522090
5020 522090
5025 522091
5030 522091
5035 522091
5040 522092
5045 522092
5050 522092
5055 522093
5060 522093
5065 522093
5070 522094
5075 522094
5080 522094
5085 522095
5090 522095
5095 522096
5100 522096
5105 522096
5110 522097
5115 522097
5120 522097
5125 522098
5130 522098
5135 522098
5140 522099
5145 522099
5150 522099
5155 522100
5160 522100
5165 522100
5170 522101
5175 522101
5180 522101
5185 522102
5190 522102
5195 522102
5200 522103
5205 522103
5210 522104
5215 522104
5220 522104
5225 522105
5230 522105
5235 522105
5240 522106
5245 522106
5250 522106
5255 522107
5260 522107
5265 522107
5270 522108
5275 522108
5280 522108
5285 522109
5290 522109
5295 522109
5300 522110
5305 522110
5310 522110
5315 522111
5320 522111
5325 522112
5330 522112
5335 522112
5340 522113
5345 522113
5350 522113
5355 522114
5360 522114
5365 522114
5370 522115
5375 522115
5380 522116
5385 522116
5390 522116
5395 522117
5400 522117
};

\addlegendentry{Baseline}
\addplot [semithick, color2, mark=triangle*, mark size=3, mark options={solid,fill=white,draw=red},  mark repeat={180}]
table[y expr=(\thisrowno{1}-573544)/1000] {%
0 573544
5 573545
10 573545
15 573545
20 573546
25 573546
30 573547
35 573547
40 573547
45 573548
50 573548
55 573548
60 573549
65 573549
70 573549
75 573550
80 573550
85 573550
90 573551
95 573551
100 573551
105 573552
110 573552
115 573552
120 573553
125 573553
130 573554
135 573554
140 573554
145 573555
150 573555
155 573555
160 573556
165 573556
170 573557
175 573557
180 573557
185 573558
190 573558
195 573559
200 573559
205 573560
210 573561
215 573561
220 573562
225 573563
230 573563
235 573564
240 573565
245 573565
250 573566
255 573567
260 573567
265 573568
270 573569
275 573569
280 573570
285 573571
290 573571
295 573572
300 573572
305 573573
310 573574
315 573574
320 573575
325 573576
330 573576
335 573577
340 573578
345 573578
350 573579
355 573580
360 573580
365 573581
370 573582
375 573582
380 573583
385 573584
390 573584
395 573585
400 573586
405 573586
410 573587
415 573588
420 573588
425 573589
430 573589
435 573590
440 573591
445 573591
450 573592
455 573593
460 573593
465 573594
470 573595
475 573595
480 573596
485 573597
490 573597
495 573598
500 573599
505 573599
510 573600
515 573601
520 573601
525 573602
530 573603
535 573603
540 573604
545 573605
550 573605
555 573606
560 573607
565 573607
570 573608
575 573609
580 573609
585 573610
590 573611
595 573611
600 573612
605 573613
610 573613
615 573614
620 573615
625 573615
630 573616
635 573617
640 573617
645 573618
650 573618
655 573619
660 573620
665 573620
670 573621
675 573622
680 573622
685 573623
690 573623
695 573624
700 573625
705 573625
710 573626
715 573626
720 573627
725 573628
730 573628
735 573629
740 573630
745 573630
750 573631
755 573632
760 573632
765 573633
770 573634
775 573634
780 573635
785 573636
790 573636
795 573637
800 573638
805 573638
810 573639
815 573640
820 573640
825 573641
830 573642
835 573642
840 573643
845 573644
850 573644
855 573645
860 573646
865 573646
870 573647
875 573647
880 573648
885 573649
890 573649
895 573650
900 573651
905 573651
910 573652
915 573653
920 573653
925 573654
930 573655
935 573655
940 573656
945 573657
950 573657
955 573658
960 573659
965 573659
970 573660
975 573661
980 573661
985 573662
990 573663
995 573663
1000 573664
1005 573665
1010 573665
1015 573666
1020 573667
1025 573667
1030 573668
1035 573669
1040 573669
1045 573670
1050 573671
1055 573671
1060 573672
1065 573673
1070 573673
1075 573674
1080 573675
1085 573675
1090 573676
1095 573676
1100 573677
1105 573678
1110 573678
1115 573679
1120 573680
1125 573680
1130 573681
1135 573682
1140 573682
1145 573683
1150 573684
1155 573684
1160 573685
1165 573686
1170 573686
1175 573687
1180 573688
1185 573688
1190 573689
1195 573690
1200 573690
1205 573691
1210 573692
1215 573692
1220 573693
1225 573694
1230 573694
1235 573695
1240 573696
1245 573696
1250 573697
1255 573698
1260 573698
1265 573699
1270 573700
1275 573700
1280 573701
1285 573702
1290 573702
1295 573703
1300 573704
1305 573704
1310 573705
1315 573706
1320 573706
1325 573707
1330 573708
1335 573708
1340 573709
1345 573710
1350 573710
1355 573711
1360 573712
1365 573712
1370 573713
1375 573714
1380 573714
1385 573715
1390 573716
1395 573716
1400 573717
1405 573718
1410 573718
1415 573719
1420 573720
1425 573720
1430 573721
1435 573722
1440 573722
1445 573723
1450 573724
1455 573724
1460 573725
1465 573726
1470 573726
1475 573727
1480 573728
1485 573728
1490 573729
1495 573730
1500 573730
1505 573731
1510 573732
1515 573732
1520 573733
1525 573734
1530 573734
1535 573735
1540 573736
1545 573736
1550 573737
1555 573738
1560 573738
1565 573739
1570 573740
1575 573740
1580 573741
1585 573742
1590 573742
1595 573743
1600 573744
1605 573744
1610 573745
1615 573746
1620 573746
1625 573747
1630 573748
1635 573748
1640 573749
1645 573750
1650 573750
1655 573751
1660 573751
1665 573752
1670 573753
1675 573753
1680 573754
1685 573755
1690 573755
1695 573756
1700 573757
1705 573757
1710 573758
1715 573759
1720 573759
1725 573760
1730 573761
1735 573761
1740 573762
1745 573763
1750 573763
1755 573764
1760 573765
1765 573765
1770 573766
1775 573767
1780 573767
1785 573768
1790 573769
1795 573769
1800 573770
1805 573771
1810 573771
1815 573772
1820 573773
1825 573773
1830 573774
1835 573775
1840 573775
1845 573776
1850 573777
1855 573777
1860 573778
1865 573779
1870 573779
1875 573780
1880 573780
1885 573781
1890 573782
1895 573782
1900 573783
1905 573784
1910 573784
1915 573785
1920 573786
1925 573786
1930 573787
1935 573788
1940 573788
1945 573789
1950 573790
1955 573790
1960 573791
1965 573791
1970 573792
1975 573793
1980 573793
1985 573794
1990 573795
1995 573795
2000 573796
2005 573797
2010 573797
2015 573798
2020 573799
2025 573799
2030 573800
2035 573801
2040 573801
2045 573802
2050 573803
2055 573803
2060 573804
2065 573805
2070 573805
2075 573806
2080 573807
2085 573807
2090 573808
2095 573808
2100 573809
2105 573810
2110 573810
2115 573811
2120 573812
2125 573812
2130 573813
2135 573814
2140 573814
2145 573815
2150 573816
2155 573816
2160 573817
2165 573818
2170 573818
2175 573819
2180 573820
2185 573820
2190 573821
2195 573822
2200 573822
2205 573823
2210 573824
2215 573824
2220 573825
2225 573826
2230 573826
2235 573827
2240 573828
2245 573828
2250 573829
2255 573830
2260 573830
2265 573831
2270 573831
2275 573832
2280 573833
2285 573833
2290 573834
2295 573835
2300 573835
2305 573836
2310 573837
2315 573837
2320 573838
2325 573839
2330 573839
2335 573840
2340 573841
2345 573841
2350 573842
2355 573843
2360 573843
2365 573844
2370 573845
2375 573845
2380 573846
2385 573847
2390 573847
2395 573848
2400 573849
2405 573849
2410 573850
2415 573851
2420 573851
2425 573852
2430 573853
2435 573853
2440 573854
2445 573854
2450 573855
2455 573855
2460 573856
2465 573857
2470 573857
2475 573858
2480 573858
2485 573859
2490 573859
2495 573860
2500 573860
2505 573860
2510 573861
2515 573861
2520 573861
2525 573862
2530 573862
2535 573862
2540 573863
2545 573863
2550 573863
2555 573864
2560 573864
2565 573865
2570 573865
2575 573865
2580 573866
2585 573866
2590 573866
2595 573867
2600 573867
2605 573867
2610 573868
2615 573868
2620 573868
2625 573869
2630 573869
2635 573870
2640 573870
2645 573870
2650 573871
2655 573871
2660 573871
2665 573872
2670 573872
2675 573872
2680 573873
2685 573873
2690 573873
2695 573874
2700 573874
2705 573875
2710 573875
2715 573875
2720 573876
2725 573876
2730 573876
2735 573877
2740 573877
2745 573877
2750 573878
2755 573878
2760 573878
2765 573879
2770 573879
2775 573879
2780 573880
2785 573880
2790 573881
2795 573881
2800 573881
2805 573882
2810 573882
2815 573882
2820 573883
2825 573883
2830 573883
2835 573884
2840 573884
2845 573884
2850 573885
2855 573885
2860 573885
2865 573886
2870 573886
2875 573887
2880 573887
2885 573887
2890 573888
2895 573888
2900 573888
2905 573889
2910 573889
2915 573889
2920 573890
2925 573890
2930 573890
2935 573891
2940 573891
2945 573891
2950 573892
2955 573892
2960 573893
2965 573893
2970 573893
2975 573894
2980 573894
2985 573894
2990 573895
2995 573895
3000 573895
3005 573896
3010 573896
3015 573896
3020 573897
3025 573897
3030 573897
3035 573898
3040 573898
3045 573898
3050 573899
3055 573899
3060 573900
3065 573900
3070 573900
3075 573901
3080 573901
3085 573902
3090 573902
3095 573902
3100 573903
3105 573903
3110 573903
3115 573904
3120 573904
3125 573905
3130 573905
3135 573905
3140 573906
3145 573906
3150 573906
3155 573907
3160 573907
3165 573907
3170 573908
3175 573908
3180 573908
3185 573909
3190 573909
3195 573909
3200 573910
3205 573910
3210 573910
3215 573911
3220 573911
3225 573912
3230 573912
3235 573912
3240 573913
3245 573913
3250 573913
3255 573914
3260 573914
3265 573914
3270 573915
3275 573915
3280 573915
3285 573916
3290 573916
3295 573916
3300 573917
3305 573917
3310 573918
3315 573918
3320 573918
3325 573919
3330 573919
3335 573919
3340 573920
3345 573920
3350 573920
3355 573921
3360 573921
3365 573921
3370 573922
3375 573922
3380 573922
3385 573923
3390 573923
3395 573924
3400 573924
3405 573924
3410 573925
3415 573925
3420 573925
3425 573926
3430 573926
3435 573926
3440 573927
3445 573927
3450 573927
3455 573928
3460 573928
3465 573929
3470 573929
3475 573929
3480 573930
3485 573930
3490 573930
3495 573931
3500 573931
3505 573931
3510 573932
3515 573932
3520 573932
3525 573933
3530 573933
3535 573933
3540 573934
3545 573934
3550 573934
3555 573935
3560 573935
3565 573936
3570 573936
3575 573936
3580 573937
3585 573937
3590 573937
3595 573938
3600 573938
3605 573938
3610 573939
3615 573939
3620 573939
3625 573940
3630 573940
3635 573940
3640 573941
3645 573941
3650 573941
3655 573942
3660 573942
3665 573943
3670 573943
3675 573943
3680 573943
3685 573944
3690 573944
3695 573944
3700 573945
3705 573945
3710 573945
3715 573946
3720 573946
3725 573947
3730 573947
3735 573947
3740 573948
3745 573948
3750 573948
3755 573949
3760 573949
3765 573949
3770 573950
3775 573950
3780 573950
3785 573951
3790 573951
3795 573951
3800 573952
3805 573952
3810 573952
3815 573953
3820 573953
3825 573954
3830 573954
3835 573954
3840 573955
3845 573955
3850 573955
3855 573956
3860 573956
3865 573956
3870 573957
3875 573957
3880 573957
3885 573958
3890 573958
3895 573958
3900 573959
3905 573959
3910 573960
3915 573960
3920 573960
3925 573961
3930 573961
3935 573961
3940 573962
3945 573962
3950 573962
3955 573963
3960 573963
3965 573963
3970 573964
3975 573964
3980 573964
3985 573965
3990 573965
3995 573965
4000 573966
4005 573966
4010 573967
4015 573967
4020 573967
4025 573968
4030 573968
4035 573968
4040 573969
4045 573969
4050 573969
4055 573970
4060 573970
4065 573970
4070 573971
4075 573971
4080 573971
4085 573972
4090 573972
4095 573973
4100 573973
4105 573973
4110 573974
4115 573974
4120 573974
4125 573975
4130 573975
4135 573975
4140 573976
4145 573976
4150 573976
4155 573977
4160 573977
4165 573977
4170 573978
4175 573978
4180 573979
4185 573979
4190 573979
4195 573980
4200 573980
4205 573980
4210 573981
4215 573981
4220 573981
4225 573982
4230 573982
4235 573982
4240 573983
4245 573983
4250 573983
4255 573984
4260 573984
4265 573984
4270 573985
4275 573985
4280 573985
4285 573986
4290 573986
4295 573987
4300 573987
4305 573987
4310 573988
4315 573988
4320 573988
4325 573989
4330 573989
4335 573989
4340 573990
4345 573990
4350 573990
4355 573991
4360 573991
4365 573991
4370 573992
4375 573992
4380 573993
4385 573993
4390 573993
4395 573994
4400 573994
4405 573994
4410 573995
4415 573995
4420 573995
4425 573996
4430 573996
4435 573996
4440 573997
4445 573997
4450 573997
4455 573998
4460 573998
4465 573998
4470 573999
4475 573999
4480 573999
4485 574000
4490 574000
4495 574001
4500 574001
4505 574001
4510 574002
4515 574002
4520 574002
4525 574003
4530 574003
4535 574003
4540 574004
4545 574004
4550 574004
4555 574005
4560 574005
4565 574005
4570 574006
4575 574006
4580 574007
4585 574007
4590 574007
4595 574008
4600 574008
4605 574008
4610 574009
4615 574009
4620 574009
4625 574010
4630 574010
4635 574010
4640 574011
4645 574011
4650 574011
4655 574012
4660 574012
4665 574012
4670 574013
4675 574013
4680 574014
4685 574014
4690 574014
4695 574015
4700 574015
4705 574015
4710 574016
4715 574016
4720 574016
4725 574017
4730 574017
4735 574017
4740 574018
4745 574018
4750 574018
4755 574019
4760 574019
4765 574019
4770 574020
4775 574020
4780 574020
4785 574021
4790 574021
4795 574022
4800 574022
4805 574022
4810 574023
4815 574023
4820 574023
4825 574024
4830 574024
4835 574024
4840 574025
4845 574025
4850 574025
4855 574026
4860 574026
4865 574026
4870 574027
4875 574027
4880 574028
4885 574028
4890 574028
4895 574029
4900 574029
4905 574029
4910 574030
4915 574030
4920 574030
4925 574031
4930 574031
4935 574031
4940 574032
4945 574032
4950 574032
4955 574033
4960 574033
4965 574034
4970 574034
4975 574034
4980 574035
4985 574035
4990 574035
4995 574036
5000 574036
5005 574036
5010 574037
5015 574037
5020 574037
5025 574038
5030 574038
5035 574038
5040 574039
5045 574039
5050 574040
5055 574040
5060 574040
5065 574041
5070 574041
5075 574041
5080 574042
5085 574042
5090 574042
5095 574043
5100 574043
5105 574043
5110 574044
5115 574044
5120 574044
5125 574045
5130 574045
5135 574045
5140 574046
5145 574046
5150 574047
5155 574047
5160 574047
5165 574048
5170 574048
5175 574048
5180 574049
5185 574049
5190 574049
5195 574050
5200 574050
5205 574050
5210 574051
5215 574051
5220 574051
5225 574052
5230 574052
5235 574052
5240 574053
5245 574053
5250 574053
5255 574054
5260 574054
5265 574055
5270 574055
5275 574055
5280 574056
5285 574056
5290 574056
5295 574057
5300 574057
5305 574057
5310 574058
5315 574058
5320 574058
5325 574059
5330 574059
5335 574059
5340 574060
5345 574060
5350 574060
5355 574061
5360 574061
5365 574062
5370 574062
5375 574062
5380 574063
5385 574063
5390 574063
5395 574064
5400 574064

};
\addlegendentry{Different sockets}
\addplot[semithick, color3, mark=square*, mark size=3, mark options={solid,fill=white,draw=red},  mark repeat={180}]
table[y expr=(\thisrowno{1}-572976)/1000] {%
0 572976
5 572977
10 572977
15 572978
20 572978
25 572978
30 572979
35 572979
40 572979
45 572980
50 572980
55 572980
60 572981
65 572981
70 572982
75 572982
80 572982
85 572983
90 572983
95 572983
100 572984
105 572984
110 572984
115 572985
120 572985
125 572986
130 572986
135 572986
140 572987
145 572987
150 572987
155 572988
160 572988
165 572989
170 572989
175 572989
180 572990
185 572990
190 572991
195 572992
200 572992
205 572993
210 572994
215 572994
220 572995
225 572995
230 572996
235 572996
240 572997
245 572997
250 572998
255 572998
260 572999
265 572999
270 573000
275 573000
280 573001
285 573001
290 573002
295 573002
300 573003
305 573003
310 573004
315 573004
320 573005
325 573005
330 573006
335 573006
340 573007
345 573007
350 573008
355 573008
360 573009
365 573009
370 573010
375 573010
380 573011
385 573011
390 573012
395 573012
400 573013
405 573013
410 573014
415 573014
420 573015
425 573016
430 573016
435 573017
440 573017
445 573018
450 573018
455 573019
460 573019
465 573019
470 573020
475 573020
480 573021
485 573021
490 573022
495 573022
500 573023
505 573023
510 573024
515 573024
520 573025
525 573025
530 573026
535 573026
540 573027
545 573027
550 573028
555 573028
560 573029
565 573029
570 573030
575 573030
580 573031
585 573031
590 573032
595 573032
600 573033
605 573033
610 573034
615 573034
620 573035
625 573035
630 573036
635 573036
640 573037
645 573037
650 573038
655 573038
660 573039
665 573039
670 573040
675 573040
680 573041
685 573041
690 573042
695 573042
700 573043
705 573043
710 573044
715 573044
720 573045
725 573045
730 573046
735 573046
740 573047
745 573047
750 573048
755 573048
760 573049
765 573049
770 573050
775 573050
780 573051
785 573051
790 573052
795 573053
800 573053
805 573053
810 573054
815 573054
820 573055
825 573055
830 573056
835 573057
840 573057
845 573058
850 573058
855 573059
860 573059
865 573060
870 573060
875 573061
880 573061
885 573062
890 573062
895 573063
900 573063
905 573064
910 573064
915 573065
920 573065
925 573066
930 573066
935 573067
940 573067
945 573068
950 573068
955 573069
960 573069
965 573070
970 573070
975 573071
980 573071
985 573072
990 573072
995 573073
1000 573073
1005 573074
1010 573074
1015 573075
1020 573075
1025 573076
1030 573076
1035 573077
1040 573077
1045 573078
1050 573078
1055 573079
1060 573079
1065 573080
1070 573081
1075 573081
1080 573082
1085 573082
1090 573083
1095 573083
1100 573084
1105 573084
1110 573085
1115 573085
1120 573086
1125 573086
1130 573087
1135 573087
1140 573088
1145 573088
1150 573089
1155 573089
1160 573090
1165 573090
1170 573091
1175 573091
1180 573092
1185 573092
1190 573093
1195 573093
1200 573094
1205 573094
1210 573095
1215 573095
1220 573096
1225 573096
1230 573097
1235 573097
1240 573098
1245 573098
1250 573099
1255 573099
1260 573100
1265 573101
1270 573101
1275 573101
1280 573102
1285 573102
1290 573103
1295 573103
1300 573104
1305 573104
1310 573105
1315 573105
1320 573106
1325 573107
1330 573107
1335 573108
1340 573108
1345 573109
1350 573109
1355 573110
1360 573110
1365 573111
1370 573111
1375 573112
1380 573112
1385 573113
1390 573113
1395 573114
1400 573114
1405 573115
1410 573115
1415 573116
1420 573116
1425 573117
1430 573117
1435 573118
1440 573118
1445 573119
1450 573119
1455 573120
1460 573120
1465 573121
1470 573121
1475 573122
1480 573122
1485 573123
1490 573123
1495 573124
1500 573124
1505 573125
1510 573125
1515 573126
1520 573126
1525 573127
1530 573127
1535 573128
1540 573128
1545 573129
1550 573129
1555 573130
1560 573130
1565 573131
1570 573131
1575 573132
1580 573132
1585 573133
1590 573133
1595 573134
1600 573134
1605 573135
1610 573135
1615 573136
1620 573136
1625 573137
1630 573137
1635 573138
1640 573138
1645 573139
1650 573139
1655 573140
1660 573140
1665 573141
1670 573142
1675 573142
1680 573143
1685 573143
1690 573144
1695 573144
1700 573145
1705 573145
1710 573146
1715 573146
1720 573147
1725 573147
1730 573148
1735 573148
1740 573149
1745 573149
1750 573150
1755 573150
1760 573151
1765 573151
1770 573152
1775 573152
1780 573153
1785 573153
1790 573154
1795 573154
1800 573155
1805 573155
1810 573156
1815 573156
1820 573157
1825 573157
1830 573158
1835 573158
1840 573159
1845 573159
1850 573160
1855 573160
1860 573161
1865 573162
1870 573162
1875 573162
1880 573163
1885 573164
1890 573164
1895 573165
1900 573165
1905 573165
1910 573166
1915 573167
1920 573167
1925 573168
1930 573168
1935 573169
1940 573169
1945 573170
1950 573170
1955 573171
1960 573171
1965 573172
1970 573172
1975 573173
1980 573173
1985 573174
1990 573174
1995 573175
2000 573175
2005 573176
2010 573176
2015 573177
2020 573177
2025 573178
2030 573178
2035 573179
2040 573179
2045 573180
2050 573180
2055 573181
2060 573181
2065 573182
2070 573182
2075 573183
2080 573183
2085 573184
2090 573184
2095 573185
2100 573185
2105 573186
2110 573186
2115 573187
2120 573187
2125 573188
2130 573188
2135 573189
2140 573189
2145 573190
2150 573190
2155 573191
2160 573191
2165 573192
2170 573192
2175 573193
2180 573193
2185 573194
2190 573194
2195 573195
2200 573195
2205 573196
2210 573196
2215 573197
2220 573198
2225 573198
2230 573198
2235 573199
2240 573199
2245 573200
2250 573200
2255 573201
2260 573202
2265 573202
2270 573203
2275 573204
2280 573204
2285 573204
2290 573205
2295 573205
2300 573206
2305 573206
2310 573207
2315 573207
2320 573208
2325 573208
2330 573209
2335 573210
2340 573210
2345 573211
2350 573211
2355 573212
2360 573212
2365 573213
2370 573213
2375 573214
2380 573214
2385 573215
2390 573215
2395 573216
2400 573216
2405 573217
2410 573217
2415 573218
2420 573218
2425 573219
2430 573219
2435 573220
2440 573220
2445 573221
2450 573221
2455 573222
2460 573222
2465 573223
2470 573223
2475 573224
2480 573224
2485 573225
2490 573225
2495 573226
2500 573226
2505 573227
2510 573227
2515 573228
2520 573228
2525 573229
2530 573229
2535 573230
2540 573230
2545 573231
2550 573231
2555 573232
2560 573232
2565 573233
2570 573233
2575 573234
2580 573234
2585 573235
2590 573235
2595 573236
2600 573236
2605 573237
2610 573237
2615 573238
2620 573238
2625 573239
2630 573239
2635 573240
2640 573240
2645 573241
2650 573241
2655 573242
2660 573242
2665 573243
2670 573243
2675 573244
2680 573244
2685 573245
2690 573245
2695 573246
2700 573246
2705 573247
2710 573247
2715 573248
2720 573248
2725 573249
2730 573250
2735 573250
2740 573250
2745 573251
2750 573251
2755 573252
2760 573252
2765 573253
2770 573254
2775 573254
2780 573255
2785 573255
2790 573256
2795 573256
2800 573257
2805 573257
2810 573258
2815 573258
2820 573259
2825 573259
2830 573259
2835 573260
2840 573261
2845 573261
2850 573262
2855 573262
2860 573262
2865 573263
2870 573263
2875 573264
2880 573264
2885 573265
2890 573265
2895 573266
2900 573266
2905 573267
2910 573267
2915 573268
2920 573268
2925 573269
2930 573269
2935 573270
2940 573270
2945 573271
2950 573271
2955 573272
2960 573272
2965 573273
2970 573273
2975 573274
2980 573274
2985 573275
2990 573275
2995 573276
3000 573276
3005 573277
3010 573277
3015 573278
3020 573278
3025 573279
3030 573279
3035 573280
3040 573280
3045 573281
3050 573281
3055 573282
3060 573282
3065 573283
3070 573283
3075 573284
3080 573284
3085 573285
3090 573285
3095 573286
3100 573286
3105 573287
3110 573288
3115 573288
3120 573289
3125 573289
3130 573290
3135 573290
3140 573290
3145 573291
3150 573291
3155 573292
3160 573293
3165 573293
3170 573294
3175 573294
3180 573295
3185 573295
3190 573296
3195 573296
3200 573297
3205 573297
3210 573298
3215 573298
3220 573299
3225 573299
3230 573300
3235 573300
3240 573301
3245 573301
3250 573302
3255 573302
3260 573303
3265 573303
3270 573304
3275 573304
3280 573305
3285 573305
3290 573306
3295 573306
3300 573307
3305 573307
3310 573308
3315 573308
3320 573309
3325 573309
3330 573310
3335 573310
3340 573311
3345 573311
3350 573312
3355 573312
3360 573313
3365 573313
3370 573314
3375 573314
3380 573315
3385 573315
3390 573316
3395 573316
3400 573317
3405 573317
3410 573318
3415 573318
3420 573319
3425 573319
3430 573320
3435 573320
3440 573321
3445 573321
3450 573322
3455 573322
3460 573323
3465 573324
3470 573324
3475 573325
3480 573325
3485 573326
3490 573326
3495 573327
3500 573327
3505 573328
3510 573328
3515 573329
3520 573329
3525 573330
3530 573330
3535 573331
3540 573331
3545 573332
3550 573332
3555 573333
3560 573333
3565 573334
3570 573334
3575 573335
3580 573335
3585 573336
3590 573336
3595 573337
3600 573337
3605 573338
3610 573338
3615 573339
3620 573340
3625 573340
3630 573340
3635 573341
3640 573342
3645 573342
3650 573342
3655 573343
3660 573343
3665 573344
3670 573344
3675 573345
3680 573346
3685 573346
3690 573347
3695 573347
3700 573348
3705 573348
3710 573349
3715 573349
3720 573350
3725 573350
3730 573351
3735 573351
3740 573352
3745 573352
3750 573353
3755 573353
3760 573354
3765 573354
3770 573355
3775 573355
3780 573356
3785 573356
3790 573357
3795 573357
3800 573358
3805 573358
3810 573359
3815 573359
3820 573360
3825 573360
3830 573361
3835 573361
3840 573362
3845 573362
3850 573363
3855 573363
3860 573364
3865 573364
3870 573365
3875 573365
3880 573366
3885 573366
3890 573367
3895 573367
3900 573368
3905 573368
3910 573369
3915 573369
3920 573370
3925 573370
3930 573371
3935 573371
3940 573372
3945 573372
3950 573373
3955 573373
3960 573374
3965 573374
3970 573375
3975 573375
3980 573376
3985 573376
3990 573377
3995 573377
4000 573378
4005 573378
4010 573379
4015 573379
4020 573379
4025 573380
4030 573380
4035 573381
4040 573381
4045 573381
4050 573382
4055 573382
4060 573382
4065 573383
4070 573383
4075 573383
4080 573384
4085 573384
4090 573384
4095 573385
4100 573385
4105 573386
4110 573386
4115 573386
4120 573387
4125 573387
4130 573387
4135 573388
4140 573388
4145 573388
4150 573389
4155 573389
4160 573389
4165 573390
4170 573390
4175 573390
4180 573391
4185 573391
4190 573392
4195 573392
4200 573392
4205 573393
4210 573393
4215 573393
4220 573394
4225 573394
4230 573394
4235 573395
4240 573395
4245 573395
4250 573396
4255 573396
4260 573397
4265 573397
4270 573397
4275 573398
4280 573398
4285 573398
4290 573399
4295 573399
4300 573399
4305 573400
4310 573400
4315 573400
4320 573401
4325 573401
4330 573401
4335 573402
4340 573402
4345 573403
4350 573403
4355 573403
4360 573404
4365 573404
4370 573404
4375 573405
4380 573405
4385 573405
4390 573406
4395 573406
4400 573406
4405 573407
4410 573407
4415 573407
4420 573408
4425 573408
4430 573408
4435 573409
4440 573409
4445 573410
4450 573410
4455 573410
4460 573411
4465 573411
4470 573411
4475 573412
4480 573412
4485 573412
4490 573413
4495 573413
4500 573413
4505 573414
4510 573414
4515 573415
4520 573415
4525 573415
4530 573416
4535 573416
4540 573416
4545 573417
4550 573417
4555 573417
4560 573418
4565 573418
4570 573418
4575 573419
4580 573419
4585 573419
4590 573420
4595 573420
4600 573420
4605 573421
4610 573421
4615 573422
4620 573422
4625 573422
4630 573423
4635 573423
4640 573423
4645 573424
4650 573424
4655 573424
4660 573424
4665 573425
4670 573425
4675 573425
4680 573426
4685 573426
4690 573426
4695 573427
4700 573427
4705 573428
4710 573428
4715 573428
4720 573429
4725 573429
4730 573429
4735 573430
4740 573430
4745 573430
4750 573431
4755 573431
4760 573431
4765 573432
4770 573432
4775 573432
4780 573433
4785 573433
4790 573433
4795 573434
4800 573434
4805 573435
4810 573435
4815 573435
4820 573436
4825 573436
4830 573436
4835 573437
4840 573437
4845 573437
4850 573438
4855 573438
4860 573438
4865 573439
4870 573439
4875 573440
4880 573440
4885 573440
4890 573441
4895 573441
4900 573441
4905 573442
4910 573442
4915 573442
4920 573443
4925 573443
4930 573443
4935 573444
4940 573444
4945 573444
4950 573445
4955 573445
4960 573446
4965 573446
4970 573446
4975 573447
4980 573447
4985 573447
4990 573448
4995 573448
5000 573448
5005 573449
5010 573449
5015 573449
5020 573450
5025 573450
5030 573450
5035 573451
5040 573451
5045 573452
5050 573452
5055 573452
5060 573453
5065 573453
5070 573453
5075 573454
5080 573454
5085 573454
5090 573455
5095 573455
5100 573455
5105 573456
5110 573456
5115 573456
5120 573457
5125 573457
5130 573457
5135 573458
5140 573458
5145 573459
5150 573459
5155 573459
5160 573460
5165 573460
5170 573460
5175 573461
5180 573461
5185 573461
5190 573462
5195 573462
5200 573462
5205 573463
5210 573463
5215 573464
5220 573464
5225 573464
5230 573465
5235 573465
5240 573465
5245 573466
5250 573466
5255 573466
5260 573466
5265 573467
5270 573467
5275 573467
5280 573468
5285 573468
5290 573469
5295 573469
5300 573469
5305 573470
5310 573470
5315 573470
5320 573471
5325 573471
5330 573471
5335 573472
5340 573472
5345 573472
5350 573473
5355 573473
5360 573474
5365 573474
5370 573474
5375 573475
5380 573475
5385 573475
5390 573476
5395 573476
5400 573476

};
\addlegendentry{Same socket}
\addplot [semithick, color2,mark=triangle*, mark size=3, mark options={solid,fill=red,draw=red}]
table[y expr=(\thisrowno{1}-573544)/1000]{
2525 573862
};
\addplot [semithick, color1, mark=square*, mark size=3, mark options={solid,fill=red,draw=red}]
table[y expr=(\thisrowno{1}-572976)/1000]{
4035 573381
};
\end{axis}
\node[text width=8cm] at (3,-1.5) {Filled markers indicate the end of jobs};

\end{tikzpicture}

%% file: results/power_intel_seq.tex
\begin{tikzpicture}[font=\Large]

\definecolor{color0}{rgb}{0.12156862745098,0.466666666666667,0.705882352941177}
\definecolor{color1}{rgb}{1,0.498039215686275,0.0549019607843137}
\definecolor{color2}{rgb}{0.172549019607843,0.627450980392157,0.172549019607843}
\definecolor{color3}{rgb}{0.83921568627451,0.152941176470588,0.156862745098039}
\definecolor{color4}{rgb}{0.580392156862745,0.403921568627451,0.741176470588235}
\definecolor{color5}{rgb}{0,0,0}
\definecolor{color6}{rgb}{0.07, 0.04, 0.56}

\begin{axis}[
legend cell align={left},
legend columns=3,
legend style={fill opacity=0.8, draw opacity=1, text opacity=1, at={(1.05,1.15)}, anchor=east, draw=white!80.0!black},
tick align=outside,
tick pos=left,
x grid style={white!69.01960784313725!black},
xlabel={Time (SEC)},
xmin=0, xmax=4800,
xtick={0,1200,2400,3600,4800},
xtick style={color=black},
y grid style={white!69.01960784313725!black},
ylabel={Power consumption (W)},
ymin=150, ymax=500,
ytick={150,200,250,300,350,400,450,500},
xmajorgrids,
ymajorgrids,
ytick style={color=black}
]
\addplot [thick, olive,mark=star, mark size=3, mark options={solid,fill=white,draw=red},  mark repeat={180}]
table {%
0 253
5 262
10 287
15 302
20 311
25 323
30 311
35 301
40 260
45 254
50 254
55 253
60 255
65 254
70 254
75 253
80 253
85 255
90 253
95 253
100 255
105 254
110 265
115 254
120 305
125 306
130 305
135 306
140 305
145 292
150 294
155 289
160 289
165 312
170 307
175 307
180 309
185 305
190 469
195 462
200 463
205 465
210 465
215 466
220 466
225 467
230 468
235 466
240 468
245 467
250 468
255 469
260 470
265 471
270 470
275 470
280 469
285 469
290 469
295 469
300 469
305 469
310 469
315 469
320 469
325 469
330 469
335 469
340 469
345 469
350 469
355 469
360 469
365 470
370 468
375 471
380 470
385 472
390 470
395 470
400 471
405 470
410 471
415 470
420 471
425 472
430 471
435 470
440 471
445 470
450 471
455 471
460 472
465 472
470 471
475 471
480 472
485 472
490 472
495 471
500 472
505 472
510 471
515 472
520 472
525 471
530 472
535 472
540 472
545 472
550 472
555 472
560 472
565 472
570 472
575 473
580 472
585 473
590 473
595 473
600 473
605 473
610 473
615 471
620 473
625 473
630 474
635 473
640 474
645 474
650 475
655 474
660 475
665 475
670 473
675 475
680 475
685 474
690 474
695 474
700 475
705 475
710 474
715 475
720 474
725 474
730 475
735 473
740 473
745 473
750 473
755 473
760 473
765 473
770 474
775 473
780 474
785 473
790 473
795 473
800 474
805 473
810 474
815 473
820 474
825 473
830 474
835 473
840 473
845 473
850 472
855 473
860 472
865 473
870 472
875 472
880 472
885 472
890 472
895 473
900 472
905 471
910 472
915 471
920 471
925 471
930 471
935 471
940 471
945 472
950 471
955 471
960 471
965 469
970 471
975 471
980 471
985 471
990 471
995 471
1000 472
1005 472
1010 471
1015 470
1020 470
1025 471
1030 471
1035 470
1040 472
1045 470
1050 471
1055 470
1060 470
1065 472
1070 468
1075 471
1080 471
1085 470
1090 472
1095 470
1100 471
1105 472
1110 471
1115 472
1120 472
1125 471
1130 471
1135 472
1140 473
1145 472
1150 472
1155 472
1160 472
1165 472
1170 472
1175 472
1180 472
1185 472
1190 472
1195 472
1200 473
1205 473
1210 470
1215 469
1220 469
1225 469
1230 469
1235 469
1240 469
1245 469
1250 469
1255 469
1260 469
1265 470
1270 470
1275 469
1280 469
1285 469
1290 469
1295 468
1300 470
1305 469
1310 469
1315 468
1320 468
1325 472
1330 473
1335 472
1340 472
1345 472
1350 471
1355 471
1360 471
1365 472
1370 472
1375 471
1380 471
1385 472
1390 471
1395 471
1400 471
1405 471
1410 472
1415 471
1420 471
1425 472
1430 471
1435 471
1440 471
1445 473
1450 471
1455 471
1460 472
1465 471
1470 471
1475 471
1480 471
1485 471
1490 471
1495 471
1500 471
1505 471
1510 471
1515 471
1520 472
1525 473
1530 473
1535 473
1540 472
1545 473
1550 472
1555 473
1560 473
1565 473
1570 473
1575 473
1580 473
1585 474
1590 473
1595 473
1600 473
1605 474
1610 472
1615 472
1620 470
1625 470
1630 470
1635 469
1640 469
1645 471
1650 470
1655 473
1660 474
1665 474
1670 473
1675 474
1680 474
1685 473
1690 473
1695 473
1700 473
1705 473
1710 474
1715 473
1720 473
1725 474
1730 474
1735 473
1740 473
1745 473
1750 473
1755 474
1760 473
1765 473
1770 473
1775 474
1780 474
1785 473
1790 473
1795 473
1800 474
1805 473
1810 473
1815 473
1820 474
1825 473
1830 473
1835 474
1840 473
1845 473
1850 473
1855 473
1860 473
1865 474
1870 473
1875 473
1880 473
1885 473
1890 474
1895 473
1900 473
1905 473
1910 474
1915 471
1920 474
1925 474
1930 473
1935 474
1940 474
1945 474
1950 474
1955 475
1960 475
1965 475
1970 475
1975 474
1980 475
1985 475
1990 475
1995 474
2000 474
2005 474
2010 475
2015 475
2020 474
2025 474
2030 474
2035 475
2040 475
2045 475
2050 474
2055 475
2060 474
2065 474
2070 474
2075 474
2080 474
2085 474
2090 474
2095 475
2100 474
2105 475
2110 474
2115 475
2120 474
2125 474
2130 475
2135 474
2140 476
2145 475
2150 474
2155 474
2160 475
2165 476
2170 474
2175 474
2180 474
2185 474
2190 474
2195 474
2200 474
2205 474
2210 474
2215 475
2220 473
2225 474
2230 473
2235 473
2240 474
2245 473
2250 474
2255 472
2260 473
2265 472
2270 472
2275 473
2280 473
2285 474
2290 473
2295 473
2300 473
2305 472
2310 472
2315 474
2320 472
2325 473
2330 473
2335 472
2340 472
2345 472
2350 472
2355 472
2360 472
2365 472
2370 471
2375 471
2380 472
2385 471
2390 472
2395 471
2400 471
2405 472
2410 471
2415 471
2420 471
2425 471
2430 472
2435 471
2440 471
2445 472
2450 472
2455 471
2460 471
2465 471
2470 471
2475 472
2480 472
2485 472
2490 472
2495 472
2500 472
2505 473
2510 473
2515 472
2520 472
2525 474
2530 472
2535 472
2540 472
2545 473
2550 474
2555 472
2560 473
2565 473
2570 473
2575 474
2580 473
2585 473
2590 473
2595 474
2600 473
2605 474
2610 473
2615 474
2620 474
2625 474
2630 474
2635 474
2640 474
2645 474
2650 474
2655 475
2660 474
2665 474
2670 474
2675 474
2680 475
2685 474
2690 474
2695 474
2700 474
2705 475
2710 475
2715 475
2720 475
2725 472
2730 474
2735 474
2740 476
2745 474
2750 474
2755 475
2760 474
2765 475
2770 475
2775 474
2780 474
2785 475
2790 474
2795 473
2800 473
2805 473
2810 474
2815 473
2820 474
2825 474
2830 472
2835 473
2840 473
2845 473
2850 473
2855 472
2860 473
2865 472
2870 473
2875 473
2880 473
2885 473
2890 472
2895 472
2900 473
2905 472
2910 472
2915 472
2920 472
2925 472
2930 472
2935 471
2940 471
2945 471
2950 471
2955 471
2960 472
2965 471
2970 472
2975 471
2980 472
2985 471
2990 472
2995 471
3000 472
3005 471
3010 471
3015 471
3020 471
3025 471
3030 471
3035 471
3040 472
3045 471
3050 472
3055 472
3060 471
3065 471
3070 472
3075 471
3080 471
3085 471
3090 472
3095 472
3100 472
3105 472
3110 473
3115 472
3120 472
3125 472
3130 473
3135 473
3140 472
3145 472
3150 473
3155 473
3160 473
3165 472
3170 472
3175 472
3180 473
3185 474
3190 473
3195 473
3200 472
3205 473
3210 473
3215 473
3220 473
3225 472
3230 473
3235 473
3240 473
3245 473
3250 474
3255 474
3260 474
3265 475
3270 474
3275 474
3280 474
3285 474
3290 474
3295 474
3300 474
3305 474
3310 474
3315 474
3320 474
3325 475
3330 474
3335 474
3340 475
3345 474
3350 475
3355 474
3360 474
3365 474
3370 475
3375 475
3380 475
3385 475
3390 474
3395 474
3400 469
3405 470
3410 471
3415 470
3420 471
3425 472
3430 470
3435 472
3440 470
3445 475
3450 474
3455 476
3460 474
3465 472
3470 475
3475 475
3480 475
3485 475
3490 473
3495 475
3500 473
3505 474
3510 473
3515 474
3520 473
3525 474
3530 474
3535 474
3540 473
3545 473
3550 474
3555 473
3560 473
3565 474
3570 473
3575 472
3580 472
3585 473
3590 474
3595 473
3600 473
3605 473
3610 472
3615 473
3620 472
3625 473
3630 473
3635 472
3640 474
3645 472
3650 472
3655 472
3660 472
3665 472
3670 472
3675 472
3680 473
3685 473
3690 472
3695 472
3700 472
3705 472
3710 472
3715 472
3720 472
3725 472
3730 472
3735 472
3740 472
3745 473
3750 472
3755 473
3760 473
3765 472
3770 470
3775 472
3780 472
3785 473
3790 474
3795 473
3800 474
3805 473
3810 474
3815 474
3820 473
3825 474
3830 473
3835 473
3840 473
3845 475
3850 475
3855 474
3860 474
3865 474
3870 474
3875 474
3880 475
3885 473
3890 474
3895 475
3900 474
3905 474
3910 475
3915 474
3920 474
3925 474
3930 473
3935 472
3940 475
3945 474
3950 474
3955 475
3960 474
3965 473
3970 474
3975 473
3980 474
3985 474
3990 473
3995 473
4000 473
4005 474
4010 472
4015 468
4020 430
4025 398
4030 398
4035 395
4040 421
4045 408
4050 385
4055 317
4060 288
4065 287
4070 287
4075 290
4080 288
4085 274
4090 267
4095 263
4100 260
4105 258
4110 261
4115 260
4120 260
4125 261
4130 260
4135 259
4140 259
4145 261
4150 261
4155 260
4160 263
4165 262
4170 259
4175 261
4180 261
4185 262
4190 261
4195 263
4200 262
4205 261
4210 260
4215 261
4220 261
4225 262
4230 259
4235 261
4240 262
4245 260
4250 260
4255 263
4260 262
4265 260
4270 259
4275 258
4280 257
4285 257
4290 254
4295 257
4300 257
4305 257
4310 260
4315 258
4320 257
4325 256
4330 259
4335 258
4340 258
4345 258
4350 259
4355 259
4360 259
4365 256
4370 259
4375 257
4380 259
4385 259
4390 259
4395 258
4400 258
4405 258
4410 258
4415 257
4420 257
4425 260
4430 259
4435 259
4440 259
4445 260
4450 259
4455 259
4460 258
4465 259
4470 257
4475 258
4480 258
4485 258
4490 259
4495 260
4500 258
4505 260
4510 258
4515 258
4520 257
4525 259
4530 259
4535 258
4540 257
4545 259
4550 258
4555 258
4560 259
4565 259
4570 259
4575 259
4580 259
4585 259
4590 258
4595 259
4600 259
4605 258
4610 259
4615 259
4620 258
4625 260
4630 259
4635 258
4640 259
4645 259
4650 260
4655 257
4660 259
4665 258
4670 258
4675 260
4680 258
4685 258
4690 259
4695 258
4700 259
4705 259
4710 259
4715 258
4720 258
4725 259
4730 259
4735 258
4740 258
4745 258
4750 259
4755 259
4760 258
4765 259
4770 259
4775 258
4780 258
4785 258
4790 259
4795 259
4800 259
4805 258
4810 260
4815 258
4820 258
4825 259
4830 259
4835 258
4840 258
4845 258
4850 258
4855 259
4860 259
4865 258
4870 258
4875 259
4880 256
4885 258
4890 258
4895 258
4900 258
4905 259
4910 258
4915 261
4920 260
4925 259
4930 258
4935 259
4940 258
4945 258
4950 258
4955 258
4960 259
4965 262
4970 259
4975 259
4980 259
4985 259
4990 259
4995 259
5000 258
5005 257
5010 259
5015 259
5020 259
5025 258
5030 259
5035 258
5040 259
5045 258
5050 258
5055 259
5060 259
5065 257
5070 258
5075 259
5080 259
5085 259
5090 259
5095 259
5100 260
5105 257
5110 258
5115 259
5120 259
5125 258
5130 258
5135 259
5140 258
5145 257
5150 257
5155 258
5160 258
5165 257
5170 258
5175 257
5180 259
5185 258
5190 258
5195 258
5200 259
5205 257
5210 259
5215 259
5220 258
5225 258
5230 258
5235 256
5240 258
5245 260
5250 258
5255 258
5260 258
5265 258
5270 258
5275 258
5280 258
5285 257
5290 258
5295 258
5300 257
5305 257
5310 257
5315 257
5320 258
5325 256
5330 257
5335 255
5340 257
5345 258
5350 258
5355 257
5360 257
5365 258
5370 257
5375 258
5380 258
5385 258
5390 258
5395 259
};
\addlegendentry{4 jobs in parallel}
\addplot [semithick, color5]
table{
0 254
5 262
10 252
15 253
20 253
25 253
30 253
35 254
40 253
45 253
50 254
55 254
60 253
65 254
70 254
75 254
80 254
85 254
90 254
95 254
100 254
105 254
110 253
115 254
120 254
125 254
130 254
135 253
140 253
145 253
150 253
155 254
160 254
165 253
170 254
175 254
180 253
185 254
190 254
195 254
200 253
205 254
210 254
215 254
220 253
225 254
230 253
235 253
240 253
245 252
250 253
255 253
260 254
265 254
270 254
275 253
280 254
285 253
290 253
295 253
300 254
305 253
310 254
315 253
320 253
325 253
330 254
335 254
340 253
345 253
350 253
355 254
360 253
365 253
370 254
375 254
380 254
385 254
390 254
395 254
400 253
405 253
410 253
415 254
420 253
425 254
430 253
435 253
440 254
445 254
450 254
455 254
460 253
465 254
470 253
475 253
480 252
485 254
490 254
495 253
500 253
505 254
510 254
515 254
520 253
525 253
530 254
535 254
540 253
545 254
550 253
555 253
560 253
565 253
570 254
575 253
580 254
585 253
590 254
595 254
600 253
605 254
610 253
615 254
620 254
625 253
630 254
635 253
640 253
645 254
650 254
655 253
660 254
665 253
670 253
675 254
680 253
685 254
690 254
695 254
700 254
705 253
710 254
715 254
720 253
725 254
730 253
735 253
740 253
745 254
750 254
755 254
760 254
765 254
770 254
775 253
780 254
785 254
790 254
795 253
800 254
805 253
810 253
815 253
820 254
825 254
830 253
835 253
840 253
845 253
850 253
855 253
860 254
865 253
870 253
875 252
880 254
885 254
890 253
895 253
900 254
905 253
910 254
915 254
920 253
925 254
930 254
935 254
940 253
945 254
950 253
955 254
960 254
965 253
970 254
975 254
980 254
985 252
990 253
995 253
1000 253
1005 253
1010 253
1015 253
1020 252
1025 254
1030 253
1035 254
1040 252
1045 254
1050 254
1055 254
1060 253
1065 254
1070 254
1075 254
1080 253
1085 253
1090 253
1095 254
1100 254
1105 253
1110 253
1115 254
1120 253
1125 254
1130 254
1135 253
1140 253
1145 253
1150 253
1155 254
1160 254
1165 254
1170 254
1175 254
1180 253
1185 253
1190 254
1195 253
1200 253
1205 254
1210 254
1215 254
1220 254
1225 254
1230 251
1235 253
1240 254
1245 253
1250 254
1255 253
1260 254
1265 254
1270 254
1275 254
1280 254
1285 253
1290 253
1295 253
1300 253
1305 254
1310 254
1315 254
1320 254
1325 254
1330 254
1335 253
1340 255
1345 254
1350 254
1355 253
1360 254
1365 254
1370 254
1375 253
1380 254
1385 254
1390 253
1395 254
1400 255
1405 253
1410 254
1415 254
1420 254
1425 254
1430 253
1435 254
1440 254
1445 253
1450 253
1455 253
1460 253
1465 253
1470 253
1475 253
1480 254
1485 254
1490 254
1495 253
1500 254
1505 253
1510 252
1515 254
1520 254
1525 254
1530 253
1535 254
1540 253
1545 254
1550 253
1555 254
1560 252
1565 254
1570 254
1575 254
1580 254
1585 254
1590 253
1595 254
1600 251
1605 254
1610 253
1615 254
1620 253
1625 254
1630 253
1635 254
1640 254
1645 253
1650 253
1655 254
1660 254
1665 253
1670 253
1675 253
1680 253
1685 254
1690 254
1695 254
1700 254
1705 253
1710 253
1715 254
1720 254
1725 252
1730 252
1735 253
1740 254
1745 253
1750 254
1755 253
1760 253
1765 253
1770 254
1775 252
1780 253
1785 253
1790 252
1795 253
1800 253
1805 253
1810 253
1815 253
1820 254
1825 252
1830 254
1835 253
1840 253
1845 253
1850 252
1855 252
1860 253
1865 253
1870 253
1875 254
1880 253
1885 252
1890 253
1895 253
1900 253
1905 252
1910 254
1915 252
1920 254
1925 253
1930 254
1935 253
1940 253
1945 254
1950 253
1955 254
1960 253
1965 253
1970 254
1975 254
1980 252
1985 253
1990 254
1995 253
2000 253
2005 253
2010 254
2015 253
2020 253
2025 253
2030 253
2035 253
2040 254
2045 254
2050 253
2055 254
2060 253
2065 253
2070 254
2075 253
2080 253
2085 254
2090 253
2095 254
2100 254
2105 253
2110 254
2115 254
2120 252
2125 254
2130 254
2135 253
2140 253
2145 253
2150 254
2155 254
2160 254
2165 253
2170 251
2175 253
2180 253
2185 254
2190 254
2195 253
2200 253
2205 254
2210 253
2215 254
2220 254
2225 254
2230 253
2235 254
2240 253
2245 254
2250 253
2255 254
2260 254
2265 253
2270 254
2275 254
2280 253
2285 253
2290 254
2295 254
2300 253
2305 253
2310 254
2315 253
2320 253
2325 253
2330 253
2335 254
2340 254
2345 254
2350 254
2355 253
2360 253
2365 253
2370 253
2375 254
2380 254
2385 253
2390 253
2395 254
2400 253
2405 253
2410 253
2415 253
2420 254
2425 253
2430 253
2435 253
2440 253
2445 254
2450 254
2455 254
2460 254
2465 251
2470 254
2475 253
2480 254
2485 253
2490 253
2495 254
2500 254
2505 253
2510 254
2515 254
2520 254
2525 253
2530 253
2535 251
2540 252
2545 253
2550 254
2555 253
2560 253
2565 254
2570 253
2575 253
2580 253
2585 253
2590 253
2595 253
2600 253
2605 253
2610 253
2615 253
2620 253
2625 261
2630 252
2635 254
2640 253
2645 254
2650 253
2655 254
2660 253
2665 253
2670 254
2675 254
2680 253
2685 253
2690 253
2695 253
2700 253
2705 254
2710 254
2715 254
2720 254
2725 253
2730 253
2735 253
2740 253
2745 254
2750 254
2755 253
2760 254
2765 252
2770 254
2775 254
2780 254
2785 254
2790 254
2795 253
2800 254
2805 253
2810 253
2815 253
2820 253
2825 253
2830 254
2835 254
2840 254
2845 254
2850 254
2855 254
2860 253
2865 254
2870 254
2875 253
2880 254
2885 251
2890 253
2895 254
2900 253
2905 253
2910 254
2915 253
2920 249
2925 252
2930 253
2935 252
2940 251
2945 252
2950 252
2955 253
2960 252
2965 252
2970 252
2975 252
2980 252
2985 251
2990 252
2995 252
3000 251
3005 252
3010 252
3015 252
3020 251
3025 251
3030 252
3035 251
3040 252
3045 252
3050 251
3055 252
3060 252
3065 252
3070 252
3075 252
3080 252
3085 252
3090 253
3095 251
3100 252
3105 252
3110 251
3115 252
3120 252
3125 251
3130 252
3135 253
3140 252
3145 252
3150 251
3155 251
3160 252
3165 252
3170 252
3175 251
3180 252
3185 251
3190 251
3195 253
3200 251
3205 251
3210 251
3215 252
3220 251
3225 252
3230 252
3235 252
3240 252
3245 251
3250 252
3255 251
3260 251
3265 252
3270 252
3275 251
3280 252
3285 251
3290 251
3295 251
3300 251
3305 252
3310 251
3315 253
3320 251
3325 251
3330 251
3335 252
3340 251
3345 251
3350 252
3355 252
3360 252
3365 252
3370 252
3375 252
3380 252
3385 253
3390 251
3395 253
3400 251
3405 252
3410 252
3415 252
3420 252
3425 251
3430 251
3435 253
3440 252
3445 251
3450 252
3455 251
3460 251
3465 251
3470 253
3475 253
3480 254
3485 253
3490 254
3495 254
3500 253
3505 254
3510 253
3515 254
3520 254
3525 254
3530 254
3535 253
3540 254
3545 253
3550 254
3555 254
3560 254
3565 253
3570 253
3575 254
3580 253
3585 254
3590 254
3595 253
3600 252
3605 253
3610 253
3615 254
3620 253
3625 255
3630 254
3635 254
3640 253
3645 253
3650 254
3655 253
3660 254
3665 254
3670 253
3675 252
3680 252
3685 254
3690 253
3695 253
3700 254
3705 253
3710 254
3715 253
3720 253
3725 253
3730 253
3735 252
3740 252
3745 254
3750 253
3755 250
3760 253
3765 253
3770 253
3775 254
3780 253
3785 253
3790 253
3795 254
3800 252
3805 254
3810 254
3815 253
3820 253
3825 253
3830 253
3835 254
3840 253
3845 252
3850 254
3855 253
3860 253
3865 253
3870 253
3875 253
3880 253
3885 253
3890 253
3895 254
3900 252
3905 254
3910 253
3915 253
3920 253
3925 253
3930 254
3935 254
3940 253
3945 253
3950 252
3955 253
3960 253
3965 252
3970 253
3975 254
3980 253
3985 254
3990 252
3995 253
4000 254
4005 253
4010 254
4015 253
4020 253
4025 254
4030 253
4035 254
4040 254
4045 254
4050 253
4055 254
4060 253
4065 254
4070 254
4075 254
4080 254
4085 254
4090 253
4095 254
4100 253
4105 253
4110 254
4115 253
4120 253
4125 254
4130 253
4135 253
4140 254
4145 254
4150 254
4155 253
4160 253
4165 254
4170 253
4175 253
4180 253
4185 254
4190 254
4195 253
4200 253
4205 254
4210 253
4215 253
4220 254
4225 253
4230 253
4235 253
4240 253
4245 254
4250 253
4255 253
4260 254
4265 253
4270 254
4275 254
4280 253
4285 253
4290 253
4295 254
4300 254
4305 254
4310 254
4315 253
4320 254
4325 254
4330 254
4335 253
4340 263
4345 253
4350 254
4355 254
4360 254
4365 254
4370 254
4375 254
4380 254
4385 254
4390 253
4395 253
4400 254
4405 253
4410 254
4415 254
4420 253
4425 253
4430 254
4435 253
4440 253
4445 254
4450 254
4455 253
4460 254
4465 253
4470 254
4475 253
4480 254
4485 253
4490 254
4495 254
4500 253
4505 253
4510 254
4515 253
4520 254
4525 253
4530 253
4535 253
4540 254
4545 254
4550 253
4555 253
4560 253
4565 253
4570 253
4575 254
4580 254
4585 253
4590 254
4595 254
4600 254
4605 254
4610 253
4615 254
4620 253
4625 253
4630 253
4635 253
4640 254
4645 253
4650 254
4655 254
4660 254
4665 253
4670 254
4675 254
4680 254
4685 253
4690 253
4695 254
4700 254
4705 254
4710 254
4715 254
4720 254
4725 253
4730 254
4735 253
4740 254
4745 253
4750 253
4755 254
4760 254
4765 253
4770 254
4775 254
4780 253
4785 254
4790 253
4795 253
4800 254
4805 253
4810 254
4815 254
4820 254
4825 254
4830 254
4835 253
4840 254
4845 254
4850 254
4855 253
4860 254
4865 253
4870 254
4875 253
4880 254
4885 253
4890 254
4895 254
4900 254
4905 253
4910 253
4915 254
4920 253
4925 254
4930 254
4935 254
4940 253
4945 254
4950 253
4955 254
4960 254
4965 252
4970 252
4975 254
4980 254
4985 254
4990 254
4995 254
5000 253
5005 254
5010 253
5015 254
5020 253
5025 254
5030 254
5035 254
5040 254
5045 252
5050 255
5055 253
5060 253
5065 253
5070 254
5075 254
5080 253
5085 254
5090 251
5095 253
5100 254
5105 254
5110 254
5115 254
5120 253
5125 254
5130 254
5135 254
5140 254
5145 254
5150 254
5155 254
5160 254
5165 253
5170 254
5175 253
5180 254
5185 254
5190 253
5195 254
5200 253
5205 253
5210 253
5215 252
5220 253
5225 254
5230 254
5235 253
5240 253
5245 253
5250 253
5255 253
5260 253
5265 253
5270 253
5275 252
5280 254
5285 254
5290 252
5295 253
5300 253
5305 254
5310 253
5315 254
5320 254
5325 254
5330 253
5335 254
5340 254
5345 253
5350 253
5355 254
5360 254
5365 253
5370 253
5375 253
5380 254
5385 254
5390 253
5395 253
};
\addlegendentry{Baseline}
\addplot [semithick, color1, mark=diamond*, mark size=3, mark options={solid,fill=white,draw=red},  mark repeat={180}]
table{%
0 265
5 262
10 266
15 269
20 270
25 266
30 262
35 255
40 253
45 254
50 253
55 254
60 253
65 253
70 253
75 254
80 254
85 254
90 254
95 253
100 254
105 256
110 266
115 266
120 266
125 267
130 262
135 263
140 263
145 261
150 260
155 270
160 266
165 267
170 259
175 267
180 357
185 360
190 361
195 360
200 361
205 361
210 361
215 361
220 359
225 360
230 360
235 361
240 361
245 361
250 359
255 358
260 359
265 360
270 361
275 362
280 360
285 363
290 360
295 362
300 362
305 362
310 363
315 363
320 362
325 362
330 363
335 364
340 362
345 363
350 363
355 364
360 364
365 363
370 364
375 364
380 364
385 363
390 363
395 364
400 364
405 364
410 364
415 365
420 364
425 364
430 364
435 365
440 365
445 364
450 365
};
\addlegendentry{1 job}
\addplot [semithick, color2, mark=square*, mark size=3, mark options={solid,fill=white,draw=red},  mark repeat={180}]
table {%
455 367
460 374
465 378
470 381
475 382
480 367
485 376
490 366
495 366
500 365
505 365
510 364
515 366
520 366
525 366
530 367
535 371
540 367
545 369
550 366
555 367
560 378
565 379
570 382
575 379
580 372
585 370
590 370
595 371
600 370
605 381
610 388
615 389
620 383
625 376
630 465
635 464
640 465
645 466
650 466
655 467
660 468
665 468
670 469
675 469
680 469
685 468
690 469
695 470
700 470
705 470
710 470
715 471
720 471
725 472
730 471
735 471
740 472
745 470
750 472
755 470
760 470
765 471
770 470
775 470
780 470
785 471
790 471
795 471
800 471
805 471
810 468
815 470
820 472
825 471
830 472
835 471
840 470
845 471
850 471
855 471
860 471
865 468
870 471
875 471
880 471
885 472
890 471
895 471
900 471
};
\addlegendentry{2 jobs}
\addplot [semithick, color3, mark=triangle*, mark size=3, mark options={solid,fill=white,draw=red},  mark repeat={180}]
table {%
905 475
910 472
915 471
920 471
925 474
930 472
935 469
940 470
945 470
950 470
955 470
960 472
965 470
970 471
975 470
980 469
985 470
990 469
995 470
1000 470
1005 470
1010 470
1015 469
1020 470
1025 471
1030 471
1035 471
1040 472
1045 472
1050 471
1055 471
1060 470
1065 470
1070 471
1075 471
1080 471
1085 470
1090 468
1095 470
1100 471
1105 465
1110 464
1115 471
1120 472
1125 472
1130 472
1135 472
1140 473
1145 471
1150 472
1155 470
1160 472
1165 472
1170 473
1175 472
1180 472
1185 473
1190 473
1195 472
1200 472
1205 474
1210 473
1215 473
1220 473
1225 472
1230 473
1235 473
1240 473
1245 472
1250 473
1255 473
1260 472
1265 473
1270 471
1275 472
1280 473
1285 473
1290 473
1295 472
1300 474
1305 474
1310 476
1315 475
1320 473
1325 474
1330 473
1335 473
1340 473
1345 472
1350 473
};
\addlegendentry{3 jobs}
\addplot [semithick, color4, mark=o, mark size=3, mark options={solid,fill=white,draw=red},  mark repeat={180}]
table {%
1355 474
1360 474
1365 474
1370 475
1375 475
1380 474
1385 475
1390 472
1395 472
1400 471
1405 471
1410 481
1415 469
1420 473
1425 472
1430 473
1435 475
1440 475
1445 475
1450 475
1455 475
1460 476
1465 477
1470 477
1475 476
1480 476
1485 475
1490 475
1495 475
1500 475
1505 473
1510 473
1515 476
1520 475
1525 475
1530 476
1535 475
1540 477
1545 475
1550 475
1555 475
1560 475
1565 475
1570 474
1575 474
1580 474
1585 474
1590 474
1595 471
1600 473
1605 474
1610 473
1615 474
1620 473
1625 475
1630 473
1635 474
1640 474
1645 474
1650 474
1655 473
1660 473
1665 474
1670 474
1675 473
1680 473
1685 474
1690 474
1695 474
1700 474
1705 474
1710 474
1715 473
1720 474
1725 473
1730 474
1735 474
1740 471
1745 474
1750 473
1755 473
1760 474
1765 473
1770 474
1775 475
1780 474
1785 473
1790 474
1795 475
1800 475
1805 472
1810 474
1815 475
1820 474
1825 474
1830 475
1835 475
1840 475
1845 475
1850 474
1855 475
1860 475
1865 475
1870 474
1875 474
1880 473
1885 474
1890 474
1895 474
1900 475
1905 474
1910 474
1915 473
1920 474
1925 474
1930 474
1935 473
1940 474
1945 471
1950 472
1955 472
1960 472
1965 473
1970 473
1975 472
1980 472
1985 473
1990 472
1995 473
2000 472
2005 473
2010 472
2015 472
2020 471
2025 472
2030 470
2035 472
2040 472
2045 472
2050 471
2055 471
2060 471
2065 472
2070 471
2075 471
2080 472
2085 472
2090 472
2095 472
2100 472
2105 472
2110 472
2115 472
2120 472
2125 471
2130 471
2135 473
2140 472
2145 472
2150 472
2155 472
2160 472
2165 473
2170 473
2175 473
2180 472
2185 473
2190 472
2195 473
2200 473
2205 473
2210 473
2215 472
2220 474
2225 474
2230 474
2235 473
2240 473
2245 474
2250 474
2255 473
2260 474
2265 473
2270 473
2275 473
2280 474
2285 474
2290 473
2295 474
2300 474
2305 475
2310 474
2315 474
2320 473
2325 473
2330 475
2335 473
2340 475
2345 473
2350 474
2355 475
2360 475
2365 475
2370 475
2375 475
2380 475
2385 474
2390 475
2395 475
2400 475
2405 474
2410 475
2415 475
2420 475
2425 475
2430 475
2435 475
2440 475
2445 474
2450 474
2455 475
2460 475
2465 475
2470 475
2475 474
2480 475
2485 475
2490 474
2495 475
2500 475
2505 461
2510 463
2515 472
2520 471
2525 471
2530 471
2535 471
2540 471
2545 471
2550 472
2555 471
2560 472
2565 471
2570 471
2575 471
2580 471
2585 471
2590 471
2595 471
2600 471
2605 470
2610 471
2615 470
2620 471
2625 471
2630 470
2635 471
2640 471
2645 471
2650 471
2655 471
2660 471
2665 470
2670 470
2675 470
2680 469
2685 470
2690 469
2695 473
2700 473
2705 472
2710 474
2715 473
2720 472
2725 474
2730 472
2735 473
2740 473
2745 472
2750 473
2755 473
2760 472
2765 472
2770 472
2775 472
2780 472
2785 473
2790 471
2795 473
2800 472
2805 474
2810 473
2815 472
2820 472
2825 473
2830 473
2835 473
2840 473
2845 473
2850 472
2855 472
2860 472
2865 472
2870 473
2875 473
2880 472
2885 472
2890 473
2895 472
2900 473
2905 473
2910 473
2915 473
2920 473
2925 473
2930 473
2935 472
2940 473
2945 473
2950 473
2955 474
2960 473
2965 473
2970 474
2975 474
2980 474
2985 474
2990 474
2995 474
3000 474
3005 474
3010 472
3015 472
3020 473
3025 474
3030 474
3035 473
3040 472
3045 473
3050 472
3055 473
3060 473
3065 473
3070 473
3075 473
3080 473
3085 473
3090 472
3095 472
3100 472
3105 472
3110 472
3115 471
3120 471
3125 471
3130 471
3135 472
3140 472
3145 472
3150 472
3155 471
3160 472
3165 472
3170 471
3175 472
3180 471
3185 471
3190 470
3195 471
3200 471
3205 470
3210 471
3215 471
3220 470
3225 471
3230 471
3235 469
3240 471
3245 471
3250 469
3255 471
3260 470
3265 471
3270 472
3275 472
3280 473
3285 472
3290 472
3295 472
3300 472
3305 472
3310 473
3315 472
3320 473
3325 472
3330 472
3335 472
3340 473
3345 472
3350 471
3355 470
3360 471
3365 470
3370 471
3375 472
3380 472
3385 472
3390 471
3395 473
3400 474
3405 474
3410 475
3415 473
3420 473
3425 474
3430 472
3435 473
3440 472
3445 473
3450 474
3455 474
3460 475
3465 475
3470 474
3475 473
3480 474
3485 475
3490 474
3495 472
3500 473
3505 473
3510 471
3515 471
3520 472
3525 472
3530 473
3535 473
3540 472
3545 473
3550 473
3555 473
3560 472
3565 473
3570 472
3575 472
3580 473
3585 472
3590 472
3595 472
3600 472
3605 471
3610 473
3615 472
3620 472
3625 473
3630 473
3635 473
3640 472
3645 471
3650 472
3655 473
3660 472
3665 473
3670 472
3675 472
3680 472
3685 472
3690 472
3695 473
3700 472
3705 473
3710 472
3715 473
3720 471
3725 471
3730 472
3735 473
3740 472
3745 472
3750 472
3755 473
3760 473
3765 471
3770 472
3775 471
3780 473
3785 473
3790 473
3795 473
3800 470
3805 473
3810 473
3815 472
3820 473
3825 473
3830 474
3835 474
3840 475
3845 476
3850 474
3855 475
3860 476
3865 476
3870 476
3875 476
3880 476
3885 476
3890 475
3895 476
3900 475
3905 474
3910 474
3915 473
3920 473
3925 473
3930 472
3935 475
3940 473
3945 472
3950 473
3955 471
3960 472
3965 473
3970 472
};
\addlegendentry{4 jobs}
\addplot [semithick, color3, mark=triangle*, mark size=3, mark options={solid,fill=white,draw=red},  mark repeat={180}]
table {%
3970 472
3975 466
3980 449
3980 449
3985 462
3990 449
3995 448
4000 447
4005 447
4010 467
4015 467
4020 467
4025 468
4030 468
4035 468
4040 468
4045 468
4050 468
4055 468
4060 468
4065 469
4070 469
4075 469
4080 469
4085 468
4090 470
4095 469
4100 470
4105 469
4110 470
4115 470
4120 470
4125 470
4130 469
4135 471
4140 471
4145 471
4150 471
4155 471
4160 470
4165 470
4170 470
4175 472
4180 472
4185 472
4190 469
4195 472
4200 473
4205 471
4210 472
4215 472
4220 472
4225 471
4230 472
4235 472
4240 472
4245 471
4250 471
4255 472
4260 471
4265 471
4270 473
4275 471
4280 472
4285 472
4290 473
4295 473
4300 472
4305 473
4310 472
4315 472
4320 473
4325 472
4330 473
4335 473
4340 472
4345 473
4350 473
4355 472
4360 471
4365 472
4370 471
4375 471
4380 471
4385 471
4390 472
4395 471
4400 471
};
\addplot [semithick, color2, mark=square*, mark size=3, mark options={solid,fill=white,draw=red},  mark repeat={180}]
table {%
4400 471
4405 465
4410 424
4410 424
4415 406
4420 386
4425 381
4430 377
4435 380
4440 379
4445 374
4450 367
4455 366
4460 367
4465 364
4470 364
4475 364
4480 364
4485 364
4490 365
4495 365
4500 364
4505 364
4510 366
4515 364
4520 364
4525 365
4530 365
4535 364
4540 365
4545 366
4550 365
4555 366
4560 364
4565 364
4570 366
4575 365
4580 365
4585 366
4590 365
4595 365
4600 366
4605 365
4610 366
4615 365
4620 364
4625 366
4630 365
4635 365
4640 366
4645 366
4650 366
4655 366
4660 365
4665 367
4670 366
4675 366
4680 366
4685 366
4690 367
};
\addplot [semithick, color1, mark=diamond*, mark size=3, mark options={solid,fill=white,draw=red},  mark repeat={180}]
table {%
4690 367
4695 340
4700 312
4705 299
4710 285
4715 285
4720 272
4725 272
4730 271
4735 271
4740 258
4745 258
4750 257
4755 257
4760 257
4765 257
4770 256
4775 257
4780 256
4785 255
4790 255
4795 255
4800 256
};

\end{axis}

\end{tikzpicture}

%% file: results/energy_intel_seq.tex
\begin{tikzpicture}[font=\Large]

\definecolor{color0}{rgb}{0.12156862745098,0.466666666666667,0.705882352941177}
\definecolor{color1}{rgb}{1,0.498039215686275,0.0549019607843137}
\definecolor{color2}{rgb}{0.172549019607843,0.627450980392157,0.172549019607843}
\definecolor{color3}{rgb}{0.83921568627451,0.152941176470588,0.156862745098039}
\definecolor{color4}{rgb}{0.580392156862745,0.403921568627451,0.741176470588235}
\definecolor{color5}{rgb}{0,0,0}
\definecolor{color6}{rgb}{0.07, 0.04, 0.56}

\begin{axis}[
legend cell align={left},
legend columns=3,
legend style={fill opacity=0.8, draw opacity=1, text opacity=1, at={(1.05,1.18)}, anchor=east, draw=white!80.0!black},
tick align=outside,
tick pos=left,
x grid style={white!69.01960784313725!black},
xlabel={Time (SEC)},
xmin=0, xmax=4800,
xtick={0,1200,2400,3600,4800},
xtick style={color=black},
y grid style={white!69.01960784313725!black},
ylabel={Total energy consumption (kWh)},
ymin=0, ymax=0.6,
xmajorgrids,
ymajorgrids,
ytick style={color=black},
ytick={0.05,0.1,0.15,0.2,0.25,0.3,0.35,0.4,0.45,0.5,0.55,0.6},
yticklabel style={
        /pgf/number format/fixed,
        /pgf/number format/precision=2
},
 y label style={at={(axis description cs:-0.05,.5)}}
]
\addplot [semithick, olive, mark=star, mark size=3, mark options={solid,fill=white,draw=red},  mark repeat={180}]
table[y expr=(\thisrowno{1}-422665)/1000] {%
0 422665
5 422665
10 422666
15 422666
20 422666
25 422667
30 422667
35 422668
40 422668
45 422668
50 422669
55 422669
60 422669
65 422670
70 422670
75 422670
80 422671
85 422671
90 422672
95 422672
100 422672
105 422673
110 422673
115 422673
120 422674
125 422674
130 422675
135 422675
140 422675
145 422676
150 422676
155 422677
160 422677
165 422677
170 422678
175 422678
180 422679
185 422679
190 422680
195 422680
200 422681
205 422681
210 422682
215 422683
220 422684
225 422684
230 422685
235 422685
240 422686
245 422686
250 422687
255 422688
260 422688
265 422689
270 422690
275 422690
280 422691
285 422692
290 422692
295 422693
300 422693
305 422694
310 422695
315 422695
320 422696
325 422697
330 422697
335 422698
340 422699
345 422699
350 422700
355 422701
360 422701
365 422702
370 422703
375 422703
380 422704
385 422705
390 422705
395 422706
400 422707
405 422707
410 422708
415 422709
420 422709
425 422710
430 422711
435 422711
440 422712
445 422712
450 422713
455 422714
460 422714
465 422715
470 422716
475 422716
480 422717
485 422718
490 422718
495 422719
500 422720
505 422720
510 422721
515 422722
520 422722
525 422723
530 422724
535 422724
540 422725
545 422725
550 422726
555 422727
560 422727
565 422728
570 422729
575 422729
580 422730
585 422731
590 422731
595 422732
600 422733
605 422733
610 422734
615 422735
620 422735
625 422736
630 422737
635 422737
640 422738
645 422739
650 422739
655 422740
660 422741
665 422741
670 422742
675 422743
680 422743
685 422744
690 422745
695 422745
700 422746
705 422747
710 422747
715 422748
720 422748
725 422749
730 422750
735 422750
740 422751
745 422752
750 422753
755 422753
760 422754
765 422755
770 422755
775 422756
780 422756
785 422757
790 422758
795 422758
800 422759
805 422760
810 422760
815 422761
820 422762
825 422762
830 422763
835 422763
840 422764
845 422764
850 422765
855 422766
860 422766
865 422767
870 422768
875 422768
880 422769
885 422770
890 422770
895 422771
900 422772
905 422772
910 422773
915 422774
920 422774
925 422775
930 422776
935 422776
940 422777
945 422777
950 422778
955 422779
960 422780
965 422780
970 422781
975 422782
980 422782
985 422783
990 422784
995 422784
1000 422785
1005 422785
1010 422786
1015 422787
1020 422787
1025 422788
1030 422789
1035 422789
1040 422790
1045 422791
1050 422791
1055 422792
1060 422793
1065 422793
1070 422794
1075 422795
1080 422795
1085 422796
1090 422796
1095 422797
1100 422798
1105 422798
1110 422799
1115 422800
1120 422800
1125 422801
1130 422802
1135 422802
1140 422803
1145 422804
1150 422804
1155 422805
1160 422806
1165 422806
1170 422807
1175 422808
1180 422808
1185 422809
1190 422810
1195 422810
1200 422811
1205 422812
1210 422812
1215 422813
1220 422814
1225 422814
1230 422815
1235 422815
1240 422816
1245 422817
1250 422817
1255 422818
1260 422819
1265 422819
1270 422820
1275 422821
1280 422821
1285 422822
1290 422823
1295 422823
1300 422824
1305 422825
1310 422825
1315 422826
1320 422827
1325 422827
1330 422828
1335 422828
1340 422829
1345 422830
1350 422830
1355 422831
1360 422832
1365 422832
1370 422833
1375 422834
1380 422834
1385 422835
1390 422836
1395 422836
1400 422837
1405 422838
1410 422838
1415 422839
1420 422840
1425 422840
1430 422841
1435 422841
1440 422842
1445 422843
1450 422843
1455 422844
1460 422845
1465 422845
1470 422846
1475 422847
1480 422847
1485 422848
1490 422848
1495 422849
1500 422850
1505 422850
1510 422851
1515 422852
1520 422852
1525 422853
1530 422854
1535 422854
1540 422855
1545 422856
1550 422856
1555 422857
1560 422858
1565 422858
1570 422859
1575 422860
1580 422860
1585 422861
1590 422862
1595 422862
1600 422863
1605 422864
1610 422864
1615 422865
1620 422866
1625 422866
1630 422867
1635 422868
1640 422868
1645 422869
1650 422870
1655 422870
1660 422871
1665 422871
1670 422872
1675 422873
1680 422873
1685 422874
1690 422875
1695 422875
1700 422876
1705 422877
1710 422877
1715 422878
1720 422879
1725 422879
1730 422880
1735 422881
1740 422881
1745 422882
1750 422883
1755 422883
1760 422884
1765 422885
1770 422885
1775 422886
1780 422887
1785 422887
1790 422888
1795 422889
1800 422889
1805 422890
1810 422891
1815 422891
1820 422892
1825 422892
1830 422893
1835 422894
1840 422894
1845 422895
1850 422896
1855 422896
1860 422897
1865 422898
1870 422898
1875 422899
1880 422900
1885 422900
1890 422901
1895 422902
1900 422902
1905 422903
1910 422904
1915 422904
1920 422905
1925 422906
1930 422906
1935 422907
1940 422908
1945 422908
1950 422909
1955 422909
1960 422910
1965 422911
1970 422911
1975 422912
1980 422913
1985 422914
1990 422914
1995 422915
2000 422915
2005 422916
2010 422917
2015 422917
2020 422918
2025 422919
2030 422919
2035 422921
2040 422921
2045 422922
2050 422923
2055 422923
2060 422924
2065 422924
2070 422925
2075 422926
2080 422926
2085 422927
2090 422928
2095 422928
2100 422929
2105 422930
2110 422930
2115 422931
2120 422932
2125 422932
2130 422933
2135 422934
2140 422934
2145 422935
2150 422936
2155 422936
2160 422937
2165 422938
2170 422938
2175 422939
2180 422940
2185 422940
2190 422941
2195 422942
2200 422942
2205 422943
2210 422944
2215 422944
2220 422945
2225 422946
2230 422946
2235 422947
2240 422948
2245 422948
2250 422949
2255 422949
2260 422950
2265 422951
2270 422951
2275 422952
2280 422953
2285 422953
2290 422954
2295 422955
2300 422955
2305 422956
2310 422957
2315 422957
2320 422958
2325 422959
2330 422959
2335 422960
2340 422961
2345 422961
2350 422962
2355 422963
2360 422963
2365 422964
2370 422965
2375 422965
2380 422966
2385 422967
2390 422967
2395 422968
2400 422969
2405 422969
2410 422970
2415 422970
2420 422971
2425 422972
2430 422972
2435 422973
2440 422974
2445 422974
2450 422975
2455 422976
2460 422976
2465 422977
2470 422978
2475 422978
2480 422979
2485 422980
2490 422980
2495 422981
2500 422982
2505 422982
2510 422983
2515 422984
2520 422984
2525 422985
2530 422986
2535 422986
2540 422987
2545 422988
2550 422988
2555 422989
2560 422989
2565 422990
2570 422991
2575 422991
2580 422992
2585 422993
2590 422993
2595 422994
2600 422995
2605 422995
2610 422996
2615 422997
2620 422997
2625 422998
2630 422999
2635 423000
2640 423000
2645 423001
2650 423002
2655 423002
2660 423003
2665 423004
2670 423004
2675 423005
2680 423006
2685 423006
2690 423007
2695 423008
2700 423008
2705 423009
2710 423010
2715 423010
2720 423011
2725 423012
2730 423012
2735 423013
2740 423014
2745 423014
2750 423015
2755 423016
2760 423016
2765 423017
2770 423018
2775 423018
2780 423019
2785 423019
2790 423020
2795 423021
2800 423021
2805 423022
2810 423023
2815 423023
2820 423024
2825 423025
2830 423025
2835 423026
2840 423027
2845 423027
2850 423028
2855 423029
2860 423029
2865 423030
2870 423031
2875 423031
2880 423032
2885 423033
2890 423033
2895 423034
2900 423035
2905 423035
2910 423036
2915 423036
2920 423037
2925 423038
2930 423039
2935 423039
2940 423040
2945 423040
2950 423041
2955 423042
2960 423042
2965 423043
2970 423044
2975 423044
2980 423045
2985 423046
2990 423046
2995 423047
3000 423048
3005 423048
3010 423049
3015 423050
3020 423050
3025 423051
3030 423052
3035 423052
3040 423053
3045 423054
3050 423054
3055 423055
3060 423056
3065 423056
3070 423057
3075 423057
3080 423058
3085 423059
3090 423059
3095 423060
3100 423061
3105 423061
3110 423062
3115 423063
3120 423063
3125 423064
3130 423065
3135 423065
3140 423066
3145 423067
3150 423067
3155 423068
3160 423069
3165 423069
3170 423070
3175 423071
3180 423071
3185 423072
3190 423073
3195 423073
3200 423074
3205 423074
3210 423075
3215 423076
3220 423076
3225 423077
3230 423078
3235 423079
3240 423080
3245 423080
3250 423081
3255 423082
3260 423082
3265 423083
3270 423084
3275 423084
3280 423085
3285 423086
3290 423086
3295 423087
3300 423088
3305 423088
3310 423089
3315 423089
3320 423090
3325 423091
3330 423091
3335 423092
3340 423093
3345 423093
3350 423094
3355 423095
3360 423095
3365 423096
3370 423097
3375 423097
3380 423098
3385 423099
3390 423099
3395 423100
3400 423101
3405 423101
3410 423102
3415 423103
3420 423103
3425 423104
3430 423105
3435 423105
3440 423106
3445 423107
3450 423107
3455 423108
3460 423108
3465 423109
3470 423110
3475 423110
3480 423111
3485 423112
3490 423112
3495 423113
3500 423114
3505 423114
3510 423115
3515 423116
3520 423116
3525 423117
3530 423118
3535 423118
3540 423119
3545 423120
3550 423120
3555 423121
3560 423122
3565 423122
3570 423123
3575 423124
3580 423124
3585 423125
3590 423126
3595 423126
3600 423127
3605 423128
3610 423128
3615 423129
3620 423129
3625 423130
3630 423131
3635 423131
3640 423132
3645 423133
3650 423133
3655 423134
3660 423135
3665 423135
3670 423136
3675 423137
3680 423137
3685 423138
3690 423139
3695 423139
3700 423140
3705 423141
3710 423141
3715 423142
3720 423143
3725 423143
3730 423144
3735 423145
3740 423145
3745 423146
3750 423147
3755 423147
3760 423148
3765 423148
3770 423149
3775 423150
3780 423150
3785 423151
3790 423152
3795 423152
3800 423153
3805 423154
3810 423154
3815 423155
3820 423156
3825 423156
3830 423156
3835 423157
3840 423158
3845 423159
3850 423159
3855 423160
3860 423160
3865 423161
3870 423162
3875 423162
3880 423163
3885 423164
3890 423164
3895 423165
3900 423166
3905 423166
3910 423167
3915 423168
3920 423168
3925 423169
3930 423170
3935 423170
3940 423171
3945 423172
3950 423172
3955 423173
3960 423174
3965 423174
3970 423175
3975 423176
3980 423176
3985 423177
3990 423178
3995 423178
4000 423179
4005 423180
4010 423180
4015 423181
4020 423181
4025 423182
4030 423182
4035 423183
4040 423184
4045 423184
4050 423185
4055 423185
4060 423186
4065 423186
4070 423186
4075 423187
4080 423187
4085 423188
4090 423188
4095 423188
4100 423189
4105 423189
4110 423189
4115 423190
4120 423190
4125 423191
4130 423191
4135 423191
4140 423192
4145 423192
4150 423192
4155 423193
4160 423193
4165 423193
4170 423194
4175 423194
4180 423195
4185 423195
4190 423195
4195 423196
4200 423196
4205 423196
4210 423197
4215 423197
4220 423197
4225 423198
4230 423198
4235 423199
4240 423199
4245 423199
4250 423200
4255 423200
4260 423200
4265 423201
4270 423201
4275 423201
4280 423202
4285 423202
4290 423202
4295 423203
4300 423203
4305 423204
4310 423204
4315 423204
4320 423205
4325 423205
4330 423205
4335 423206
4340 423206
4345 423206
4350 423207
4355 423207
4360 423207
4365 423208
4370 423208
4375 423208
4380 423209
4385 423209
4390 423210
4395 423210
4400 423210
4405 423211
4410 423211
4415 423211
4420 423212
4425 423212
4430 423212
4435 423213
4440 423213
4445 423214
4450 423214
4455 423214
4460 423215
4465 423215
4470 423215
4475 423216
4480 423216
4485 423216
4490 423217
4495 423217
4500 423217
4505 423218
4510 423218
4515 423218
4520 423219
4525 423219
4530 423220
4535 423220
4540 423220
4545 423221
4550 423221
4555 423221
4560 423222
4565 423222
4570 423222
4575 423223
4580 423223
4585 423223
4590 423224
4595 423224
4600 423225
4605 423225
4610 423225
4615 423226
4620 423226
4625 423226
4630 423227
4635 423227
4640 423227
4645 423228
4650 423228
4655 423228
4660 423229
4665 423229
4670 423230
4675 423230
4680 423230
4685 423231
4690 423231
4695 423231
4700 423232
4705 423232
4710 423232
4715 423233
4720 423233
4725 423233
4730 423234
4735 423234
4740 423235
4745 423235
4750 423235
4755 423236
4760 423236
4765 423236
4770 423237
4775 423237
4780 423237
4785 423238
4790 423238
4795 423238
4800 423239
4805 423239
4810 423240
4815 423240
4820 423240
4825 423241
4830 423241
4835 423241
4840 423242
4845 423242
4850 423242
4855 423243
4860 423243
4865 423243
4870 423244
4875 423244
4880 423245
4885 423245
4890 423245
4895 423246
4900 423246
4905 423246
4910 423247
4915 423247
4920 423247
4925 423248
4930 423248
4935 423249
4940 423249
4945 423249
4950 423250
4955 423250
4960 423250
4965 423251
4970 423251
4975 423251
4980 423252
4985 423252
4990 423252
4995 423253
5000 423253
5005 423253
5010 423254
5015 423254
5020 423255
5025 423255
5030 423256
5035 423256
5040 423256
5045 423257
5050 423257
5055 423257
5060 423258
5065 423258
5070 423258
5075 423259
5080 423259
5085 423259
5090 423260
5095 423260
5100 423260
5105 423261
5110 423261
5115 423262
5120 423262
5125 423262
5130 423263
5135 423263
5140 423263
5145 423264
5150 423264
5155 423264
5160 423265
5165 423265
5170 423265
5175 423266
5180 423266
5185 423266
5190 423267
5195 423267
5200 423268
5205 423268
5210 423268
5215 423269
5220 423269
5225 423269
5230 423270
5235 423270
5240 423270
5245 423271
5250 423271
5255 423272
5260 423272
5265 423272
5270 423273
5275 423273
5280 423273
5285 423274
5290 423274
5295 423274
5300 423275
5305 423275
5310 423275
5315 423276
5320 423276
5325 423276
5330 423277
5335 423277
5340 423278
5345 423278
5350 423278
5355 423279
5360 423279
5365 423279
5370 423280
5375 423280
5380 423280
5385 423281
5390 423281
5395 423282
5400 423282
};
\addlegendentry{4 jobs in parallel}
\addplot [semithick, color5]
table [y expr=(\thisrowno{1}-521738)/1000] {%
0 521738
5 521738
10 521739
15 521739
20 521739
25 521740
30 521740
35 521740
40 521741
45 521741
50 521741
55 521742
60 521742
65 521743
70 521743
75 521743
80 521744
85 521744
90 521744
95 521745
100 521745
105 521745
110 521746
115 521746
120 521746
125 521747
130 521747
135 521747
140 521748
145 521748
150 521748
155 521749
160 521749
165 521749
170 521750
175 521750
180 521751
185 521751
190 521751
195 521752
200 521752
205 521752
210 521753
215 521753
220 521753
225 521754
230 521754
235 521754
240 521755
245 521755
250 521755
255 521756
260 521756
265 521757
270 521757
275 521757
280 521758
285 521758
290 521758
295 521759
300 521759
305 521759
310 521760
315 521760
320 521760
325 521761
330 521761
335 521761
340 521762
345 521762
350 521762
355 521763
360 521763
365 521763
370 521764
375 521764
380 521765
385 521765
390 521765
395 521766
400 521766
405 521766
410 521767
415 521767
420 521767
425 521768
430 521768
435 521768
440 521769
445 521769
450 521769
455 521770
460 521770
465 521771
470 521771
475 521771
480 521772
485 521772
490 521772
495 521773
500 521773
505 521773
510 521774
515 521774
520 521774
525 521775
530 521775
535 521775
540 521776
545 521776
550 521776
555 521777
560 521777
565 521778
570 521778
575 521778
580 521779
585 521779
590 521779
595 521780
600 521780
605 521780
610 521781
615 521781
620 521781
625 521782
630 521782
635 521782
640 521783
645 521783
650 521784
655 521784
660 521784
665 521785
670 521785
675 521785
680 521786
685 521786
690 521786
695 521787
700 521787
705 521787
710 521788
715 521788
720 521788
725 521789
730 521789
735 521790
740 521790
745 521790
750 521791
755 521791
760 521791
765 521792
770 521792
775 521792
780 521793
785 521793
790 521793
795 521794
800 521794
805 521794
810 521795
815 521795
820 521795
825 521796
830 521796
835 521796
840 521797
845 521797
850 521798
855 521798
860 521798
865 521799
870 521799
875 521799
880 521800
885 521800
890 521800
895 521801
900 521801
905 521801
910 521802
915 521802
920 521802
925 521803
930 521803
935 521804
940 521804
945 521804
950 521805
955 521805
960 521805
965 521806
970 521806
975 521806
980 521807
985 521807
990 521807
995 521808
1000 521808
1005 521808
1010 521809
1015 521809
1020 521809
1025 521810
1030 521810
1035 521810
1040 521811
1045 521811
1050 521812
1055 521812
1060 521812
1065 521813
1070 521813
1075 521813
1080 521814
1085 521814
1090 521814
1095 521815
1100 521815
1105 521815
1110 521816
1115 521816
1120 521816
1125 521817
1130 521817
1135 521817
1140 521818
1145 521818
1150 521819
1155 521819
1160 521819
1165 521820
1170 521820
1175 521820
1180 521821
1185 521821
1190 521821
1195 521822
1200 521822
1205 521823
1210 521823
1215 521823
1220 521824
1225 521824
1230 521824
1235 521825
1240 521825
1245 521825
1250 521826
1255 521826
1260 521826
1265 521827
1270 521827
1275 521827
1280 521828
1285 521828
1290 521828
1295 521829
1300 521829
1305 521830
1310 521830
1315 521830
1320 521831
1325 521831
1330 521831
1335 521832
1340 521832
1345 521832
1350 521833
1355 521833
1360 521833
1365 521834
1370 521834
1375 521834
1380 521835
1385 521835
1390 521835
1395 521836
1400 521836
1405 521837
1410 521837
1415 521837
1420 521838
1425 521838
1430 521838
1435 521839
1440 521839
1445 521839
1450 521840
1455 521840
1460 521840
1465 521841
1470 521841
1475 521841
1480 521842
1485 521842
1490 521842
1495 521843
1500 521843
1505 521843
1510 521844
1515 521844
1520 521845
1525 521845
1530 521845
1535 521846
1540 521846
1545 521846
1550 521847
1555 521847
1560 521847
1565 521848
1570 521848
1575 521848
1580 521849
1585 521849
1590 521849
1595 521850
1600 521850
1605 521851
1610 521851
1615 521851
1620 521852
1625 521852
1630 521852
1635 521853
1640 521853
1645 521853
1650 521854
1655 521854
1660 521854
1665 521855
1670 521855
1675 521855
1680 521856
1685 521856
1690 521856
1695 521857
1700 521857
1705 521858
1710 521858
1715 521858
1720 521859
1725 521859
1730 521859
1735 521860
1740 521860
1745 521860
1750 521861
1755 521861
1760 521861
1765 521862
1770 521862
1775 521863
1780 521863
1785 521863
1790 521864
1795 521864
1800 521864
1805 521865
1810 521865
1815 521865
1820 521866
1825 521866
1830 521866
1835 521867
1840 521867
1845 521867
1850 521868
1855 521868
1860 521868
1865 521869
1870 521869
1875 521869
1880 521870
1885 521870
1890 521871
1895 521871
1900 521871
1905 521872
1910 521872
1915 521872
1920 521873
1925 521873
1930 521873
1935 521874
1940 521874
1945 521874
1950 521875
1955 521875
1960 521875
1965 521876
1970 521876
1975 521876
1980 521877
1985 521877
1990 521878
1995 521878
2000 521878
2005 521879
2010 521879
2015 521879
2020 521880
2025 521880
2030 521880
2035 521881
2040 521881
2045 521881
2050 521882
2055 521882
2060 521882
2065 521883
2070 521883
2075 521883
2080 521884
2085 521884
2090 521885
2095 521885
2100 521885
2105 521886
2110 521886
2115 521886
2120 521887
2125 521887
2130 521887
2135 521888
2140 521888
2145 521888
2150 521889
2155 521889
2160 521889
2165 521890
2170 521890
2175 521890
2180 521891
2185 521891
2190 521892
2195 521892
2200 521892
2205 521893
2210 521893
2215 521893
2220 521894
2225 521894
2230 521894
2235 521895
2240 521895
2245 521895
2250 521896
2255 521896
2260 521896
2265 521897
2270 521897
2275 521898
2280 521898
2285 521898
2290 521899
2295 521899
2300 521899
2305 521900
2310 521900
2315 521900
2320 521901
2325 521901
2330 521901
2335 521902
2340 521902
2345 521902
2350 521903
2355 521903
2360 521904
2365 521904
2370 521904
2375 521905
2380 521905
2385 521905
2390 521906
2395 521906
2400 521906
2405 521907
2410 521907
2415 521907
2420 521908
2425 521908
2430 521908
2435 521909
2440 521909
2445 521909
2450 521910
2455 521910
2460 521911
2465 521911
2470 521911
2475 521912
2480 521912
2485 521912
2490 521913
2495 521913
2500 521913
2505 521914
2510 521914
2515 521914
2520 521915
2525 521915
2530 521915
2535 521916
2540 521916
2545 521916
2550 521917
2555 521917
2560 521918
2565 521918
2570 521918
2575 521919
2580 521919
2585 521919
2590 521920
2595 521920
2600 521920
2605 521921
2610 521921
2615 521921
2620 521922
2625 521922
2630 521922
2635 521923
2640 521923
2645 521924
2650 521924
2655 521924
2660 521925
2665 521925
2670 521925
2675 521926
2680 521926
2685 521926
2690 521927
2695 521927
2700 521927
2705 521928
2710 521928
2715 521928
2720 521929
2725 521929
2730 521929
2735 521930
2740 521930
2745 521931
2750 521931
2755 521931
2760 521932
2765 521932
2770 521932
2775 521933
2780 521933
2785 521933
2790 521934
2795 521934
2800 521934
2805 521935
2810 521935
2815 521935
2820 521936
2825 521936
2830 521936
2835 521937
2840 521937
2845 521938
2850 521938
2855 521938
2860 521939
2865 521939
2870 521939
2875 521940
2880 521940
2885 521940
2890 521941
2895 521941
2900 521941
2905 521942
2910 521942
2915 521942
2920 521943
2925 521943
2930 521943
2935 521944
2940 521944
2945 521944
2950 521945
2955 521945
2960 521946
2965 521946
2970 521946
2975 521947
2980 521947
2985 521947
2990 521948
2995 521948
3000 521948
3005 521949
3010 521949
3015 521950
3020 521950
3025 521950
3030 521951
3035 521951
3040 521951
3045 521952
3050 521952
3055 521952
3060 521953
3065 521953
3070 521953
3075 521954
3080 521954
3085 521954
3090 521955
3095 521955
3100 521955
3105 521956
3110 521956
3115 521957
3120 521957
3125 521957
3130 521958
3135 521958
3140 521958
3145 521959
3150 521959
3155 521959
3160 521960
3165 521960
3170 521960
3175 521961
3180 521961
3185 521961
3190 521962
3195 521962
3200 521962
3205 521963
3210 521963
3215 521964
3220 521964
3225 521964
3230 521965
3235 521965
3240 521965
3245 521966
3250 521966
3255 521966
3260 521967
3265 521967
3270 521967
3275 521968
3280 521968
3285 521968
3290 521969
3295 521969
3300 521969
3305 521970
3310 521970
3315 521970
3320 521971
3325 521971
3330 521971
3335 521972
3340 521972
3345 521973
3350 521973
3355 521973
3360 521974
3365 521974
3370 521974
3375 521975
3380 521975
3385 521975
3390 521976
3395 521976
3400 521976
3405 521977
3410 521977
3415 521977
3420 521978
3425 521978
3430 521978
3435 521979
3440 521979
3445 521979
3450 521980
3455 521980
3460 521980
3465 521981
3470 521981
3475 521982
3480 521982
3485 521982
3490 521983
3495 521983
3500 521983
3505 521984
3510 521984
3515 521984
3520 521985
3525 521985
3530 521985
3535 521986
3540 521986
3545 521986
3550 521987
3555 521987
3560 521988
3565 521988
3570 521989
3575 521989
3580 521989
3585 521990
3590 521990
3595 521990
3600 521991
3605 521991
3610 521991
3615 521992
3620 521992
3625 521992
3630 521993
3635 521993
3640 521993
3645 521994
3650 521994
3655 521995
3660 521995
3665 521995
3670 521996
3675 521996
3680 521996
3685 521997
3690 521997
3695 521997
3700 521998
3705 521998
3710 521998
3715 521999
3720 521999
3725 521999
3730 522000
3735 522000
3740 522000
3745 522001
3750 522001
3755 522001
3760 522002
3765 522002
3770 522003
3775 522003
3780 522003
3785 522004
3790 522004
3795 522004
3800 522005
3805 522005
3810 522005
3815 522006
3820 522006
3825 522006
3830 522007
3835 522007
3840 522007
3845 522008
3850 522008
3855 522008
3860 522009
3865 522009
3870 522010
3875 522010
3880 522010
3885 522011
3890 522011
3895 522011
3900 522012
3905 522012
3910 522012
3915 522013
3920 522013
3925 522013
3930 522014
3935 522014
3940 522014
3945 522015
3950 522015
3955 522015
3960 522016
3965 522016
3970 522016
3975 522017
3980 522017
3985 522018
3990 522018
3995 522018
4000 522019
4005 522019
4010 522019
4015 522020
4020 522020
4025 522020
4030 522021
4035 522021
4040 522021
4045 522022
4050 522022
4055 522022
4060 522023
4065 522023
4070 522024
4075 522024
4080 522024
4085 522025
4090 522025
4095 522025
4100 522026
4105 522026
4110 522026
4115 522027
4120 522027
4125 522027
4130 522028
4135 522028
4140 522028
4145 522029
4150 522029
4155 522030
4160 522030
4165 522030
4170 522031
4175 522031
4180 522031
4185 522032
4190 522032
4195 522032
4200 522033
4205 522033
4210 522033
4215 522034
4220 522034
4225 522035
4230 522035
4235 522035
4240 522036
4245 522036
4250 522036
4255 522037
4260 522037
4265 522037
4270 522038
4275 522038
4280 522038
4285 522039
4290 522039
4295 522039
4300 522040
4305 522040
4310 522040
4315 522041
4320 522041
4325 522042
4330 522042
4335 522042
4340 522043
4345 522043
4350 522043
4355 522044
4360 522044
4365 522044
4370 522045
4375 522045
4380 522045
4385 522046
4390 522046
4395 522046
4400 522047
4405 522047
4410 522047
4415 522048
4420 522048
4425 522048
4430 522049
4435 522049
4440 522050
4445 522050
4450 522050
4455 522051
4460 522051
4465 522051
4470 522052
4475 522052
4480 522052
4485 522053
4490 522053
4495 522053
4500 522054
4505 522054
4510 522054
4515 522055
4520 522055
4525 522056
4530 522056
4535 522056
4540 522057
4545 522057
4550 522057
4555 522058
4560 522058
4565 522058
4570 522059
4575 522059
4580 522059
4585 522060
4590 522060
4595 522060
4600 522061
4605 522061
4610 522061
4615 522062
4620 522062
4625 522063
4630 522063
4635 522063
4640 522064
4645 522064
4650 522064
4655 522065
4660 522065
4665 522065
4670 522066
4675 522066
4680 522066
4685 522067
4690 522067
4695 522067
4700 522068
4705 522068
4710 522068
4715 522069
4720 522069
4725 522069
4730 522070
4735 522070
4740 522071
4745 522071
4750 522071
4755 522072
4760 522072
4765 522072
4770 522073
4775 522073
4780 522073
4785 522074
4790 522074
4795 522074
4800 522075
4805 522075
4810 522075
4815 522076
4820 522076
4825 522077
4830 522077
4835 522077
4840 522078
4845 522078
4850 522078
4855 522079
4860 522079
4865 522079
4870 522080
4875 522080
4880 522080
4885 522081
4890 522081
4895 522081
4900 522082
4905 522082
4910 522083
4915 522083
4920 522083
4925 522084
4930 522084
4935 522084
4940 522085
4945 522085
4950 522085
4955 522086
4960 522086
4965 522086
4970 522087
4975 522087
4980 522087
4985 522088
4990 522088
4995 522089
5000 522089
5005 522089
5010 522090
5015 522090
5020 522090
5025 522091
5030 522091
5035 522091
5040 522092
5045 522092
5050 522092
5055 522093
5060 522093
5065 522093
5070 522094
5075 522094
5080 522094
5085 522095
5090 522095
5095 522096
5100 522096
5105 522096
5110 522097
5115 522097
5120 522097
5125 522098
5130 522098
5135 522098
5140 522099
5145 522099
5150 522099
5155 522100
5160 522100
5165 522100
5170 522101
5175 522101
5180 522101
5185 522102
5190 522102
5195 522102
5200 522103
5205 522103
5210 522104
5215 522104
5220 522104
5225 522105
5230 522105
5235 522105
5240 522106
5245 522106
5250 522106
5255 522107
5260 522107
5265 522107
5270 522108
5275 522108
5280 522108
5285 522109
5290 522109
5295 522109
5300 522110
5305 522110
5310 522110
5315 522111
5320 522111
5325 522112
5330 522112
5335 522112
5340 522113
5345 522113
5350 522113
5355 522114
5360 522114
5365 522114
5370 522115
5375 522115
5380 522116
5385 522116
5390 522116
5395 522117
5400 522117
};
\addlegendentry{Baseline}
\addplot [semithick, color1, mark=diamond*, mark size=3, mark options={solid,fill=white,draw=red},  mark repeat={180}]
table[y expr=(\thisrowno{1}-381590)/1000] {%
0 381590
5 381590
10 381591
15 381591
20 381592
25 381592
30 381592
35 381593
40 381593
45 381593
50 381594
55 381594
60 381594
65 381595
70 381595
75 381595
80 381596
85 381596
90 381597
95 381597
100 381597
105 381598
110 381598
115 381598
120 381599
125 381599
130 381599
135 381600
140 381600
145 381600
150 381601
155 381601
160 381602
165 381602
170 381602
175 381603
180 381603
185 381604
190 381604
195 381605
200 381605
205 381606
210 381606
215 381607
220 381607
225 381608
230 381608
235 381609
240 381609
245 381610
250 381610
255 381611
260 381611
265 381612
270 381612
275 381613
280 381613
285 381614
290 381614
295 381615
300 381615
305 381616
310 381616
315 381617
320 381617
325 381618
330 381618
335 381619
340 381619
345 381620
350 381620
355 381621
360 381621
365 381622
370 381622
375 381623
380 381623
385 381624
390 381624
395 381625
400 381625
405 381626
410 381626
415 381627
420 381627
425 381628
430 381628
435 381629
440 381629
445 381630
450 381630
};
\addlegendentry{1 job}
\addplot [semithick, color2, mark=square*, mark size=3, mark options={solid,fill=white,draw=red},  mark repeat={180}]
table[y expr=(\thisrowno{1}-381590)/1000] {%
450 381630
455 381631
460 381631
465 381632
470 381632
475 381633
480 381634
485 381634
490 381635
495 381635
500 381636
505 381636
510 381637
515 381637
520 381638
525 381638
530 381639
535 381639
540 381640
545 381640
550 381641
555 381641
560 381642
565 381642
570 381643
575 381643
580 381644
585 381644
590 381645
595 381645
600 381646
605 381646
610 381647
615 381647
620 381648
625 381648
630 381649
635 381650
640 381650
645 381651
650 381652
655 381652
660 381653
665 381653
670 381654
675 381655
680 381655
685 381656
690 381657
695 381657
700 381658
705 381659
710 381659
715 381660
720 381661
725 381661
730 381662
735 381663
740 381663
745 381664
750 381665
755 381665
760 381666
765 381667
770 381667
775 381668
780 381668
785 381669
790 381670
795 381671
800 381671
805 381672
810 381672
815 381673
820 381674
825 381674
830 381675
835 381675
840 381676
845 381677
850 381677
855 381678
860 381679
865 381679
870 381680
875 381680
880 381681
885 381682
890 381682
895 381683
900 381684
};
\addlegendentry{2 jobs}
\addplot [semithick, color3, mark=triangle*, mark size=3, mark options={solid,fill=white,draw=red},  mark repeat={180}]
table[y expr=(\thisrowno{1}-381590)/1000] {%
905 381684
910 381685
915 381686
920 381686
925 381687
930 381688
935 381688
940 381689
945 381690
950 381690
955 381691
960 381692
965 381692
970 381693
975 381694
980 381694
985 381695
990 381696
995 381696
1000 381697
1005 381697
1010 381698
1015 381699
1020 381699
1025 381700
1030 381701
1035 381701
1040 381702
1045 381703
1050 381703
1055 381704
1060 381705
1065 381705
1070 381706
1075 381707
1080 381707
1085 381708
1090 381709
1095 381709
1100 381710
1105 381710
1110 381711
1115 381712
1120 381712
1125 381713
1130 381714
1135 381714
1140 381715
1145 381716
1150 381716
1155 381717
1160 381718
1165 381718
1170 381719
1175 381720
1180 381720
1185 381721
1190 381722
1195 381722
1200 381723
1205 381724
1210 381724
1215 381725
1220 381726
1225 381726
1230 381727
1235 381728
1240 381728
1245 381729
1250 381729
1255 381730
1260 381731
1265 381731
1270 381732
1275 381733
1280 381733
1285 381734
1290 381735
1295 381735
1300 381736
1305 381737
1310 381737
1315 381738
1320 381739
1325 381739
1330 381740
1335 381741
1340 381741
1345 381742
1350 381743
};
\addlegendentry{3 jobs}
\addplot [semithick, color4, mark=o, mark size=3, mark options={solid,fill=white,draw=red},  mark repeat={180}]
table[y expr=(\thisrowno{1}-381590)/1000] {%
1350 381743
1355 381743
1360 381744
1365 381745
1370 381745
1375 381746
1380 381746
1385 381747
1390 381748
1395 381748
1400 381749
1405 381750
1410 381750
1415 381751
1420 381752
1425 381752
1430 381753
1435 381753
1440 381754
1445 381755
1450 381755
1455 381756
1460 381757
1465 381757
1470 381758
1475 381759
1480 381759
1485 381760
1490 381761
1495 381761
1500 381762
1505 381763
1510 381763
1515 381764
1520 381765
1525 381765
1530 381766
1535 381767
1540 381767
1545 381768
1550 381769
1555 381769
1560 381770
1565 381771
1570 381771
1575 381772
1580 381772
1585 381773
1590 381774
1595 381774
1600 381775
1605 381776
1610 381776
1615 381777
1620 381778
1625 381778
1630 381779
1635 381780
1640 381780
1645 381781
1650 381782
1655 381782
1660 381783
1665 381784
1670 381784
1675 381785
1680 381786
1685 381786
1690 381787
1695 381788
1700 381788
1705 381789
1710 381790
1715 381790
1720 381791
1725 381791
1730 381792
1735 381793
1740 381794
1745 381794
1750 381795
1755 381796
1760 381796
1765 381797
1770 381798
1775 381798
1780 381799
1785 381800
1790 381800
1795 381801
1800 381801
1805 381802
1810 381803
1815 381803
1820 381804
1825 381805
1830 381805
1835 381806
1840 381807
1845 381807
1850 381808
1855 381809
1860 381809
1865 381810
1870 381811
1875 381811
1880 381812
1885 381813
1890 381813
1895 381814
1900 381815
1905 381815
1910 381816
1915 381817
1920 381817
1925 381818
1930 381819
1935 381819
1940 381820
1945 381821
1950 381821
1955 381822
1960 381823
1965 381823
1970 381824
1975 381824
1980 381825
1985 381826
1990 381826
1995 381827
2000 381828
2005 381828
2010 381829
2015 381830
2020 381830
2025 381831
2030 381832
2035 381833
2040 381833
2045 381834
2050 381835
2055 381835
2060 381836
2065 381837
2070 381837
2075 381838
2080 381838
2085 381839
2090 381840
2095 381840
2100 381841
2105 381842
2110 381842
2115 381843
2120 381844
2125 381844
2130 381845
2135 381846
2140 381846
2145 381847
2150 381848
2155 381848
2160 381849
2165 381850
2170 381850
2175 381851
2180 381852
2185 381852
2190 381853
2195 381854
2200 381854
2205 381855
2210 381856
2215 381856
2220 381857
2225 381858
2230 381858
2235 381859
2240 381860
2245 381860
2250 381861
2255 381862
2260 381862
2265 381863
2270 381863
2275 381864
2280 381865
2285 381865
2290 381866
2295 381867
2300 381867
2305 381868
2310 381869
2315 381869
2320 381870
2325 381871
2330 381871
2335 381872
2340 381873
2345 381873
2350 381874
2355 381875
2360 381875
2365 381876
2370 381877
2375 381877
2380 381878
2385 381879
2390 381879
2395 381880
2400 381881
2405 381881
2410 381882
2415 381882
2420 381883
2425 381884
2430 381884
2435 381885
2440 381886
2445 381886
2450 381887
2455 381888
2460 381888
2465 381889
2470 381890
2475 381890
2480 381891
2485 381892
2490 381892
2495 381893
2500 381894
2505 381894
2510 381895
2515 381896
2520 381896
2525 381897
2530 381898
2535 381898
2540 381899
2545 381900
2550 381900
2555 381901
2560 381901
2565 381902
2570 381903
2575 381903
2580 381904
2585 381905
2590 381905
2595 381906
2600 381907
2605 381907
2610 381908
2615 381909
2620 381909
2625 381910
2630 381910
2635 381911
2640 381911
2645 381912
2650 381913
2655 381913
2660 381914
2665 381915
2670 381915
2675 381916
2680 381917
2685 381917
2690 381918
2695 381919
2700 381919
2705 381920
2710 381921
2715 381921
2720 381922
2725 381923
2730 381923
2735 381924
2740 381925
2745 381925
2750 381926
2755 381927
2760 381927
2765 381928
2770 381929
2775 381929
2780 381930
2785 381931
2790 381931
2795 381932
2800 381933
2805 381933
2810 381934
2815 381934
2820 381935
2825 381936
2830 381936
2835 381937
2840 381938
2845 381938
2850 381939
2855 381940
2860 381940
2865 381941
2870 381942
2875 381942
2880 381943
2885 381944
2890 381944
2895 381945
2900 381946
2905 381946
2910 381947
2915 381948
2920 381948
2925 381949
2930 381949
2935 381950
2940 381951
2945 381951
2950 381952
2955 381953
2960 381953
2965 381954
2970 381955
2975 381955
2980 381956
2985 381957
2990 381957
2995 381958
3000 381959
3005 381959
3010 381960
3015 381961
3020 381961
3025 381962
3030 381963
3035 381963
3040 381964
3045 381965
3050 381965
3055 381966
3060 381966
3065 381967
3070 381968
3075 381968
3080 381969
3085 381970
3090 381970
3095 381971
3100 381972
3105 381972
3110 381973
3115 381974
3120 381974
3125 381975
3130 381976
3135 381976
3140 381977
3145 381978
3150 381978
3155 381979
3160 381980
3165 381980
3170 381981
3175 381982
3180 381982
3185 381983
3190 381983
3195 381984
3200 381985
3205 381985
3210 381986
3215 381987
3220 381987
3225 381988
3230 381988
3235 381989
3240 381990
3245 381990
3250 381991
3255 381992
3260 381992
3265 381993
3270 381994
3275 381994
3280 381995
3285 381996
3290 381996
3295 381997
3300 381997
3305 381998
3310 381999
3315 381999
3320 382000
3325 382001
3330 382001
3335 382002
3340 382003
3345 382003
3350 382004
3355 382005
3360 382005
3365 382006
3370 382007
3375 382007
3380 382008
3385 382009
3390 382009
3395 382010
3400 382011
3405 382011
3410 382012
3415 382013
3420 382013
3425 382014
3430 382015
3435 382015
3440 382016
3445 382017
3450 382017
3455 382018
3460 382019
3465 382019
3470 382020
3475 382020
3480 382021
3485 382022
3490 382022
3495 382023
3500 382024
3505 382024
3510 382025
3515 382026
3520 382026
3525 382027
3530 382028
3535 382028
3540 382029
3545 382030
3550 382030
3555 382031
3560 382032
3565 382032
3570 382033
3575 382034
3580 382034
3585 382035
3590 382036
3595 382036
3600 382037
3605 382037
3610 382038
3615 382039
3620 382039
3625 382040
3630 382041
3635 382041
3640 382042
3645 382043
3650 382043
3655 382044
3660 382045
3665 382045
3670 382046
3675 382047
3680 382047
3685 382048
3690 382049
3695 382049
3700 382050
3705 382051
3710 382051
3715 382052
3720 382053
3725 382053
3730 382054
3735 382055
3740 382055
3745 382056
3750 382056
3755 382057
3760 382058
3765 382058
3770 382059
3775 382060
3780 382060
3785 382061
3790 382062
3795 382062
3800 382063
3805 382064
3810 382064
3815 382065
3820 382065
3825 382066
3830 382067
3835 382067
3840 382068
3845 382069
3850 382069
3855 382070
3860 382071
3865 382071
3870 382072
3875 382073
3880 382073
3885 382074
3890 382075
3895 382075
3900 382076
3905 382077
3910 382077
3915 382078
3920 382078
3925 382079
3930 382080
3935 382080
3940 382081
3945 382082
3950 382082
3955 382083
3960 382084
3965 382084
3970 382085
3975 382086
3980 382086
};
\addlegendentry{4 jobs}
\addplot [semithick, color3, mark=triangle*, mark size=3, mark options={solid,fill=white,draw=red},  mark repeat={180}]
table[y expr=(\thisrowno{1}-381590)/1000] {%
3980 382086
3985 382087
3990 382088
3995 382088
4000 382089
4005 382090
4010 382090
4015 382091
4020 382091
4025 382092
4030 382093
4035 382093
4040 382094
4045 382095
4050 382095
4055 382096
4060 382097
4065 382097
4070 382098
4075 382099
4080 382099
4085 382100
4090 382101
4095 382101
4100 382102
4105 382103
4110 382103
4115 382104
4120 382104
4125 382105
4130 382106
4135 382106
4140 382107
4145 382108
4150 382108
4155 382109
4160 382110
4165 382110
4170 382111
4175 382112
4180 382112
4185 382113
4190 382114
4195 382114
4200 382115
4205 382116
4210 382116
4215 382117
4220 382118
4225 382118
4230 382119
4235 382119
4240 382120
4245 382121
4250 382121
4255 382122
4260 382123
4265 382123
4270 382124
4275 382125
4280 382125
4285 382126
4290 382127
4295 382127
4300 382128
4305 382129
4310 382129
4315 382130
4320 382131
4325 382131
4330 382132
4335 382133
4340 382133
4345 382134
4350 382135
4355 382135
4360 382136
4365 382137
4370 382137
4375 382138
4380 382139
4385 382139
4390 382140
4395 382140
4400 382141
4405 382142
4410 382142
};
\addplot [semithick, color2, mark=square*, mark size=3, mark options={solid,fill=white,draw=red},  mark repeat={180}]
table[y expr=(\thisrowno{1}-381590)/1000] {%
4410 382142
4415 382143
4420 382144
4425 382144
4430 382145
4435 382145
4440 382146
4445 382146
4450 382147
4455 382147
4460 382148
4465 382148
4470 382149
4475 382149
4480 382150
4485 382150
4490 382151
4495 382151
4500 382152
4505 382152
4510 382153
4515 382153
4520 382154
4525 382154
4530 382155
4535 382155
4540 382156
4545 382156
4550 382157
4555 382157
4560 382158
4565 382158
4570 382159
4575 382159
4580 382160
4585 382160
4590 382161
4595 382161
4600 382162
4605 382162
4610 382163
4615 382163
4620 382164
4625 382164
4630 382165
4635 382165
4640 382166
4645 382166
4650 382167
4655 382167
4660 382168
4665 382168
4670 382169
4675 382169
4680 382170
4685 382171
4690 382171
4695 382171
};
\addplot [semithick, color1, mark=diamond*, mark size=3, mark options={solid,fill=white,draw=red},  mark repeat={180}]
table[y expr=(\thisrowno{1}-381590)/1000] {%
4695 382171
4700 382172
4705 382172
4710 382173
4715 382173
4720 382173
4725 382174
4730 382174
4735 382175
4740 382175
4745 382175
4750 382176
4755 382176
4760 382176
4765 382177
4770 382177
4775 382177
4780 382178
4785 382178
4790 382179
4795 382179
4800 382179
};
\addplot [semithick, color2,mark=triangle*, mark size=3, mark options={solid,fill=red,draw=red}]
table[y expr=(\thisrowno{1}-381590)/1000]{
3980 382086
};

\addplot [semithick, color2,mark=square*, mark size=3, mark options={solid,fill=red,draw=red}]
table[y expr=(\thisrowno{1}-381590)/1000] {%
4410 382142
};

\addplot [semithick, color2,mark=diamond*, mark size=3, mark options={solid,fill=red,draw=red}]
table[y expr=(\thisrowno{1}-381590)/1000] {%
4695 382171
};

\end{axis}
\node[text width=8cm] at (3.5,-1.5) {Filled markers indicate the end of jobs};

\end{tikzpicture}

%% file: results/power_amd_seq.tex
\begin{tikzpicture}[font=\Large]

\definecolor{color0}{rgb}{0.12156862745098,0.466666666666667,0.705882352941177}
\definecolor{color1}{rgb}{1,0.498039215686275,0.0549019607843137}
\definecolor{color2}{rgb}{0.172549019607843,0.627450980392157,0.172549019607843}
\definecolor{color3}{rgb}{0.83921568627451,0.152941176470588,0.156862745098039}
\definecolor{color4}{rgb}{0.580392156862745,0.403921568627451,0.741176470588235}
\definecolor{color5}{rgb}{0,0,0}
\definecolor{color6}{rgb}{0.07, 0.04, 0.56}

\begin{axis}[
legend cell align={left},
legend columns=3,
legend style={fill opacity=0.8, draw opacity=1, text opacity=1, at={(1.05,1.15)}, anchor=east, draw=white!80.0!black},
tick align=outside,
tick pos=left,
x grid style={white!69.01960784313725!black},
xlabel={Time (SEC)},
xmin=0, xmax=5310,
xtick style={color=black},
y grid style={white!69.01960784313725!black},
ylabel={Power consumption (W)},
ymin=150, ymax=350,
ytick={150,200,250,300,350},
xmajorgrids,
ymajorgrids,
ytick style={color=black}
]
\addplot [thick, olive, mark=star, mark size=3, mark options={solid,fill=white,draw=red},  mark repeat={180}]
table {%
0 215
5 215
10 216
15 217
20 211
25 212
30 208
35 206
40 206
45 206
50 205
55 205
60 205
65 205
70 205
75 206
80 205
85 205
90 205
95 205
100 206
105 205
110 206
115 205
120 207
125 206
130 205
135 206
140 205
145 206
150 204
155 205
160 206
165 206
170 206
175 205
180 206
185 206
190 207
195 206
200 208
205 205
210 206
215 209
220 207
225 220
230 220
235 220
240 220
245 219
250 214
255 215
260 213
265 214
270 214
275 213
280 219
285 219
290 211
295 212
300 220
305 273
310 291
315 293
320 293
325 294
330 294
335 294
340 294
345 296
350 296
355 296
360 297
365 298
370 300
375 300
380 301
385 302
390 302
395 302
400 302
405 302
410 302
415 302
420 302
425 303
430 304
435 304
440 304
445 304
450 304
455 304
460 304
465 304
470 304
475 304
480 306
485 303
490 306
495 304
500 306
505 304
510 303
515 302
520 306
525 306
530 306
535 306
540 305
545 303
550 304
555 306
560 306
565 304
570 304
575 306
580 304
585 306
590 307
595 306
600 307
605 304
610 304
615 304
620 304
625 306
630 305
635 305
640 305
645 306
650 304
655 306
660 306
665 306
670 304
675 304
680 303
685 306
690 305
695 307
700 304
705 304
710 305
715 306
720 306
725 306
730 306
735 306
740 306
745 306
750 304
755 304
760 306
765 306
770 306
775 304
780 304
785 306
790 306
795 306
800 307
805 306
810 306
815 305
820 304
825 304
830 304
835 304
840 306
845 306
850 306
855 305
860 305
865 306
870 306
875 306
880 304
885 304
890 304
895 305
900 306
905 305
910 305
915 304
920 304
925 302
930 303
935 306
940 306
945 306
950 304
955 304
960 304
965 304
970 307
975 306
980 304
985 304
990 306
995 306
1000 306
1005 304
1010 304
1015 305
1020 306
1025 306
1030 306
1035 306
1040 304
1045 304
1050 304
1055 304
1060 304
1065 307
1070 306
1075 307
1080 306
1085 306
1090 304
1095 304
1100 306
1105 305
1110 306
1115 306
1120 304
1125 304
1130 304
1135 304
1140 304
1145 304
1150 306
1155 306
1160 305
1165 304
1170 304
1175 306
1180 307
1185 306
1190 306
1195 304
1200 305
1205 302
1210 302
1215 303
1220 304
1225 304
1230 304
1235 306
1240 306
1245 306
1250 304
1255 304
1260 304
1265 304
1270 306
1275 306
1280 304
1285 304
1290 306
1295 304
1300 304
1305 304
1310 306
1315 306
1320 306
1325 305
1330 302
1335 304
1340 303
1345 306
1350 305
1355 304
1360 304
1365 304
1370 306
1375 306
1380 304
1385 304
1390 306
1395 302
1400 304
1405 304
1410 304
1415 304
1420 304
1425 304
1430 306
1435 306
1440 306
1445 303
1450 304
1455 304
1460 303
1465 304
1470 306
1475 306
1480 304
1485 304
1490 306
1495 306
1500 306
1505 306
1510 306
1515 304
1520 304
1525 304
1530 306
1535 306
1540 304
1545 304
1550 304
1555 306
1560 306
1565 304
1570 306
1575 303
1580 302
1585 305
1590 304
1595 302
1600 303
1605 304
1610 305
1615 306
1620 306
1625 306
1630 302
1635 302
1640 302
1645 303
1650 306
1655 304
1660 304
1665 304
1670 305
1675 309
1680 307
1685 306
1690 306
1695 304
1700 304
1705 306
1710 304
1715 304
1720 304
1725 304
1730 304
1735 307
1740 307
1745 306
1750 307
1755 305
1760 302
1765 304
1770 304
1775 304
1780 306
1785 304
1790 304
1795 304
1800 304
1805 304
1810 306
1815 306
1820 304
1825 304
1830 307
1835 306
1840 304
1845 305
1850 306
1855 306
1860 306
1865 306
1870 304
1875 304
1880 304
1885 306
1890 304
1895 304
1900 304
1905 306
1910 307
1915 306
1920 305
1925 304
1930 307
1935 307
1940 304
1945 304
1950 304
1955 305
1960 305
1965 306
1970 306
1975 306
1980 306
1985 304
1990 306
1995 306
2000 303
2005 304
2010 304
2015 306
2020 306
2025 303
2030 306
2035 306
2040 306
2045 304
2050 304
2055 303
2060 303
2065 304
2070 306
2075 306
2080 306
2085 306
2090 305
2095 306
2100 306
2105 306
2110 307
2115 304
2120 306
2125 306
2130 306
2135 304
2140 305
2145 304
2150 304
2155 304
2160 304
2165 306
2170 306
2175 306
2180 306
2185 304
2190 304
2195 303
2200 306
2205 306
2210 306
2215 307
2220 307
2225 306
2230 306
2235 306
2240 305
2245 304
2250 306
2255 306
2260 304
2265 304
2270 304
2275 306
2280 306
2285 306
2290 306
2295 306
2300 304
2305 303
2310 306
2315 306
2320 306
2325 306
2330 306
2335 306
2340 306
2345 306
2350 304
2355 304
2360 304
2365 307
2370 306
2375 306
2380 303
2385 304
2390 304
2395 306
2400 306
2405 306
2410 304
2415 306
2420 304
2425 304
2430 305
2435 306
2440 306
2445 304
2450 306
2455 306
2460 306
2465 306
2470 306
2475 306
2480 304
2485 304
2490 306
2495 306
2500 304
2505 304
2510 306
2515 304
2520 305
2525 306
2530 306
2535 304
2540 304
2545 306
2550 306
2555 304
2560 304
2565 304
2570 306
2575 306
2580 306
2585 306
2590 306
2595 306
2600 304
2605 304
2610 304
2615 304
2620 306
2625 302
2630 304
2635 306
2640 306
2645 304
2650 304
2655 306
2660 306
2665 304
2670 306
2675 306
2680 306
2685 304
2690 304
2695 304
2700 304
2705 306
2710 304
2715 306
2720 306
2725 304
2730 304
2735 304
2740 306
2745 304
2750 306
2755 304
2760 307
2765 306
2770 305
2775 304
2780 304
2785 304
2790 304
2795 304
2800 304
2805 305
2810 304
2815 306
2820 306
2825 306
2830 306
2835 304
2840 304
2845 306
2850 305
2855 304
2860 306
2865 306
2870 306
2875 304
2880 306
2885 306
2890 306
2895 304
2900 306
2905 307
2910 306
2915 306
2920 306
2925 306
2930 304
2935 306
2940 306
2945 304
2950 304
2955 306
2960 306
2965 304
2970 304
2975 306
2980 307
2985 306
2990 306
2995 304
3000 304
3005 306
3010 307
3015 306
3020 306
3025 306
3030 304
3035 306
3040 306
3045 306
3050 306
3055 304
3060 304
3065 306
3070 306
3075 306
3080 304
3085 306
3090 306
3095 306
3100 306
3105 306
3110 306
3115 306
3120 304
3125 304
3130 306
3135 304
3140 305
3145 306
3150 306
3155 306
3160 307
3165 306
3170 306
3175 304
3180 305
3185 306
3190 306
3195 306
3200 306
3205 304
3210 306
3215 302
3220 304
3225 306
3230 306
3235 306
3240 305
3245 304
3250 304
3255 304
3260 304
3265 306
3270 306
3275 306
3280 306
3285 306
3290 306
3295 304
3300 306
3305 304
3310 304
3315 306
3320 306
3325 304
3330 306
3335 306
3340 306
3345 306
3350 304
3355 304
3360 304
3365 306
3370 304
3375 302
3380 305
3385 297
3390 303
3395 305
3400 303
3405 302
3410 306
3415 306
3420 306
3425 306
3430 306
3435 305
3440 304
3445 302
3450 302
3455 303
3460 305
3465 304
3470 303
3475 302
3480 304
3485 304
3490 304
3495 304
3500 302
3505 303
3510 306
3515 306
3520 306
3525 303
3530 302
3535 302
3540 303
3545 302
3550 304
3555 304
3560 303
3565 302
3570 302
3575 304
3580 305
3585 302
3590 303
3595 302
3600 305
3605 303
3610 304
3615 302
3620 302
3625 302
3630 303
3635 302
3640 303
3645 303
3650 300
3655 296
3660 292
3665 282
3670 268
3675 262
3680 258
3685 257
3690 248
3695 238
3700 227
3705 223
3710 213
3715 213
3720 212
3725 213
3730 212
3735 209
3740 205
3745 205
3750 206
3755 206
3760 205
3765 206
3770 206
3775 205
3780 207
3785 206
3790 205
3795 206
3800 205
3805 205
3810 206
3815 205
3820 205
3825 206
3830 205
3835 206
3840 205
3845 205
3850 205
3855 205
3860 205
3865 206
3870 205
3875 204
3880 205
3885 205
3890 208
3895 205
3900 205
3905 205
3910 207
3915 206
3920 206
3925 206
3930 205
3935 206
3940 205
3945 206
3950 206
3955 205
3960 205
3965 205
3970 205
3975 205
3980 205
3985 206
3990 205
3995 205
4000 205
4005 205
4010 209
4015 206
4020 206
4025 206
4030 206
4035 205
4040 206
4045 206
4050 206
4055 206
4060 205
4065 205
4070 206
4075 206
4080 206
4085 205
4090 206
4095 205
4100 206
4105 206
4110 205
4115 205
4120 206
4125 205
4130 206
4135 206
4140 206
4145 206
4150 205
4155 205
4160 205
4165 206
4170 206
4175 206
4180 205
4185 207
4190 208
4195 206
4200 207
4205 207
4210 205
4215 206
4220 206
4225 205
4230 206
4235 206
4240 205
4245 205
4250 206
4255 205
4260 206
4265 206
4270 205
4275 205
4280 206
4285 206
4290 206
4295 205
4300 206
4305 205
4310 206
4315 206
4320 205
4325 206
4330 206
4335 205
4340 205
4345 206
4350 206
4355 206
4360 206
4365 206
4370 206
4375 205
4380 206
4385 206
4390 205
4395 206
4400 206
4405 206
4410 206
4415 206
4420 205
4425 206
4430 205
4435 206
4440 205
4445 205
4450 206
4455 206
4460 205
4465 205
4470 206
4475 206
4480 206
4485 205
4490 206
4495 205
4500 207
4505 206
4510 205
4515 205
4520 205
4525 206
4530 206
4535 206
4540 206
4545 206
4550 206
4555 205
4560 205
4565 206
4570 206
4575 206
4580 205
4585 206
4590 206
4595 205
4600 205
4605 206
4610 206
4615 205
4620 206
4625 205
4630 205
4635 206
4640 205
4645 205
4650 206
4655 205
4660 206
4665 206
4670 206
4675 205
4680 206
4685 205
4690 206
4695 205
4700 206
4705 206
4710 206
4715 205
4720 206
4725 205
4730 205
4735 206
4740 206
4745 205
4750 205
4755 205
4760 206
4765 207
4770 205
4775 205
4780 206
4785 206
4790 205
4795 205
4800 205
4805 206
4810 206
4815 206
4820 205
4825 205
4830 206
4835 205
4840 205
4845 205
4850 206
4855 206
4860 205
4865 205
4870 206
4875 205
4880 205
4885 206
4890 206
4895 204
4900 205
4905 205
4910 206
4915 205
4920 205
4925 205
4930 205
4935 205
4940 205
4945 206
4950 207
4955 205
4960 206
4965 205
4970 205
4975 206
4980 205
4985 205
4990 206
4995 205
5000 205
5005 206
5010 205
5015 205
5020 206
5025 205
5030 206
5035 206
5040 205
5045 206
5050 205
5055 206
5060 205
5065 205
5070 205
5075 205
5080 206
5085 206
5090 205
5095 205
5100 205
5105 206
5110 205
5115 205
5120 206
5125 205
5130 205
5135 205
5140 205
5145 205
5150 206
5155 206
5160 205
5165 205
5170 206
5175 205
5180 205
5185 205
5190 206
5195 205
5200 206
5205 206
5210 206
5215 206
5220 205
5225 205
5230 205
5235 205
5240 205
5245 205
5250 205
5255 205
5260 205
5265 205
5270 205
5275 206
5280 205
5285 205
5290 205
5295 205
5300 206
5305 206
5310 206
5315 205
5320 206
5325 205
5330 205
5335 205
5340 206
5345 206
5350 206
5355 205
5360 205
5365 205
5370 205
5375 205
5380 206
5385 205
5390 205
5395 205
5400 206
};
\addlegendentry{4 jobs in parallel}
\addplot [semithick, color5]
table{
0 205
5 205
10 204
15 205
20 205
25 204
30 205
35 204
40 204
45 204
50 205
55 205
60 204
65 205
70 204
75 205
80 204
85 205
90 205
95 205
100 205
105 204
110 204
115 205
120 205
125 204
130 205
135 205
140 204
145 204
150 205
155 205
160 205
165 204
170 205
175 204
180 204
185 205
190 205
195 204
200 204
205 206
210 206
215 206
220 206
225 206
230 206
235 206
240 205
245 206
250 205
255 206
260 205
265 206
270 205
275 206
280 206
285 205
290 205
295 205
300 205
305 205
310 205
315 206
320 205
325 205
330 205
335 205
340 206
345 205
350 205
355 205
360 205
365 205
370 206
375 206
380 205
385 205
390 205
395 206
400 205
405 206
410 206
415 206
420 206
425 206
430 206
435 206
440 206
445 205
450 208
455 205
460 205
465 206
470 206
475 206
480 205
485 205
490 206
495 205
500 205
505 205
510 205
515 205
520 205
525 206
530 205
535 206
540 205
545 206
550 205
555 205
560 205
565 205
570 205
575 206
580 206
585 206
590 205
595 206
600 205
605 206
610 206
615 206
620 205
625 206
630 206
635 206
640 205
645 206
650 206
655 206
660 205
665 206
670 206
675 205
680 206
685 206
690 205
695 205
700 205
705 205
710 205
715 205
720 205
725 206
730 205
735 206
740 207
745 207
750 206
755 205
760 205
765 205
770 206
775 206
780 205
785 206
790 206
795 206
800 205
805 206
810 205
815 205
820 206
825 205
830 206
835 205
840 205
845 206
850 206
855 206
860 205
865 206
870 205
875 206
880 205
885 206
890 205
895 206
900 206
905 206
910 205
915 206
920 206
925 205
930 205
935 205
940 206
945 206
950 205
955 206
960 205
965 206
970 206
975 206
980 205
985 206
990 206
995 206
1000 206
1005 205
1010 206
1015 205
1020 205
1025 206
1030 206
1035 205
1040 206
1045 206
1050 205
1055 205
1060 205
1065 205
1070 206
1075 205
1080 206
1085 206
1090 205
1095 205
1100 205
1105 205
1110 205
1115 205
1120 206
1125 205
1130 205
1135 206
1140 205
1145 206
1150 205
1155 205
1160 206
1165 206
1170 206
1175 205
1180 206
1185 205
1190 205
1195 205
1200 205
1205 205
1210 205
1215 205
1220 206
1225 206
1230 206
1235 205
1240 206
1245 206
1250 206
1255 205
1260 205
1265 206
1270 205
1275 205
1280 205
1285 205
1290 205
1295 205
1300 205
1305 205
1310 205
1315 205
1320 206
1325 206
1330 206
1335 206
1340 206
1345 205
1350 206
1355 206
1360 206
1365 205
1370 206
1375 205
1380 206
1385 206
1390 205
1395 205
1400 206
1405 206
1410 206
1415 206
1420 206
1425 205
1430 205
1435 206
1440 205
1445 206
1450 206
1455 205
1460 206
1465 206
1470 206
1475 205
1480 205
1485 206
1490 206
1495 206
1500 205
1505 206
1510 206
1515 205
1520 205
1525 205
1530 206
1535 206
1540 206
1545 206
1550 205
1555 206
1560 206
1565 206
1570 206
1575 206
1580 205
1585 206
1590 206
1595 205
1600 206
1605 206
1610 206
1615 205
1620 206
1625 206
1630 205
1635 206
1640 206
1645 207
1650 206
1655 206
1660 206
1665 205
1670 205
1675 205
1680 205
1685 206
1690 206
1695 206
1700 206
1705 205
1710 205
1715 206
1720 206
1725 206
1730 206
1735 205
1740 205
1745 206
1750 206
1755 206
1760 206
1765 205
1770 205
1775 205
1780 205
1785 205
1790 205
1795 205
1800 206
1805 205
1810 206
1815 205
1820 205
1825 206
1830 206
1835 205
1840 206
1845 206
1850 206
1855 206
1860 205
1865 205
1870 206
1875 205
1880 205
1885 205
1890 205
1895 206
1900 206
1905 205
1910 206
1915 205
1920 206
1925 205
1930 205
1935 206
1940 205
1945 205
1950 205
1955 204
1960 206
1965 206
1970 205
1975 205
1980 205
1985 205
1990 205
1995 205
2000 206
2005 205
2010 205
2015 205
2020 205
2025 205
2030 205
2035 205
2040 205
2045 205
2050 205
2055 205
2060 205
2065 206
2070 205
2075 206
2080 206
2085 205
2090 205
2095 205
2100 206
2105 205
2110 205
2115 206
2120 205
2125 205
2130 205
2135 205
2140 205
2145 206
2150 206
2155 205
2160 205
2165 205
2170 205
2175 206
2180 205
2185 206
2190 205
2195 205
2200 205
2205 205
2210 205
2215 205
2220 206
2225 205
2230 205
2235 206
2240 206
2245 206
2250 205
2255 205
2260 205
2265 205
2270 205
2275 205
2280 205
2285 205
2290 205
2295 205
2300 206
2305 206
2310 206
2315 205
2320 205
2325 205
2330 205
2335 205
2340 205
2345 205
2350 206
2355 206
2360 205
2365 205
2370 205
2375 204
2380 205
2385 206
2390 205
2395 205
2400 205
2405 205
2410 204
2415 205
2420 205
2425 205
2430 206
2435 206
2440 205
2445 206
2450 206
2455 205
2460 205
2465 205
2470 206
2475 205
2480 206
2485 205
2490 206
2495 205
2500 205
2505 206
2510 205
2515 205
2520 205
2525 206
2530 206
2535 205
2540 206
2545 206
2550 205
2555 206
2560 206
2565 205
2570 206
2575 205
2580 206
2585 205
2590 206
2595 205
2600 205
2605 206
2610 206
2615 206
2620 205
2625 206
2630 206
2635 205
2640 205
2645 205
2650 206
2655 205
2660 206
2665 206
2670 206
2675 205
2680 206
2685 206
2690 206
2695 206
2700 205
2705 206
2710 206
2715 206
2720 205
2725 206
2730 205
2735 205
2740 206
2745 205
2750 205
2755 205
2760 205
2765 206
2770 207
2775 205
2780 206
2785 205
2790 205
2795 206
2800 205
2805 200
2810 205
2815 205
2820 205
2825 205
2830 205
2835 205
2840 205
2845 206
2850 205
2855 206
2860 205
2865 205
2870 205
2875 204
2880 204
2885 204
2890 205
2895 204
2900 205
2905 205
2910 206
2915 205
2920 205
2925 205
2930 206
2935 205
2940 204
2945 205
2950 205
2955 206
2960 205
2965 205
2970 205
2975 205
2980 206
2985 205
2990 204
2995 206
3000 205
3005 205
3010 205
3015 204
3020 205
3025 205
3030 204
3035 204
3040 205
3045 205
3050 205
3055 205
3060 205
3065 205
3070 204
3075 205
3080 205
3085 205
3090 205
3095 205
3100 205
3105 205
3110 205
3115 205
3120 205
3125 205
3130 205
3135 205
3140 205
3145 207
3150 205
3155 205
3160 205
3165 205
3170 205
3175 205
3180 205
3185 204
3190 205
3195 205
3200 205
3205 205
3210 204
3215 205
3220 205
3225 206
3230 205
3235 205
3240 205
3245 205
3250 205
3255 205
3260 205
3265 205
3270 205
3275 204
3280 205
3285 205
3290 204
3295 204
3300 205
3305 205
3310 205
3315 205
3320 205
3325 206
3330 205
3335 205
3340 205
3345 204
3350 205
3355 205
3360 205
3365 205
3370 205
3375 205
3380 204
3385 205
3390 204
3395 205
3400 205
3405 205
3410 205
3415 204
3420 205
3425 205
3430 205
3435 205
3440 205
3445 208
3450 205
3455 205
3460 204
3465 205
3470 205
3475 205
3480 206
3485 205
3490 205
3495 205
3500 205
3505 205
3510 205
3515 204
3520 204
3525 205
3530 205
3535 205
3540 205
3545 204
3550 205
3555 205
3560 205
3565 206
3570 204
3575 205
3580 205
3585 205
3590 205
3595 205
3600 205
3605 205
3610 205
3615 205
3620 205
3625 205
3630 205
3635 204
3640 204
3645 205
3650 205
3655 204
3660 204
3665 205
3670 204
3675 205
3680 205
3685 205
3690 205
3695 205
3700 205
3705 205
3710 205
3715 205
3720 206
3725 204
3730 205
3735 204
3740 204
3745 206
3750 205
3755 205
3760 205
3765 205
3770 205
3775 205
3780 205
3785 205
3790 205
3795 204
3800 204
3805 205
3810 206
3815 205
3820 204
3825 205
3830 204
3835 205
3840 205
3845 204
3850 205
3855 204
3860 205
3865 205
3870 205
3875 205
3880 205
3885 205
3890 205
3895 205
3900 205
3905 204
3910 205
3915 205
3920 205
3925 205
3930 204
3935 205
3940 205
3945 204
3950 205
3955 205
3960 205
3965 205
3970 205
3975 205
3980 205
3985 204
3990 204
3995 205
4000 205
4005 204
4010 204
4015 205
4020 205
4025 205
4030 205
4035 204
4040 205
4045 205
4050 205
4055 205
4060 204
4065 205
4070 204
4075 205
4080 205
4085 205
4090 205
4095 205
4100 205
4105 205
4110 205
4115 205
4120 205
4125 205
4130 205
4135 205
4140 204
4145 205
4150 205
4155 205
4160 205
4165 205
4170 204
4175 205
4180 205
4185 205
4190 205
4195 205
4200 204
4205 205
4210 204
4215 205
4220 205
4225 205
4230 205
4235 205
4240 205
4245 204
4250 204
4255 205
4260 205
4265 205
4270 205
4275 205
4280 204
4285 204
4290 205
4295 205
4300 204
4305 206
4310 205
4315 205
4320 205
4325 205
4330 205
4335 205
4340 205
4345 205
4350 204
4355 205
4360 205
4365 205
4370 205
4375 205
4380 205
4385 205
4390 205
4395 205
4400 205
4405 205
4410 205
4415 205
4420 205
4425 204
4430 205
4435 206
4440 205
4445 205
4450 204
4455 205
4460 205
4465 205
4470 205
4475 205
4480 204
4485 205
4490 206
4495 204
4500 205
4505 205
4510 204
4515 204
4520 204
4525 205
4530 204
4535 205
4540 205
4545 205
4550 205
4555 205
4560 205
4565 205
4570 205
4575 206
4580 205
4585 205
4590 205
4595 205
4600 205
4605 205
4610 205
4615 205
4620 205
4625 205
4630 204
4635 205
4640 205
4645 205
4650 205
4655 205
4660 204
4665 204
4670 205
4675 205
4680 205
4685 205
4690 205
4695 205
4700 205
4705 205
4710 205
4715 205
4720 205
4725 205
4730 204
4735 204
4740 205
4745 205
4750 205
4755 205
4760 204
4765 205
4770 205
4775 205
4780 204
4785 205
4790 204
4795 205
4800 205
4805 205
4810 204
4815 204
4820 205
4825 205
4830 205
4835 204
4840 205
4845 204
4850 205
4855 205
4860 204
4865 206
4870 205
4875 205
4880 204
4885 204
4890 204
4895 205
4900 205
4905 205
4910 205
4915 205
4920 205
4925 205
4930 205
4935 205
4940 204
4945 205
4950 205
4955 205
4960 205
4965 205
4970 205
4975 205
4980 205
4985 205
4990 206
4995 205
5000 205
5005 205
5010 205
5015 205
5020 204
5025 205
5030 205
5035 205
5040 205
5045 205
5050 204
5055 204
5060 205
5065 204
5070 205
5075 205
5080 206
5085 205
5090 205
5095 205
5100 205
5105 205
5110 205
5115 205
5120 205
5125 206
5130 205
5135 205
5140 205
5145 205
5150 205
5155 205
5160 205
5165 204
5170 205
5175 204
5180 204
5185 205
5190 204
5195 204
5200 205
5205 205
5210 204
5215 205
5220 205
5225 205
5230 205
5235 205
5240 205
5245 204
5250 205
5255 204
5260 205
5265 205
5270 205
5275 205
5280 204
5285 205
5290 205
5295 205
5300 205
5305 205
5310 204
5315 205
5320 204
5325 204
5330 205
5335 205
5340 205
5345 205
5350 204
5355 205
5360 205
5365 205
5370 205
5375 204
5380 205
5385 205
5390 205
5395 205
5400 204
5405 205
5410 205
5415 205
5420 204
5425 205
5430 204
5435 205
5440 205
5445 205
5450 205
5455 205
5460 205
5465 204
5470 205
5475 204
5480 205
5485 205
5490 205
5495 204
5500 204
5505 205
5510 205
5515 204
5520 204
5525 205
5530 205
5535 206
5540 205
5545 204
5550 204
5555 206
5560 205
5565 204
5570 205
5575 204
5580 205
5585 205
5590 205
5595 205
5600 206
5605 205
5610 205
5615 205
5620 205
5625 205
5630 205
5635 205
5640 205
5645 205
5650 204
5655 205
5660 205
5665 205
5670 205
5675 204
5680 204
5685 205
5690 205
5695 205
5700 206
5705 205
5710 205
5715 205
5720 205
5725 205
5730 204
5735 205
5740 205
5745 205
5750 205
5755 205
5760 204
5765 205
5770 205
5775 205
5780 205
5785 205
5790 204
5795 205
5800 205
5805 206
5810 205
5815 205
5820 205
5825 205
5830 204
5835 206
5840 206
5845 205
5850 205
5855 205
5860 204
5865 205
5870 206
5875 205
5880 205
5885 205
5890 205
5895 205
5900 205
5905 205
5910 204
5915 205
5920 206
5925 205
5930 205
5935 205
5940 205
5945 206
5950 205
5955 205
5960 205
5965 205
5970 205
5975 204
5980 205
5985 205
5990 204
5995 205
6000 205
6005 204
6010 205
6015 205
6020 205
6025 205
6030 205
6035 205
6040 205
6045 205
6050 205
6055 205
6060 205
6065 205
6070 204
6075 206
6080 206
6085 204
6090 205
6095 204
6100 204
6105 205
6110 205
6115 205
6120 205
6125 206
6130 205
6135 205
6140 206
6145 205
6150 206
6155 206
6160 204
6165 205
6170 205
6175 205
6180 205
6185 205
6190 206
6195 205
6200 204
6205 205
6210 205
6215 205
6220 205
6225 205
6230 205
6235 205
6240 205
6245 205
6250 205
6255 204
6260 204
6265 204
6270 203
6275 204
6280 204
6285 204
6290 204
6295 204
6300 204
6305 203
6310 203
6315 204
6320 204
6325 204
6330 204
6335 204
6340 203
6345 204
6350 204
6355 204
6360 203
6365 205
6370 203
6375 204
6380 204
6385 204
6390 204
6395 204
6400 204
6405 204
6410 204
6415 203
6420 205
6425 205
6430 204
6435 204
6440 205
6445 204
6450 204
6455 204
6460 204
6465 204
6470 204
6475 203
6480 204
6485 204
6490 204
6495 203
6500 204
6505 204
6510 204
6515 204
6520 204
6525 204
6530 204
6535 203
6540 204
6545 204
6550 204
6555 204
6560 204
6565 204
6570 203
6575 204
6580 204
6585 205
6590 203
6595 203
6600 204
6605 204
6610 203
6615 204
6620 205
6625 203
6630 204
6635 204
6640 204
6645 204
6650 204
6655 204
6660 203
6665 204
6670 203
6675 203
6680 203
6685 204
6690 204
6695 204
6700 204
6705 204
6710 204
6715 204
6720 204
6725 204
6730 204
6735 204
6740 206
6745 204
6750 204
6755 204
6760 204
6765 203
6770 204
6775 204
6780 204
6785 203
6790 203
6795 203
6800 204
6805 203
6810 204
6815 204
6820 204
6825 204
6830 203
6835 204
6840 204
6845 204
6850 204
6855 203
6860 204
6865 204
6870 204
6875 204
6880 203
6885 204
6890 204
6895 204
6900 203
6905 204
6910 204
6915 204
6920 204
6925 204
6930 204
6935 204
6940 203
6945 203
6950 204
6955 203
6960 205
6965 203
6970 204
6975 204
6980 204
6985 204
6990 204
6995 204
7000 204
7005 204
7010 204
7015 204
7020 204
7025 203
7030 204
7035 203
7040 207
7045 203
7050 203
7055 204
7060 204
7065 204
7070 204
7075 203
7080 203
7085 204
7090 203
7095 204
7100 204
7105 204
7110 203
7115 204
7120 204
7125 204
7130 204
7135 203
7140 204
7145 204
7150 205
7155 204
7160 204
7165 204
7170 204
7175 204
7180 204
7185 204
7190 203
7195 203
7200 204
};
\addlegendentry{Baseline}
\addplot [semithick, color1, mark=diamond*, mark size=3, mark options={solid,fill=white,draw=red},  mark repeat={180}]
table{%
0 205
5 210
10 209
15 211
20 210
25 211
30 207
35 211
40 205
45 206
50 205
55 204
60 205
65 205
70 205
75 205
80 205
85 205
90 205
95 205
100 205
105 205
110 209
115 210
120 210
125 211
130 211
135 211
140 209
145 209
150 209
155 209
160 209
165 210
170 211
175 211
180 211
185 211
190 211
195 211
200 210
205 210
210 210
215 211
220 211
225 210
230 212
235 210
240 211
245 210
250 211
255 211
260 211
265 211
270 210
275 211
280 211
285 211
290 211
295 211
300 210
305 211
310 211
315 211
320 211
325 211
330 211
335 211
340 211
345 210
350 211
355 210
360 210
365 211
370 210
375 210
380 211
385 210
390 210
395 211
400 210
405 210
410 210
415 210
420 211
425 210
430 210
435 210
440 210
445 211
450 211
};
\addlegendentry{1 job}
\addplot [semithick, color2, mark=square*, mark size=3, mark options={solid,fill=white,draw=red},  mark repeat={180}]
table {%
455 211
460 215
465 214
470 215
475 215
480 215
485 215
490 211
495 210
500 211
505 210
510 210
515 210
520 211
525 211
530 212
535 211
540 211
545 210
550 210
555 211
560 212
565 215
570 214
575 215
580 215
585 214
590 213
595 213
600 213
605 213
610 213
615 212
620 214
625 214
630 214
635 219
640 215
645 246
650 247
655 247
660 246
665 247
670 247
675 246
680 246
685 247
690 247
695 247
700 247
705 248
710 248
715 248
720 247
725 248
730 248
735 248
740 248
745 248
750 249
755 248
760 249
765 247
770 248
775 248
780 249
785 249
790 248
795 248
800 249
805 248
810 249
815 249
820 249
825 249
830 249
835 249
840 249
845 249
850 248
855 249
860 249
865 249
870 249
875 248
880 249
885 249
890 249
895 250
900 248
};
\addlegendentry{2 jobs}
\addplot [semithick, color3, mark=triangle*, mark size=3, mark options={solid,fill=white,draw=red},  mark repeat={180}]
table {%
905 249
910 249
915 251
920 251
925 250
930 251
935 252
940 252
945 248
950 248
955 247
960 248
965 248
970 247
975 247
980 248
985 248
990 248
995 248
1000 247
1005 247
1010 248
1015 248
1020 251
1025 251
1030 252
1035 252
1040 251
1045 251
1050 250
1055 250
1060 249
1065 249
1070 249
1075 254
1080 253
1085 255
1090 252
1095 254
1100 280
1105 284
1110 284
1115 285
1120 287
1125 287
1130 286
1135 284
1140 285
1145 285
1150 287
1155 287
1160 287
1165 285
1170 286
1175 287
1180 285
1185 283
1190 284
1195 285
1200 284
1205 284
1210 287
1215 284
1220 283
1225 286
1230 286
1235 285
1240 285
1245 285
1250 286
1255 286
1260 287
1265 286
1270 286
1275 287
1280 286
1285 286
1290 285
1295 285
1300 285
1305 284
1310 285
1315 286
1320 285
1325 286
1330 286
1335 285
1340 286
1345 286
1350 286
};
\addlegendentry{3 jobs}
\addplot [semithick, color4, mark=o, mark size=3, mark options={solid,fill=white,draw=red},  mark repeat={180}]
table {%
1355 285
1360 286
1365 286
1370 286
1375 286
1380 285
1385 285
1390 285
1395 285
1400 285
1405 280
1410 281
1415 282
1420 282
1425 282
1430 287
1435 287
1440 289
1445 289
1450 286
1455 286
1460 286
1465 284
1470 285
1475 284
1480 286
1485 286
1490 286
1495 285
1500 284
1505 286
1510 290
1515 289
1520 290
1525 293
1530 291
1535 290
1540 290
1545 290
1550 289
1555 290
1560 284
1565 287
1570 287
1575 288
1580 288
1585 289
1590 305
1595 307
1600 307
1605 303
1610 304
1615 302
1620 304
1625 305
1630 305
1635 305
1640 307
1645 308
1650 307
1655 308
1660 308
1665 308
1670 308
1675 306
1680 307
1685 308
1690 306
1695 306
1700 305
1705 305
1710 308
1715 306
1720 305
1725 309
1730 308
1735 308
1740 308
1745 307
1750 305
1755 306
1760 308
1765 308
1770 303
1775 304
1780 305
1785 307
1790 308
1795 307
1800 307
1805 308
1810 307
1815 305
1820 305
1825 306
1830 305
1835 305
1840 306
1845 308
1850 308
1855 305
1860 307
1865 307
1870 307
1875 308
1880 307
1885 305
1890 305
1895 304
1900 307
1905 306
1910 306
1915 307
1920 307
1925 306
1930 306
1935 305
1940 305
1945 307
1950 307
1955 307
1960 306
1965 306
1970 307
1975 305
1980 304
1985 304
1990 305
1995 304
2000 308
2005 305
2010 306
2015 305
2020 305
2025 306
2030 308
2035 305
2040 306
2045 305
2050 305
2055 306
2060 307
2065 306
2070 306
2075 304
2080 305
2085 305
2090 305
2095 304
2100 306
2105 305
2110 305
2115 305
2120 304
2125 308
2130 306
2135 304
2140 304
2145 306
2150 307
2155 303
2160 304
2165 304
2170 304
2175 302
2180 304
2185 308
2190 305
2195 304
2200 305
2205 305
2210 308
2215 306
2220 302
2225 302
2230 304
2235 304
2240 302
2245 302
2250 302
2255 302
2260 306
2265 305
2270 305
2275 305
2280 308
2285 305
2290 305
2295 305
2300 304
2305 307
2310 306
2315 306
2320 304
2325 308
2330 305
2335 305
2340 307
2345 307
2350 304
2355 305
2360 304
2365 306
2370 306
2375 305
2380 304
2385 305
2390 303
2395 305
2400 305
2405 303
2410 304
2415 305
2420 304
2425 303
2430 304
2435 304
2440 308
2445 306
2450 305
2455 308
2460 307
2465 305
2470 305
2475 305
2480 305
2485 308
2490 306
2495 306
2500 306
2505 304
2510 304
2515 304
2520 304
2525 304
2530 304
2535 305
2540 304
2545 306
2550 304
2555 304
2560 304
2565 307
2570 306
2575 304
2580 304
2585 304
2590 305
2595 305
2600 304
2605 304
2610 305
2615 304
2620 302
2625 302
2630 301
2635 302
2640 302
2645 301
2650 301
2655 302
2660 301
2665 301
2670 302
2675 301
2680 301
2685 304
2690 305
2695 304
2700 304
2705 304
2710 305
2715 304
2720 304
2725 304
2730 305
2735 305
2740 304
2745 304
2750 304
2755 306
2760 304
2765 304
2770 304
2775 305
2780 305
2785 304
2790 305
2795 308
2800 305
2805 302
2810 302
2815 305
2820 304
2825 304
2830 304
2835 304
2840 304
2845 304
2850 304
2855 305
2860 304
2865 306
2870 308
2875 304
2880 305
2885 305
2890 304
2895 308
2900 306
2905 305
2910 304
2915 304
2920 304
2925 304
2930 304
2935 304
2940 304
2945 304
2950 304
2955 305
2960 304
2965 305
2970 308
2975 306
2980 306
2985 306
2990 304
2995 304
3000 304
3005 304
3010 305
3015 304
3020 306
3025 306
3030 306
3035 304
3040 304
3045 304
3050 307
3055 306
3060 304
3065 305
3070 305
3075 307
3080 306
3085 306
3090 308
3095 304
3100 305
3105 305
3110 304
3115 304
3120 304
3125 305
3130 308
3135 305
3140 304
3145 308
3150 308
3155 305
3160 306
3165 306
3170 305
3175 306
3180 307
3185 308
3190 307
3195 306
3200 306
3205 305
3210 305
3215 304
3220 304
3225 306
3230 307
3235 308
3240 304
3245 306
3250 306
3255 304
3260 306
3265 304
3270 304
3275 304
3280 305
3285 306
3290 306
3295 304
3300 307
3305 304
3310 304
3315 304
3320 304
3325 304
3330 306
3335 304
3340 304
3345 304
3350 304
3355 305
3360 306
3365 305
3370 304
3375 304
3380 305
3385 306
3390 308
3395 306
3400 305
3405 304
3410 305
3415 305
3420 306
3425 305
3430 304
3435 303
3440 305
3445 306
3450 304
3455 306
3460 304
3465 307
3470 305
3475 306
3480 304
3485 304
3490 301
3495 301
3500 301
3505 301
3510 302
3515 302
3520 302
3525 302
3530 304
3535 305
3540 304
3545 304
3550 304
3555 304
3560 304
3565 305
3570 306
3575 304
3580 304
3585 303
3590 303
3595 303
3600 304
3605 304
3610 304
3615 304
3620 304
3625 305
3630 305
3635 304
3640 305
3645 302
3650 302
3655 304
3660 303
3665 304
3670 304
3675 303
3680 304
3685 304
3690 305
3695 304
3700 304
3705 308
3710 305
3715 304
3720 304
3725 307
3730 307
3735 305
3740 304
3745 304
3750 304
3755 304
3760 304
3765 304
3770 304
3775 304
3780 304
3785 305
3790 304
3795 308
3800 306
3805 304
3810 305
3815 306
3820 306
3825 304
3830 304
3835 306
3840 304
3845 304
3850 304
3855 305
3860 304
3865 305
3870 305
3875 308
3880 304
3885 305
3890 305
3895 305
3900 305
3905 308
3910 305
3915 308
3920 305
3925 304
3930 304
3935 302
3940 303
3945 305
3950 304
3955 304
3960 305
3965 307
3970 303
3975 299
3980 299
};
\addlegendentry{4 jobs}
\addplot [semithick, color3, mark=triangle*, mark size=3, mark options={solid,fill=white,draw=red},  mark repeat={180}]
table {%
3980 299
3985 296
3990 291
3995 292
4000 290
4000 290
4005 289
4010 289
4015 289
4020 288
4025 285
4030 283
4035 285
4040 285
4045 285
4050 285
4055 285
4060 285
4065 285
4070 285
4075 285
4080 285
4085 285
4090 285
4095 285
4100 285
4105 285
4110 283
4115 285
4120 285
4125 285
4130 285
4135 285
4140 285
4145 286
4150 285
4155 286
4160 285
4165 285
4170 285
4175 285
4180 285
4185 285
4190 285
4195 285
4200 285
4205 285
4210 286
4215 285
4220 285
4225 285
4230 285
4230 285
4235 284
4240 286
4245 286
4250 285
4255 285
4260 285
4265 286
4270 285
4275 287
4280 285
4285 286
4290 285
4295 286
4300 286
4305 286
4310 287
4315 286
4320 285
4325 285
4330 285
4335 286
4340 285
4345 285
4350 284
4355 285
4360 285
4365 286
4370 286
4375 287
4380 286
4385 287
4390 287
4395 287
4400 286
4405 287
4410 285
4415 275
4420 275
};
\addplot [semithick, color2, mark=square*, mark size=3, mark options={solid,fill=white,draw=red},  mark repeat={180}]
table {%
4420 275
4425 266
4430 265
4435 261
4440 256
4440 256
4445 257
4450 257
4455 254
4460 256
4465 255
4470 250
4475 249
4480 249
4485 250
4490 249
4495 250
4500 249
4505 248
4510 252
4515 250
4520 251
4525 251
4530 251
4535 252
4540 252
4545 251
4550 252
4555 250
4560 252
4565 251
4570 251
4575 252
4580 252
4585 252
4590 252
4595 251
4600 251
4605 251
4610 252
4615 251
4620 251
4625 251
4630 251
4635 252
4640 251
4645 251
4650 252
4655 251
4660 251
4665 251
4670 251
4675 251
4680 252
4685 251
4690 251
4695 251
4700 251
4705 251
4710 251
4715 251
4720 250
4725 253
4730 251
4735 251
4740 252
4745 252
4750 251
4755 250
4760 250
4765 251
4770 251
4775 251
4780 251
4785 251
4790 251
4795 250
4800 252
4805 250
4810 251
4815 249
4820 251
4825 250
4830 251
};
\addplot [semithick, color1, mark=diamond*, mark size=3, mark options={solid,fill=white,draw=red},  mark repeat={180}]
table {%
4835 251
4840 241
4840 241
4845 233
4850 229
4855 227
4860 219
4865 219
4870 218
4875 219
4880 219
4885 217
4890 213
4895 212
4900 212
4905 213
4910 213
4915 213
4920 212
4925 214
4930 212
4935 213
4940 212
4945 212
4950 212
4955 212
4960 212
4965 211
4970 212
4975 212
4980 213
4985 212
4990 212
4995 212
5000 212
5005 212
5010 211
5015 211
5020 211
5025 212
5030 211
5035 211
5040 212
5045 211
5050 211
5055 211
5060 210
5065 212
5070 211
5075 211
5080 211
5085 212
5090 211
5095 211
5100 212
5105 212
5110 211
5115 211
5120 211
5125 212
5130 212
5135 210
5140 211
5145 211
5150 212
5155 211
5160 210
5165 211
5170 210
5175 211
5180 211
5185 211
5190 210
5195 211
5200 210
5205 211
5210 211
5215 211
5220 211
5225 211
5230 211
5235 211
5240 211
5245 211
5250 210
5255 210
5260 210
5265 211
5270 211
5275 211
5280 211
5285 211
5290 210
5295 211
5300 210
5305 211
5310 205
};

\end{axis}

\end{tikzpicture}

%% file: results/energy_amd_seq.tex
\begin{tikzpicture}[font=\Large]

\definecolor{color0}{rgb}{0.12156862745098,0.466666666666667,0.705882352941177}
\definecolor{color1}{rgb}{1,0.498039215686275,0.0549019607843137}
\definecolor{color2}{rgb}{0.172549019607843,0.627450980392157,0.172549019607843}
\definecolor{color3}{rgb}{0.83921568627451,0.152941176470588,0.156862745098039}
\definecolor{color4}{rgb}{0.580392156862745,0.403921568627451,0.741176470588235}
\definecolor{color5}{rgb}{0,0,0}
\definecolor{color6}{rgb}{0.07, 0.04, 0.56}

\begin{axis}[
legend cell align={left},
legend columns=3,
legend style={fill opacity=0.8, draw opacity=1, text opacity=1, at={(1.05,1.18)}, anchor=east, draw=white!80.0!black},
tick align=outside,
tick pos=left,
x grid style={white!69.01960784313725!black},
xlabel={Time (SEC)},
xmin=0, xmax=5310,
xtick style={color=black},
y grid style={white!69.01960784313725!black},
xmajorgrids,
ymajorgrids,
ylabel={Total energy consumption (kWh)},
ymin=0, ymax=0.4,
ytick style={color=black},
ytick={0.05, 0.1,0.15,0.2,0.25,0.3,0.35,0.4},
yticklabel style={
        /pgf/number format/fixed,
        /pgf/number format/precision=2
},
 y label style={at={(axis description cs:-0.05,.5)}}
]
\addplot [semithick, olive, mark=star, mark size=3, mark options={solid,fill=white,draw=red},  mark repeat={180}]
table[y expr=(\thisrowno{1}-278785)/1000] {%
0 278785
5 278785
10 278785
15 278786
20 278786
25 278786
30 278787
35 278787
40 278787
45 278788
50 278788
55 278788
60 278788
65 278788
70 278789
75 278789
80 278790
85 278790
90 278790
95 278790
100 278790
105 278790
110 278792
115 278792
120 278792
125 278792
130 278792
135 278792
140 278793
145 278793
150 278794
155 278794
160 278794
165 278794
170 278795
175 278795
180 278795
185 278795
190 278796
195 278796
200 278796
205 278797
210 278797
215 278797
220 278797
225 278798
230 278798
235 278799
240 278799
245 278799
250 278799
255 278799
260 278800
265 278801
270 278801
275 278801
280 278801
285 278801
290 278801
295 278803
300 278803
305 278803
310 278803
315 278803
320 278804
325 278805
330 278805
335 278805
340 278805
345 278806
350 278807
355 278807
360 278807
365 278807
370 278808
375 278809
380 278809
385 278809
390 278809
395 278811
400 278811
405 278811
410 278811
415 278812
420 278813
425 278813
430 278813
435 278813
440 278814
445 278815
450 278815
455 278815
460 278816
465 278816
470 278817
475 278817
480 278817
485 278818
490 278819
495 278819
500 278819
505 278819
510 278820
515 278821
520 278821
525 278821
530 278821
535 278822
540 278823
545 278823
550 278823
555 278824
560 278824
565 278825
570 278825
575 278825
580 278826
585 278827
590 278827
595 278827
600 278827
605 278827
610 278829
615 278829
620 278829
625 278829
630 278830
635 278831
640 278831
645 278831
650 278831
655 278832
660 278833
665 278833
670 278833
675 278833
680 278834
685 278835
690 278835
695 278835
700 278836
705 278837
710 278837
715 278837
720 278837
725 278838
730 278839
735 278839
740 278839
745 278839
750 278840
755 278841
760 278841
765 278841
770 278842
775 278842
780 278843
785 278843
790 278843
795 278844
800 278845
805 278845
810 278845
815 278846
820 278846
825 278847
830 278847
835 278847
840 278848
845 278848
850 278849
855 278849
860 278849
865 278850
870 278850
875 278851
880 278851
885 278852
890 278852
895 278853
900 278853
905 278853
910 278854
915 278854
920 278855
925 278855
930 278855
935 278856
940 278856
945 278857
950 278857
955 278858
960 278858
965 278858
970 278859
975 278859
980 278860
985 278860
990 278861
995 278861
1000 278861
1005 278862
1010 278862
1015 278863
1020 278863
1025 278864
1030 278864
1035 278864
1040 278865
1045 278865
1050 278866
1055 278866
1060 278866
1065 278867
1070 278868
1075 278868
1080 278868
1085 278869
1090 278869
1095 278870
1100 278870
1105 278870
1110 278871
1115 278871
1120 278872
1125 278872
1130 278872
1135 278873
1140 278874
1145 278874
1150 278874
1155 278874
1160 278875
1165 278876
1170 278876
1175 278876
1180 278876
1185 278878
1190 278878
1195 278878
1200 278878
1205 278879
1210 278880
1215 278880
1220 278880
1225 278880
1230 278881
1235 278882
1240 278882
1245 278882
1250 278882
1255 278884
1260 278884
1265 278884
1270 278884
1275 278885
1280 278886
1285 278886
1290 278886
1295 278886
1300 278887
1305 278888
1310 278888
1315 278888
1320 278889
1325 278889
1330 278890
1335 278890
1340 278890
1345 278891
1350 278892
1355 278892
1360 278892
1365 278892
1370 278893
1375 278894
1380 278894
1385 278894
1390 278895
1395 278895
1400 278896
1405 278896
1410 278896
1415 278897
1420 278897
1425 278898
1430 278898
1435 278898
1440 278899
1445 278899
1450 278900
1455 278900
1460 278901
1465 278901
1470 278902
1475 278902
1480 278902
1485 278903
1490 278903
1495 278904
1500 278904
1505 278905
1510 278905
1515 278905
1520 278906
1525 278906
1530 278907
1535 278907
1540 278907
1545 278908
1550 278908
1555 278909
1560 278909
1565 278910
1570 278910
1575 278910
1580 278911
1585 278911
1590 278912
1595 278912
1600 278913
1605 278913
1610 278913
1615 278914
1620 278914
1625 278915
1630 278915
1635 278915
1640 278916
1645 278917
1650 278917
1655 278917
1660 278917
1665 278918
1670 278919
1675 278919
1680 278919
1685 278920
1690 278920
1695 278921
1700 278921
1705 278921
1710 278922
1715 278923
1720 278923
1725 278923
1730 278923
1735 278924
1740 278925
1745 278925
1750 278925
1755 278925
1760 278926
1765 278927
1770 278927
1775 278927
1780 278927
1785 278929
1790 278929
1795 278929
1800 278929
1805 278931
1810 278931
1815 278931
1820 278931
1825 278931
1830 278933
1835 278933
1840 278933
1845 278933
1850 278934
1855 278935
1860 278935
1865 278935
1870 278935
1875 278936
1880 278937
1885 278937
1890 278937
1895 278937
1900 278938
1905 278939
1910 278939
1915 278939
1920 278940
1925 278941
1930 278941
1935 278941
1940 278941
1945 278942
1950 278943
1955 278943
1960 278943
1965 278944
1970 278944
1975 278945
1980 278945
1985 278945
1990 278946
1995 278946
2000 278947
2005 278947
2010 278947
2015 278948
2020 278949
2025 278949
2030 278949
2035 278950
2040 278950
2045 278951
2050 278951
2055 278951
2060 278952
2065 278952
2070 278953
2075 278953
2080 278953
2085 278954
2090 278954
2095 278955
2100 278955
2105 278956
2110 278956
2115 278957
2120 278957
2125 278957
2130 278958
2135 278958
2140 278959
2145 278959
2150 278959
2155 278960
2160 278960
2165 278961
2170 278961
2175 278962
2180 278962
2185 278963
2190 278963
2195 278963
2200 278964
2205 278964
2210 278965
2215 278965
2220 278965
2225 278966
2230 278966
2235 278967
2240 278967
2245 278968
2250 278968
2255 278968
2260 278969
2265 278969
2270 278970
2275 278970
2280 278970
2285 278971
2290 278972
2295 278972
2300 278972
2305 278973
2310 278973
2315 278974
2320 278974
2325 278974
2330 278975
2335 278975
2340 278976
2345 278976
2350 278976
2355 278977
2360 278978
2365 278978
2370 278978
2375 278978
2380 278979
2385 278980
2390 278980
2395 278980
2400 278980
2405 278982
2410 278982
2415 278982
2420 278982
2425 278983
2430 278984
2435 278984
2440 278984
2445 278984
2450 278985
2455 278986
2460 278986
2465 278986
2470 278986
2475 278988
2480 278988
2485 278988
2490 278988
2495 278989
2500 278990
2505 278990
2510 278990
2515 278990
2520 278991
2525 278992
2530 278992
2535 278992
2540 278992
2545 278993
2550 278994
2555 278994
2560 278994
2565 278995
2570 278996
2575 278996
2580 278996
2585 278996
2590 278997
2595 278998
2600 278998
2605 278998
2610 278999
2615 278999
2620 279000
2625 279000
2630 279000
2635 279001
2640 279001
2645 279002
2650 279002
2655 279002
2660 279003
2665 279004
2670 279004
2675 279004
2680 279005
2685 279005
2690 279006
2695 279006
2700 279006
2705 279007
2710 279007
2715 279008
2720 279008
2725 279008
2730 279009
2735 279009
2740 279010
2745 279010
2750 279011
2755 279011
2760 279011
2765 279012
2770 279012
2775 279013
2780 279013
2785 279014
2790 279014
2795 279014
2800 279015
2805 279015
2810 279016
2815 279016
2820 279017
2825 279017
2830 279017
2835 279018
2840 279018
2845 279019
2850 279019
2855 279020
2860 279020
2865 279021
2870 279021
2875 279021
2880 279022
2885 279022
2890 279023
2895 279023
2900 279023
2905 279024
2910 279024
2915 279025
2920 279025
2925 279025
2930 279026
2935 279027
2940 279027
2945 279027
2950 279027
2955 279028
2960 279029
2965 279029
2970 279029
2975 279029
2980 279031
2985 279031
2990 279031
2995 279031
3000 279031
3005 279033
3010 279033
3015 279033
3020 279033
3025 279034
3030 279035
3035 279035
3040 279035
3045 279035
3050 279037
3055 279037
3060 279037
3065 279037
3070 279038
3075 279039
3080 279039
3085 279039
3090 279039
3095 279040
3100 279041
3105 279041
3110 279041
3115 279041
3120 279042
3125 279043
3130 279043
3135 279043
3140 279044
3145 279044
3150 279045
3155 279045
3160 279045
3165 279046
3170 279047
3175 279047
3180 279047
3185 279048
3190 279048
3195 279049
3200 279049
3205 279049
3210 279050
3215 279050
3220 279051
3225 279051
3230 279051
3235 279052
3240 279052
3245 279053
3250 279053
3255 279054
3260 279054
3265 279055
3270 279055
3275 279055
3280 279056
3285 279056
3290 279057
3295 279057
3300 279058
3305 279058
3310 279058
3315 279059
3320 279059
3325 279060
3330 279060
3335 279060
3340 279061
3345 279061
3350 279062
3355 279062
3360 279063
3365 279063
3370 279063
3375 279064
3380 279064
3385 279065
3390 279065
3395 279066
3400 279066
3405 279066
3410 279067
3415 279067
3420 279068
3425 279068
3430 279069
3435 279069
3440 279070
3445 279070
3450 279070
3455 279071
3460 279071
3465 279072
3470 279072
3475 279072
3480 279073
3485 279073
3490 279074
3495 279074
3500 279074
3505 279075
3510 279076
3515 279076
3520 279076
3525 279076
3530 279077
3535 279078
3540 279078
3545 279078
3550 279079
3555 279079
3560 279080
3565 279080
3570 279080
3575 279080
3580 279082
3585 279082
3590 279082
3595 279082
3600 279083
3605 279084
3610 279084
3615 279084
3620 279084
3625 279085
3630 279086
3635 279086
3640 279086
3645 279087
3650 279088
3655 279088
3660 279088
3665 279088
3670 279089
3675 279089
3680 279090
3685 279090
3690 279090
3695 279091
3700 279091
3705 279091
3710 279092
3715 279092
3720 279092
3725 279093
3730 279093
3735 279093
3740 279093
3745 279094
3750 279094
3755 279094
3760 279095
3765 279095
3770 279095
3775 279095
3780 279095
3785 279096
3790 279097
3795 279097
3800 279097
3805 279097
3810 279097
3815 279097
3820 279098
3825 279099
3830 279099
3835 279099
3840 279099
3845 279099
3850 279099
3855 279100
3860 279101
3865 279101
3870 279101
3875 279101
3880 279101
3885 279102
3890 279102
3895 279102
3900 279103
3905 279103
3910 279103
3915 279103
3920 279104
3925 279104
3930 279104
3935 279104
3940 279105
3945 279105
3950 279106
3955 279106
3960 279106
3965 279106
3970 279106
3975 279106
3980 279108
3985 279108
3990 279108
3995 279108
4000 279108
4005 279108
4010 279109
4015 279110
4020 279110
4025 279110
4030 279110
4035 279110
4040 279111
4045 279111
4050 279111
4055 279112
4060 279112
4065 279112
4070 279112
4075 279113
4080 279113
4085 279113
4090 279113
4095 279114
4100 279114
4105 279115
4110 279115
4115 279115
4120 279115
4125 279115
4130 279115
4135 279117
4140 279117
4145 279117
4150 279117
4155 279117
4160 279117
4165 279117
4170 279118
4175 279118
4180 279118
4185 279119
4190 279119
4195 279119
4200 279119
4205 279120
4210 279120
4215 279120
4220 279120
4225 279121
4230 279121
4235 279122
4240 279122
4245 279122
4250 279122
4255 279122
4260 279122
4265 279124
4270 279124
4275 279124
4280 279124
4285 279124
4290 279124
4295 279125
4300 279125
4305 279126
4310 279126
4315 279126
4320 279126
4325 279126
4330 279127
4335 279127
4340 279128
4345 279128
4350 279128
4355 279128
4360 279129
4365 279129
4370 279129
4375 279129
4380 279130
4385 279130
4390 279131
4395 279131
4400 279131
4405 279131
4410 279131
4415 279131
4420 279133
4425 279133
4430 279133
4435 279133
4440 279133
4445 279133
4450 279133
4455 279134
4460 279135
4465 279135
4470 279135
4475 279135
4480 279135
4485 279136
4490 279136
4495 279136
4500 279137
4505 279137
4510 279137
4515 279138
4520 279138
4525 279138
4530 279138
4535 279139
4540 279139
4545 279139
4550 279140
4555 279140
4560 279140
4565 279140
4570 279140
4575 279141
4580 279142
4585 279142
4590 279142
4595 279142
4600 279142
4605 279142
4610 279143
4615 279144
4620 279144
4625 279144
4630 279144
4635 279144
4640 279145
4645 279145
4650 279145
4655 279146
4660 279146
4665 279146
4670 279146
4675 279147
4680 279147
4685 279147
4690 279147
4695 279148
4700 279148
4705 279149
4710 279149
4715 279149
4720 279149
4725 279149
4730 279149
4735 279151
4740 279151
4745 279151
4750 279151
4755 279151
4760 279151
4765 279152
4770 279152
4775 279152
4780 279152
4785 279153
4790 279153
4795 279153
4800 279154
4805 279154
4810 279154
4815 279154
4820 279154
4825 279155
4830 279156
4835 279156
4840 279156
4845 279156
4850 279156
4855 279156
4860 279157
4865 279158
4870 279158
4875 279158
4880 279158
4885 279158
4890 279158
4895 279159
4900 279160
4905 279160
4910 279160
4915 279160
4920 279160
4925 279161
4930 279161
4935 279161
4940 279162
4945 279162
4950 279162
4955 279162
4960 279163
4965 279163
4970 279163
4975 279163
4980 279164
4985 279164
4990 279165
4995 279165
5000 279165
5005 279165
5010 279165
5015 279165
5020 279167
5025 279167
5030 279167
5035 279167
5040 279167
5045 279167
5050 279168
5055 279169
5060 279169
5065 279169
5070 279169
5075 279169
5080 279169
5085 279170
5090 279170
5095 279171
5100 279171
5105 279171
5110 279171
5115 279172
5120 279172
5125 279172
5130 279172
5135 279173
5140 279173
5145 279174
5150 279174
5155 279174
5160 279174
5165 279174
5170 279174
5175 279176
5180 279176
5185 279176
5190 279176
5195 279176
5200 279176
5205 279176
5210 279177
5215 279178
5220 279178
5225 279178
5230 279178
5235 279178
5240 279179
5245 279179
5250 279179
5255 279180
5260 279180
5265 279180
5270 279181
5275 279181
5280 279181
5285 279181
5290 279182
5295 279182
5300 279183
5305 279183
5310 279183
5315 279183
5320 279183
5325 279183
5330 279184
5335 279185
5340 279185
5345 279185
5350 279185
5355 279185
5360 279185
5365 279186
5370 279186
5375 279186
5380 279187
5385 279187
5390 279187
5395 279187
5400 279188
};
\addlegendentry{4 jobs in parallel}
\addplot [semithick, color5, mark=.]
table[y expr=(\thisrowno{1}-350880)/1000] {%
0 350880
5 350880
10 350880
15 350880
20 350880
25 350880
30 350881
35 350882
40 350882
45 350882
50 350882
55 350882
60 350883
65 350883
70 350883
75 350884
80 350884
85 350884
90 350884
95 350885
100 350885
105 350885
110 350885
115 350886
120 350886
125 350887
130 350887
135 350887
140 350887
145 350887
150 350888
155 350889
160 350889
165 350889
170 350889
175 350889
180 350889
185 350889
190 350891
195 350891
200 350891
205 350891
210 350891
215 350891
220 350892
225 350892
230 350893
235 350893
240 350893
245 350893
250 350894
255 350894
260 350894
265 350894
270 350895
275 350895
280 350895
285 350896
290 350896
295 350896
300 350896
305 350896
310 350897
315 350898
320 350898
325 350898
330 350899
335 350899
340 350899
345 350900
350 350900
355 350900
360 350900
365 350900
370 350901
375 350901
380 350902
385 350902
390 350902
395 350902
400 350902
405 350902
410 350904
415 350904
420 350904
425 350904
430 350904
435 350904
440 350905
445 350905
450 350906
455 350906
460 350906
465 350906
470 350907
475 350907
480 350907
485 350907
490 350908
495 350908
500 350908
505 350909
510 350909
515 350909
520 350909
525 350910
530 350910
535 350911
540 350911
545 350911
550 350911
555 350911
560 350911
565 350913
570 350913
575 350913
580 350913
585 350913
590 350913
595 350913
600 350914
605 350915
610 350915
615 350915
620 350915
625 350915
630 350916
635 350916
640 350916
645 350917
650 350917
655 350917
660 350918
665 350918
670 350918
675 350918
680 350918
685 350919
690 350920
695 350920
700 350920
705 350920
710 350920
715 350920
720 350920
725 350922
730 350922
735 350922
740 350922
745 350922
750 350922
755 350923
760 350923
765 350924
770 350924
775 350924
780 350924
785 350925
790 350925
795 350925
800 350925
805 350926
810 350926
815 350927
820 350927
825 350927
830 350927
835 350927
840 350928
845 350928
850 350929
855 350929
860 350929
865 350929
870 350929
875 350929
880 350931
885 350931
890 350931
895 350931
900 350931
905 350931
910 350932
915 350932
920 350932
925 350932
930 350933
935 350933
940 350933
945 350934
950 350934
955 350934
960 350934
965 350934
970 350935
975 350936
980 350936
985 350936
990 350936
995 350936
1000 350936
1005 350936
1010 350938
1015 350938
1020 350938
1025 350938
1030 350938
1035 350938
1040 350939
1045 350940
1050 350940
1055 350940
1060 350940
1065 350940
1070 350941
1075 350941
1080 350941
1085 350942
1090 350942
1095 350942
1100 350943
1105 350943
1110 350943
1115 350943
1120 350943
1125 350944
1130 350944
1135 350945
1140 350945
1145 350945
1150 350945
1155 350945
1160 350945
1165 350947
1170 350947
1175 350947
1180 350947
1185 350947
1190 350947
1195 350948
1200 350948
1205 350949
1210 350949
1215 350949
1220 350949
1225 350949
1230 350950
1235 350950
1240 350950
1245 350951
1250 350951
1255 350951
1260 350952
1265 350952
1270 350952
1275 350952
1280 350952
1285 350953
1290 350954
1295 350954
1300 350954
1305 350954
1310 350954
1315 350954
1320 350955
1325 350956
1330 350956
1335 350956
1340 350956
1345 350956
1350 350956
1355 350957
1360 350958
1365 350958
1370 350958
1375 350958
1380 350958
1385 350959
1390 350959
1395 350959
1400 350960
1405 350960
1410 350960
1415 350961
1420 350961
1425 350961
1430 350961
1435 350961
1440 350962
1445 350963
1450 350963
1455 350963
1460 350963
1465 350963
1470 350963
1475 350963
1480 350965
1485 350965
1490 350965
1495 350965
1500 350965
1505 350965
1510 350966
1515 350966
1520 350966
1525 350967
1530 350967
1535 350967
1540 350968
1545 350968
1550 350968
1555 350968
1560 350968
1565 350969
1570 350969
1575 350970
1580 350970
1585 350970
1590 350970
1595 350970
1600 350970
1605 350972
1610 350972
1615 350972
1620 350972
1625 350972
1630 350972
1635 350972
1640 350973
1645 350974
1650 350974
1655 350974
1660 350974
1665 350974
1670 350975
1675 350975
1680 350975
1685 350976
1690 350976
1695 350976
1700 350977
1705 350977
1710 350977
1715 350977
1720 350977
1725 350978
1730 350979
1735 350979
1740 350979
1745 350979
1750 350979
1755 350979
1760 350979
1765 350981
1770 350981
1775 350981
1780 350981
1785 350981
1790 350981
1795 350982
1800 350982
1805 350983
1810 350983
1815 350983
1820 350983
1825 350984
1830 350984
1835 350984
1840 350985
1845 350985
1850 350985
1855 350986
1860 350986
1865 350986
1870 350986
1875 350986
1880 350987
1885 350987
1890 350988
1895 350988
1900 350988
1905 350988
1910 350988
1915 350988
1920 350990
1925 350990
1930 350990
1935 350990
1940 350990
1945 350990
1950 350991
1955 350991
1960 350992
1965 350992
1970 350992
1975 350992
1980 350992
1985 350993
1990 350993
1995 350993
2000 350994
2005 350994
2010 350994
2015 350995
2020 350995
2025 350995
2030 350995
2035 350995
2040 350996
2045 350997
2050 350997
2055 350997
2060 350997
2065 350997
2070 350997
2075 350997
2080 350999
2085 350999
2090 350999
2095 350999
2100 350999
2105 350999
2110 351000
2115 351000
2120 351000
2125 351001
2130 351001
2135 351001
2140 351002
2145 351002
2150 351002
2155 351002
2160 351002
2165 351003
2170 351004
2175 351004
2180 351004
2185 351004
2190 351004
2195 351004
2200 351004
2205 351006
2210 351006
2215 351006
2220 351006
2225 351006
2230 351006
2235 351007
2240 351007
2245 351008
2250 351008
2255 351008
2260 351008
2265 351009
2270 351009
2275 351009
2280 351009
2285 351010
2290 351010
2295 351010
2300 351011
2305 351011
2310 351011
2315 351011
2320 351012
2325 351012
2330 351013
2335 351013
2340 351013
2345 351013
2350 351013
2355 351013
2360 351015
2365 351015
2370 351015
2375 351015
2380 351015
2385 351015
2390 351016
2395 351016
2400 351017
2405 351017
2410 351017
2415 351017
2420 351017
2425 351018
2430 351018
2435 351018
2440 351019
2445 351019
2450 351019
2455 351020
2460 351020
2465 351020
2470 351020
2475 351020
2480 351021
2485 351022
2490 351022
2495 351022
2500 351022
2505 351022
2510 351022
2515 351022
2520 351024
2525 351024
2530 351024
2535 351024
2540 351024
2545 351024
2550 351025
2555 351026
2560 351026
2565 351026
2570 351026
2575 351026
2580 351027
2585 351027
2590 351027
2595 351027
2600 351028
2605 351028
2610 351028
2615 351029
2620 351029
2625 351029
2630 351029
2635 351030
2640 351030
2645 351031
2650 351031
2655 351031
2660 351031
2665 351031
2670 351031
2675 351033
2680 351033
2685 351033
2690 351033
2695 351033
2700 351033
2705 351034
2710 351034
2715 351034
2720 351034
2725 351035
2730 351035
2735 351035
2740 351036
2745 351036
2750 351036
2755 351036
2760 351036
2765 351037
2770 351038
2775 351038
2780 351038
2785 351038
2790 351038
2795 351038
2800 351038
2805 351040
2810 351040
2815 351040
2820 351040
2825 351040
2830 351040
2835 351041
2840 351042
2845 351042
2850 351042
2855 351042
2860 351042
2865 351043
2870 351043
2875 351043
2880 351044
2885 351044
2890 351044
2895 351045
2900 351045
2905 351045
2910 351045
2915 351045
2920 351046
2925 351046
2930 351047
2935 351047
2940 351047
2945 351047
2950 351047
2955 351047
2960 351049
2965 351049
2970 351049
2975 351049
2980 351049
2985 351049
2990 351050
2995 351050
3000 351051
3005 351051
3010 351051
3015 351051
3020 351052
3025 351052
3030 351052
3035 351052
3040 351053
3045 351053
3050 351053
3055 351054
3060 351054
3065 351054
3070 351054
3075 351054
3080 351055
3085 351056
3090 351056
3095 351056
3100 351056
3105 351056
3110 351056
3115 351058
3120 351058
3125 351058
3130 351058
3135 351058
3140 351058
3145 351058
3150 351059
3155 351060
3160 351060
3165 351060
3170 351060
3175 351060
3180 351061
3185 351061
3190 351061
3195 351062
3200 351062
3205 351062
3210 351063
3215 351063
3220 351063
3225 351063
3230 351063
3235 351064
3240 351064
3245 351065
3250 351065
3255 351065
3260 351065
3265 351065
3270 351065
3275 351067
3280 351067
3285 351067
3290 351067
3295 351067
3300 351067
3305 351068
3310 351068
3315 351068
3320 351068
3325 351069
3330 351069
3335 351070
3340 351070
3345 351070
3350 351070
3355 351070
3360 351070
3365 351071
3370 351072
3375 351072
3380 351072
3385 351072
3390 351072
3395 351072
3400 351074
3405 351074
3410 351074
3415 351074
3420 351074
3425 351074
3430 351075
3435 351075
3440 351076
3445 351076
3450 351076
3455 351076
3460 351076
3465 351077
3470 351077
3475 351077
3480 351078
3485 351078
3490 351078
3495 351079
3500 351079
3505 351079
3510 351079
3515 351079
3520 351080
3525 351081
3530 351081
3535 351081
3540 351081
3545 351081
3550 351081
3555 351082
3560 351083
3565 351083
3570 351083
3575 351083
3580 351083
3585 351083
3590 351084
3595 351085
3600 351085
3605 351085
3610 351085
3615 351085
3620 351086
3625 351086
3630 351086
3635 351087
3640 351087
3645 351087
3650 351088
3655 351088
3660 351088
3665 351088
3670 351088
3675 351089
3680 351090
3685 351090
3690 351090
3695 351090
3700 351090
3705 351090
3710 351090
3715 351092
3720 351092
3725 351092
3730 351092
3735 351092
3740 351092
3745 351093
3750 351094
3755 351094
3760 351094
3765 351094
3770 351094
3775 351095
3780 351095
3785 351095
3790 351096
3795 351096
3800 351096
3805 351096
3810 351097
3815 351097
3820 351097
3825 351097
3830 351098
3835 351098
3840 351099
3845 351099
3850 351099
3855 351099
3860 351099
3865 351099
3870 351101
3875 351101
3880 351101
3885 351101
3890 351101
3895 351101
3900 351102
3905 351102
3910 351102
3915 351102
3920 351103
3925 351103
3930 351103
3935 351104
3940 351104
3945 351104
3950 351104
3955 351104
3960 351105
3965 351106
3970 351106
3975 351106
3980 351106
3985 351106
3990 351106
3995 351106
4000 351108
4005 351108
4010 351108
4015 351108
4020 351108
4025 351108
4030 351109
4035 351109
4040 351110
4045 351110
4050 351110
4055 351110
4060 351111
4065 351111
4070 351111
4075 351112
4080 351112
4085 351112
4090 351112
4095 351113
4100 351113
4105 351113
4110 351113
4115 351114
4120 351114
4125 351115
4130 351115
4135 351115
4140 351115
4145 351115
4150 351115
4155 351117
4160 351117
4165 351117
4170 351117
4175 351117
4180 351117
4185 351118
4190 351118
4195 351119
4200 351119
4205 351119
4210 351119
4215 351119
4220 351120
4225 351120
4230 351121
4235 351121
4240 351121
4245 351121
4250 351122
4255 351122
4260 351122
4265 351122
4270 351123
4275 351123
4280 351124
4285 351124
4290 351124
4295 351124
4300 351124
4305 351124
4310 351125
4315 351126
4320 351126
4325 351126
4330 351126
4335 351126
4340 351126
4345 351127
4350 351128
4355 351128
4360 351128
4365 351128
4370 351128
4375 351129
4380 351129
4385 351130
4390 351130
4395 351130
4400 351130
4405 351130
4410 351131
4415 351131
4420 351131
4425 351132
4430 351132
4435 351132
4440 351133
4445 351133
4450 351133
4455 351133
4460 351133
4465 351134
4470 351134
4475 351135
4480 351135
4485 351135
4490 351135
4495 351135
4500 351135
4505 351136
4510 351136
4515 351136
4520 351137
4525 351137
4530 351137
4535 351137
4540 351138
4545 351138
4550 351138
4555 351138
4560 351139
4565 351140
4570 351140
4575 351140
4580 351140
4585 351140
4590 351140
4595 351142
4600 351142
4605 351142
4610 351142
4615 351142
4620 351142
4625 351142
4630 351143
4635 351144
4640 351144
4645 351144
4650 351144
4655 351144
4660 351145
4665 351145
4670 351145
4675 351146
4680 351146
4685 351146
4690 351146
4695 351147
4700 351147
4705 351147
4710 351147
4715 351148
4720 351148
4725 351149
4730 351149
4735 351149
4740 351149
4745 351149
4750 351150
4755 351150
4760 351151
4765 351151
4770 351151
4775 351151
4780 351151
4785 351151
4790 351153
4795 351153
4800 351153
4805 351153
4810 351153
4815 351153
4820 351154
4825 351154
4830 351155
4835 351155
4840 351155
4845 351155
4850 351155
4855 351156
4860 351156
4865 351157
4870 351157
4875 351157
4880 351157
4885 351158
4890 351158
4895 351158
4900 351158
4905 351159
4910 351159
4915 351159
4920 351160
4925 351160
4930 351160
4935 351160
4940 351161
4945 351161
4950 351162
4955 351162
4960 351162
4965 351162
4970 351162
4975 351162
4980 351164
4985 351164
4990 351164
4995 351164
5000 351164
5005 351164
5010 351164
5015 351165
5020 351166
5025 351166
5030 351166
5035 351166
5040 351166
5045 351167
5050 351167
5055 351167
5060 351168
5065 351168
5070 351168
5075 351168
5080 351169
5085 351169
5090 351169
5095 351170
5100 351170
5105 351170
5110 351171
5115 351171
5120 351172
5125 351172
5130 351172
5135 351172
5140 351172
5145 351173
5150 351173
5155 351174
5160 351174
5165 351175
5170 351175
5175 351175
5180 351175
5185 351175
5190 351175
5195 351177
5200 351177
5205 351177
5210 351177
5215 351177
5220 351177
5225 351177
5230 351178
5235 351179
5240 351179
5245 351179
5250 351179
5255 351179
5260 351180
5265 351180
5270 351181
5275 351181
5280 351181
5285 351181
5290 351181
5295 351182
5300 351182
5305 351182
5310 351183
5315 351183
5320 351183
5325 351184
5330 351184
5335 351184
};
\addlegendentry{Baseline}
\addplot [semithick, color1, mark=diamond*, mark size=3, mark options={solid,fill=white,draw=red},  mark repeat={180}]
table[y expr=(\thisrowno{1}-346049)/1000] {%
0 346049
5 346049
10 346049
15 346050
20 346050
25 346050
30 346051
35 346051
40 346051
45 346051
50 346052
55 346052
60 346052
65 346052
70 346053
75 346053
80 346054
85 346054
90 346054
95 346054
100 346054
105 346054
110 346056
115 346056
120 346056
125 346056
130 346056
135 346056
140 346057
145 346057
150 346058
155 346058
160 346058
165 346058
170 346058
175 346059
180 346059
185 346060
190 346060
195 346060
200 346060
205 346061
210 346061
215 346061
220 346061
225 346062
230 346062
235 346063
240 346063
245 346064
250 346064
255 346064
260 346064
265 346065
270 346065
275 346065
280 346066
285 346066
290 346066
295 346067
300 346067
305 346067
310 346067
315 346067
320 346068
325 346069
330 346069
335 346069
340 346069
345 346069
350 346069
355 346071
360 346071
365 346071
370 346071
375 346071
380 346071
385 346072
390 346072
395 346073
400 346073
405 346073
410 346073
415 346074
420 346074
425 346074
430 346075
435 346075
440 346075
445 346075
450 346076
};
\addlegendentry{1 job}
\addplot [semithick, color2, mark=square*, mark size=3, mark options={solid,fill=white,draw=red},  mark repeat={180}]
table[y expr=(\thisrowno{1}-346049)/1000] {%
450 346076
455 346076
460 346076
465 346077
470 346077
475 346077
480 346078
485 346078
490 346078
495 346078
500 346078
505 346079
510 346080
515 346080
520 346080
525 346080
530 346080
535 346080
540 346082
545 346082
550 346082
555 346082
560 346082
565 346082
570 346083
575 346083
580 346084
585 346084
590 346084
595 346084
600 346084
605 346085
610 346085
615 346086
620 346086
625 346086
630 346086
635 346087
640 346087
645 346087
650 346088
655 346088
660 346089
665 346089
670 346089
675 346089
680 346090
685 346090
690 346091
695 346091
700 346091
705 346091
710 346091
715 346093
720 346093
725 346093
730 346093
735 346093
740 346093
745 346095
750 346095
755 346095
760 346095
765 346095
770 346096
775 346097
780 346097
785 346097
790 346097
795 346098
800 346098
805 346099
810 346099
815 346099
820 346099
825 346100
830 346100
835 346100
840 346101
845 346101
850 346102
855 346102
860 346102
865 346102
870 346103
875 346103
880 346103
885 346103
890 346104
895 346104
900 346105
};
\addlegendentry{2 jobs}
\addplot [semithick, color3, mark=triangle*, mark size=3, mark options={solid,fill=white,draw=red},  mark repeat={180}]
table[y expr=(\thisrowno{1}-346049)/1000] {%
905 346105
910 346105
915 346106
920 346106
925 346106
930 346106
935 346107
940 346108
945 346108
950 346108
955 346108
960 346108
965 346109
970 346110
975 346110
980 346110
985 346110
990 346110
995 346112
1000 346112
1005 346112
1010 346112
1015 346112
1020 346112
1025 346114
1030 346114
1035 346114
1040 346114
1045 346114
1050 346115
1055 346115
1060 346116
1065 346116
1070 346116
1075 346117
1080 346117
1085 346117
1090 346118
1095 346118
1100 346119
1105 346119
1110 346119
1115 346120
1120 346120
1125 346121
1130 346121
1135 346121
1140 346121
1145 346122
1150 346123
1155 346123
1160 346123
1165 346123
1170 346124
1175 346125
1180 346125
1185 346125
1190 346125
1195 346126
1200 346127
1205 346127
1210 346127
1215 346127
1220 346128
1225 346129
1230 346129
1235 346129
1240 346129
1245 346130
1250 346131
1255 346131
1260 346131
1265 346131
1270 346133
1275 346133
1280 346133
1285 346133
1290 346133
1295 346134
1300 346135
1305 346135
1310 346135
1315 346135
1320 346136
1325 346137
1330 346137
1335 346137
1340 346137
1345 346138
1350 346139
};
\addlegendentry{3 jobs}
\addplot [semithick, color4, mark=o, mark size=3, mark options={solid,fill=white,draw=red},  mark repeat={180}]
table[y expr=(\thisrowno{1}-346049)/1000] {%
1350 346139
1355 346139
1360 346139
1365 346139
1370 346140
1375 346141
1380 346141
1385 346141
1390 346141
1395 346142
1400 346143
1405 346143
1410 346143
1415 346143
1420 346144
1425 346144
1430 346144
1435 346145
1440 346146
1445 346146
1450 346146
1455 346146
1460 346147
1465 346148
1470 346148
1475 346148
1480 346148
1485 346150
1490 346150
1495 346150
1500 346150
1505 346150
1510 346152
1515 346152
1520 346152
1525 346152
1530 346152
1535 346154
1540 346154
1545 346154
1550 346154
1555 346154
1560 346156
1565 346156
1570 346156
1575 346156
1580 346156
1585 346157
1590 346158
1595 346158
1600 346158
1605 346158
1610 346160
1615 346160
1620 346160
1625 346160
1630 346161
1635 346162
1640 346162
1645 346162
1650 346162
1655 346163
1660 346164
1665 346164
1670 346164
1675 346165
1680 346166
1685 346166
1690 346166
1695 346166
1700 346167
1705 346168
1710 346168
1715 346168
1720 346168
1725 346169
1730 346170
1735 346170
1740 346170
1745 346171
1750 346171
1755 346172
1760 346172
1765 346172
1770 346173
1775 346173
1780 346174
1785 346174
1790 346175
1795 346175
1800 346176
1805 346176
1810 346176
1815 346177
1820 346177
1825 346178
1830 346178
1835 346178
1840 346179
1845 346179
1850 346180
1855 346180
1860 346181
1865 346181
1870 346181
1875 346182
1880 346182
1885 346183
1890 346183
1895 346184
1900 346184
1905 346184
1910 346185
1915 346185
1920 346186
1925 346186
1930 346187
1935 346187
1940 346187
1945 346188
1950 346188
1955 346189
1960 346189
1965 346190
1970 346190
1975 346190
1980 346191
1985 346191
1990 346192
1995 346192
2000 346193
2005 346193
2010 346193
2015 346194
2020 346194
2025 346195
2030 346196
2035 346196
2040 346196
2045 346197
2050 346197
2055 346198
2060 346198
2065 346198
2070 346199
2075 346199
2080 346200
2085 346201
2090 346201
2095 346201
2100 346201
2105 346203
2110 346203
2115 346203
2120 346203
2125 346204
2130 346205
2135 346205
2140 346205
2145 346205
2150 346206
2155 346207
2160 346207
2165 346207
2170 346207
2175 346209
2180 346209
2185 346209
2190 346209
2195 346209
2200 346211
2205 346211
2210 346211
2215 346211
2220 346212
2225 346213
2230 346213
2235 346213
2240 346213
2245 346215
2250 346215
2255 346215
2260 346215
2265 346215
2270 346217
2275 346217
2280 346217
2285 346217
2290 346218
2295 346219
2300 346219
2305 346219
2310 346219
2315 346221
2320 346221
2325 346221
2330 346221
2335 346221
2340 346223
2345 346223
2350 346223
2355 346223
2360 346223
2365 346225
2370 346225
2375 346225
2380 346225
2385 346226
2390 346227
2395 346227
2400 346227
2405 346227
2410 346228
2415 346229
2420 346229
2425 346229
2430 346230
2435 346231
2440 346231
2445 346231
2450 346231
2455 346232
2460 346233
2465 346233
2470 346233
2475 346233
2480 346234
2485 346235
2490 346235
2495 346235
2500 346236
2505 346237
2510 346237
2515 346237
2520 346237
2525 346238
2530 346239
2535 346239
2540 346239
2545 346239
2550 346240
2555 346241
2560 346241
2565 346241
2570 346242
2575 346242
2580 346243
2585 346243
2590 346243
2595 346244
2600 346245
2605 346245
2610 346245
2615 346245
2620 346246
2625 346246
2630 346247
2635 346247
2640 346248
2645 346248
2650 346248
2655 346249
2660 346249
2665 346250
2670 346250
2675 346250
2680 346251
2685 346252
2690 346252
2695 346252
2700 346252
2705 346254
2710 346254
2715 346254
2720 346254
2725 346255
2730 346256
2735 346256
2740 346256
2745 346256
2750 346257
2755 346258
2760 346258
2765 346258
2770 346258
2775 346260
2780 346260
2785 346260
2790 346260
2795 346260
2800 346262
2805 346262
2810 346262
2815 346262
2820 346263
2825 346264
2830 346264
2835 346264
2840 346264
2845 346266
2850 346266
2855 346266
2860 346266
2865 346266
2870 346268
2875 346268
2880 346268
2885 346268
2890 346269
2895 346270
2900 346270
2905 346270
2910 346270
2915 346272
2920 346272
2925 346272
2930 346272
2935 346272
2940 346274
2945 346274
2950 346274
2955 346274
2960 346274
2965 346276
2970 346276
2975 346276
2980 346276
2985 346277
2990 346278
2995 346278
3000 346278
3005 346278
3010 346280
3015 346280
3020 346280
3025 346280
3030 346281
3035 346282
3040 346282
3045 346282
3050 346282
3055 346283
3060 346284
3065 346284
3070 346284
3075 346284
3080 346285
3085 346286
3090 346286
3095 346286
3100 346287
3105 346288
3110 346288
3115 346288
3120 346288
3125 346289
3130 346290
3135 346290
3140 346290
3145 346291
3150 346291
3155 346292
3160 346292
3165 346292
3170 346293
3175 346294
3180 346294
3185 346294
3190 346294
3195 346295
3200 346296
3205 346296
3210 346296
3215 346297
3220 346297
3225 346297
3230 346298
3235 346298
3240 346299
3245 346299
3250 346299
3255 346300
3260 346301
3265 346301
3270 346301
3275 346301
3280 346303
3285 346303
3290 346303
3295 346303
3300 346304
3305 346305
3310 346305
3315 346305
3320 346305
3325 346306
3330 346307
3335 346307
3340 346307
3345 346307
3350 346309
3355 346309
3360 346309
3365 346309
3370 346310
3375 346311
3380 346311
3385 346311
3390 346311
3395 346312
3400 346313
3405 346313
3410 346313
3415 346313
3420 346315
3425 346315
3430 346315
3435 346315
3440 346315
3445 346317
3450 346317
3455 346317
3460 346317
3465 346319
3470 346319
3475 346319
3480 346319
3485 346319
3490 346321
3495 346321
3500 346321
3505 346321
3510 346321
3515 346323
3520 346323
3525 346323
3530 346323
3535 346323
3540 346325
3545 346325
3550 346325
3555 346325
3560 346326
3565 346327
3570 346327
3575 346327
3580 346327
3585 346329
3590 346329
3595 346329
3600 346329
3605 346330
3610 346331
3615 346331
3620 346331
3625 346331
3630 346332
3635 346333
3640 346333
3645 346333
3650 346333
3655 346334
3660 346335
3665 346335
3670 346335
3675 346336
3680 346336
3685 346337
3690 346337
3695 346337
3700 346338
3705 346339
3710 346339
3715 346339
3720 346339
3725 346340
3730 346341
3735 346341
3740 346341
3745 346342
3750 346342
3755 346343
3760 346343
3765 346343
3770 346344
3775 346345
3780 346345
3785 346345
3790 346346
3795 346346
3800 346347
3805 346347
3810 346347
3815 346348
3820 346348
3825 346348
3830 346349
3835 346349
3840 346350
3845 346350
3850 346351
3855 346351
3860 346352
3865 346352
3870 346352
3875 346353
3880 346354
3885 346354
3890 346354
3895 346354
3900 346355
3905 346356
3910 346356
3915 346356
3920 346356
3925 346358
3930 346358
3935 346358
3940 346358
3945 346359
3950 346360
3955 346360
3960 346360
3965 346360
3970 346362
3975 346362
3980 346362
};
\addlegendentry{4 jobs}
\addplot [semithick, color3, mark=triangle*, mark size=3, mark options={solid,fill=white,draw=red},  mark repeat={180}]
table[y expr=(\thisrowno{1}-346049)/1000] {%
3980 346362
3985 346362
3990 346362
3995 346364
4000 346364
4005 346364
4010 346364
4015 346364
4020 346366
4025 346366
4030 346366
4035 346366
4040 346366
4045 346368
4050 346368
4055 346368
4060 346368
4065 346368
4070 346370
4075 346370
4080 346370
4085 346370
4090 346370
4095 346371
4100 346372
4105 346372
4110 346372
4115 346372
4120 346373
4125 346374
4130 346374
4135 346374
4140 346375
4145 346375
4150 346376
4155 346376
4160 346376
4165 346377
4170 346377
4175 346378
4180 346378
4185 346378
4190 346379
4195 346379
4200 346380
4205 346380
4210 346380
4215 346381
4220 346381
4225 346381
4230 346382
4235 346382
4240 346383
4245 346383
4250 346383
4255 346384
4260 346384
4265 346385
4270 346385
4275 346385
4280 346386
4285 346386
4290 346387
4295 346387
4300 346387
4305 346388
4310 346389
4315 346389
4320 346389
4325 346389
4330 346390
4335 346391
4340 346391
4345 346391
4350 346391
4355 346392
4360 346393
4365 346393
4370 346393
4375 346393
4380 346393
4385 346395
4390 346395
4395 346395
4400 346395
4405 346395
4410 346397
4415 346397
4420 346397
};
\addplot [semithick, color2, mark=square*, mark size=3, mark options={solid,fill=white,draw=red},  mark repeat={180}]
table[y expr=(\thisrowno{1}-346049)/1000] {%
4420 346397
4425 346398
4430 346398
4435 346399
4440 346399
4445 346399
4450 346399
4455 346399
4460 346401
4465 346401
4470 346401
4475 346401
4480 346402
4485 346402
4490 346403
4495 346403
4500 346403
4505 346404
4510 346404
4515 346404
4520 346405
4525 346405
4530 346405
4535 346406
4540 346406
4545 346406
4550 346407
4555 346407
4560 346408
4565 346408
4570 346408
4575 346408
4580 346408
4585 346410
4590 346410
4595 346410
4600 346410
4605 346410
4610 346411
4615 346412
4620 346412
4625 346412
4630 346412
4635 346413
4640 346413
4645 346414
4650 346414
4655 346414
4660 346414
4665 346415
4670 346415
4675 346416
4680 346416
4685 346416
4690 346417
4695 346417
4700 346417
4705 346417
4710 346418
4715 346419
4720 346419
4725 346419
4730 346419
4735 346419
4740 346421
4745 346421
4750 346421
4755 346421
4760 346421
4765 346421
4770 346423
4775 346423
4780 346423
4785 346423
4790 346423
4795 346424
4800 346425
4805 346425
4810 346425
4815 346425
4820 346426
4825 346426
4830 346426
};
\addplot [semithick, color1, mark=diamond*, mark size=3, mark options={solid,fill=white,draw=red},  mark repeat={180}]
table[y expr=(\thisrowno{1}-346049)/1000] {%
4830 346426
4835 346427
4840 346427
4845 346427
4850 346428
4855 346428
4860 346428
4865 346429
4870 346429
4875 346429
4880 346430
4885 346430
4890 346430
4895 346430
4900 346431
4905 346431
4910 346432
4915 346432
4920 346432
4925 346432
4930 346432
4935 346432
4940 346434
4945 346434
4950 346434
4955 346434
4960 346434
4965 346434
4970 346435
4975 346436
4980 346436
4985 346436
4990 346436
4995 346436
5000 346437
5005 346437
5010 346437
5015 346437
5020 346437
5025 346437
5030 346439
5035 346439
5040 346439
5045 346439
5050 346439
5055 346439
5060 346439
5065 346440
5070 346441
5075 346441
5080 346442
5085 346442
5090 346442
5095 346442
5100 346442
5105 346443
5110 346444
5115 346444
5120 346444
5125 346444
5130 346444
5135 346444
5140 346444
5145 346446
5150 346446
5155 346446
5160 346446
5165 346446
5170 346446
5175 346447
5180 346448
5185 346448
5190 346448
5195 346448
5200 346448
5205 346449
5210 346449
5215 346449
5220 346450
5225 346450
5230 346450
5235 346451
5240 346451
5245 346451
5250 346451
5255 346452
5260 346452
5265 346453
5270 346453
5275 346453
5280 346453
5285 346453
5290 346453
5295 346454
5300 346455
5305 346455
5310 346455
};
\addplot [semithick, color2,mark=triangle*, mark size=3, mark options={solid,fill=red,draw=red}]
table[y expr=(\thisrowno{1}-346049)/1000] {%
3980 346362
};

\addplot [semithick, color2,mark=square*, mark size=3, mark options={solid,fill=red,draw=red}]
table[y expr=(\thisrowno{1}-346049)/1000] {%
4420 346397
};

\addplot [semithick, color2,mark=diamond*, mark size=3, mark options={solid,fill=red,draw=red}]
table[y expr=(\thisrowno{1}-346049)/1000] {%
4830 346426
};

\end{axis}
\node[text width=8cm] at (3,-1.5) {Filled markers indicate the end of jobs};

\end{tikzpicture}

%% file: results/time_cores.tex
\begin{tikzpicture}[font=\Large]

\begin{axis}[
legend cell align={left},
legend columns=1,
legend style={font=\Large,at={(0.6, 1)}},
legend image post style={scale=2},
tick align=outside,
tick pos=left,
x grid style={white!69.01960784313725!black},
xlabel={Number of cores},
xmajorgrids,
xmin=0, xmax=4.845,
xtick style={color=black},
xtick={0.45,1.45,2.45,3.45,4.45},
xticklabels={1,4,8,16,16$^{*}$},
y grid style={white!69.01960784313725!black},
ylabel={Execution time (SEC)},
ymajorgrids,
ymin=0, ymax=21000,
ytick style={color=black},
ytick={0,2000,4000,6000, 20000},
tick label style={font=\large}
]
\addlegendimage{ybar,ybar legend,preaction={fill, green},pattern=crosshatch dots};

\draw[preaction={fill, green},pattern=crosshatch dots] (axis cs:0.15,0) rectangle (axis cs:0.45,18291);

\addlegendentry{Intel}
\draw[preaction={fill, green},pattern=crosshatch dots] (axis cs:1.15,0) rectangle (axis cs:1.45,4678);
\draw[preaction={fill, green},pattern=crosshatch dots] (axis cs:2.15,0) rectangle (axis cs:2.45,2553);
\draw[preaction={fill, green},pattern=crosshatch dots] (axis cs:3.15,0) rectangle (axis cs:3.45,1842);
\draw[preaction={fill, green},pattern=crosshatch dots] (axis cs:4.15,0) rectangle (axis cs:4.45,1235);

\addlegendimage{ybar,ybar legend, preaction={fill, red},pattern=north west lines};
\addlegendentry{AMD}

\draw[preaction={fill, red},pattern=north west lines, pattern color=black] (axis cs:0.45,0) rectangle (axis cs:0.75,20689);
\draw[preaction={fill, red},pattern=north west lines, pattern color=black] (axis cs:1.45,0) rectangle (axis cs:1.75,5662);
\draw[preaction={fill, red},pattern=north west lines, pattern color=black] (axis cs:2.45,0) rectangle (axis cs:2.75,2792);
\draw[preaction={fill, red},pattern=north west lines, pattern color=black] (axis cs:3.45,0) rectangle (axis cs:3.75,1540);
\draw[preaction={fill, red},pattern=north west lines, pattern color=black] (axis cs:4.45,0) rectangle (axis cs:4.75,1300);
\end{axis}
\node[text width=4cm] at (4,-1.25) {$*$: 32 threads};

\end{tikzpicture}

%% file: results/time_jobs.tex
\begin{tikzpicture}[font=\Large]

\begin{axis}[
legend cell align={left},
legend columns=1,
legend style={font=\Large,at={(0.7, 1)}},
legend image post style={scale=2},
tick align=outside,
tick pos=left,
x grid style={white!69.01960784313725!black},
xlabel={Number of jobs},
xmajorgrids,
xmin=0, xmax=2.845,
xtick style={color=black},
xtick={0.45,1.45,2.45},
xticklabels={1,2,4},
y grid style={white!69.01960784313725!black},
ylabel={Execution time (SEC)},
y label style={at={(axis description cs:-0.1,0.5)},anchor=north},
ymajorgrids,
ymin=0, ymax=4200,
ytick style={color=black},
ytick={0,1000,2000,3000,4000},
tick label style={font=\large}
]
\draw[preaction={fill, green},pattern=crosshatch dots] (axis cs:0.15,0) rectangle (axis cs:0.45,2553);
\addlegendimage{ybar,ybar legend,preaction={fill, green},pattern=crosshatch dots};
\addlegendentry{Intel}

\draw[preaction={fill, green},pattern=crosshatch dots] (axis cs:1.15,0) rectangle (axis cs:1.45,2535);
\draw[preaction={fill, green},pattern=crosshatch dots] (axis cs:2.15,0) rectangle (axis cs:2.45,4092);
\addlegendimage{ybar,ybar legend,preaction={fill, red},pattern=north west lines, pattern color=black};
\addlegendentry{AMD}

\draw[preaction={fill, red},pattern=north west lines, pattern color=black] (axis cs:0.45,0) rectangle (axis cs:0.75,2792);
\draw[preaction={fill, red},pattern=north west lines, pattern color=black] (axis cs:1.45,0) rectangle (axis cs:1.75,2844);
\draw[preaction={fill, red},pattern=north west lines, pattern color=black] (axis cs:2.45,0) rectangle (axis cs:2.75,3783);
\end{axis}

\end{tikzpicture}

%% file: results/time_energy_intel.tex
\definecolor{darkgreen}{rgb}{0.0, 0.5, 0.0}
\begin{tikzpicture}[font=\large,spy using outlines= {circle, magnification=2, connect spies}]
\begin{axis}[
legend cell align={left},
legend columns=2,
legend style={fill opacity=0.8, draw opacity=1, text opacity=1, at={(1.05,1.22)}, anchor=east, draw=white!80.0!black},
tick align=outside,
tick pos=left,
x grid style={white!69.01960784313725!black},
xlabel={Execution time per job (SEC)},
xmin=0, xmax=6000,
xtick={0,1000,2000,3000,4000,5000,6000},
xtick style={color=black},
y grid style={white!69.01960784313725!black},
ylabel style={align=center},
ylabel={Total energy consumption \\ per job ($\times 10^{-3}$kWh)},
ymin=0, ymax=400,
ytick={0,100,200,300,400},
ytick style={color=black},
xmajorgrids,
ymajorgrids,
 y label style={at={(axis description cs:-0.03,.5)}}
]
 \pgfplotstableread{
x y  
18291 1351 
4678   392  
2553   251   
1842   202 
1235   147 
1267   157  
1023 130  
            }\datatable

\addplot[scatter/classes={ b={mark=x,draw=red}, c={mark=pentagon*,darkgreen}, d={mark=diamond*,purple}, e={mark=square*,olive}, f={mark=triangle*,black}, g={mark=*,gray}},
        scatter, only marks, mark size=3,  
        scatter src=explicit symbolic,
        ]%
table[meta=class] {
x y class 
18291 1351 a 
4678   392  b 
2553   251  c 
1842   202 d
1235   147 e
1267   157 f 
1023 130 g 
    };
    \legend{1 job 4 cores,1 job 8 cores, 1 job 16 cores,1 job 16* core (32 threads), 2 jobs 8 cores, 4 jobs 8 cores}

\addplot[thick, shorten >= -4cm,
                shorten <= -2cm,orange, forget plot]
table [x = x, y = {create col/linear regression={y=y}},
            ] {\datatable};
            
\addlegendimage{empty legend}
    \addlegendentry{}
\addlegendimage{no markers,orange}
\addlegendentry{
$ y =
    \pgfmathprintnumber{\pgfplotstableregressiona}
    \cdot x
    \pgfmathprintnumber[print sign]{\pgfplotstableregressionb}$
};
\coordinate (spypoint)at (axis cs:1000,150);
\coordinate (magnifyglass)at (axis cs:1150,300);;
\end{axis}
\spy [brown, size=2.5cm] on (spypoint) in node[fill=white] at (magnifyglass); 
\end{tikzpicture}

%% file: results/time_energy_intel_ex_baseline.tex
\definecolor{darkgreen}{rgb}{0.0, 0.5, 0.0}

\begin{tikzpicture}[font=\large]
\begin{axis}[
legend cell align={left},
legend columns=2,
legend style={fill opacity=0.8, draw opacity=1, text opacity=1, at={(1.05,1.22)}, anchor=east, draw=white!80.0!black},
tick align=outside,
tick pos=left,
x grid style={white!69.01960784313725!black},
xlabel={Execution time per job (SEC)},
xmin=0, xmax=6000,
xtick={0,1000,2000,3000,4000,5000,6000},
xtick style={color=black},
y grid style={white!69.01960784313725!black},
ylabel style={align=center},
ylabel={Total energy consumption \\ per job ($\times 10^{-3}$kWh)},
ymin=0, ymax=80,
ytick={10,20,30,40,50,60,70,80},
ytick style={color=black},
xmajorgrids,
ymajorgrids,
 y label style={at={(axis description cs:-0.03,.5)}}
]
 \pgfplotstableread{
x y  
18291 80.22 
4678   66.5  
2553   75.63  
1842   74.03 
1235   61.01 
1267   71.81 
1023 59.50 
            }\datatable
\addplot[scatter/classes={ b={mark=x,draw=red}, c={mark=pentagon*,darkgreen}, d={mark=diamond*,purple}, e={mark=square*,olive}, f={mark=triangle*,black}, g={mark=*,gray}},
        scatter, only marks, mark size=3,  
        scatter src=explicit symbolic,
        ]%
table[meta=class] {
x y class 
4678   66.5  b 
2553   75.63  c 
1842   74.03 d
1235   61.01 e
1267   71.81 f 
1023 59.50 g 
    };
    \legend{1 job 4 cores,1 job 8 cores, 1 job 16 cores,1 job 16* core (32 threads), 2 jobs 8 cores, 4 jobs 8 cores}
            \addplot[
                thick,                
                shorten >= -4cm,
                shorten <= -2cm,
                orange, forget plot
            ]
            table [
                x = x,
                 y = {create col/linear regression={y=y}},
            ] {\datatable};
\addlegendimage{empty legend}
    \addlegendentry{}
\addlegendimage{no markers,orange}
\addlegendentry{
 $ y =
    \pgfmathprintnumber{\pgfplotstableregressiona}
    \cdot x
    \pgfmathprintnumber[print sign]{\pgfplotstableregressionb}$
};
\end{axis}
\end{tikzpicture}

%% file: results/time_energy_amd.tex
\definecolor{darkgreen}{rgb}{0.0, 0.5, 0.0}

\begin{tikzpicture}[font=\large,spy using outlines= {circle, magnification=2, connect spies}]
\begin{axis}[
legend cell align={left},
legend columns=2,
legend style={fill opacity=0.8, draw opacity=1, text opacity=1, at={(1.05,1.22)}, anchor=east, draw=white!80.0!black},
tick align=outside,
tick pos=left,
x grid style={white!69.01960784313725!black},
xlabel={Execution time per job (SEC)},
xmin=0, xmax=6000,
xtick={0,1000,2000,3000,4000,5000,6000},
xtick style={color=black},
y grid style={white!69.01960784313725!black},
ylabel style={align=center},
ylabel={Total energy consumption \\ per job ($\times 10^{-3}$kWh)},
ymin=0, ymax=400,
ytick={0,100,200,300,400},
ytick style={color=black},
xmajorgrids,
ymajorgrids,
 y label style={at={(axis description cs:-0.03,.5)}}
]
 \pgfplotstableread{
x y  
5662   351   
2792   189  
1540   118 
1300   104 
1422   111 
945 77 
            }\datatable
\addplot[scatter/classes={ b={mark=x,draw=red}, c={mark=pentagon*,darkgreen}, d={mark=diamond*,purple}, e={mark=square*,olive}, f={mark=triangle*,black}, g={mark=*,gray}},
        scatter, only marks, mark size=3,  
        scatter src=explicit symbolic,
        ]%
table[meta=class] {
x y class 
5662   351  b 
2792   189  c 
1540   118 d
1300   104 e
1422   111 f 
945 77 g 
    };
    \legend{1 job 4 cores,1 job 8 cores, 1 job 16 cores,1 job 16* core (32 threads), 2 jobs 8 cores, 4 jobs 8 cores}

\coordinate (spypoint)at (axis cs:1200,100);
\coordinate (magnifyglass)at (axis cs:1250,300);;
            \addplot[
                thick,
                shorten >= -4cm,
                shorten <= -2cm,
                orange, forget plot
            ]
            table [
                x = x,
                 y = {create col/linear regression={y=y}},
            ] {\datatable};
\addlegendimage{empty legend}
    \addlegendentry{}
\addlegendimage{no markers,orange}
\addlegendentry{
 $ y =
    \pgfmathprintnumber{\pgfplotstableregressiona}
    \cdot x
    \pgfmathprintnumber[print sign]{\pgfplotstableregressionb}$
};
\end{axis}
\spy [brown, size=2.5cm] on (spypoint) in node[fill=white] at (magnifyglass); 
\end{tikzpicture}

%% file: results/time_energy_amd_ex_baseline.tex
\definecolor{darkgreen}{rgb}{0.0, 0.5, 0.0}
\begin{tikzpicture}[font=\large,spy using outlines= {circle, magnification=2, connect spies}]
\begin{axis}[
legend cell align={left},
legend columns=2,
legend style={fill opacity=0.8, draw opacity=1, text opacity=1, at={(1.05,1.22)}, anchor=east, draw=white!80.0!black},
tick align=outside,
tick pos=left,
x grid style={white!69.01960784313725!black},
xlabel={Execution time per job (SEC)},
xmin=0, xmax=6000,
xtick={0,1000,2000,3000,4000,5000,6000},
xtick style={color=black},
y grid style={white!69.01960784313725!black},
ylabel style={align=center},
ylabel={Total energy consumption \\ per job ($\times 10^{-3}$kWh)},
ymin=0, ymax=80,
ytick={10,20,30,40,50,60,70,80},
ytick style={color=black},
xmajorgrids,
ymajorgrids,
 y label style={at={(axis description cs:-0.03,.5)}}
]
 \pgfplotstableread{
x y  
20689 34.48 
5662   28.73   
2792   29.39   
1540   29.86 
1300   30.08 
1422   29.75 
945 28.50 
            }\datatable
\addplot[scatter/classes={ b={mark=x,draw=red}, c={mark=pentagon*,darkgreen}, d={mark=diamond*,purple}, e={mark=square*,olive}, f={mark=triangle*,black}, g={mark=*,gray}},
        scatter, only marks, mark size=3, 
        scatter src=explicit symbolic,
        ]%
table[meta=class] {
x y class 
5662   28.73  b 
2792   29.39  c 
1540   29.86 d
1300   30.08 e
1422   29.75 f 
945 28.50 g 
    };
    \legend{1 job 4 cores,1 job 8 cores, 1 job 16 cores,1 job 16* core (32 threads), 2 jobs 8 cores, 4 jobs 8 cores}
            \addplot[
                thick,
                shorten >= -4cm,
                shorten <= -2cm,
                orange, forget plot
            ]
            table [
                x = x,
                 y = {create col/linear regression={y=y}}
            ] {\datatable};
\addlegendimage{empty legend}
    \addlegendentry{}
\addlegendimage{no markers,orange}
\addlegendentry{
 $ y =
    \pgfmathprintnumber{\pgfplotstableregressiona}
    \cdot x
    \pgfmathprintnumber[print sign]{\pgfplotstableregressionb}$
};

\coordinate (spypoint)at (axis cs:1180,30);
\coordinate (magnifyglass)at (axis cs:3800,55);;
\end{axis}
\spy [brown, size=2.5cm] on (spypoint) in node[fill=white] at (magnifyglass); 
\end{tikzpicture}

%% file: ref.bib
@IEEEtranBSTCTL{IEEEexample:BSTcontrol,
CTLdash_repeated_names= "no",
CTLuse_forced_etal       = "yes",
	CTLmax_names_forced_etal = "3",
	CTLnames_show_etal       = "1"
}

@article{malmodin2024ict,
  title={ICT sector electricity consumption and greenhouse gas emissions--2020 outcome},
  author={Malmodin, Jens and L{\"o}vehagen, Nina and Bergmark, Pernilla and Lund{\'e}n, Dag},
  journal={Telecommunications Policy},
  volume={48},
  number={3},
  pages={102701},
  year={2024},
  publisher={Elsevier}
}

@misc{ stfcreport,
    title = {STFC Sustainability and Carbon Management
},
    note = {\url{https://www.she.stfc.ac.uk/Pages/Sustainability-and-Carbon-Management.aspx}}
}

@misc{ gpuconsumption,
    title = {{ENERGETIC}: Final Report},
author = {Bane, M. K. and Brown, O. and Bhowmik, D. and Quinn, J. and Ali, T. and Quinn, J.,and Stansby, D.},
    note = {\url{https://zenodo.org/records/7692272}},
year={2023}
}

@misc{ STFC,
    title = {Science and Technology Facilities Council },
    note = {\url{https://www.ukri.org/councils/stfc/}}
}

@misc{ docker,
    title = {Docker},
    note = {\url{https://www.docker.com/}}
}

@inproceedings{rimoldi2004simulation,
  title={The simulation for the ATLAS experiment. Present status and outlook},
  author={Rimoldi, A and Dell'Acqua, A and Gallas, M and Nairz, A and Boudreau, J and Tsulaia, V and Costanzo, D},
  booktitle={IEEE Symposium Conference Record Nuclear Science 2004.},
  volume={3},
  pages={1886--1890},
  year={2004},
  organization={IEEE}
}

@article{jin2020review,
  title={A review of power consumption models of servers in data centers},
  author={Jin, Chaoqiang and Bai, Xuelian and Yang, Chao and Mao, Wangxin and Xu, Xin},
  journal={applied energy},
  volume={265},
  pages={114806},
  year={2020},
  publisher={Elsevier}
}

@inproceedings{kaffes2020leveraging,
  title={Leveraging application classes to save power in highly-utilized data centers},
  author={Kaffes, Kostis and Sbirlea, Dragos and Lin, Yiyan and Lo, David and Kozyrakis, Christos},
  booktitle={Proceedings of the 11th ACM Symposium on Cloud Computing},
  pages={134--149},
  year={2020}
}

@inproceedings{jin2012energy,
  title={Energy efficiency and server virtualization in data centers: An empirical investigation},
  author={Jin, Yichao and Wen, Yonggang and Chen, Qinghua},
  booktitle={2012 Proceedings IEEE INFOCOM Workshops},
  pages={133--138},
  year={2012},
  organization={IEEE}
}

@inproceedings{zhang2020estimating,
  title={Estimating power consumption of containers and virtual machines in data centers},
  author={Zhang, Xusheng and Shen, Ziyu and Xia, Bin and Liu, Zheng and Li, Yun},
  booktitle={2020 IEEE International Conference on Cluster Computing (CLUSTER)},
  pages={288--293},
  year={2020},
  organization={IEEE}
}

@article{dayarathna2015data,
  title={Data center energy consumption modeling: A survey},
  author={Dayarathna, Miyuru and Wen, Yonggang and Fan, Rui},
  journal={IEEE Communications Surveys \& Tutorials},
  volume={18},
  number={1},
  pages={732--794},
  year={2015},
  publisher={IEEE}
}

@article{brady2013case,
  title={A case study and critical assessment in calculating power usage effectiveness for a data centre},
  author={Brady, Gemma A and Kapur, Nikil and Summers, Jonathan L and Thompson, Harvey M},
  journal={Energy Conversion and Management},
  volume={76},
  pages={155--161},
  year={2013},
  publisher={Elsevier}
}

@article{aroca2015measurement,
  title={A measurement-based characterization of the energy consumption in data center servers},
  author={Aroca, Jordi Arjona and Chatzipapas, Angelos and Anta, Antonio Fern{\'a}ndez and Mancuso, Vincenzo},
  journal={IEEE Journal on selected areas in communications},
  volume={33},
  number={12},
  pages={2863--2877},
  year={2015},
  publisher={IEEE}
}

@article{varasteh2015server,
  title={Server consolidation techniques in virtualized data centers: A survey},
  author={Varasteh, Amir and Goudarzi, Maziar},
  journal={IEEE Systems Journal},
  volume={11},
  number={2},
  pages={772--783},
  year={2015},
  publisher={IEEE}
}

@article{capozzoli2015cooling,
  title={Cooling systems in data centers: state of art and emerging technologies},
  author={Capozzoli, Alfonso and Primiceri, Giulio},
  journal={Energy Procedia},
  volume={83},
  year={2015},
  publisher={Elsevier}
}

@article{shuja2014survey,
  title={Survey of techniques and architectures for designing energy-efficient data centers},
  author={Shuja, Junaid and Bilal, Kashif and Madani, Sajjad A and Othman, Mazliza and Ranjan, Rajiv and Balaji, Pavan and Khan, Samee U},
  journal={IEEE Systems Journal},
  year={2014},
  publisher={IEEE}
}

@article{lo2014towards,
  title={Towards energy proportionality for large-scale latency-critical workloads},
  author={Lo, David and Cheng, Liqun and Govindaraju, Rama and Barroso, Luiz Andr{\'e} and Kozyrakis, Christos},
  journal={ACM SIGARCH Computer Architecture News},
  volume={42},
  number={3},
  pages={301--312},
  year={2014},
  publisher={ACM New York, NY, USA}
}

@article{sprecher2014recycling,
  title={Recycling potential of neodymium: the case of computer hard disk drives},
  author={Sprecher, Benjamin and Kleijn, Rene and Kramer, Gert Jan},
  journal={Environmental science \& technology},
  volume={48},
  number={16},
  pages={9506--9513},
  year={2014},
  publisher={ACS Publications}
}

@inproceedings{narayanan2009migrating,
  title={Migrating server storage to SSDs: analysis of tradeoffs},
  author={Narayanan, Dushyanth and Thereska, Eno and Donnelly, Austin and Elnikety, Sameh and Rowstron, Antony},
  booktitle={Proceedings of the 4th ACM European conference on Computer systems},
  pages={145--158},
  year={2009}
}

@inproceedings{shi2013cpt,
  title={CPT: An energy-efficiency model for multi-core computer systems},
  author={Shi, Weisong and Wang, Shinan and Luo, Bing},
  booktitle={Proc. 5th Workshop Energy-Efficient Des.},
  pages={1--6},
  year={2013}
}

@misc{ cquandri,
    title = {{CQuanDRI} scripts and results},
    note = {\url{https://github.com/ox-computing/project-cquandri}}
}

@inproceedings{arslan2023green,
  title={{Green With Envy: Unfair Congestion Control Algorithms Can Be More Energy Efficient}},
  author={Arslan, Serhat and Renganathan, Sundararajan and Spang, Bruce},
  booktitle={Proceedings of the 22nd ACM Workshop on Hot Topics in Networks},
  pages={220--228},
  year={2023}
}

@inproceedings{ciko2023going,
  title={{Going Dark: A Software “Light Switch” for Internet Servers}},
  author={Ciko, Kristjon and Welzl, Michael and Teymoori, Peyman},
  booktitle={2023 IEEE 29th International Symposium on Local and Metropolitan Area Networks (LANMAN)},
  pages={1--6},
  year={2023},
  organization={IEEE}
}

@misc{cpupower,
    title = {cpupower(1) - Linux man page
},
    note = {\url{https://linux.die.net/man/1/cpupower}}
}

@misc{iea,
    author = {{IEA}},
    title={Data Centres and Data Transmission Networks},
    year={2023},
    note = {\url{https://www.iea.org/energy-system/buildings/data-centres-and-data-transmission-networks}}
}

@inproceedings{guenter2011managing,
  title={Managing cost, performance, and reliability tradeoffs for energy-aware server provisioning},
  author={Guenter, Brian and Jain, Navendu and Williams, Charles},
  booktitle={2011 Proceedings IEEE INFOCOM},
  pages={1332--1340},
  year={2011},
  organization={IEEE}
}

@article{perumal2014power,
  title={Power-conservative server consolidation based resource management in cloud},
  author={Perumal, Varalakshmi and Subbiah, Sankari},
  journal={International Journal of Network Management},
  volume={24},
  number={6},
  pages={415--432},
  year={2014},
  publisher={Wiley Online Library}
}

@inproceedings{song2013unified,
  title={Unified performance and power modeling of scientific workloads},
  author={Song, Shuaiwen Leon and Barker, Kevin and Kerbyson, Darren},
  booktitle={Proceedings of the 1st International Workshop on Energy Efficient Supercomputing},
  year={2013}
}

@article{aguado2008cvmfs,
  title={CVMFS-a file system for the CernVM virtual appliance},
  author={Aguado Sanchez, Carlos and Bloomer, J and Buncic, P and Franco, L and Klemer, S and Mato, P},
  journal={XII Advanced Computing and Analysis Techniques in Physics Research},
  pages={52},
  year={2008}
}

@article{barroso2018datacenter,
  title={The datacenter as a computer: Designing warehouse-scale machines},
  author={Barroso, Luiz Andr{\'e} and H{\"o}lzle, Urs and Ranganathan, Parthasarathy},
  journal={Synthesis Lectures on Computer Architecture},
  volume={13},
  number={3},
  pages={i--189},
  year={2018},
  publisher={Morgan \& Claypool Publishers}
}

@article{krawczyk2022ethernet,
  title={Ethernet for high-throughput computing at CERN},
  author={Krawczyk, Rafal and Colombo, Tommaso and Neufeld, Niko and Pisani, Flavio and Valat, S{\'e}bastien},
  journal={IEEE Transactions on Parallel and Distributed Systems},
  volume={33},
  number={12},
  pages={3640--3650},
  year={2022},
  publisher={IEEE}
}

@misc{iea2024,
  title={Electricity 2024},
  author={{IEA}},
  year={2024},
  note={\url{https://www.iea.org/reports/electricity-2024}}
}

@inproceedings{tsaregorodtsev2004dirac,
  title={DIRAC: A scalable lightweight architecture for high throughput computing},
  author={Tsaregorodtsev, Andrei and Garonne, Vincent and Stokes-Rees, Ian},
  booktitle={Fifth IEEE/ACM International Workshop on Grid Computing},
  year={2004},
  organization={IEEE}
}
